\documentclass[a4paper, 12pt]{book}           
\usepackage[pdftex]{graphicx} 
\usepackage{euler}
\usepackage{palatino}
\usepackage{amsmath}
\usepackage{amsfonts}
\usepackage{amssymb}
\usepackage{slashed}
\usepackage[utf8]{inputenc}
\usepackage[T1]{fontenc}
\usepackage{wrapfig}
\usepackage{booktabs}
\usepackage{sidecap}
\usepackage{phdthesis}
\usepackage{pdflscape}
\usepackage{hyperref}
\usepackage{rotating}

\usepackage[square, numbers, sort&compress]{natbib}

\title{
\bf Measurement of double polarized asymmetries \\ 
in quasi-elastic processes ${}^3\vec{\mathrm{He}}\left(
\vec{\mathrm{e}},\mathrm{e'} \mathrm{d }\right)$ and 
${}^3\vec{\mathrm{He}}\left(\vec{\mathrm{e}},\mathrm{e'} 
\mathrm{p}\right)$}
\author{Miha Mihovilovi\v{c}}   
\date{\today}                           
      
\frontmatter          

\begin{document}   
\thispagestyle{empty}
\begin{center}
UNIVERSITY OF LJUBLJANA\\
FACULTY OF MATHEMATICS AND PHYSICS\\
DEPARTMENT OF PHYSICS
\end{center}
\vspace*{6cm}

\begin{center}
{\large Miha Mihovilovi\v{c}}
\end{center}
\begin{center}
{\LARGE \bf Measurement of double polarized asymmetries \\ 
in quasi-elastic processes ${}^3\vec{\mathrm{He}}\left(
\vec{\mathrm{e}},\mathrm{e'} \mathrm{d }\right)$ and 
${}^3\vec{\mathrm{He}}\left(\vec{\mathrm{e}},\mathrm{e'} 
\mathrm{p}\right)$}
\end{center}
\begin{center}
{\large Doctoral thesis}
\end{center}
\vspace*{3cm}
\begin{center}
{\large ADVISER: assoc.~prof.~dr.~Simon \v{S}irca\\[3mm]
COADVISER: dr.~Douglas~W.~Higinbotham }
\end{center}
\vfill

\begin{center}
Ljubljana, 2012
\end{center}

\newpage
\thispagestyle{empty}
\vspace*{5cm}
{}
\newpage


\thispagestyle{empty}
\begin{center}
UNIVERZA V LJUBLJANI\\
FAKULTETA ZA MATEMATIKO IN FIZIKO\\
ODDELEK ZA FIZIKO
\end{center}
\vspace*{6cm}

\begin{center}
{\large Miha Mihovilovi\v{c}}
\end{center}
\begin{center}
{\LARGE \bf Meritev dvojnopolarizacijskih asimetrij v \\ kvazielasti\v{c}nih procesih 
${}^3\vec{\mathrm{He}}\left(\vec{\mathrm{e}},\mathrm{e'} \mathrm{d }\right)$ in 
${}^3\vec{\mathrm{He}}\left(\vec{\mathrm{e}},\mathrm{e'} \mathrm{p}\right)$} 
\end{center}
\begin{center}
{\large Doktorska disertacija}
\end{center}
\vspace*{3cm}
\begin{center}
{\large MENTOR: izr.~prof.~dr.~Simon \v{S}irca\\[3mm]
SOMENTOR: dr.~Douglas~W.~Higinbotham }
\end{center}
\vfill

\begin{center}
Ljubljana, 2012
\end{center}

\newpage
\thispagestyle{empty}
\vspace*{5cm}
{}
\newpage

\thispagestyle{empty}
\vspace*{10cm}
\begin{flushright}
To my parents for all their love.
\end{flushright}
\newpage
\thispagestyle{empty}
\vspace*{5cm}
{}
\newpage

\thispagestyle{empty}
\begin{flushright}
\parskip3mm  
{\bf \huge Acknowledgments}\\
\vspace*{1cm}
I want to express my deepest gratitude to my mentor assoc.~prof.~dr.~Simon~\v{S}irca for 
believing in me and giving me the opportunity to participate in the experiments at Jefferson Lab
and be a part of the E05-102 collaboration. He has been a wonderful mentor, who has 
given me complete intellectual freedom and let me perform my research in any way I wanted. 
He has always been available for my questions and helped me 
solve countless problems with infinite patience. I was able to learn so many things 
from him, not only about physics and computer science, but also about literature, 
history, opera and classical music. I find all this experience priceless. 
In the end, I would like to thank him for sacrificing many hours reading my 
thesis and correcting all "the" mistakes that I made.  

I am grateful also to my second adviser dr.~Douglas~Higinbotham for his help and 
guidance throughout the data analysis. I have always enjoyed fruitful 
discussions with him and found his advices invaluable. I would also like to thank 
him for enabling me to  spend time at Jefferson Lab. While working
there, I was able to communicate with scientists on-site and learned 
from their experience. These interactions have helped my research tremendously 
and allowed me to proceed much faster with the analysis. 

I want to thank  dr.~Vincent~Sulkosky, dr.~Bryan~Moffit and dr.~Yi~Qiang for 
helping me before and during the experiment. At the time of the experiment I was
a complete novice to Hall-A  equipment and really appreciated 
their advice and supervision. I also need to thank them for writing numerous 
e-mails answering all my questions regarding the data analysis. 
Many thanks also go to my fellow students Elena~Long, Yawei~Zhang and 
dr.~Ge~Jin, for spending long nights with me in the Hall A counting house 
collecting experimental data. At this point I would also like to thank to the 
rest of the E05-102 collaboration, without whom this experiment 
would not have been possible.

My special thanks go to dr.~Kees~de~Jager, assist.~prof.~dr.~Matej~Lipoglav\v{s}ek
and prof.~dr.~Andrej~Likar, for kindly providing the funds for my trips to the 
United States.

I would like to thank to colleagues and friends at the Jo\v{z}ef Stefan Institute. I have
really enjoyed working with you and will never forget all the fun that we had and
all the challenges that we faced together.

Finally, I owe gratitude to my family for their love and support during all my 
years of schooling, and to Helena for encouraging me and patiently 
waiting, while I was away in the office writing my thesis.

\end{flushright}
\newpage
\thispagestyle{empty}
\vspace*{5cm}
{}
\newpage

\begin{center}
{\huge \bf Abstract}
\end{center}
The ${}^3\mathrm{He}$ nucleus is a subject of considerable interest. It is an 
excellent system for testing our understanding of nuclear forces and the structure 
of its ground state, as well as to study the reaction mechanisms involved in 
electron-induced hadron knockout. It is a precisely calculable nuclear system 
where theoretical predictions can be compared to experimental data to ever 
increasing accuracy. The ${}^3\mathrm{He}$ nucleus has been extensively used 
as an effective neutron target in the experiments dedicated to measuring properties 
of the neutron. The interpretation of those measurements depends directly on the 
quality of this substitution, which makes a detailed knowledge of the structure 
of ${}^3\mathrm{He}$ even more important.
\newline
\newline
This thesis is dedicated to a study of a spin-isospin structure of the 
polarized ${}^3\mathrm{He}$. First, an introduction to the spin structure of 
${}^3\mathrm{He}$ is given, followed by a brief overview of past experiments. 
The main focus of the thesis is the E05-102 experiment at Jefferson Lab, in 
which the reactions ${}^3\vec{\mathrm{He}}\left(\vec{\mathrm{e}},\mathrm{e'} \mathrm{d }\right)$
and ${}^3\vec{\mathrm{He}}\left(\vec{\mathrm{e}},\mathrm{e'} \mathrm{p}\right)$ 
in the quasi-elastic region were studied. The purpose of this experiment 
was to better understand the effects of 
the S'- and D-state contributions to the ${}^3\mathrm{He}$ ground-state wave-functions by a 
precise measurement of beam-target asymmetries $A_x$ and $A_z$ in the range of recoil 
momenta from $0$ to about $300\,\mathrm{MeV}/c$. The experimental equipment utilized in 
these measurements is described, with special attention devoted to the 
calibration of the hadron spectrometer, BigBite.  Results on the measured 
asymmetries are presented, together with first attempts at their comparison 
to the state-of-the art Faddeev calculations. The remaining open problems 
and challenges for future work are also discussed.
\vspace*{1cm}
\begin{flushleft}
{\bf PACS (2010)}: 29.30.Aj,\hspace*{1mm} 29.85.Fj,\hspace*{1mm} 25.30.-c
\end{flushleft}

\newpage
\thispagestyle{empty}
\vspace*{5cm}
{}
\newpage

\begin{center}
{\huge \bf Povzetek}
\end{center}

V svojem doktorskem delu sem raziskoval spinsko-izospinsko zgradbo polariziranih
jeder ${}^3\mathrm{He}$. Jedro ${}^3\mathrm{He}$ je zelo zanimivo, saj ga je 
kljub kompleksni notranji zgradbi mo\v{c} ra\v{c}unsko obvladati. To nam omogo\v{c}a, 
da teoreti\v{c}ne napovedi o strukturi jedra natan\v{c}no primerjamo 
z izmerjenimi podatki in tako preverimo na\v{s}e razumevanje 
jedrskih sil med nukleoni in reakcijskih mehanizmov, ki spremljajo jedrske procese.
V eksperimentih jedro ${}^3\mathrm{He}$ pogosto uporabljamo tudi kot efektivno 
polarizirano nevtronsko tar\v{c}o za posredno raziskovanje lastnosti nevtrona. 
Kakovost tako pridobljenih informacij o nevtronu je neposredno odvisna od tega, 
kako natan\v{c}no poznamo lastnosti ${}^3\mathrm{He}$, kar le \v{s}e dodatno motivira na\v{s}e 
raziskave.
\newline
\newline
V disertaciji najprej predstavim  osnovne lastnosti jedra 
${}^3\mathrm{He}$. Temu sledi kratek pregled preteklih poskusov 
posve\v{c}enih raziskovanju tega jedra in prikaz njihovih najpomembnej\v{s}ih 
ugotovitev. Nato se osredoto\v{c}im na eksperiment E05-102, ki smo ga
izvedli v okviru kolaboracije Hall-A v Thomas Jefferson National Accelerator Facility v 
Zdru\v{z}enih dr\v{z}avah Amerike. Opravili smo meritev
dvojnopolarizacijskih asimetrij $A_x$ in $A_z$ v kvazielasti\v{c}nih procesih 
${}^3\vec{\mathrm{He}}\left(\vec{\mathrm{e}},\mathrm{e'} \mathrm{d }\right)$
in ${}^3\vec{\mathrm{He}}\left(\vec{\mathrm{e}},\mathrm{e'} \mathrm{p}\right)$,
na obmo\v{c}ju gibalnih koli\v{c}in nedetektiranih delcev $0-300\,\mathrm{MeV}/c$ 
z \v{z}eljo, da bi bolje razumeli prisotnost stanj S' in D v valovni funkciji
osnovnega stanja ${}^3\mathrm{He}$. V poglavjih, ki sledijo, zato natan\v{c}no 
opi\v{s}em postopke ter aparature, ki smo jih uporabili.
Pri tem posebno pozornost namenim umeritvi hadronskega spektrometra BigBite. 
Zatem predstavim potek analize meritev, ter podam rezultate. Na koncu izmerjene 
asimetrije soo\v{c}im s teoreti\v{c}nimi izra\v{c}uni, ki slonijo na re\v{s}evanju 
zahtevnih nerelativisti\v{c}nih integralskih ena\v{c}b Faddeeva. Izpostavim tudi 
nere\v{s}ene probleme ter predstavim izzive, s katerimi se bomo spopadli v prihodnje.


\vspace*{1cm}
\begin{flushleft}
{\bf Klju\v{c}ne besede}: Jefferson Lab, polarizirano jedro helija-3, 
spinsko-izospinska zgradba helija-3, dvojnopolarizacijske asimetrije, magnetni
spektrometri, BigBite, opti\v{c}na kalibracija, rekonstrukcija trajektorij delcev, 
ena\v{c}be Faddeeva.\\[3mm]
{\bf PACS (2010)}: 29.30.Aj,\hspace*{1mm} 29.85.Fj,\hspace*{1mm} 25.30.-c
\end{flushleft}

\newpage

\tableofcontents                        
\mainmatter
\parskip3mm    
\chapter{Physics Motivation}

\section{Why ${}^3\mathrm{He}$?}

The ${}^3\mathrm{He}$ nucleus is the subject of considerable current 
interest. Being a three nucleon system makes it an ideal case for testing 
our understanding of the nuclear forces between nucleons, the structure of 
the nuclear ground state and the reaction mechanisms. It is complex enough 
to exhibit all important features that are present in the reactions of heavier 
nuclei. On the other hand it is small enough to be an exactly calculable 
nuclear system, where theoretical predictions of its nuclear structure 
can be compared with the experimental data to an increasingly accurate 
degree~\cite{e05102}. This way it is also an ideal testing ground 
for studying effects such as Final State Interactions (FSI) and the 
Meson Exchange Currents (MEC).

Consisting of three nucleons (two protons and a neutron), 
${}^3\mathrm{He}$ is also an ideal candidate for studying 
the effects of three-nucleon forces (3NF). A need for inclusion of 3NF 
has been clearly demonstrated in calculations of binding energies 
of ${}^3\mathrm{H}$ and ${}^3\mathrm{He}$~\cite{nogga2003}, where 
correct values are achieved only if 3NF is added. When using only
nucleon-nucleon forces, binding energies of both nuclei are 
underestimated.

As an alternative to ${}^3\mathrm{He}$, ${}^3\mathrm{H}$ nuclei 
could be considered for testing few-body theories. Since triton 
has exactly opposite structure than helium 
(${}^3\mathrm{He(ppn)}\rightarrow{}^3\mathrm{H(nnp)}$), comparison 
of the two would give even further insight into the underlying theories. 
Unfortunately ${}^3\mathrm{H}$ is radioactive, which prevents it from being 
explored in modern experiments, due to the safety concerns.  

\vspace*{-0.2cm}

\subsection{Structure of the ground-state wave-function}
\label{sec:he3WF}
\begin{figure}[!hb]
\begin{center}
\includegraphics[width=0.7\textwidth]{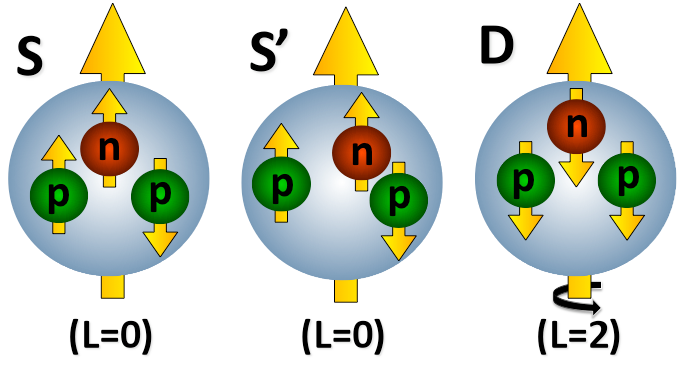}
\caption{Three dominant components (S, S' and D) of the ${}^3\mathrm{He}$
ground-state wave-function. Arrows show the spin orientations of 
the nucleons and nuclei, while $L$ denotes the orbital angular momentum. 
\label{fig_He3States}}
\end{center}
\end{figure}

Theoretical calculations~\cite{blankleider, afnan} of the three-body 
bound state predict, that three components dominate the ${}^3\mathrm{He}$ 
ground-state wave-function. See Fig.~\ref{fig_He3States} and 
Table~\ref{table_He3WaveFunctionComponents}. 
The dominant component of 
the ${}^3\mathrm{He}$ wave function is a spatially symmetric 
S state, in which the proton spins are in the spin-singlet state 
(anti-parallel) and the ${}^3\mathrm{He}$ spin is predominantly carried by 
the neutron. This configuration accounts for $\approx 90\,\%$ of the 
spin-averaged wave function. The dominance of this state is supported by the 
fact, that the magnetic momentum of the ${}^3\mathrm{He}$ is very close
to the magnetic momentum of the neutron~\cite{anderson49}:
\vspace*{-0.2cm}
\begin{eqnarray}
 \frac{\mu_{{}^3\mathrm{He}}}{\mu_{\mathrm{n}}} = \frac{-2.131}{-1.913} \approx 1 \nonumber
\end{eqnarray}

\begin{table}[!hb]
 \begin{center}
\caption{The partial wave components (Faddeev channels) 
of the three-nucleon 
($\alpha$, $\beta$, $\gamma$) wave function within the 
Derrick-Blatt scheme~\cite{blankleider}. $S$ and $L$ are the 
spin and the angular momentum of ${}^3\mathrm{He}$. 
$l_{\alpha}$ is the relative orbital angular momentum 
of the pair ($\beta\gamma$), while $L_{\alpha}$ represents
the orbital angular momentum of the ($\beta\gamma$) center
of mass relative to $\alpha$, with $\vec{L} = 
\vec{l}_{\alpha} + \vec{L}_{\alpha}$. $P$ and $K$ label 
the basis vectors $|PK\rangle$ of the irreducible
representations of the permutation group $S_{3}$, which
are considered for description of the three-nucleon 
spin states. \label{table_He3WaveFunctionComponents}}
\begin{tabular}{ccccccccc}
\toprule
 Channel & & & & & & & Probability & WF\\[-4pt]
 number  & $L$ & $S$ & $l_{\alpha}$ & $L_{\alpha}$ 
         & $P$ & $K$ & (\%)        & State\\ 
\midrule
  1& 0 & 0.5 & 0 & 0 & A & 1 & 87.44 & S\\
  2& 0 & 0.5 & 2 & 2 & A & 1 & 1.20 & S\\
\hline
  3& 0 & 0.5 & 0 & 0 & M & 2 & 0.74 & S'\\
  4& 0 & 0.5 & 1 & 1 & M & 1 & 0.74 & S'\\
  5& 0 & 0.5 & 2 & 2 & M & 2 & 0.06 & S'\\
\hline
  6& 1 & 0.5 & 1 & 1 & M & 1 & 0.01 & P\\
  7& 1 & 0.5 & 2 & 2 & A & 1 & 0.01 & P\\
  8& 1 & 0.5 & 2 & 2 & M & 2 & 0.01 & P\\
  9& 1 & 1.5 & 1 & 1 & M & 1 & 0.01 & P\\
 10& 1 & 1.5 & 2 & 2 & M & 2 & 0.01 & P\\
\hline
 11& 2 & 1.5 & 0 & 2 & M & 2 & 1.08 & D\\
 12& 2 & 1.5 & 1 & 1 & M & 1 & 2.63 & D\\
 13& 2 & 1.5 & 1 & 3 & M & 1 & 1.05 & D\\
 14& 2 & 1.5 & 2 & 0 & M & 2 & 3.06 & D\\
 15& 2 & 1.5 & 2 & 2 & M & 2 & 0.18 & D\\
 16& 2 & 1.5 & 3 & 1 & M & 1 & 0.37 & D\\
\bottomrule
 \end{tabular}
\end{center}
\end{table}

An additional $\approx 8\,\%$  of the spin-averaged wave-function can be 
attributed to the D state generated by the tensor component of the 
nucleon-nucleon force. In this case, the three nucleon spins are predominantly 
oriented opposite to the ${}^3\mathrm{He}$ nuclear spin. The remaining 
$\approx 2\,\%$ originate from a mixed-symmetry configuration of the nucleons, 
the S' state. It arises because of the differences between the isoscalar ($T=0$)
and isovector ($T=1$) forces and hence reflects (spin-isospin)-space 
correlations~\cite{blatt}. The results of Faddeev calculations predict~\cite{nogga2003,friar85}, 
that the probability for S' state depends on the binding energy of the nuclei and
scales approximately as $P_{\mathrm{S'}} \approx (E_{\mathrm{Binding}}/\mathrm{MeV})^{-2.1}$. 
The S' state does not exist for ${}^2\mathrm{H}$, whereas for ${}^4\mathrm{He}$ and heavier nuclei it is 
expected to be strongly suppressed, with $P_{\mathrm{S'}} < 0.1\,\mathrm{\%}$. 
This makes ${}^3\mathrm{He}$ and ${}^3\mathrm{H}$ the only two nuclei, where $S'$ state 
can be observed. Estimated probabilities for finding nuclei in this state 
are $P_{{}^3\mathrm{He}} = 1.24\,\mathrm{\%}$ and $P_{{}^3\mathrm{H}} = 1.05\,\mathrm{\%}$. 
Hence, different probabilities for the $S'$ state in  ${}^3\mathrm{He}$ and ${}^3\mathrm{H}$
explain the bulk of the difference between their charge 
radii ($r_{{}^3\mathrm{He}} \approx 1.7\,\mathrm{fm}$, 
$r_{{}^3\mathrm{H}} \approx 1.5\,\mathrm{fm}$)~\cite{friar87}.

The contributions from other components of 
the ${}^3\mathrm{He}$ ground-state wave-function (e.g. P-state) are estimated to be 
very small and can be neglected.

Understanding the role of the D and S' states  
in ${}^3\mathrm{He}$ is a very important aspect of the few-body theory. In 
particular, the observables that are sensitive to the S' state constitute a 
stringent test of the quality of the theoretical calculations.

\subsection{${}^3\vec{\mathrm{He}}$ as an effective $\vec{n}$ target}

\begin{figure}[!ht]
\begin{center}
\begin{minipage}[!ct]{0.44\textwidth}
\includegraphics[width=1\textwidth]{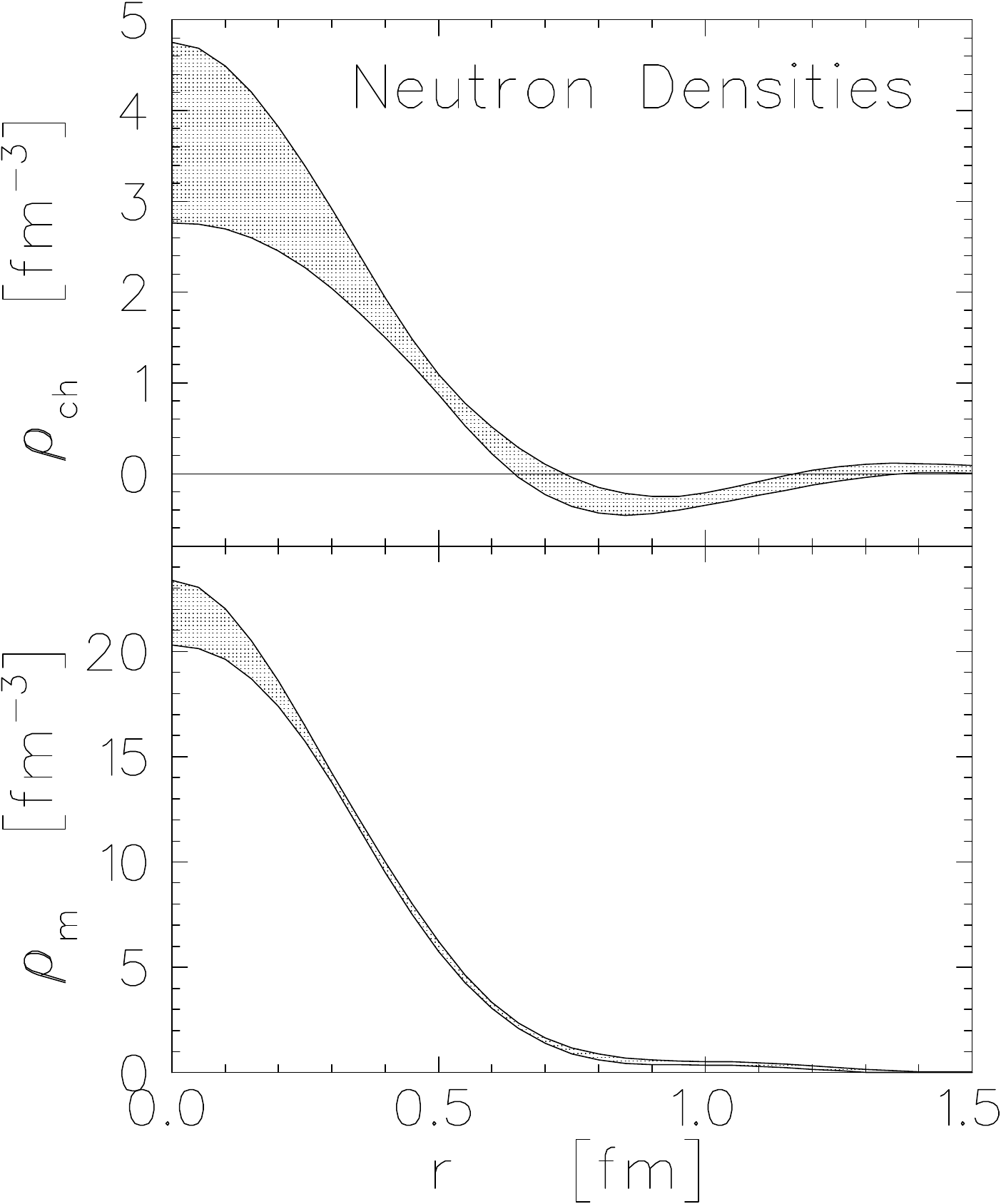}
\end{minipage}
\hfill
\begin{minipage}[!ct]{0.53\textwidth}
\includegraphics[angle=90,width=1\textwidth]{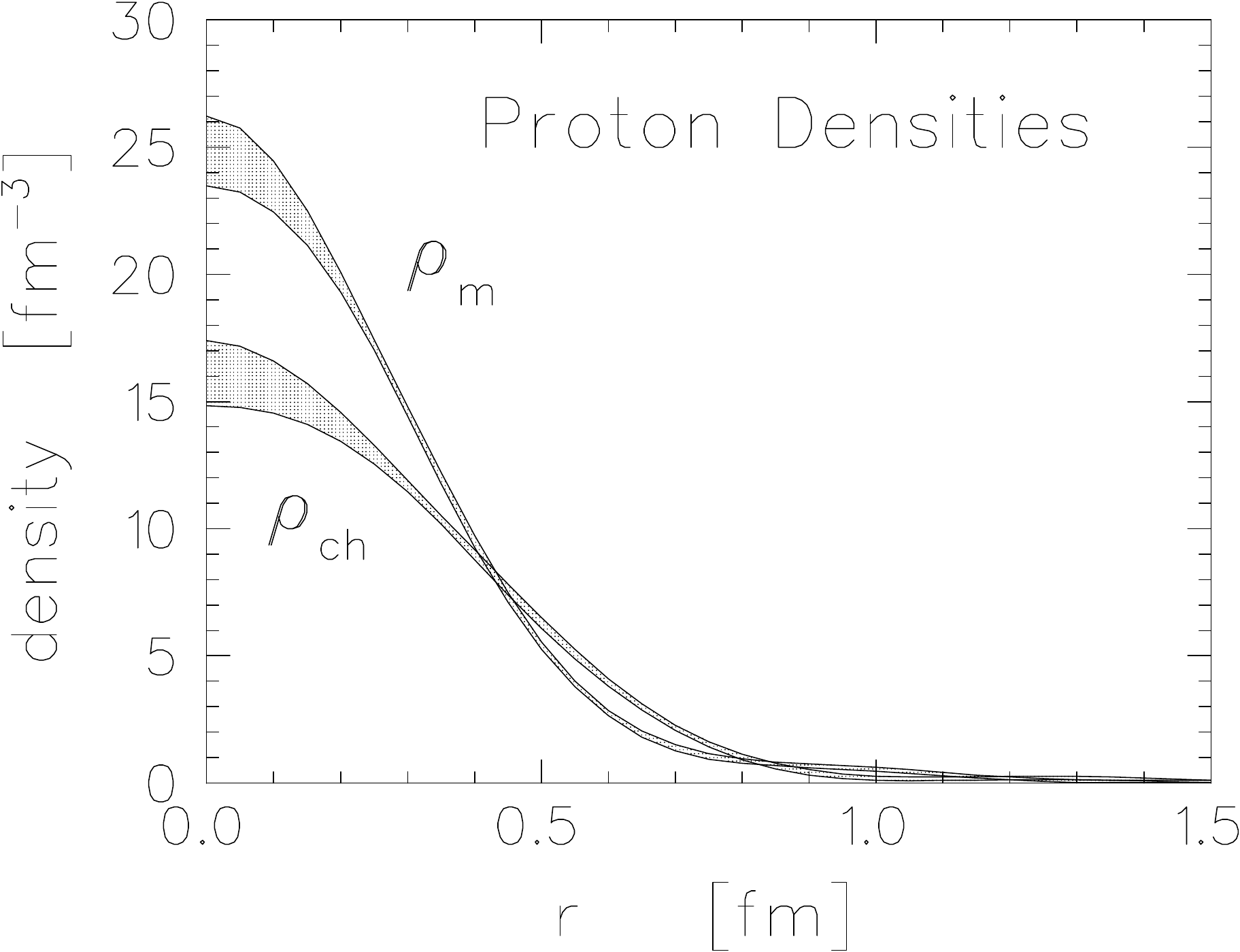}
\caption{Charge ($\rho_{\mathrm{cm}}$) and magnetization ($\rho_{\mathrm{m}}$) 
densities for the neutron (left) and proton (right)~\cite{kelly}.
\label{fig_NucleonDensities}}
\end{minipage}
\end{center}
\end{figure}

A detailed knowledge of the ground-state spin structure of ${}^3\mathrm{He}$ is 
crucial also for extracting precise information on neutron structure.
While the properties of the proton are nowadays well known and precisely measured, 
the structure of the neutron is not yet understood to a desirable accuracy. 
The most uncertain is the information about the neutron
charge distribution. Fig.~\ref{fig_NucleonDensities} shows the calculated 
values of the charge distributions for both proton and neutron~\cite{kelly}. 

The proton charge density is determined with an accuracy better than 
$8\,\mathrm{\%}$, while neutron charge density is known only with an accuracy 
of about $25\,\mathrm{\%}$. There is a continuous effort among the nuclear 
society to measure neutron property more precisely. However, the problem is,
that direct measurements are not possible, because there is no neutron target.  
The structure of the neutron must therefore be determined indirectly. For that we 
use scattering experiments on the deuterium target, where we can assume 
that the neutron behaves almost as a free particle due to the small binding energy 
of the deuteron. As an effective polarized neutron target, a polarized 
${}^3\mathrm{He}$ can be used, by exploiting the fact, that 
the spin of the ${}^3\mathrm{He}$ is essentially carried by the neutron.

\subsubsection{The electric form factor of the neutron}
\begin{figure}[!ht]
\begin{center}
\includegraphics[width=1\textwidth]{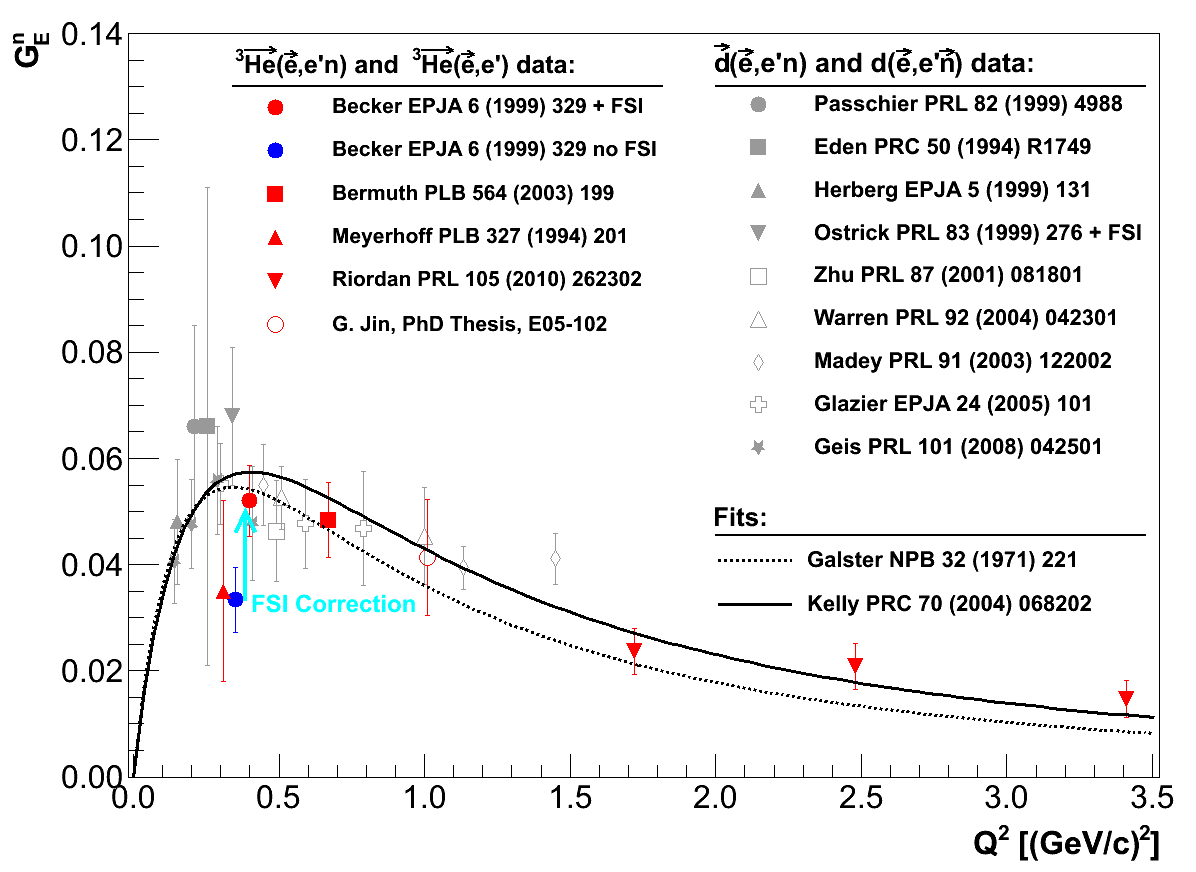}
\caption{Current status of extractions of $\mathrm{G_E^n}$.  Color points 
show results obtained from the cross-section asymmetries from polarized 
${}^3{\vec{\mathrm{He}}}$ target. The gray points are from the analysis of
the cross-section asymmetries using a polarized ${}^2{\vec{\mathrm{H}}}$ 
target. Full and dashed line represent the Kelly and Galster fit to $\mathrm{G_E^n}$.
Full red and blue circle represent the data point from the same measurement before and
after the correction for FSI. The arrow shows the size of the theoretical correction needed
to properly interpret the datum. \label{fig_GEnPlot}}
\end{center}
\end{figure}

The neutron charge distribution is determined through the measurement of the 
neutron electric form factor ($\mathrm{G_E^n}$). Fig.~\ref{fig_GEnPlot} shows the 
majority of the available data for the $\mathrm{G_E^n}$, obtained from
the experiments using polarized deuterium (${}^2\vec{\mathrm{H}}$) and polarized
helium (${}^{3}\vec{\mathrm{He}}$) targets.
 
When utilizing polarized ${}^3\mathrm{He}$ 
target, $\mathrm{G_E^n}$ is determined from the measurements of double polarized
asymmetries in quasi-elastic processes ${}^3{\vec{\mathrm{He}}}\left(
\vec{\mathrm{e}},\mathrm{e' n}\right)$, ${}^3{\vec{\mathrm{He}}}\left(
\vec{\mathrm{e}},\mathrm{e'}\right)$. To extract a precise information on 
the neutron electromagnetic form factors from these data, it is crucial to 
understand the ground-state spin structure of the $\mathrm{{}^3He}$
nucleus in details. The importance of accurate theoretical description 
is illustrated in Fig.~\ref{fig_GEnPlot}. The cyan arrow shows the size of a
theoretical correction needed to properly interpret the Becker 
${}^3{\vec{\mathrm{He}}}\left( \vec{\mathrm{e}},\mathrm{e' n}\right)$ 
datum at $Q^2 = 0.35\,(\mathrm{GeV}/c)^2$ as effective neutron 
data~\cite{becker}. This clearly shows, that a satisfactory description 
of the scattering process ${}^3{\vec{\mathrm{He}}}\left( 
\vec{\mathrm{e}},\mathrm{e' n}\right)$ can no longer be provided by the
plane-wave calculation, where ${}^3\mathrm{He}$ is assumed to be only in 
a S-state and spin of the nuclei carried completely by the neutron. 
Instead, state-of-the-art Faddeev calculations are used, which consider 
full $\mathrm{{}^3He}$ ground-state wave function, together with
the reaction-mechanism effects such as final-state interactions (FSI) and
meson-exchange currents (MEC). The differences between the plane-wave 
approximation and Faddeev calculations are significant, especially at low 
values of the $Q^2$, where the effects of the FSI for this process are 
most prominent. According to Fig.~\ref{fig_GEnPlot} the discrepancy
exceeds the presently achievable experimental uncertainties 
by almost factor of three. 

\subsubsection{The magnetic form factor of the neutron}

\begin{figure}[htp]
\begin{center}
\includegraphics[width=0.9\textwidth]{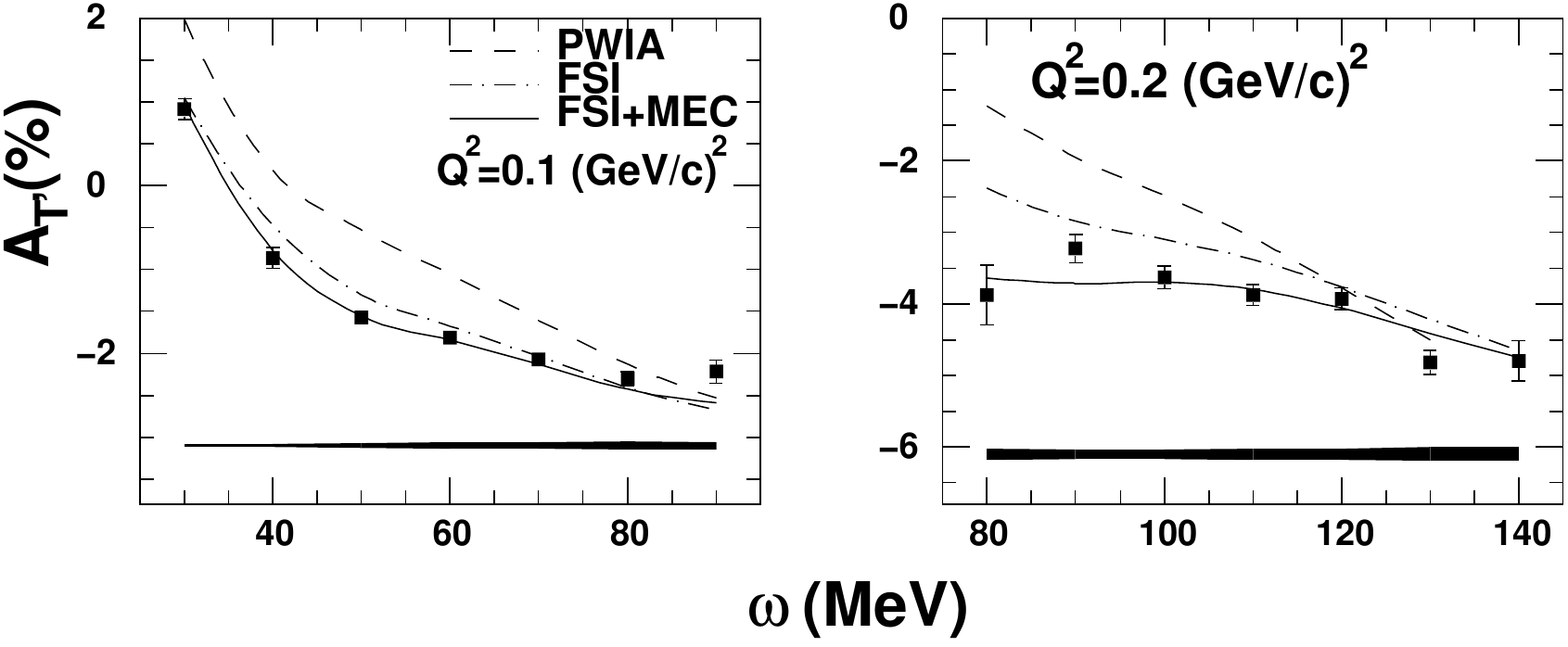}
\caption{ The measured transverse asymmetry $A_{\mathrm{T'}}$ in 
${}^3{\vec{\mathrm{He}}}\left( \vec{\mathrm{e}},\mathrm{e'}\right)$ at 
$Q^2=0.1,\,0.2\,(\mathrm{GeV}/c)^2$~\cite{xu2000}. PWIA calculations 
are shown as dashed lines. The Faddeev calculations with FSI only and 
with both FSI and MEC are shown as dash-dotted and solid lines, 
respectively. \label{fig_XuGMn}}
\end{center}
\end{figure}

The ${}^3\mathrm{He}$ target was extensively used also for determination of the 
magnetic form-factor of the neutron ($G_{\mathrm{M}}^n$), which is intimately 
connected to the magnetization distribution inside the neutron 
(see Fig.~\ref{fig_NucleonDensities}). The $G_{\mathrm{M}}^n$ is extracted 
from the measurement of the double-polarization asymmetry $A_{\mathrm{T'}}$ in the
inclusive ${}^3{\vec{\mathrm{He}}}\left( \vec{\mathrm{e}},\mathrm{e'}\right)$
reaction. Fig.~\ref{fig_XuGMn} shows results of such measurement performed
at Jefferson Lab~\cite{xu2000}.

These experiments again relay on a fact, that in the ground state 
of ${}^3\mathrm{He}$, proton spins cancel each other out, and polarized
${}^3\mathrm{He}$ behaves as an  effective neutron target. However, 
the results show, that PWIA alone does not provide an adequate descriptions. 
Corrections for FSI and MEC have to be applied, which once more require a
precise theoretical insight into the ${}^3\mathrm{He}$ reaction mechanism.

\subsubsection{Polarized quark structure functions}

\begin{figure}[!ht]
\begin{center}
\includegraphics[width=0.46\textwidth]{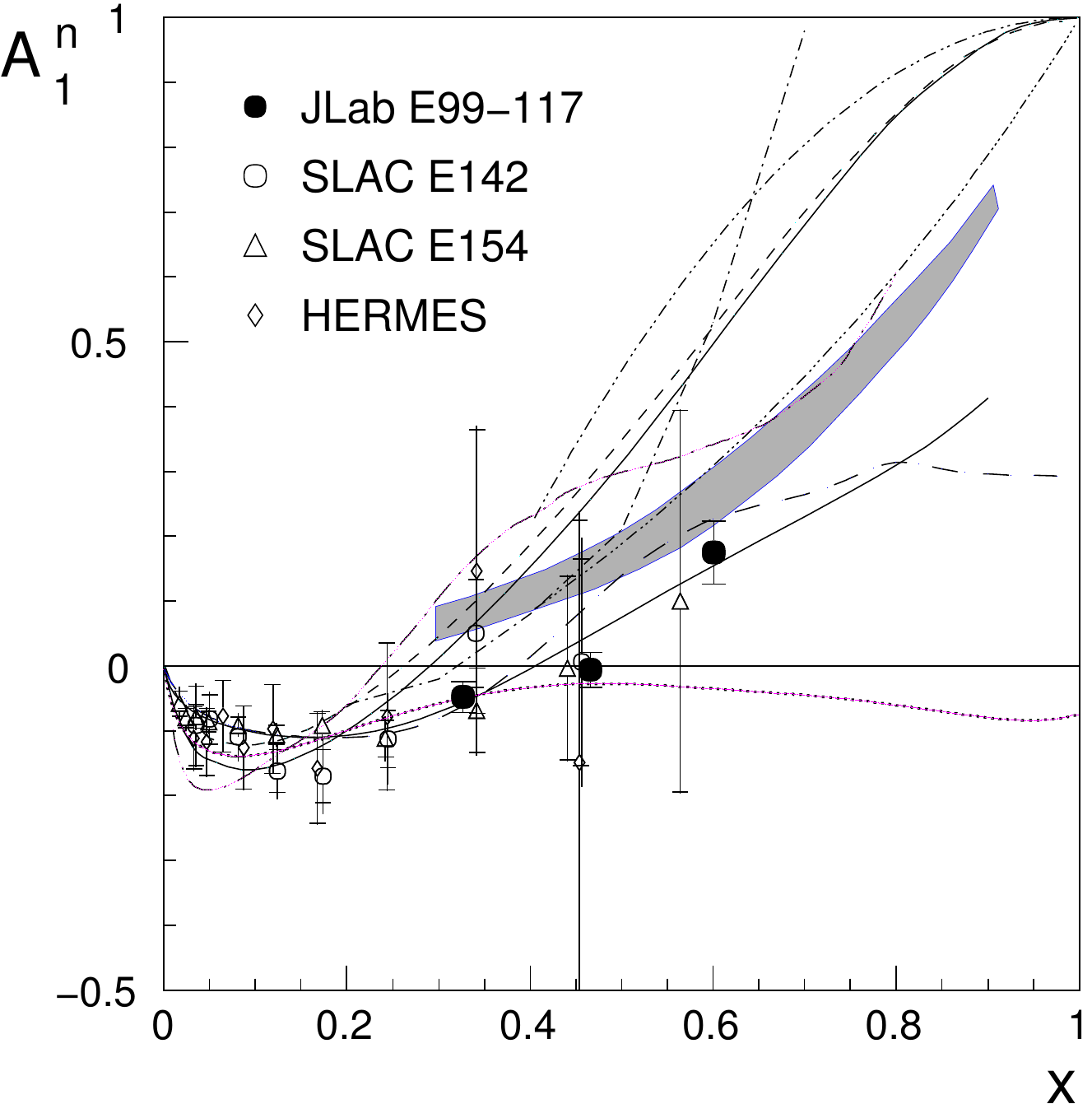}
\hfill
\includegraphics[width=0.49\textwidth]{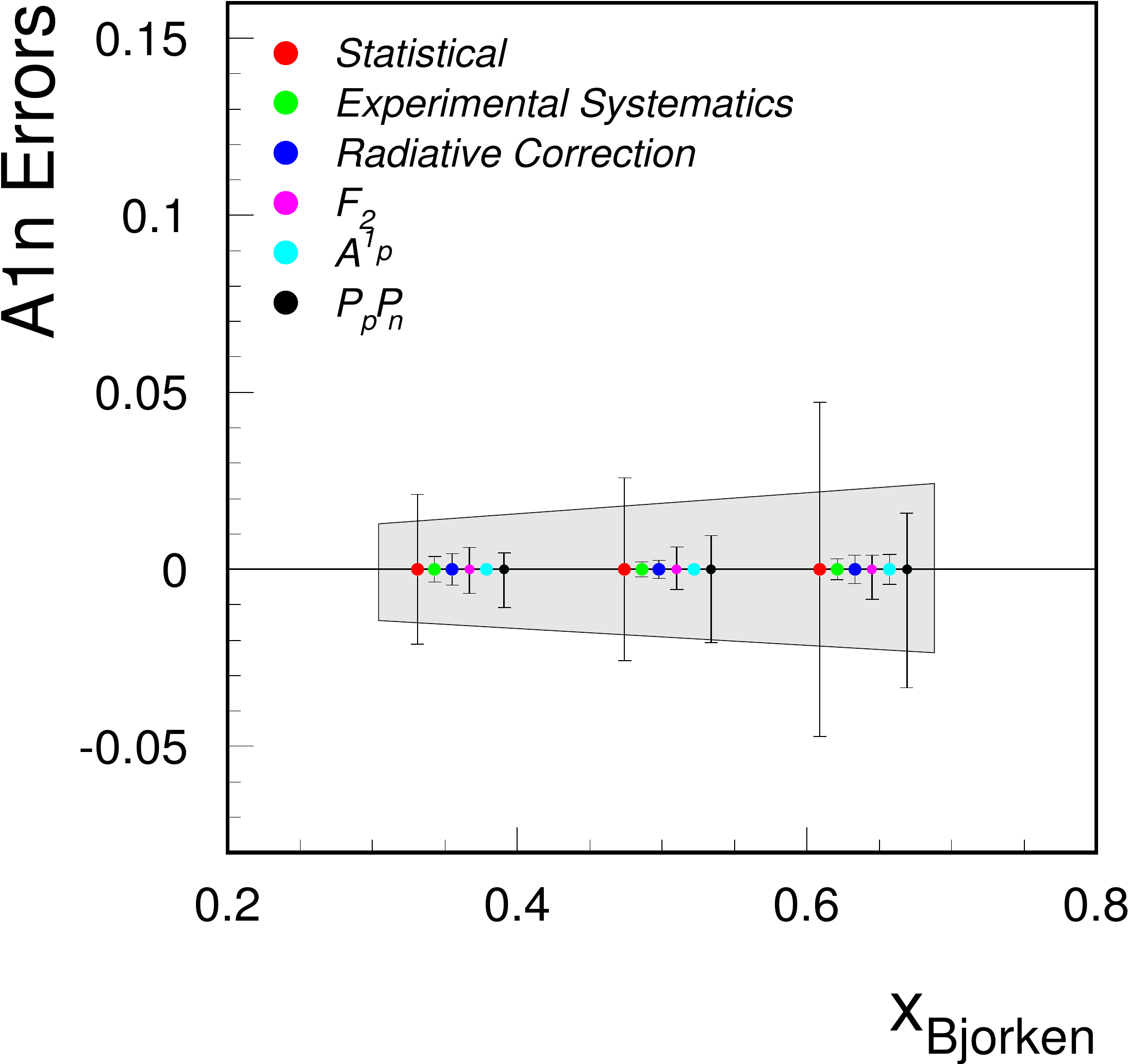}
\caption{[Left] The results of the $A_{1}^{n}(x)$ measurements, compared to
the theoretical predictions~\cite{xiaochaoPRL}. The measurements
show a zero crossing around $x=0.47$ and a significantly positive value at
$x=0.60$. The results at high $x$ agree with the predictions of the 
constituent quark model but disagree with the predictions of the leading-order
perturbative QCD (PQCD). This might indicate, that quarks have non-zero
orbital angular momentum~\cite{xiaochaoPRC}. [Right] Summary of uncertainties 
in the experiment E99-117. The statistical uncertainty dominates. 
$P_{\mathrm{p}} P_{\mathrm{n}}$ denotes the error caused by the uncertainty 
of the polarization of the proton and neutron in ${}^3\vec{\mathrm{He}}$. The
light-blue band represents the predicted statistical error of the upcoming
experiment E12-06-122. The uncertainty becomes comparable to or even smaller than 
the error in $P_{\mathrm{p}} P_{\mathrm{n}}$.\label{fig_A1n}}
\end{center}
\end{figure}

A detailed knowledge of the ground-state spin structure of ${}^3\mathrm{He}$ is 
essential also for other types of experiments that are considering ${}^3\mathrm{He}$ as
an effective polarized neutron target. An example of such experiment is the
measurement of the neutron spin asymmetry $A_1^n$, which is important for 
understanding the spin structure of the neutron (see Fig.~\ref{fig_A1n}). 
In particular, it provides a definitive information about the spin carried 
by the quarks and gives an insight in to the continuing question of 
the role of the quark orbital angular momentum in the nucleon 
wave function~\cite{e12-06-122}. Fig~\ref{fig_A1n} shows the error budget 
of the experiment E99-117~\cite{e05102,xiaochaoPRL,xiaochaoPRC} in which 
asymmetry $A_1^n$ was extracted. The two largest sources of error are the 
statistical uncertainty and the uncertainty of the polarization of proton 
($P_{\mathrm{p}}$) and neutron ($P_{\mathrm{n}}$) inside  
the ${}^3\mathrm{He}$. These polarizations depend directly
on structure of the nuclei and three components of the ground-state wave 
function.

Although the statistical error dominates over the rest of the error 
contributions in the E99-117, it is estimated that it will 
become comparable or even smaller than the uncertainty in $P_{\mathrm{p}}$
and $P_{\mathrm{n}}$ for the upcoming $A_{1}^{n}$ experiment E12-06-122,
which will be utilizing a polarized ${}^3\mathrm{He}$ with the 
$11\,\mathrm{GeV}$ electron beam. At that point any corrections to 
$P_{\mathrm{p}}$ and $P_{\mathrm{n}}$ would result in a shift of 
all points up or down and consequently change the interpretation 
of the zero crossing of $A_{1}^{n}(x)$.

With the increasing statistics, the precision of the current and 
upcoming double-polarized experiments is reaching a level, that can only be matched 
by the best theoretical models of the ${}^3\mathrm{He}$ nucleus~\cite{e05102}. 
These models therefore require progressively more accurate input to adjust their 
parameters like the ground-state wave-function components, and a complete 
understating of the spin and isospin dependence of final-state interactions 
and meson-exchange currents. To achieve this a direct measurement devoted to a better
understanding of the ${}^3\mathrm{He}$ itself is needed. Without a significant 
improvement of this understanding, future experiments on ${}^3\mathrm{He}$ will 
be seriously impaired~\cite{e05102}. The properties of the ${}^3\mathrm{He}$ 
need to be studied on a broad kinematic range to create enough lever to constrain
the theories. 

\section{Previous experiments involving ${}^3\mathrm{He}$}

\subsection{Unpolarized experiments}

\begin{figure}[!ht]
\begin{center}
\begin{minipage}[t]{0.42\textwidth}
\hrule height 0pt
\includegraphics[width=\textwidth]{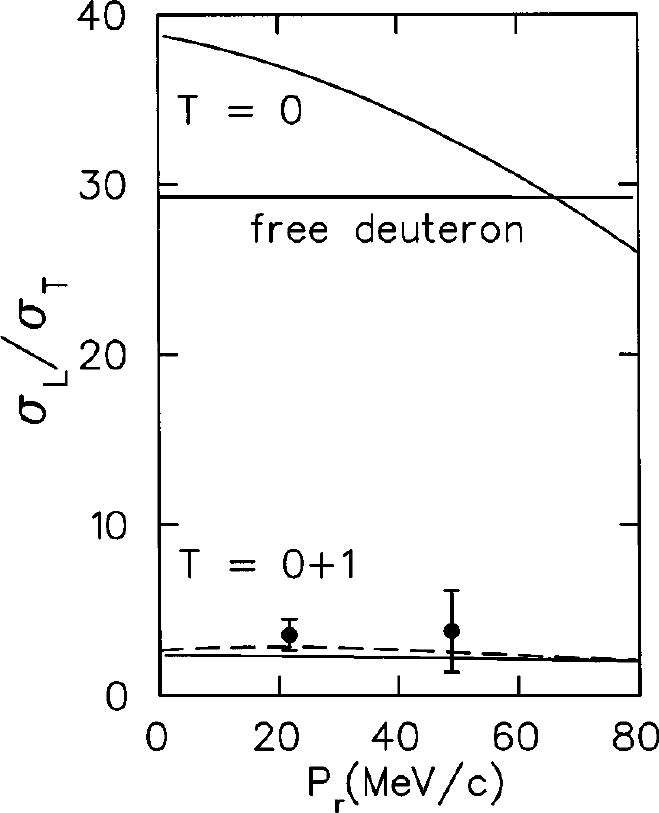}
\end{minipage}
\hspace*{1cm}
\begin{minipage}[t]{0.45\textwidth}
\hrule height 0pt
\vskip -3mm
\caption{The ratio of the longitudinal and transverse response
functions $\sigma_{\mathrm{L}}/\sigma_{\mathrm{T}}$ as a function of recoil
momentum $P_{r}$~\cite{tripp96}. The theoretical calculations considering
only $T=0$ currents and both $T=0$ and $T=1$ currents are shown with
dashed and solid line, respectively. The ratio for the free deuteron is also
shown. 
\label{fig_Tripp}}
\end{minipage}
\end{center}
\end{figure}

A high level of interest and motivation for understanding the structure and properties 
of ${}^3\mathrm{He}$ can be recognized in  an extensive theoretical and experimental
effort to study unpolarized processes ${}^3\mathrm{He}(\mathrm{e,e'd})$ and
${}^3\mathrm{He}(\mathrm{e,e'p})$. The MIT-Bates experiment~\cite{tripp96}
measured reaction ${}^3\mathrm{He}(\mathrm{e,e'd})\mathrm{p}$ at four-momentum 
transfer $q = 420\,\mathrm{MeV}/c$, for two different proton recoil momenta 
$p_r = 22,$ and $54\,\mathrm{MeV}/c$, in parallel deuteron kinematics. In the region of 
low proton recoil momentum ($p_r \approx 0$) one would naively expect that the 
cross-section follows that for the elastic scattering of a free deuteron. Since 
elastic scattering of a free deuteron chooses isoscalar currents ($T=0$), 
one would think that the dominant mechanism in the two-body ($p+d$) breakup 
would also involve the interaction with a correlated $T=0$ pair, known as a 
quasi-deuteron model (QDM). However, the two-body currents in the 
${}^3\mathrm{He}(\mathrm{e,e'd})\mathrm{p}$ reaction contain also the isovector
($T=1$) components. The Bates experiment has shown, that the $T=1$ currents play 
an important role and contributes substantially to the final value of the
cross-section (see Fig.~\ref{fig_Tripp}). 
Hence, ${}^3\mathrm{He}(\mathrm{e,e'd})\mathrm{p}$ can not be 
adequately explained by the QDM, but reaction mechanism containing both isoscalar
and isovector currents must be considered. 

A Jefferson Lab experiment E89-044 also contributed an important new
insight into the characteristics of the ${}^3\mathrm{He}$ breakup process.
In particular, they found an evidence of NN-correlations and demonstrated
the importance of the FSI.
The ${}^3\mathrm{He}(\mathrm{e,e'p})\mathrm{d}$ reaction~\cite{marat05} and  
${}^3\mathrm{He}(\mathrm{e,e'p})\mathrm{pn}$ reaction~\cite{fatiha05} were
measured at fixed energy and momentum transfers ($\omega = 840\,\mathrm{MeV}$
and $|\vec{q}| = 1502\,\mathrm{MeV}/c$), and covered a tremendous range of 
missing momenta up to $1\,\mathrm{GeV}/c$, for missing energies up to the pion 
threshold.  These benchmark measurements were much higher in statistics than any previous
previous measurement~\cite{marchand88}. The experimental data  for both, 
two-body breakup (2bbu) and three-body breakup (3bbu), channels
were well described by the approach of Laget~\cite{laget2005}.

\begin{figure}[!ht]
\begin{center}
\includegraphics[width=0.49\textwidth]{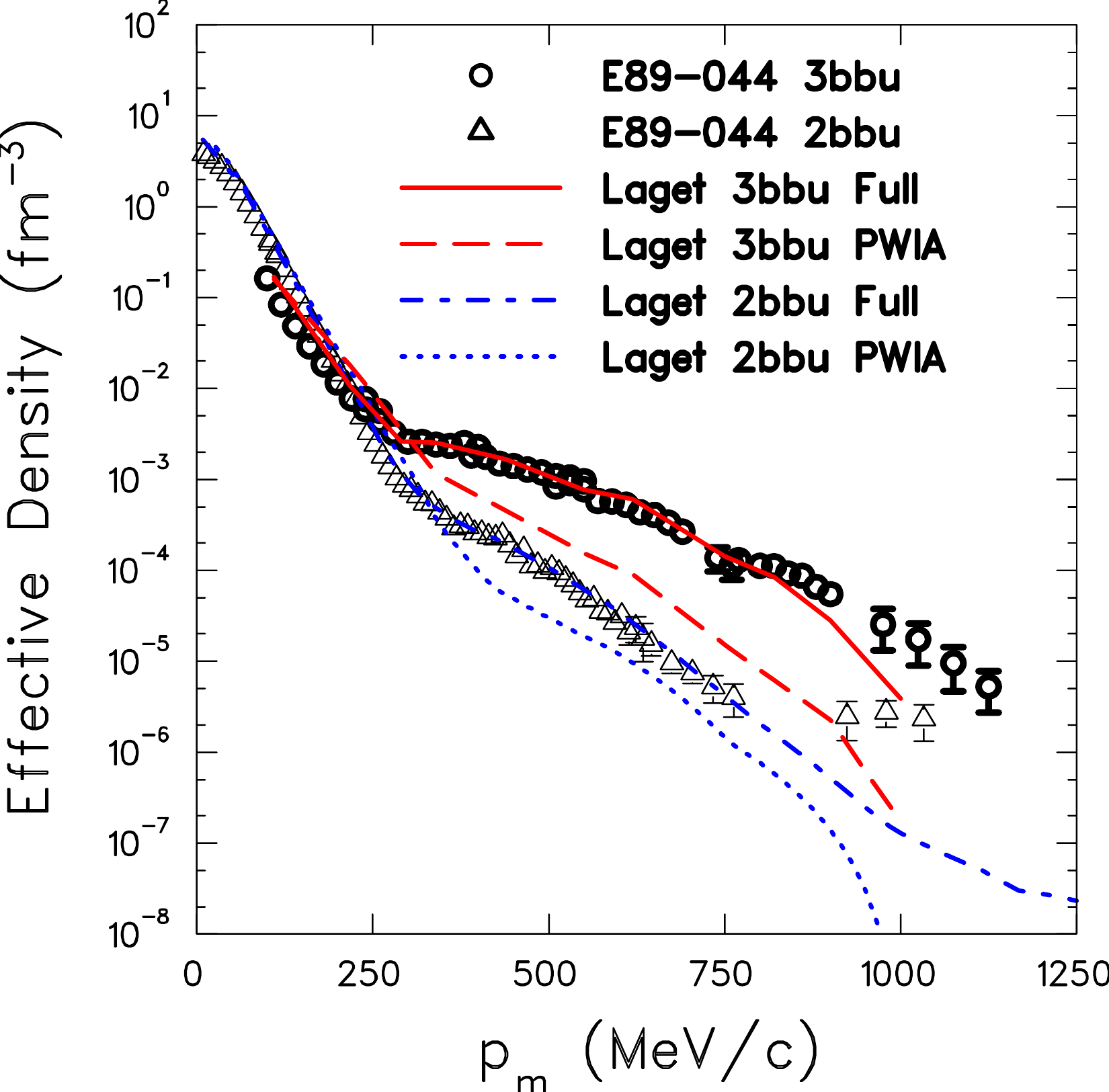}
\hfill
\includegraphics[width=0.435\textwidth]{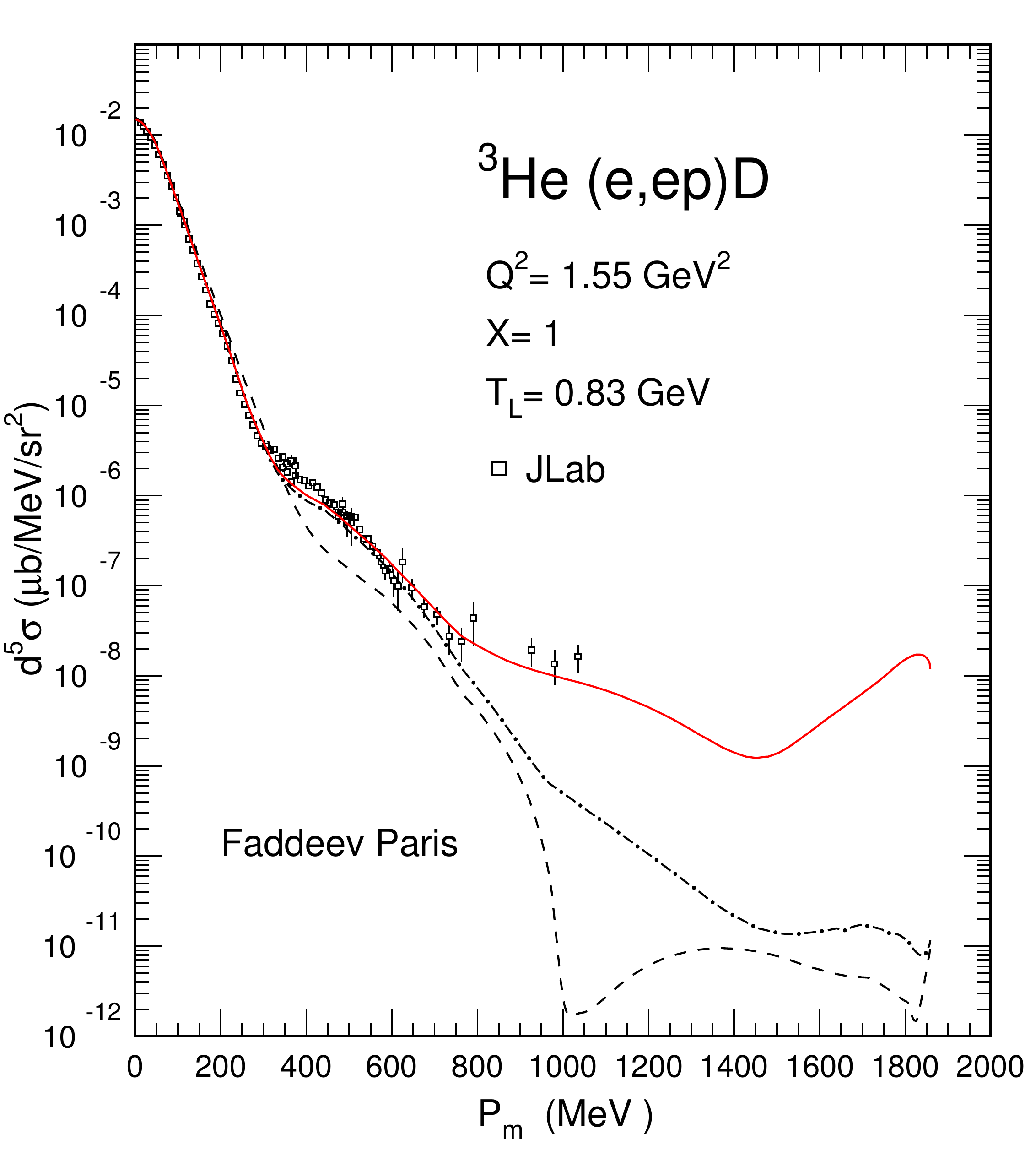}
\caption{ 
[Left] Proton effective momentum density distributions in ${}^3\mathrm{He}$
extracted from ${}^3\mathrm{He}(\mathrm{e,e'p})\mathrm{d}$ and  
${}^3\mathrm{He}(\mathrm{e,e'p})\mathrm{pn}$~\cite{fatiha05}. Experimental 
results are compared to the calculations by Laget~\cite{laget2005}. PWIA calculations 
and full Faddeev calculations for each reaction channel are shown with 
dashed and solid line, respectively. [Right] Theoretical predictions by 
Laget for the momentum distribution in 
reaction ${}^3\mathrm{He}(\mathrm{e,e'p})\mathrm{d}$ at $Q^2=1.55\,(\mathrm{GeV}/c)^2$ 
compared to the JLab data~\cite{marat05}. Dashed line shows the PWIA calculations.
Dash-dotted line considers two-body FSI, MEC and $\Delta$ production. Full line
represents full calculations including also three-body processes. 
\label{fig_FatihaLaget}}
\end{center}
\end{figure}

One-body mechanisms, where electron interacts with 
a single nucleon and deuteron is just a spectator (PWIA), is sufficient to describe 
${}^3\mathrm{He}(\mathrm{e,e'p})$ reactions below recoil (missing) momenta 
$p_{r} \approx 300\,\mathrm{MeV}/c$. In the $p_{r}$ region between $300\,\mathrm{MeV}/c$
and $700\,\mathrm{MeV}/c$, nucleon-nucleon final-state interactions become
important. Fig.~\ref{fig_FatihaLaget} shows that the 3bbu cross-section in this region
is up to three orders of magnitude larger than the corresponding 2bbu cross section. 
This significant difference is caused by a much larger role of FSI and NN-correlations 
in the 3bbu than in the 2bbu, because of the reduced probability for the two
undetected nucleons to recombine and form the ejected deuteron at high $p_r$. 
The comparison of PWIA calculations for both reaction channels reveals only one 
order of magnitude enhancement of the 3bbu over the 2bbu, due to the NN-correlations.
The rest is contributed by the FSI and is represented in Fig.~\ref{fig_FatihaLaget}
as a difference between the dashed-line (PWIA calculation) and solid line (full calculations).

The two-orders of magnitude correction to the cross-section contributed by the FSI 
indicates a great importance of the FSI in the 3bbu of ${}^3\mathrm{He}$. 
On the other hand, the two-body processes 
like meson-exchange currents and formation of $\Delta$, contribute only at level 
of $\approx 20\,\mathrm{\%}$. The flattening of the 2bbu cross-section 
in the $p_r$ region between $700\,\mathrm{MeV}/c$ and $1000\,\mathrm{MeV}/c$ was 
explained by the three-body mechanism which dominates there~\cite{laget2005}. 
A virtual photon is absorbed by a nucleon at rest. This nucleon 
emits meson, which is then absorbed by the the remaining two nucleons~\cite{sircaWL2009}.
From all this we can see, that experiment E89-044 enabled a simultaneous study and interpretation of 
one-, two- and three-body mechanisms and significantly enriched our knowledge
on the ${}^3\mathrm{He}$ system.

\begin{figure}[!ht]
\begin{center}
\begin{minipage}[t]{0.49\textwidth}
\hrule height 0pt
\includegraphics[width=\textwidth]{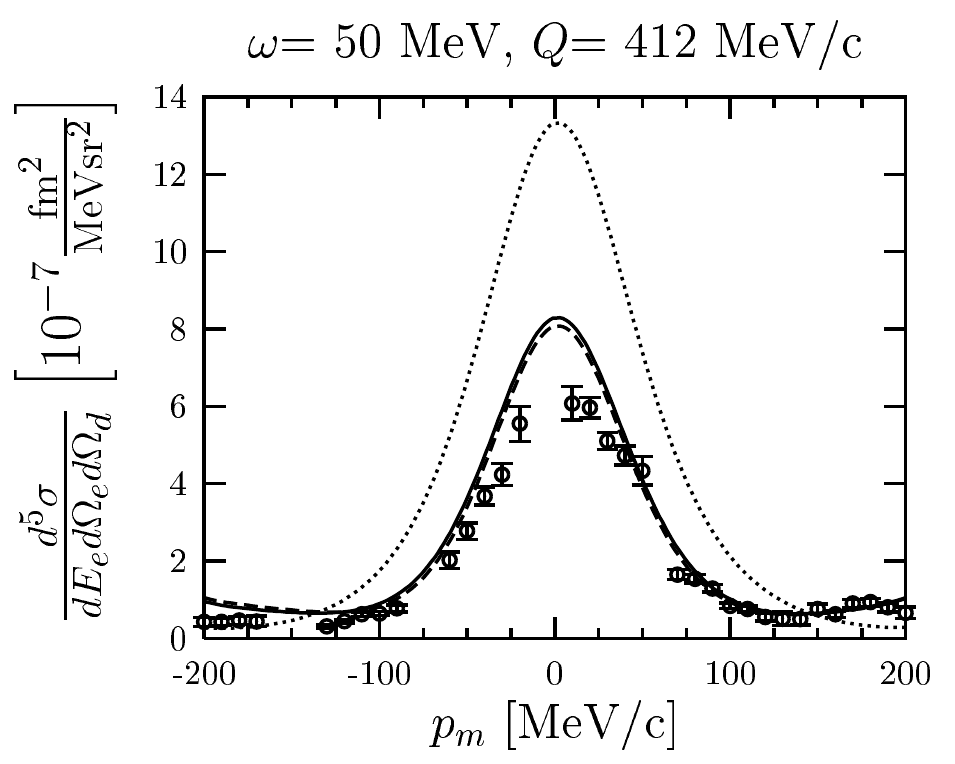}
\end{minipage}
\hspace*{1cm}
\begin{minipage}[t]{0.4\textwidth}
\hrule height 0pt
\caption{The measured cross-section for ${}^3\mathrm{He}(\mathrm{e,e'd})\mathrm{p}$ 
as a function of proton recoil momentum~\cite{spaltro2002}. Dotted curve represents 
the results of the PWIAS calculations. Dashed line shows full calculations without
MEC. Predictions shown with solid line include also MEC. The PWIA 
calculations are not shown~\cite{glockle2004}. 
\label{fig_spaltro}}
\end{minipage}
\end{center}
\end{figure}

In the case of the deuteron knockout, things are unfortunately not that well 
understood. An important puzzle is related to the results of the NIKHEF 
experiment~\cite{spaltro2002}, where they measured unpolarized cross-section for the
${}^3\mathrm{He}(\mathrm{e,e'd})\mathrm{p}$ reaction as a function of 
recoil momentum $p_r$ in $(q,\omega)$-constant kinematics. An example of their 
measurements is shown in Fig.~\ref{fig_spaltro}. 

To the date, the theory
was unable to adequately describe these data. In spite continuous theoretical 
efforts the inconsistency remains. Even most sophisticated Faddeev calculations, 
which employ the AV18 nucleon-nucleon interaction and include MEC overestimate
the measured cross sections. At present is not clear if this discrepancy 
is caused by an error in the measurements or by an inadequate theoretical description.
Therefore further measurements with greater precision and sensitivity to the theoretical 
ingredients are needed to resolve these issues~\cite{glockle2004,sircaWL2009}. 
Beam-target asymmetries seem to be very promising candidates for such observables.

\subsection{Double-polarization experiments}
The results from the unpolarized ${}^3\mathrm{He}$ experiments and corresponding 
theoretical calculations have revealed important information about the structure
and properties of the ${}^3\mathrm{He}$ system and have proven the need for full 
Faddeev calculations. Regrettably, the measurements of the unpolarized cross-section 
do not have the strength to isolate small components ($S'$ and $D$) of 
the ${}^3\mathrm{He}$ ground-state wave-function which have the ability
to further constrain theoretical models. This gives double-polarization 
observables an important advantage. It has been demonstrated by various 
theoretical groups~\cite{nagorny96, nagorny98, golak2005, sauer93}, that measurements of 
beam-target asymmetries in ${}^3\vec{\mathrm{He}}\left(\vec{\mathrm{e}},
\mathrm{e'} \mathrm{d }\right)$ and ${}^3\vec{\mathrm{He}}\left(\vec{\mathrm{e}},
\mathrm{e'} \mathrm{p}\right)$ can provide precise information on both
$S'$ and $D$ components of the ${}^3\mathrm{He}$ ground-state wave-function.

\subsubsection{Experiments with polarized ${}^2\mathrm{H}$ at NIKHEF}

\begin{figure}[!ht]
\begin{center}
\begin{minipage}[t]{0.5\textwidth}
\hrule height 0pt
\includegraphics[width=\textwidth]{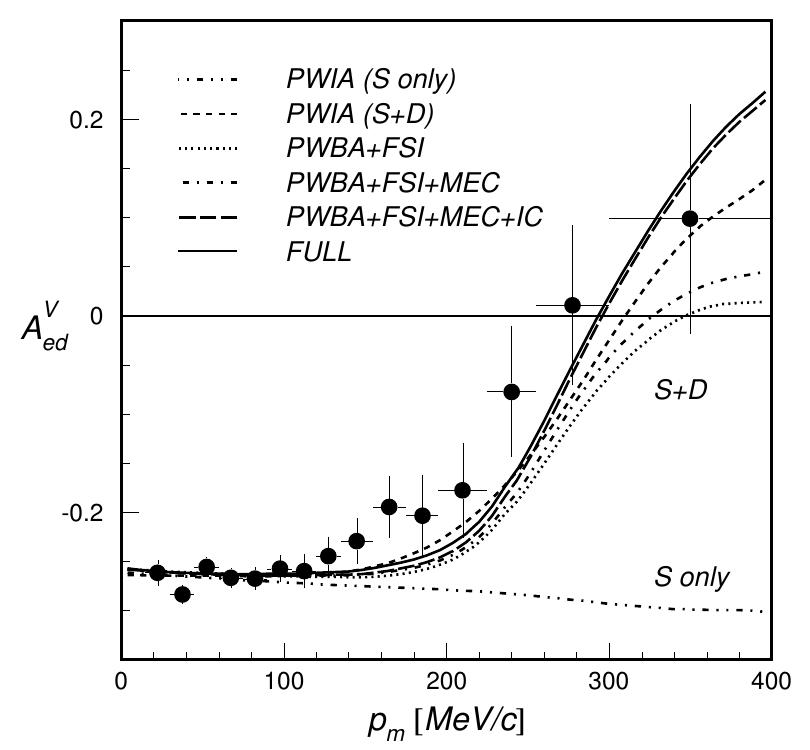}
\end{minipage}
\hfill
\begin{minipage}[t]{0.46\textwidth}
\hrule height 0pt
\caption{ 
Spin correlation parameter $A_{ed}^{V}$ in reaction ${}^2\vec{\mathrm{H}}\left(
\vec{\mathrm{e}},\mathrm{e'} \mathrm{p}\right)\mathrm{n}$ as a function of
recoil (neutron) momentum at $Q^2=0.21\,(\mathrm{GeV}/c)^2$. 
Curves show predictions of different theoretical models. The need for 
inclusion of the D state into the predictions is clearly demonstrated, since
the PWIA with S-state only does not change sign at higher recoil (missing)
momenta. Incorporation of FSI, MEC  and IC have tendency to move theoretical 
predictions closer to the experimental values, but do not have power to 
significantly change results.
\label{fig_Passchier}}
\end{minipage}
\end{center}
\end{figure}

The use of double-polarization observables has already proven to be very successful
in experiments with polarized ${}^2\mathrm{H}$. Fig~\ref{fig_Passchier} shows 
the results of a NIKHEF experiment~\cite{passcier2002}, where they used deuteron
as a benchmark for testing nuclear theory. 
They measured spin-momentum correlation parameter $A_{ed}^{V}$ for the 
${}^2\vec{\mathrm{H}}\left(\vec{\mathrm{e}},\mathrm{e'} \mathrm{p}\right)\mathrm{n}$ 
reaction at $Q^2 = 0.21\,(\mathrm{GeV}/c)^2$. The measured data 
give precise information about the deuteron spin structure and are in good 
agreement with different theoretical models. Theory predicts that 
${}^2\mathrm{H}$ ground-state wave functions is principally combined of 
two major components.  In the dominant $S$-component are spins of both, proton 
and neutron, aligned with the spin of the nuclei, while in the $D$ component they 
are oriented in the opposite direction. There is a $\approx 90\,\mathrm{\%}$
probability of finding deuteron in the $S$-state and $\approx 10\,\mathrm{\%}$
probability of finding it in the $D$ state. Fig.~\ref{fig_Passchier} shows that 
contribution of $D$ state is essential for obtaining proper description of the 
experimental data at higher missing momenta. Hence, if ground-state wave 
function would consist of only $S$-state, spin correlation parameter $A_{ed}^{V}$ 
would have to have significantly different shape. 

One can also see, that final-state effects become important only at high 
recoil momenta. Unfortunately, the accuracy of the data in this region becomes 
poor and obstructs the study of such effects. This speaks in favor of 
double-polarization measurements with polarized ${}^3\mathrm{He}$, where significant 
contributions final-state effects are measurable already at smaller recoil 
momenta. In addition, ${}^2\mathrm{H}$ does not contain the $S'$ state, 
which is dominating the region of small recoil momenta and has a potential to 
further constrain theoretical models. Such effects can be studied 
only with ${}^3\mathrm{He}$, which makes ${}^3\mathrm{He}$ even more exciting 
playground to test nuclear dynamics.

\subsubsection{The proton scattering experiments at IUCF}

An important milestone in study of polarization degrees of freedom in ${}^3\mathrm{He}$
was set by the experiment at Indiana University Cyclotron Facility (IUCF)~\cite{milner1996}.
They determined for the first time spin asymmetries in the momentum distributions of 
the neutron and proton in ${}^{3}\mathrm{He}$. The measured asymmetries in quasi-elastic processes  
${}^3\vec{\mathrm{He}}(\vec{p},2p)$ and ${}^3\vec{\mathrm{He}}(\vec{p},pn)$ were compared
to the PWIA calculations. They observed a good agreement of the measurements with the theory.
A $100\,\mathrm{\%}$ asymmetry in the momentum distribution of the neutron at low momenta 
demonstrated a strong dominance of the $S$ state in the ${}^3\mathrm{He}$ ground-state 
wave-function and provided confidence, that ${}^3\mathrm{He}$ can be used as an effective 
polarized neutron target for scattering experiments in nuclear and particle physics. 
On the other hand, the observed negative asymmetry ($\approx -15\,\mathrm{\%}$) in the 
momentum distribution of a proton was interpreted as an indication for the presence 
of the $S'$-state. However, this interpretation was later refuted by noticing that 
the major reason for the asymmetry are the relative differences between the two-body
breakup and three-body breakup cross-sections.

The IUCF experiment also revealed the weaknesses inherent to the use of hadronic probes 
and helped to initiate a study of spin-dependent momentum distributions in ${}^3\mathrm{He}$ 
with the use of  electrons. being a point-like particles, electrons represent the cleanest 
probe for testing
nuclear structure and can provide much better resolving power than hadrons. 
Hence, the effort for disentangle the effects of small wave-function components has 
shifted to electro-disintegration of polarized ${}^3\mathrm{He}$.

\subsubsection{First measurement of electron induced beam-target asymmetries}

The pilot measurement of beam-target asymmetries, utilizing a ${}^3\mathrm{He}$ target in 
combination with polarized electron beam, was performed at NIKHEF~\cite{poolman},
where they determined the $A_z$ asymmetry in ${}^3\vec{\mathrm{He}}\left(\vec{\mathrm{e}},
\mathrm{e'} \mathrm{p}\right)$ and ${}^3\vec{\mathrm{He}}\left(\vec{\mathrm{e}},
\mathrm{e'} \mathrm{n}\right)$ reactions at beam energy of $442\,\mathrm{MeV}$ and
at $Q^2 = 0.16\,(\mathrm{GeV}/c)^2$. They obtained a small value of the 
asymmetry in the proton channel ($A_z = 0.15\pm 0.11$), but a large value in the
neutron channel ($A_z = -0.56\pm 0.18$). This was in agreement with the theoretical 
predictions where the main contribution of the ${}^3\mathrm{He}$ wave-function 
represents the spatially symmetric $S$-state, where the protons occupy a spin-singlet
state and neutron caries the majority of ${}^3\mathrm{He}$ spin. Although this measurement
was low in statistics, it already demonstrated the feasibility of the double-polarization
experiments at medium beam energies ($\approx 1\,\mathrm{GeV}$), indicated 
the need for full Faddeev calculations and laid the ground work for further experiments. 

\subsubsection{Study of two-body and three-body breakup processes at Mainz}

\begin{figure}[!ht]
\begin{center}
\begin{minipage}[t]{0.55\textwidth}
\hrule height 0pt
\includegraphics[width=\textwidth]{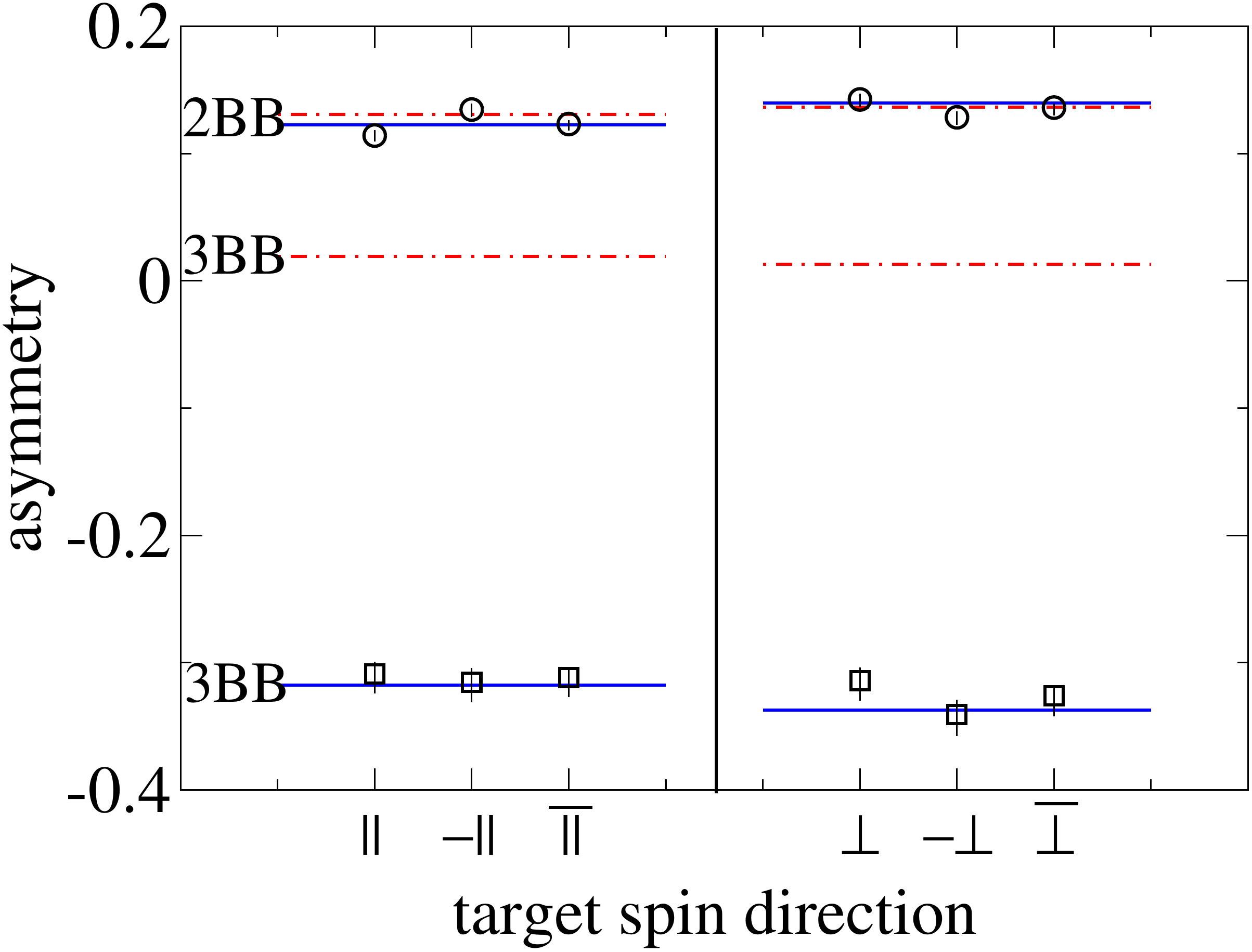}
\end{minipage}
\hfill
\begin{minipage}[t]{0.42\textwidth}
\hrule height 0pt
\caption{ 
Measured longitudinal $A_z$ and transverse $A_{x}$ asymmetries for 2BBU reaction channel
${}^3\vec{\mathrm{He}}\left(\vec{\mathrm{e}},\mathrm{e'} \mathrm{p}\right)\mathrm{d}$ and
3BBU reaction channel ${}^3\vec{\mathrm{He}}\left(\vec{\mathrm{e}},\mathrm{e'} 
\mathrm{p}\right)\mathrm{pn}$. Symbols $\parallel$, -$\parallel$, $\perp$ and 
-$\perp$ denote four target spin orientations: parallel to $\vec q$, anti-parallel, 
perpendicular and anti-perpendicular. 
The symbols $\overline{\parallel}$ and $\overline{\perp}$ represent 
the mean value of the asymmetry in each direction. Dash-dotted and full line represent PWIA 
and full theoretical predictions for 2BBU and 3BBU, respectively~\cite{achenbach2008}. 
\label{fig_MAINZeep}}
\end{minipage}
\end{center}
\end{figure}

Polarization degrees of freedom were successfully utilized in the Mainz 
experiment~\cite{achenbach2008} for studying meson-exchange currents and final-state 
interactions in the ${}^3\mathrm{He}$ breakup process. They measured 
beam-target asymmetries $A_{z}$ and $A_{x}$  in both ${}^3\vec{\mathrm{He}}\left(\vec{\mathrm{e}},
\mathrm{e'} \mathrm{p}\right)\mathrm{d}$ and 
${}^3\vec{\mathrm{He}}\left(\vec{\mathrm{e}},\mathrm{e'} \mathrm{p}\right)\mathrm{pn}$ reactions. 
The measurements were performed
at the top of the quasi-elastic peak at $Q^2 = 0.3\,(\mathrm{GeV}/c)^2$, with
$\omega = 135\,\mathrm{MeV}$ and $|\vec q| = 570\,\mathrm{MeV}/c$. According to Faddeev 
calculations for this kinematic conditions, MEC effects should be small. On the other 
hand, calculation predict very distinct roles of FSI in the 2bbu and 3bbu processes.
In the 2bbu channel, the PWIA asymmetry and the asymmetry 
including FSI are almost identical. However, in the 3bbu case, the PWIA asymmetry is 
almost exactly zero, implying the usual picture of a spin-singlet proton-proton pair
(see Fig.~\ref{fig_He3States}), while the asymmetry including FSI is large and 
negative~\cite{sircaWL2009}.  The measured results are shown in Fig.~\ref{fig_MAINZeep}
and they agree well with the computed values.

Mainz experiment unfortunately provided data only at low recoil momenta 
($p_r \leq 120\,\mathrm{MeV}/c$). They also performed no binning of the 
datum the in the $p_r$ variable. All their data were collected in one 
bin with the mean value of $p_r\approx 40\,\mathrm{MeV}/c$. 
Furthermore, they measured asymmetries only in ${}^3\vec{\mathrm{He}}\left(\vec{\mathrm{e}},
\mathrm{e'} \mathrm{p}\right)$ reaction, while the deuteron channel 
${}^3\vec{\mathrm{He}}\left(\vec{\mathrm{e}},\mathrm{e'} \mathrm{d}\right)$ remains 
unexplored. Hence, new measurements are required, similar to those 
from NIKHEF~\cite{passcier2002}, to obtain double-polarization asymmetries 
as a function of recoil momentum in all reaction channels, which could reveal 
the presence of $S'$- and $D$-state in the ${}^3\mathrm{He}$ ground-state wave-function.
The first attempt of such measurement was performed in NIKHEF~\cite{sixPhD}. However, the
statistical accuracy of those measurements was insufficient to resolve the role of $S'$ 
component at low recoil momenta. Therefore their results were never
published. This way E05-102 is the first experiment, where $S'$- and $D$-wave contributions 
to the ${}^3\mathrm{He}$ wave-function will be inspected in most direct manor. With 
these ground breaking measurements we will be able to confirm or reject theoretical 
predictions on spin and iso-spin structure of the nuclei and re-examine our understanding 
of the meson exchange currents and final state interactions.

\section{The E05-102 experiment}

\begin{figure}[!ht]
\begin{center}
\includegraphics[width=0.49\textwidth]{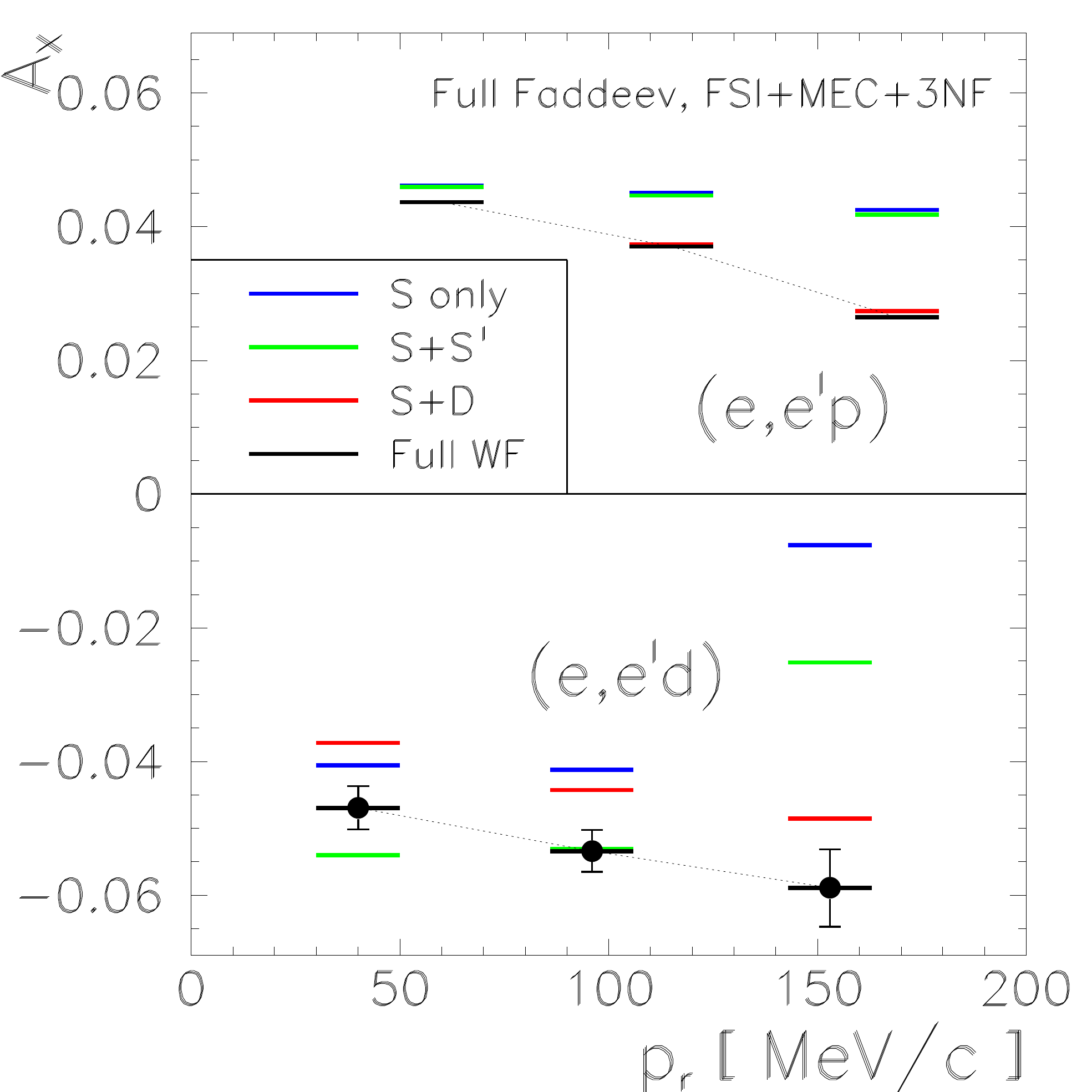}
\includegraphics[width=0.49\textwidth]{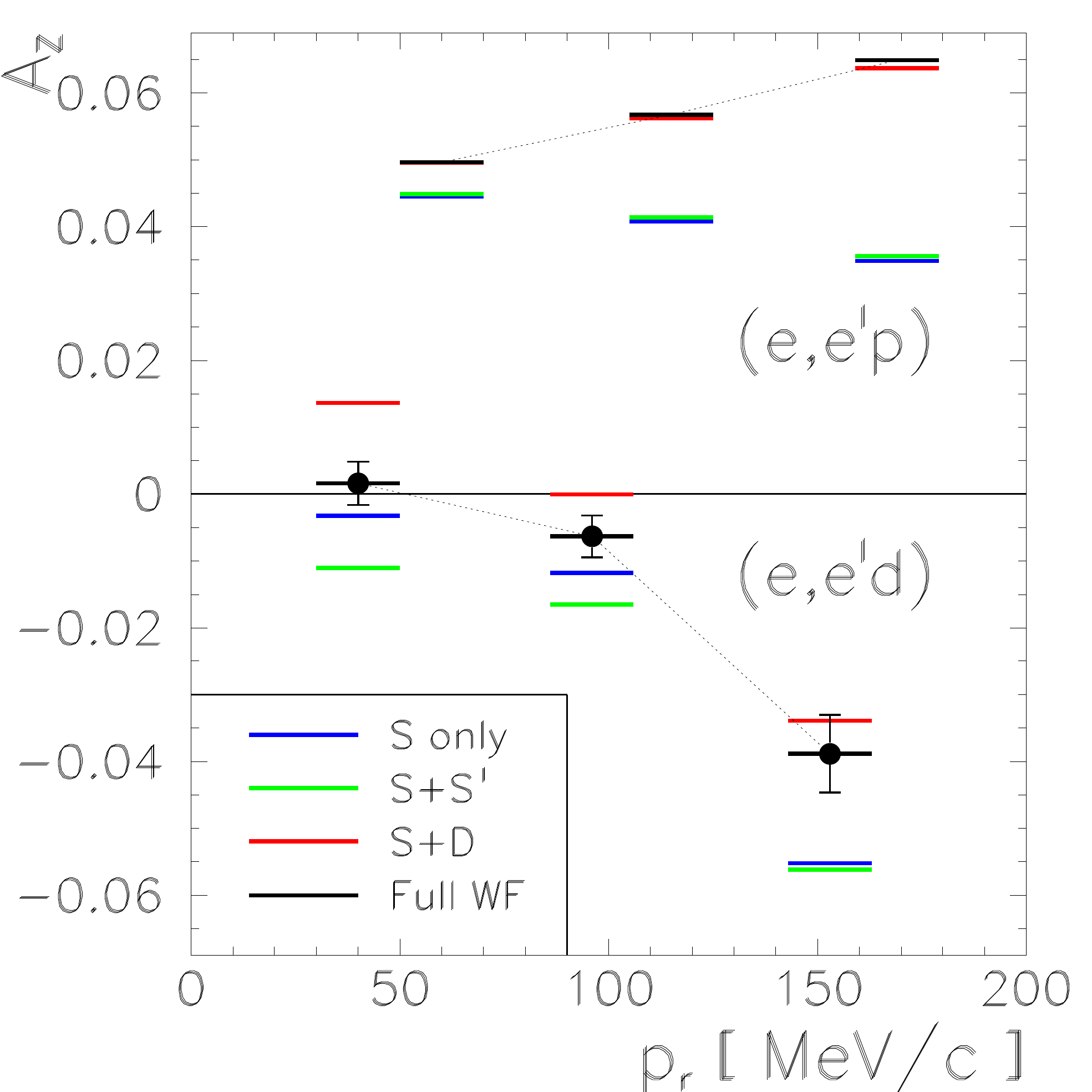}
\caption{ Asymmetries $Ax$ and $Az$ in processes 
${}^3\vec{\mathrm{He}}\left(\vec{\mathrm{e}},\mathrm{e'} \mathrm{d }\right)\mathrm{p}$ 
and ${}^3\vec{\mathrm{He}}\left(\vec{\mathrm{e}},\mathrm{e'} \mathrm{p}\right)\mathrm{d}$
for a beam energy of $2.4\,\mathrm{GeV}$ and $|\vec{q}| = 620\,\mathrm{MeV}/c$ predicted
by Golak~\cite{e05102}. Asymmetries depend on the $S'$ and $D$
components of the ${}^3\mathrm{He}$ ground-state wave function. The relative influence of each 
component changes with the recoil momentum $p_{r}$. Measurement of 
double-polarized asymmetries therefore represents a valuable technique for studying 
properties of ${}^3\mathrm{He}$.
\label{fig_eed_eep}}
\end{center}
\end{figure}

Experiment E05-102 is the only polarized ${}^3\mathrm{He}$ experiment carried out at
Jefferson Lab which is seeking to better understand the ${}^3\mathrm{He}$ system,
by measuring beam-target asymmetries $A_{x}$ and $A_{z}$ in reactions 
${}^3\vec{\mathrm{He}}\left(\vec{\mathrm{e}},\mathrm{e'} \mathrm{d }\right)$ and 
${}^3\vec{\mathrm{He}}\left(\vec{\mathrm{e}},\mathrm{e'} \mathrm{p}\right)$.
The asymmetries were measured for recoil momenta $p_{Miss}$ between 
$0$ and almost $300\,\mathrm{GeV}/c$. In this kinematics range $(e,e'd)$ channel is
predicted to be uniquely sensitive to the effects of both $S'$ and $D$ components 
of the ${}^3\mathrm{He}$ ground-state wave function. Fig.~\ref{fig_eed_eep} shows
the calculated values of the asymmetries, made for kinematic settings very similar
to those during the E05-102 experiment. An important role of the $S'$-state is 
evident at small recoil momenta, where asymmetry seems to be relatively 
flat and independent of the $p_{r}$. In this region a potential absence of the 
$S'$-state would cause almost a $30\,\mathrm{\%}$ change in the asymmetry. 
The contribution of the $D$-component becomes significant at larger 
$p_{r}>120\,\mathrm{MeV}/c$, with asymmetry starting to change dramatically.  

When looking at the double-polarized asymmetries in 
${}^3\vec{\mathrm{He}}\left(\vec{\mathrm{e}},\mathrm{e'} \mathrm{p}\right)$,
one has to consider that two reaction channels ${}^3\vec{\mathrm{He}}\left(
\vec{\mathrm{e}},\mathrm{e'} \mathrm{p}\right)\mathrm{d}$ and ${}^3\vec{\mathrm{He}}
\left(\vec{\mathrm{e}},\mathrm{e'} \mathrm{p}\right)\mathrm{pn}$ are possible.
The experimental setup of the experiment E05-102 allowed  simultaneous
measurement of both channels. Measured asymmetries are this way linear combinations 
of contributions from both channels. This has to be acknowledged
when data are compared to the theoretical calculations. The predicted asymmetries
for the two-body-breakup (2BBU) are presented in Fig.~\ref{fig_eed_eep}.  
It is believed that this channel is sensitive only to the $D$-state,
while the contributions of the $S'$ state are negligible. 
Here, asymmetry is also predicted to be reasonably flat for small recoil momenta
and starts decreasing when $p_{r}>120\,\mathrm{MeV}/c$. This behavior is 
again governed by the $D$-state.

\begin{figure}[!ht]
\begin{center}
\begin{minipage}[t]{0.36\textwidth}
\hrule height 0pt
\includegraphics[width=\textwidth]{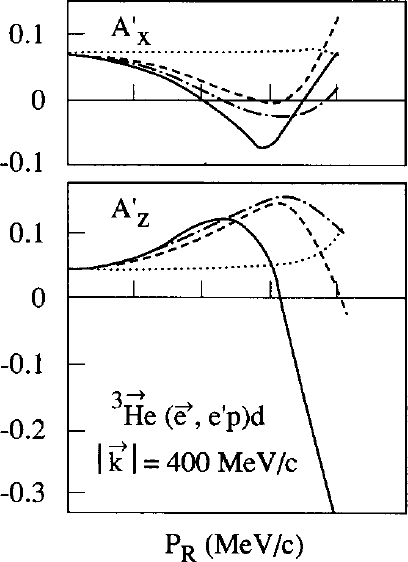}
\end{minipage}
\hspace*{1cm}
\begin{minipage}[t]{0.45\textwidth}
\hrule height 0pt
\caption{Asymmetries $A_{x}$ and $A_{z}$ in the reaction 
${}^3\vec{\mathrm{He}}\left( \vec{\mathrm{e}},\mathrm{e'} 
\mathrm{p}\right)\mathrm{d}$ as a function of recoil momentum $p_r$ 
for parallel kinematics ($\vec{p_r} \parallel \vec{q}$) at a beam energy 
of $880\,\mathrm{MeV}$. The dotted lines and dashed lines correspond to 
the PWIA when only S-wave and both S- and D-waves are respectively 
taken into account. The dash-dotted lines include FSI, while full lines 
include also MEC~\cite{laget92}. 
\label{fig_Laget1}}
\end{minipage}
\end{center}
\end{figure}

The properties of final state interactions and meson exchange currents
are also unmasked through the measurement of $A_{x}$ and $A_{z}$ asymmetries.
This was indicated already by Laget~\cite{laget92}. Within the framework
of plain-wave impulse approximation (PWIA) he demonstrated, that
inclusion of such effects can dramatically change the final value of the
asymmetries in $\mathrm{(e,e'p)d}$ and $\mathrm{(e,e'p)pn}$ reactions. 
See Fig.~\ref{fig_Laget1} for the details.
FSI and MEC contributions are different in each reaction channel,
and in general can not be neglected. This gives an excellent opportunity
to study FSI and MEC via double-polarized asymmetries on ${}3\mathrm{He}$
target.

\begin{figure}[!ht]
\begin{center}
\includegraphics[width=0.49\textwidth]{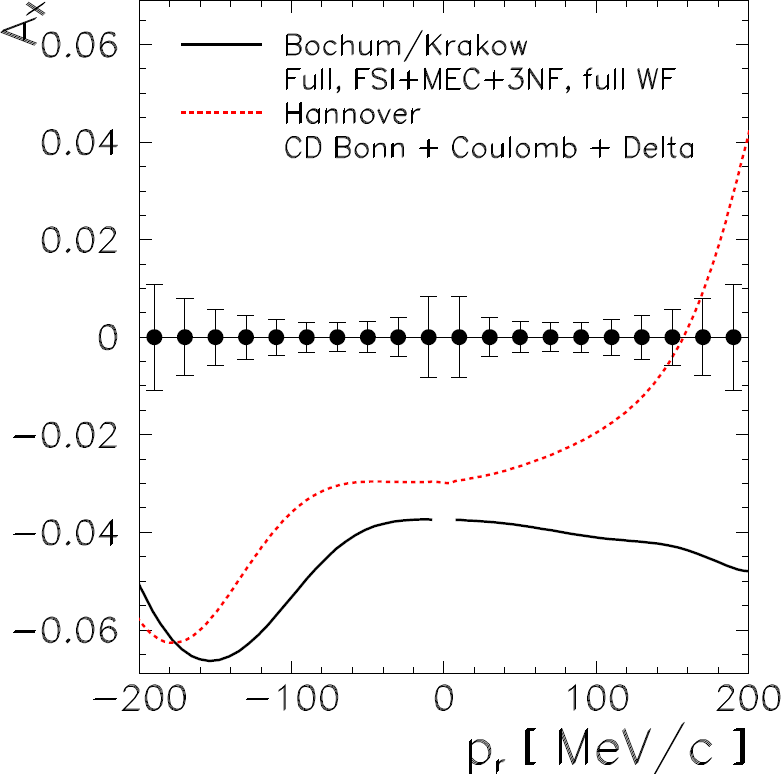}
\includegraphics[width=0.49\textwidth]{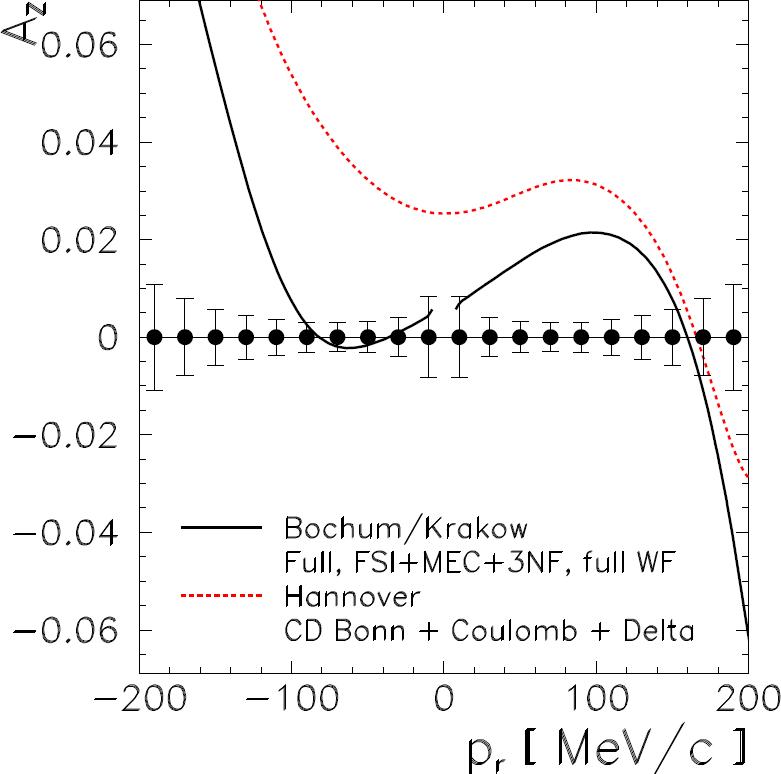}
\caption{Comparison of predicted double-polarized asymmetries $A_{x}$
and $A_{z}$ provided by Bochum/Krakow and Hannover group. A significant 
discrepancy between two calculations is observed for the longitudinal 
asymmetry $A_{z}$. Figure taken from~\cite{e05102}.
\label{fig_HannoverKrakowComparison}}
\end{center}
\end{figure}

The predictions for observables $A_{x}$ and $A_{z}$ based on the 
non-relativistic Faddeev calculations are provided by various 
theoretical groups. In the last few years, all groups  made major 
theoretical advances, especially in the treatment of the 
meson exchange currents and three nucleon force. Their 
calculations have been cross-checked in many instances~\cite{e05102}.
However, in some aspects there are still significant differences 
between them. Fig.~\ref{fig_HannoverKrakowComparison} shows
the comparison of the state-of-the-art predictions of Bochum/Krakow
and Hannover groups for ${}^3\vec{\mathrm{He}}\left(\vec{\mathrm{e}},
\mathrm{e'} \mathrm{d }\right)\mathrm{p}$. Differences in predictions 
are marked in both asymmetries, especially in the case of 
longitudinal asymmetry. The conclusive results from the direct measurements
of $A_{x}$ and $A_{z}$ will put these competing theoretical calculations to 
the test and help significantly in diminishing the differences between them.

\section{The layout of this thesis}

In this thesis I will study the spin-isospin structure of the polarized 
${}^3\mathrm{He}$, through the analysis of double-polarization asymmetries, 
measured in the experiment E05-102. The data for all three reaction channels
${}^3\vec{\mathrm{He}}\left(\vec{\mathrm{e}}, \mathrm{e'} \mathrm{d }\right)\mathrm{p}$,
${}^3\vec{\mathrm{He}}\left(\vec{\mathrm{e}},\mathrm{e'} \mathrm{p }\right)\mathrm{d}$ and
${}^3\vec{\mathrm{He}}\left(\vec{\mathrm{e}},
\mathrm{e'} \mathrm{p}\right)\mathrm{pn}$ will be inspected in order to get 
new insight into the properties of the ${}^3\mathrm{He}$. Within the limits of 
this work all open questions, discusses in this chapter, will not answered. However, 
the results obtained from these ground breaking measurements with contribute extensively to our 
knowledge on the structure of the ${}^3\mathrm{He}$ and accompanying reactions effects.
The high precision data, covering wide range  of recoil momenta (up to 
$\approx 300\,\mathrm{MeV}/c$) at different $Q^2$, will allow us to confirm or reject 
the theoretical predictions on the spin and iso-spin structure of the nuclei and 
check our understanding of the meson exchange currents and final state interactions. 
Without this knowledge all future experiments on ${}^3\mathrm{He}$ at low $Q^2$ will 
be seriously impaired.

The thesis will be divided in following sections. First, the underlying theoretical 
formalism will be briefly explained. Numerical
predictions provided by different theoretical groups will be presented and compared. 
Next, the apparatus utilized for the experiment E05-102 will be described, followed
by the chapter about the calibration of spectrometers and detectors. 
A special attention will be dedicated to the optical calibration of the BigBite spectrometer,
which I commit a lot of time to. After the calibration chapters, systematic and 
statistic uncertainties will be determined. This will be followed by the main chapter of
this thesis, where final experimental asymmetries will be presented and compared to the 
theoretical calculations. Pursuing chapter will be devoted to interpretation and 
discussion of obtained results. Finally, summary of the main findings will be presented, 
followed by the outlook for future work.

\chapter{General Formalism}

This chapter presents a brief overview of the theory of spin-dependent 
quasi-elastic electron scattering. First, kinematic variables used in the 
analysis of such reactions are defined. Then, the derivation
of the double-polarization asymmetries from spin-dependent cross-sections
is presented in the most general manner, followed by a more detailed formalism 
developed within the Born approximation. In the following section, an introduction 
to Faddeev calculations is made, which are considered for calculation
of the nuclear matrix elements. The predictions of a  relativistic gauge 
invariant approach of Nagorny are also presented.

\section{Electron scattering on ${}^3\vec{\mathrm{He}}$}

\begin{figure}[!ht]
\begin{center}
\begin{minipage}[t]{0.66\textwidth}
\hrule height 0pt
\includegraphics[width=1\textwidth]{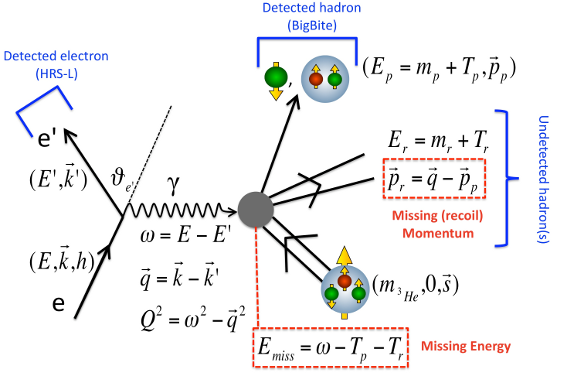}
\end{minipage}
\hfill
\begin{minipage}[t]{0.33\textwidth}
\hrule height 0pt
\caption{Kinematics of the reactions ${}^3\mathrm{He}(e,e'N)$ in the plane-wave Born approximation. 
In the experiment E05-102 a scattered electron and only one of the reaction
products were detected. The remaining hadrons were left undetected.
\label{fig_theory_Kinematics}}
\end{minipage}
\end{center}
\end{figure}

The kinematics of semi-exclusive quasi-elastic scattering process is 
presented in Fig.~\ref{fig_theory_Kinematics}. In such a reaction the incident
electron $e$ with energy $E$, momentum $\vec{k}$ and helicity $h$
interacts with the target ${}^3\mathrm{He}$ nucleus at rest. The 
${}^3\mathrm{He}$ nucleus has mass $m_{{}^3\mathrm{He}}$ and nuclear 
spin $\vec{s}$. The scattered electron $e'$ has the energy $E'$, 
momentum $\vec{k'}$ and scattering angle $\theta_{e'}$.
In this process, the electron gives up a part of its energy to the ${}^3\mathrm{He}$ nucleus
by emitting a virtual photon $\gamma$ with energy $\omega = E - E'$ and momentum 
transfer vector $\vec{q} = \vec{k} - \vec{k'}$. The $Q^2$ represents the square of the
momentum-transfer four-vector:
\begin{eqnarray}
  Q^2 = \left|q^{\mu}\right|^2 = \left|(\omega, \vec{q})\right|^2 = \omega^2 - \vec{q}^2\,. \nonumber 
\end{eqnarray}
In the reaction the ${}^3\mathrm{He}$ nucleus breaks 
into two or three reaction products. One of them (proton or deuteron) with mass $m_{p}$ 
is then detected by the hadron spectrometer. There, its identity together 
with its energy $E_{p}$ and momentum $\vec{p}_{p}$ are determined. The rest of the 
reaction products with the adjoined mass $m_r$  are left undetected. 
They are all together assigned a recoil momentum 
(or missing momentum) $\vec{p_r}$ and corresponding recoil energy $E_r$, defined as:
\begin{eqnarray}
\vec{p}_r = \vec{q} - \vec{p}_p\,,\qquad
E_{r} = \sqrt{m_{r}^2 + \left(\vec{p}_r\right)^2}\,, \nonumber
\end{eqnarray}
where in the calculation of the recoil energy, a two-body breakup is assumed. 
Additionally, missing energy is defined as:
\begin{eqnarray}
  E_{\mathrm{miss}} = \omega - T_{p} - T_{r}\,.\nonumber
\end{eqnarray}
The $T_p$ and $T_d$ are the kinetic energies of the detected hadron and undetected part, respectively,
and for particle momenta $\lesssim 1\,\mathrm{GeV}/c$ can be approximated as:
\begin{equation}
 T_{p} = \sqrt{m_p^2 + \vec{p}_{p}^2} - m_p \approx \frac{\vec{p}_{p}^2}{2m_p}\,,\qquad
 T_{r} = \sqrt{m_r^2 + \vec{p}_{r}^2} - m_r \approx \frac{\vec{p}_{r}^2}{2m_r}\,.\nonumber
\end{equation}
Commonly used are also the invariant mass $W$ and the Bjorken scaling variable $x$, which are defined
as:
\begin{eqnarray}
  W^2 = m_{{}^3\mathrm{He}}^2 + 2m_{{}^3\mathrm{He}}\omega - Q^2\,, \qquad x = \frac{Q^2}{2m_{{}^3\mathrm{He}}\omega}\,. \nonumber
\end{eqnarray}
For the elastic scattering $W = m_{{}^3\mathrm{He}}$ and $x=1$, while for the inelastic scattering
$W>m_{{}^3\mathrm{He}}$ and $0 < x < 1$. For a typical kinematical setting during the
E05-102 experiment, $Q^2 = 0.25\,\left(\mathrm{GeV}/c\right)^2$ and $\omega \approx 0.1\,\mathrm{GeV}$
were obtained, resulting in $W \approx 2.86\,\mathrm{GeV}$ and $x \approx 0.45$.

\begin{figure}[!ht]
\begin{center}
\begin{minipage}[t]{0.6\textwidth}
\hrule height 0pt
\includegraphics[width=1\textwidth]{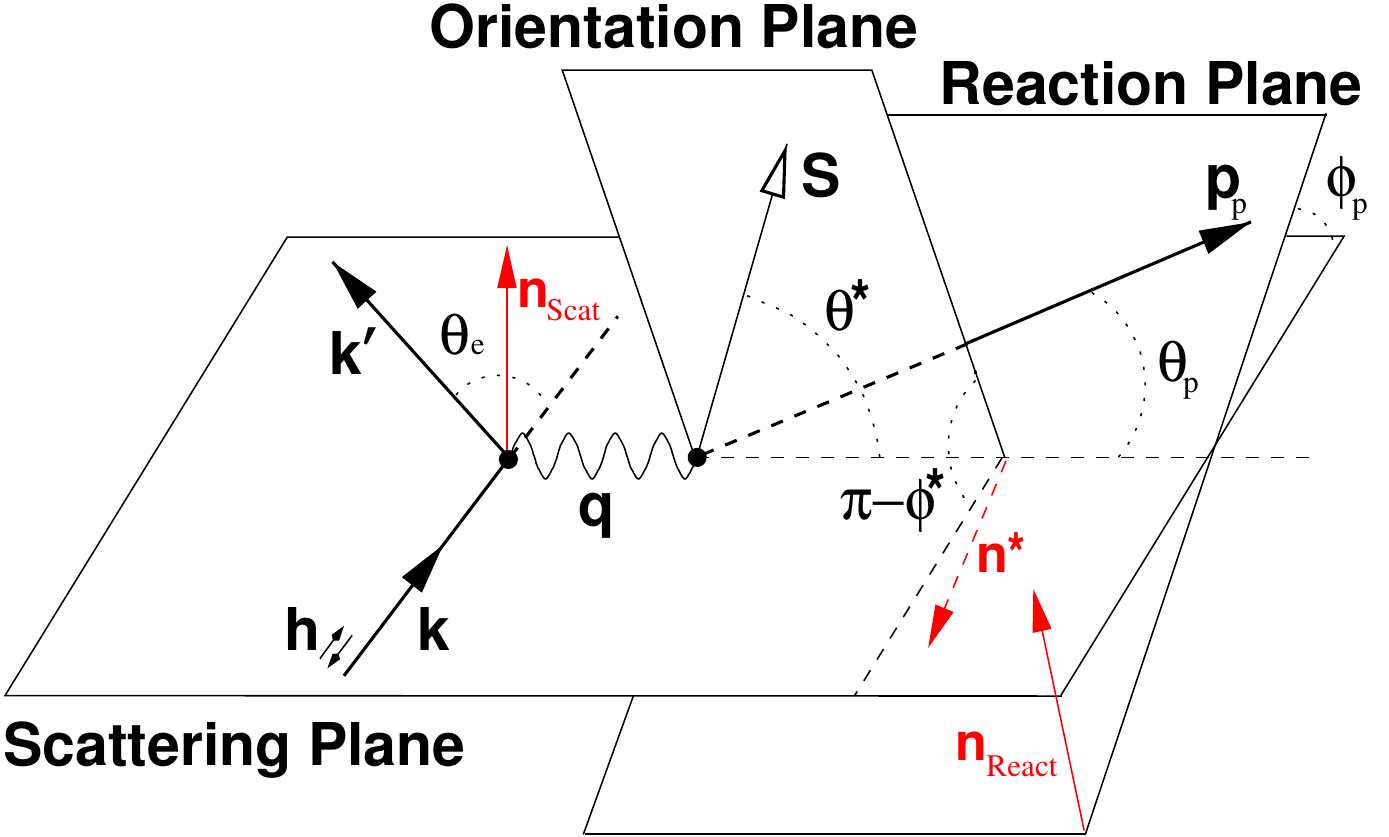}
\end{minipage}
\hfill
\begin{minipage}[t]{0.37\textwidth}
\hrule height 0pt
\caption{Coordinate system used to describe the scattering of polarized electrons from a vector-polarized
nuclear target. 
Figure from after Ref.~\cite{sixPhD}.
\label{fig_theory_planes}}
\end{minipage}
\end{center}
\end{figure}

The kinematic variables in the scattering of polarized electrons from a polarized nuclear target,
 shown in  Fig.~\ref{fig_theory_Kinematics}, define three distinct planes. The momenta of the 
incident and scattered electrons define the scattering plane, as demonstrated in 
Fig.~\ref{fig_theory_planes}. By definition the vector $\vec{q}$ also lies within this plane. The reaction
plane is determined by the momentum  of the detected hadron and the vector $\vec{q}$, while the
orientation plane is defined by the vector $\vec q$ and the target spin orientation vector $\vec{S}$.
To each of these three planes, an unit vector perpendicular to the plane (normal vector) can be assigned.
They can be calculated as follows:
\begin{eqnarray}
\mathrm{Scattering\>plane:}\qquad \hat{n}_{\mathrm{Scat}} &=& \frac{\vec{k} \times \vec{k'}}{|\vec{k}|\> |\vec{k'}|} = 
                      \frac{\vec{q} \times \vec{k}}{|\vec{q}|\> |\vec{k}|}\,,\nonumber \\
\mathrm{Reaction\>plane:}\qquad \hat{n}_{\mathrm{React}} &=& \frac{\vec{q} \times \vec{p}_p}{|\vec{q}|\> |\vec{p}_p|}\,, 
\label{eq_theory_normalvec}\\
\mathrm{Orientation\>plane:}\qquad \hat{n}^{*} &=& \frac{\vec{q} \times \vec{S}}{|\vec{q}|\> |\vec{S}|}\,. \nonumber
\end{eqnarray}

Certain observables measured in such scattering experiments depend also on two pairs of angles 
$(\theta^{*}, \phi^{*})$ and $(\theta_{p}, \phi_{p})$. The polar angle $\theta^{*}$ is the 
angle between the spin vector $\vec{S}$ and the vector $\vec{q}$, while $\theta_{p}$ is the 
angle between the vectors $\vec{p}_{p}$ and $\vec{q}$: 
\begin{eqnarray}
  \theta^{*} = \arccos\left(\frac{\vec{q}\cdot\vec{S}}{|\vec{q}|\> |\vec{S}|}\right)\,,\qquad 
  \theta_{p} = \arccos\left(\frac{\vec{q}\cdot\vec{p}_{p}}{|\vec{q}|\> |\vec{p}_p|}\right)\,. \nonumber
\end{eqnarray}
The  angles $\phi^{*}$ and $\phi_{p}$ can be calculated by using normal vectors 
defined by Eqs.~(\ref{eq_theory_normalvec}):
\begin{eqnarray}
  \phi^{*} = \arccos\left(\hat{n}_{\mathrm{Scat}}\cdot \hat{n}^{*}\right)\,,\qquad 
  \phi_{p} = \arccos\left(\hat{n}_{\mathrm{Scat}}\cdot \hat{n}_{\mathrm{React}}\right)\,. \nonumber
\end{eqnarray}
As demonstrated in Fig.~\ref{fig_theory_planes}, $\phi^{*}$ represents the angle between the scattering 
plane and the orientation plane, while $\phi_{\mathrm{React}}$ corresponds to the angle between the 
scattering plane and the reaction plane.

\section{Spin dependent cross-section and the asymmetries }
The cross-section for  semi-exclusive quasi-elastic reactions $(e,e'N)$, where a longitudinally 
polarized electron beam is used in conjunction with a polarized target of spin $S=1/2$, 
has the following form~\cite{laget92}:
\begin{eqnarray}
  \frac{d^6\sigma(h,\vec{S})}{d\Omega_e\,dE'\,d\Omega_p\,dp_p} = \frac{d^6\sigma_0(h,\vec{S})}{d\Omega_e\,dE'\,d\Omega_p\,dp_p}
  \left[ 1 + \vec{S}\cdot\vec{A^0} + h\left(A_e + \vec{S}\cdot\vec{A'}\right) \right]\,. \label{eq_theory_gcs}
\end{eqnarray} 
Here, $h$ represents the helicity of the incident electron, $\vec S$ is the spin of the target and $\sigma_0$ the 
unpolarized cross-section. The $\vec{A^0}$ and $A_e$ indicate asymmetries generated by the polarization of only the 
target or only the beam. On the other hand $\vec{A}'$ is the asymmetry when both beam and  target are polarized. 
The target spin and the asymmetry vector, given in Eq.~(\ref{eq_theory_gcs}) are defined in the reference frame in which
the quantization axis $z$ lies in the direction of the momentum transfer $\vec{q}$, the $x$-axis is perpendicular
to it and lies in the reaction plane, while the $y$-axis is normal to it.

We will focus on the double-polarization asymmetry $\vec{A}'$. In coplanar geometry, where the spin lies in 
the scattering plane, only $A_x'$ and $A_z'$ components survive. They can be determined through the
measurement of the cross-section ratios:
\begin{eqnarray}
	A_{x,z}' = \frac{\left(d\sigma_{++}+d\sigma_{- -}\right) - \left(d\sigma_{+ -}+d\sigma_{- -}\right)}
                       {\left(d\sigma_{++}+d\sigma_{- -}\right) - \left(d\sigma_{+ -}+d\sigma_{- -}\right)}\,.\label{eq_theory_doubleasymm}
\end{eqnarray}
For the measurement of the asymmetry $A_z'$ the target spin must be oriented along the $z$-axis, while for the 
extraction of $A_x'$ the target spin should be pointing along the $x$-axis. The $(\pm,\pm)$ signs represent 
the beam helicities and the projections (parallel, anti-parallel) of the target spin along the 
quantization axis. Using this four-fold spin flip sequence gives us a direct insight into the double-polarization 
asymmetries, since the contributions of the single spin asymmetries $\vec{A}^0$ and $A_e$ cancel.

However, the measurement of the cross-sections for all four orientations of the two spins, as presented in 
Eq.~(\ref{eq_theory_doubleasymm}), is not obligatory in order to extract $A_x'$ and $A_z'$. Under certain
conditions, it is enough to measure only the beam-helicity asymmetry with a fixed target orientation 
to determine $A_x$ and $A_z$. 

When the emitted nucleon lies in the electron scattering plane, it was 
determined~\cite{laget92} that $A_x^0 = 0$, $A_z^0 = 0$ and $A_y' = 0$. The component 
$A_y^0 \neq 0$ due to the effects of the FSI and MEC. However, since the target is not being 
polarized in the $y$-direction, the term $\vec{S}\cdot \vec{A}^0=0$. 
The parity violating asymmetry $A_e$ is also expected~\cite{zafar} to be much smaller than 
the asymmetries $A_x'$ and $A_z'$. 
Consequently, the contribution of the $A_e$ in Eq.~(\ref{eq_theory_gcs}) can be neglected,
and only the  double-polarization term remains. 
Hence, the asymmetries $A_x'$ and $A_z'$ can be extracted from the measurements where only beam 
helicity is flipped, while the target spin orientation remains fixed in any of 
two principal directions (along or perpendicular to $\vec{q}$):
\begin{eqnarray}
  A_{x,z}'(+S_{x,z}) = \frac{d\sigma_{++} - d\sigma_{-+}}{d\sigma_{++}+d\sigma_{-+}} \,,\qquad 
  A_{x,z}'(-S_{x,z}) = \frac{d\sigma_{+-} - d\sigma_{--}}{d\sigma_{+-}+d\sigma_{--}}\,, \label{eq_theory_asymmetry}
\end{eqnarray}	
where $(+S_{x,z})$ and $(-S_{x,z})$ denote the parallel and anti-parallel orientation of the target
spin with respect to the quantization axis. Considering also the identities $(d\sigma_{++}=d\sigma_{--})$
and $(d\sigma_{+-} = d\sigma_{-+})$, the following relation is obtained:
\begin{eqnarray}
A_{x,z}'(+S_{x,z}) = - A_{x,z}'(-S_{x,z})\,,\label{eq_theory_SpinFlipAsymmetry}
\end{eqnarray} 
According to this equation, the asymmetries ${A'}_{x,z}$, measured at two opposite orientation of the
target spin, should differ only in sign. This can be exploited to search for false asymmetries.

\section{Born approximation}
Quasi-elastic scattering of electrons on ${}^3\mathrm{He}$ nuclei at intermediate to high energies,
can be adequately described by the plane-wave Born approximation (PWBA), where a single photon is 
exchanged between the electron and the nucleus, and the electrons are treated as plane-waves.
Within this framework, the differential cross-section in the laboratory frame for the two-body 
breakup channels ${}^3\mathrm{He}(e,e'd)p$ and ${}^3\mathrm{He}(e,e'p)d$ can be written 
as~\cite{donnelly86, donnelly89}:
\begin{eqnarray}
  d\sigma = \frac{m_e }{E |\vec{v}_e|}\left| \mathcal{M}_{fi}\right|^2 \left[\frac{m_e}{E'}\frac{d^3\vec{k}'}{(2\pi)^3} 
            \frac{m_p}{E_p}\frac{d^3\vec{p}_p}{(2\pi)^3} \frac{m_r}{E_r}\frac{d^3\vec{p}_r}{(2\pi)^3} \right]
             (2\pi)^4\delta^{(4)}\left(k+p_{{}^3\mathrm{He}}-k'-p_p-p_{r}\right),\,{} \label{eq_theory_PWBAcs}
\end{eqnarray}
where $|\vec{v}_e| = |\vec{p}_e|/E$ is the velocity of the incident electron, $\left| \mathcal{M}_{fi}\right|^2$ is
the complex square of the invariant matrix element for the process under consideration, and the $\delta^{(4)}$ function
represents the overall momentum and energy conservation. The terms within the square brackets are the phase
spaces of all outgoing particles. In the considered semi-exclusive reactions only the electron and one of the
hadrons is detected. Since the recoiling particle (see Fig.~\ref{fig_theory_Kinematics}) is not detected,  
the integration of Eq.~(\ref{eq_theory_PWBAcs}) over the recoil momentum $\vec{p_r}$ is performed. Additionally,
the integration over the $|\vec{p}_p|$ is then carried out, which introduces a recoil factor $f_{\mathrm{rec}}$
into $d\sigma$, resulting in the following expression for the cross-section~\cite{donnelly89,sircaPhD}:
\begin{eqnarray}
  \frac{d^5 \sigma}{dE'\,d\Omega_{e'}\,d\Omega_{p}} = \frac{m_e^2\,m_{p}\,m_{r}}{(2\pi)^5}\frac{|\vec{p}_p|}{m_{{}^3\mathrm{He}}} 
    \frac{1}{f_{\mathrm{rec}}} \left| \mathcal{M}_{fi}\right|^2\,, \label{eq_theory_PWBAcs2}
\end{eqnarray}
where
\begin{eqnarray}
f_{\mathrm{rec}} = \left| 1 + \frac{\omega |\vec{p}_{p}| - E_p|\vec{q}|\cos\theta_{p}}{M_{{}^3\mathrm{He}}\,|\vec{p}_{p}|} \right| \nonumber
\end{eqnarray}
and $\theta_{p}$ is the angle between $\vec{p}_p$ and $\vec{q}$ as demonstrated in Fig.~\ref{fig_theory_planes}. Spherical
angles $\Omega_{e'}$ and $\Omega_{p}$ of both detected particles were introduced into Eq.~(\ref{eq_theory_PWBAcs2}),
by using $d^3\vec{p} = |\vec{p}|\,E\,dE\,d\Omega$.

The Lorentz invariant matrix element $\mathcal{M}_{fi}$, which describes the interaction of the electron with the nucleus, can 
be written~\cite{donnelly89, halzen-martin} as a product of the electron electro-magnetic current, 
the photon propagator and the hadron electro-magnetic current:
\begin{eqnarray}
 \mathcal{M}_{fi} = -e^2 j_e^{\mu}\left[ \frac{-g_{\mu\nu}}{q^2}\right] J^{\nu}(q)\,.\label{eq_theory_me}
\end{eqnarray}
Two tensors are then usually defined, the leptonic and the hadronic tensor, both depending on the corresponding 
electro-magnetic currents. The leptonic tensor $L^{\mu\nu}$ describes the electron part of the investigated
process. Because the electron is a point Dirac particle, the tensor can be exactly expressed as~\cite{donnelly86, donnelly89}:
\begin{eqnarray}
	L^{\mu\nu} = j_e^{\mu} {j_e^{\nu}}^* = 
\left[\overline{u}_e(k',s')\gamma^\mu u_e(k,s) \right] \left[\overline{u}_e(k',s')\gamma^\nu u_e(k,s) \right]^*\,.\nonumber
\end{eqnarray} 
Here $u_e(k,s)$ represents the standard Dirac spinor for an electron with four-momentum $k$ and spin $s$. Considering an experiment
where only the incident electron is polarized with helicity $h$, the leptonic tensor in the extreme
relativistic limit ($E,E' \gg m_e$) becomes~\cite{donnelly89}:
\begin{eqnarray}
L^{\mu \nu} = \frac{1}{4m_e^2}\left(2\left(k^\mu {k'}^\nu - {k'}^\mu k^\nu \right) + g^{\mu\nu}Q^2 - 
              2hi\epsilon^{\mu \nu\alpha\beta} k_\alpha {k'}_{\beta} \right)\,, \label{eq_theory_leptensor}
\end{eqnarray}
where $\epsilon^{\mu \nu\alpha\beta}$ represents the four dimensional Levi-Civita symbol. 
Similarly, the hadronic tensor $W^{\mu\nu}$ describes the hadron part of the interaction. It contains all information 
on the nuclear structure and dynamics, and is defined as:
\begin{eqnarray} 
W^{\mu \nu} = J^{\mu}(q){J^\nu}^{*}(q)\,.\label{eq_theory_hadrtensor}
\end{eqnarray} 
Consequently, the square of the invariant matrix element can be written as a contraction of leptonic and 
hadronic tensors. Using Eqs.~(\ref{eq_theory_leptensor}) and (\ref{eq_theory_hadrtensor}) one gets:
\begin{eqnarray}
\left| \mathcal{M}_{fi}\right|^2 = \frac{e^4}{Q^4} L^{\mu\nu}W_{\mu \nu} = \frac{e^4}{Q^4} \frac{1}{4m_e^2} \left[
\left| K_\mu J^\mu \right|^2 + Q^2 {J^\mu}^{*}J_\mu 
- 2ih\epsilon_{\mu\nu\alpha\beta}k^{\alpha}{k'}^{\beta}{J^\mu}^*J^{\nu} \right]\,, \label{eq_theory_SME}
\end{eqnarray}
where $K_\mu = k_\mu + {k'}_\mu$. Considering the conservation of the hadronic transition
current $\partial_\mu J^\mu = q_\mu(\rho, J_x, J_y, J_z)^\mu =0$, Eq.~(\ref{eq_theory_SME}) can be further
expressed as:
\begin{eqnarray}
  \left| \mathcal{M}_{fi}\right|^2 = \frac{e^4}{Q^4} \frac{1}{4m_e^2} v_0\left[\left( v_L R_{fi}^L +v_T R_{fi}^T + 
	v_{TT} R_{fi}^{TT} + v_{TL} R_{fi}^{TL} \right) + h \left( v_{T'} R_{fi}^{T'} + v_{TL'} R_{fi}^{TL'} \right)
 \right]\,, \label{eq_theory_SME2}
\end{eqnarray}
where $v_\kappa$ are  kinematic factors, which in the laboratory system have the forms:
\begin{eqnarray}
 	v_L &=& \left( \frac{Q^2}{q^2} \right)^2\,, \nonumber\\
        v_T &=&  -\frac{1}{2}\frac{Q^2}{q^2} + \tan^2 \frac{\theta_e}{2}\,, \nonumber\\
        v_{TT} &=& \frac{1}{2}\frac{Q^2}{q^2}\,,\nonumber\\
        v_{TL} &=& \frac{1}{\sqrt{2}} \frac{Q^2}{q^2} \sqrt{-\frac{Q^2}{q^2} + 
               \tan^2\frac{\theta_e}{2}}\,, \label{eq_theory_kinfac}\\
        v_{T'} &=& \sqrt{-\frac{Q^2}{q^2} + \tan^2\frac{\theta_e}{2} }\tan \frac{\theta_e}{2}\,, \nonumber\\
        v_{TL'} &=& \frac{1}{\sqrt{2}}\frac{Q^2}{q^2}\tan\frac{\theta_e}{2}\,.\nonumber
\end{eqnarray}
Note that these factors depend only on the electron kinematics.
The functions $R_{fi}^\kappa$ are the nuclear response functions~\cite{sixPhD,donnelly86} and store the physical content 
of the process under consideration.  They can be directly expressed in terms of the nuclear magnetic 
charge $\rho(\vec{q})$ and the transverse currents $J(\vec{q}; \pm 1) = \mp \left(J_{x}(\vec q) \pm iJ_{y}(\vec{q})\right)/\sqrt{2}$:
\begin{eqnarray}
 	R_{fi}^L &=& \left| \rho(\vec q) \right|^2\,,\nonumber\\
        R_{fi}^T &=& \left| J(\vec q; 1)\right|^2 + \left| J(\vec q; -1)\right|^2\,,\nonumber\\
        R_{fi}^{TT} &=& 2Re\left[ J^*(\vec q; 1)J(\vec q; -1)\right]\,,\nonumber\\
        R_{fi}^{TL} &=& -2Re\left[ \rho^*(\vec q)\left( J(\vec q; 1) - J(\vec q; -1) \right) \right]\,,\nonumber\\
        R_{fi}^{T'} &=& \left| J(\vec q; 1)\right|^2 - \left| J(\vec q; -1)\right|^2\,, \nonumber\\
        R_{fi}^{TL'} &=& -2Re\left[ \rho^*(\vec q)\left( J(\vec q; 1) + J(\vec q; -1) \right) \right]\,. \label{eq_theory_responses}
\end{eqnarray} 
By introducing Eq.~(\ref{eq_theory_SME2}) to  Eq.~(\ref{eq_theory_PWBAcs2}), the cross-section for two-body breakup of 
${}^3\mathrm{He}$ can be written in terms of the response functions as:
\begin{eqnarray}
&&\frac{d^5 \sigma}{dE'\,d\Omega_{e'}\,d\Omega_{p}} =\frac{m_{p}\,m_{r}}{(2\pi)^3}\frac{|\vec{p}_p|}{m_{{}^3\mathrm{He}}} 
    \frac{1}{f_{\mathrm{rec}}} \left[\frac{\alpha \cos\frac{\theta_e}{2}}{2E\sin^2\frac{\theta_e}{2}}\right]^2 \nonumber\\
         &&\times \left\{\left( v_L R_{fi}^L +v_T R_{fi}^T + 
	v_{TT} R_{fi}^{TT} + v_{TL} R_{fi}^{TL} \right)
  + h \left( v_{T'} R_{fi}^{T'} + v_{TL'} R_{fi}^{TL'} \right) \right\}\,, \label{eq_theory_finalcs}
\end{eqnarray}
where we have considered that in the extreme relativistic limit $Q^2 = 4E\,E'\sin^{2}\frac{\theta_e}{2}$ and 
$v_0 = 4E\,E'\cos^{2}\frac{\theta_e}{2}$.  The term in the square brackets represents the Mott cross-section, and
$\alpha$ is the fine-structure constant. This cross-section can now be used to determine the experimentally
interesting asymmetry for the two-body (pd) breakup of ${}^3\mathrm{He}$. By inserting Eq.~(\ref{eq_theory_finalcs}) 
into Eq.~(\ref{eq_theory_asymmetry}) one gets 
\begin{eqnarray}
A_{\mathrm{pd}} = \frac{v_{T'}R_{T'} + v_{TL'}R_{TL'}}{v_{L}R_{L}+v_{T}R_{T} + v_{TT}R_{TT} + v_{TL}R_{TL}}\,.\label{eq_theory_asymmetry_pd}
\end{eqnarray}

If the quantization axis of ${}^3\vec{\mathrm{He}}$ is not along $\vec{q}$ but in 
the direction given by the angles $(\theta^*, \phi^*)$, then the ${}^3\mathrm{He}$ state can be written as:
\begin{eqnarray}
	 |\Psi_{{}^3\mathrm{He}}(m,\theta^*,\phi^*)\rangle = 
             \sum_{m'}D_{m'm}^{(1/2)}(\phi^*, \theta^*, 0)|\Psi_{{}^3\mathrm{He}}(m')\rangle\,, \nonumber
\end{eqnarray}
where $|\Psi_{{}^3\mathrm{He}}(m')\rangle$ is quantized along $\vec{q}$, and $D_{m'm}^{(1/2)}(\phi^*, \theta^*, 0)$
are the Wigner D-matrices~\cite{edmonds}. Considering this in the calculation of the matrix elements for the 
nuclear transition currents, one obtains an explicit $(\theta^*, \phi^*)$ dependence  of the following structure
functions~\cite{ishikawa98}:
\begin{eqnarray}
	R_{fi}^{TL} &=& \tilde{R}_{fi}^{TL}\sin{\theta^*}\sin{\phi^*}\,,\nonumber\\
	R_{fi}^{T'} &=& \tilde{R}_{fi}^{T'}\cos{\theta^*}\,,\label{eq_theory_reduced_responses}\\
	R_{fi}^{TL'} &=& \tilde{R}_{fi}^{TL'}\sin{\theta^*}\cos{\phi^*}\,,\nonumber
\end{eqnarray}  
where $\tilde{R}_{fi}^{TL}$, $\tilde{R}_{fi}^{T'}$ and $\tilde{R}_{fi}^{TL'}$ represent the reduced nuclear 
structure functions, which no longer depend on the target orientation. 
The nuclear structure functions $R_{fi}^{T}$, $R_{fi}^{L}$ and $R_{fi}^{TT}$ remain independent of the $(\theta^*, \phi^*)$.
Considering Eqs.~(\ref{eq_theory_reduced_responses}) in Eq.~(\ref{eq_theory_asymmetry_pd}), the asymmetry for 
the two-body breakup can be expressed as:
\begin{eqnarray}
  A_{\mathrm{pd}}(\theta^*, \phi^*) = \frac{v_{T'}\tilde{R}_{fi}^{T'}\cos{\theta^*} + v_{TL'}\tilde{R}_{fi}^{TL'}\sin{\theta^*}\cos{\phi^*}}
{v_{L}R_{L}+v_{T}R_{T} + v_{TT}R_{TT} + v_{TL}\tilde{R}_{fi}^{TL}\sin{\theta^*}\sin{\phi^*}}\,.\label{eq_theory_finalasymmetry_pd}
\end{eqnarray}

An analogous approach can be utilized to obtain the asymmetry for the three-body breakup of ${}^3\mathrm{He}$, where
the initial nucleus fragments into two protons and a neutron. In this case two reaction products remain undetected. Consequently, 
an additional integration over the direction of the relative momentum of the two undetected nucleons $\hat{p}_\mathrm{pn}$ 
is required in the expression for the asymmetry~\cite{golak2005}:
\begin{eqnarray}
 A_{ppn} = \frac{\int d\hat{p}_\mathrm{pn}\,\left(v_{T'}R_{T'} + v_{TL'}R_{TL'}\right)}
{\int d\hat{p}_\mathrm{pn}\,\left(v_{L}R_{L}+v_{T}R_{T} + v_{TT}R_{TT} + v_{TL}R_{TL}\right)}\,. \label{eq_theory_finalasymmetry_ppn}
\end{eqnarray}
In order to predict the behavior of the asymmetries $A_{pd}$ and $A_{ppn}$, the response functions given by 
Eqs.~(\ref{eq_theory_responses}) must be known. Hence, the transition nuclear currents $J_\mu$ must be determined, 
which describe the transition of the hadronic system from the initial state (${}^3\mathrm{He}$) to the final hadronic 
states (pd, ppn), after the interaction with the virtual photon.  There are various approaches for obtaining these 
currents~\cite{nagorny96, nagorny98, golak2005}. The predictions made for the E05-102 experiment are based on the Faddeev 
calculations~\cite{golak2005}. The calculations were performed for  both longitudinal and transverse spin orientations, 
providing us with predictions for both $A_x$ and $A_z$ asymmetries.

Expression (\ref{eq_theory_finalasymmetry_pd}) also shows, that, in general, asymmetry relation (\ref{eq_theory_SpinFlipAsymmetry}) holds
only in the approximation, where target spin vector lays within the scattering plane $(\phi^* = 0^\circ,180^\circ)$. In this case,
$\theta^*$ dependence in the denominator disappears and asymmetry changes sign, when spin is flipped for $180^\circ$.
However, the theory~\cite{golak2005} predicts, that response functions $R_{TT}$ and $\tilde{R}_{TL}$ are at least 
an order of magnitude smaller than the longitudinal and transverse response functions $R_{L}$ and $R_{T}$.
Considering also the experimental conditions of the E05-102 experiment, where $\theta^*\approx 70^\circ, 160^\circ$, 
the term $\sin{\theta^*}\sin{\phi^*}$ reduces their influence even further, resulting in a few percent correction 
to the denominator. Since such small corrections could not be recognized with a given statistical accuracy of the 
experiment, contributions of $R_{TT}$ and $\tilde{R}_{TL}$ will be neglected and we will pretend, that 
relation (\ref{eq_theory_SpinFlipAsymmetry}) holds for all $\phi^*$.

\section{Faddeev equations}

The most essential part in the description of the photon-induced breakup of ${}^3\mathrm{He}$ are the nuclear
nuclear transition currents~\cite{glockle2004, golak2005}:
\begin{eqnarray}
	J^\mu = \langle\Psi_f|\hat{O}^\mu|\Psi_{{}^3\mathrm{He}}(\theta^*, \phi^*)\rangle\,, \label{eq_theory_ME}
\end{eqnarray} 
where $|\Psi_f\rangle$ is final three-nucleon scattering state, $\hat{O}^\mu$ is the photon absorption 
operator, and $|\Psi_{{}^3\mathrm{He}}\rangle$ is the initial polarized ${}^3\mathrm{He}$ target state.
For a two-body breakup, the final state is combined of a proton and a deuteron (pd), while for the 
three-body breakup, the final state consists of three unbound nucleons (ppn). The initial ${}^3\mathrm{He}$ spin 
direction is  determined by the angles $\theta^*$ and $\phi^*$.

\begin{figure}[!ht]
\begin{center}
\begin{minipage}[t]{0.6\textwidth}
\hrule height 0pt
\includegraphics[width=\textwidth]{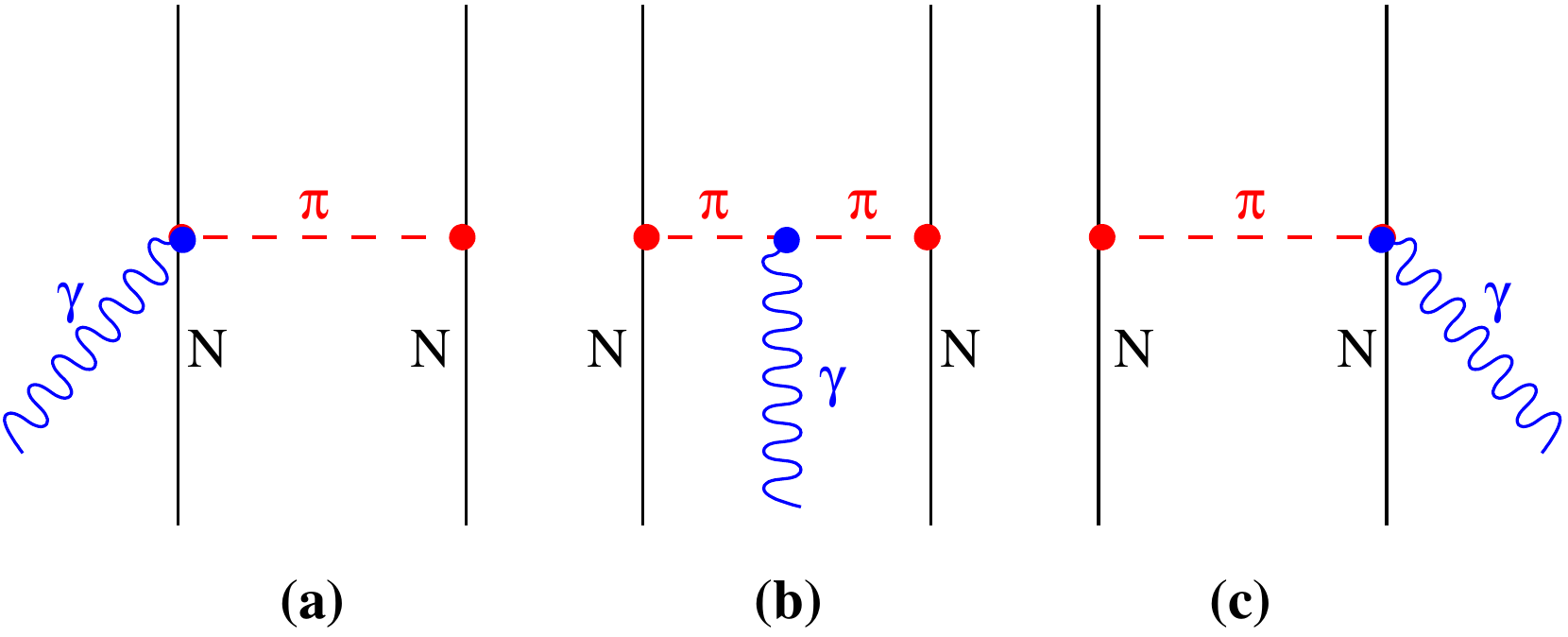}
\end{minipage}
\hfill
\begin{minipage}[t]{0.36\textwidth}
\hrule height 0pt
\caption{ Feynman diagrams for the two-body currents associated with 
pion exchange. Figure adopted from Ref.~\cite{carlson}.
\label{fig_theory_MEC}}
\end{minipage}
\end{center}
\end{figure}

The virtual photon can interact individually with any of the three nucleons. The interaction with the i-th nucleon
is described by a standard single-nucleon electro-magnetic current operator 
$\hat{J}_{\mathrm{SN}(i)}^\mu = (J_{\mathrm{SN}(i)}^0,\vec{J}_{\mathrm{SN}(i)})$, which can be written 
as~\cite{golak2005, halzen-martin}:
\begin{eqnarray}
   J_{\mathrm{SN}(i)}^0 &=& \left[F_1 + \frac{\vec{q}^2}{2m_N}F_2\right]\left(1 -  \frac{\vec{q}^2}{8m_N^2} \right) + 
   i\left[4m_N F_1 - F_2\left(1 -  \frac{\vec{q}^2}{8m_N^2} \right)\right]\left( \vec\sigma\cdot\left(\vec{k'} \times \vec{k}\right)\right)\,,\nonumber\\
   \vec{J}_{\mathrm{SN}(i)} &=& F_1\frac{\vec{k}+\vec{k'}}{2m_N} + \frac{i}{2m_N}\left[F_1 + 2m_N F_2 \right]\left( \vec\sigma\cdot\left(\vec{k'} - \vec{k}\right)\right)\,,\nonumber 
\end{eqnarray}
where $F_1$ and $F_2$ are the nucleon (proton or neutron) form factors~\cite{halzen-martin, arrington}. In addition to
the single-nucleon photon absorption, many-body interactions also need to be considered (see Fig.~\ref{fig_theory_MEC}). 
The interaction of  two or more nucleons with a photon can be modeled by a meson exchange~\cite{schiavilla, carlson} 
and leads to the effective meson-exchange current operators $\hat{J}_{\mathrm{MEC}(i)}^\mu$. Hence, the photon 
absorption operator can be written as:
\begin{eqnarray}
	\hat{O}^\mu = \sum_{i=1}^{3} \hat{J}_{\mathrm{SN}(i)}^\mu + \hat{J}_{\mathrm{MEC}(i)}^\mu\,. \nonumber
\end{eqnarray}

\begin{figure}[!ht]
\begin{center}
\includegraphics[width=0.85\textwidth]{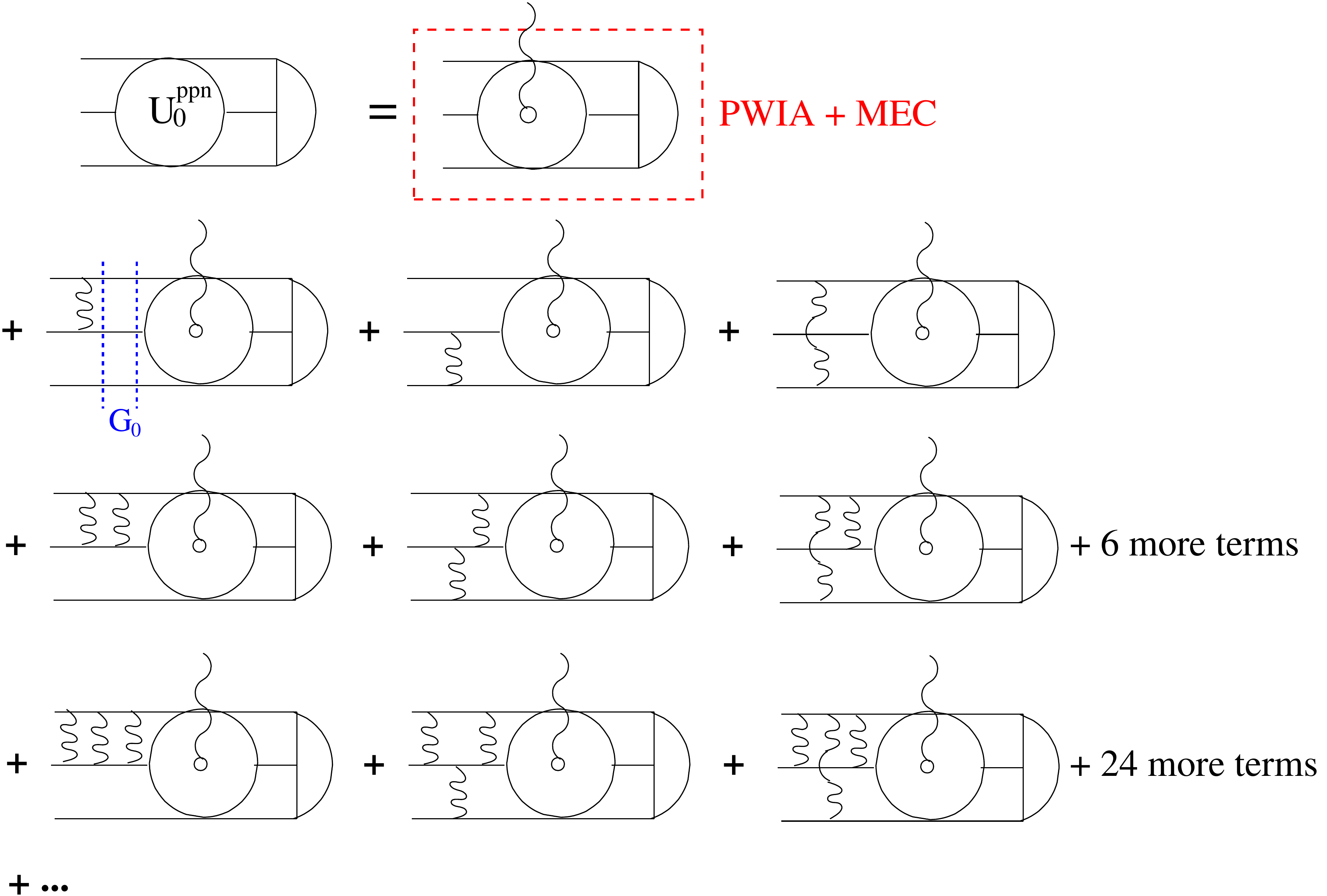}
\caption{ Diagrams for the complete 3N-breakup amplitude
due to photon absorption. The half-circle on the right represents the initial 
${}^3\mathrm{He}$ state, the circle with the wiggly line in the center represents the one-photon
absorption process. The wiggly lines between two horizontal lines represent the NN-interactions 
between the nucleons. Three-nucleon forces (3NF) are neglected. Between the dashed vertical lines
three nucleons propagate freely (denoted by the free 3N propagator $G_0$)
Three horizontal line on the left represent three final states (ppn). Figure adopted 
from Ref.~\cite{golak2005}. 
\label{fig_theory_diagrams_ppn}}
\end{center}
\end{figure}

\begin{figure}[!ht]
\begin{center}
\begin{minipage}[t]{0.65\textwidth}
\hrule height 0pt
\includegraphics[width=\textwidth]{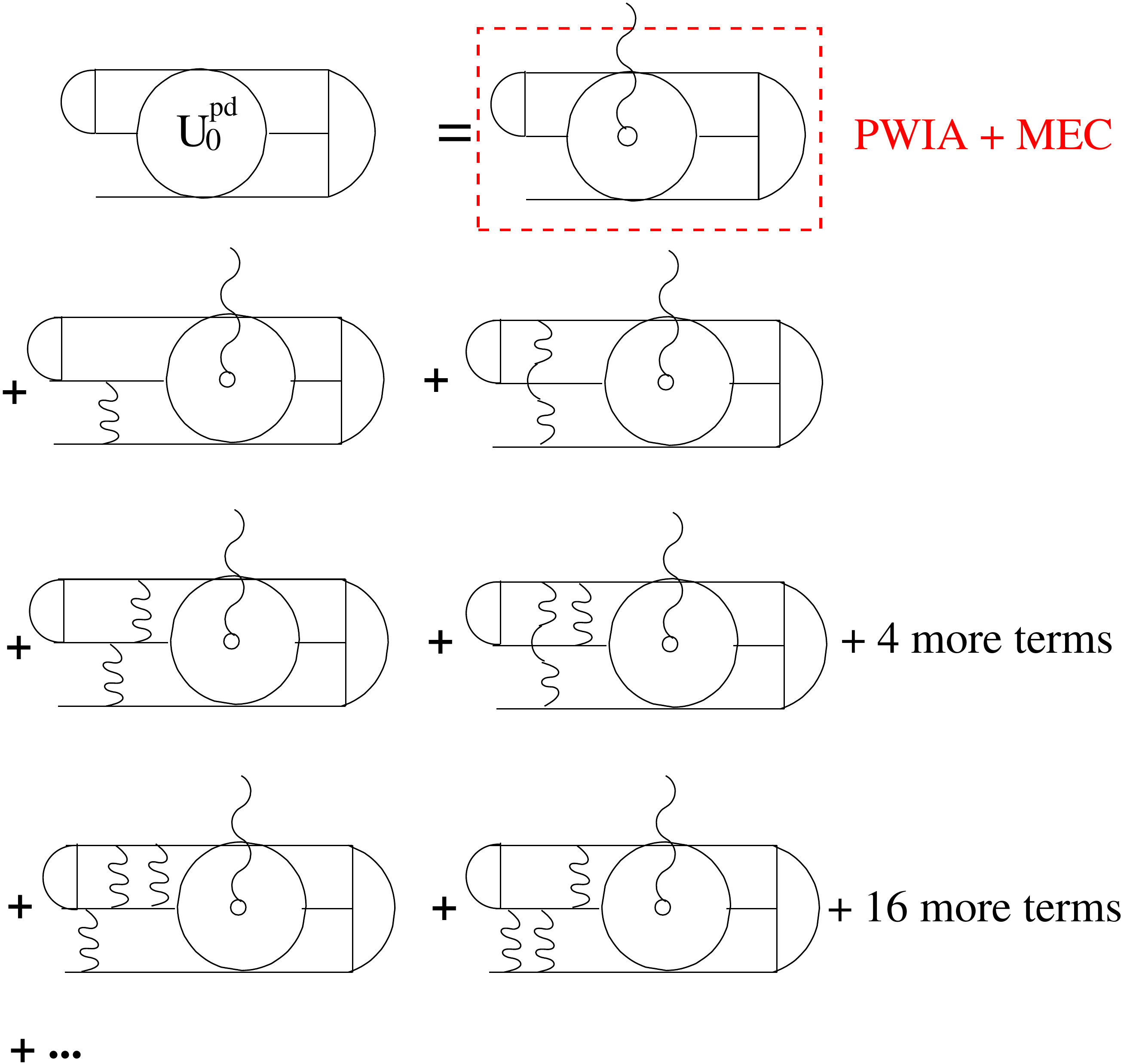}
\end{minipage}
\hfill
\begin{minipage}[t]{0.33\textwidth}
\hrule height 0pt
\caption{ Diagrams for the pd-breakup amplitude due to photon absorption. 
The smaller half-moon and the horizontal line on the left correspond to the final 
deuteron and proton, respectively. Other notation as in Fig.~\ref{fig_theory_diagrams_ppn}.
Figure adopted from Ref.~\cite{golak2005}. 
\label{fig_theory_diagrams_pd}}
\end{minipage}
\end{center}
\end{figure}

After the absorption of a photon, three nucleons may further interact.
Figs.~\ref{fig_theory_diagrams_ppn} and~\ref{fig_theory_diagrams_pd} summarize everything 
that can happen in photo-disintegration of ${}^3\mathrm{He}$. In the 
absence of MEC effects the first diagram corresponds to the plane-wave impulse 
approximation, since the nucleons do not interact after the breakup of the nucleus. 
In the rest of the diagrams, the nucleons interact by pairwise forces. The re-scatterings
of the nucleons are known as final state interactions (FSI).

To account for FSI in the calculation of the matrix elements, auxiliary
states $|U_f^\mu\rangle$ are introduced as:
\begin{eqnarray}
  \langle\Psi_f|\hat{O}^\mu|\Psi_{{}^3\mathrm{He}}(\theta^*, \phi^*)\rangle \stackrel{\mathrm{FSI}}{\longrightarrow}
   \langle\Psi_f|U_f^\mu\rangle\,. \nonumber
\end{eqnarray}

\begin{figure}[!ht]
\begin{center}
\begin{minipage}[t]{0.7\textwidth}
\hrule height 0pt
\includegraphics[width=\textwidth]{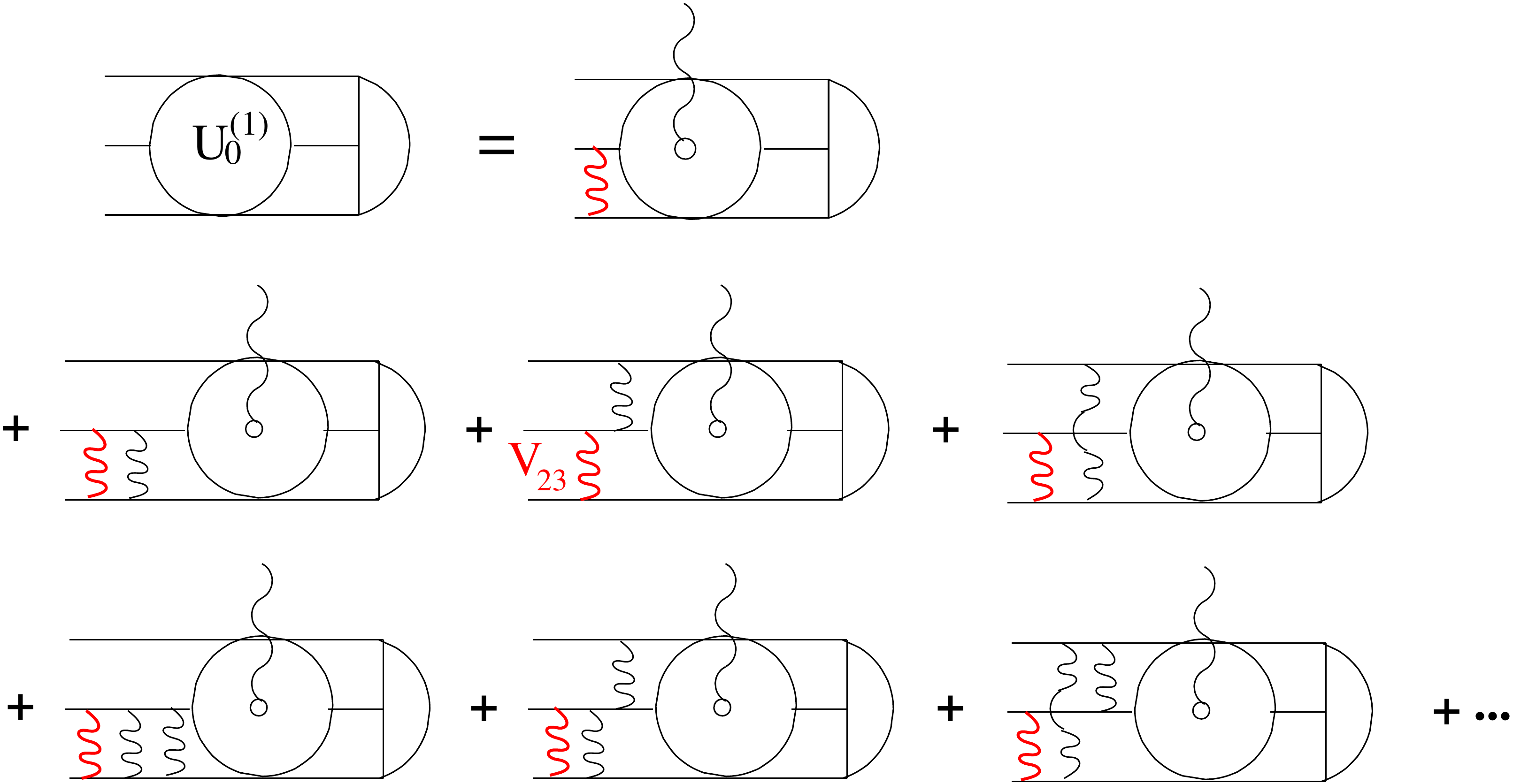}
\end{minipage}
\hfill
\begin{minipage}[t]{0.29\textwidth}
\hrule height 0pt
\caption{ A subset of  diagrams which contribute to the FSI in  
the (ppn) and (pd) breakup, which end with the interaction of the nucleons 2 and 3. 
Symbols as in Figs.~\ref{fig_theory_diagrams_ppn} 
and~\ref{fig_theory_diagrams_pd}. Figure adopted from Ref.~\cite{golak2005}. 
\label{fig_theory_diagrams_T23}}
\end{minipage}
\end{center}
\end{figure}

Diagrams in Figs.~\ref{fig_theory_diagrams_ppn} and~\ref{fig_theory_diagrams_pd} with interacting 
nucleons can be divided into three subsets. The first set contains graphs in which nucleons 2 and 3 
interact last (see Fig.~\ref{fig_theory_diagrams_T23}), the second set includes graphs which end 
with interaction of nucleons 1 and 3, while the last subset combines diagrams where nucleons 1 and 2 interact
last. 

Hence, the auxiliary states for the (ppn) and (pd) breakup can be written as:
\begin{eqnarray}
  |U_{ppn}^\mu\rangle &=& \hat{O}^\mu|\Psi_{{}^3\mathrm{He}}\rangle + |U_1^\mu\rangle + |U_2^\mu\rangle + |U_3^\mu\rangle\,, \nonumber\\
  |U_{pd}^\mu\rangle &=& \hat{O}^\mu|\Psi_{{}^3\mathrm{He}}\rangle + |U_1^\mu\rangle + |U_2^\mu\rangle\,. \label{eq_theory_aux}
\end{eqnarray}
Note that by keeping just the first terms of these expansions, one reinstates PWIA. The terms 
$|U_{1,2,3}^\mu\rangle$ correspond to each subset of the FSI diagrams. In the case of the (pd) breakup, 
the last interaction between nucleons 1 and 2 does not represent the FSI, but corresponds to the interaction
between nucleons within the compound final state (deuteron). Consequently $|U_{3}^\mu\rangle$ is not 
considered in the auxiliary state for this breakup. Considering the diagrams of Fig~\ref{fig_theory_diagrams_T23}, 
a recursion formula can be written for the first subset:
\begin{eqnarray}
|U_1^\mu\rangle = \hat{V}_{23}\hat{G}_{0}\hat{O}^\mu|\Psi_{{}^3\mathrm{He}}\rangle + 
    \hat{V}_{23}\hat{G}_{0}\left(|U_1^\mu\rangle + |U_2^\mu\rangle + |U_3^\mu\rangle \right)\,.\label{eq_theory_u1}
\end{eqnarray} 
Here, $\hat{G}_0$ is the free three-nucleon propagator~\cite{golak2005}, and the potential operator $\hat{V}_{ij}$ 
describes the nucleon-nucleon (NN) force between nucleons $i$ and $j$. By re-arranging the terms in 
Eq.~(\ref{eq_theory_u1}) and introducing the NN t-operator as
\begin{eqnarray}
  \hat{t}_{ij} = \left( \mathbb{I}- \hat{V}_{ij}\hat{G}_0\right)^{-1}\hat{V}_{23}\,, \nonumber
\end{eqnarray}
one obtains:
\begin{eqnarray}
  |U_1^\mu\rangle  = \hat{t}_{23}\hat{G}_0 \hat{O}^\mu |\Psi_{{}^3\mathrm{He}}\rangle + 
\hat{t}_{23}\hat{G}_0\left(|U_2^\mu\rangle + |U_3^\mu\rangle \right)\,. \label{eq_theory_u1a}
\end{eqnarray} 
Assuming that three nucleons are identical in sense of isospin symmetry, prescriptions for the remaining two 
subsets of diagrams can be expressed in terms of $|U_1^\mu\rangle$ as:
\begin{eqnarray}
  |U_2^\mu\rangle = \hat{P}_{12}\hat{P}_{23}|U_1^\mu\rangle\,,\qquad |U_3^\mu\rangle = \hat{P}_{13}\hat{P}_{23}|U_1^\mu\rangle\,,
\end{eqnarray}
where $\hat{P}_{ij}$ is a permutation operator~\cite{GreinerSymmetry} that interchanges nucleons $i$ and $j$. By defining 
$\hat{P} \equiv   \hat{P}_{12}\hat{P}_{23} + \hat{P}_{13}\hat{P}_{23}$, Eq.~(\ref{eq_theory_u1a}) can be written as:
\begin{eqnarray}
  |U_1^\mu\rangle  = \hat{t}_{23}\hat{G}_0 \hat{O}^\mu |\Psi_{{}^3\mathrm{He}}\rangle + 
\hat{t}_{23}\hat{G}_0 \hat{P} |U_1^\mu\rangle \,. \label{eq_theory_faddeev_eq}
\end{eqnarray}
This is the Faddeev integral equation~\cite{golak2005, glockle2004}. By iterating this equation, an expansion for 
$|U_1^\mu\rangle$ is obtained, now formulated in terms  of the NN t-operators:
\begin{eqnarray}
  |U_1^\mu\rangle  = \hat{t}_{23}\hat{G}_0 \hat{O}^\mu |\Psi_{{}^3\mathrm{He}}\rangle +  \hat{t}_{23}\hat{G}_0 \hat{P}\hat{t}_{23}\hat{G}_0 \hat{O}^\mu |\Psi_{{}^3\mathrm{He}}\rangle + \hat{t}_{23}\hat{G}_0 \hat{P}\hat{t}_{23} \hat{t}_{23}\hat{G}_0\hat{P}\hat{t}_{23}\hat{G}_0 \hat{O}^\mu |\Psi_{{}^3\mathrm{He}}\rangle + \cdots\,,\nonumber
\end{eqnarray}
Using this formalism in Eqs.~(\ref{eq_theory_ME}) and~(\ref{eq_theory_aux}),  the final expressions for the
nuclear transition currents can be obtained:
\begin{eqnarray}
	J_{ppn}^\mu &=& \langle \Psi_{ppn} | U_{ppn}^\mu \rangle = \langle \Psi_{ppn}|\hat{O}|\Psi_{{}^3\mathrm{He}}\rangle 
                    + \langle \Psi_{ppn}|(1+\hat{P})|U_1^\mu\rangle \,,\nonumber \\
	J_{pd}^\mu &=& \langle \Psi_{pd} | U_{pd}^\mu \rangle = \langle \Psi_{pd}|\hat{O}|\Psi_{{}^3\mathrm{He}}\rangle 
                    + \langle \Psi_{pd}|(1+\hat{P}_{12}\hat{P}_{23})|U_1^\mu\rangle \,.\nonumber\,
\end{eqnarray}
In order to obtain the matrix elements for the two- and three-body breakup of ${}^3\mathrm{He}$, the Faddeev integral equations
given in Eq.~(\ref{eq_theory_faddeev_eq}) must be solved. The calculation of the full auxiliary states is not trivial and need 
to be calculated on  super-computers. For the experiment E05-102, most advanced calculations were performed by the theory
group from Krakow and Bochum~\cite{golakcalculations}. They employ a modern nucleon-nucleon potential AV18~\cite{wiringa} for 
the description of the two-nucleon forces. In addition to the two-body interactions used in this short overview, 
they consider also three-nucleon forces (3NF) which describe the interactions of three nucleons. For the implementation of 3NF they utilize 
the UrbanaIX~\cite{urbanaIX} model. 

However, to be able to perform any kind of matrix-element calculations, the ground-state wave-function of ${}^3\mathrm{He}$
must first be understood. Nowadays, the $|\Psi_{{}^3\mathrm{He}}(\theta^*, \phi^*)\rangle$ state can also be  obtained as a 
solution of the Faddeev equation, using an approach very similar to the one discussed above~\cite{nogga2003, glockle2004}.

\section{Relativistic gauge invariant approach}
\label{sec:nagorny}

\begin{figure}[!ht]
\begin{center}
\includegraphics[width=\textwidth]{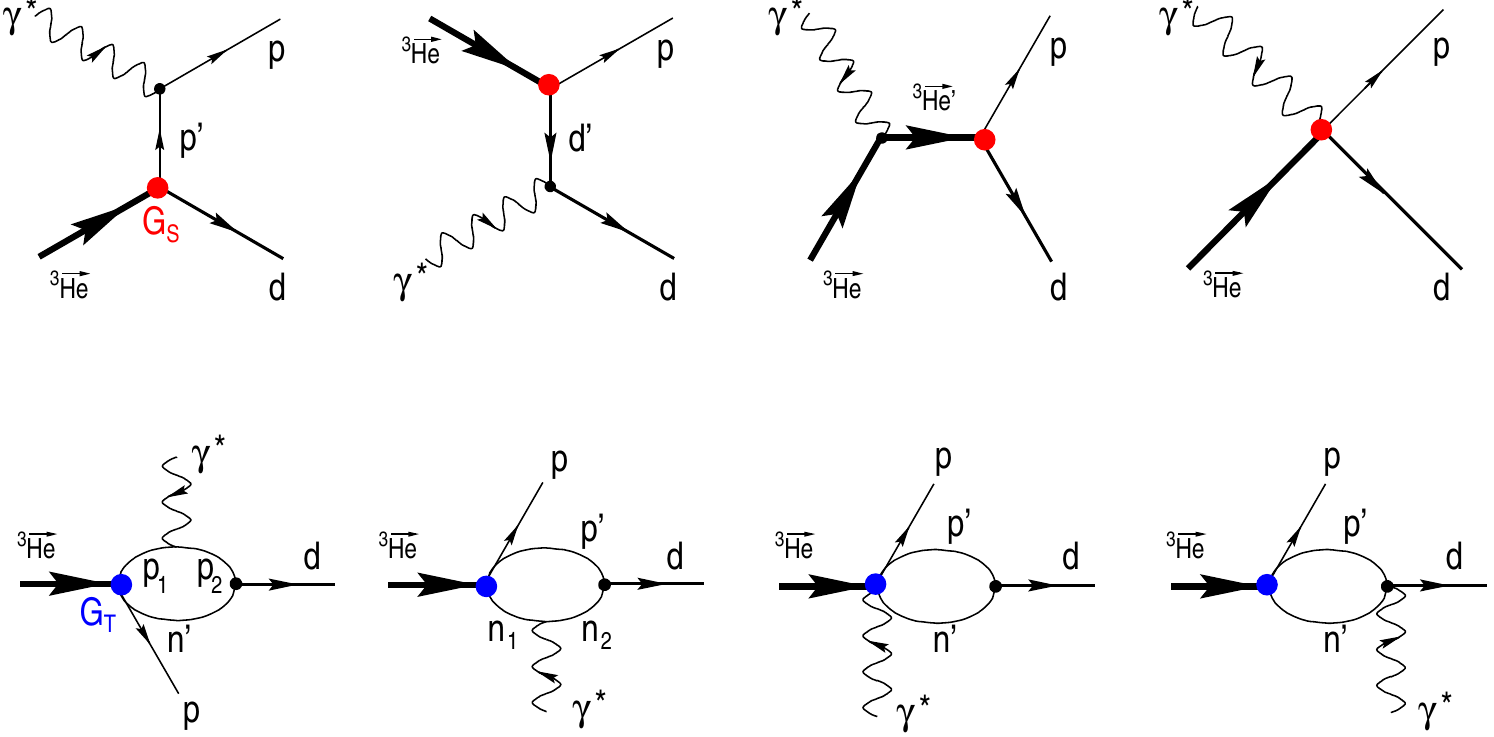}
\caption{ The minimal set of covariant diagrams which provides the conservation of the isoscalar (top)
and isovector (bottom) currents. Figure adopted after Refs.~\cite{sixPhD, nagorny96}.
\label{fig_theory_nagorny_2bbu}}
\end{center}
\end{figure}

This sections briefly summarizes the predictions of Nagorny~\cite{nagorny96, nagorny98, nagorny94}, who considered 
the relativistic gauge invariant approach that combines the requirements of covariance and current conservation
with accounting of nuclear structure, final state interactions (FSI) and meson exchange currents. 
Fig.~\ref{fig_theory_nagorny_2bbu} shows the minimal set of diagrams for the two-body breakup channels 
${}^3\mathrm{He}(e,e'p)d$ and ${}^3\mathrm{He}(e,e'd)p$, which provides nuclear current conservation and a 
good enough description of the unpolarized cross-section for the two-body disintegration of ${}^3\mathrm{He}$.

The first diagram (proton pole) represents the plane-wave impulse approximation (PWIA). The diagram 
with the deuteron pole corresponds to the quasi-deuteron model. The third diagram with the ${}^3\mathrm{He}$
pole contributes to the FSI, while the last diagram in the top row ensures, that the isoscalar current is conserved. 
Such contact terms account for the effects of meson exchange (MEC). 
In all these diagrams, the internal state of the pn-pair before 
and after the interaction with the photon remains the same: spin=1 and isospin=0. Since the
isospin of the pn-pair does not change during the photo-absorption, the nuclear current corresponding to 
these diagrams is isoscalar.

On the other-hand, the diagrams in the bottom row of Fig.~\ref{fig_theory_nagorny_2bbu} correspond to
the isovector current. In this case the internal state of the pn-pair changes in the interaction 
with a photon from (spin=0, isospin=1) to (spin=1, isospin=0). Hence, the isospin state of the pn-pair
changes with the photo-absorption, for which an isovector current is required.  

For numerical calculations Nagorny considered the ${}^3\mathrm{He}$ wave-function obtained 
from Faddeev calculations with the Reid Soft Core potential. As discussed in 
Sec.~\ref{sec:he3WF} the ${}^3\mathrm{He}$ ground-state wave-function predominantly consists of 
the S-wave and D-wave parts:
\begin{eqnarray}
	\Psi({}^3\mathrm{He}) = \Psi_{S-wave} + \Psi_{D-wave}\,,\nonumber
\end{eqnarray}	 
where the S-wave part can be further represented as:
\begin{eqnarray}
  \Psi_{S-wave} = -\psi^{s}\xi^{a} + \psi'\xi'' - \psi''\xi'\,.
\end{eqnarray}
Here $\psi^s$ represents the fully symmetric (S-wave) space component (see Fig.~\ref{fig_He3States}), while 
$\psi'$ and $\psi''$ correspond to the space components with mixed symmetry (S'-state). The spin-isospin 
parts of the wave-function are the fully antisymmetric $\xi^a$ and the mixed symmetry configurations 
$\xi'$, $\xi''$. In terms of these components, the vertices $G_{s,t}$ of the ${}^3\mathrm{He}$ breakup
with pn-pairs in the singlet and triplet spin states can be expressed as:
\begin{eqnarray}
	G_s = -\frac{\psi^s + \psi''}{\sqrt{2}}\,,\qquad G_t = \frac{\psi^s - \psi''}{\sqrt{2}}\,. \nonumber
\end{eqnarray}

Although only full currents from Fig.~\ref{fig_theory_nagorny_2bbu} are conserved, only particular 
diagrams remain relevant, with a proper choice of kinematics, while the rest are suppressed. 
For the $(e,e'p)d$ reaction in quasi-elastic kinematics 
the proton-pole graph (see Fig.~\ref{fig_theory_nagorny_2bbu}) dominates at low recoil momenta $p_r$.   
The electron scatters off a quasi-free proton, while leaving the deuteron as a
spectator practically at rest. The amplitude for this process can be factorized exactly and 
depends only of the $G_s$ vertex. Consequently all information about the nuclear structure
is canceled when the ratio of the cross-sections is formed. In this case, the polarized 
${}^3\mathrm{He}$ acts like a polarized proton target. Hence, the asymmetry $A^{\vec{e}\>{}^3\vec{\mathrm{He}}}$ equals the 
asymmetry  $A^{\vec{e}\vec{p}}$ for elastic scattering of polarized electrons on polarized 
protons~\cite{gao} (corrected for the sign, since the spin of the ejected proton is opposite 
to the spin of ${}^3\mathrm{He}$):
\begin{eqnarray}
 A(\theta^*, \phi^*) = - A^{\vec{e}\vec{p}} = \frac{2\tau v_{T'}\cos{\theta^*} {G_{M}^{p}}^2  - 
                      2\sqrt{2\tau(1+\tau)}v_{TL'}\sin{\theta^*} \cos{\phi^*}G_M^p G_{E}^p }
                      {(1+\tau)v_L {G_E^p}^2 + 2\tau v_T {G_{M}^p}^2}\,. \label{eq_theory_ElasticProtonAsymmetry}
\end{eqnarray}
Here $\tau = \frac{Q^2}{2M_p}$, while $G_M^p$ and $G_E^p$ are the Sachs form-factors~\cite{arrington}, 
and $v_{\kappa}$ are the kinematical factors given by Eqs.~(\ref{eq_theory_kinfac}).

However, for a description of the deuteron channel $(e,e'd)$ in quasi-elastic kinematics, 
considering only the deuteron-pole diagram  in the limit of low recoil momenta, 
by analogy to the proton channel, is insufficient. 
Both isoscalar and isovector transitions (see Fig.~\ref{fig_Tripp}) must be taken into 
account. Hence, amplitudes with both singlet vertices $G_S$ and triplet vertices $G_T$ 
contribute. Due to the interference of the isoscalar and isovector currents, the terms which 
contain the information about the nuclear dynamics do not cancel when the 
asymmetry is formed, making the ${}^3\vec{\mathrm{He}}(\vec{e},e'd)p$ channel 
at low recoil momenta sensitive to the S'-components.

The $(e,e'd)p$ channel is also a valuable source of information on the D-components of
the ${}^3\mathrm{He}$ ground-state wave-function, which become important at large
recoil momenta. In the D-state the spins of all three nucleons are oriented in the opposite
direction to the nuclear spin (see Fig.~\ref{fig_He3States}), while in the S-state
the spin of the neutron and one of the protons are pointing in the direction of the nuclear spin.  
Therefore, the triplet pn-pairs  in the S- and D-states have different spin orientations 
in relation to the nuclear spin direction, which means that their contributions to
the asymmetry will have opposite signs. Hence, due to the presence of the D-state, 
the asymmetry is expected to change sign at high recoil momenta. 

On the other hand, in the $(e,e'p)d$ channel, the presence of the D-state is not 
that prominent. Since the spin of the knocked-out proton is opposite to
the nuclear spin in both S- and D-states, no sign change is expected in this channel
at high recoil momenta. 

The calculations were performed also for the three-body breakup channel $(e,e'p)pn$. 
At low recoil momenta this process is dominated by the proton-pole diagram~\cite{nagorny98} 
with either singlet or triplet spectator pn-pairs and their amplitudes are determined
by the $G_s$ and $G_t$ vertices. The asymmetry for the $(e,e'p)pn$ channel, calculated
from these currents, becomes:
\begin{eqnarray}
	A = \kappa_p(p_r) \times A^{\vec{e}\vec{p}}\,.\label{eq_theory_asymmetry_ppn}
\end{eqnarray}
The parameter $\kappa_p(p_r)$ can be interpreted as an effective proton polarization 
in ${}^3\vec{\mathrm{He}}$ and contains all information on the nuclear structure and
dynamics~\cite{nagorny98}:
\begin{eqnarray}
  \kappa_p = \frac{G_s^2 - G_t^2}{G_s^2 + 3G_t^2}\,. \label{eq_theory_kappa}
\end{eqnarray} 
Considering the predominant S-configuration of the nucleus, the production 
of a singlet pn-pair corresponds to the absorption of a photon by the proton, whose spin 
oriented in the same direction as ${}^3\vec{\mathrm{He}}$ spin. On the other hand, to generate
the triplet pn-pair the photon must be absorbed by a proton, whose spin is oriented in 
the opposite direction to the nuclear spin. In the PWIA, the squares of the amplitudes for 
these two processes differ only by a sign, which results in a zero asymmetry 
of  ${}^3\vec{\mathrm{He}}(\vec{e},e'p)np$, when only a fully symmetric configuration 
of ${}^3\mathrm{He}$ is considered.  This reflects the fact that, unlike the neutron, 
the protons in the S-state of ${}^3\mathrm{He}$ are unpolarized. 
It also means that in PWIA at low recoil momenta, the 
asymmetry given by Eq.~(\ref{eq_theory_asymmetry_ppn}) may arise only due to 
the presence of the S'-state, which would make this channel ideal for the investigation 
of the mixed-symmetry states. However, the magnitude of the asymmetry should be very small, 
since the S'-state represents only $\approx 2\,\mathrm{\%}$ of the spin averaged wave-function:
\begin{eqnarray}
  \kappa_p \stackrel{PWIA}{\longrightarrow}\frac{\psi''}{\psi^S} \ll 1\,.
\end{eqnarray} 

However, experiments~\cite{achenbach2008} have measured significantly larger asymmetries 
at low recoil momenta as predicted by the PWIA model (see Fig.~\ref{fig_MAINZeep}). 
This findings can not be contributed to the stronger influence of the S'-state in 
the ${}^3\mathrm{He}$ wave-function. The observed large asymmetry arises 
due to the final state interactions. At low recoil momenta and high $|\vec{q}|$,
the major FSI are between the nucleons in the spectator pn-pair, since their relative
momentum will be small due to energy and momentum conservation, while the momentum
of the stuck proton with respect to the pn-pair will be large enough 
($|\vec{p}_p|\sim|\vec{q}|$) so that their interaction may be neglected.
In Nagorny's calculations, FSI were considered by replacing the 
vertex functions $G_{s,t}$ in Eq.~(\ref{eq_theory_kappa}) with re-normalized 
ones, which include additional loops to account for such effects.


\chapter{Experimental Setup}
The experiment E05-102 was conducted in Hall A at Thomas Jefferson National 
Accelerator Facility between May 12 and June 15, 2009. 

\section{Jefferson Lab}
Thomas Jefferson National Accelerator Facility (TJNAF), known also as 
Jefferson Lab (JLab), is one of 17 national laboratories funded by 
the U.S. Department of Energy (DOE). It is located in Newport News,
Virginia and is managed and operated by the Jefferson Science 
Associates, LLC. Its $6\,\mathrm{GeV}$ polarized continuous electron beam 
accelerator together with the three experimental halls makes it 
one of one of the world's leading medium energy nuclear laboratories,
ideal for the research of the structure of nuclei, hadrons and underlying
fundamental interactions.

\begin{figure}[!ht]
\begin{center}
\begin{minipage}[t]{0.5\textwidth}
\hrule height 0pt
\includegraphics[width=\textwidth]{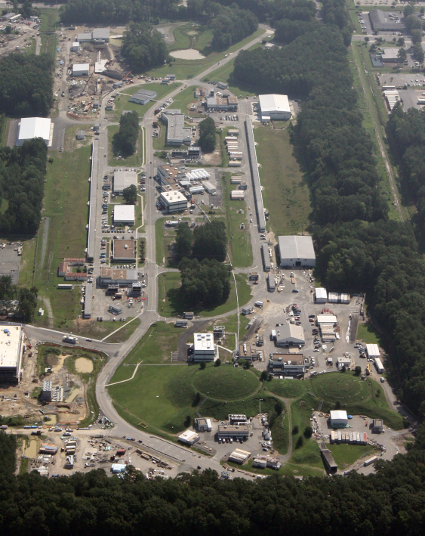}
\end{minipage}
\hfill
\begin{minipage}[t]{0.45\textwidth}
\hrule height 0pt
\caption{ Aerial photo of Jefferson Lab. The accelerator ring together 
with three round experimental halls (at the bottom) are clearly 
visible. The construction site for the new experimental Hall D can also 
be seen in the upper right corner.  \label{fig_JeffersonLab}}
\end{minipage}
\end{center}
\end{figure}

The accelerator was originally designed~\cite{alcorn} to accelerate 
electrons up to $4.4\,\mathrm{GeV}$, by recirculating the electron beam 
four times through two linear accelerators (South and North LINAC), 
producing an energy gain of $400\,\mathrm{MeV}$ per pass. Each LINAC consists 
of 20 cryo-modules with an accelerating gradient of $5\,\mathrm{MeV/m}$,
made of niobium cavities, cooled by liquid helium to $2\,\mathrm{K}$. 
Continuous research and improvements to the accelerator resulted in 
a higher mean gradient of $7\,\mathrm{MeV/m}$, which made it possible
to accelerate electrons to $5.7\,\mathrm{GeV}$. The maximum 
continuous-wave electron beam current is $200\,\mathrm{\mu A}$ and can
be split arbitrarily between three interleaved $499\,\mathrm{MHz}$
bunch trains. One such bunch train can be peeled off after each LINAC 
pass to any one of the experimental halls using radio-frequency (RF)
separators and septum magnets.

\begin{figure}[!ht]
\begin{center}
\begin{minipage}[t]{0.6\textwidth}
\hrule height 0pt
\includegraphics[width=\textwidth]{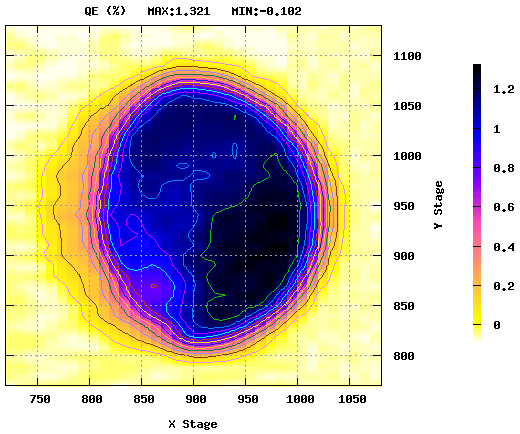}
\end{minipage}
\hfill
\begin{minipage}[t]{0.35\textwidth}
\hrule height 0pt
\caption{The photocathode quantum efficiency during experiment E05-102. 
During the operation, the QE starts to decrease where illuminated with 
high power lasers. The deterioration of the QE when moving to the right 
side is clearly visible. Therefore QE must be constantly monitored and
laser spot moved to provide beam with high polarization. When QE becomes
to low, photocathode is taken through recovery procedure, to reactive it
(increase QE).
 \label{fig_cathodeQE}}
\end{minipage}
\end{center}
\end{figure}

Polarized electrons are injected into the accelerator from a DC 
photo-gun~\cite{sinclair}. There a strained GaAs photocathode
is illuminated with a circularly polarized infrared laser light
with a wavelength of $860\,\mathrm{nm}$. This light has just enough
energy to excite the electrons from the $m_j = \pm 3/2$ sub-states
of the $P_{3/2}$ valence-band  to the  $m_j = \pm 1/2$ states of the 
$S_{1/2}$ conducting band. The polarization
of the electrons depends on the helicity of the laser light. When 
using light with positive helicity, only electrons with 
spin $m_j = -1/2$ are ejected and vice versa. This procedure in principle
allows $100\,\mathrm{\%}$ polarization of the beam. In reality, the achieved
polarization is between $70\,\mathrm{\%}$ and $90\,\mathrm{\%}$. The number 
of the emitted electrons directly depends on the number of photons incident on the
photocathode and is given phenomenologically by:
\begin{eqnarray}
  I = \frac{\lambda}{124 \left(\mathrm{\frac{W\,nm}{mA}}\right)}\cdot P \cdot QE\,, \nonumber 
\end{eqnarray}
where $I$ represents the electron current, $P$ is the power of the laser light
and $QE$ represents the quantum efficiency. QE is the percentage of the photons 
of a particular wavelength $\lambda$ that eject an electron.  The $QE$ strongly depends on the 
material used for the photocathode. Furthermore, the operational lifetime of a photocathode,
which is defined as the time required for the QE to fall below $1/e$ of its
original value, strongly depends  on the generated electron current and the quality of 
the vacuum. Fig.~\ref{fig_cathodeQE} shows the measured QE of the 
photocathode during the E05-102 experiment. 
The lifetimes of the high-polarization photogun at Jefferson 
Lab  are among the longest lifetimes~\cite{stutzman2008} measured at 550 beam hours 
at an average current of $100\,\mathrm{\mu A}$. To provide an independent 
control of the beam current to all three experimental halls, the photocathode
is illuminated by three $499\,\mathrm{Mhz}$ gain-switched diode lasers.

Polarized electrons ejected from the photo-cathode are then accelerated in a linear 
accelerator to $45\,\mathrm{MeV}$ before reaching the main acceleration ring. 
The polarization of the beam can be oriented with the Wien filter. The 
polarization at the injector is measured by using the Mott 
polarimeter.

Experiments at Jefferson Lab are carried out in three experimental 
halls (A,B, and C). Hall C has been operational since November 1995, Hall A 
since May 1997 and Hall B since December 1997. The energy of the electrons 
delivered to  each of the experimental halls can be changed in multiples 
of $1/5$ of the maximum energy ($1.2\,\mathrm{GeV}, 2.4\,\mathrm{GeV}$, 
$3.6\,\mathrm{GeV}$, $4.8\,\mathrm{GeV}$, $5.7\,\mathrm{GeV}$). 
Maximum energy beam can be delivered simultaneously to all three halls,
at three different beam currents. Hall B with its large acceptance 
spectrometer (CLAS) requires a current as low as $1\,\mathrm{nA}$, 
while $100\,\mathrm{\mu A}$ beam is usually required in Halls A and C. 

\begin{figure}[htp]
\begin{center}
\includegraphics[width=0.8\textwidth]{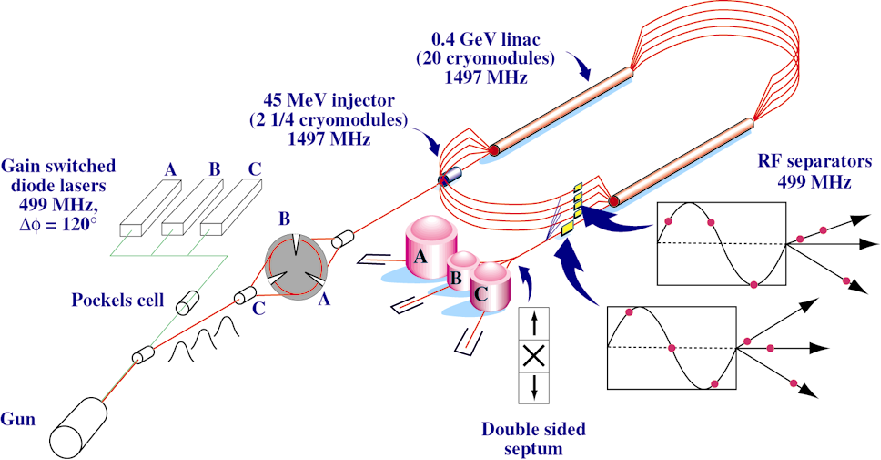}
\caption{Schematic drawing of the Jefferson Lab's $6\,\mathrm{GeV}$ 
accelerator operation: In the injector, light from three lasers (one 
for each experimental hall) hits the photocathode and emits polarized 
electrons, which are then accelerated to $45\,\mathrm{MeV}$ and injected 
into the main accelerator ring. There, electrons get accelerated to up to
$5.7\,\mathrm{GeV}$ by recirculating them five times through the two linear
accelerators. In the end, the beam gets extracted from the ring by using
RF separators and septum magnets, and is simultaneously delivered to 
all three experimental halls. \label{fig_CEBAF6GeV}}
\end{center}
\end{figure}

In 2012, after 16 years of very successful operations, Jefferson Lab will
be upgraded to $12\,\mathrm{GeV}$. The renovation will
take place between May 2012 and May 2013. When the accelerator 
will be upgraded with new and improved cryo-modules. In addition, five 
more cavities will be added to each LINAC, which will increase the 
energy from $1.1\,\mathrm{GeV}$ to $2.2\,\mathrm{GeV}$ per recirculation. 
A maximum energy of $11\,\mathrm{GeV}$ will then become available in halls A, B, and C. 
The intensity of these beams will be $85\,\mathrm{\mu A}$ with polarization
up to $85\%$. An additional arc will be added to the accelerator for
the beam to reach the new experimental Hall D with the maximum energy of 
$12\,\mathrm{GeV}$. With the new apparatus, Jefferson Lab wants to continue
its quest in understanding the structure of nuclear matter. Various 
experiments are proposed that would give a detailed insight to the
physical origin of quark confinement, spin and flavor structure of
the proton and neutron and the quark structure of the nucleons. New high 
precision parity violating experiments will allow stringent tests of the 
Standard model which could lead to the discovery of new physics beyond 
this model.

\section{Experimental Hall A}

\begin{figure}[!ht]
\begin{center}
\includegraphics[width=0.95\textwidth]{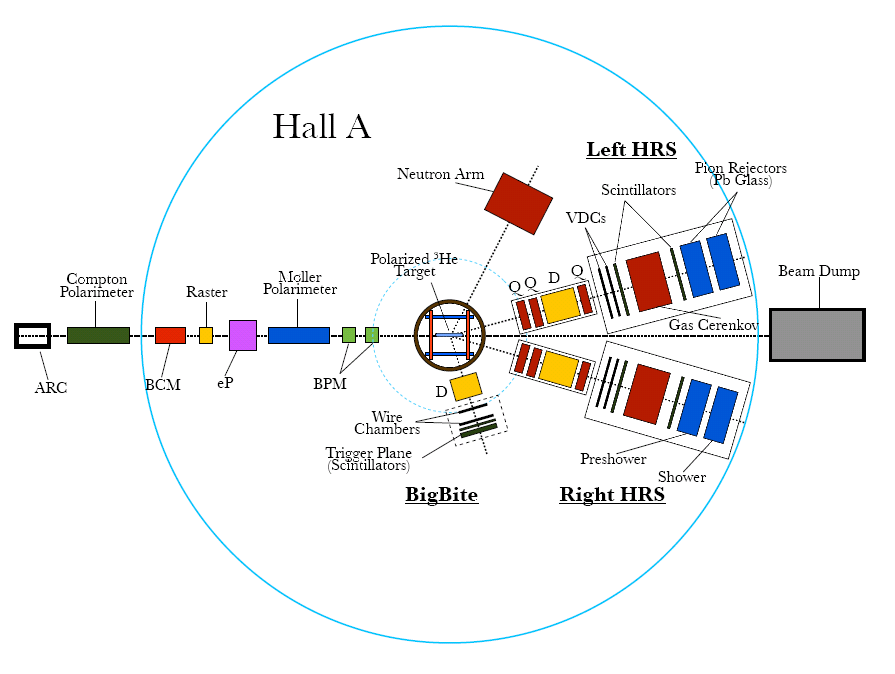}
\vspace*{-5mm}
\caption{Experimental Hall A during the E05-102 experiment. The electron
beam comes into the hall from the left, passing beam quality 
monitors before hitting the target at the center of the hall. Various
targets are possible. In this particular experiment a polarized ${}^3\mathrm{He}$
target was used. Scattered electrons were detected by two High Resolution
Spectrometers (Left-HRS and Right-HRS). In addition to the standard equipment, a
smaller spectrometer BigBite and a Neutron detector were used to detect 
hadrons. \label{fig_HALLA}}
\end{center}
\end{figure}

Hall A~\cite{alcorn} is the largest of the experimental halls. Most of the hall is 
under ground. It has a circular shape, measuring $53\,\mathrm{m}$ in diameter.  
Figure~\ref{fig_HALLA} shows the schematics drawing of the hall during
the E05-102 experiment. After entering the hall enclosure, the electron 
beam first passes through various diagnostics instruments before
hitting the target. With these instruments we determine the beam current,
its position at the target and the degree of polarization. Particles that do 
not interact with the target material are then stopped at the beam dump, 
which is located at the opposite side of the hall. The experimental target is 
located at the center of the hall. Various cryogenic, solid and gaseous target are 
being used. For the E05-102 experiment a polarized ${}^3\mathrm{He}$ target
was used.

The reaction products are detected by two High Resolution Spectrometers, known as 
HRS-Left and HRS-Right. Their momentum resolution is better than $2\times10^{-4}$ and 
horizontal angular resolution is better than $2\,\mathrm{mrad}$, rendering Hall-A 
ideal for the experiments which require high luminosity and high momentum and/or angular 
resolution for at least one of the reaction products. Beside the HRS spectrometer, a smaller 
BigBite spectrometer and a Neutron detector (HAND) can be employed for the  detection of 
the scattered particles. They are mostly being used for the  double or triple coincidence 
experiments.

\section{Beam line}
The hall A beam line starts at the beam switchyard, where the beam is extracted from
the accelerator, and ends at the Hall A beam dump~\cite{alcorn,osp}. There are
many control and measurement devices placed along the $200\,\mathrm{m}$-long
beam line, which are necessary to transport the beam with required properties 
onto the target and into the beam dump, and to simultaneously measure the properties
of the beam. The most relevant elements (see figure~\ref{fig_HALLA}) are the 
arc magnets (positioned after the switchyard) that are used for the beam 
energy measurement, Compton polarimeter, beam current monitors, Unser monitor, 
a fast raster, eP device for the beam energy determination (not used during E05-102),
M{\o}ller polarimeter and beam position monitors. With the use of these devices, 
the properties of the beam can be precisely determined (see table~\ref{table_BeamProperties}).
\begin{table}[!ht]
 \begin{center}
\begin{minipage}[t]{0.6\textwidth}
\hrule height 0pt
\begin{tabular}{ll}
\toprule
Parameter & Accuracy\\
\midrule
Energy & $1\times 10^{-4}$ (relative)\\
Current ($\geq 1\,\mathrm{\mu A}$) & $\leq 5 \times 10^{-3}$ (relative)\\
Position (at target) & $10\,\mathrm{\mu m}$\\
Direction (at target) & $30\,\mathrm{\mu m}$\\
Polarization &  $\approx 3\,\%$ \\
\bottomrule
 \end{tabular}
\end{minipage}
\begin{minipage}[t]{0.35\textwidth}
\hrule height 0pt
\caption{Overview of the available accuracy for the 
determination of beam parameters~\cite{alcorn}.
\label{table_BeamProperties}}
\end{minipage}
\end{center}
\end{table}

\subsection{Measurement of Beam Energy}
Beam energy is measured absolutely by the Arc method~\cite{alcorn}. 
This measurement is based of the principle, that an electron in 
a constant magnetic field moves in a circle, the 
radius of which depends on the density of the magnetic field and
the momentum (energy) of the particle. By measuring the 
magnetic field integral and the total angular deflection $\phi$ of 
the electron beam inside the magnetic field, the electron momentum 
can then be determined by
\begin{eqnarray}
  p = k \frac{\int_{0}^{l} B_{\perp}dl}{\phi}\,\,\,, \nonumber
\end{eqnarray}
where $k = 0.299792\,\mathrm{GeV~T^{-1} m^{-1}}/c$~\cite{alcorn}, $l$ represents the 
total path-length of the particle through the magnetic field, while 
$B_{\perp}$ is the component of the magnetic field density, 
perpendicular to the particle track.

\begin{figure}[!ht]
\begin{center}
\includegraphics[width=0.7\textwidth]{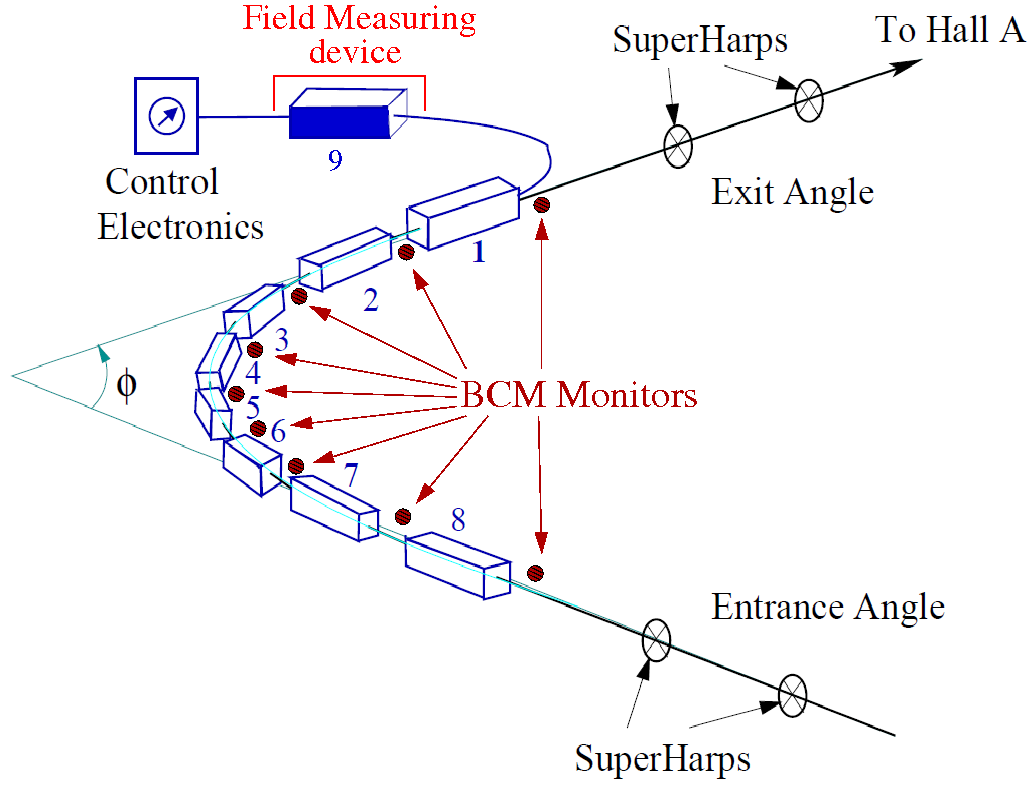}
\caption{ Drawing of the arc section of the Hall A beam line~\cite{ZhengPhD}. 
Eight magnets are used to deflect the beam by approximately $\phi=34.3^\mathrm{o}$. The 
position of the beam before and after the arc is determined by the wire scanners
(SuperHarps). The measurement of the magnetic field integral of the eight
bending magnets is based on the direct field measurement in the ninth reference
magnet. \label{fig_HALLA_Arc}}
\end{center}
\end{figure}

\vspace*{-10mm}
Figure~\ref{fig_HALLA_Arc} presents the setup used for the Arc method. 
Measurement is done in the arc section of the beam line by using 
nine identical dipole magnets ($3\,\mathrm{m}$ long), powered in 
series. This ensures the same current through all the magnets. First eight 
magnets are used to steer the beam through the arc. The measurement of the 
deflection angle, $\phi$, of the beam is based on a set of wire scanners 
called superHarps~\cite{osp}. These scanners are moved across the beam path. 
When the beam strikes a
wire, a Photomultiplier tube (PMT) records a signal due to the electromagnetic
shower induced in the wire. Recording the scanner position and the PMT output 
voltage allows us to determine the beam position at each scanner location. 
Then, by knowing the positions of the scanners, we can deduce the effective bending 
angle through the arc. The mean bending angle is $\overline{\phi} = 34.3^\mathrm{o}$. 
The measurement of the magnetic field integral is done by using the last 
ninth dipole magnet, which is positioned outside the beam line. A special
magnetic flux meter is attached to this magnet, which moves back and forth through
the magnet and measures the inside magnetic field as a function of distance.  
The field maps of the first eight magnets can then be calibrated relatively to this one, 
since all magnets are identical and run at the same conditions.

\begin{figure}[!ht]
\begin{center}
\includegraphics[width=1\textwidth]{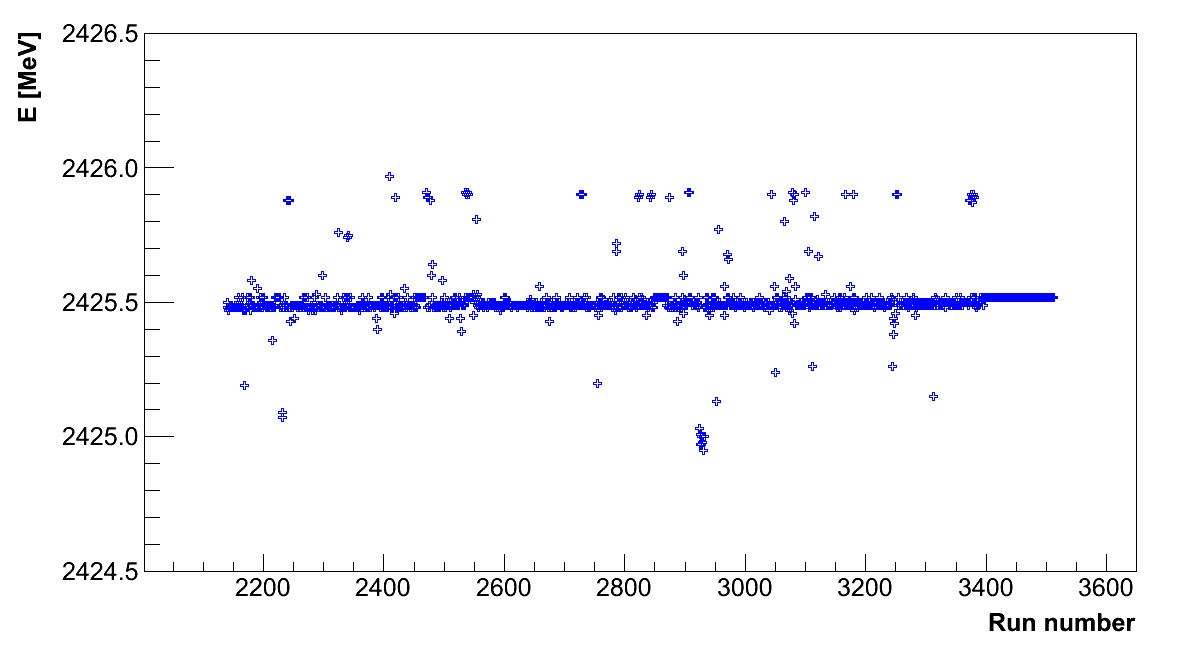}
\caption{ Beam energy during the E05-102 experiment, measured at the 
beginning of each collected dataset (run) by the non-invasive Tiefenbach method. 
The resolution of this method is   
$dE_{\mathrm{Beam}}/E_{\mathrm{Beam}}\approx 5\times10^{-4}$.
\label{fig_TiefenbachEnergy}}
\end{center}
\end{figure}

The precision of the Arc method is 
$dE_{\mathrm{Beam}}/E_{\mathrm{Beam}}\approx 2\times10^{-4}$. 
Unfortunately this is an invasive technique (electron beam is affected) and
can not be used to monitor the beam energy during the production running
of the experiment. For that, a non-invasive Tiefenbach method was considered.
Instead of superHarps, this method uses the Hall A arc beam position monitors in
combination with the current values of the arc magnetic field integral to
determine the beam energy. The resolution of this approach is 
$dE_{\mathrm{Beam}}/E_{\mathrm{Beam}}\approx 5\times10^{-4}$. 
The measurements of the beam energy during the E05-102 experiment 
are shown in Fig.~\ref{fig_TiefenbachEnergy}.

\subsection{Measurement of Beam Position}
\label{sec_BPMs}
\begin{figure}[!htpb]
\begin{center}
\includegraphics[width=0.7\textwidth]{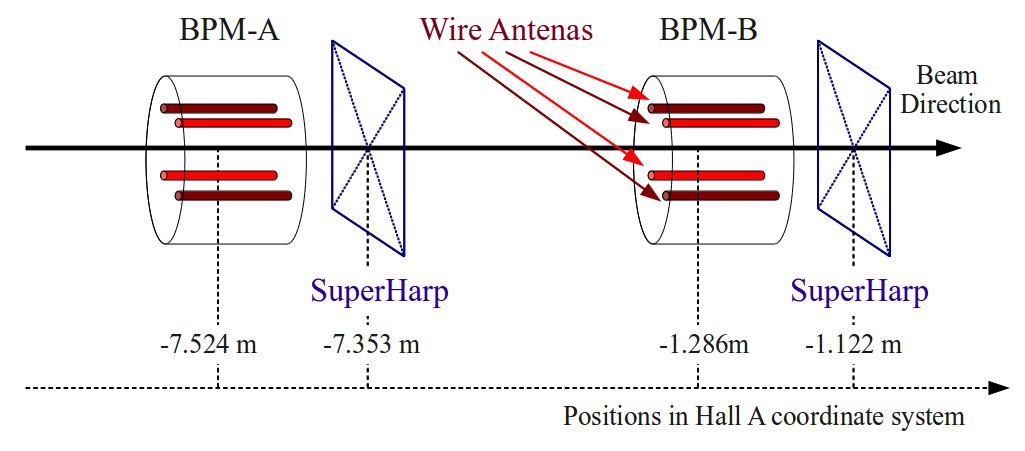}
\caption{Beam position monitors are located upstream from the 
target (center of Hall A). Each BPM consists of four wire 
antennas. The relative position of the beam is determined by using
the difference-over-sum technique between two opposite wires 
(first pair: light-red wires, second pair: dark-red wires.) 
For the absolute positions they have to be calibrated 
against SuperHarps. \label{fig_HALLA_BPMs}}
\end{center}
\end{figure}

For non-destructive determination of the position and direction of 
the beam at the target location, two beam position monitors (BPMs) are 
employed~\cite{BPM_Wright,alcorn}. They are located $7.524\,\mathrm{m}$ 
and $1.286\,\mathrm{m}$ upstream of the target (see Fig~\ref{fig_HALLA_BPMs}).  
Each BPM cavity is a four wire antenna array. The detection method is
based on comparing the signals induced by the beam passing two opposite 
antennas. From the recorded signals from two such pairs, beam
coordinates at the center of the BPM module can be reconstructed. 
By combining this information from both modules, a relative position of 
the beam at the target  can be determined to within 
$100\,\mathrm{\mu m}$ for currents above $1\,\mathrm{\mu A}$.

\begin{figure}[!ht]
\begin{center}
\includegraphics[width=0.9\textwidth]{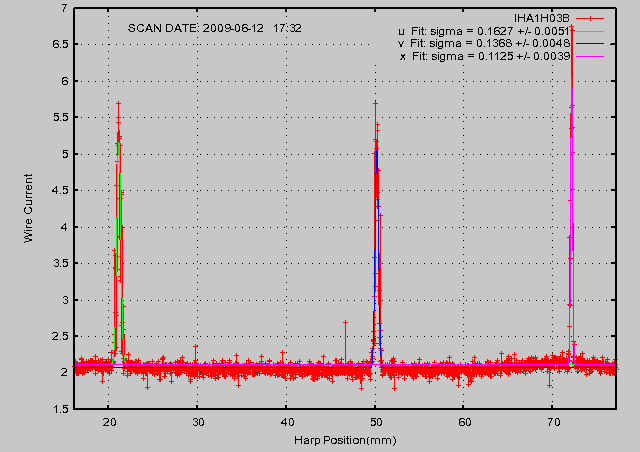}
\caption{A typical Harp-scan measurement performed during the E05-102 experiment to
determine the transverse position and spot size of a beam. 
Each of the wire scanners consists of three thin wires that are
moved across the beam path. The first into the beam is a vertical wire
(labeled as ``x``). The next two wires are at $\pm 45$ degrees to the 
vertical (labeled as ``u`` and ''v''). When the beam hits a wire, a signal 
is detected by the PMT's positioned next to the scanner.  From the combination 
of the size of the signal and the position of the scanner, a precise beam 
position and its transverse size are determined. The figure shows the results for the SuperHarp, 
which is positioned right after the second BPM.  The profile of the 
electron beam has an ellipsoidal shape with a spot size of 
$\sigma_{x,y} < 200\,\mathrm{\mu m}$.
\label{fig_HARP}}
\end{center}
\end{figure}

Before being able to use BPMs, they have to be calibrated. For that we
use wire scanners (superHarps) adjacent to each of 
the BPMs. This invasive calibration procedure is called ``Bull's eye``. 
The wire scanners are surveyed with respect to the Hall A coordinates
with the accuracy of less than $200\,\mathrm{\mu m}$. An example of a SuperHarp
measurement is shown in figure~\ref{fig_HARP}.

The BPM information is recorded in two different ways. Event-by-event 
information from BPMs is recorded in the CODA data 
stream from each of the 8 BCM antennas. In addition, the averaged position over 
$0.3\,\mathrm{s}$ is logged every second into the EPICS database and 
injected asynchronously into the CODA data-stream every few seconds
(see Sec.~\ref{sec:CODA}). 

\subsection{Beam Raster}  
The beam is rastered at the target with the amplitude of the several
millimeters at $25\,\mathrm{kHz}$ to prevent damaging the target cell.
The main concern is the overheating of the target gas which could
lead to the explosion  of the target. The raster is a pair of horizontal
and vertical air-core dipoles located $23\,\mathrm{m}$ upstream
of the target~\cite{osp}. Since 2003 a linear beam raster
system is being considered~\cite{yan_linear}. The system generates  a
rectangular raster pattern with a highly uniform  distribution
on the target. It allows  the use of beam currents up to
$200\,\mathrm{\mu A}$.

\begin{figure}[!htpb]
\begin{center}
\begin{minipage}[t]{0.6\textwidth}
\hrule height 0pt
\includegraphics[width=1\textwidth]{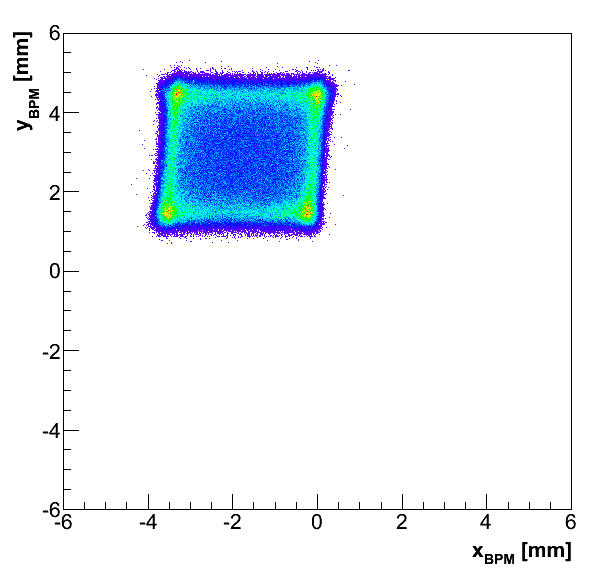}
\end{minipage}
\hfill
\begin{minipage}[t]{0.35\textwidth}
\hrule height 0pt
\caption{A typical beam spot at the target with a
$4\times4\,\mathrm{mm^2}$ raster during
production running. The position of the beam
at the target at any given moment was reconstructed 
by the beam position monitors (BPMs). \label{fig_raster}}
\end{minipage}
\end{center}
\end{figure}

During the E05-102 experiment an electron beam with $4\times4\,\mathrm{mm^2}$
raster  was used (see Fig.~\ref{fig_raster}). The size of the beam spot
at the target was monitored with the BPMs.
In addition, $6\times6\,\mathrm{mm^2}$ and
$2\times2\,\mathrm{mm^2}$ raster sizes were used during beam recovery
procedures to check if the beam is pointing to the center of the cell.
When the electron
beam was hitting the edge of the target, detected event rates changed
dramatically when the raster size was increased. For the
purpose of the optics calibration, an unrastered beam was also
considered, but only in the combination with the solid carbon target.
The typical size of an unrastered beam (see Fig.~\ref{fig_HARP}) 
is $\sigma_{x,y} < 200\,\mathrm{\mu m}$.

\subsection{Measurement of Beam Current}
\begin{figure}[!ht]
\begin{center}
\includegraphics[width=0.8\textwidth]{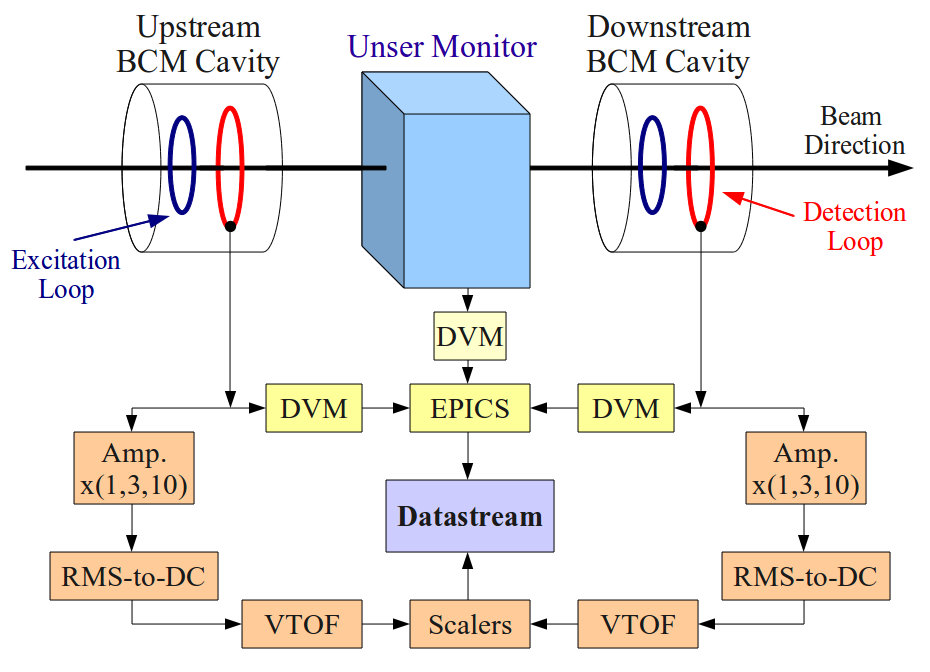}
\caption{Schematic of the Hall A beam current measurement system. Two 
BCM cavities are employed for the continuous monitoring of the beam current. 
Measurements are recorded to data stream as a part of EPICS information 
(sampled data) and as a scaler information (integrated data).
\label{fig_BCM}}
\end{center}
\end{figure}

The current of the electron beam entering the Hall A is determined by the
beam current monitor (BCM), which provides a stable, low-noise and 
non-intercepting measurement~\cite{alcorn, osp}. It is located
$24.5\,\mathrm{m}$ upstream of the target location and consists of an Unser 
monitor and two RF cavities enclosed in a box to improve magnetic shielding
and temperature stabilization (see Fig.~\ref{fig_BCM}). The Unser 
monitor~\cite{unser} is a Parametric Current Transformer and is used 
as an absolute reference. It can not be used for continuous monitoring of
the beam current because its output signal drifts significantly with time. 
Instead, two resonant RF cavity monitors on either side of the Unser monitor 
are used to measure the beam current. Both cavities 
are stainless steel cylindrical waveguides with high quality factor. They are
tuned to the frequency of the beam ($1.497\,\mathrm{MHz}$) resulting in  
voltage levels at their output proportional to the beam current.
The output from each cavity is divided into two components. The first component is 
processed by a high-precision AC voltmeter, which effectively measures the 
RMS of the input every second. The resulting number is 
stored to the data stream as an EPICS variable every few seconds.

The second component of the output signal is amplified and sent to 
the RMS-to-DC converters which produce an analog DC voltage level. 
This level drives a Voltage-to-Frequency converter (VTOF) whose output frequency 
is proportional to the input voltage level. The output signals are then led
to the scaler modules and injected into the data stream. During each run, 
scalers simply accumulate, resulting at the end in a number 
proportional to the time-integrated beam current, i.e. the accumulated beam charge.

The RMS-to-DC output is linear for currents from $5\,\mathrm{\mu A}$ to 
above $200\,\mathrm{\mu A}$. Since it is non-linear at very low currents, 
two additional amplifiers were used in front of the RMS-to-DC module with 
different gains ($\times3$ and $\times 10$), to push the non-linear region to 
smaller currents at the expense of saturation at very high currents.  
Hence there are three signals coming from each BCM ($\mathrm{BCM\times}1$,
$\mathrm{BCM\times}3$, $\mathrm{BCM\times}10$). All signals are 
fed into scaler modules of each spectrometer (HRS-L, HRS-R and BigBite) 
and recorded in the data-stream. This way we have a redundancy of 
18 scaler outputs for determining the beam current and charge during a run. 
When running at low currents, the scaler information obtained from the BCM channel with
a higher gain amplifiers should be used to ensure linearity. When running 
at high currents, the scaler information from BCM channels with a lower gain amplifiers 
should be considered to avoid saturation. Each of the scaler outputs is 
being calibrated during the calibration running to find a relation 
between the raw BCM scaler reading and the true accumulated charge. 

\subsection{Beam helicity and Beam half-wave plate}
\label{sec:HWP}
\begin{figure}[!ht]
\begin{center}
\includegraphics[width=0.8\textwidth]{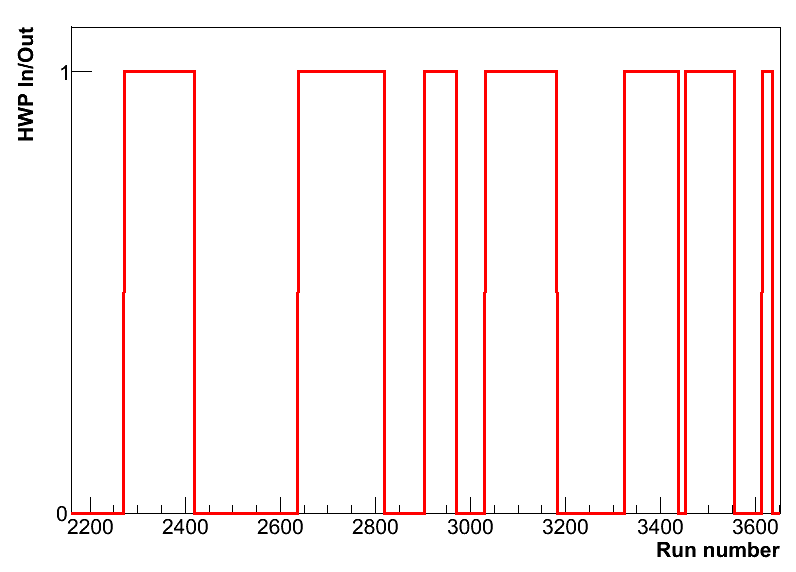}
\vspace*{-3mm}
\caption{Position of the Beam half-wave plate (HWP) during the E05-102 experiment. 
Half of the experimental statistics was collected with the 
HWP inserted. \label{fig_BHWP}}
\end{center}
\end{figure}

The helicity of the electron beam is determined by the polarization of the 
laser light that hits the photo-cathode at the injector gun. The helicity 
sign can be flipped at the rate by changing the polarization of the light,
by using a Pockels cell. In addition, a removable half-wave plate~\cite{sinclair}, 
known as a ''beam half-wave plate'', can be inserted in front of the Pockels cell 
to reverse the sense of  the polarization determined by the Pockels cell. 
When the beam half-wave plate is inserted, the helicity of the outgoing electron 
beam is opposite to the case when it is not inserted. This way it provides
a powerful way to test for any false asymmetries, since the physics asymmetries
measured with and without the beam half-wave plate should have opposite 
signs~\cite{ZhengPhD}. During the E05-102 experiment half of the
statistics was collected with the inserted beam half-wave plate 
(see Fig.~\ref{fig_BHWP}). This allowed us to continuously control 
false asymmetries. 

\subsection{Measurement of Beam Polarization}
\label{sec:Moller}
For the experiments conducted in Hall A a longitudinally polarized 
electron beam is being used with typical beam polarization of 
$75\mathrm{-}85\,\mathrm{\%}$. To measure the polarization of the
beam  Compton and M\o{}ller polarimeters are being employed.

The Compton polarimeter~\cite{alcorn} is positioned at the entrance to 
the hall (see Fig.~\ref{fig_HALLA}), after the beam comes out of 
the arc. It exploits the process of Compton scattering, where 
circularly polarized photon is being scattered off a polarized electron. 
The polarimeter measures the beam helicity asymmetry for this process, 
which is directly proportional to the beam polarization. Unfortunately, 
this polarimeter was not working properly during the E05-102 experiment. 
Therefore we have not considered its measurements in our analysis.

\begin{figure}[!ht]
\begin{center}
\includegraphics[width=1\textwidth]{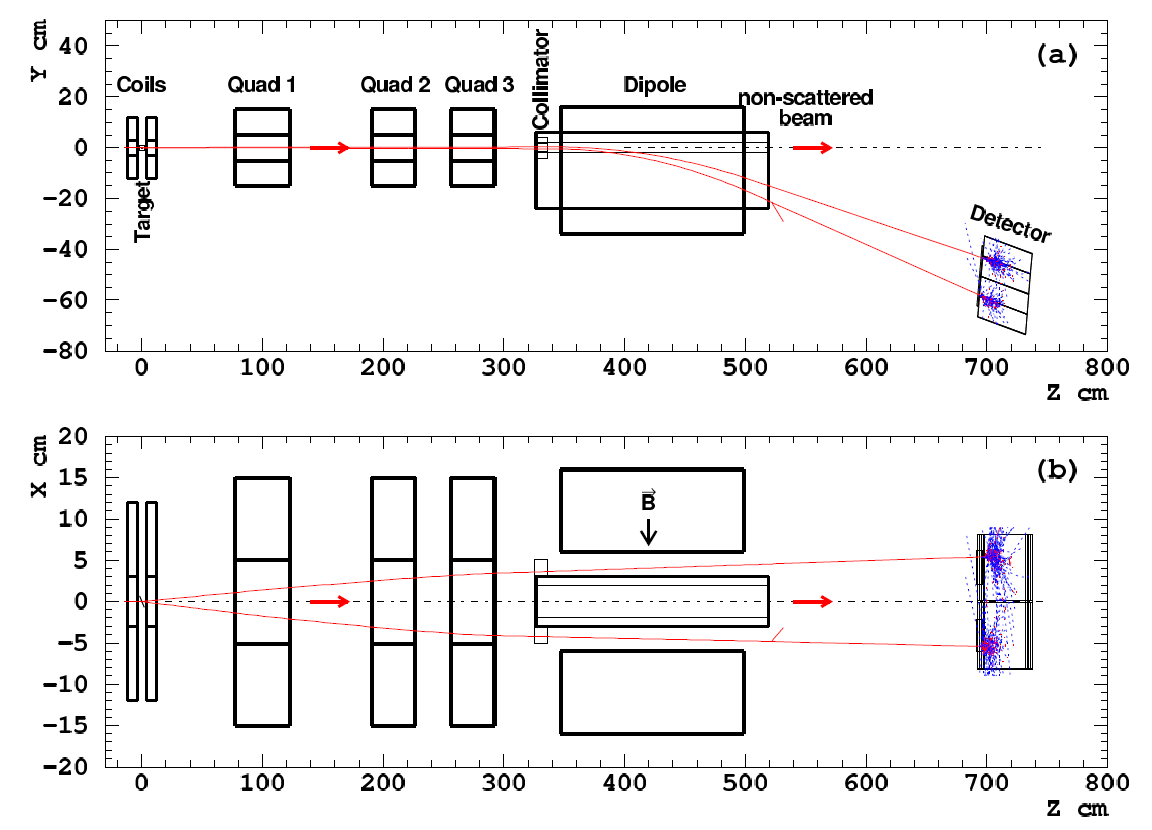}
\caption{Schematic drawing of the M\o{}ller polarimeter. It consists
of a magnetized solid target, four sequential magnets (three quadrupoles
and a dipole) and a segmented calorimeter, for the detection of 
scattered electrons. The z-direction is along the beam line,
x is pointing to the left of the beam and y is pointing upwards.
The horizontal arrows show direction of the non-scattered beam.
\vspace*{5mm}
\label{fig_Moller}}
\end{center}
\end{figure}

The M\o{}ller polarimeter is based on M\o{}ller scattering of
polarized electrons off polarized atomic electrons in a magnetized 
foil~\cite{alcorn,osp}. The cross-section for the process
$\vec{e}^- + \vec{e}^- \longrightarrow e^{-} + e^{-}$ is proportional 
to the target polarization $P_i^{\mathrm{Target}}$ and beam polarization
$P_{i}^{\mathrm{Beam}}$:
\begin{eqnarray}
\sigma \propto \left( 1+ \sum_{i = X,Y,Z}A_{ii}P_i^{\mathrm{Target}}
P_{i}^{Beam} \right)\,, \nonumber
\end{eqnarray}
where $i \in \{X,Y,Z\}$ defines the projections of the polarizations. The 
analyzing power depends on the scattering angle in the 
center-of-mass coordinate system ($\theta_{\mathrm{CMS}}$). 
Assuming that the beam direction is along
the Z-axis and that the scattering occurs in the XZ-plane (see Fig.~\ref{fig_Moller}),
the components of the analyzing power can be written as:
\begin{eqnarray}
A_{ZZ} = -\frac{sin^2{\theta_{\mathrm{CMS}}\left(7+
\cos^2{\theta_{CMS}}\right)}}{\left( 3+ \cos^2{\theta_{\mathrm{CMS}}}\right)}
\,,\,\,\,\,\,\,\,
A_{XX} = -A_{YY} = -\frac{\sin^4{\theta_{\mathrm{CMS}}}}
{\left( 3+ \cos^2{\theta_{\mathrm{CMS}}}\right)}\,. \nonumber
\end{eqnarray}

The longitudinal beam polarization is determined via the measurement of 
the helicity asymmetry:
\begin{eqnarray}
P_{Z}^{\mathrm{Beam}} = \frac{1}{P_{Z}^{\mathrm{Target}} \langle A_{ZZ}\rangle}
\left(\frac{N_{+} - N_{-}}{N_{+} + N_{-}}\right)\,,
\end{eqnarray}
where $N_{+}$ and $N_{-}$ are the measured counting rates for two 
opposite orientations of the beam helicity and $P_{Z}^{\mathrm{Target}}$ is 
the target polarization along the beam line.  $\langle A_{ZZ}\rangle$ 
represents the mean analyzing power, obtained by averaging $A_{ZZ}$ over 
all available scattering angles, using a Monte-Carlo calculation of 
the M\o{}ller spectrometer acceptance. 

The M\o{}ller polarimeter is installed between the eP detector and 
the beam position monitors, approximately $17.5\,\mathrm{m}$ upstream of
the target (see Fig.~\ref{fig_HALLA}). As a target of polarized 
electrons it uses a ferromagnetic foil which is magnetized in a 
magnetic field of $24\,\mathrm{mT}$ along its main symmetry axis. Five different 
foils, made of Aluminum, Iron or Supermendur alloy, can be used 
as a target. The target polarization is derived from the foil
magnetization measurements and is usually around $8\,\mathrm{\%}$. 
The secondary electrons are detected by the magnetic spectrometer
which consists of three sequential quadrupoles, a dipole, and a 
lead-glass calorimeter. The detector is divided into two segments
in order to detect two scattered electrons in coincidence. The
coincidence measurement significantly reduces the noise caused by 
 non-M\o{}ller sources. 

\begin{figure}[!ht]
\begin{center}
\begin{minipage}[t]{0.67\textwidth}
\hrule height 0pt
\includegraphics[width=1\textwidth]{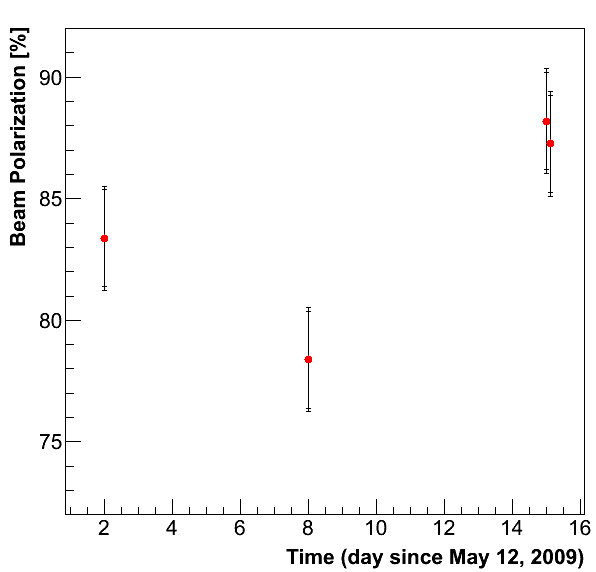}
\end{minipage}
\hfill
\begin{minipage}[t]{0.3\textwidth}
\hrule height 0pt
\caption{Results of the beam polarization measurements
during the E05-102 experiment with the M{\o}ller 
polarimeter. The first error-bar on each data-point represents
the systematic error of approx. $2\,\mathrm{\%}$.
The second (smaller) error-bar represents the statistical 
uncertainty which is estimated to be around $0.2\,\mathrm{\%}$   
\label{fig_MollerMeasurement}}
\end{minipage}
\end{center}
\end{figure}

The Hall A M\o{}ller polarimeter can be used at beam energies from 
$0.8\,\mathrm{GeV}$ to $6\,\mathrm{GeV}$. The M\o{}ller polarization 
measurements are invasive and can not be performed 
during the experiment. The measurements are usually made with 
a beam current of $0.5\,\mathrm{\mu A}$ and typically take one hour, 
resulting in the statistical accuracy of about $0.2\,\mathrm{\%}$ 
and with a $2\,\mathrm{\%}$ systematical uncertainty. During 
the experiment E05-102 four M\o{}ller measurements were performed.
Results are shown in Fig.~\ref{fig_MollerMeasurement}.

\section{Target System}
\label{sec:TargetSysyem}

The polarized ${}^3\vec{\mathrm{He}}$ target has been used extensively
at SLAC, Mainz, MIT-Bates, DESY and Jefferson Lab, mostly for
studying the spin structure of the neutron. At Jefferson Lab, 
the polarized ${}^3\mathrm{He}$ target was first utilized in 1998 for 
the E94-010 experiment~\cite{mezziani}.

The original target consisted of three basic 
parts. The main component was a glass target cell, 
filled with ${}^3\mathrm{He}$ gas at high pressure, nitrogen and a small 
amount of alkali vapors.  
The second important component were the high power lasers, which provided 
intense circularly polarized $795\,\mathrm{nm}$ light, required for 
polarizing the  ${}^3\mathrm{He}$ by 
spin exchange optical pumping.
The third constituent part were the two pairs of Helmholtz coils, which were 
employed to rotate and hold the polarization in any in-plane  direction
(parallel to the floor of the experimental hall). 

\begin{figure}[!ht]
\begin{center}
\begin{minipage}[t]{0.67\textwidth}
\hrule height 0pt
\includegraphics[width=1\textwidth]{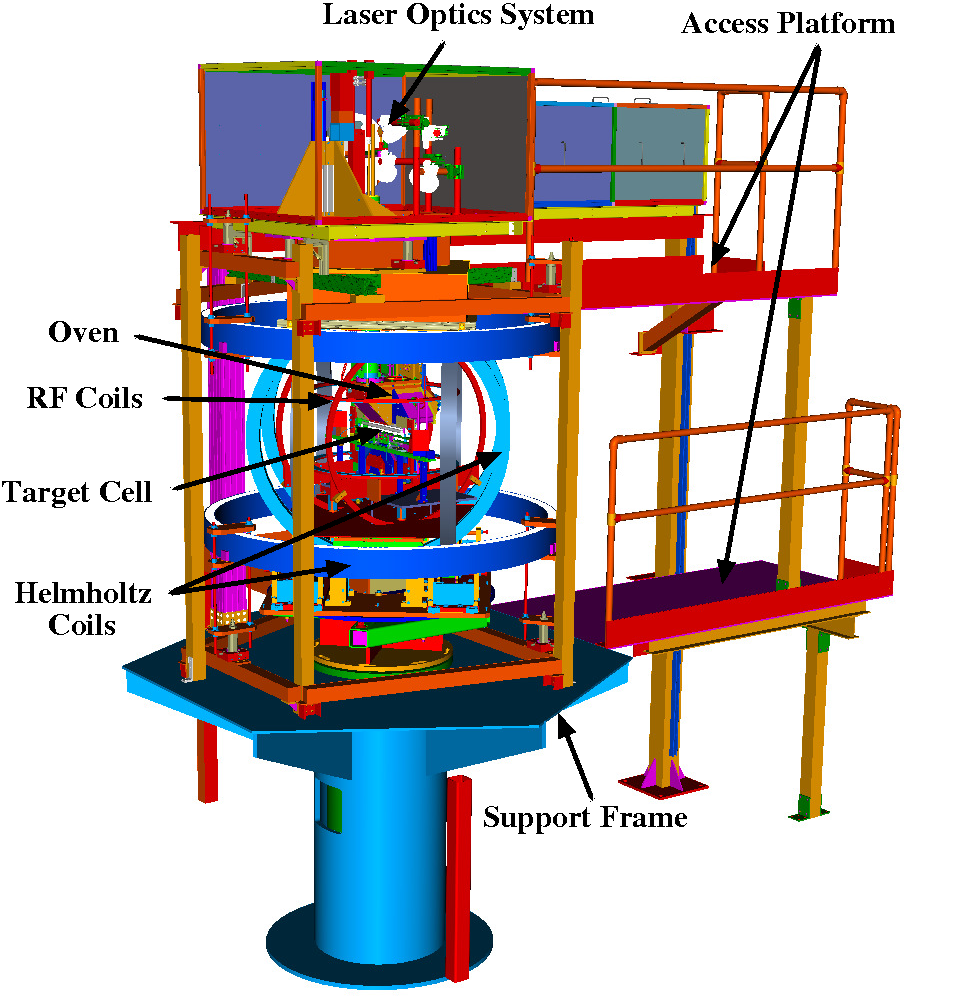}
\end{minipage}
\hfill
\begin{minipage}[t]{0.3\textwidth}
\hrule height 0pt
\caption{The ${}^3\mathrm{He}$ target system used in the 
E05-102 experiment. The electron beam enters at the
back of the picture and exits at the front. 
\label{fig_TargetSetup}}
\end{minipage}
\end{center}
\end{figure}

The original target system 
was upgraded~\cite{yi_menu2010} in 2008 to satisfy the more demanding needs 
of the Big Family experiments~\cite{e06010, e05102, e06014}. First, a larger 
target cell was considered. Then, a third set of Helmholtz coils was added to
provide  additional magnetic field in the vertical direction. The optical system 
together with all the lasers was extensively modified to allow pumping in all 
three directions. For that, a bigger oven, where target gets 
heated also had to be redesigned. Finally, new control and diagnostics
software was written. The structure of the target setup used in the
E05-102 experiment is shown in Fig.~\ref{fig_TargetSetup}.

\subsection{Spin-Exchange Optical Pumping}
\label{sec:SEOP}
The Hall A ${}^3\mathrm{He}$-target is polarized by the spin-exchange optical
pumping (SEOP) method, developed at SLAC~\cite{SLAC_optical_pumping}. This
is a two step method~\cite{huangPhD}, where atoms of alkali metal are first polarized
to produce a source of polarized electrons. In the second step, these polarized electrons 
collide and exchange their spin with the ${}^3\mathrm{He}$ nuclei.

\subsubsection{Optical pumping}
\begin{figure}[!hb]
\begin{center}
\begin{minipage}[t]{0.55\textwidth}
\hrule height 0pt
\includegraphics[width=1\textwidth]{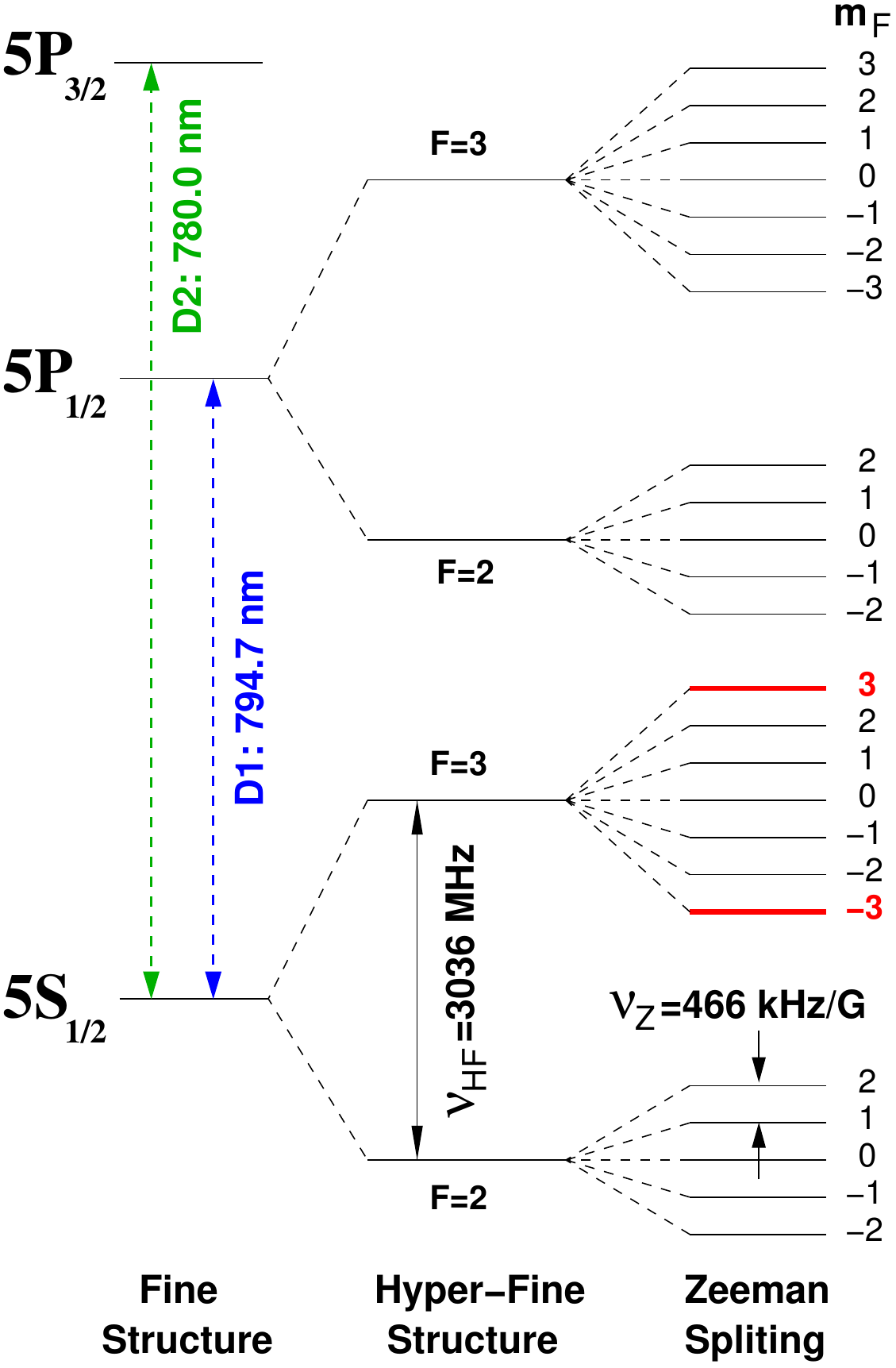}
\end{minipage}
\hfill
\begin{minipage}[t]{0.40\textwidth}
\hrule height 0pt
\caption{The energy levels of ${}^{85}\mathrm{Rb}$ with $I=\frac{5}{2}$. The valence band 
$5S_{\frac{1}{2}}$ and the first empty band $5P_{\frac{1}{2}}$ are shown. The degenerated 
energy levels are separated due to hyper-fine interaction and Zeeman effect. The sizes of both
splittings are given by $\nu_{\mathrm{HF}}$ and $\nu_{Z}$. The electrons from the
ground state are excited into the $5P_{\frac{1}{2}}$ by using circularly polarized light with
wave-length tuned to the rubidium $D_1$ transition. 
Depending on the orientation of the laser light polarization, 
selection rules ensure that either sub-level $F=3\,,m_{F}=3$ or sub-level $F=3\,\,m_{F}=-3$
get saturated. 
\label{fig_RbStates}}
\end{minipage}
\end{center}
\end{figure}

The polarized electrons are generated by optically pumping rubidium (Rb) atoms. 
The  Rubidium is an alkali metal with a single electron in the outer shell $(5S_{\frac{1}{2}})$. 
In this state the electron has intrinsic spin $S=\frac{1}{2}$ and angular momentum $L=0$ (S-state).
Its total angular momentum ($|L-S|\leq J \leq |L+S|$) is therefore also $J = \frac{1}{2}$.
In the naturally occurring rubidium there is approximately $72\,\mathrm{\%}$ of ${}^{85}\mathrm{Rb}$ 
and approximately $28\,\mathrm{\%}$ of ${}^{87}\mathrm{Rb}$. Our polarization technique is
based on the ${}^{85}\mathrm{Rb}$ with the nuclear spin $I = \frac{5}{2}$. The hyper-fine
interaction between nucleus and valence electron constrains the resulting angular momentum 
of a rubidium atom ($|I-J|\leq F \leq |I+J|$) and splits the degenerated energy 
states into two sub-levels $F=2$ and to $F=3$ (see Fig.~\ref{fig_RbStates}).
When the external magnetic field $\vec{B}$, is applied to the Rb atoms, two energy sub-levels get further
separated due to the Zeeman splitting into five and seven substates. The energy level of each 
substate can be obtained from the Breit-Rabi Hamiltonian:
\begin{eqnarray}
   \hat{H} = A_{\mathrm{HF}} \vec{I}\cdot\vec{S} + \mu_{e}\vec{S}\cdot\vec{B} - \mu_{I}\vec{I}\cdot\vec{B}\,,\nonumber  
\end{eqnarray}
where $\mu_e$ and $\mu_I$ are the electron and nuclear magnetic moments and $A_{\mathrm{HF}}$ is the 
hyper-fine coupling constant.

In the process of optical pumping, Rb is first heated to $\approx 513\,\mathrm{K}$ to produce alkali 
vapors. The vapor is then exposed to left-handed circularly polarized laser light with a wave-length of 
$794.7\,\mathrm{nm}$, which gives valence electrons exactly enough energy, to excite them from the 
$5S_{\frac{1}{2}}$ sublevels to the $5P_{\frac{1}{2}}$ sublevels (a $D_{1}$ transition). 
When the angular momentum of the incident photon is pointing in the same direction 
as the external magnetic field, the conservation of the
angular momentum sets the selection rule for the z-component of the total angular momentum 
$\Delta m_{F} = +1$, since in this configuration the photon brings exactly one unit of angular momentum. 
In this case, electrons from all sublevels can be excited to $5P_{\frac{1}{2}}$, except those in 
the $F=3, m_{F}=3$ state, since there is no $m_{F}=4$ state available for the $D_1$ excitation. 

\begin{figure}[!hb]
\begin{center}
\begin{minipage}[t]{0.55\textwidth}
\hrule height 0pt
\includegraphics[width=1\textwidth]{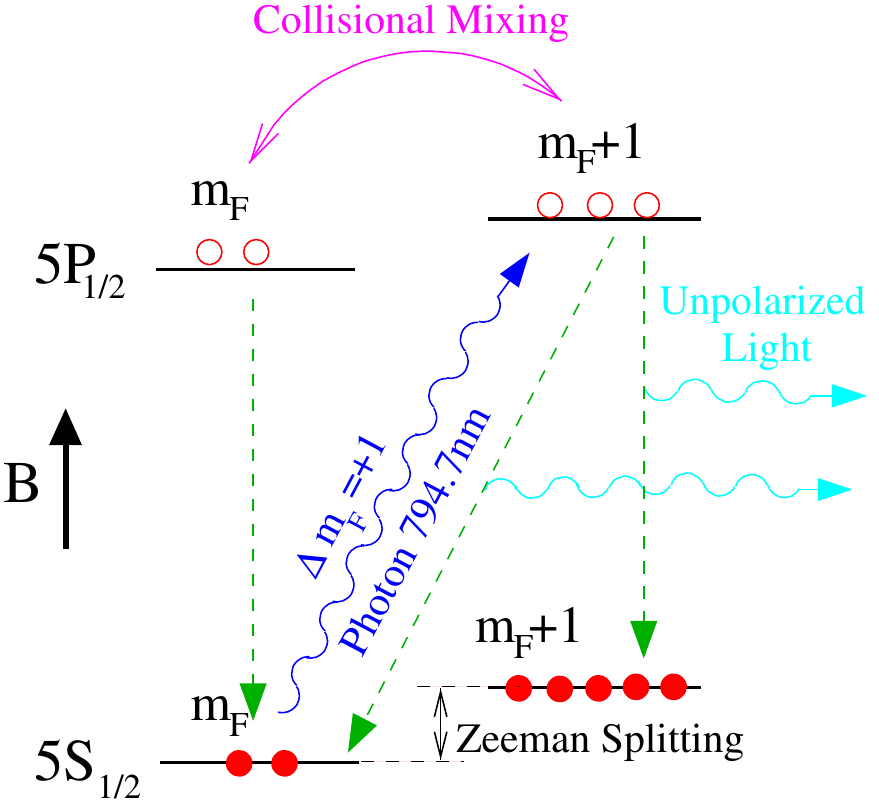}
\end{minipage}
\hfill
\begin{minipage}[t]{0.40\textwidth}
\hrule height 0pt
\caption{Optical pumping of rubidium by circularly polarized laser light. The 
selection rule allows only $\Delta m_{F} = +1$ transitions from the ground state to
the excited state. The selection rule does not apply for the relaxation of the excited
states, where transitions $\Delta m_{F} = 0,\> \pm1$ are allowed. The emitted unpolarized
light can depolarize Rb. Therefore nitrogen gas is used for quenching.  
\label{fig_OpticalPumping}}
\end{minipage}
\end{center}
\end{figure}

The excited electrons spontaneously decay from $5P_{\frac{1}{2}}$ states back to the 
$5S_{\frac{1}{2}}$ states. The relaxation happens through radiative and non-radiative transitions
to all sublevels  of the ground-state, also $F=3, m_{F}=3$. See Fig.~\ref{fig_OpticalPumping}. Once 
they are back in the $5S_{\frac{1}{2}}$, they can absorb another photon and get excited again.
By repeating this process, electrons will start gathering in the $F=3, m_{F}=3$ state, while
the other sublevels will get depopulated. The spin of the electrons in the $F=3, m_{F}=3$ state
is pointing in the direction of the external magnetic field and this is how Rb atoms get polarized.

However, when electrons in the $m_{F}$ excited state decay back to the $m_{F}-1$ level 
of the $5S_{\frac{1}{2}}$ states, they emit photons with the same wave-length as the pumping lasers. The emitted 
photons are no longer polarized along the quantization axis determined by the magnetic field. Therefore, 
the selection rule $\Delta m_{F} = 1$ no longer applies and these photons have the ability to excite 
the electrons from the $F=3\,,\>m_{F}=3$ ground state and depopulate it. This consequently decreases the pumping 
efficiency. To minimize this effect, a small amount of nitrogen gas ($\mathrm{N_2}$) is added to the gas
mixture. Nitrogen enables electrons to decay to the ground state without emitting 
photons~\cite{ZhengPhD, huangPhD}, by loosing energy 
through collisions of rubidium atoms with nitrogen molecules. The energy absorbed by nitrogen 
is then distributed among its rotational and vibrational degrees of freedom.  
This is known as non-radiative quenching and results in only about $5\,\mathrm{\%}$ of 
the excited electrons to decay by emitting a photon~\cite{ZhengPhD, SliferPhD}. 

The procedure described above can also be performed with laser light with a right-handed circular 
polarization. In this case the selection rule allows only $\Delta m_{F} = -1$ transition to 
the $5P_{\frac{1}{2}}$ excited states, resulting in the saturation of electrons in the
$F=3,\> m_{F} = -3$ substate. This option is chosen when we want to polarize the target 
in the opposite direction.


\subsubsection{Spin exchange} 

After Rb atoms are polarized, they transfer their polarization to
the ${}^3\mathrm{He}$ nuclei in the spin exchange process. This effect
was first observed in 1960 by Bouchiat~et~al.~\cite{bouchiat} They discovered
that the Rb atoms in the presence of the ${}^3\mathrm{He}$ gas
relax from the optically polarized  state to the depolarized state by flipping 
the ${}^3\mathrm{He}$ nuclei to a polarized condition. The spin transfer is
governed by the binary collisions between atoms~\cite{huangPhD, ZhengPhD}. In these 
collisions, spin-exchange occurs via the hyper-fine interaction between the Rb 
electron and ${}^3\mathrm{He}$ nucleus. The efficiency of this process at 
$190^\circ\mathrm{C}$ was determined to be $\approx2\,\mathrm{\%}$, which means that 
50 collisions are required to polarize one ${}^3\mathrm{He}$ nucleus~\cite{babcock2003}.
This procedure was utilized in the original Hall-A ${}^3\mathrm{He}$-target. 

\begin{figure}[!ht]
\begin{center}
\begin{minipage}[t]{0.60\textwidth}
\hrule height 0pt
\includegraphics[width=1\textwidth]{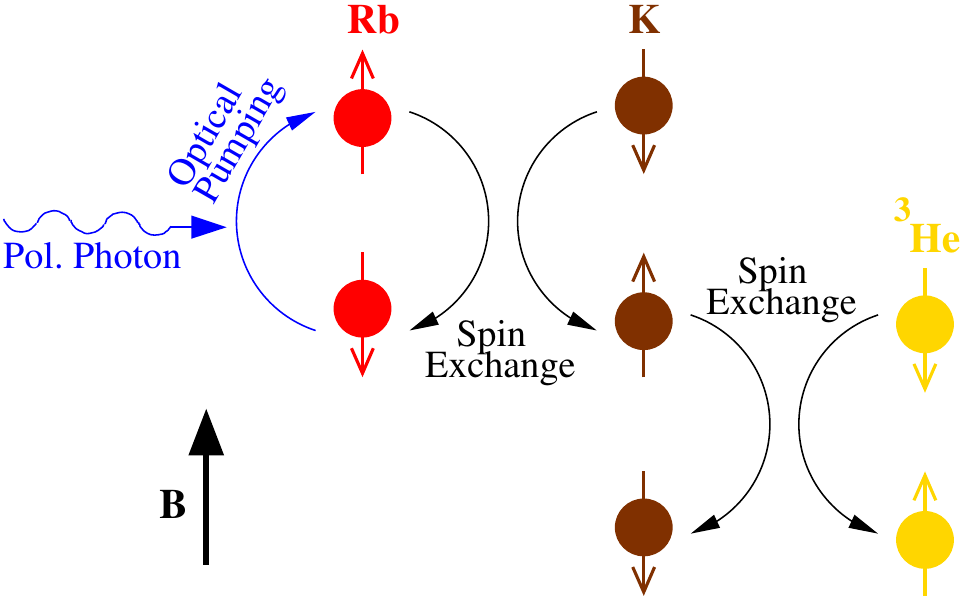}
\end{minipage}
\hfill
\begin{minipage}[t]{0.35\textwidth}
\hrule height 0pt
\caption{Hybrid spin exchange process in the $\mathrm{Rb}$-$\mathrm{K}$-${}^3\mathrm{He}$
gas mixture. Polarized Rb atoms quickly transfer their spin to K atoms, which then
interact with ${}^3\mathrm{He}$ nuclei and polarize it. Depolarized Rb atoms are
polarized again by optical pumping and the whole process is repeated. The spin 
of the polarized ${}^3\mathrm{He}$ nuclei is  parallel to the magnetic field.
\label{fig_SpinExchangeProcess}}
\end{minipage}
\end{center}
\end{figure}

\begin{figure}[!ht]
\begin{center}
\begin{minipage}[t]{0.55\textwidth}
\hrule height 0pt
\includegraphics[width=1\textwidth]{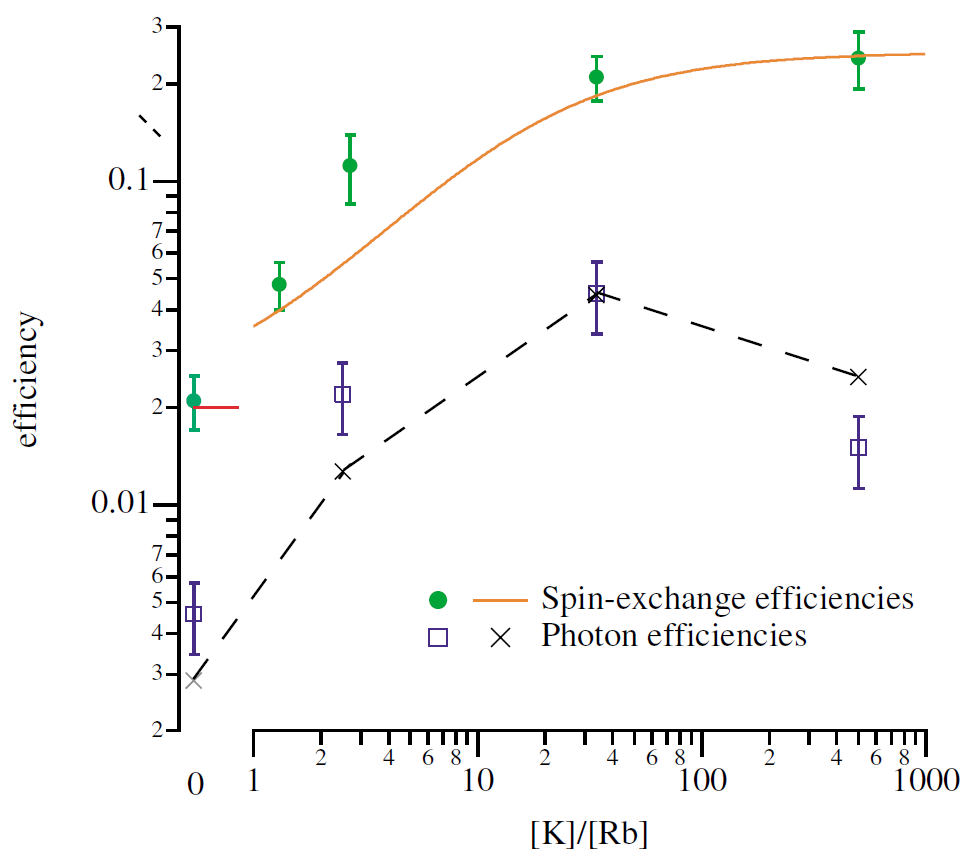}
\end{minipage}
\hfill
\begin{minipage}[t]{0.40\textwidth}
\hrule height 0pt
\caption{The spin exchange efficiency is the ratio between the rate
at which spin is transfered to the ${}^3\mathrm{He}$ and the rate at which
alkali atoms loose their angular momentum. This ratio is shown as a function 
of the K-Rb density ratio, at a temperature of 
$190\,\mathrm{{}^\circ C}$. The photon efficiency is the ratio between the number 
of polarized ${}^3\mathrm{He}$ nuclei and the number of photons absorbed
by the Rb vapors. Figure is taken from~\cite{babcock2003}.
\label{fig_SpinExchangeEfficiency}}
\end{minipage}
\end{center}
\end{figure}

However, it was later determined~\cite{baranga1998, babcock2003} that the efficiency for 
polarizing the ${}^3\mathrm{He}$ increases for an order of magnitude 
if potassium (K) is added to the gas mixture. See Fig.~\ref{fig_SpinExchangeEfficiency}.
In this case, Rb atoms first transfer their spin to K, which happens very quickly
and efficiently. The polarized K atoms then undergo the spin-exchange process with 
${}^3\mathrm{He}$. This is demonstrated in Fig.~\ref{fig_SpinExchangeProcess}.
This time only two collisions are necessary to polarize one
${}^3\mathrm{He}$ nucleus. Consequently, the target can be polarized much faster. This 
hybrid spin exchange optical pumping process has been incorporated in the
new design of the target which was used during the E05-102 experiment. 

\subsection{${}^3He$ Target Cell}
\begin{figure}[!ht]
\begin{center}
\begin{minipage}[t]{0.55\textwidth}
\hrule height 0pt
\includegraphics[width=1\textwidth]{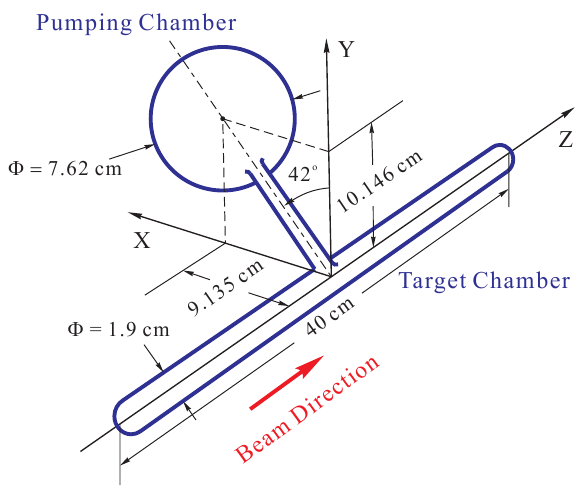}
\end{minipage}
\hfill
\begin{minipage}[t]{0.40\textwidth}
\hrule height 0pt
\caption{ Schematics of the target cell. 
It consist of two main parts: the cigar-shaped main target cell, which is exposed 
to the beam and where the reactions occur, and a spherical pumping chamber 
(positioned outside the beam line) where incident laser light hits the 
${}^3He$ gas and polarizes it.
\label{fig_3HeTargetCell}}
\end{minipage}
\end{center}
\end{figure}

The target cell used for the  E05-102 experiment is made of 
GE-180 glass~\cite{UVa_target_Group} and consists of two major
parts: the scattering chamber and the pumping chamber (see Fig.~\ref{fig_3HeTargetCell}).
In the scattering chamber polarized ${}^3\mathrm{He}$ nuclei interact with the 
electrons from the beam. It is $398.78\,\mathrm{mm}$ long and 
$18.75\,\mathrm{mm}$ wide~\cite{yawei_3HeTarget}. 
The average  thickness of the glass in 
the scattering chamber is $1.638\,\mathrm{mm}$, except at the 
entrance and the exit windows, where the glass is only 
$0.14\,\mathrm{mm}$ thick. To prevent the electron beam from breaking the cell,
the entrance and exit windows are constantly cooled with 
${}^4\mathrm{He}$ gas. In the pumping chamber, the polarization of 
the ${}^3\mathrm{He}$ gas takes place. It has a spherical shape with
$\approx 76.6\,\mathrm{mm}$ diameter. The pumping chamber resides in 
the heater oven, where it is heated to $\approx 270\,\mathrm{{}^\circ C}$,
and illuminated by polarized laser light. Both chambers are connected via
the transfer tube. This tube is $93.98\,\mathrm{mm}$ long with the diameter of 
$12.47\,\mathrm{mm}$ and is tilted with the respect to the vertical axis for 
$42^\circ$. 

The cell was constructed and tested in the University of Virginia Spin Physics 
Lab~\cite{UVa_target_Group} and was named Moss. Before the cell was sealed, it
was filled with the ${}^3He$ gas, five-to-one K/Rb mixture and a small amount of
nitrogen. The number density of the ${}^3He$ gas inside the target cell was 
measured to be $7.96\,\mathrm{amg}$\footnote{An amagat (amg) is a unit of number 
density. It is defined as the number of ideal gas molecules per unit volume 
at $p_0=101.325\,\mathrm{kPa}$ and $T_0 = 273.15\,\mathrm{K}$. The number density
of an ideal gas at pressure $p$ and temperature $T$ is calculated as: 
$n = \frac{p}{p_0}\frac{T_0}{T}\,\mathrm{amg}$.}~\cite{yawei_3HeTarget}.

\subsection{Targets for calibration}
\label{sec:ref_targets}

\begin{figure}[!ht]
\begin{center}
\includegraphics[width=0.85\textwidth]{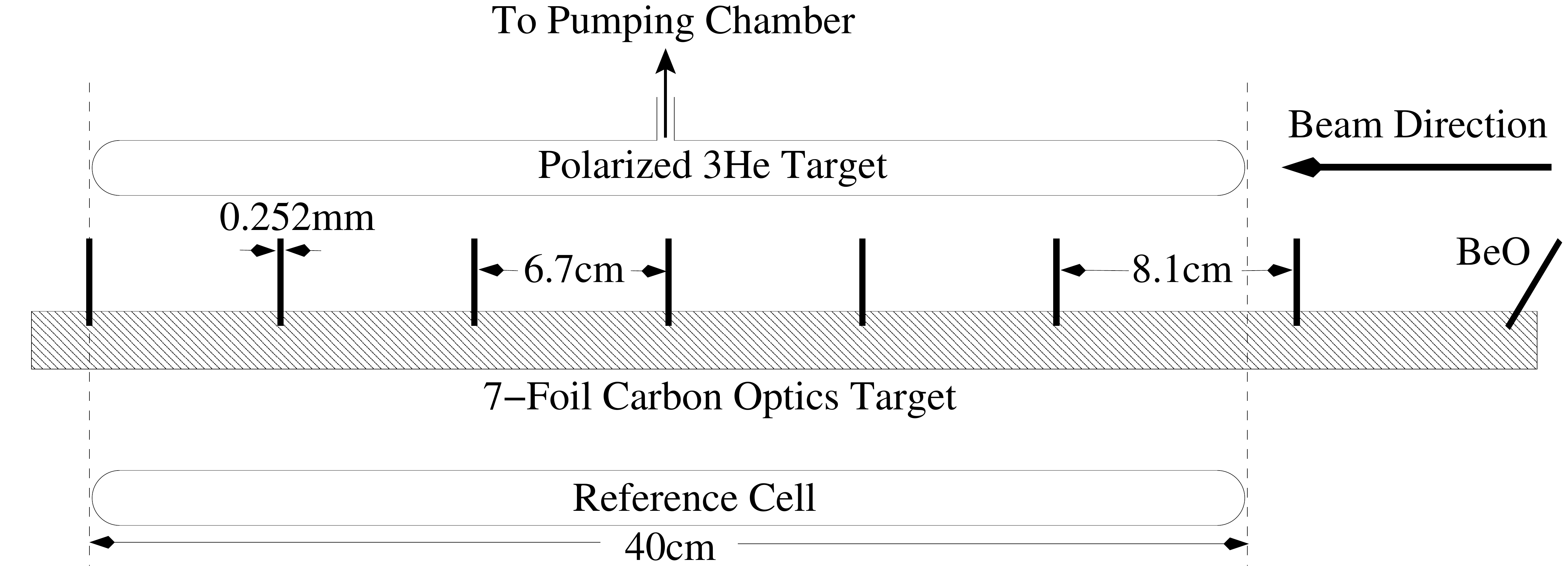}
\caption{The target system including the polarized
${}^3\mathrm{He}$ cell at the top, the multi-foil carbon optics target,
and the reference cell at the bottom.  The slanted BeO foil
is used for visual inspection of the beam impact point.\label{fig_tg_TGSYSTEM}}
\end{center}
\end{figure}

\begin{figure}[!ht]
\begin{center}
  \includegraphics[width=0.9\linewidth]{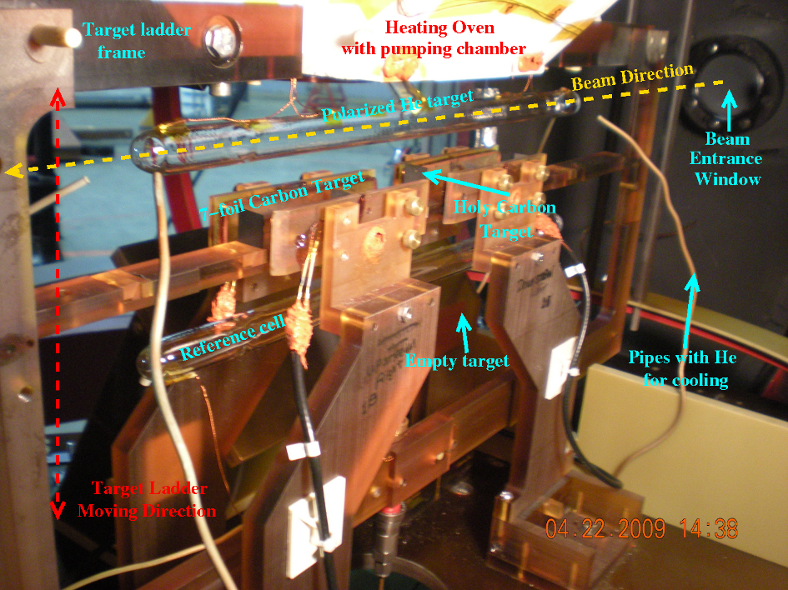}
\end{center}
\caption{The photograph of the target ladder assembly. For details, see text.
\label{fig_TargetLadder}}
\end{figure}

In addition to the polarized ${}^3He$ target cell, other targets were utilized 
for the calibration purposes (see Fig.~\ref{fig_tg_TGSYSTEM}). 
First is a  $40\,\mathrm{cm}$-long multi-foil 
carbon target, which consists of seven  $0.252\,\mathrm{mm}$-thick carbon foils 
mounted to a plastic frame which are preceded by a single slanted BeO foil, which 
served for visual inspection of the beam impact point. The carbon foil at 
the center is slightly higher with a hole inside this upper section 
(see Fig~\ref{fig_TargetLadder}). This is know as a ``holy target`` and
was used to test beam alignment.
Next to the multi-foil target, a dummy (reference) cell 
was installed. It consists of a glass tube, very similar to the scattering
chamber of the polarized ${}^3He$ target and can be either evacuated or 
filled with hydrogen, deuterium, nitrogen or unpolarized ${}^3He$ at a desired 
pressure. 


These extra targets are mounted together with the polarized ${}^3He$ target
on a plastic target ladder which can move in vertical direction by means 
of a remotely-controlled electro-motor. This gives us the ability to quickly
change between targets. At the bottom of the target ladder one slot is left
empty (known as ''empty target'').

\subsection{Holding Magnetic Field}
\label{Target_HoldingField}

\begin{figure}[!ht]
\begin{center}
\begin{minipage}[t]{0.55\textwidth}
\hrule height 0pt
\includegraphics[width=1\textwidth]{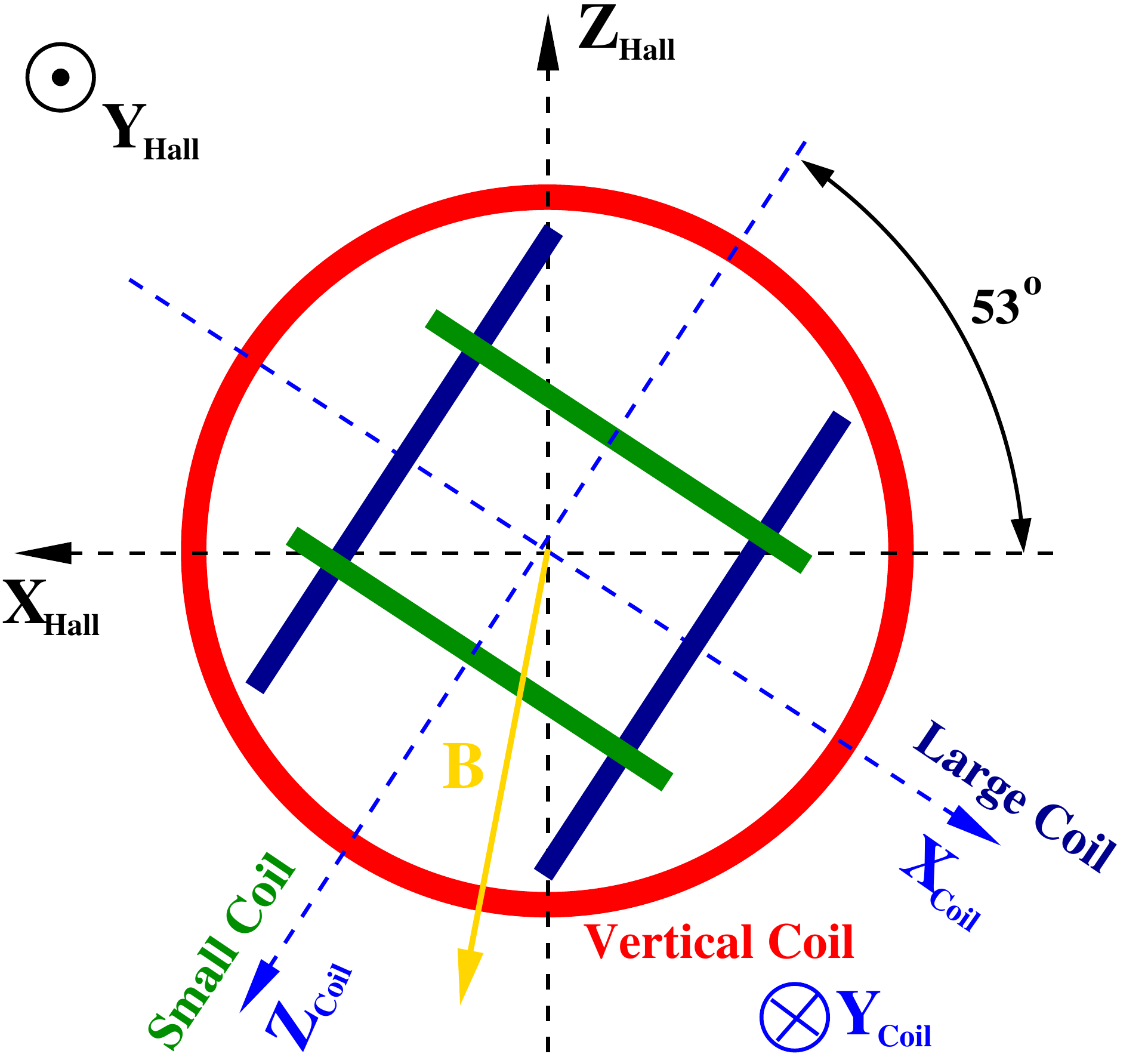}
\end{minipage}
\hfill
\begin{minipage}[t]{0.39\textwidth}
\hrule height 0pt
\caption{ Schematics of the Helmholtz coil orientations 
with respect to the beam direction.
\label{fig_HoldingField}}
\end{minipage}
\end{center}
\end{figure}

The E05-102 experiment required a flexible alignment of the target polarization vector 
parallel to (for $A_z$ asymmetry) and perpendicular to (for $A_x$ asymmetry) 
the direction of the momentum transfer $\vec q$. To achieve that, three Helmholtz 
coils were employed, which can rotate and hold the polarization in any given 
direction. See Fig.~\ref{fig_HoldingField}. The small and the large coils  
generate uniform magnetic fields in horizontal directions, while the vertical coils
generate vertical holding fields~\cite{HallA_wiki_3HeTarget}. The average strength
of the generated field is $25\,\mathrm{Gauss}$. The characteristics of these coils 
are listed in table~\ref{table_HoldingCoils}. The fields from these 
three coils are orthogonal and define the Coil Coordinate system, which is rotated 
for $37^\circ$ with respect to the the Hall System.

\begin{figure}[!ht]
\begin{center}
\includegraphics[width=\textwidth]{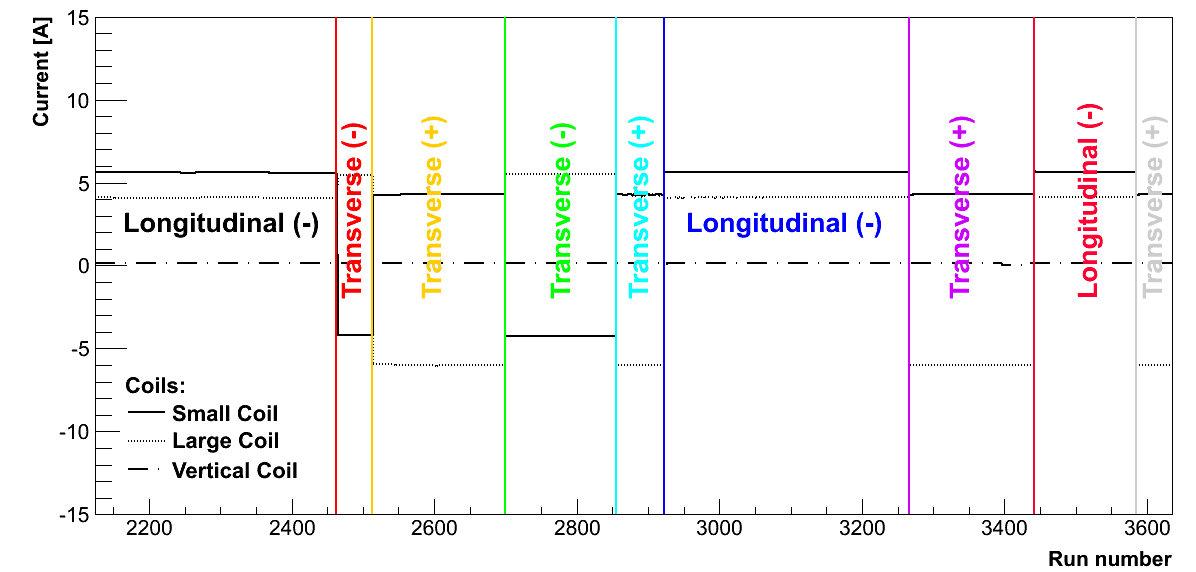}
\caption{The EPICS readouts of the currents in all three pairs of Helmholtz coils during 
the experiment. These current settings provided magnetic fields pointing in three
main (in-plane) orientations. The direction of the magnetic field is exactly opposite to the
target spin orientation due to negative magnetic momentum of the ${}^3\mathrm{He}$. 
\label{fig_TargetFieldOrientation}}
\end{center}
\end{figure}

Unfortunately, we were able to optically pump the target in only three directions: 
along the beam line (called longitudinal polarization), vertically, or perpendicularly 
(in-plane) to the beam line (called transverse direction). To produce the target 
polarization in an arbitrary direction, the target would have to be polarized first 
in one of the three primary directions and then rotated to a particular 
direction. Unfortunately, when exposed to the beam, the target polarization
decreases rapidly  without constant pumping and can be used for only a few hours. 
Afterwards, the polarization would drop too much, and the whole
procedure would have to be repeated. This takes approximately four to eight hours 
and in the meanwhile the polarized target can not be used. Hence our 
asymmetry measurements were performed with the target polarized in these
three primary directions. To provide a holding magnetic field in the longitudinal, 
transverse and vertical direction, precise currents must be set for each
pair of coils. Fig.~\ref{fig_TargetFieldOrientation} shows how the current settings for
each coil were changing during the experiment. The appropriate current values for 
each field orientation were determined with the Compass-calibration procedure and 
are gathered in table~\ref{table_HoldingCoils}. 

\begin{table}[!ht]
\begin{center} 
\caption{[Left] Main characteristics of the Helmholtz coils 
used to generate the holding field for the polarized ${}^3\mathrm{He}$
target. [Right] Coil current settings for generating 
the magnetic field in three principal directions. 
\label{table_HoldingCoils}}
\begin{minipage}[t]{0.35\textwidth}
\hrule height 0pt
\begin{tabular}{lcc}
\toprule
\multicolumn{3}{c}{{\bf Helmholtz coils properties}}\\
\midrule
Coil & Diameter & Number\\[-2pt]
     & $[m]$ & of turns\\
\midrule
Small & 1.27 & 256\\
Large & 1.45 & 272\\
Vertical & 1.83 & 355\\
\bottomrule
 \end{tabular}
\end{minipage}
\hfill
\begin{minipage}[t]{0.52\textwidth}
\hrule height 0pt
\begin{tabular}{lccc}
\toprule
\multicolumn{4}{c}{{\bf Coil current settings}}\\
\midrule
Field & $I_{\mathrm{Small}}$ & $I_{\mathrm{Vertical}}$ & $I_{\mathrm{Large}}$\\[-1pt]
Direction & $[A]$ & $[A]$ & $[A]$ \\
\midrule
Vertical & 0.028 & 13.020 & $-$0.212\\
Longitudinal & 5.593 & 0.168 & 4.061\\
Transverse $+90^\circ$ & 4.286 & 0.168 & $-$6.000 \\
Transverse $-90^\circ$ & $-$4.238 & 0.168 & 5.527\\
\bottomrule
 \end{tabular}
\end{minipage}
\end{center}
\end{table}

\subsection{Laser System}
\begin{figure}[!ht]
\begin{center}
\begin{minipage}[t]{0.65\textwidth}
\hrule height 0pt
\includegraphics[width=1\textwidth]{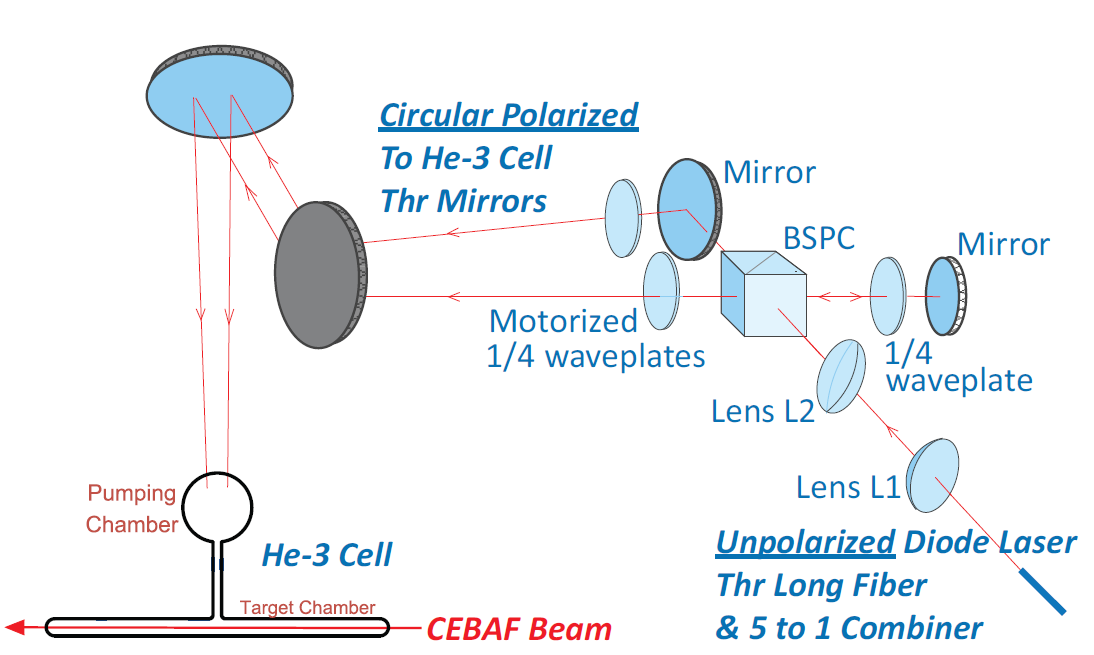}
\end{minipage}
\hfill
\begin{minipage}[t]{0.3\textwidth}
\hrule height 0pt
\caption{ Schematic diagram of the optics setup for the vertical pumping laser line.
Laser light from three IR-laser is first gathered by two lenses, then split and 
linearly polarized by BSCP. Finally series of mirrors and 1/4 wave-plates are used to 
circularly polarize light and transfer it to pumping chamber. 
Figure is taken from~\cite{huangPhD}.
\label{fig_LaserOpticsSystem}}
\end{minipage}
\end{center}
\end{figure}

A $\lambda=749.7\,\mathrm{nm}$ laser light needed for the optical pumping of the Rb vapors 
was provided by three Newport/Spectra-Physics COMET lasers with a line width of 
$\Delta \lambda = 0.2\,\mathrm{nm}$ for $90\,\mathrm{\%}$ of the total power~\cite{huangPhD}. 
Each laser provided a power of $\approx 30\,\mathrm{W}$.
They were installed in a dedicated room outside the experimental hall. A set of 
$75\,\mathrm{m}$ long optical fibers was utilized to transport the generated laser light
to the hall, where it was introduced to three optics setups mounted on the top
of the target system (see Fig.~\ref{fig_TargetSetup}). The purpose of these optics setups
was to produce circularly polarized light and deliver it to the pumping chamber  to polarize 
the target in one of the three possible pumping directions: vertical, horizontal transverse-to-beam
and longitudinal. The schematics of one such system is demonstrated in Fig.~\ref{fig_LaserOpticsSystem}.
It consists of a beam splitting  polarizing cube (BSPC), a series of lenses and mirrors, and motorized
quarter-wave plates. They allowed us to remotely flip the circular polarization of the light, thus changing 
flipping the rubidium states of $m_F = \pm 3$ and changing the polarization direction of ${}^3\mathrm{He}$.
The mirrors and the lenses were positioned such, that the size of the spot, hitting the
pumping chamber covered a large fraction of its surface area in order to maximize the pumping efficiency. 

The optical system and the whole target enclosure had to be optically isolated from the rest of the
experiment to prevent injury or damage caused by the high power infrared laser light. In addition, 
both the laser room and the optical setup enclosure were equipped with sensors and interlocked.

\subsection{Target Polarimetry}
\label{sec:NMR}

\begin{figure}[!ht]
\begin{center}
\includegraphics[width=\textwidth]{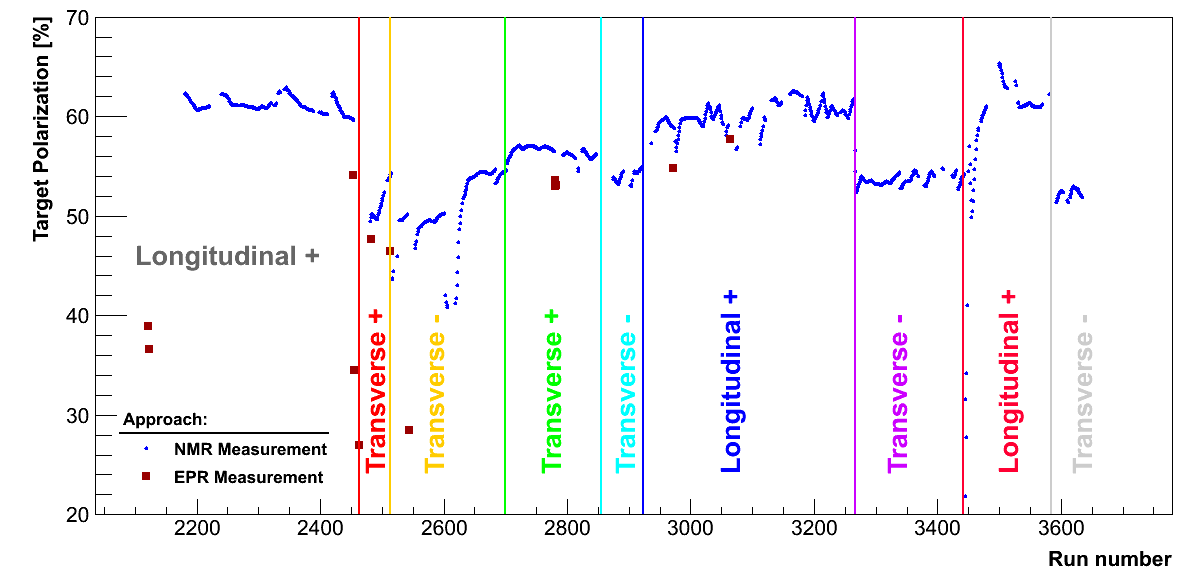}
\caption{ The results of the NMR and EPR polarimetry during the E05-102 experiment. 
NMR measurements were performed every four hours, while the EPR measurements were
done only after the change of the target spin orientation. The results show the polarization 
of the ${}^3\mathrm{He}$ nuclei inside the pumping chamber for each target spin orientation. To determine
the physically relevant polarization of the gas inside the scattering chamber, one 
needs to understand very accurately how the polarization falls off when going from the
pumping chamber down to the target cell. The analysis of the NMR and EPR data was performed by
Yawei Zhang~\cite{yawei_3HeTarget}. 
\label{fig_TargetPolarizationPlot}}
\end{center}
\end{figure}

During the experiment the degree of polarization of the target was monitored periodically.
Two different techniques were employed to measure the polarization: the nuclear magnetic resonance
(NMR) polarimetry and the electron paramagnetic resonance (EPR) polarimetry. The results of the 
polarization measurements are shown in Fig.~\ref{fig_TargetPolarizationPlot}. The NMR 
measurements were performed every four hours, or approximately after every four collected data sets. 
The polarization of the target for the data between two sequential measurements was determined by 
 linear interpolation.  The EPR measurements were not performed regularly, but were done only 
when the target spin orientation was changed. When flipping the target spin, the polarization of the 
target usually drops and requires some time to recover to its previous value. Since the EPR polarimetry 
was performed right after the spin flip, the measurements show smaller values of the 
polarization.
 
In the process 
of SEOP, a $\mathrm{{}^3\mathrm{He}}$ gas in the pumping chamber is being polarized. The maximum 
polarization then diffuses when moving down towards the scattering chamber through a thin pipe. 
Hence, the target polarization in the scattering chamber $P_{\mathrm{sc}}$ is smaller than 
the polarization in the pumping chamber $P_{\mathrm{pc}}$. 

The NMR and EPR polarimetry measure only the polarization of the gas inside the pumping chamber, while
the scattering processes are happening in the scattering chamber. Therefore a polarization loss factor 
$P_{\mathrm{tc}}/P_{\mathrm{pc}}$ is required to determine the physically relevant target 
polarization from the direct measurements inside the pumping chamber. For the considered target 
cell the polarization loss factor was estimated~\cite{huangPhD, yawei_3HeTarget} to be 
around $\left(89 \pm 5\right)\,\mathrm{\%}$.

\subsubsection{NMR Polarimetry}
With the NMR polarimetry, the target polarization is determined from the measurements of the ${}^3\mathrm{He}$ 
NMR signal during the spin reversal of the ${}^3\mathrm{He}$ nuclei through the adiabatic fast passage 
technique (AFP)~\cite{huangPhD}.

\begin{figure}[!ht]
\begin{center}
\begin{minipage}[t]{0.63\textwidth}
\hrule height 0pt
\includegraphics[width=1\textwidth]{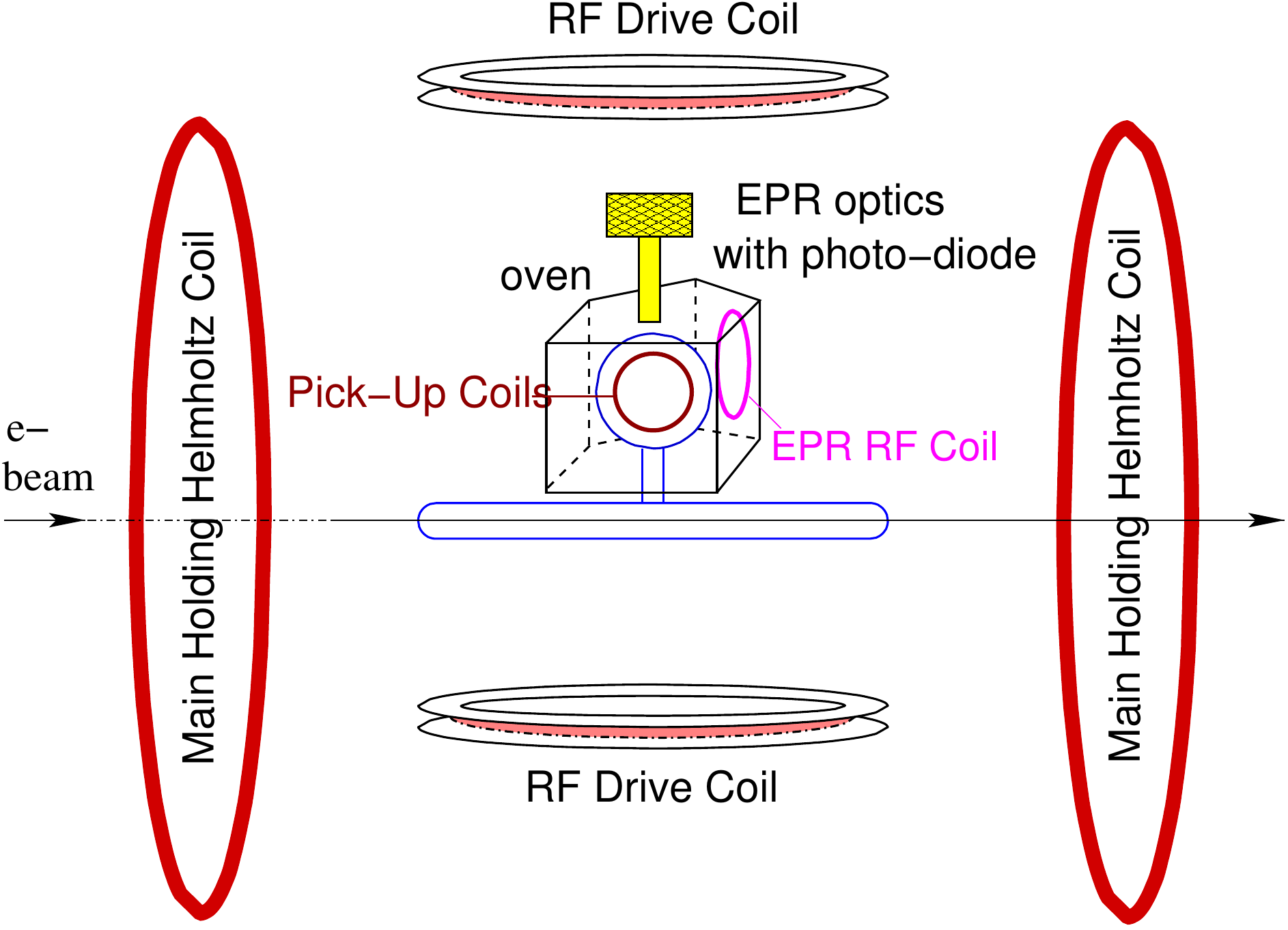}
\end{minipage}
\hfill
\begin{minipage}[t]{0.33\textwidth}
\hrule height 0pt
\caption{ The schematics of the apparatus used for the NMR and 
EPR polarimetry. The holding field is provided by the main holding Helmholtz coils. 
For the AFP spin flip, RF drive coils are used. The NMR signal is detected by the pick-up 
coils mounted on the scattering chamber. To depolarize Rb atoms in the process of the EPR 
polarimetry, EPR RF coil is considered, mounted in the vicinity of the pumping chamber. 
The emitted $D_2$ light is detected by the photo-diode. Figure is taken from~\cite{ZhengPhD}.
\label{fig_Target_NMR_EPR}}
\end{minipage}
\end{center}
\end{figure}

When a free particle with spin $\vec{S}$ is put into magnetic field $\vec{B}_{0}$, it experiences
a magnetic torque $\vec{T}_{\mathrm{M}}$,
\begin{eqnarray}
  \vec{T}_{M} = \gamma \vec{p}_m \times \vec{B}_{0} = \frac{d\vec{S}}{dt} = \frac{1}{\gamma}\frac{d\vec{p}_m}{dt}\>, 
  \label{eq_NMR_1} 
\end{eqnarray}
where $\vec{p}_m = \gamma \vec{S}$ is the magnetic moment of ${}^3\mathrm{He}$ and $\gamma = g_{{}^3\mathrm{He}}\mu_{N}/\hbar$.
Here $g_{{}^3\mathrm{He}}$ is the gyro-magnetic ratio for ${}^3\mathrm{He}$, $\mu_N$ is the nuclear magneton and $\hbar$ is the Planck's constant.
A magnetic field $\vec{B}_1$ perpendicular to $\vec{B}_0$ is added, which rotates with frequency $\vec{\omega}_0$ in the 
opposite direction to $\vec{B}_0$. In this case Eq.~(\ref{eq_NMR_1}) becomes
\begin{eqnarray}
  \frac{d\vec{p}_m}{dt} = \gamma \vec{p}_m \times \left(\vec{B}_0 + \vec{B}_1\right)\,. \label{eq_NMR_2}
\end{eqnarray}
In the cylindrical coordinate system which rotates synchronously with $\vec{B}_1$, Eq.~(\ref{eq_NMR_2}) can be 
rewritten as~\cite{goldstein}:
\begin{eqnarray}
  \frac{d\vec{p}_m}{dt} = \gamma \vec{p}_m \times \left(\left(B_0 - \frac{\omega_0}{\gamma}\right)\hat{z} + B_1\hat{r}\right)\>, \nonumber
\end{eqnarray}
where the unit vector $\hat{z}$ is pointing along the holding field $\vec{B}_0$, and the unit vector $\hat{r}$ is perpendicular to $\hat{z}$. 
In the case of the ${}^3\mathrm{He}$ target, $\vec{B}_0$ represents the holding field, provided by the three main Helmholtz coils, while 
the rotational transverse field with magnitude $B_1$ is provided by the two RF-Coils (see Fig.~\ref{fig_Target_NMR_EPR}).  
These coils generate  linearly oscillating magnetic field $B_{\mathrm{RF}} = 2B_1\cos(\omega_0 t)$.
This field can be decomposed into two components with opposite rotations $(\pm \omega_0\hat{z})$. In this
procedure we use negatively oriented component $-\omega_0\hat{z}$ of rotation. The positively
oriented component $+\omega_0 \hat{z}$ does not play a role during the AFP and will not be considered here.

The target spin $\vec{S}$ is primarily oriented parallel to the holding field $\vec{B}_0$. When doing NMR polarimetry with
the AFP, the magnetic field $\left(B_0 - \frac{\omega_0}{\gamma}\right)$ is changed from 
$\left(B_0 - \frac{\omega_0}{\gamma}\right) \ll -B_1$, passing zero, to $\left(B_0 - \frac{\omega_0}{\gamma}\right) \gg B_1$.
Consequently, the magnetic moment of the ${}^3\mathrm{He}$ nucleus rotates with the field. The field must be changing slowly to keep the target 
spin in the same quantum state (adiabatic limit), but must be fast enough to prevent the spin relaxation of ${}^3\mathrm{He}$. 
The rotating magnetic moment induces a signal in the pick-up coils. These coils are oriented perpendicularly 
 to the RF coils (see Fig.~\ref{fig_Target_NMR_EPR})  to minimize the detection of the RF signal which could 
distort the measurement.  The signal induced in the pick-up coils is proportional to the average 
polarization $P_{{}^3\mathrm{He}}$ of the ${}^3\mathrm{He}$ target and is given by:
\begin{eqnarray}
  S_{\mathrm{NMR}} = A_{\mathrm{NMR}}^{\mathrm{Calib}}\cdot P_{{}^3\mathrm{He}}\cdot\frac{B_1}{\sqrt{\left(B_{0} - \frac{\omega_0}{\gamma}\right)^2 + B_{1}^2}}\,.\nonumber
\end{eqnarray}
Unfortunately the NMR measurement provides only a relative value of the target polarization. To compute the absolute
polarization, the calibration constant  $A_{\mathrm{NMR}}^{\mathrm{Calib}}$ needs to be determined. It is
obtained from the cross calibration with the EPR measurements. 

The sweep of the magnetic field $\left(B_0 - \frac{\omega}{\gamma}\right)$ can be achieved by either changing 
the holding field $B_0$ or by changing the frequency of the rotating field $\omega_0$. The first technique
is knows as the AFP field sweep, while the second one is called the AFP frequency sweep. Both approaches were considered
in the E05-102 experiment. However, for the regular polarization measurement only the frequency sweep measurements
were preformed due to their lower signal to noise ratio. 

\subsubsection{EPR Polarimetry}

In the presence of external holding magnetic field $B_0$, the energy sub-levels of rubidium
are separated due to the Zeeman effect (see Fig.~\ref{fig_RbStates}). The energy (frequency) 
splitting is directly proportional to the size of the magnetic fields:
\begin{eqnarray}
  \nu_{Z} = \kappa_{Z}B_0\,, \nonumber
\end{eqnarray}
where $\kappa_Z = 0.466\,\mathrm{MHz/G}$. When the target is polarized, the ${}^3\mathrm{He}$ spins create an 
additional magnetic field $\delta B_{{}^3\mathrm{He}}$. This field is small but strong 
enough to create a detectable change $\Delta \nu_{z}$ in the energy splitting of the Rb sub-levels.
The change in the frequency depends on the relative orientation of the target spin with the respect 
to the holding field:
\begin{eqnarray}
  \nu_{Z}^{\pm} = \nu_{Z} \pm \Delta \nu_{Z} = \kappa_{Z}(B_0 \pm \delta B_{{}^3\mathrm{He}})\,. \nonumber
\end{eqnarray}
The main idea of the EPR polarimeter is to measure this deviation $\Delta \nu_Z$, which 
is directly proportional to the target polarization $P_{{}^3\mathrm{He}}$~\cite{yawei_EPR_report}:
\begin{eqnarray}
  \Delta \nu_{Z} = \frac{\mu_0}{3}\frac{d\nu_{\mathrm{EPR}}}{dB}\kappa_{0}p_{m}
  n_{\mathrm{pc}}P_{{}^3\mathrm{He}}\,. \nonumber
\end{eqnarray}
Here $\mu_0$ is the magnetic permeability of the vacuum, $\vec{p}_{m}$ is the magnetic moment of ${}^3\mathrm{He}$,
$n_{\mathrm{pc}}$ is the ${}^3\mathrm{He}$ number density in the pumping chamber, while $\frac{d\nu_{\mathrm{EPR}}}{dB}$ 
and $\kappa_{0}$ are parameters obtained from atomic physics experiments~\cite{yawei_EPR_report}.

The frequency difference $\Delta \nu_{Z}$ is determined 
by measuring the Zeeman frequencies $\nu_Z^{\pm}$ for both positive and negative spin orientations and subtracting 
the results ($\Delta \mu_F = \frac{1}{2}\left(\nu_{Z}^{+} - \nu_{Z}^{-}\right)$). To flip the target spin, 
the AFP technique is again utilized, using the NMR RF coils. Typically four spin flips were performed in each EPR
measurement.

When the target is being polarized in positive the direction, the vast majority of the Rb atoms is found in 
the ($F=3$, $m_{F} = 3$) state. By introducing a small oscillating magnetic field, 
the electrons from ($F=3$, $m_{F} = 3$) sub-level
can be kicked to the ($F=3$, $m_{F} = 2$) state and consequently depolarize rubidium. This perturbation
magnetic field is provided by a small EPR-RF-coil attached to the pumping chamber (see Fig.~\ref{fig_Target_NMR_EPR}). 
The electrons in the state ($F=3$, $m_{F} = 2$) are then excited again in order to re-polarize rubidium. In the
process of de-excitation of these electrons light is emitted. The light output is detected with a 
photo-diode. Since a small amount of light emitted from the $D_1$ transition could not be detected due to the
intense incoming laser light, the diode was configured to detect the $D_2$ transition light with 
$\lambda = 780\,\mathrm{nm}$. The light from $D_2$ transition corresponds to the Rb atom decaying from the 
$5P_{\frac{3}{2}}$ to the $5S_{\frac{1}{2}}$ state, and the light emission is most intense when
the frequency of the perturbation field corresponds to the frequency of the Zeeman splitting $\nu_Z$.
These transitions are possible because the electrons from the excited states $5P_{\frac{1}{2}}$ can be 
thermally excited to the $5P_{\frac{3}{2}}$ states~\cite{SliferPhD}.

Instead of using the EPR RF magnetic field to directly depolarize Rb atoms, a magnetic field with different frequency 
could be utilized to first depolarize K atoms. In this case, rubidium would become depolarized through the fast 
spin exchange between both alkali metals. In the E05-102 experiment both approaches were considered.

Since the EPR polarization measurement determines the absolute value of the ${}^3\mathrm{He}$ target polarization,
it could also be used to calibrate the NMR polarimeter. Hence, for each AFP pass during the EPR measurements,
an NMR signal was also recorded at the same running conditions. By comparing the size of the NMR signal to
the polarization determined with the EPR polarimeter, the NMR calibration constant was
determined~\cite{yawei_3HeTarget}.

\section{High Resolution Spectrometers}
The core components of the Hall A equipment are two almost identical 
High Resolution Spectrometers (HRS)~\cite{alcorn}. The first spectrometer is positioned
on the left side of the beam line and is accordingly called Left HRS (HRS-L), 
while the  other one is positioned on the right side
(Right HRS, HRS-R). See Fig.~\ref{fig_HALLA} for details. 
Experiment E05-102 employed only HRS-L, in coincidence with the BigBite 
spectrometer, for detection of scattered electrons. 

\begin{figure}[!ht]
\begin{center}
\begin{minipage}[t]{0.6\textwidth}
\hrule height 0pt
\includegraphics[width=1\textwidth]{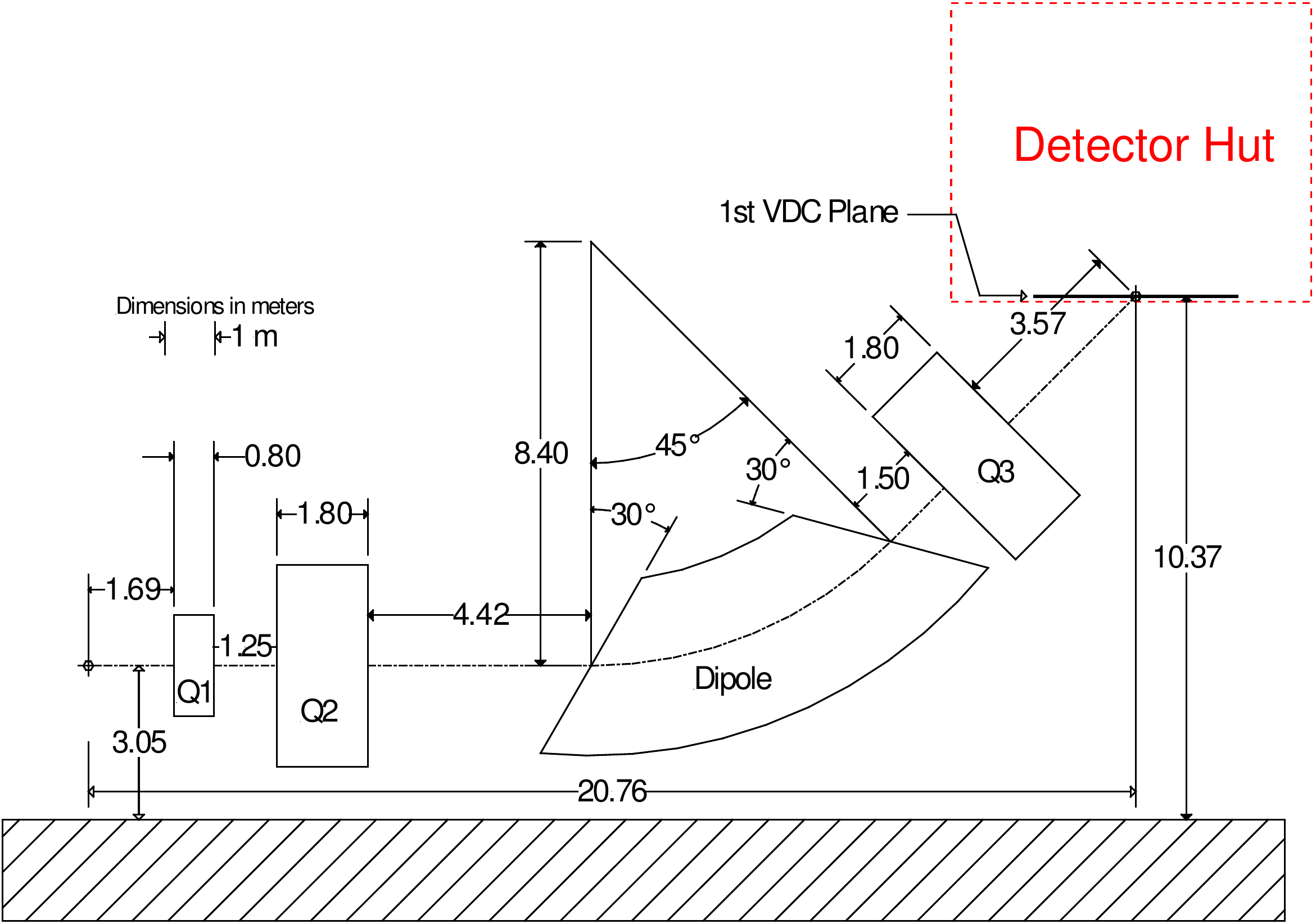}
\end{minipage}
\hfill
\begin{minipage}[t]{0.35\textwidth}
\hrule height 0pt
\caption{The layout of the High Resolution Spectrometer, showing the dimensions
and positions of three quadrupole magnets and the dipole~\cite{alcorn}. 
The scheme also shows the position of the detector hut with the detector package. 
After passing the magnets, the particles first hit the VDCs (tracking detectors). 
\label{fig_HRS_layout}}
\end{minipage}
\end{center}
\end{figure}

The basic layout of the spectrometer is shown in Fig.~\ref{fig_HRS_layout}.
It consists of three quadrupole and one dipole magnet in a QQDQ configuration,
with a central bending angle of $45^\circ$ in the vertical direction and an
optical length of $23.4\,\mathrm{m}$. The selected magnet configuration 
enables an extensive momentum range of the spectrometer from $0.8$ to 
$4.0\,\mathrm{GeV}/c$, large acceptance in both angle and momentum, and good 
position, angular and momentum resolutions. The main design characteristics of 
HRS-L are gathered in Table~\ref{table_HRSL}.
\begin{table}[!ht]
\begin{center} 
\begin{minipage}[t]{0.7\textwidth}
\hrule height 0pt
\begin{tabular}{ll}
\toprule
\multicolumn{2}{c}{\textbf{HRS-L main characteristics}} \\
\midrule
Configuration  & QQDQ\\
Optical length & $23.4\,\mathrm{m}$\\
Bending angle  & $45\,{}^{\mathrm{o}}$ (vertical direction)\\[4pt]
Momentum range & $0.3-4.0\,\mathrm{GeV}/c$ \\ 
Momentum acceptance & $-4.5\,\mathrm{\%} \leq \delta_{Tg} \leq 4.5\,\mathrm{\%}  $\\ 
Momentum resolution & $2.0\times 10^{-4}$\\[4pt]
Angular range & $12.5-150^{\circ}$\\[4pt]
Angular acceptance &  \\
\,\,\,\,\,\,Horizontal     & $\pm 30\,\mathrm{mrad}$ \\
\,\,\,\,\,\,Vertical       & $\pm 60\,\mathrm{mrad}$\\[4pt]
Angular resolution &     \\
\,\,\,\,\,\,Horizontal     & $0.5\,\mathrm{mrad}$ \\
\,\,\,\,\,\,Vertical       & $1.0\,\mathrm{mrad}$ \\[4pt]
Transverse length acceptance & $\pm 5\,\mathrm{cm}$\\
Transverse position resolution & $1\,\mathrm{mm}$\\
\bottomrule
\end{tabular}
\end{minipage}
\hfill
\begin{minipage}[t]{0.29\textwidth}
\hrule height 0pt
\caption{Main characteristics of the High Resolution Spectrometer. The 
resolutions are given in terms of their FWHM values. 
\label{table_HRSL}}
\end{minipage}
\end{center}
\end{table}

All four magnets in the HRS are superconducting and are cooled with super-critical 
helium gas to $\approx 4.5\,\mathrm{K}$. They can be 
energized with currents that exceed $1.3\,\mathrm{kA}$, producing magnetic
fields (for the dipole magnet) in excess of $1.5\,\mathrm{T}$. The magnetic field
in the dipole is monitored by three NMR field probes positioned inside
the magnet and provide a very precise field reading at the $10^{-5}$ level.
In addition to the NMR probes, the dipole magnet is also equipped with a 
Hall probe for supplementary field measurements. In particular, they are used
for low field measurements ($B<0.17\,\mathrm{T}$) where NMR probes start to fail. 
Hall probes are also  used for magnetic field measurements inside quadrupole magnets, 
since they are not equipped with the NMR sensors. Unfortunately Hall 
probes are known for their long-term instability and can not be used as an
absolute reference. Therefore, the fields in the quadrupole 
magnets are set based on their current settings and not field read-back from Hall probes.

Spectrometer magnets are mounted on a rigid metal support frame, which allows 
the spectrometer to be positioned azimuthally around the  central Hall A pivot,
and is designed such that relative positions of the magnets and the detector hut 
remain constant regardless of the spectrometer position.  The motion of 
the $10^{6}\,\mathrm{kg}$ heavy spectrometer is realized by means 
of eight-wheeled boogies, which are mounted at the bottom of the spectrometer. 
One wheel of each boogie is driven by a servomotor through a gear reducer and is 
controlled through a computer
interface. This allows the spectrometer to be remotely positioned to any angle
between $12.5^{\circ}$ and $150^{\circ}$. The maximum speed of motion is $3^{\circ}$ per
minute. The exact angular position of the spectrometer is determined from the 
floor marks underneath the HRS with a resolution better than $\pm 0.14\,\mathrm{mrad}$.

\subsection{Detector package}

The HRS detector package, together with all the Data-Acquisition electronics
is located behind the last quadrupole magnet, inside the detector hut 
at the back of the spectrometer (see Fig.~\ref{fig_HRS_layout}). 
The detector hut protects the detector package  
against damaging radiation from all directions and is made of $10\,\mathrm{cm}$
thick steel frame with a $5\,\mathrm{cm}$ lead layer inside and a layer of
concrete outside~\cite{alcorn}. Individual detectors are mounted on a 
retractable frame and can be moved outside of the detector hut for repair 
or reconfiguration. 

\begin{figure}[!ht]
\begin{center}
\includegraphics[width=0.49\textwidth]{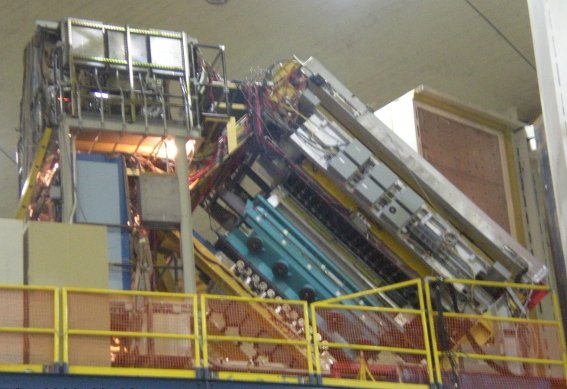}
\hfill
\includegraphics[width=0.49\textwidth]{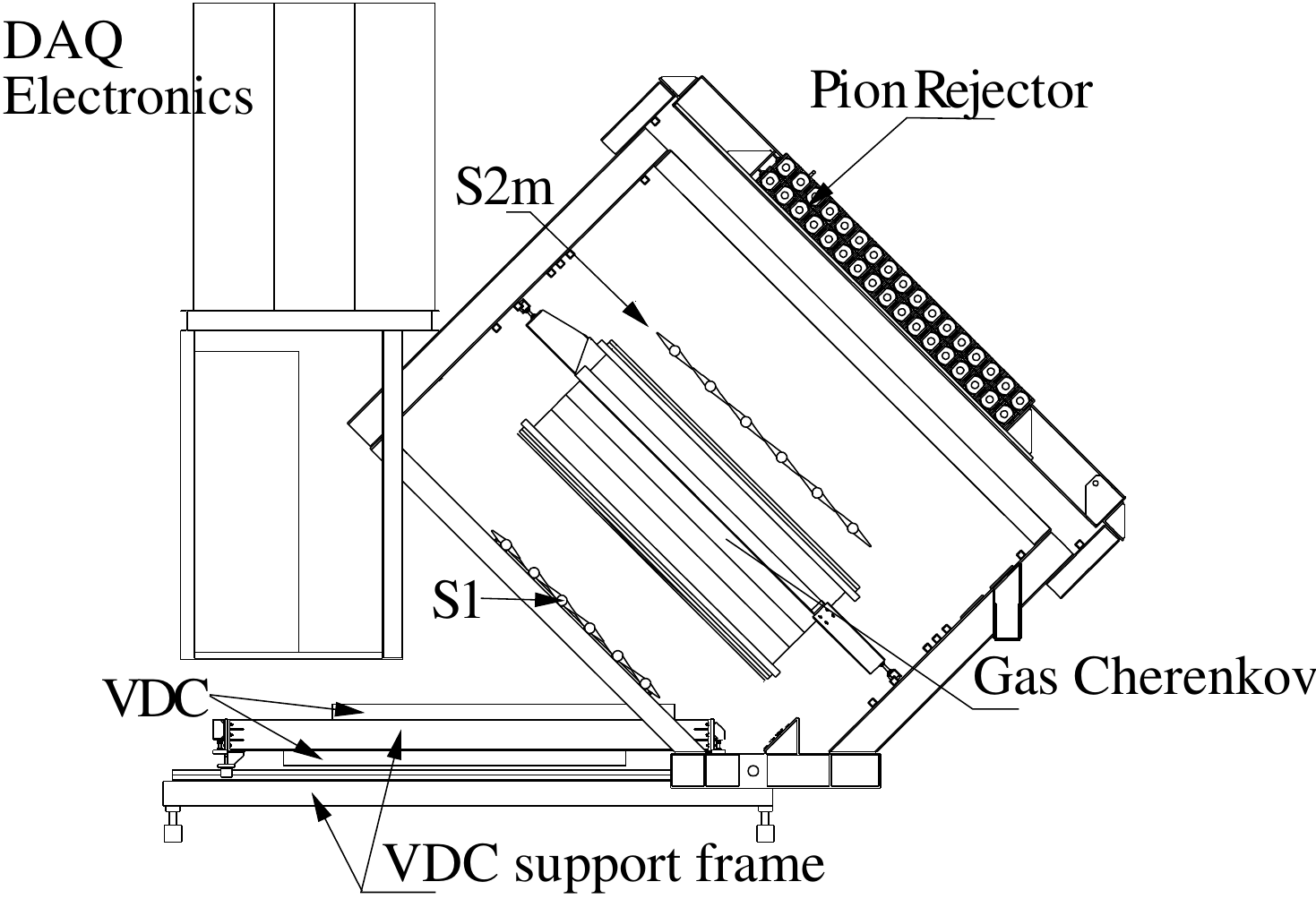}
\caption{[Left] The detector package in from of the spectrometer's
detector hut during the reconfiguration. [Right] Schematic layout of 
the detector stack considered for E05-102 experiment. Positions of
individual detectors and DAQ electronics are clearly demonstrated. 
\label{fig_HRS_DetectorPackage}}
\end{center}
\end{figure}

The purpose of the detector package is to perform various functions in the 
characterization of the charged particles passing through the spectrometer.
The configuration of the HRS-L detector package 
is not fixed, but can be customized to the needs of a particular experiment. 
For the E05-102 experiment, the detector package consisted of 
vertical drift chambers (VDCs) for tracking, two scintillation detectors
(S1 and S2m) for triggering and time-of-flight measurements, a Cherenkov 
detector for identification of electrons  and two Shower detectors for 
additional particle identification. The position of each detector 
in the detector package is shown in Fig.~\ref{fig_HRS_DetectorPackage}.

\subsubsection{Vertical Drift Chambers}

Particle tracking for HRS is provided by two Vertical Drift Chambers
(VDCs) which have the ability to reconstruct the particle's trajectory  
at the focal plane with spatial resolution $\sigma_{x,y} \approx 100\,\mathrm{\mu m}$
and angular resolution $\sigma_{\theta,\phi} \approx 0.5\,\mathrm{mrad}$.
This is essential for precise reconstruction of particle's momentum vector at the target. 

\begin{figure}[!ht]
\begin{center}
\includegraphics[width=0.8\textwidth]{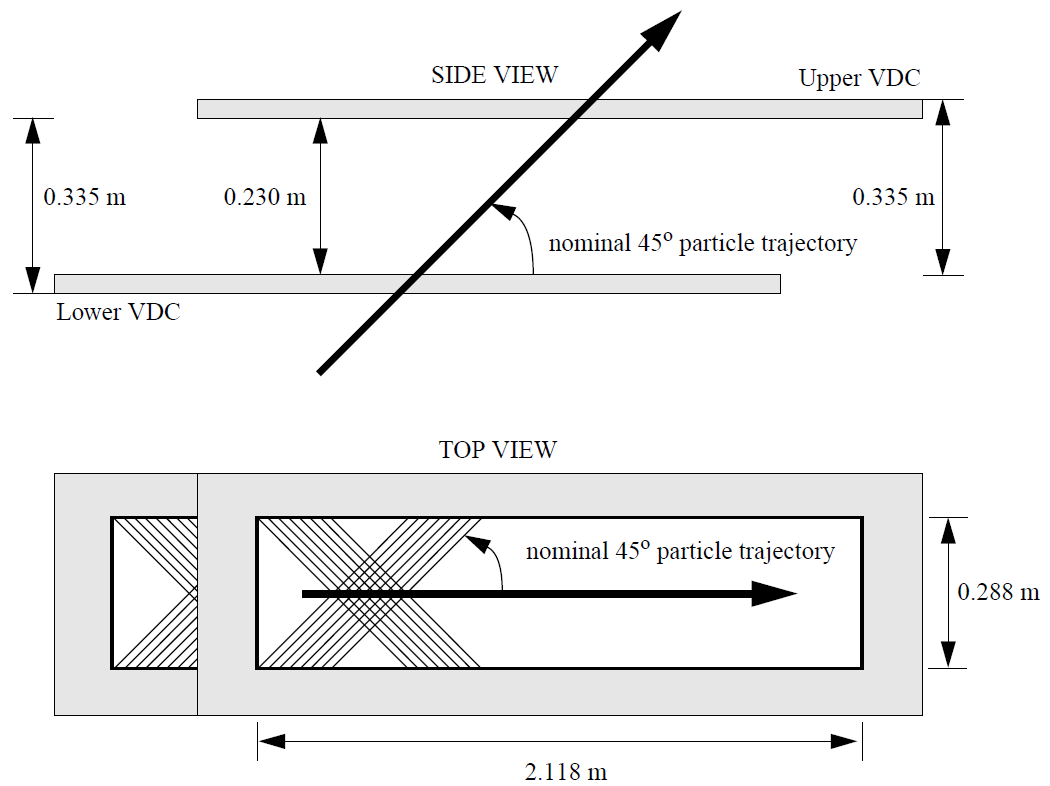}
\caption{Geometrical properties of the VDCs. The size of the rectangular aperture 
of each chamber is $2118\,\mathrm{mm} \times 288\,\mathrm{mm}$. Each chamber consists 
of two orthogonal planes of wires (U and V) that are inclined at an angle
of $45^\circ$ with respect to the spectrometer's dispersive direction. The VDCs lie
in the laboratory horizontal plane. The vertical spacing between the matching 
wire planes (U1 and U2 or V1 and V2) is $335\,\mathrm{mm}$~\cite{alcorn,fissum}.
\label{fig_HRSL_VDCs}}
\end{center}
\end{figure}

The two VDCs were designed and constructed by the Nuclear Interactions Group 
at MIT-LNS in conjunction with the Jefferson Lab~\cite{fissum}. The active
area of each chamber has a size of $2118\,\mathrm{mm} \times 288\,\mathrm{mm}$ and
consists of two planes of wires in a UV configuration~\cite{alcorn}. 
The sense wires in each plane are perpendicular to each other and are inclined
at an angle of $45^{\circ}$ with respect to both the dispersive and non-dispersive 
direction. There are a total of 368 sense wires in each plane, spaced $4.24\,\mathrm{mm}$
apart. Each chamber also has three gold-plated Mylar windows, two located
on each side of the chamber and one in-between U and V wire-planes. All three Mylar 
windows are connected to high voltage at about $-4\,\mathrm{kV}$ and together with
grounded signal wires create an almost uniform drifting field around each plane of 
wires. The VDCs are filled with a gas mixture of argon ($62\,\mathrm{\%}$)
and ethane ($38\%$), which flows through the chambers at approximately $6\,\mathrm{l}$
per hour. 

Vertical Drift Chambers are positioned at the bottom of the detector package and
are the fist detectors that particles hit after they exit the magnets. Chambers lie
in the laboratory horizontal plane and are separated by about $0.230\,\mathrm{m}$.
The lower VDC is as close as possible to the spectrometer focal plane. The central
trajectory crosses the wire planes at the angle of $45^{\circ}$.

When a particle passes the VDCs, it ionizes the gas surrounding each wire-plane. 
The electrons from the ionized gas then travel to the sense wires along the path
of least time. The drift velocity of the electrons is approximately $50\,\mathrm{\mu m/ns}$.
The drifting time is measured by the TDC modules with the resolution 
of $0.5\,\mathrm{ns/channel}$. The TDC readout is started by the signals in the hit
wires, and a common stop is provided by the triggering circuit. Typically five wires 
have non-zero TDC readings per event. The measured time is then converted into the 
vertical distance between the hit position and the signal wire. The position 
resolution in each plane is $225\,\mathrm{\mu m}$ (FWHM).  The distance information from all hit 
wires in both wire-planes is then considered in a linear fit to determine the two-dimensional 
position of a track at the entrance to the VDC. Finally, by combining the position information 
from both chambers, the position and the angles of a particle track at the focal plane can be 
reconstructed.  

\subsubsection{Scintillation Detector}
The HRS-L employs two scintillation detectors~\cite{alcorn}, called S1 and S2m, for triggering
and precise time of flight measurements ($\sigma_{\mathrm{TOF}} \approx 0.5\,\mathrm{ns}$). 
They were built by the University of Regina and assembled at TRIUMF (Canada). 
The detectors are mounted on the detector frame behind the VDCs (see Fig.~\ref{fig_HRS_DetectorPackage}) 
and are separated by a distance of about $2\,\mathrm{m}$. 

The first plane, S1, consists of six paddles made of BC408 plastic scintillator. 
Each paddle is $35.5\,\mathrm{cm}$ long, $29.5\,\mathrm{cm}$ wide and $5\,\mathrm{mm}$
thick. The neighboring paddles overlap by $10\,\mathrm{mm}$. The light signal from 
each scintillator bar is collected by two 2-inch photo-multiplier tubes mounted at
the opposite ends of a bar. On the other hand, the second, S2m detector consists of 
sixteen scintillator bars which are made of EJ-230 
plastic scintillator with dimensions of $43.2\,\mathrm{cm}$ by $14.0\,\mathrm{cm}$ by 
$5.1\,\mathrm{cm}$ thick~\cite{osp}. Here, paddles do not overlap. Each bar is again viewed
by a 2-inch photo-multiplier-tubes at each end. 
The time resolution per plane is $\sigma \approx 0.30\,\mathrm{ns}$. The signals from each
PMT are led from detectors to the front-end crates with electronics for further processing. 

\subsubsection{Cherenkov Detector}
\vspace*{2mm}
Identification of the particles inside the HRS is provided by a threshold gas
Cherenkov detector. In particular, it is utilized to separate electrons from pions and protons
with  $99\,\mathrm{\%}$ efficiency. The detector was designed and constructed for
Jefferson Lab by groups from INFN and Saclay~\cite{iodice1998}. 

The operation of the detector is based on the Cherenkov effect. The Cherenkov radiation
is emitted when a particle travels through a medium faster than the speed of light
$c_{\mathrm{Medium}} = c_{0}/n$ in that medium. Here, $c_{0}$ is the speed of light in vacuum
and $n$ is the refractive index of the medium~\cite{leo}. The threshold velocity can be transformed 
into the minimal particle momentum required for a particular particle to emit Cherenkov light:
\begin{eqnarray}
  p_{\mathrm{Threshold}} = \frac{m \frac{c_{0}}{n}}{\sqrt{1 - \frac{v^2 n^2}{c_{0}^2}}} 
  \approx \frac{mc_{0}}{\sqrt{n^2-1}}  \nonumber
\end{eqnarray}
The threshold momentum depends on the particle mass and refractive index and can be set
arbitrarily with the proper choice of the radiative medium to satisfy the needs of
a particular detector.

The HRS Cherenkov detector is $1\,\mathrm{m}$ long, with a $250\times 80\,\mathrm{cm^2}$ wide 
entrance surface. It is positioned between the S1 and S2m trigger scintillator planes. The detector 
is filled with the $\mathrm{CO_2}$ gas at atmospheric pressure with the refractive 
index $n = 1.00041$. This sets the threshold momentum for electrons to $0.017\,\mathrm{GeV}/c$,
for pions to $4.8\,\mathrm{GeV}/c$, and for protons to $32\,\mathrm{GeV}/c$. Hence, within the 
momentum acceptance of the HRS spectrometer (see table~\ref{table_HRSL}) only electrons can
emit Cherenkov light, which allows us to distinguish them from the other particles. 
This way we can use the threshold Cherenkov detector either for tagging electrons or as a veto for the 
identification of the heavier hadron components.

\begin{figure}[!ht]
\begin{center}
\includegraphics[width=0.7\textwidth]{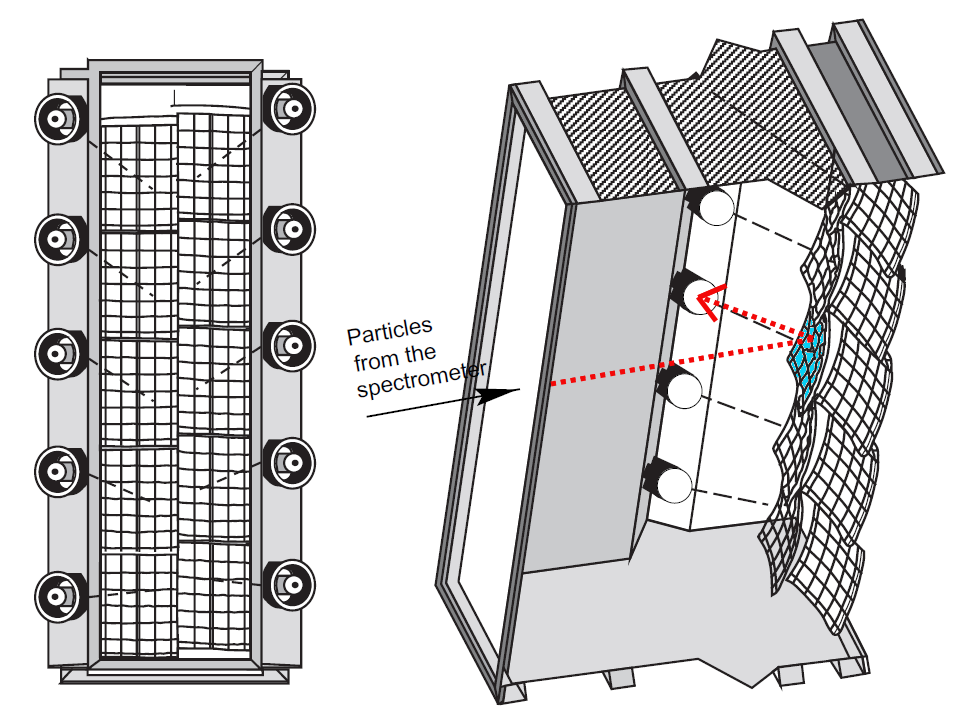}
\caption{ Front and side view of the Gas Cherenkov Detector~\cite{iodice1998}. Very
fast electrons (black arrow) produce Cherenkov radiation (dashed line) inside the
$\mathrm{CO_2}$ radiator. Emitted light is reflected from the mirrors and collected 
by the PMTs mounted on the side of the detector housing.  
\label{fig_HRS_Cherenkov}}
\end{center}
\end{figure}

The emitted Cherenkov light is collected by the mirrors mounted on the opposite 
side of the entrance window (see Fig.~\ref{fig_HRS_Cherenkov}), and then focused on ten
5-inch photo-multiplier tubes located on the side of the detector box. The signals from each
PMT are taken to the front-end crates for further processing. 

\subsubsection{Pion Rejector}

\begin{figure}[!ht]
\begin{center}
\includegraphics[width=0.75\textwidth]{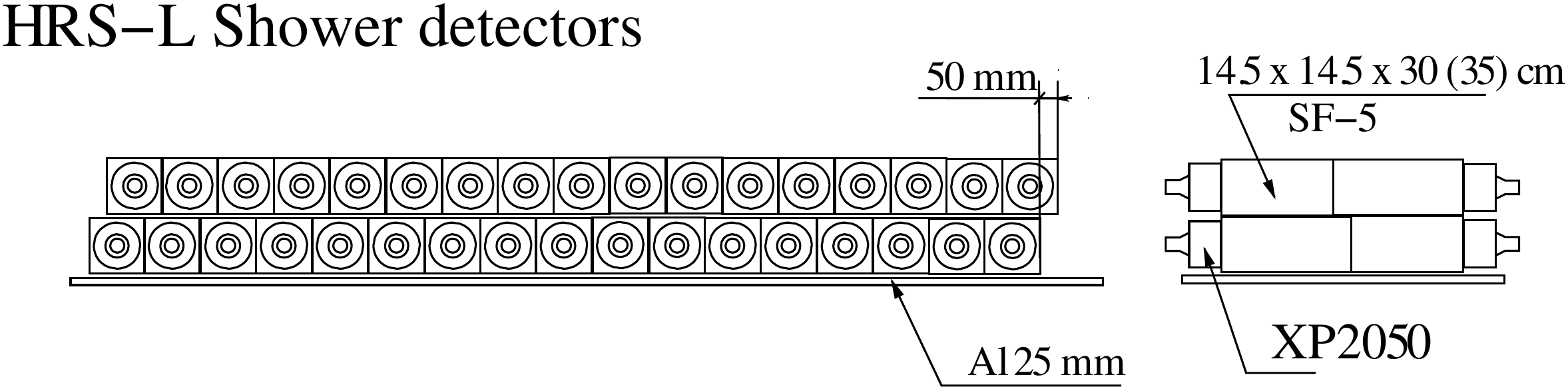}
\caption{HRS-L Pion rejector~\cite{alcorn}. Particles enter the detector
from the bottom. 
\label{fig_HRSL_PionRejector}}
\end{center}
\end{figure}

In addition to the Cherenkov detector, an electromagnetic calorimeter (also called Pion rejector) has 
been instrumented for particle identification~\cite{alcorn}. The detector is composed of 
two layers of lead glass blocks. Each layer consists of 17 long blocks of dimensions 
$15\,\mathrm{cm} \times 15\,\mathrm{cm} \times 35\,\mathrm{cm}$ and 17 short blocks of 
dimensions $15\,\mathrm{cm} \times 15\,\mathrm{cm} \times 30\,\mathrm{cm}$, which are
assembled together as shown in Fig~\ref{fig_HRSL_PionRejector}. 

When incident electrons enter the Pion rejector they produce photons through brems\-strahlung. 
These photons create new 
electron-positron pairs which again generate bremsstrahlung photons. This repetitive 
process creates an electromagnetic avalanche. The positrons and electrons in the avalanche 
are fast enough to produce Cherenkov radiation inside the lead glass which is then detected 
by the PMTs at the end of each block. The detected light yield is proportional to 
the energy lost by the incident particle. 

On the other hand, hadrons deposit much less energy in the Pion rejector, since 
hadronic showers can not develop in a such short calorimeter. This is due to the fact
that hadrons have  much larger mean free paths in lead glass than electrons. 

These differences between electromagnetic and hadronic showers help us 
to identify the incident particle. Electrons are likely to start the shower
already in the first layer and consequently produce strong signals in both layers
of the calorimeter. However, hadrons may mostly pass through the first layer without starting
a shower and produce signal only in the second layer. Hence, by comparing the energy deposits in the 
first and the second layer of the calorimeter, electrons can be clearly distinguished
from hadrons, mainly pions.

\section{BigBite Spectrometer}

\begin{figure}[!ht]
\begin{center}
\begin{minipage}[t]{0.45\textwidth}
\hrule height 0pt
\includegraphics[width=1\textwidth]{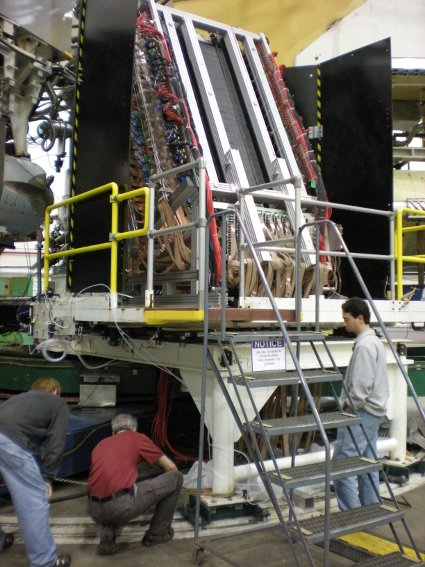}
\end{minipage}
\hspace*{1cm}
\begin{minipage}[t]{0.4\textwidth}
\hrule height 0pt
\caption{The back of the BigBite spectrometer with the hadron 
detector package during the E05-102 experiment. The spectrometer
is mounted on a metal frame that can be rotated around the 
pivot to a desirable scattering angle. The 
spectrometer is surrounded by the field clamp (black metal walls) 
to shield detectors from undesirable radiation and to minimize the
BigBite's residual magnetic field at the target.
\label{fig_BBPhoto}}
\end{minipage}
\end{center}
\end{figure}

The BigBite spectrometer is a recent acquisition in the experimental 
Hall A of the Thomas Jefferson National Accelerator Facility. 
It was previously used at the Internal Target Facility of the 
AmPS ring at NIKHEF for the detection of electrons
\cite{lange-general,lange-optics}.  At Jefferson Lab,
BigBite has been re-implemented as a versatile spectrometer
that can be instrumented with various detector packages
optimized for the particular requirements of the experiments 
(see Fig.~\ref{fig_BBPhoto}).
BigBite complements the High-Resolution Spectrometers,
which are part of the standard equipment of Hall~A \cite{alcorn}.
Adding BigBite allows one to devise more flexible experimental
setups involving double- and even triple-coincidence measurements.

In 2005, the BigBite spectrometer was first used in Hall~A 
as the hadron arm in the E01-015 experiment, which investigated
nucleon-nucleon short-range correlations \cite{subedi, shneor}.
In 2006, it was instrumented as the electron arm for the measurement
of the neutron electric form factor (experiment E02-013 \cite{riordan}). 
In 2008 and 2009, it has been used in two large groups of experiments
spanning a broad range of physics topics.  We studied near-threshold
neutral pion production on protons (experiment E04-007 \cite{e04007})
and measured single-spin asymmetries in semi-inclusive pion electro-production 
on polarized $^3\mathrm{He}$ (experiments E06-010 and E06-011
\cite{e06010,e06011,qian11,huang11}).  In the same period, we also 
measured parallel and perpendicular asymmetries on polarized 
$^3\mathrm{He}$ in order to extract the $g_2^\mathrm{n}$ polarized 
structure function in the deep-inelastic regime (experiment 
E06-014 \cite{e06014}). Finally, in May and June 2009 it was employed 
to detect protons and deuterons during the experiment E05-102.

\begin{table}
\begin{center} 
\begin{minipage}[t]{0.5\textwidth}
\hrule height 0pt
\begin{tabular}{ll}
\toprule
\multicolumn{2}{c}{\textbf{BigBite main characteristics}} \\
\midrule
Configuration  & Dipole\\
Optical length & $\approx 2.7\,\mathrm{m}$\\
Bending angle  & $25\,{}^{\mathrm{o}}$\\[4pt]
Momentum range & $200-900\,\mathrm{MeV}$ \\ 
Momentum acceptance & $-0.6 \leq \delta_{Tg} \leq 0.8  $\\ 
Momentum resolution & $1.6\,\mathrm{\%}$\\[4pt]
Angular acceptance &  \\
\,\,\,\,\,\,Horizontal     & $\approx 240\,\mathrm{mrad}$ \\
\,\,\,\,\,\,Vertical       & $\approx 500\,\mathrm{mrad}$\\[4pt]
Angular resolution &     \\
\,\,\,\,\,\,Horizontal     & $7\,\mathrm{mrad}$ \\
\,\,\,\,\,\,Vertical       & $16\,\mathrm{mrad}$ \\[4pt]
Vertex resolution & $1.2\,\mathrm{cm}$\\
\bottomrule
\end{tabular}
\end{minipage}
\hfill
\begin{minipage}[t]{0.45\textwidth}
\hrule height 0pt
\caption{Main characteristics of the BigBite spectrometer. The 
resolutions (sigma) were determined 
for $0.5\,\mathrm{GeV}/c$ protons~\cite{miha_NIM} during 
the E05-102 experiment, when BigBite was used  with 
the hadron detector package. \label{table_BBChar}}
\end{minipage}
\end{center}
\end{table}

\begin{figure}[!ht]
\begin{center}
\includegraphics[width=1\textwidth]{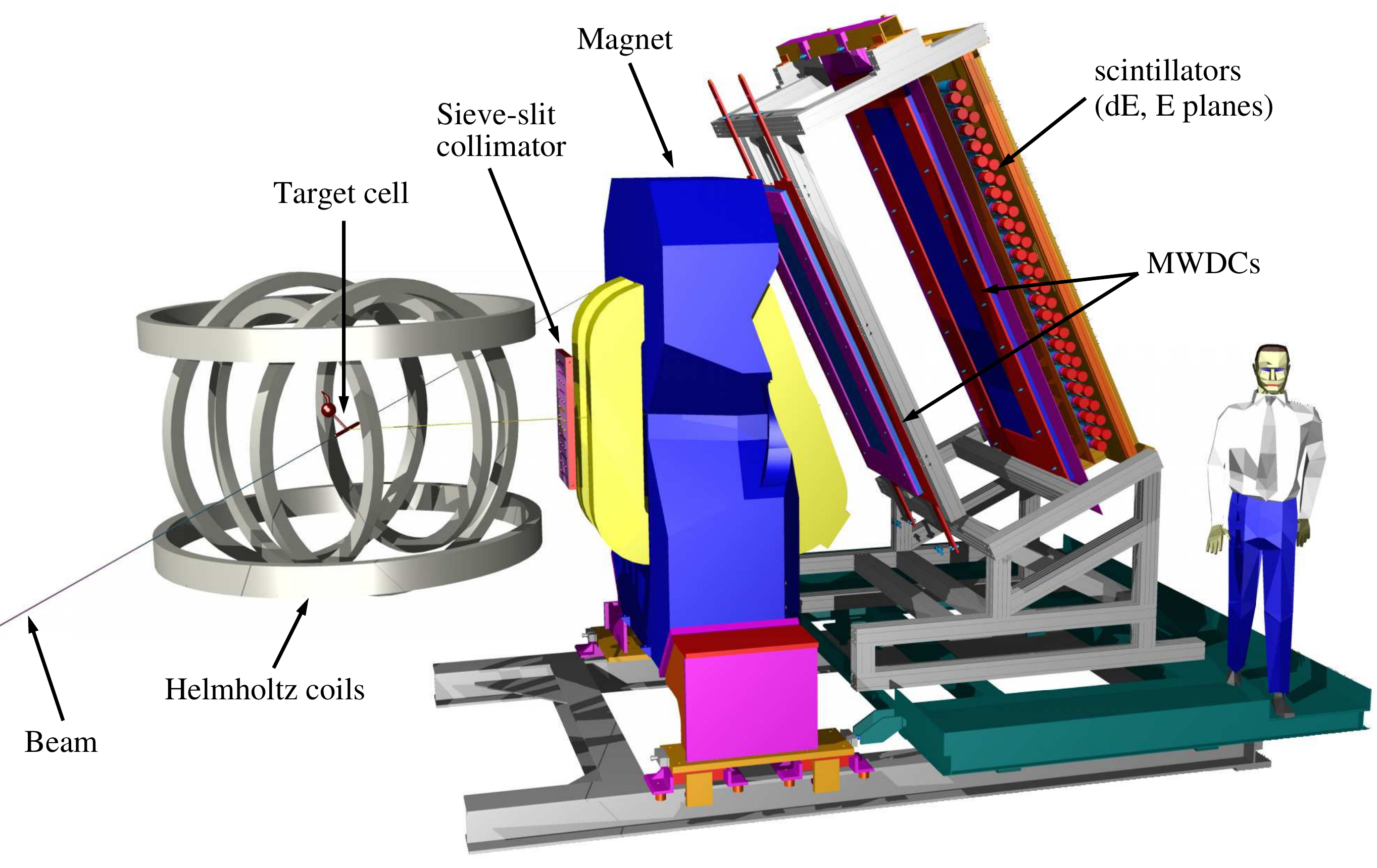}
\caption{The BigBite spectrometer on its support frame. 
BigBite consists of a dipole magnet, followed by the
detector package assembled from a pair of multi-wire drift 
chambers (MWDC) and two scintillator planes (dE and E). 
The field clamp minimizes stray magnetic fields in the vicinity 
of the target. The sieve-slit collimator can be inserted vertically
in front of the spectrometer by means of a pulley. \label{BBSpectrometer}}
\end{center}
\end{figure}

BigBite is a non-focusing spectrometer consisting of
a single dipole with large momentum and angular acceptances.
See Table~\ref{table_BBChar} for details. The magnet is followed
by a hadron detector package (see Fig.~\ref{BBSpectrometer}) consisting of two
Multi-Wire Drift Chambers (MWDC) for particle tracking and two 
planes of scintillation detectors (denoted by dE and E) for triggering,
particle identification, and energy determination. 

\subsection{BigBite Magnet}
The BigBite spectrometer utilizes a single room-temperature dipole 
magnet, shown in Fig.~\ref{BBSpectrometer}. The magnet was designed by the
Budker Institute for Nuclear Physics in Novosibirsk for the experiments
at NIKHEF~\cite{lange-general}. The gap between pole-faces measures $25\,\mathrm{cm}$
in the horizontal direction and $84\,\mathrm{cm}$ in the vertical direction.
This gives BigBite more than fifteen-times bigger acceptance compared to the 
High Resolution Spectrometers.

\begin{figure}[!ht]
\begin{center}
\includegraphics[width=0.35\textwidth]{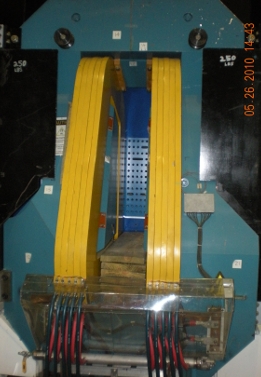}
\hfill
\includegraphics[width=0.6\textwidth]{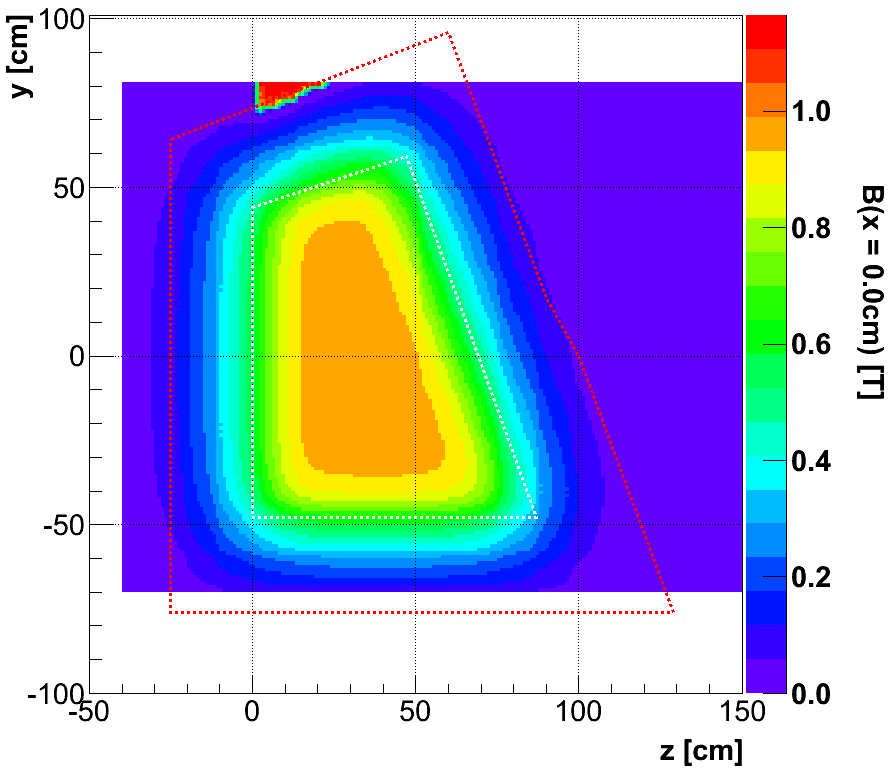}
\caption{[Left] Photograph of the BigBite magnet, taken from its back side. 
The blue yoke, the yellow coils around the pole-faces and the blue 
sieve-slit collimator at the entrance to the magnet are clearly visible. 
[Right] The simulated magnetic field of BigBite in the mid-plane
of the magnet. The white dashed line shows the position and size of 
the magnet poles. The red dashed line denotes the outer limits of the coils. 
The magnetic field is homogeneous inside the magnet, while the magnitude
decreases quickly on the outside of the pole-faces. Fringe magnetic fields 
can reach outside of the magnet. The calculation was made with 
MAFIA~\cite{MAFIA, vladimir}.
\label{fig_BBField}}
\end{center}
\end{figure}

Energizing the magnet with a current of $518\,\mathrm{A}$
results in a mean field density of $0.92\,\mathrm{T}$, corresponding to a central 
momentum of $p_\mathrm{c}=0.5\,\mathrm{GeV}/c$ and a bending angle of $25^\circ$. 
See Fig.~\ref{fig_BBField}. The central trajectory of the magnet is 
perpendicular to the vertical entrance face, while it subtends  
a $5\,{}^{\mathrm{o}}$ angle with the exit face, which is  rotated 
for $20\,{}^{\mathrm{o}}$ with respect to the vertical direction. This
enhances the field integral for particles entering the upper
region of the magnet, while it reduces it for lower, 
which makes the dispersion more uniform across the acceptance
of the spectrometer~\cite{lange-general}.

The magnet is mounted on a support frame which can be rotated around a pivot 
at the center of the target. The support also carries the detector package and 
a metal field shielding wall. See Fig.~\ref{BBSpectrometer}. The field clamp was 
installed in front of BigBite to shield detectors from unwanted radiation and 
to minimize the BigBite fringe magnetic fields at the target, which could 
distort its operations. The residual BigBite field on the target was 
estimated to a few Gauss and was taken into consideration  when configuring the target 
holding magnetic field (which is $\approx 25\,\mathrm{Gauss}$).

\subsection{Multiwire Drift Chambers}
\label{sec:MWDCs}

To reconstruct the track of a charged particle through the BigBite's detector 
package two multi-wire drift chambers (MWDC) were used. The first chamber is 
located $\approx 30\,\mathrm{cm}$ behind the the dipole magnet. The second one
is positioned $\approx 76\,\mathrm{cm}$ behind the first one. Both chambers are mounted
on an aluminum frame and are rotated for $25\,\mathrm{{}^o}$ with respect to the 
vertical direction, in order to be perpendicular to the central track through 
the spectrometer.

\begin{figure}[!ht]
\begin{center}
\includegraphics[width=0.4\textwidth]{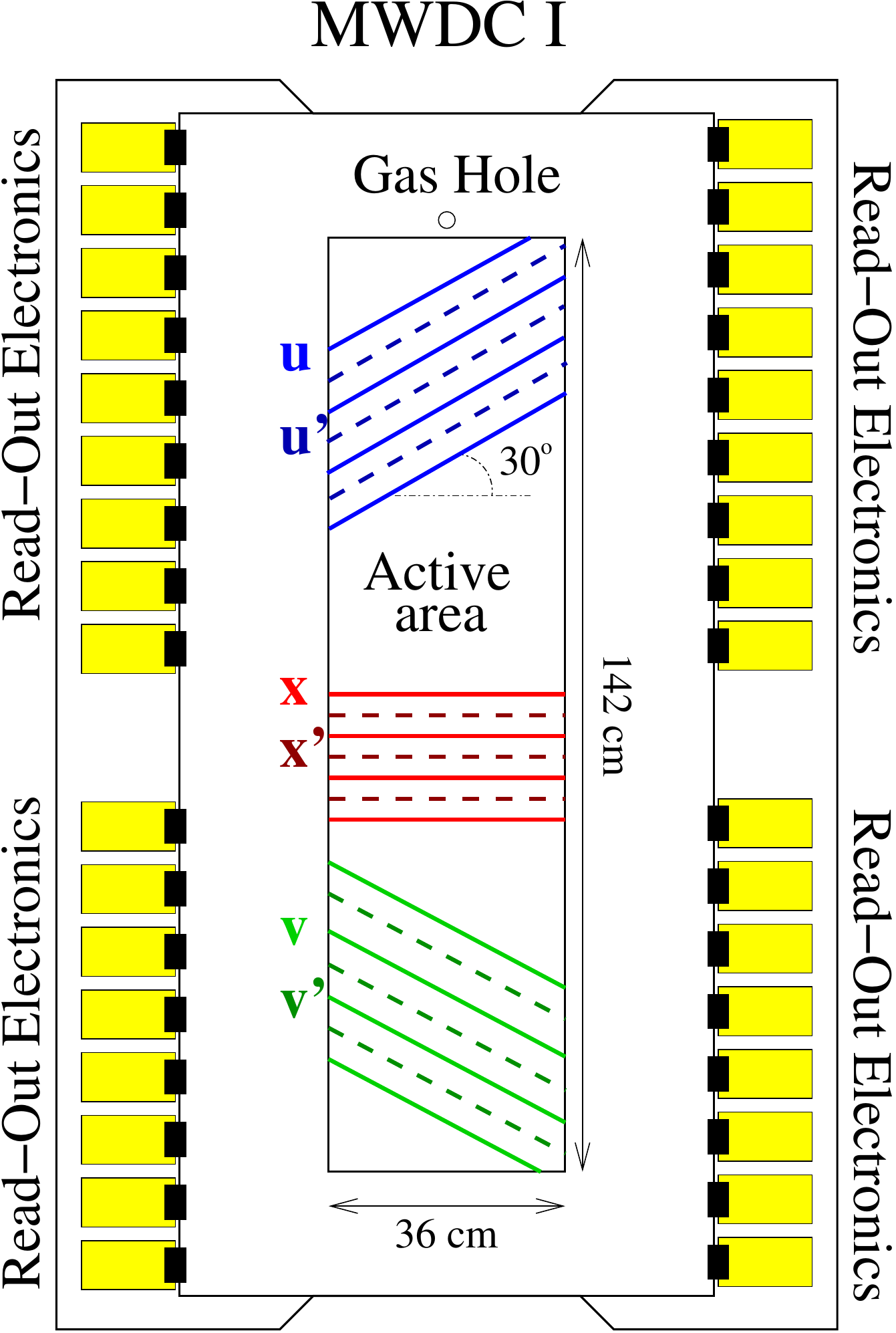}
\hspace*{1cm}
\includegraphics[width=0.46\textwidth]{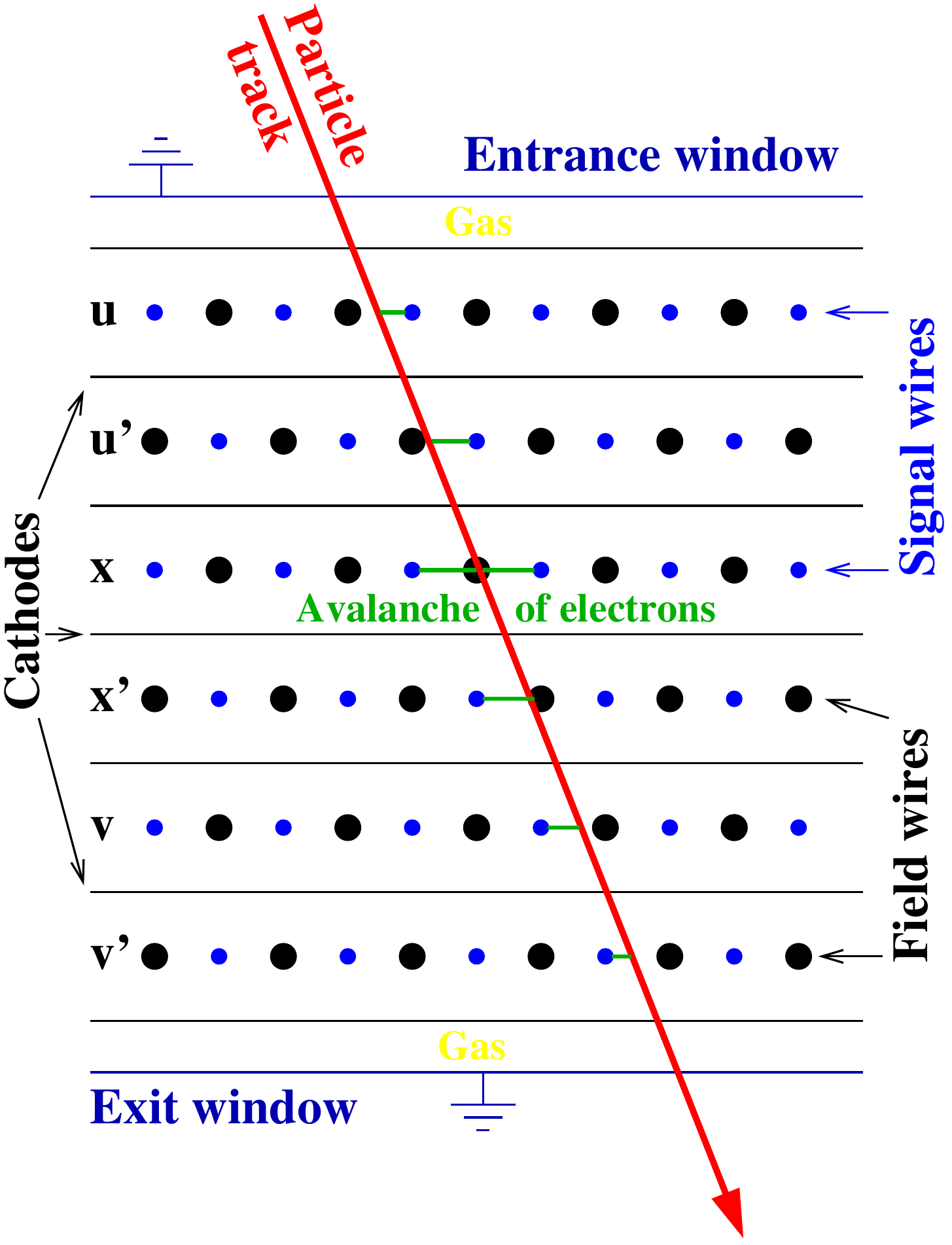}
\caption{[Left] Schematics of the first BigBite's MWDC,
showing the orientation of the wire planes. The wires in  planes (x, x') are 
aligned horizontally, the wires in planes (u, u') are oriented
at an angle of $30^\circ$ with respect to the horizontal direction and  
the wires of the planes (v, v') are oriented at $-30^\circ$. The signals from 
all wires are read out through the electronic circuit boards mounted 
on the edges of the chamber. [Right] Particle tracking through the MWDC. 
On its path, the particle ionizes the atoms in gas. Created 
electrons drift inside the homogeneous electric field to the nearest signal 
wire. From the positions of the hit wires and drift times 
particle track can then be reconstructed. 
\label{fig_MWDCScheme}}
\end{center}
\end{figure}

The chambers were constructed at the University of Virginia~\cite{chanMSc}. 
Each MWDC consists of six planes of wires. See Fig.~\ref{fig_MWDCScheme}.  
The wires in the first two planes (u, u') are oriented at an angle 
of $60^\circ$ with respect to the dispersive direction.  The wires in the 
third and fourth plane (x, x') are aligned horizontally, while the wires of 
the last two planes (v, v') are oriented at $-60^\circ$. 
See Fig.~\ref{fig_MWDCScheme} for details. In between 
every two signal wires a field wire is inserted.  
The spacing between two signal wires in a single plane is $1\,\mathrm{cm}$.
The wires in the odd planes (x',u',v') are shifted by a half of a 
signal wire spacing relative to the even planes (x, u, v) to 
resolve left/right ambiguities~\cite{huangPhD}. Two different 
types of wires are used. The signal 
wires are made of $25\,\mathrm{\mu m}$ thick gold-plated tungsten, while 
$90\,\mathrm{\mu m}$ thick copper-beryllium alloy is used for field 
wires. Each wire plane is sandwiched between two cathode planes made of 
copper-coated ($0.12\,\mathrm{\mu m}$) DuPont mylar ($0.2\,\mathrm{mm}$). The field 
wires and the cathode planes were set at a voltage of about $-1600\,\mathrm{V}$
during the experiment. 

The active  
area of the first (smaller) chamber is $140\,\mathrm{cm} \times 35\,\mathrm{cm}$ 
and contains 141 signal and field wires for the u, u', v and v' planes and 
142 wires for the x and x' planes. The active area of the second (larger) chamber 
is $200\,\mathrm{cm} \times 35\,\mathrm{cm}$ and contains 200 signal and 
field wires for the u, u', v and v' planes and 202 wires for the x and x' 
planes. The chamber entrance and exit windows are made of $0.4\,\mathrm{mm}$ 
thick aluminized kapton. Both windows are grounded and are separated from 
the outermost cathodes by gas pockets. The gas that was kept flowing through
the MWDC system was a $50\,\mathrm{\%}/50\,\mathrm{\%}$ mixture of argon and
ethane.

When a charged particles passes through the wire plane, it ionizes the gas that
fills the volume between the wires, producing ions and electrons 
(see Fig.~\ref{fig_MWDCScheme}). A created avalanche of electrons 
then drifts to the nearest signal wire, where it produces
an electronic signal. The time that electrons need to come from the track position 
to the signal wire is proportional to the distance traveled. The drift velocity
depends on the considered gas and the applied electric field~\cite{leo} and has 
typical values of $\approx 50\,\mathrm{\mu m/ns}$. A signal from every wire
is read through the read-out circuit boards mounted on the edges of the 
MWDCs, and recorded by the time-digital converters (TDCs).

\begin{figure}[!ht]
\begin{center}
\includegraphics[width=0.49\textwidth]{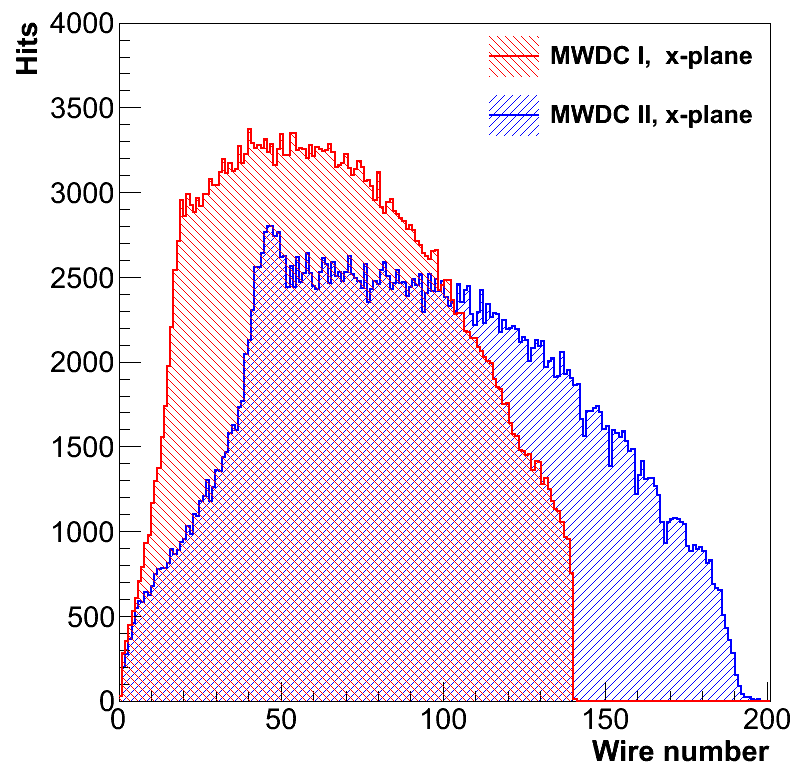}
\includegraphics[width=0.49\textwidth]{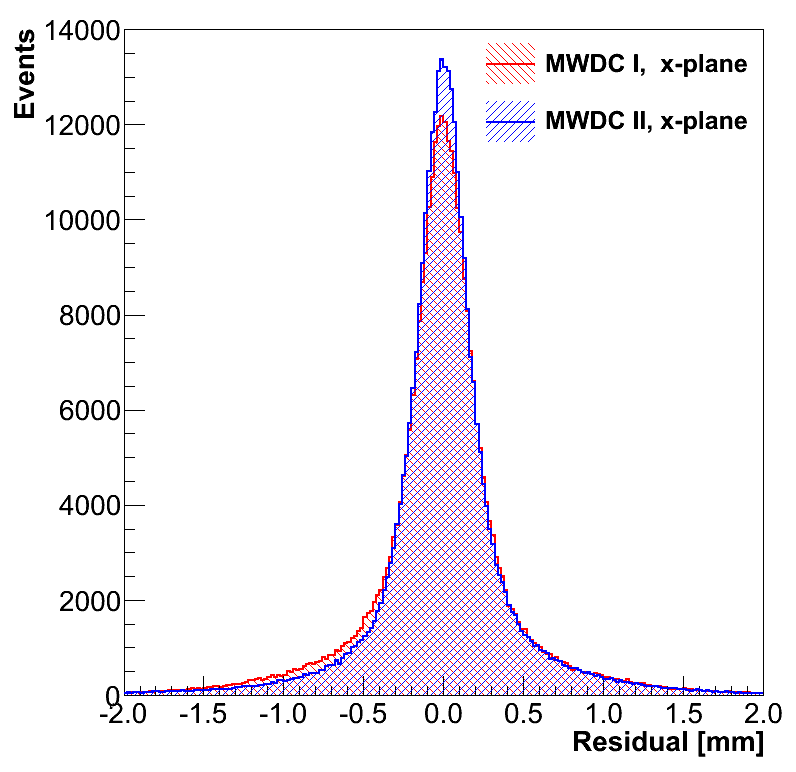}
\caption{ Typical results of the analysis of the MWDC performance. [Left]
Response of wire-plane x as a function of the wire number in the first and
second MWDCs. [Right] Spatial resolution of the wire-plane x for both MWDC. 
The residual is defined as a distance between the hit position and the 
track projection on the hit wire plane. The spatial resolution of each 
wire-plane was determined~\cite{GeJinMWDC} to be better 
than $\mathrm{200\,\mathrm{\mu m}}$.  
\label{fig_MWDCHits}}
\end{center}
\end{figure}

The spatial resolution of the MWDCs per wire-plane was determined 
by comparing the position of the hit (determined from the drift time) 
and the projection of the track to the hit wire plane (see Fig.~\ref{fig_MWDCHits}). 
The  resolution of each wire-plane was determined~\cite{GeJinMWDC} to be better 
than $\mathrm{200\,\mathrm{\mu m}}$.

\begin{figure}[!ht]
\begin{center}
\includegraphics[width=0.8\textwidth]{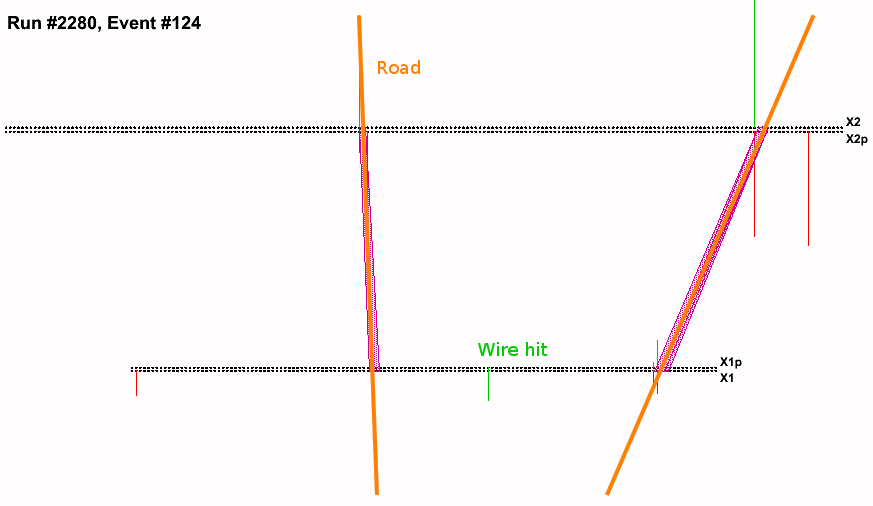}
\caption{ Reconstructed two-dimensional tracks (roads) in the x-projection 
of MWDCs. Each projection combines four wire planes (e.g. x1, x1p, x2, x2p). 
Wire hits in each plane are shown with red and green vertical lines. 
The length of a line corresponds to the distance (not in scale) from the wire.
Hits from all wires are introduced to the Pattern Match Tree Search
algorithm which finds possible roads. The roads from all three projections 
(x, u, v) are then combined into three-dimensional tracks.  
\label{fig_MWDC_roads}}
\end{center}
\end{figure}

\begin{figure}[!hb]
\begin{center}
\begin{minipage}[t]{0.5\textwidth}
\hrule height 0pt
\includegraphics[width=1\textwidth]{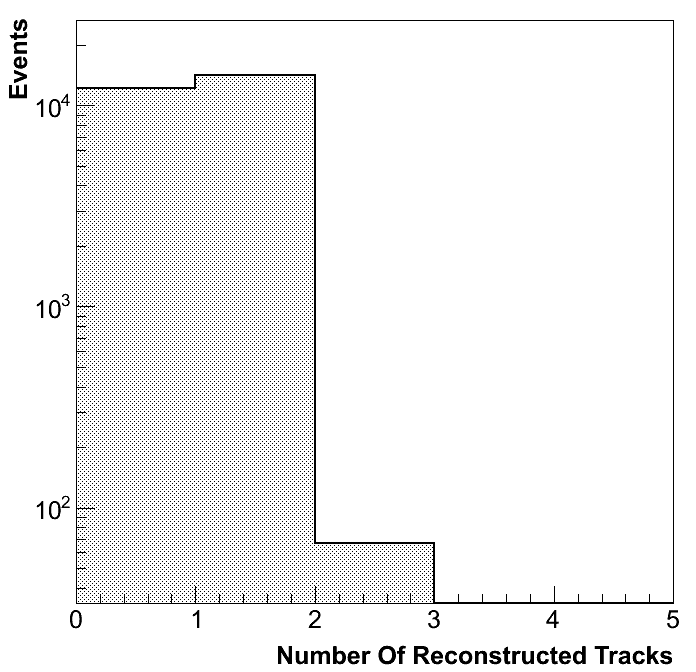}
\end{minipage}
\hspace*{1cm}
\begin{minipage}[t]{0.4\textwidth}
\hrule height 0pt
\caption{ Number of reconstructed tracks in the BigBite MWDCs. There
are events for which no track could be reconstructed. These are mostly
cosmic events. The majority of coincidence events are with one reconstructed track.
Less than $1\,\mathrm{\%}$ of the events have more than one track. 
\label{fig_BigBite_NumberOfTracks}}
\end{minipage}
\end{center}
\end{figure}

A one-dimensional positional information from all the wire planes of 
both MWDCs is then used to reconstruct the full track of a particle that 
is flying through the detector package. Each track is determined by four 
coordinates: two positions ($x_{\mathrm{Det}},y_{\mathrm{Det}}$) at the 
entrance to the first MWDC and two angles 
($\theta_{\mathrm{Det}}, \phi_{\mathrm{Det}}$). They are determined
by a dedicated analyzing software~\cite{OleThreeSearch}. The program
first divides hits into three groups (or projections) according to their 
orientation. Within each projection two dimensional tracks (or roads) are 
then reconstructed. A search for roads is done by a  Pattern Match Tree Search
algorithm~\cite{treesearch}. An example of reconstructed roads for 
a given hit pattern in a $x$-projection of MWDCs is demonstrated 
in Fig.~\ref{fig_MWDC_roads}. In the last step, roads for all projections
are combined into a full three-dimensional track. The coordinates
of the full track are determined by solving a set of equations
by using Cholesky decomposition~\cite{nrc}:
\begin{eqnarray}
    \xi_i = \left( x_{\mathrm{Det}} + z_i 
            \cdot \tan{\theta_{\mathrm{Det}}}\right)\cos{\alpha_i} +  
            \left( y_{\mathrm{Det}} + z_i\cdot \tan{\phi_{\mathrm{Det}}} 
            \right) \sin{\alpha_i}\,,\nonumber
\end{eqnarray}
where $\xi_i$ is the directly measured linear coordinate in the $i$-th wire plane, which 
is positioned at $z_i$ and $\alpha_i$ is the angle of the $i$-th wire plane 
with respect to the horizontal (x) axis. The number of successfully 
reconstructed tracks per event is demonstrated in Fig.~\ref{fig_BigBite_NumberOfTracks}.

The particle track parameters are 
obtained  together with the corresponding uncertainties, which are directly 
related to the MWDC wire-plane spatial resolution. The resolution 
for the dispersive coordinate $x_{\mathrm{Det}}$ was estimated to
$\sigma_{x_{\mathrm{Det}}} = 91\,\mathrm{\mu m}$, while for the non-dispersive
coordinate $y_{\mathrm{Det}}$ it was shown to be 
$\sigma_{x_{\mathrm{Det}}} = 200\,\mathrm{\mu m}$. The resolution for the 
$x_{\mathrm{Det}}$ is expectingly better because of the wire orientations.
Only two planes of wires can be used to determine $y_{\mathrm{Det}}$, 
while three are considered for $x_{\mathrm{Det}}$. The corresponding
angular resolutions are 
$\sigma_{\theta_{\mathrm{Det}}} = 0.16\,\mathrm{mrad}$ and
$\sigma_{\phi_{\mathrm{Det}}} = 0.35\,\mathrm{mrad}$. Again for the same 
reasons, the out-of-plane angle $\theta_{\mathrm{Det}}$ is determined 
with better accuracy than the in-plane angle $\phi_{\mathrm{Det}}$.

In the calculation of these errors, the uncertainty in the relative positions
of the wire-chambers was not considered. It was estimated to be 
$\approx 0.5\,\mathrm{\%}$ and would represent the dominant part of  the error. 
However, since MWDC were not moving during the experiment, it affects
only the absolute values of the track angles, which can be
compensated by the appropriate correction in the optics matrix.

\subsection{Scintillation detector}

\begin{figure}[!ht]
\begin{center}
\includegraphics[width=1\textwidth]{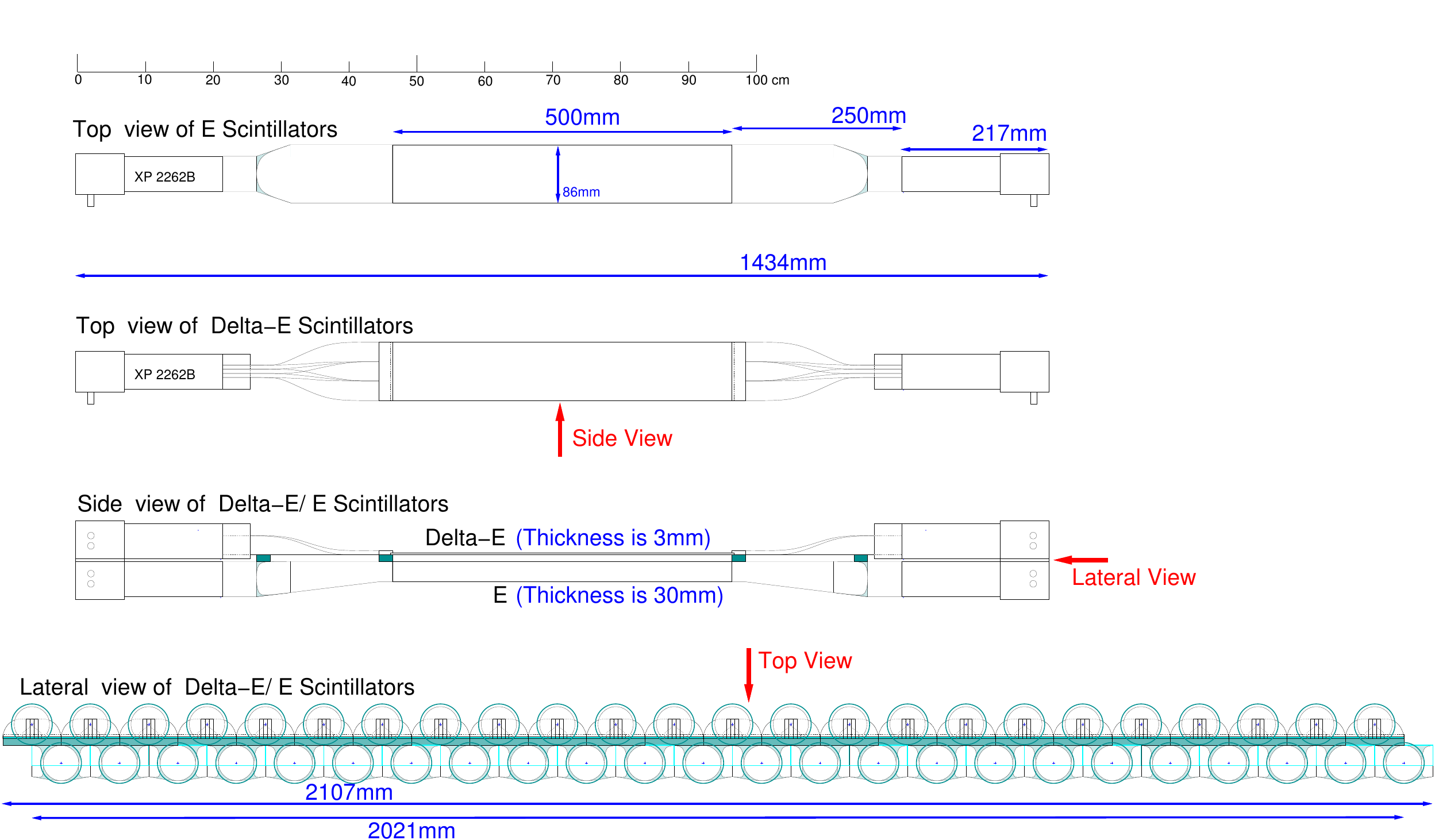}
\caption{Schematics of the BigBite scintillation detector. It consists
of two detector planes (dE and E), each made of 24 scintillator paddles.
The first dE-plane is $3\,\mathrm{mm}$ thick while the second E-plane 
is made of $3\,\mathrm{cm}$ thick material. Light signal from each 
paddle is read out with two photomultiplier tubes (PMTs) mounted on each
end of the paddle and connected to the scintillator with the plastic
light guides. \label{fig_BBScintillators}}
\end{center}
\end{figure}

The particle identification (PID) and timing information was provided 
by the scintillation detector (also called the Trigger plane), which 
was built by University of Glasgow~\cite{annand}. It consists 
of two individual layers of scintillation bars (called dE-plane and 
E-plane) separated by $8\,\mathrm{mm}$. See Fig.~\ref{fig_BBScintillators}
for details. They are mounted at the back of the BigBite hadron detector 
package, approximately $10\,\mathrm{cm}$ behind the second MWDC. The dE- and 
E-planes each consist of $24$ scintillator bars, made of EJ-204 plastic. 
The bars and are $50\,\mathrm{cm}$ long and $8.6\,\mathrm{cm}$ wide. For 
the dE-plane, thinner bars ($0.3\,\mathrm{cm}$) were used to detect low-energy 
particles, while for the E-plane, a thickness of $3\,\mathrm{cm}$ was chosen to 
allow for the detection of more energetic particles.  The light pulses in each 
bar were detected by Photonis XP2262B PMTs mounted at 
each end of the bar. The PMTs were coupled to the scintillator paddles through 
the fish-tail light guides made of UV transmitting plexiglass BC800. To double 
the  spatial and momentum resolution, the bars in the E-plane are offset from 
those in the dE-plane by one half of the bar width ($4.3\,\mathrm{cm}$).

A signal from each PMT is first amplified ten times and then divided into three
copies. The first copy goes to the trigger electronics to form a trigger pulse,
which starts the data reading from all the detectors. The second and the third
copy are lead to analog-to-digital converters (ADC) and time-to-digital converters 
(TDC), where timing and amplitude information is recorded. 

\begin{figure}[!ht]
\begin{center}
\includegraphics[width=0.49\textwidth]{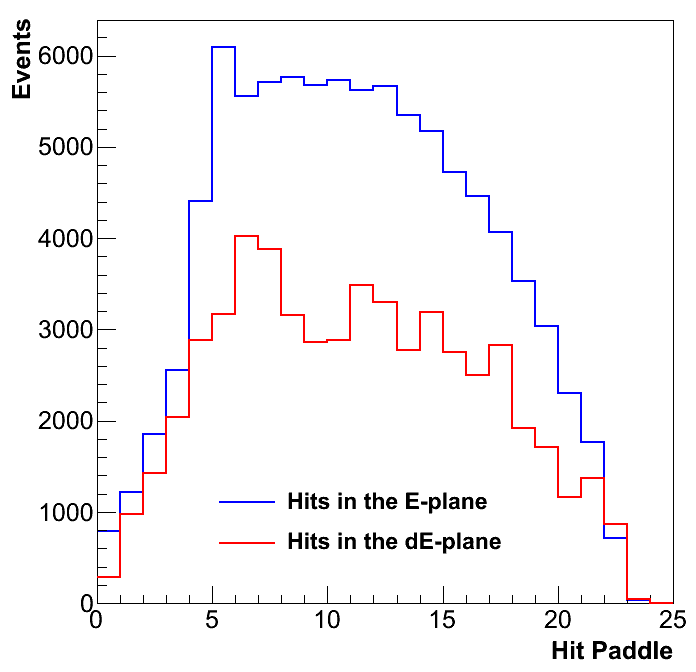}
\hfill
\includegraphics[width=0.49\textwidth]{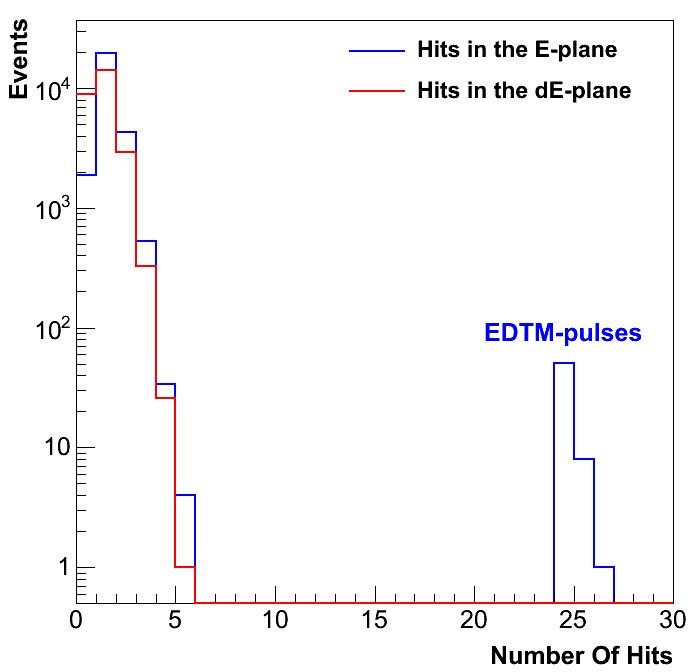}
\caption{ [Left] Distribution of particle hits in the two BigBite scintillation
detector planes as a function of the paddle number. [Right] Number of 
fully reconstructed hits in the E- and dE-planes for a single event. 
Events with zero hits
in the detectors correspond to HRS-L only events (not coincidences). EDTM events,
where all 24 scintillation paddles are triggered simultaneously, are also visible 
for the E-plane. EDTM events are not visible in the dE-plane TDC spectrum, 
because of the structure of the dE-plane triggering circuit, where EDTM pulses
are injected to the circuit after the TDCs are read.      
\label{fig_BigBite_NumberOfEdEHits}}
\end{center}
\end{figure}

A typical distribution of particle hits in the scintillation detector is demonstrated 
in figure Fig.~\ref{fig_BigBite_NumberOfEdEHits}. The majority of the particles hit
the central part of the detector. The least populated is the bottom part of the detector, 
where very high momentum particles are usually detected. A smaller number of hits in the 
dE-plane is a consequence of its smaller detection power for high momentum particles 
and set detection thresholds. The distribution of the hits in the scintillation detector 
of course also resembles the distribution of hits in the wire-chambers, shown 
in Fig.~\ref{fig_MWDCHits}. In the majority of the coincidence events detected in the
E05-102 experiment, the scintillation detector is hit only once, while only 
$\approx 10\,\mathrm{\%}$ of the events have more than one valid hit.
This can significantly simplify the data analysis. By limiting the analysis only to
the events with one hit per event, the doubt about which hit in the BigBite
detector package corresponds to the coincidence event is automatically removed. 
However, random coincidences are still possible. 

\section{Data Acquisition (DAQ)}
\label{sec:CODA}

Hall A uses CEBAF Online Data Acquisition system (CODA) to collect the data taken
during the experiment. CODA~\cite{CODA} was developed by the Jefferson Lab
Data Acquisition Group and combines various hardware modules and software packages 
for the acquisition, monitoring and storage of the experimental data. 
It is designed to be modular and scalable.

The raw signals from detectors are first amplified and then sent to the 
analog-to-digital converters (ADCs) and via discriminators to the time-to-digital 
converters (TDCs) and scalers. These modules are installed on the front-end crates 
and are used to convert electronic signals (amplitude, time and counts) into 
digitized information. The operation of all modules in one crate is controlled 
by the Readout Controller (ROC). ROCs are single-board computers mounted
at the beginning of each crate and are running the VxWorks real-time kernel. 
Each ROC is loaded with a dedicated programming script which specifies the positions of the 
modules in a crate, their type and properties (such as number of channels). ROC also manages 
the communication of the crate though the Ethernet network, which is considered for the 
transport of data from the digitizing modules to the CODA Event Builder (EB). 
EB~\cite{MonaghanPhD} is a program that collects pieces of information from all ROCs 
and constructs a single data structure. We call this structure an event. The combined datum 
is then sent to CODA Event Recorder (ER), which writes it to a disk.

Which events are recorded and which are rejected is decided by the Trigger Supervisor (TS). 
It couples the experiment specific triggering system (see Sec.~\ref{sec:TriggerSetup}) 
with the CODA system. The trigger signals from the trigger circuit are accepted 
through eight input channels, T1 to T8. According to a set of 
scaling factors (called pre-scale factors) it decides which trigger should 
be accepted. It can accept multiple triggers. When a particular trigger is accepted, 
TS returns a Level One Accept (L1A) pulse, which tells the ROCs to start reading the data 
from the digitizing modules. During readout the TS sets the {\tt busy} flag, which 
prevents any additional triggers to be accepted before the  ROCs have finished 
processing data. When maximum event processing rate is comparable to the 
trigger rate, this can lead to DAQ dead-time. Dead time tells us the percentage 
of the lost triggers (good events) due to the limitations of the DAQ, and can 
be determined by comparing the number of recorded CODA events to the number of 
scaler events (see Sec.~\ref{sec:ScalerMeasurements}).

\begin{figure}[!ht]
\begin{center}
\begin{minipage}[t]{0.45\textwidth}
\hrule height 0pt
\includegraphics[width=1\textwidth]{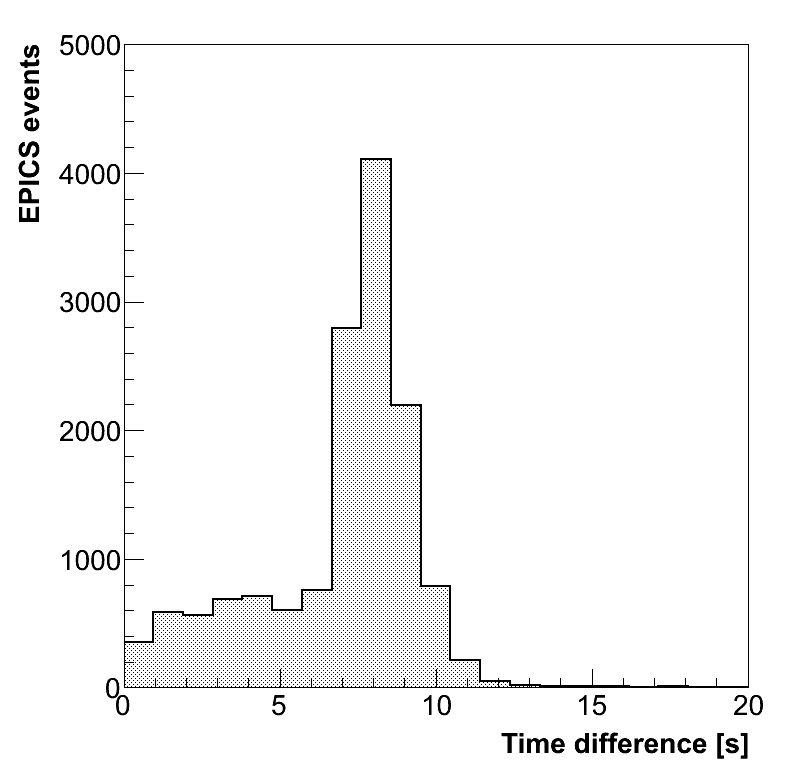}
\end{minipage}
\hspace*{1cm}
\begin{minipage}[t]{0.4\textwidth}
\hrule height 0pt
\caption{ The distribution of time difference between two sequential 
EPICS entries. EPICS information is inserted into the data stream during 
the data taking process. 
\label{fig_EPICSTime}}
\end{minipage}
\end{center}
\end{figure}

In addition to the CODA system, Experimental Physics and Industrial Control System 
(EPICS) is employed to record various assisting information at a slower update 
rate. EPICS is used to record the information regarding spectrometer
magnets, information on beam position, current and energy and various target 
information such as target polarization, magnetic field orientation and cell 
temperature. It also records the value of the beam current at the injector 
and the beam parameters delivered to halls B and C. The data are typically inserted  
into the raw data file in ASCII form every few seconds 
(see Fig.~\ref{fig_EPICSTime}). A typical set of collected data includes 
approximately 60 EPICS entries.  

The data taking process is controlled by the operators in the Hall A counting 
house via the CODA graphical user interface (GUI) known as the Run Control. First, 
the GUI is used to set the experimental configuration. At this point all 
considered parts of the DAQ are loaded with configuration scripts
for proper readout of detectors. After CODA is properly set up, the GUI can be used
to start and stop the data acquisition. During data taking, the CODA GUI serves
for checking the rate of data recording, the dead time and to monitor 
the  CODA components (ROCs, EB, EC). The raw data collected between each 
start and stop of CODA is called a run. Each run is assigned a sequential 
run number together with the name of the experiment (e.g. \texttt{e05102\_1234.dat.0}) 
and is written to a local disk array. Recorded runs are later sent to a tape 
silo called Mass Storage System (MSS) for long-term storage. The 
data are transported to the MSS approximately once per day.

Front-end crates with electronics modules for acquisition of data from the 
HRS-L detectors are installed next to the detectors inside the HRS-L detector 
hut. The electronics for collecting data from the BigBite, together with the
power supplies for all the detectors, is located in a weldment located on the 
left side of Hall A. The weldment is shielded from radiation 
by a thick lead wall. The detectors of the BigBite spectrometer, which is
positioned on the opposite side of the hall (with respect to the target),
were connected to the electronics by means of $\approx 30\,\mathrm{m}$ long 
cables. For the E05-102 experiment, the triggering system and the TS were also 
located in the BigBite weldment. 

\subsection{Trigger System} 
\label{sec:TriggerSetup}
Triggers are electronic pulses that are formed when a particle hits the detector 
or a detector package in a spectrometer. From the combination of these signals 
at a given moment we decide  whether they correspond to a certain physical 
process and whether they should be recorded or not. Which detectors need to be 
hit simultaneously to produce a specific trigger is determined by the trigger 
circuit. For construction of the trigger circuit combination of Nuclear Instrument
Modules (NIM) and Computer Automated Measurement And Control modules (CAMAC) 
was employed. After triggers are created they are introduced to the TS, which  
decides which trigger to accept and starts downloading data from ADC and TDC modules.
In the E05-102 experiment eight different triggers were considered for detection of 
coincident events with spectrometers HRS-L and BigBite.

\subsubsection{Single BigBite trigger T1}
\begin{figure}[!hb]
\begin{center}
 \includegraphics[height=0.9\linewidth, angle=90]{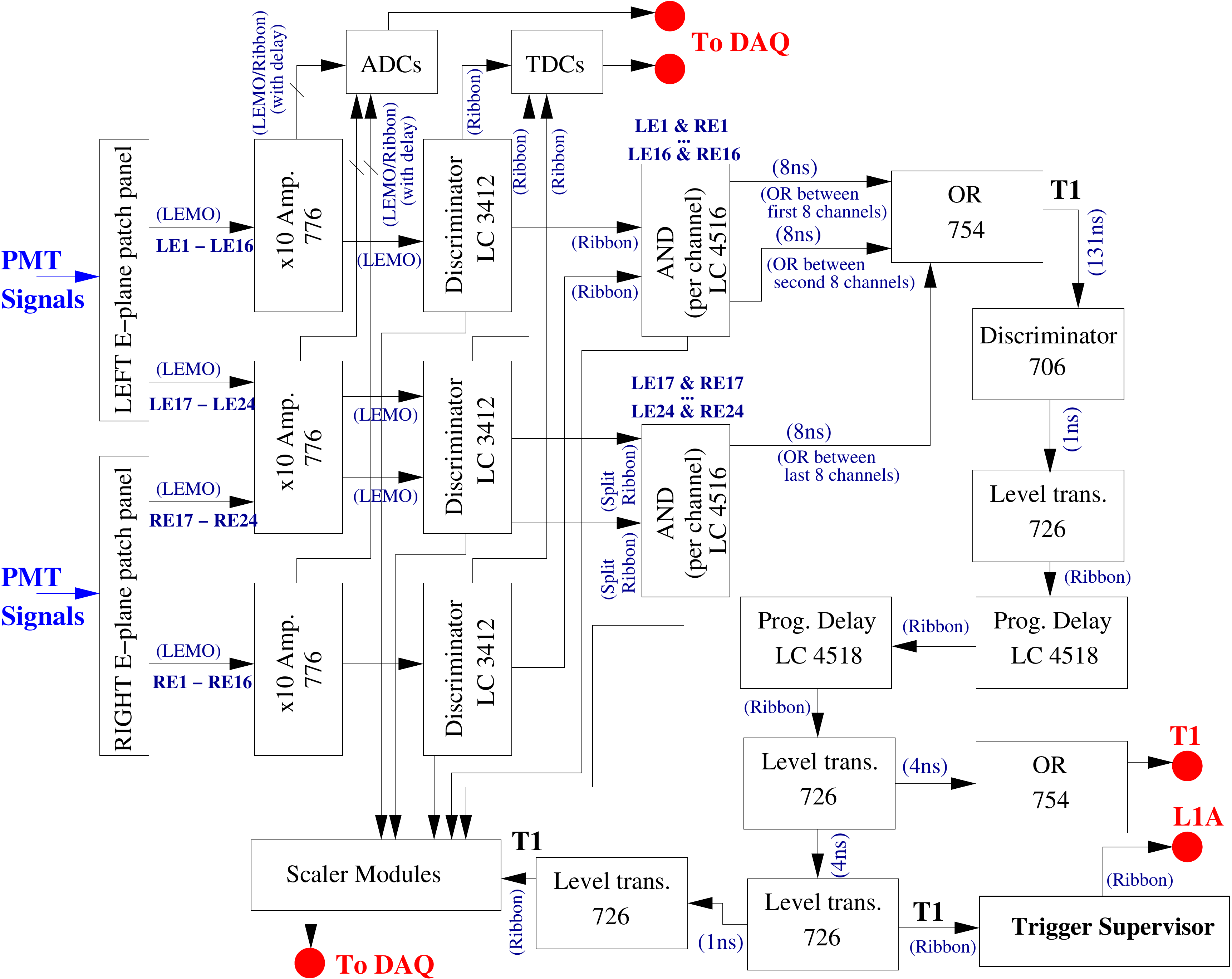}
 \caption{Electronics scheme of the BigBite main trigger T1.
 \label{fig_BBT1trigger}}
\end{center}
\end{figure}

Trigger T1 is the BigBite main trigger. It is formed whenever there is a coincidence
hit (both PMTs see a valid signal) in one of the paddles of the E-scintillation 
plane. Its complete electronics scheme is shown in Figure~\ref{fig_BBT1trigger}.
The signals from the PMTs are led from the detector patch panel (PP) to 
the BigBite weldment using $30\,\mathrm{m}$ coaxial cables. There all 48 signals 
are first amplified by a factor of $10$ by Phillips analog 
modules PS-776. Amplified signals were then taken through $8\,\mathrm{ns}$ LEMO 
cables to the discriminators. We used LeCroy LC-3412 modules with electronically 
controlled threshold levels which could be set remotely. During the experiment, the threshold
levels were adjusted to separate  protons and deuterons from minimally ionizing particles 
that were kept below threshold. This significantly reduced the number of recorded 
random events.

Twisted-pair ribbon cables were used to connect the discriminators with the 
LeCroy units LC-4516 where a logical AND between the left and right PMT 
signal for each scintillation paddle is formed. This way we specify that a
valid hit in the E-plane requires both left and right PMT to measure a pulse
that exceeds the threshold level. Since the detector consists of 24 paddles (48 PMTs) 
we require two LC-4516 modules with sixteen channels each. On the output of these 
two modules we get three signals, each representing an OR between eight 
combined (L AND R) single paddle signals. These three signals are guided over 
$8\,\mathrm{ns}$ LEMO cables to the PS-754 module where a final logical OR between 
them is performed. The single signal on the output represents the T1 trigger.

Ultimately we are interested in coincidence triggers between HRS-L and 
BigBite spectrometers. Due to the shorter cable distances, trigger pulses from 
BigBite arrive earlier to the TS than triggers from HRS-L. Therefore an additional
delay has to be applied to BigBite triggers in order for all triggers to come
simultaneously to the TS and form coincidence triggers. For that purpose, two 
electronically programmable delay modules were utilized, each being able to delay 
the signal by up to $32\,\mathrm{ns}$. The precise amount of required delay was 
determined during the commissioning phase of the experiment. Level-translation 
modules PS-726 were placed before and after the programmable delay modules to 
translate LEMO-cable signals to twisted-pair signals. For the rest of the needed 
delay a $115\,\mathrm{ns}$ cable delay was exploited. To refresh the signal 
after it comes out of a long cable, a PS-706 discriminator was used. 

\subsubsection{Single BigBite trigger T2}

\begin{figure}[!ht]
\begin{center}
 \includegraphics[height=0.9\linewidth, angle=90]{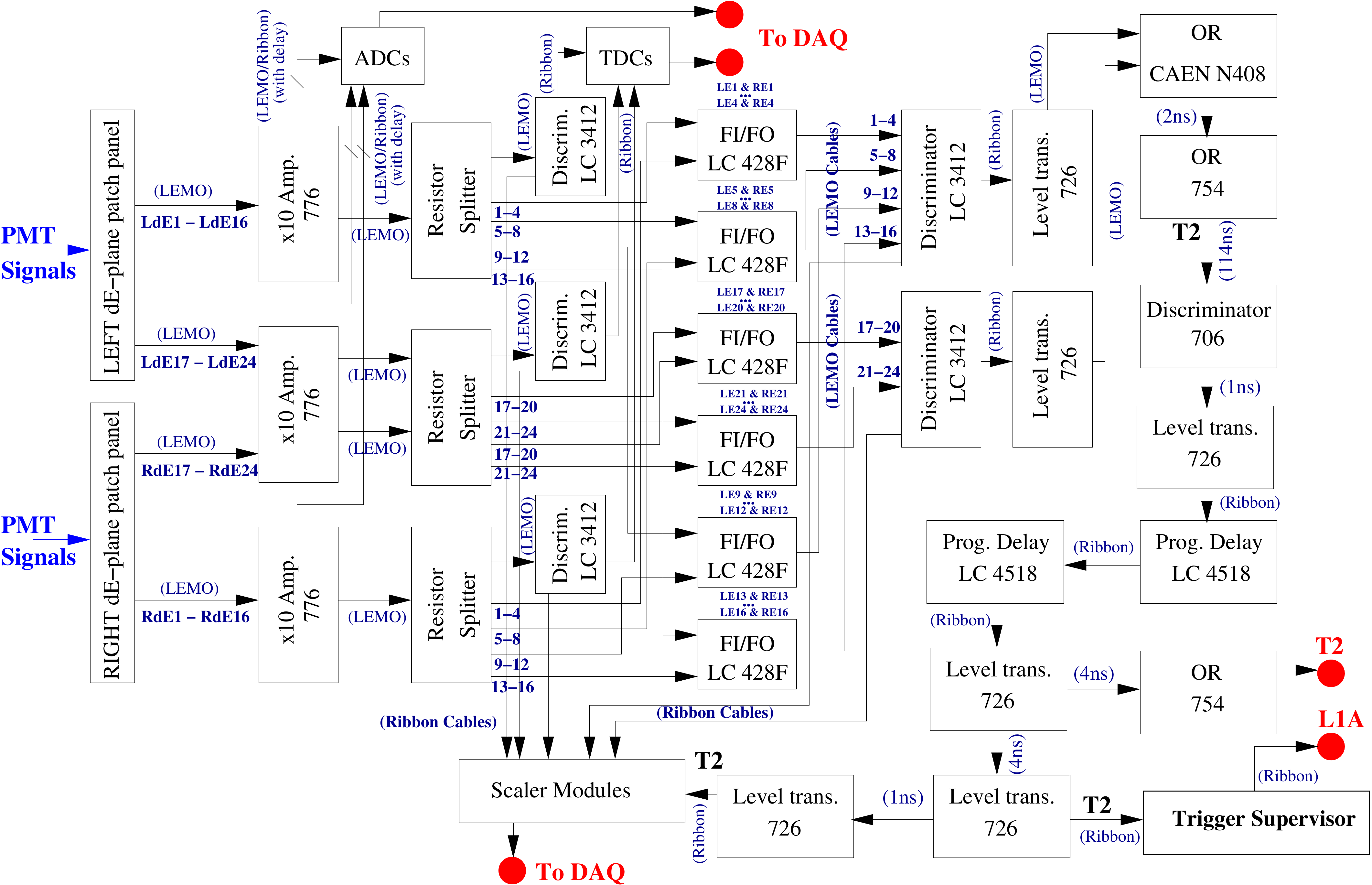}
 \caption{The electronics scheme of the BigBite secondary trigger T2.
\label{fig_BBT2trigger}}
\end{center}
\end{figure}

Trigger T2 is a secondary BigBite trigger. It is designed to select events that 
hit the dE-scintillation plane. The electronics scheme of the  T2 trigger is presented 
in Fig.~\ref{fig_BBT2trigger}. It shows that the T2 trigger is constructed a bit 
differently than T1. 
After the amplification the detector signals get divided into two parts using a simple 
resistor splitter. The first copy leads to the discriminators and then to the TDCs. 
The remaining copy is used for triggering and is connected to a LeCroy module 428F.
There an analog sum of the signals from the left and right PMTs for each paddle
is performed. The LC-428F modules have capacited and non-capacited outputs. We 
used capacited outputs to eliminate the $60\,\mathrm{Hz}$ noise in the signals 
caused by the power supply cables. After the summation, the signals from all paddles get 
discriminated using modules LC3412. With the discrimination performed only after the
summation, we accept also events with a signal in only one PMT per paddle, as long as 
the detected signal is high enough to come through the discriminator. The main reason for 
this is to detect particles with very small momenta.  From the discriminators, 24 signals
are taken to two PS-726 modules, where signals are translated from 
Ribbon cables to LEMO cables. These are then attached to the CAEN module N408 to 
form a logical OR between all of them in order to form the T2 trigger.

After T2 is created it is also introduced to some additional delay to wait for 
triggers from HRS-L. The same circuit as for T1 has been used. 
The only difference is a $10\,\mathrm{ns}$ shorter delay cable, because  
T2 requires few $\mathrm{ns}$ more to be generated. The time difference 
between T1 and T2 is caused by additional electronics and cables that 
was required for the construction of T2.  

\subsubsection{Single HRS-L triggers T3 and T4}

\begin{figure}[!ht]
\begin{center}
\includegraphics[height=\linewidth,angle=-90]{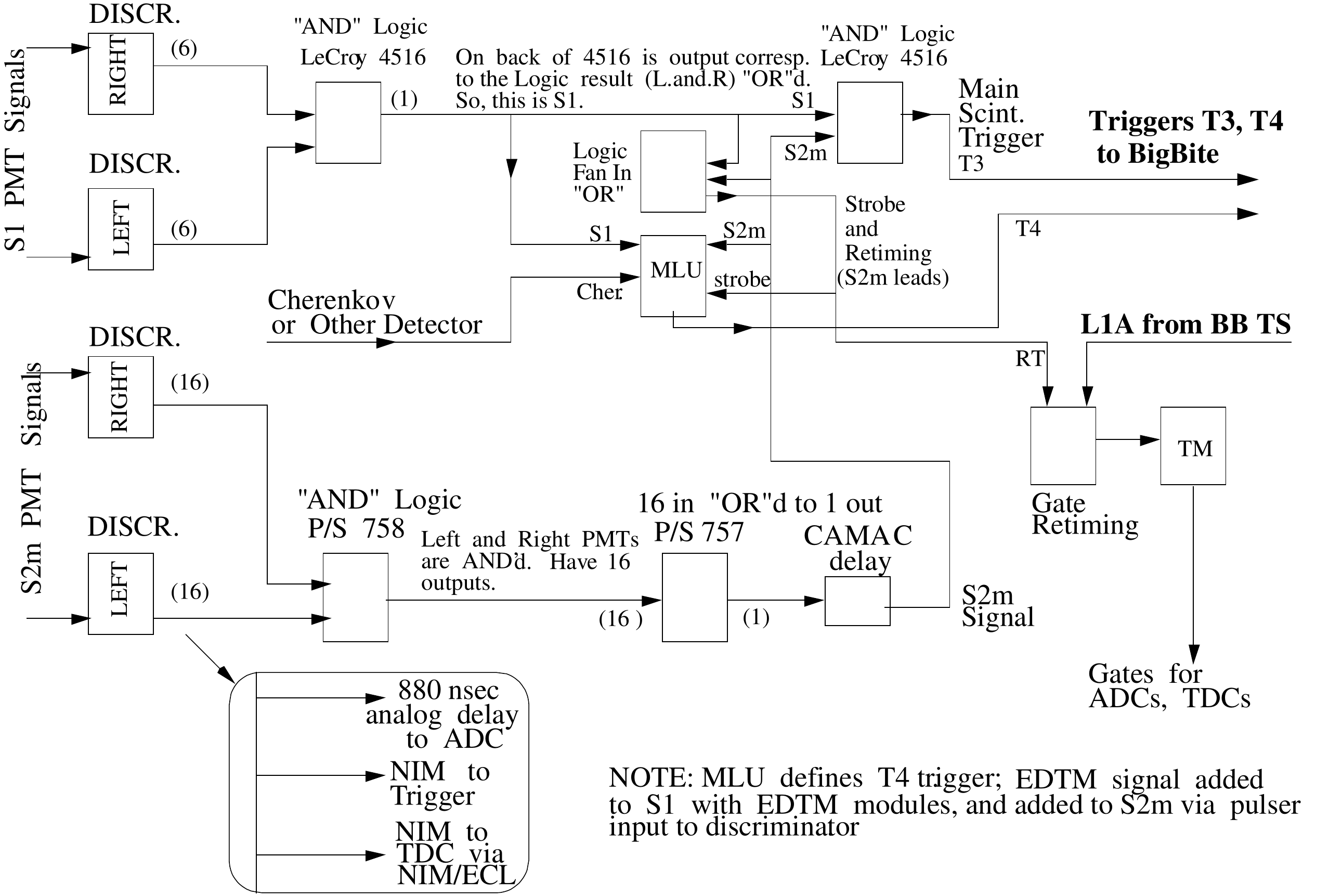}
\caption{The electronics scheme of the HRS-L single triggers T3 and T4. This
trigger scheme is a modified version of the original drawing made by Robert
Michaels~\cite{BobTriggerScheme}. 
\label{fig_BBT3trigger}}
\end{center}
\end{figure}

Triggers T3 and T4 are HRS-L triggers~\cite{alcorn}. Trigger T3 is the main trigger and is 
formed when both scintillator planes S1 and S2m have hits on both sides of a paddle 
(see Fig.~\ref{fig_BBT3trigger}).   
Trigger T4 is a supplementary HRS-L trigger and is used to measure trigger efficiency.
It is designed to select electrons which cause a hit in either the first (S1) or the second (S2m) 
scintillator plane and a hit in the Cherenkov detector which is positioned between the two 
planes. Hence, T4 is an exclusive trigger with respect to T3, since T4 is not formed 
when both S1 and S2m are hit. After T3 and T4 are formed, they are sent over 
$60\,\mathrm{m}$ long cable to the BigBite weldment to be processed by the triggering system. 
After BigBite TS accepts a trigger, the L1A signal is sent back up to the HRS-L to initiate 
the readout of the ADCs and TDCs. In order to have a constant reference point for the 
TDC information, the readout of the digitizing modules is tied to the leading edge
of the right side PMT signal of the S2m scintillator bars~\cite{huangPhD}.

\subsubsection{Coincidence triggers T5 and T6}

The coincidence triggers T5 and T6 were the main triggers for the experiment 
E05-102. Electrons and hadrons (p, d) need to be detected simultaneously in
order to measure the asymmetries for semi-exclusive reactions 
${}^3\vec{\mathrm{He}}\left(\vec{\mathrm{e}},\mathrm{e'} \mathrm{d }\right)$ and 
${}^3\vec{\mathrm{He}}\left(\vec{\mathrm{e}},\mathrm{e'} 
\mathrm{p}\right)$. The T5 trigger was constructed by overlapping T1 and T3 triggers 
in time. The bulk of the recorded coincidence events were based on this trigger. 
The secondary coincidence trigger T6 is created by overlaying triggers T2 and
T3. This trigger is important for the detection of particles with very low momenta
which do not have enough energy to reach the E-scintillation plane, but are completely
stopped already in the dE-layer of the scintillators.

The construction of the coincidence trigger needs to be precisely planned.
Details such as time delays in electronics modules and cables, as well as the 
flight time of particles through spectrometers need to be considered in order to assemble
a capable coincidence trigger. The construction of triggers T5 and T6 
started very ambitiously, with features such as programmable width and position of 
the coincidence window. However, when we made a simulation of the experiment, 
we realized that due to the long cables between HRS-L and BigBite, 
the T3 trigger arrived very late to the coincidence trigger circuit. 
Consequently, the L1A pulse from TS, which starts the readout of the digitizing 
modules, was formed even later, and by that time the raw detector signals 
that were waiting in BigBite's ADCs and TDCs to be read out, could 
already be gone. To make up for this delay, we had to construct a much simpler 
version of the coincidence trigger with as little electronics as possible 
and without any extra features. 

\begin{figure}[!ht]
\begin{center}
  \includegraphics[width=1\linewidth]{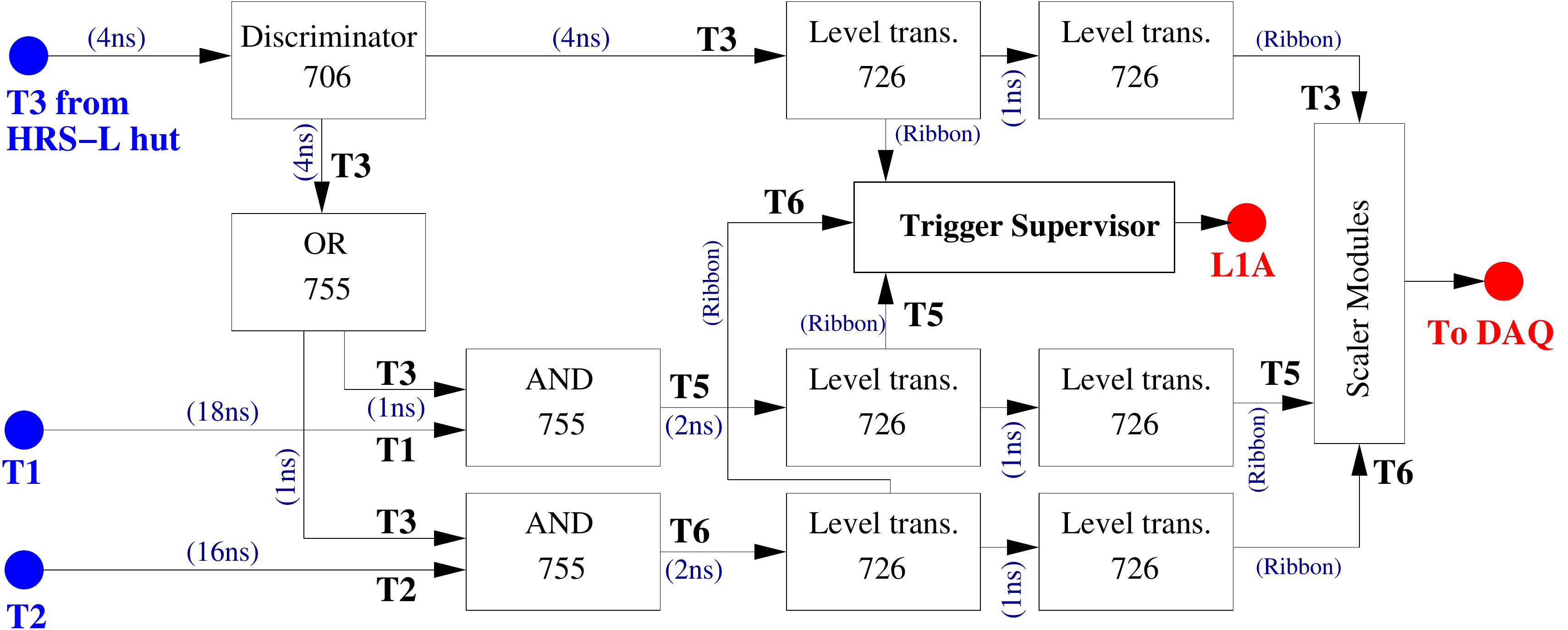}
  \caption{Electronics scheme of the coincidence triggers T5 and T6.
  \label{fig_BBCoincTrigger}}
\end{center}
\end{figure}

Figure~\ref{fig_BBCoincTrigger} shows the electronics scheme of our final 
coincidence trigger. The signal T3 from the HRS-L goes first 
into the discriminator to refresh the pulses. The output signal 
is then led to the  PS-755 module  where a coincidence window is formed. 
We have set the coincidence window to be approximately $80\,\mathrm{ns}$ wide. 
Two copies of the coincidence window signal are then sent to two PS-755 modules to perform
a logical AND with triggers T1 and T2, respectively. The outputs from these two 
modules are the coincidence triggers T5 and T6.  The cable delays were set such 
that T1 and T2 came to PS-755 modules approximately $25\,\mathrm{ns}$ after the  
T3 coincidence window. This ensures that the timing of T5 and T6 was defined by the leading 
edges of T1 and T2 triggers. After coincidence triggers were created
they were fed into the trigger supervisor via the level-translating modules 
PS-726 for further decisioning.

\subsubsection{Triggers T7 and T8}
\label{sec_BB_cosmic_trigger}

When no  beam was available, we were 
using cosmic rays to test and calibrate the BigBite detectors. For that purpose we 
devised a cosmics trigger T7.  A $2\,\mathrm{m}$ long scintillation bar 
(called HAPPEX paddle) was added to the 
BigBite detector package and was positioned  between the two wire chambers. 
This bar was long enough to vertically cover both MWDCs and scintillation planes. 
The purpose of the paddle was to ensure that the particle which hit the dE- and E-planes 
also managed to come though the MWDCs. The light signal from the paddle was 
read out by two PMTs mounted at its edges. The signals from the PMTs 
were led to the BigBite weldment where they were amplified and logical ORed.
The output signal ($\mathrm{T_{Happex}}$) was lead to a PS-755 module  to meet 
the T1 trigger. A logical AND between T1 and $\mathrm{T_{Happex}}$ defines the cosmics 
trigger T7. 
Before the experiment started, the HAPPEX paddle was removed and T7 disconnected from 
the trigger supervisor. Consequently, T7 was not available during production data
taking.

Trigger T8 represents the $1024\,\mathrm{Hz}$ pulser. It serves as a reference point
and for testing the performance of the DAQ system.

\subsubsection{BigBite re-timing}

\begin{figure}[!ht]
\begin{center}
   \includegraphics[width=\linewidth]{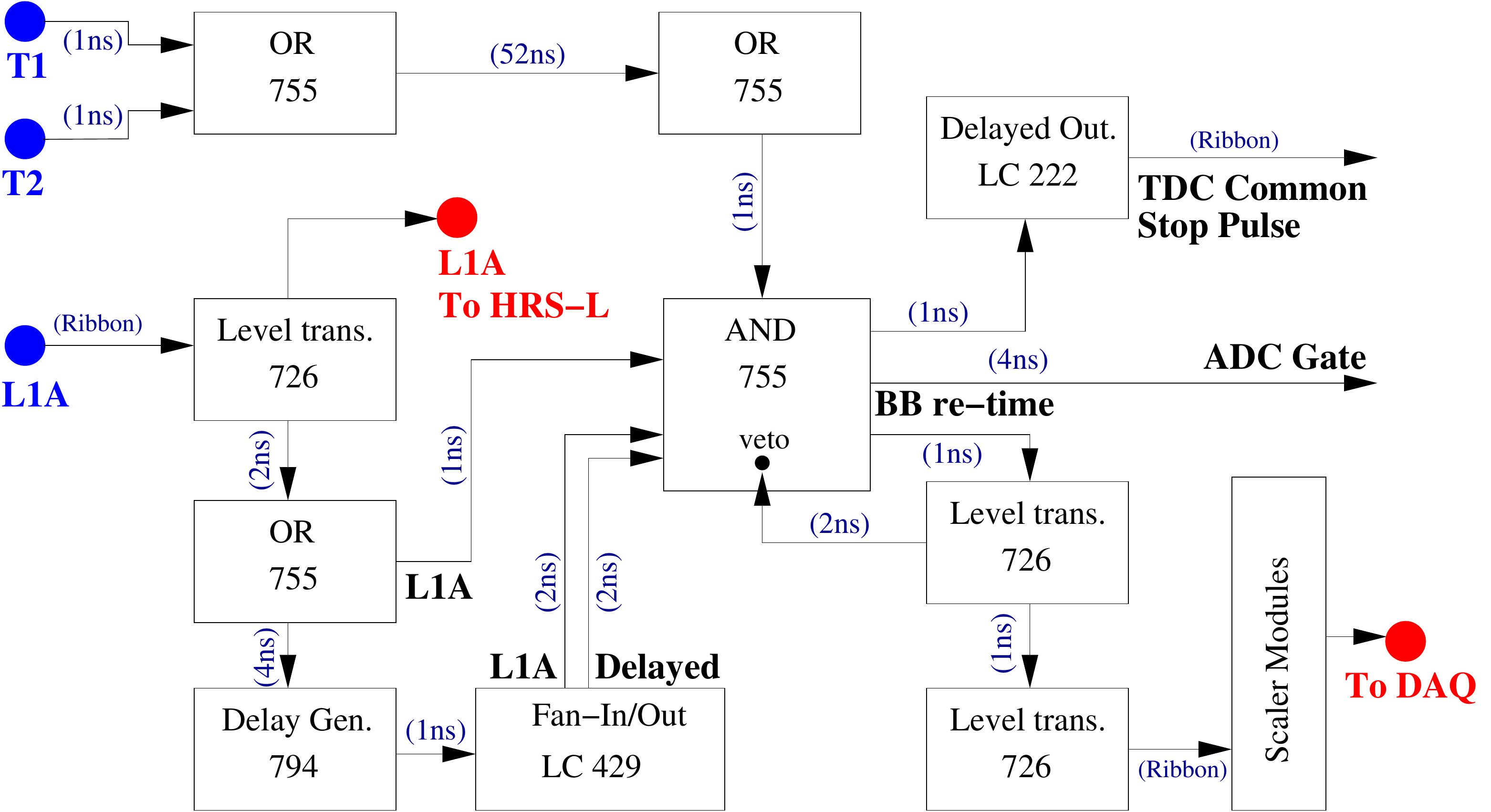}
   \caption{BigBite re-timing circuit.\label{fig_BBRetiming}}
\end{center}
\end{figure}

When the trigger supervisor accepts a trigger, it generates a L1A pulse which can
be directly used to form the gates for BigBite ADCs and TDCs. However, we want the 
gate pulse to be tied to the spectrometer triggers T1 and T2. For that purpose, a 
L1A pulse needs to be re-timed with respect to the local triggers, similarly
as it is being done for the HRS-L spectrometer (see Fig.~\ref{fig_BBT3trigger}). 
This is achieved by the BigBite re-timing circuit, which is shown in 
Fig.~\ref{fig_BBRetiming}. 
First, a PS-755 module is utilized to form a logical OR between T1 and T2. 
Delay cables are set in a way that T1 comes to the module $\approx 2\,\mathrm{ns}$ 
before T2. If the primary trigger (T1) is present, then this time difference ensures 
that T1 will be used for re-timing. If for a particular event only the secondary 
trigger (T2) is available, then T2 will be considered for it.
Then we perform a logic AND between this combined signal (T1+T2) and the L1A 
pulse, which must come to the module first in order for the circuit to work. 
Since TS requires some time to produce L1A from the input triggers, 
we need to take the (T1+T2) pulse through a $\mathrm{50\,\mathrm{ns}}$ delay.
The pulse at the output represents the BigBite re-time signal which is used to 
initiate the readout from BigBite's ADCs and TDCs.

It can happen that for a given event both T1 and T2 are absent. In order 
to record also such events, where only HRS-L triggers (T3, T4) or 
pulser (T8) are present, two additional modules (PS-794 and PS-729) are 
added to the circuit. Their task is to delay the L1A signal by 
approximately $90\,\mathrm{ns}$, which is a bit more than the 
width ($\approx 80\,\mathrm{ns}$) of the L1A pulse. If by that time T1 and T2
are not formed, the circuit decides to accept the delayed 
un-timed L1A pulse. The Phillips module PS-755, where the final logical AND 
is realized,  is set to setting No.~2, which means that it makes an AND between 
any two input signals. For a good event, these two signal are 
L1A and delayed (T1+T2). In the case where neither T1 nor T2 are present, these 
two signals are two identical copies of the delayed L1A pulse. However, a delayed L1A signal 
is always formed, also when T1 and T2 are present, and comes $90\,\mathrm{ns}$ 
after the non-delayed one. Therefore, to prevent double pulsing and recordings of bogus 
events, we loop a BigBite re-timing pulse back to the logical module as a
VETO, to disable the module's input and prevent the reading of a secondary 
L1A pulse.

\subsubsection{Prescale factors}

The trigger rates for the eight triggers can be very different and depend
strongly on the properties (cross-section) of the observed physical process and 
on the experimental conditions, such as beam current and spectrometer acceptances.
Usually raw trigger rates for single triggers (T1, T2, T3, T4) are much higher 
than the rates for the coincidence triggers (T5, T6). Since we are mostly interested  
in coincidence events, we do not want to record too many single-arm events. 
The rate at which we accept particular trigger and record the corresponding events 
can be reduced by setting proper prescale factors ($PS = \mathrm{(Raw Rate)/(Recording Rate)}$). 
Prescale factors ($PS$) are integers which are stored to the trigger supervisor 
before each run. During data taking the trigger supervisor then accepts every 
$PS^{\mathrm{th}}$ trigger. Prescale factors also allow us to properly set the 
total data collecting rate in order to keep the dead time within acceptable limits 
(typically below $10\,\mathrm{\%}$). The maximum recording rate was limited to $2.5\,\mathrm{k}$ events per 
second or approximately $5.5\,\mathrm{MB}$ per second. Typical trigger rates and considered 
prescale factors for the production data taking are shown in Table.~\ref{table_prescales}.

\begin{table}[!ht]
 \begin{center}
\caption{Prescale factors with raw and prescaled trigger rates for two kinematical 
settings considered in experiment E05-102. Presented values for each kinematics 
correspond to the mean values determined at the end of a typical run. We collected 
as many coincidence events as possible, together with some HRS-L single events. 
BigBite single events were neglected. 
\label{table_prescales}}
\begin{tabular}{c|ccccccccc}
\multicolumn{9}{c}{}\\
\multicolumn{9}{c}{$\mathrm{Run\>Number = 2300}$, $\theta_{HRSL} = 14.5^\circ$, 
$\theta_{BB} = -75.0^\circ$ }\\
\toprule
Trigger & T1 & T2 & T3 & T4 & T5 & T6 & T7 & T8 \\
\hline \\[-3.5mm]
Prescale & 16777215 & 16777215 & 3 & 4 & 1 & 1 & 65535 & 100  \\[-1.5mm]
factor & & & & & & & & \\
Raw rates & 301437.4 & 255878.6 & 8190.8 & 182.7 & 909.6 & 436.4 & 0 & 1024.0 \\[-1.5mm]
$[Hz]$ & & & & & & & & \\
Prescaled & 0.0 & 0.0 & 2730.3 & 45.7 & 909.6 & 436.4 & 0 & 10.2 \\[-1.5mm]
rates $[Hz]$ & & & & & & & & \\
\bottomrule
 \end{tabular}
\begin{tabular}{c|ccccccccc}
\multicolumn{9}{c}{}\\
\multicolumn{9}{c}{}\\
\multicolumn{9}{c}{$\mathrm{Run\>Number = 3081}$, $\theta_{HRSL} = 12.5^\circ$, 
$\theta_{BB} = -75.0^\circ$ }\\
\toprule
Trigger & T1 & T2 & T3 & T4 & T5 & T6 & T7 & T8 \\
\hline \\[-3.5mm]
Prescale & 16777215 & 16777215 & 100 & 5 & 1 & 1 & 65535 & 100  \\[-1.5mm]
factor & & & & & & & & \\
Raw rates & 282125.8 & 244688.1 & 22510.9 & 432.5 & 2512.0 & 1464.9 & 0 & 1024.0\\[-1.5mm]
$[Hz]$ & & & & & & & & \\
Prescaled &  0.0 & 0.0 & 225.1 & 86.5 & 2512.0 & 1464.9 & 0 & 10.2 \\[-1.5mm]
rates $[Hz]$ & & & & & & & & \\
\bottomrule
 \end{tabular}
\end{center}
\end{table}

\subsection{BigBite ADCs}

\begin{figure}[!ht]
\begin{center}
\begin{minipage}[t]{0.6\textwidth}
\hrule height 0pt
    \includegraphics[width=\linewidth]{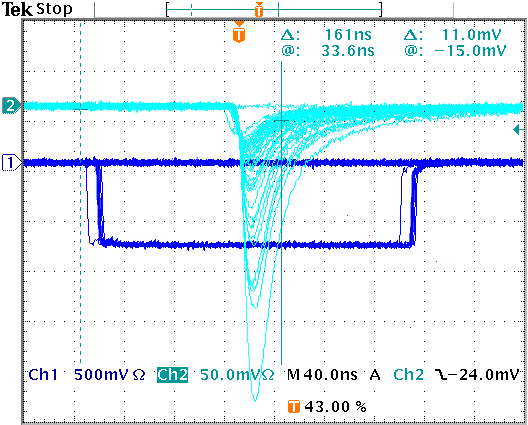}
\end{minipage}
\hfill
\begin{minipage}[t]{0.32\textwidth}
\hrule height 0pt
\caption{Scope plot of an amplified and delayed PMT signal, 
relative to the ADC gate. A point (in vertical direction) where all 
analog signals intersect, represents the discriminator (leading edge) 
threshold level considered for the T1 trigger.
The ADC gate is approximately $250\,\mathrm{ns}$ wide. 
\label{fig_BBADCGate}}
\end{minipage}
\end{center}
\end{figure}

In order to record the information about the energy deposition inside the scintillation 
detectors, a copy of the amplified PMT signals from modules PS-776 is introduced 
to the 12-bit CAEN V792 ADC modules, where the analog signals get digitized. 
See Figs.~\ref{fig_BBT1trigger} and~\ref{fig_BBT2trigger} for details. For the 
transport of the signals to the ADCs, special ribbon cables made of small coaxial cables
were used, which reduce the crosstalk in cabling. In addition, 
hi-pass filters at the input to the ADC modules were used to suppress the low 
frequency (60Hz) noise.  Before analog signals were put into the ribbon 
cables, they were first taken through long delay cables, where they had to wait
for triggers to be accepted and ADC gate opened by the BigBite re-timing pulse.
It is crucial to have properly chosen delay cables so that signals do not miss
the ADC-window. The total amount of delay from the output of the amplifier to 
the input of the ADC module was measured to be $\approx500\,\mathrm{ns}$. 
Figure~\ref{fig_BBADCGate} shows the scope plot of the ADC gate window and 
analog pulses from the E-scintillation plane just before entering the ADC modules. 
The ADC gate is approximately $250\,\mathrm{ns}$ wide. For the ADC gate we utilized 
the BigBite re-timing pulse (see figure~\ref{fig_BBRetiming}), where we increased the 
length of the outgoing pulse on the PS-755 module to the proper value. 

\subsection{BigBite TDCs}
The time information on the hits in the scintillation detectors was obtained
by means of the F1-TDC modules, which were developed at Jefferson Lab. 
These modules have a resolution of $60\,\mathrm{ps}$  and were  operating in 
 common-stop mode. The F1-TDCs have a $\approx 850\,\mathrm{ns}$ long
acceptance window where data are constantly flying through. When 
a common-stop signals appears the module reads and digitizes the data that 
are inside the window at that particular moment. For the common-stop 
pulse we once again used the BigBite re-timing pulse, which was approximately 
delayed by using the LC-222 module.

For the E-plane, the signals for the TDCs were extracted from the 
secondary output of the LC-3412 discriminator modules and transported to
the TDCs using twisted-pair ribbon cables. The signals for 
the dE-plane are connected to TDC modules somewhat differently. 
See Fig.~\ref{fig_BBT2trigger}. After the resistor splitters, the analog signals 
are lead to the discriminators and from there to the TDC modules 
via ribbon cables. Because of the resistor splitters, the signals going 
to the dE-plane TDCs have half the initial amplitude. This fact needs to be considered 
when setting the thresholds for these discriminators. They should be 
set to approximately one half of the setting used for the
dE-plane trigger discriminators.

During the experiment we also recorded the time information of all eight
triggers. For this we employed FastBus TDCs with a resolution of $500\,\mathrm{ps}$.
These data were extensively used during the experiment for monitoring of 
trigger performance. In the off-line analysis, these data can be used for 
extraction of coincidence events, background estimation (random coincidences)
and even as a part of a particle identification.

\subsection{Event Dead Time Monitor}
\label{sec:EDTM}

To test if the trigger electronics was properly assembled and configured, 
artificial signals which simulate real physical processes were considered.
To emulate such signals, we utilized two Event Dead Time Monitors (EDTMs), 
one configured as master and the other as slave. The master was
situated in the HRS-L while the slave was positioned in the BigBite 
weldment. The master EDTM sends generated pulses to the HRS-L electronics 
and to the slave EDTM, which then sends out delayed pulses to the 
BigBite electronics. The delay corresponds to the real physical 
delay, and encompasses the differences in flight paths and flight times of 
protons and electrons, as well as the length of the cables 
connecting both spectrometers.

In the HRS-L spectrometer the ETDM pulses are utilized to simulate the pulses
coming from the scintillator detectors S1 and S2m. To achieve this, we  
connected the signal cables from the S1 scintillators to the EDTM board before 
feeding signals to the discriminators. There we combined signals from the PMTs 
with the EDTM pulses. For the S2m, we attached the EDTM pulses directly to
the discriminator modules via the pulser input. See Fig.~\ref{fig_BBT3trigger}
for details.

To emulate the pulses from the BigBite E-scintillation plane, 
EDTM signals were coupled to the LC-3412 discriminators via
pulser inputs. For simulation of hits in the dE-plane, EDTM 
pulses were plugged as an extra input to the FI/FO modules LC-428F. 
The EDTM modules were configured to generate the pulses at approximately 
$20\,\mathrm{Hz}$. The findings of the study with the EDTM pulses 
are presented in Appendix~\ref{appendix:EDTM}.

\vspace*{5mm}
\subsection{Scaler modules}
\label{sec:ScalerMeasurements}

\begin{figure}[!ht]
\begin{center}
\includegraphics[width=\linewidth]{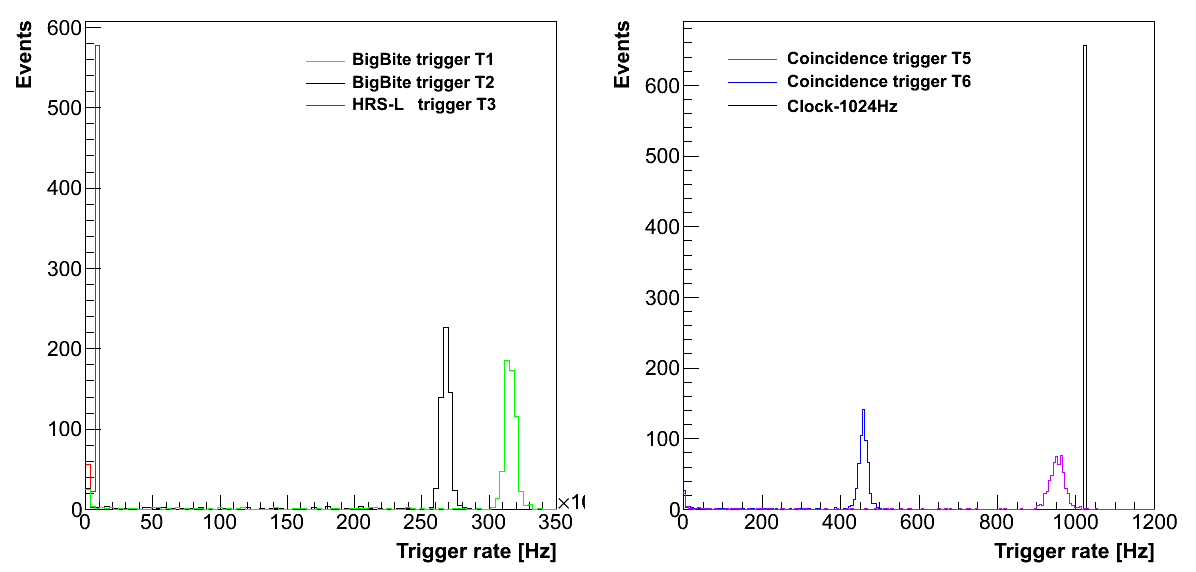}
\caption{Raw trigger rates observed during the experiment. 
Trigger rates in BigBite spectrometer are much larger than in the HRS-L 
($\approx 10\,\mathrm{kHz}$), because of the of the much larger acceptance. 
The coincidence events are detected approximately every $1\,\mathrm{ms}$.
\label{fig_scalers_trigger_rates}}
\end{center}
\end{figure}

Scaler modules are used to perform deadtime-free counting of digitalized signals and 
were utilized in the experiment to record counting rates and total number of 
counts of various important signals. The scalers were
exploited especially for real-time monitoring of the beam current, trigger rates and raw rates 
in all PMTs in all considered detectors. Typical raw trigger rates 
observed during the experiment are shown in Fig.~\ref{fig_scalers_trigger_rates}.
In the succeeding analysis the 
scaler information was used to properly normalize the data to the collected charge and
to determine the DAQ deadtime. Furthermore, they were used for finding any false 
asymmetries related to the deadtime and accumulated charge, which could affect 
the extraction of the experimental asymmetries. 

\begin{figure}[!ht]
\begin{center}
\begin{minipage}[t]{0.5\textwidth}
\hrule height 0pt
    \includegraphics[width=\linewidth]{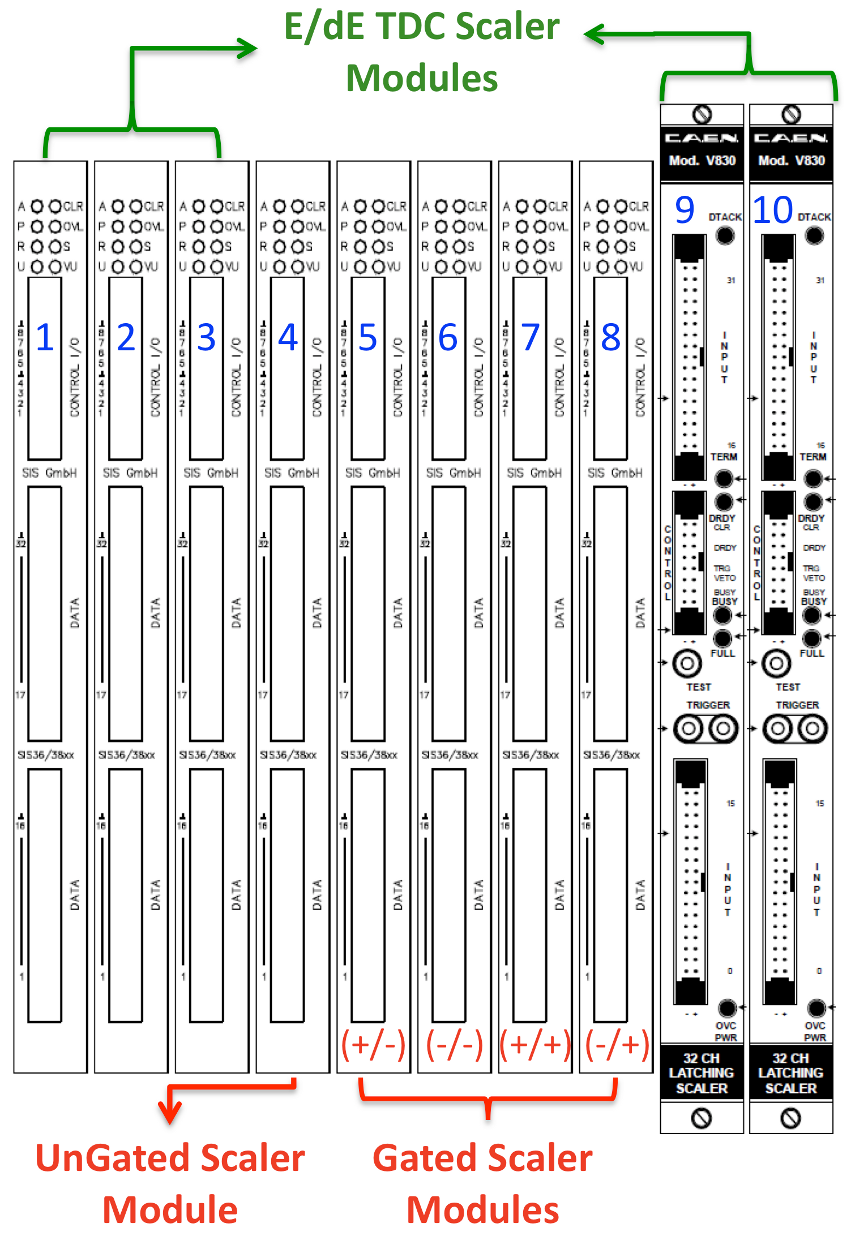}
\end{minipage}
\hfill
\begin{minipage}[t]{0.45\textwidth}
\hrule height 0pt
\caption{Setup of the scaler modules considered for the BigBite spectrometer.
Modules 1, 2 and 3 were used for recording raw digitalized pulses for each of the 
96 PMTs of the dE/E scintillation detector. Modules 9 and 10 were utilized for 
monitoring   of compound signals from each scintillator paddle. For that 
48 scaler channels are required. Modules 4, 5, 6, 7 and 8 were recording 
rates for triggers and other interesting pulses (see Table~\ref{table_scalers_list}). 
Module 4 was ungated and recorded total rates.
Modules 5, 6, 7 and 8 were gated with the beam-helicity and target-spin states. 
Each of the gated modules records rates only when matching combination 
of the (beam/target)-state is present.
\label{fig_Scalers_BigBiteConf}}
\end{minipage}
\end{center}
\end{figure}

Each spectrometer has been equipped with ten scaler modules The setup considered 
for BigBite is presented in Fig.~\ref{fig_Scalers_BigBiteConf}. The first three
modules recorded raw rates in the 96 channels of the  dE/E-scintillation detector. 
The compound signal from each scintillation paddle (see Figs.~\ref{fig_BBT1trigger} 
and~\ref{fig_BBT2trigger}) 
was also monitored using the last two CAEN modules. This way the performance of 
the scintillation detector could be inspected in detail. The remaining five 
modules in the center were daisy-chained and were utilized to record scaler 
information on triggers, beam current monitors and some other vital signals. 
The complete list of recorded signals is presented in Table~\ref{table_scalers_list}. 
Four of the modules were gated based on both helicity and target spin states. 
Hence, each module recorded identical input signal for one particular combination 
of the (beam/target) spin state: $(+/+)$, $(+/-)$, $(-/+)$, $(-/-)$. The fifth
scaler module was left ungated and recorded input information regardless of the
(beam/target) spin state. Additionally, all nine scaler modules were gated 
with a CODA run signal. This allowed scalers to record information only when 
the CODA DAQ was running. 

\begin{table}[!hb]
 \begin{center}
\caption{ Lists of crucial experimental signals which are recorded 
by the HRS-L and BigBite gated scalers.
\label{table_scalers_list}}
\vspace*{3mm}
\begin{minipage}[t]{0.49\textwidth}
\hrule height 0pt
\begin{tabular}{ll}
\multicolumn{2}{c}{{\bf BigBite scalers}}\\
\hline

\hline

\hline \\[-4mm]
\#& Recorded Signal \\[1mm]
\hline
1 & Trigger T1\\
2 & Trigger T2\\
3 & Trigger T3\\
4 & Trigger T4\\
5 & Trigger T5\\
6 & Trigger T6\\
7 & Trigger T7 (cosmics)\\
8 & Trigger T8 (1024 Hz clock)\\
9 & L1A (TS accepted triggers)\\
10 & BigBite re-timing pulse\\
11 & Trigger for the F1 TDCs\\
12 & Delayed L1A pulse\\
13 & 103.7 kHz clock\\
14 & BCM, upstream cavity, gain=1\\
15 & BCM, upstream cavity, gain=3\\
16 & BCM, upstream cavity, gain=10\\
17 & BCM, downstream cavity, gain=1\\
18 & BCM, downstream cavity, gain=3\\
19 & BCM, downstream cavity, gain=10\\
20 & Helicity state\\
21 & EDTM pulser\\
22 & Unser monitor\\
23 & ADC gate \\ 
\hline

\hline

\hline

\end{tabular}
\end{minipage}
\hfill
\begin{minipage}[t]{0.49\textwidth}
\hrule height 0pt
\begin{tabular}{ll}
\multicolumn{2}{c}{{\bf HRS-L scalers}}\\
\hline

\hline

\hline \\[-4mm]
\#& Recorded Signal \\[1mm]
\hline
1 & Trigger T1\\
2 & Trigger T2\\
3 & Trigger T3\\
4 & Trigger T4\\
5 & Trigger T5\\
6 & Trigger T6\\
7 & Trigger T7 (cosmics)\\
8 & Trigger T8 (1024 Hz clock)\\
9 & L1A (TS accepted triggers)\\
10 & BCM, upstream cavity, gain=1\\
11 & BCM, upstream cavity, gain=3\\
12 & BCM, upstream cavity, gain=10\\
13 & BCM, downstream cavity, gain=1\\
14 & BCM, downstream cavity, gain=3\\
15 & BCM, downstream cavity, gain=10\\
16 & Helicity state\\
17 & EDTM pulser\\
18 & 103.7 kHz clock\\
19 & Unser monitor\\
20 & HRS-R primary trigger (T1) \\
21 & HRS-R supplementary trigger (T2) \\ 
\hline

\hline

\hline

\end{tabular}
\end{minipage}
\end{center}
\end{table}

For the HRS-L we devised a very similar configuration. Besides the modules for recording 
the hit rates in Cherenkov and S1- and S2m-scintillation detectors, five modules 
were used to record the critical signals for all combinations of the 
(beam/target)-spin states. The list of the monitored signals is shown in 
Table~\ref{table_scalers_list}. For the redundancy and cross-checking purposes, the 
copies of the BigBite triggers and BCM rates were recorded also by the HRS-L
scalers.

Scaler information is recorded into the data stream in two different ways. 
They are attached to each recorded physical event and can be accessed through 
the \texttt{evbbite} and \texttt{evleft} scaler variables.  Since these variables 
are refreshed for each recorded event, they provide a detailed analysis of 
scalers within each run. Momentary changes in rates, caused e.g. by the beam trips, 
are best detectable through these variables.

On the other hand, the scaler rates are recorded to the data stream as 
individual events every few seconds. These special types of events 
can be reached in the analysis via variables \texttt{bbite} for BigBite 
and \texttt{left} for HRS-L scalers. Due to the low refreshing rates of these
variables they are appropriate only for the scaler analysis of full runs.

Only after the experiment it was discovered that BigBite 
scalers were not properly recorded. The problems appeared for the \texttt{evbbite}
scalers, where the information on the $(-/+)$-gated scaler, was 
overridden by the data from the last two scaler modules. This happened because 
wrong  memory addresses were assigned to those two modules. Fortunately, 
the \texttt{bbite} scalers did not suffer the same problem and could be 
considered for the extraction of the missing information. However, troubles
appeared also for the \texttt{bbite} scalers. Here, the last two CEAN-380 modules were not 
recognized by the CODA analysis library. Consequently, default values for the
scaler modules were used, which consider only first $16$ channels instead of $32$. 
This way, half of the compound E/dE signals are missing in the data stream. 
To recover the missing data, the \texttt{evbbite} scalers can be used, 
where these two modules were treated properly. Hence, by 
combining the information from both types of BigBite scalers, all observables are accessible 
in the analysis.

\chapter{Calibration of the apparatus}
The  experimental apparatus must be well understood and  properly calibrated
before the analysis of the production data could be performed. 
The HRS-L spectrometer was calibrated by Ge Jin and Yawei Zhang~\cite{GeJinPhD}. 
The latter performed also the majority of the target-related
tests and calibrated the NMR and EPR polarimeters~\cite{yawei_3HeTarget, yawei_EPR_report}. 
On the other hand, Ge Jin calibrated the BigBite MWDCs~\cite{GeJinMWDC}.

This chapter presents calibration procedures that were performed by the author,
and the corresponding results.
First, orientation of the the target magnetic field will be determined 
using the compass measurements. Then, the calibration results for the 
two beam monitors will be shown. Special attention will be dedicated 
to the analysis of the BigBite scintillation detectors, followed by the 
tests of the triggering system, which needs be understood perfectly. The 
energy losses of the ejected particles will also be investigated.
The optical calibration of the spectrometers will be
studied in a separate chapter.

\section{Magnetic Field Direction Measurement}
In the E05-102 experiment we measured the beam-target asymmetries for three different
orientation of the target spin: along the beam line (longitudinal (+)), and
two horizontal transverse-to-the-beam directions (transverse ($\pm$)).
The interpretation of the experimental results depends strongly on the direction 
of the target spin. Hence, it is essential to precisely know the orientation of the 
magnetic field that holds the spin in a particular direction. 
The magnetic field is provided by three mutually perpendicular pairs of 
Helmholtz coils (see Sec.~\ref{Target_HoldingField}).Contributions of the BigBite 
fringe-fields and Earth's magnetic field must also be considered.

The precise information on the direction of the holding magnetic field for each 
experimental setting was obtained by the compass calibration measurements. Two
different compasses were employed. The vertical compass was used to determine
the polar (vertical) angle, while the horizontal compass
was utilized to obtain the azimuthal angle of the magnetic field.

\subsection{Vertical Compass Measurement}
The magnetic field created by the Helmholtz coils is expected to be uniform 
at the target position and is proportional to the electrical current. 
By setting currents in all 
three coils, the polar angle $\phi_{\mathrm{B}}$ of the field can be expressed as:
\begin{eqnarray}
  \phi_{\mathrm{B}} = \arctan\left[\frac{\sqrt{\left(I_{S} + \alpha\right)^2 + R^2\left(I_{L} + \beta\right)^2}}
                       {K \left(I_V +E \right)} \right]\,, \label{eq_VerticalCompass_1}
\end{eqnarray}
where $I_{S}$, $I_{L}$, $I_{V}$ are the currents in the small, large and vertical coils, respectively.
Parameters $R$ and $K$ are the scaling constants. Since we are interested in the ratio
of the three currents only two such factors are needed. Parameters 
$\alpha$, $\beta$ and $E$ correspond to the residual fields, which are present even when currents
in all coils are zero. They represent the cumulative contributions of the Earth's magnetic 
field and the BigBite spectrometer in each direction. 

In order to determine the precise value of $\phi_{\mathrm{B}}$ for an arbitrary
current setting used during the experiment (demonstrated in 
Fig.~\ref{fig_TargetFieldOrientation}), the parameters $R,K,\alpha,\beta$ and $E$ need to be 
known. They were obtained from the measurements of several $\phi_{B}$ at different 
current settings using the vertical compass. 
\begin{figure}[!ht]
\begin{center}
\begin{minipage}[t]{0.6\textwidth}
\hrule height 0pt
\includegraphics[width=1\textwidth]{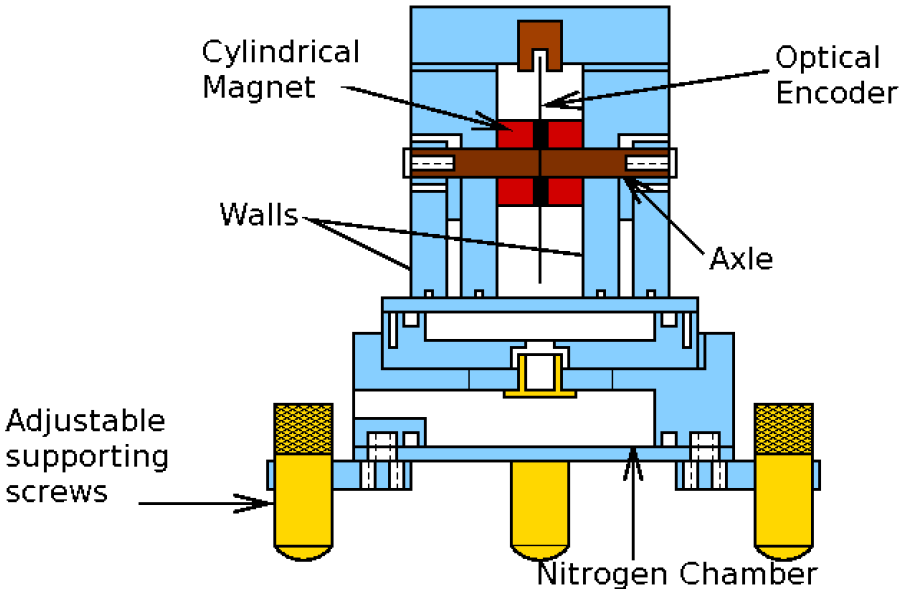}
\end{minipage}
\hfill
\begin{minipage}[t]{0.38\textwidth}
\hrule height 0pt
\caption{Schematics of the vertical compass device. The main part of the compass is a
cylindrical magnet with the digital encoder for reading out the vertical angle of the 
magnet. The datum is transfered to the computer via USB cable.   The compass is 
floating on a thin layer of nitrogen gas, which is constantly blowing from the 
adjustable platform. This enables the  compass to rotate frictionlessly in horizontal 
and vertical directions. The figure is taken from~\cite{DuttaPhD}.
\label{fig_Compass_VerticalCompass}}
\end{minipage}
\end{center}
\end{figure}

The drawing of the vertical compass is shown in Fig.~\ref{fig_Compass_VerticalCompass}.
The compass was developed at the Department of the Physics and Astronomy at the University 
of Kentucky~\cite{DuttaPhD}. The main part of the compass is a magnetic cylinder with a
digital optical encoder for reading out the vertical displacements of the magnet. The angular
accuracy of the encoder disk was $0.09\,\mathrm{deg}$. The compass was
placed on top of the adjustable platform which was installed inside the target enclosure. 
The platform was adjusted such that the center of the compass coincided with the center 
of the target. Nitrogen gas was blown into the system through the inlet and allowed the 
compass to move frictionlessly in both (horizontal and vertical) directions.  

The measurement was done in two steps. After the Helmholtz magnets were energized to the chosen currents,
the compass was left free to align itself with the magnetic field. Once aligned, the encoder reading
$N_1$ was noted.  Then the compass was rotated in the horizontal direction for $180^\circ$ and
locked at that angle. In that position the second encoder reading $N_2$ was recorded. 
The polar angle in the coil coordinate system was then determined by using:
\begin{eqnarray}
  \phi_{B}\left(I_S, I_L, I_V \right) = \frac{2000}{\pi}\cdot\frac{N_1 - N_2}{2}\,. \nonumber
\end{eqnarray}
The measurements were performed for ten different current settings. All measurements
were preformed with the BigBite magnet turned on and are
gathered in table~\ref{table_VerticalCompass}. The measured points were later fitted to
Eq.~(\ref{eq_VerticalCompass_1}) in order to determine the unknown free 
parameters~\cite{AndrejLeban}. The results are shown in table~\ref{table_VerticalCompass}. 

\begin{table}[!ht]
 \begin{center}
\caption{[Top] Vertical compass measurements performed during the commissioning 
phase of the E05-102 experiment. All measurement were performed with the BigBite magnet
turned on. The uncertainty of the measured angles was 
estimated to be $\approx 10^{-4}\,\mathrm{rad}$.
[Bottom] The resulting values of the free parameters from Eq.~(\ref{eq_VerticalCompass_1}).
The analysis was performed in collaboration with Andrej Leban~\cite{AndrejLeban}.
\label{table_VerticalCompass}}
\vspace*{3mm}
\begin{minipage}{0.49\linewidth}
\begin{tabular}{ccccc}
\multicolumn{5}{c}{{\bf Measurements for $I_L = 0.0\,\mathrm{A}$}}\\
\toprule
$I_{S}\,\mathrm{[A]}$  & $I_{V}\,\mathrm{[A]}$ & $N_1$ 
& $N_2$ & $\phi_{B}\,\mathrm{[rad]}$ \\
\midrule
7.0  & 0.0 & 1990 & 11 & 1.5543 \\ 
5.0  & 10.0 & 1462.5 & 537 & 0.7269 \\
0.0  & 14.0 & 997.5 & 1002 & 0.0035 \\
-5.0 & 10.0 & 534 & 1466 & 0.7320 \\
-7.0 & 0.0 & 7 & 1994 & 1.5606 \\
\bottomrule\\
 \end{tabular}
\end{minipage}
\hfill
\begin{minipage}{0.49\linewidth}
\begin{tabular}{ccccc}
\multicolumn{5}{c}{{\bf Measurements for $I_S = 0.0\,\mathrm{A}$}}\\
\toprule
$I_{L}\,\mathrm{[A]}$ & $I_{V}\,\mathrm{[A]}$ & $N_1$ 
& $N_2$ & $\phi_{B}\,\mathrm{[rad]}$ \\
\midrule
-7.0  & 0.0 & 3972 & 1959 & 1.5610 \\ 
-5.0  & 10.0 & 518 & 1411.5 & 0.7018 \\
0.0  & 14.0 & 983 & 947 & 0.0283 \\
5.0 & 10.0 & 1438 & 491 & 0.7438 \\
7.0 & 0.0 & 1955 & 3976 & 1.5543 \\
\bottomrule\\
 \end{tabular}
\end{minipage}
\begin{tabular}{l|ccccc}
\multicolumn{5}{c}{{\bf Parameters}}\\
\toprule
Parameter & K & R & $\alpha\,\mathrm{[A]}$ & $\beta\,\mathrm{[A]}$ & $E\,\mathrm{[A]}$ \\
\midrule
Value & 0.5504 & 0.9873 & -0.0195 & 0.1575 & 0.1681 \\
Uncertainty & 0.0081 & 0.0187 & 0.0667 & 0.0419 & 0.0605 \\ 
\bottomrule
 \end{tabular}
\end{center}
\end{table}

\subsection{Horizontal Compass Measurement}
The azimuthal angle of the magnetic field at a given combination of currents in the three pairs
of Helmholtz coils is obtained from the formula:
\begin{eqnarray}
  \theta_{B} = \arctan\left[\frac{R\left(I_{L} + \beta\right)}{\left(I_{S} + \alpha \right)} \right]\,,
    \label{eq_HorizontalCompass_1}
\end{eqnarray}
where $I_{S}$ and $I_{L}$ are the currents in the small and large vertical coils, while the
parameters $A$, $B$ and $R$ have the same meaning as in the vertical case. This time the 
horizontal compass was utilized. It is constructed of a magnetized 40-cm-long 
iron needle (dipole magnet) mounted on a support frame. The compass is again positioned at the
center of the target enclosure at the height of the beam line. In the presence of the horizontal 
field (parallel to the needle) the compass turns in the direction of the magnetic field. 
The angle of the needle with respect to the beam direction was obtained by measuring the 
absolute position of both ends of the needle. The position measurements were performed 
by the survey group using high-precision 3D laser positioning system. After the compass
rotated to a particular direction, a metal survey ball equipped with a mirror was positioned
at each end of the needle and illuminated by a laser  in order to determine their coordinates. 
Using this technique, the position of the survey ball can be determined with a very high 
precision. However, the resolution of the compass measurement is limited by the accuracy 
of placing a survey ball on the tip of the compass needle and was estimated to 
be $\approx 3\times10^{-4}\,\mathrm{rad}$, assuming that coordinates of the needle ends 
are known to $\pm 1\,\mathrm{mm}$. Furthermore, the error on the position of the needle tips 
can also lead to  non-zero offsets at the center of the compass. Fortunately this does 
not cause any issues, since the magnetic field is believed to be uniform at the target position. 

Knowing the needle
coordinates, the azimuthal angle of the magnetic field could be determined from:
\begin{eqnarray}
  \theta_{B} = \alpha_{\mathrm{Coil}} + \arctan\left(\frac{x_{2} - x_{1}}
                      {z_{2} - z_{1}} \right)\,, \label{eq_HorizontalCompass_2}
\end{eqnarray}
where the subscripts $1$ and $2$ denote the left and right end of the 
needle, respectively. In Eq.~(\ref{eq_HorizontalCompass_2}) we also had to consider
that the survey measurements were performed in the Hall coordinate system, while the angle in  
Eq.~(\ref{eq_HorizontalCompass_1}) is in the Coil coordinate system 
(see Fig.~\ref{fig_TargetFieldOrientation}). Hence, a rotation for a constant angle
$\alpha_{\mathrm{Coil}} = 143^\circ$ is required to transform the azimuthal angles 
from the Hall coordinate system to the Coil coordinate system. 

A series of horizontal compass measurements was performed at different values of $I_S$ 
and $I_L$. Furthermore, the calibration was performed with BigBite magnet on and off in order to 
estimate the influence of the BigBite fringe fields on the target field. The
measured points are shown in Table~\ref{table_HorizontalCompass1}. 
A least-square procedure~\cite{AndrejLeban} was used to determine parameters $\alpha$, $\beta$
and R by fitting  Eq.~(\ref{eq_HorizontalCompass_1}) to the measured data.
The determined parameters are shown in 
Table~\ref{table_HorizontalCompass2}. The comparison of the fitted parameters shows no 
significant difference between the cases where BigBite was on or off.
This suggests that BigBite fields clamp successfully protects the target from the unwanted 
fringe fields.

\begin{table}[!hb]
 \begin{center}
\caption{Horizontal compass measurements performed during the commissioning 
phase of the E05-102 experiment. The uncertainty of the measured angles was 
estimated to be $\leq 3\times10^{-4}\,\mathrm{rad}$.
\label{table_HorizontalCompass1}}
\vspace*{3mm}
\begin{tabular}{ccccccc}
\multicolumn{7}{c}{{\bf BigBite On}}\\
\toprule
$I_S\,\mathrm{[A]}$ & $I_L\,\mathrm{[A]}$ & $z_1\,\mathrm{[cm]}$ & 
$x_1\,\mathrm{[cm]}$ & $z_2\,\mathrm{[cm]}$ & $x_2\,\mathrm{[cm]}$ 
& $\theta_B\,\mathrm{[rad]}$ \\
\midrule
4.2 & -5.6 & -5.08 & 231.36 & 8.53 &-225.24 & -0.8952\\
-1.0 & -6.9 & 163.58 & 164.09 & -160.76 &-157.66 &-1.7144\\
-5.6 & -4.2 & 230.12 & -1.63 & -226.93 &10.15 &-2.5216\\
-6.9 & 1.0 & 124.34 & -194.29 & -136.74 &180.47 &2.8251\\
-4.2 & 5.6 & -8.92 & -235.87 & -3.86 &221.07 &2.2055\\
1.0 & 6.9 & -169.16 & -168.16 & 155.83 &153.44 &1.4259\\
5.6 & 4.2 & -234.87 & -11.25 & 222.11 &-4.57 &0.6604\\
6.9 & -1.0 & -149.41 & 170.82 & 135.21 &-186.50 &-0.2524\\
\bottomrule\\
 \end{tabular}
\begin{tabular}{ccccccc}
\multicolumn{7}{c}{{\bf BigBite Off}}\\
\toprule
$I_S\,\mathrm{[A]}$ & $I_L\,\mathrm{[A]}$ & $z_1\,\mathrm{[cm]}$ & 
$x_1\,\mathrm{[cm]}$ & $z_2\,\mathrm{[cm]}$ & $x_2\,\mathrm{[cm]}$ 
& $\theta_B\,\mathrm{[rad]}$ \\
\midrule
4.2 & -5.6 & -8.52 & 229.46 & 1.88 &-227.36 & -0.9023\\
-5.6 & -4.2 & 225.13 & 0.12 & -231.79 & 4.20 & -2.5048\\
-4.2 & 5.6 & -4.84 & -238.88 & -3.00 & 217.99 & 2.2125\\
5.6 & 4.2 & -235.88 & -3.15 & 220.98 &-3.36 & 0.6453\\
\bottomrule
 \end{tabular}
\end{center}
\end{table}

\begin{table}[!hb]
 \begin{center}
\caption{The free parameters of the horizontal compass Eq.~(\ref{eq_HorizontalCompass_1}), 
determined for the E05-102 experiment. The results show no significant difference
between BigBite-On and BigBite-Off modes. 
\label{table_HorizontalCompass2}}
\vspace*{3mm}
\begin{tabular}{l|ccc}
\multicolumn{4}{c}{{\bf Parameters}}\\
\toprule
 Parameter & R & $\alpha\,\mathrm{[A]}$ & $\beta\,\mathrm{[A]}$\\
\midrule
BigBite On & $0.9975 \pm 0.0023$ & $0.032 \pm 0.024$ &  $0.193 \pm 0.170$ \\
BigBite Off & $0.9793\pm 0.0442$ & $0.054 \pm 0.266$ &  $0.082 \pm 0.324$ \\
\bottomrule
 \end{tabular}
\end{center}
\end{table}

\subsection{Final Compass Results}

The parameters $\alpha$, $\beta$ and $R$ in Eqs.~(\ref{eq_VerticalCompass_1}) and 
(\ref{eq_HorizontalCompass_1}) were determined independently for the 
horizontal and vertical compass measurement. However, both methods are
investigating the same system, thus the parameters $\alpha$, 
$\beta$ and $R$ should be nearly identical, and we have combined 
results from  Tables~\ref{table_VerticalCompass} 
and~\ref{table_HorizontalCompass2} to calculate the mean values of the 
parameters. They are shown in Table~\ref{table_CompassParameters}.
The obtained mean values can now be applied to Eqs.~(\ref{eq_VerticalCompass_1})
and (\ref{eq_HorizontalCompass_1}) in order to determine the true orientation of 
the field for the current settings used in the actual experiment.
Together with the angles, the corresponding errors also need to be determined. 
The major contributions to the error come from the uncertainties of the 
fitted parameters. Since these parameters and their errors are not 
independent~\cite{TaylorErrorAnalysis}, the upper limit for the error can be 
estimated as:

\begin{eqnarray}
\Delta \theta_{B} &\leq& \left|\frac{d\theta_{B}}{dR}\right|\Delta R + 
                    \left|\frac{d\theta_{B}}{d\alpha}\right|\Delta \alpha +
                    \left|\frac{d\theta_{B}}{d\beta}\right|\Delta \beta\,, \nonumber \\
\nonumber \\
\Delta \phi_{B} &\leq& \left|\frac{d\phi_{B}}{dK}\right|\Delta K + 
                     \left|\frac{d\phi_{B}}{dR}\right|\Delta R + 
                     \left|\frac{d\phi_{B}}{d\alpha}\right|\Delta \alpha +
                     \left|\frac{d\phi_{B}}{d\beta}\right|\Delta \beta + 
                     \left|\frac{d\phi_{B}}{dE}\right|\Delta E\,.\nonumber 
\end{eqnarray}
The final results are gathered in Table~\ref{table_CompassResults}. 

\begin{table}[!ht]
 \begin{center}
\caption{The final values of the parameters used in Eqs.~(\ref{eq_VerticalCompass_1}) 
and (\ref{eq_HorizontalCompass_1}) to calculate the polar and azimuthal angles
of the target holding magnetic field. 
The parameters $\alpha$, $\beta$ and $R$  are the mean values of the vertical and 
horizontal compass results. Parameters $K$ and $E$ were obtained with
vertical compass only.
\label{table_CompassParameters}}
\vspace*{3mm}
\begin{tabular}{l|ccccc}
\multicolumn{5}{c}{{\bf Final Parameters}}\\
\toprule
Parameter & K & R & $\alpha\,\mathrm{[A]}$ & $\beta\,\mathrm{[A]}$ & $E\,\mathrm{[A]}$ \\
\midrule
Value & 0.5504 & 0.99745 & 0.0248 & 0.1565 & 0.1681 \\
Uncertainty & 0.0081 & 0.00130 & 0.0123 & 0.0377 & 0.0605 \\ 
\bottomrule
 \end{tabular}
\end{center}
\end{table}

\begin{table}[!ht]
 \begin{center}
\caption{ True orientation of the magnetic field, calculated for each target orientation
setup considered in the E05-102 experiment. The $\theta_B^{\mathrm{Hall}}$ is the azimuthal 
field angle, while the $\phi_B^{\mathrm{Hall}}$ is the polar angle (zero when pointing
in the vertical direction), given in the Hall coordinate system. The corresponding angles 
$\theta_{\mathrm{Spin}}^{\mathrm{Hall}}$ and $\phi_{\mathrm{Spin}}^{\mathrm{Hall}}$ 
for the target spin orientations are also calculated. The displayed errors represent the 
upper limits for the angle uncertainties. 
\label{table_CompassResults}}
\vspace*{3mm}
\begin{tabular}{cccccc|ccc}
\multicolumn{9}{c}{{\bf Field/Spin Direction}}\\
\toprule
Field & $I_S$ & $I_V$ & $I_L$ &
$\theta_{B}^{\mathrm{Hall}}$ & $\phi_{B}^{\mathrm{Hall}}$ & 
Spin & $\theta_{\mathrm{Spin}}^{\mathrm{Hall}}$ & 
$\phi_{\mathrm{Spin}}^{\mathrm{Hall}}$ \\ [-0.2mm]
Orient. &$\mathrm{[A]}$& $\mathrm{[A]}$ & $\mathrm{[A]}$ & $\mathrm{[{}^\circ]}$ &
$\mathrm{[{}^\circ]}$ & Orient. & $\mathrm{[{}^\circ]}$& $\mathrm{[{}^\circ]}$\\
\toprule
Vert. (+) & -0.130 & -13.529 & -0.065 & $282.1$ & $1.1$ & Vert. (-) & $102.1$ & 178.9\\[-1.5mm] 
             &  &  &  & $\pm15.04$ & $\pm 0.29$ &  & $\pm 15.04$& $\pm 0.29$\\[1mm] 
Vert. (-) & 0.028 & 13.020 & -0.218 & $93.7$ & $179.4$ & Vert. (+) & $-86.3$ & $0.6$\\[-1.5mm] 
             &  &  &  & $\pm24.00$ & $\pm 0.30$ &  & $\pm 24.00$& $\pm 0.30$\\ 
Long. (+) & -5.81 & 0.168 & -4.622 & $0.6$ & $91.5$ & Long. (-) & $-179.4$ & $88.5$\\[-1.5mm] 
             &  &  &  & $\pm0.33$ & $\pm 0.29$ &  & $\pm 0.33$& $\pm 0.29$\\ 
Long. (-) & 5.93 & 0.168 & 4.061 & $179.8$ & $91.5$ & Long. (+) & $-0.2$ & $88.5$\\[-1.5mm] 
             &  &  &  & $\pm0.34$ & $\pm 0.30$ &  & $\pm 0.34$& $\pm 0.30$\\ 
Trans. (-) & -4.238 & 0.168 & 5.527 & $269.6$ & $91.5$ & Trans. (+) & $89.6$ & $88.5$\\[-1.5mm] 
             &  &  &  & $\pm0.30$ & $\pm 0.30$ &  & $\pm 0.30$& $\pm 0.30$\\
Trans. (+) & 4.286 & 0.168 & -6.001 & $89.5$ & $91.5$ & Trans. (-) & $-90.5$ & $88.5$\\[-1.5mm] 
             &  &  &  & $\pm0.29$ & $\pm 0.29$ &  & $\pm 0.29$& $\pm 0.29$\\
\bottomrule
 \end{tabular}
\end{center}
\end{table}

\section{Calibration of the Beam Current Monitors}
The raw Beam Current Monitor (BCM) readout is given in terms of counts in the scaler 
modules. The total number of counts (BCM-Counts) corresponds to 
the collected charge, while the scaler count rate 
(($\mathrm{BCM-Rate}) = d(\mathrm{BCM-Counts})/dt$) corresponds to 
the beam current. The following relations are used to determine the beam current 
($I_{\mathrm{Hall-A}}$) and the collected charge ($Q_{\mathrm{Hall-A}}$) from 
each of the six BCM signals 
($x=u_{1}$, $u_{3}$,$u_{10}$, $d_{1}$, $d_{3}$, $d_{10}$):
\begin{eqnarray}
I_{\mathrm{Hall-A}} &=& \frac{dQ_{\mathrm{Hall-A}}}{dt} = 
 \frac{\mathrm{(BCM-Rate)}_{x} - O_{x}}{C_{x}}\label{BCMCurrent}  \\
Q_{\mathrm{Hall-A}} &=& \frac{\mathrm{(BCM-Counts)}_{x} - \mathrm{(Time)} 
 \times O_{x}}{C_{x}}\>, \label{BCMCharge}
\end{eqnarray}
where $C_{x}$ and $O_{x}$ are calibration constants and $\mathrm{(Time)}$ 
represents the total time of the data taking. The parameter $C_{x}$ is a 
multiplicative factor in units of ${As}^{-1}$ and transforms the 
raw scaler reading to the meaningful physical quantities. 
The offset $O_{x}$ accounts for the presence of  dark current 
in the electronics (non-zero current readings when 
the beam is turned off). 

\begin{figure}[!ht]
\begin{center}
\includegraphics[width=0.49\textwidth]{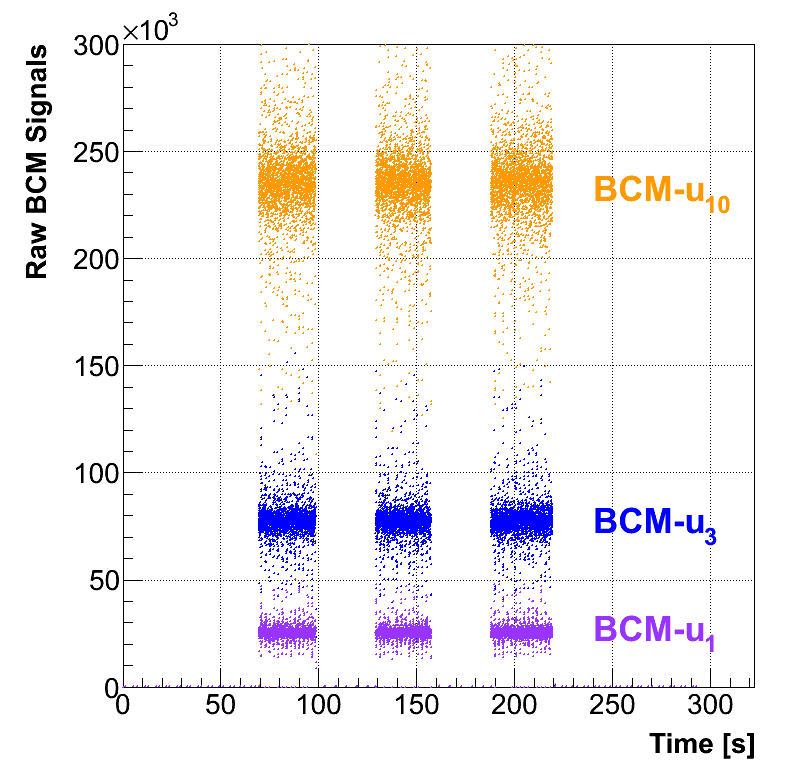}
\includegraphics[width=0.49\textwidth]{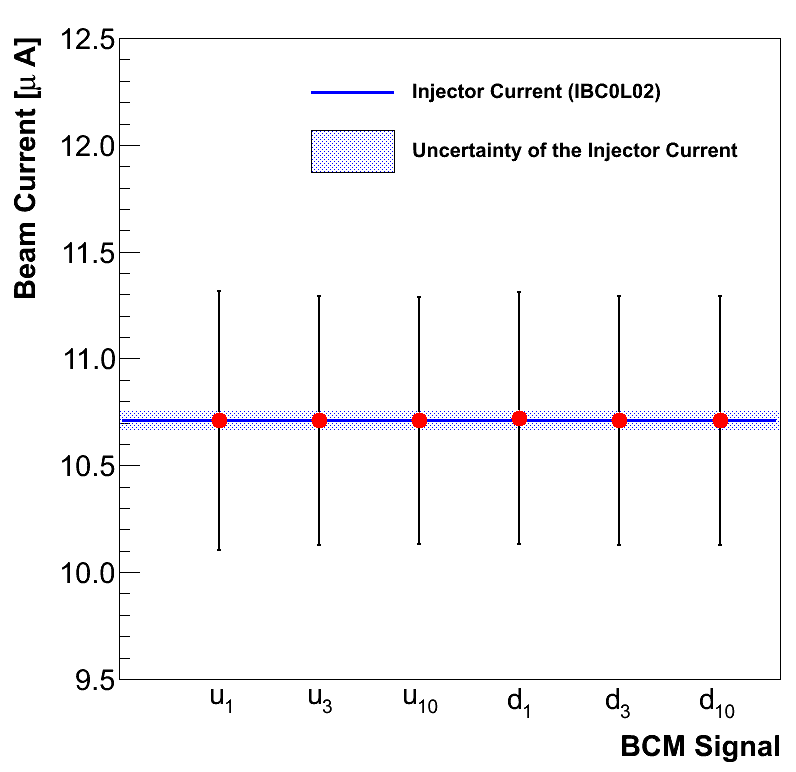}
\caption{[Left] Raw scaler reading of the beam current for three BCM signals 
$u_{1}$, $u_{3}$ and $u_{10}$ during calibration run \#2268.  The ratio of the 
scaler-rate amplitudes for two BCM signals agrees with the ratio of their 
amplification factors. Sections of zero/non-zero current are also clearly 
visible. [Right] The comparison of the reconstructed Hall A beam current with 
the current at the injector. The injector current is determined to an accuracy 
better than $\sigma_{I_{\mathrm{Injector}}}<0.05\,\mathrm{\mu A}$. With the use
of the Hall A BCM monitor, beam current can be determined with the resolution better
than  $\sigma_{I_{Hall-A}}<0.6\,\mathrm{\mu A}$, regardless of the considered
BCM signal. \label{fig_BCMCalibration}}
\end{center}
\end{figure}

The calibration constants  $C_{x}$ and $O_{x}$ for all six BCM signals were
determined using a dedicated data set (E05-102 run number \#2268). This 
particular run consists of three sequences of zero and non-zero current, 
which allows a simultaneous determination of $C_{x}$ and $O_{x}$.
Fig.~\ref{fig_BCMCalibration} shows the raw scaler reading for three
BCM signals $u_{1}$, $u_{3}$ and $u_{10}$.
During this special run, the electron beam was delivered only to  Hall A. This 
allows us to compare our BCM readings directly to the beam current at the injector. 
The value of the beam current there is precisely known ($\sigma_{I_{\mathrm{Injector}}}<
0.05\,\mathrm{\mu A}$) and is inserted into our data stream as an EPICS variable 
(IBC0L02). Combining this information with our raw BCM data in formula 
(\ref{BCMCurrent}), the calibration constants were determined. The results are gathered
in Table~\ref{table_BCM_calibration_constants}. Once knowing the calibration parameters,
we were able to independently determine the beam current that entered Hall-A, by employing each 
of the BCM signals (see Fig.~\ref{fig_BCMCalibration}). The resolution (sigma) of the 
reconstructed beam current was limited with the spread of the scaler data 
(BCM-Rate) and was estimated to $\sigma_{I_{\mathrm{Hall-A}}}<0.6\,\mathrm{\mu A}$.

\begin{table}[!ht]
 \begin{center}
\begin{minipage}[t]{0.6\textwidth}
\hrule height 0pt
\begin{tabular}{ccc}
\toprule
BCM Signal & Calibration Constant & Offset \\
\multicolumn{1}{c}{$x$}  & \multicolumn{1}{c}{$C_{x}$}&  \multicolumn{1}{c}{$O_{x}$}\\
\midrule
$\mathrm{u_1}$ & 2333.83 & 354.7 \\
$\mathrm{u_3}$ & 7183.15 & 328.5 \\
$\mathrm{u_{10}}$ & 21890.78 & 449.7 \\
\midrule
$\mathrm{d_1}$ & 2395.87 & 140.2 \\
$\mathrm{d_3}$ & 7423.52 & 124.1 \\
$\mathrm{d_{10}}$ & 23408.59 & 294.0 \\
\bottomrule
 \end{tabular}
\end{minipage}
\begin{minipage}[t]{0.39\textwidth}
\hrule height 0pt
\caption{Calibration constants $C_x$ and $O_x$ for all six
BCM signals. The coefficients are used in formulas 
(\ref{BCMCurrent}) and (\ref{BCMCharge}) to 
calculate the beam current and charge. The ratio
between the calibration constants $C_{x}$ for both BCM cavities
$u$ and $d$ approximately correspond to the ratio of the 
amplification factors for each signal. }
\label{table_BCM_calibration_constants}
\end{minipage}
\end{center}
\end{table}

\section{Determination of Beam position}

The position and direction of the incident particle beam at the target
is very important. During the experiment we had to ascertain 
that the beam has a proper size and illuminates the correct part of the 
target, otherwise it could damage the apparatus. Therefore, the position 
and size of the beam were periodically
monitored and properly adjusted when significant deviations occurred.
Furthermore, in the off-line analysis, a precise knowledge on the position of the 
beam at the target for each considered event was especially important 
for the optical calibration of the spectrometers. 

\begin{figure}[!ht]
\begin{center}
\includegraphics[width=0.49\textwidth]{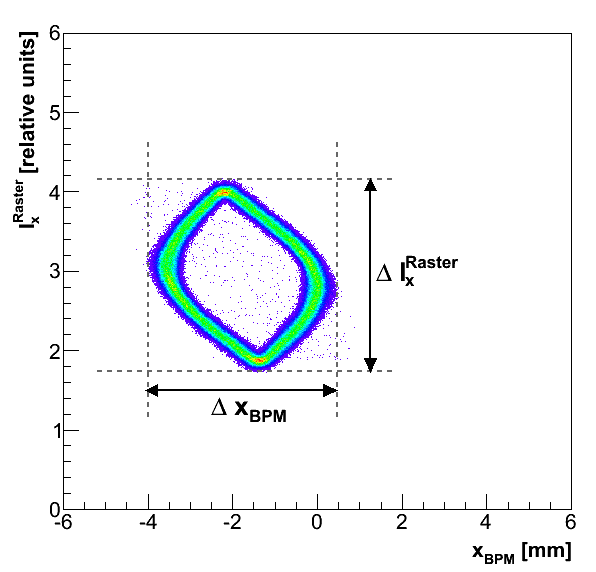}
\includegraphics[width=0.49\textwidth]{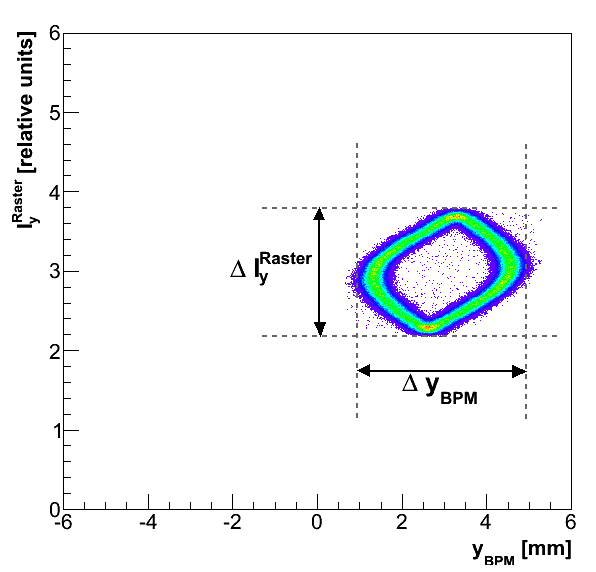}
\caption{ The horizontal ($x_{\mathrm{BPM}}$) and vertical ($y_{\mathrm{BPM}}$) position 
of the beam at the target, reconstructed by the BPMs as a function of the horizontal
($I_{\mathrm{x}}^{\mathrm{Raster}}$) and vertical ($I_{\mathrm{y}}^{\mathrm{Raster}}$) current  in the raster 
magnets. A  phase difference of almost $\pi/2$ between the two is clearly visible. 
The widths $\Delta x_{\mathrm{BPM}}$, $\Delta y_{\mathrm{BPM}}$ and 
$\Delta I_{\mathrm{x}}^{\mathrm{Raster}}$, $\Delta I_{\mathrm{y}}^{\mathrm{Raster}}$ are 
considered in the calculation of calibration constants of Eqs.~(\ref{eq_Raster_formulas}).
\label{fig_Raster_Calibration}}
\end{center}
\end{figure}

In general, the position of the beam is determined with the 
two beam position monitors (BPMs) described in  Sec.~\ref{sec_BPMs}. 
The readout from the two monitors is calibrated 
using a ``Bull's eye'' scan.  A typical beam spot on the target 
reconstructed by the BPMs is shown in 
Fig.~\ref{fig_raster}. Since rastering was used, the beam is not all
localized at one point but has an orthogonal shape. Unfortunately, when using rastered
beam, the BPMs are too slow to measure the position of the beam exactly at the instant, 
when the particle hits the target. There is a difference of few $\mathrm{\mu s}$
between the time when the particle hits the target and the time when the BPM returns 
the corresponding beam position. Consequently, the BPM information can not be used
for event by event analysis, such as optical calibration. 
Instead, current information from the beam raster was utilized to determine 
the beam position~\cite{Podd}. This is possible because raster magnets 
have a much shorter response time. The difference between the BPM measurements 
and the raster approximation is 
demonstrated in Fig.~\ref{fig_Raster_Calibration}, which compares the raster 
current in each direction with the position of the beam at the target determined 
by the BPMs. There is almost a $\pi/2$ phase difference between the two. 

When using the raster information, the position of the beam at the target at any given
moment is calculated by
\begin{eqnarray}
  x_{\mathrm{Tg}}^{\mathrm{Raster}} = O_{\mathrm{x}} + 
    A_{\mathrm{xx}}I_{\mathrm{x}}^{\mathrm{Raster}} + A_\mathrm{xy}I_{\mathrm{y}}^{\mathrm{Raster}}\,, \qquad
  y_{\mathrm{Tg}}^{\mathrm{Raster}} = O_{\mathrm{y}} + 
    A_{\mathrm{yx}}I_{\mathrm{x}}^{\mathrm{Raster}} + A_\mathrm{yy}I_{\mathrm{y}}^{\mathrm{Raster}}\,,
    \label{eq_Raster_formulas}
\end{eqnarray}
where $I_{\mathrm{x}}^{\mathrm{Raster}}$ and $I_{\mathrm{y}}^{\mathrm{Raster}}$ are the currents in the
two raster magnets.
The parameters $A_{\mathrm{xx}}$, $A_{\mathrm{xy}}$, $A_{\mathrm{yx}}$ and $A_{\mathrm{yy}}$ are 
the calibration constants required to transform currents to positions. 
In our case are $A_{\mathrm{xy}}$, $A_{\mathrm{yx}}$ identically zero, 
since raster coils are aligned with the target coordinate system. 
$A_{\mathrm{xx}}$ and $A_{\mathrm{yy}}$  are the ratios between the widths of the BPM distributions and
the current distributions (see Fig.~\ref{fig_Raster_Calibration}):
\begin{eqnarray}
  A_{\mathrm{xx}} = \frac{\Delta x_{\mathrm{BPM}}}{\Delta I_{\mathrm{x}}^{\mathrm{Raster}}}\,,\qquad
  A_{\mathrm{yy}} = \frac{\Delta y_{\mathrm{BPM}}}{\Delta I_{\mathrm{y}}^{\mathrm{Raster}}}\,. \nonumber
\end{eqnarray}
The parameters $O_{\mathrm{x}}$ and $O_{\mathrm{y}}$ represent offset corrections and are determined
by comparing the absolute position of the beam spot measured with the BPMs  to the position
reconstructed from the raster currents:
\begin{eqnarray}
  O_{\mathrm{x}} = \overline{x}_{\mathrm{BPM}} - 
        \overline{I}_{\mathrm{x}}^{\mathrm{Raster}} A_{\mathrm{xx}}\,,\qquad
  O_{\mathrm{y}} = \overline{y}_{\mathrm{BPM}} - 
        \overline{I}_{\mathrm{y}}^{\mathrm{Raster}} A_{\mathrm{yy}}\,. \nonumber
\end{eqnarray}
Here $\overline{x}_{\mathrm{BPM}}$ and $\overline{y}_{\mathrm{BPM}}$ are the mean 
positions of the beam spot at the target (determined with BPMs), while
$\overline{I}_{\mathrm{x}}$ and $\overline{I}_{\mathrm{y}}$ are the mean values
of the raster currents. 

\begin{figure}[!ht]
\begin{center}
\begin{minipage}[t]{0.6\textwidth}
\hrule height 0pt
\includegraphics[width=1\textwidth]{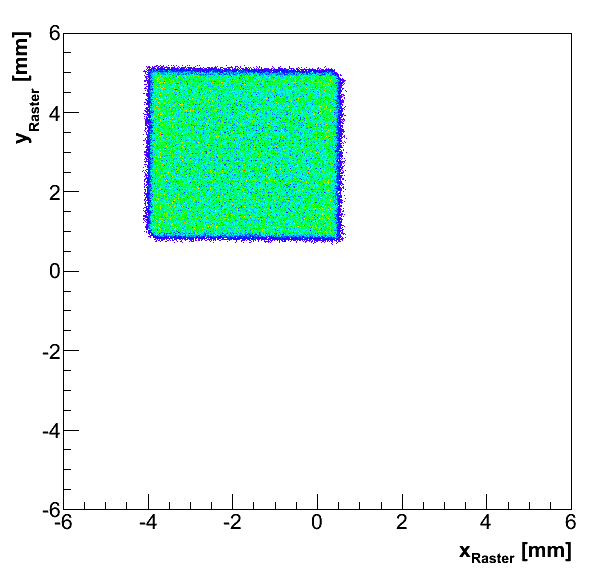}
\end{minipage}
\hfill
\begin{minipage}[t]{0.38\textwidth}
\hrule height 0pt
\caption{ The beam spot at the target position during run \#2280, 
reconstructed from the raster currents, by using Eqs.~(\ref{eq_Raster_formulas})
and calibration parameters in Table~\ref{table_Raster_parameters}. 
\label{fig_Raster_Result}}
\end{minipage}
\end{center}
\end{figure}

All free parameters required in Eqs.~(\ref{eq_Raster_formulas}) are stored in 
the \texttt{db\_rb.Raster.dat} database file and need to be determined before any 
further analysis is performed. Unfortunately the relation between the currents in 
the raster magnets and actual beam deflection and beam position 
depends strongly on the beam energy and on beam tunning.
During the  E05-102 experiment the beam energy  was very stable, while the beam
was frequently tuned. In principle, each tuning could change the calibration 
constants. Consequently, the calibration constants had to be determined for every collected
dataset (run). The calibration process was executed automatically at the beginning of 
analysis of every run and the obtained calibration constants were automatically
copied to the database file. Typical values of the calibration constants are 
gathered in Table~\ref{table_Raster_parameters}. An example of the reconstructed 
beam spot at the target by using raster currents is shown in Fig.~\ref{fig_Raster_Result}.

\section{Calibration of the BigBite TDCs}
The precise time information acquired from the BigBite scintillation detectors 
is required for the determination of particle's time-of-flight and 
formation of coincidence triggers. It can also help us in determining the horizontal (in-plane)
position of the particle hit in the detector package.

\begin{table}[!ht]
 \begin{center}
\begin{minipage}[t]{0.6\textwidth}
\hrule height 0pt
\begin{tabular}{lcccc}
\toprule
Run & $O_{\mathrm{x}}$ & $O_{\mathrm{y}}$ & $A_{\mathrm{xx}} $ & $A_{\mathrm{yy}}$ \\
\midrule
2280 & $-0.00727$ & $0.01081$ & $1.882\cdot10^{-06}$ & $-2.631\cdot10^{-06}$ \\
2480 & $-0.00738$ & $0.01099$ & $1.942\cdot10^{-06}$ & $-2.684\cdot10^{-06}$ \\
2935 & $-0.00659$ & $0.01074$ & $1.911\cdot10^{-06}$ & $-2.619\cdot10^{-06}$ \\
3078 & $-0.00725$ & $0.01081$ & $1.894\cdot10^{-06}$ & $-2.619\cdot10^{-06}$ \\
3370 & $-0.00738$ & $0.01081$ & $1.931\cdot10^{-06}$ & $-2.628\cdot10^{-06}$ \\
3488 & $-0.00716$ & $0.01072$ & $1.859\cdot10^{-06}$ & $-2.592\cdot10^{-06}$ \\
3570 & $-0.00636$ & $0.00909$ & $1.510\cdot10^{-06}$ & $-2.034\cdot10^{-06}$ \\
3606 & $-0.00623$ & $0.00903$ & $1.526\cdot10^{-06}$ & $-2.036\cdot10^{-06}$ \\
\bottomrule
 \end{tabular}
\end{minipage}
\hfill
\begin{minipage}[t]{0.29\textwidth}
\hrule height 0pt
\caption{The calibration constants required in Eqs.~(\ref{eq_Raster_formulas}) to 
reconstruct the position and size of the beam spot at the target, by using 
raster currents.}
\label{table_Raster_parameters}
\end{minipage}
\end{center}
\end{table}

The time information is obtained from the BigBite TDC modules and is always read out  
relative to the formed triggers. The triggering circuit for the E05-102 experiment 
was designed such that BigBite was self-timed. Hence, TDC signals from all scintillation 
paddles should come at the same time with respect to the trigger. Furthermore, particles
traveling through the center of the detector package should generate simultaneous pulses 
in both left and right TDCs.  This is not always true due to slightly inconsistent 
lengths of cables considered in the electronics and different gains in the PMTs (leading 
edge discriminators are used). The TDC modules  therefore have to be properly calibrated
before time information could be used for further analysis. 

The calibration of the TDC modules was performed after the experiment by 
setting the appropriate correction factors and offsets in the off-line analysis
scripts. The time information $t_i$ from each of the 96 channels is calculated from:
\begin{eqnarray}
  t_{i}^{L,R} = R_{\mathrm{TDC}}\cdot T_{i}^{L,R} - O_{i}^{L,R}\>, \qquad i = 1\,,\dots\,,48\,, \nonumber
\end{eqnarray}
where $R_{\mathrm{TDC}} = 0.0602\,\mathrm{ns/channel}$ is the resolution of 
the TDC modules, $T_i^{L,R}$ are the raw TDC readouts and $O_{i}^{L,R}$ are the offset 
corrections, obtained through the BigBite TDC calibration.
The calibration of the TDC modules was done separately for each scintillation plane,
since different physical types of events were considered for each case.  

\begin{figure}[!ht]
\begin{center}
\includegraphics[width=1\linewidth]{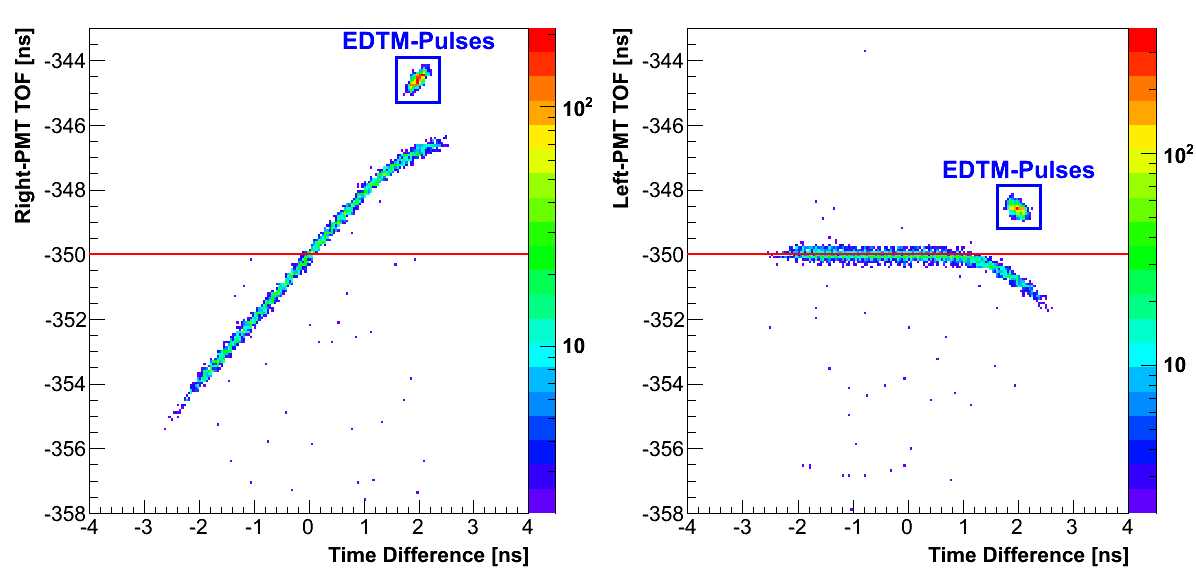}
\caption{The corrected time information $t_0^{L,R}$ for the left and 
right PMT, as a function of the time difference $(t_0^{L} - t_0^{R})$, 
obtained for the first paddle of the E-plane scintillation detector.
The signal from the left PMT almost always comes after the right PMT signal. 
Hence, the left PMT defines timing, and consequently its TDC value is
constant. The offsets for both channels were set such, that 
the signals intersect in the middle $(t_0^{L} - t_0^{R} = 0)$ and that the constant 
part of the $t_{0}^L$ is positioned at $-350\,\mathrm{ns}$. The plots 
also show EDTM pulses which 
come simultaneously to the TDC modules. Therefore, all hits accumulate
at a single point. 
\label{fig_TDC_LR_Calibration}}
\end{center}
\end{figure}

For the calibration of the E-plane, events with trigger T1 were used, 
which require valid signals in both PMTs 
in a particular scintillation paddle. As shown in Fig.~\ref{fig_BBT1trigger}, the signals 
from both PMTs are introduced to a logical AND. Consequently, the signal arriving 
last defines timing. Therefore one of the time signals is flat, while the other one
is changing and vice-versa. Which signal is arriving second depends on the lengths of 
the cables and the particle hit position. This means that it is not necessary
that one of the signals is always flat while the other one is changing. Their roles can interchange,
which slightly complicates the calibration.

The offsets $O_{i}^{L,R}$ for each scintillation 
paddle $i$ were determined in two steps. In the first step, the PMT with
the largest constant section was determined. The offset for this channel was then 
adjusted to position the time peak at $-350\,\mathrm{ns}$. The offset for
the other channel was determined in the second step of the calibration by insisting
that the time difference $(t_i^{R} - t_i^{L})$ between the left and right PMT signals 
must be zero for the events hitting the center of the paddle.  The results are shown in 
Fig.~\ref{fig_TDC_LR_Calibration}.

The TDC calibration for the dE-plane was done similarly, but now the events with
only trigger T2 were considered. If the primary trigger T1 was, the signals 
from the E-plane would define timing. Consequently, neither the left nor the right PMT signal from 
the  dE-plane scintillation bar would be sharp and TDC calibration would be impossible. 
Additionally, in contrast to the T1 trigger, only one hit per paddle in the dE-plane is 
required to form a T2 trigger. To successfully calibrate the dE-plane, one needs to impose 
constraints which demand hits in left and right PMTs.

\begin{figure}[!ht]
\begin{center}
\includegraphics[width=0.49\linewidth]{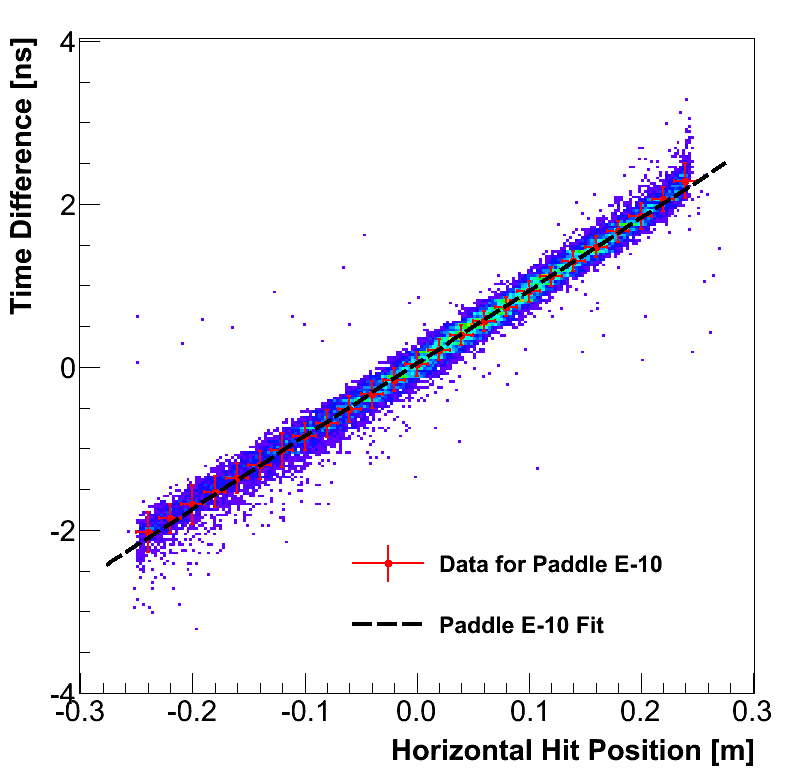}
\includegraphics[width=0.49\linewidth]{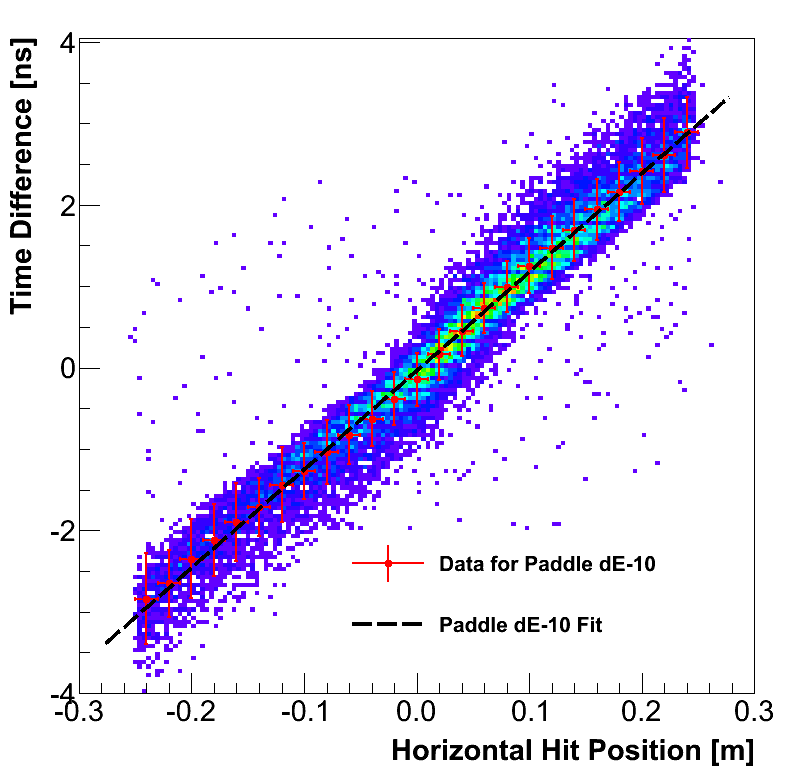}
\caption{Time difference between hits in the left and right PMTs as a function 
of the horizontal position of the particle hit. The position is obtained by 
extrapolating the particle track from the MWDCs to the scintillation paddles. 
Results are shown for $10^{\mathrm{th}}$ paddle in both E (left) and dE (right) plane. The 
slope of the obtained curve represents the inverse value of the effective 
speed-of-light in the scintillation material.
\label{fig_TDC_Velocity}}
\end{center}
\end{figure}

\begin{figure}[!ht]
\begin{center}
\includegraphics[width=0.49\linewidth]{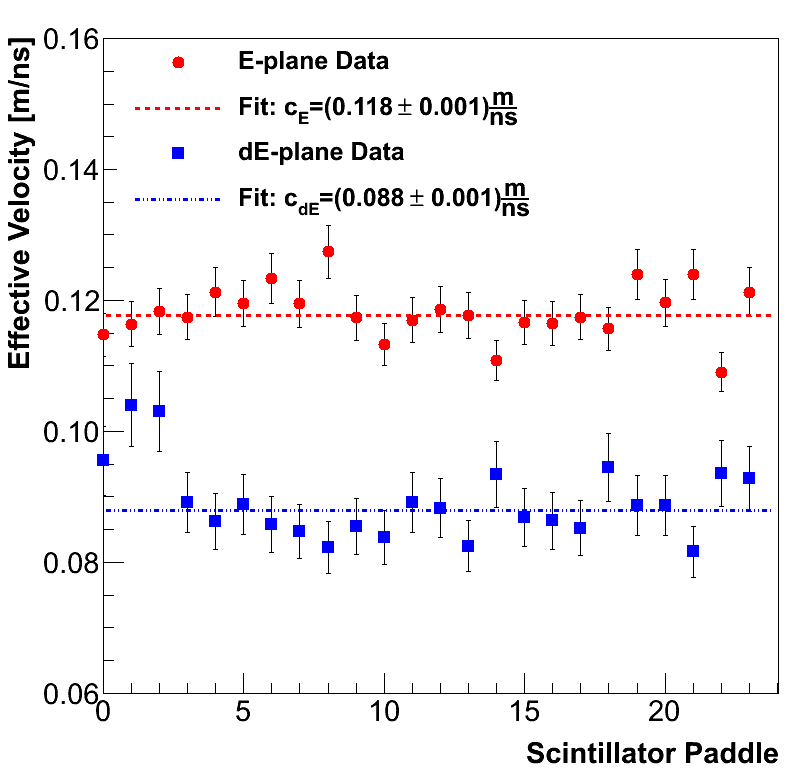}
\includegraphics[width=0.49\linewidth]{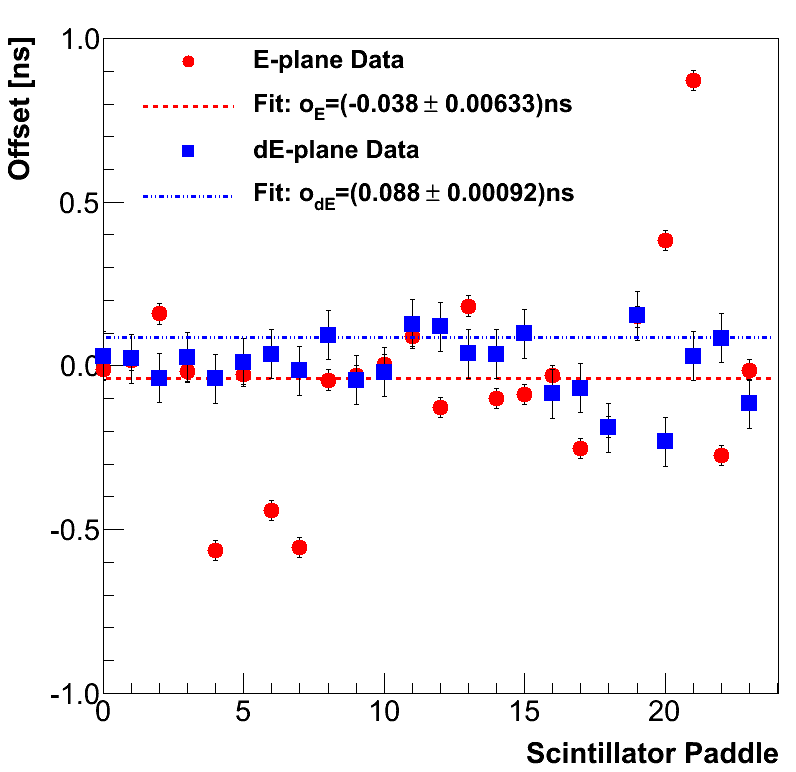}
\caption{[Left] The effective speed-of-light for all scintillation paddles in the 
dE- and E-plane. 
[Right] The time difference between the hits in the left and right PMTs for 
events  hitting the center of the paddle. The size of the offset determines
the quality of the performed TDC calibration. Ideally these offsets should be zero. 
\label{fig_TDC_FinalResults}}
\end{center}
\end{figure}

The information from the scintillation detectors can also be used to determine
the position of the particle hit in the detector. The vertical (dispersive) position, 
$x_{\mathrm{TDC}}$, is obtained from the position of the hit paddle, with the resolution 
limited to half  of the with of the paddle. On the other hand, the horizontal (non-dispersive) 
position of the particle hit, $y_{\mathrm{TDC}}$,  can be determined much more accurately by measuring 
the time difference between hits in the left and right PMTs:
\begin{eqnarray}
  y_{\mathrm{TDC}}^{E,dE} = \frac{c_{\mathrm{eff}}^{E,dE}}{2}\left(t_i^{R} -t_i^{L} \right)
  \,, \label{eq_TDC_difference}
\end{eqnarray}
where $c_{\mathrm{eff}}^{E,dE}$ represent the effective speed of light inside the paddles. 
These parameters are determined by comparing the measured time 
difference in each paddle to the horizontal position of the track, as determined by the
MWDCs, and extrapolated to the scintillation detector. Fig.~\ref{fig_TDC_Velocity} shows
such a comparison for one paddle in each scintillation plane. This comparison was performed
for all 24 bars in each scintillation plane. The results are shown in Fig.~\ref{fig_TDC_FinalResults}.

The effective velocities of light were calculated, $c_{\mathrm{E}} = 0.12\mathrm{m/ns}$ for E-plane and
$c_{\mathrm{dE}} = 0.09\,\mathrm{m/ns}$ for the dE-plane. The obtained values are smaller
than the nominal speed-of-light inside plastic scintillator 
($c = c_0/1.58 = 0.19\,\mathrm{m/ns}$). This is expected, since generated light
does not travel directly to the PMTs but undertakes many reflections before exiting 
the scintillator, so its flightpath is much longer than 
the effective distance between the position of the hit PMT.  
The number of reflections depends on the length and width of the 
scintillator~\cite{kurata94}. This also explains the difference between $c_{\mathrm{E}}$ 
and $c_{\mathrm{dE}}$. Furthermore, the reduction of the 
effective speed-of-light is also caused by the use of leading-edge discriminators.
Due to attenuations in the scintillation material, light pulses generated by a single 
hit, do not reach both PMTs with the same amplitude. Since higher pulses cross the 
discriminator threshold earlier than smaller pulses, an additional delay is introduced (walk). 
 
\section{Calibration of the BigBite ADCs}
\label{sec:ADCCalibration}
The BigBite's ADC modules record data on charge collected by the PMTs employed in 
the scintillation detectors (dE- and E-planes). The  accumulated
charge is proportional to the particle energy deposited inside the detector material, and  
represents a valuable piece of information that enables particle identification (PID). 
Additionally, it can also provide the information about the hit position and the 
momentum of the incident particle. The analysis of the E05-102 experimental data 
depends strongly on the quality of particle identification, and a precise 
calibration of these modules was imperative.

The 12-bit ADC modules return values between $0$ and $4096$ for each of the
96 monitored channels. The additional 13. bit (value $8192$) is considered for reporting overflow 
in a particular channel. The digitalized ADC values are directly proportional 
to the size (integral) of the incoming analog pulses, which correspond to the
deposited energy. However, the measurement of 
the accumulated energy is influenced by many effects which can distort 
the formation of the electronics signal read out by the ADCs. 
First, attenuation of light in the scintillator cause a smaller light signal. 
The electronics pulses generated by the PMTs then depend 
on the properties of their photocathodes, and the connected high-voltages (HV).
The output signal is introduced to further attenuations and noise in the 
long signal cables. Finally, potential inaccuracies in the amplifiers may distort
the signal before entering the ADCs. These effects were carefully studied and corrected in 
 the calibration in order to establish comparable and consistent readouts in 
all ADC channels.

\begin{figure}[!ht]
\begin{center}
\includegraphics[width=0.49\linewidth]{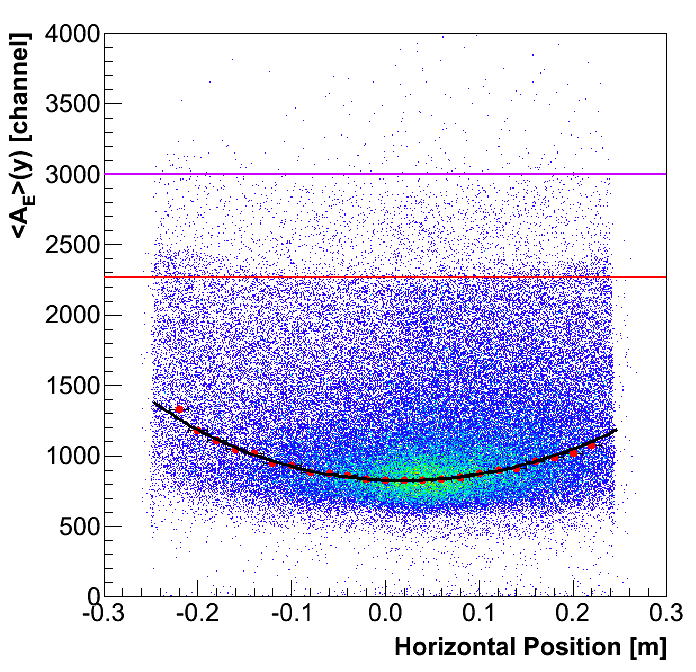}
\includegraphics[width=0.49\linewidth]{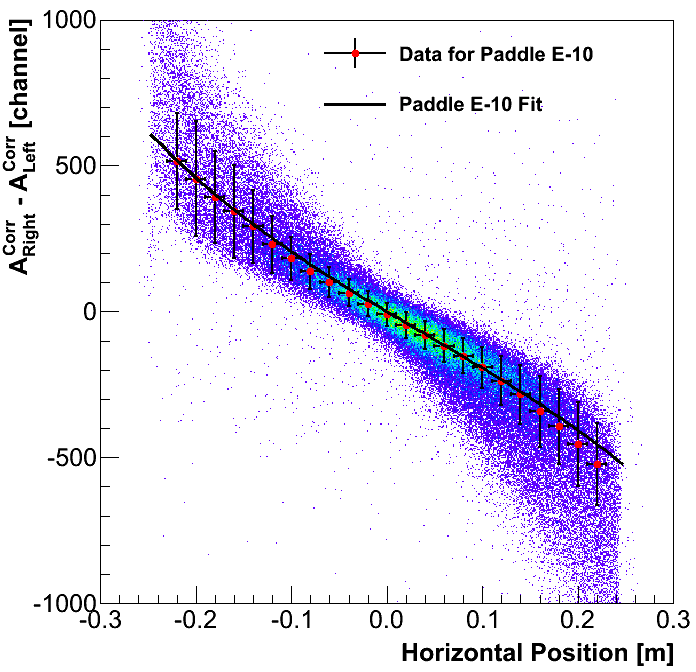}
\caption{ The ADC readings for E-plane paddle No.10. The plots show
the mean value (left) and difference (right) of the right and left PMT
signals as a function of horizontal position of the hit at the paddle.
The red dots correspond to the most probable values of the distributions 
at a given horizontal position. The mean ADC values were fitted with the 
Landau distribution function, while the Gaussian distribution was considered 
to fit the ADC difference data. Both ADC distributions  have positional 
dependence, which is roughly described by Eq.~(\ref{eq_BBADC_meanvalue}) and
Eq.~(\ref{eq_BBADC_diffvalue}). The two free parameters were determined to 
be $\lambda_{\mathrm{Left}} = 3.75$ and $\lambda_{\mathrm{Right}} = 4.35$. 
The red and violet line 
denote the proton and deuteron punch-through points, respectively.
\label{fig_BBADC_MeanDifference}}
\end{center}
\end{figure}

In a simple model, the electronic pulses generated by the two PMTs when the particle 
hits a scintillation paddle can be described as:
\begin{eqnarray}
  A_{\mathrm{Left}}(y) = G_{\mathrm{Left}}^{0}S_{0}(y)
    e^{-\lambda_{\mathrm{Left}}\left(l/2 - y\right)}\,, \qquad
A_{\mathrm{Right}}(y) = G_{\mathrm{Right}}^{0}S_{0}(y)
    e^{-\lambda_{\mathrm{Right}}\left(y +l/2\right)}\,, \label{eq_BBADC_yields}
\end{eqnarray}
where $l$ is the width of the paddle, while $\lambda_\mathrm{Left}$ 
and $\lambda_\mathrm{Right}$ represent the
effective attenuation constants describing losses on each side 
of the scintillation material. $S_0(y)$ is the light yield
produced by the incident particle and $G_{\mathrm{Left}}^{0}$, $G_{\mathrm{Right}}^{0}$
are the PMT gain factors. By requiring that scintillation light generated
in the center of the bar $(y = 0)$ must produce equivalent electric pulses on both sides
of the paddle, the gain factors must satisfy the condition
\begin{eqnarray}
  G_{\mathrm{Left}} = G_{\mathrm{Left}}^{0} e^{-\lambda_{\mathrm{Left}}\left(l/2\right)} = 
   G_{\mathrm{Right}}^{0} e^{-\lambda_{\mathrm{Right}}\left(l/2\right)} = G_{\mathrm{Right}}\,.
    \label{eq_BBADC_gains}
\end{eqnarray}
Considering Eq.~(\ref{eq_BBADC_gains}) in Eqs.~(\ref{eq_BBADC_yields}), the equations for the 
magnitude of the generated electric pulses on the output from the PMTs (or digital ADC equivalents)
are obtained:
\begin{eqnarray}
    A_{\mathrm{Left}}(y) = G_{\mathrm{Left}}S_{0}(y)
    e^{\lambda_{\mathrm{Left}}y}\,, \qquad
A_{\mathrm{Right}}(y) = G_{\mathrm{Right}}S_{0}(y)
    e^{-\lambda_{\mathrm{Right}}y}\,. \label{eq_BBADC_finalyield}
\end{eqnarray}
By dividing Eqs.~(\ref{eq_BBADC_finalyield}), one obtains the expression for 
finding the horizontal position of the particle hit from the left and right ADC readings:
\begin{eqnarray}
y = \frac{1}{\lambda_{\mathrm{Left}} + \lambda_{\mathrm{Right}}} 
  \ln\left( \frac{A_{\mathrm{Left}}(y)}{A_{\mathrm{Right}}(y) } \right)\,. \nonumber 
\end{eqnarray}
The mean value of the ADC signal obtained from a particular paddle is given by
\begin{eqnarray}
  \left\langle A \right\rangle(y) = \frac{A_{\mathrm{Right}}(y) + A_{\mathrm{Left}}(y)}{2} = 
      G_{\mathrm{Left}}\, S_0(y)\frac{ e^{\lambda_{\mathrm{Left}}y} +  
      e^{-\lambda_{\mathrm{Right}}y}}{2}\,. \label{eq_BBADC_meanvalue}
\end{eqnarray}
Similarly, the difference between the signals in the two opposite PMTs can be expressed as:
\begin{eqnarray}
  \Delta A(y) = A_{\mathrm{Right}}(y) - A_{\mathrm{Left}}(y) = 
    -2G_{\mathrm{Left}}\, S_0(y)\frac{ e^{\lambda_{\mathrm{Left}}y} -  
      e^{-\lambda_{\mathrm{Right}}y} }{2}\,. \label{eq_BBADC_diffvalue}
\end{eqnarray}
When assuming that attenuations in scintillator material are small and  attenuation factors 
identical on both sides of the bar $(\lambda_{\mathrm{Left}} = \lambda_{\mathrm{Right}}\rightarrow 0 )$,
 Eqs.~(\ref{eq_BBADC_meanvalue}) and~(\ref{eq_BBADC_diffvalue}) can be expanded in the Taylor 
series:
\begin{eqnarray}
  \left\langle A \right\rangle(y) &=& G_{\mathrm{Left}} S_0(y)\cosh{\left(\lambda_\mathrm{Left}y\right)} = 
    G_{\mathrm{Left}} S_0(y) \left[ 1+ \frac{\left(\lambda_\mathrm{Left}y\right)^2}{2!} +
      \frac{\left(\lambda_\mathrm{Left}y\right)^4}{4!} + \ldots \right]\,,\nonumber \\
  \Delta A(y) &=& -2 G_{\mathrm{Left}} S_0(y)\sinh{\left(\lambda_\mathrm{Left}y\right)} = -2
    G_{\mathrm{Left}} S_0(y) \left[ \left(\lambda_\mathrm{Left}y\right) + \frac{\left(\lambda_\mathrm{Left}y\right)^3}{3!} +
      \ldots \right]\,.\nonumber
\end{eqnarray}
In the ideal scintillator with no attenuations $(\lambda_{\mathrm{Left}} = 0)$, 
$\left\langle A \right\rangle(y)$ should be constant 
and independent of the horizontal position, while the difference $\Delta A(y)$ should always be zero.  
In a real scintillator with non-zero attenuation, the mean ADC value  results in a 
parabolic dependence of the horizontal position $y$, while the difference $\Delta A(y)$
can be explained by an odd polynomial function. The described model is in good agreement
with the measured data, as demonstrated in Fig.~\ref{fig_BBADC_MeanDifference}. When attenuation
factors differ for signals on the left and right side of the paddle, 
Eqs.~(\ref{eq_BBADC_meanvalue}) and~(\ref{eq_BBADC_diffvalue}) become more complicated. The 
$\left\langle A \right\rangle(y)$ picks up also terms with odd exponents, allowing its minimum
to move away from the center of the paddle. The difference $\Delta A(y)$ is modified 
in a similar manner.

Initial calibration of the scintillation detectors was performed already before the experiment
using cosmic events. For that purpose the cosmic trigger, described in 
Sec.~\ref{sec_BB_cosmic_trigger}, was utilized. The PMT gain matching was
done in two steps. First, high-voltages for the left PMTs were adjusted such,
that the events coming from the center of each paddle caused the same ADC readouts. 
A positional cut on the center of a paddle was performed by using
TDC information (see Eq.~(\ref{eq_TDC_difference})). This was the only possible method, 
since MWDCs were still not operational at that time. 
In the second step, high-voltages for both left and right
PMTs were changed simultaneously in order to align mean gains from all
24 paddles in both scintillation planes. The alignment was done graphically 
by cross-comparing two-dimensional ``dE/E`` plots, each showing the yield in a 
particular dE-paddle as a function of the yield in the neighboring E-paddle.

\begin{figure}[!ht]
\begin{center}
\includegraphics[width=0.49\linewidth]{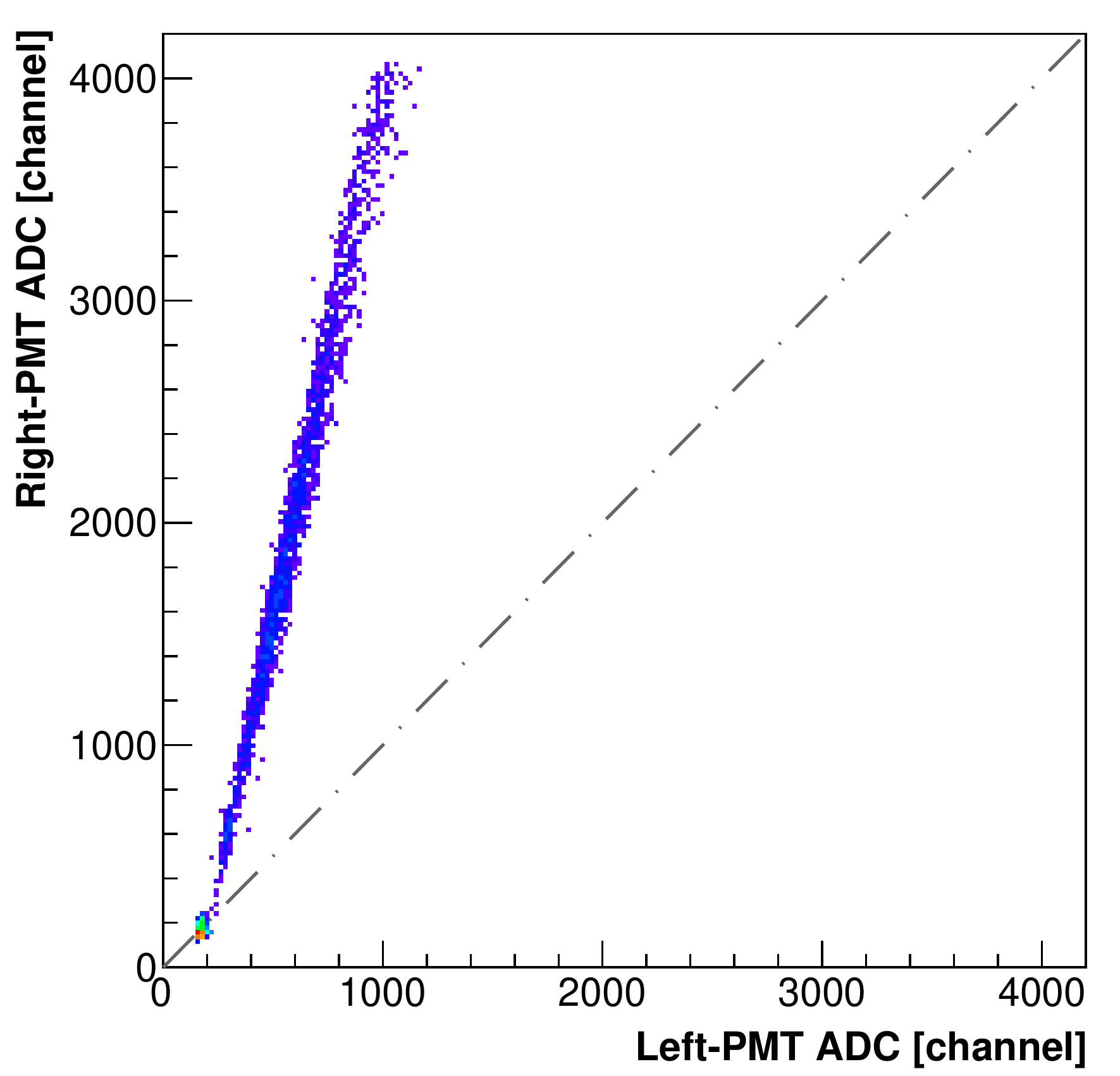}
\includegraphics[width=0.49\linewidth]{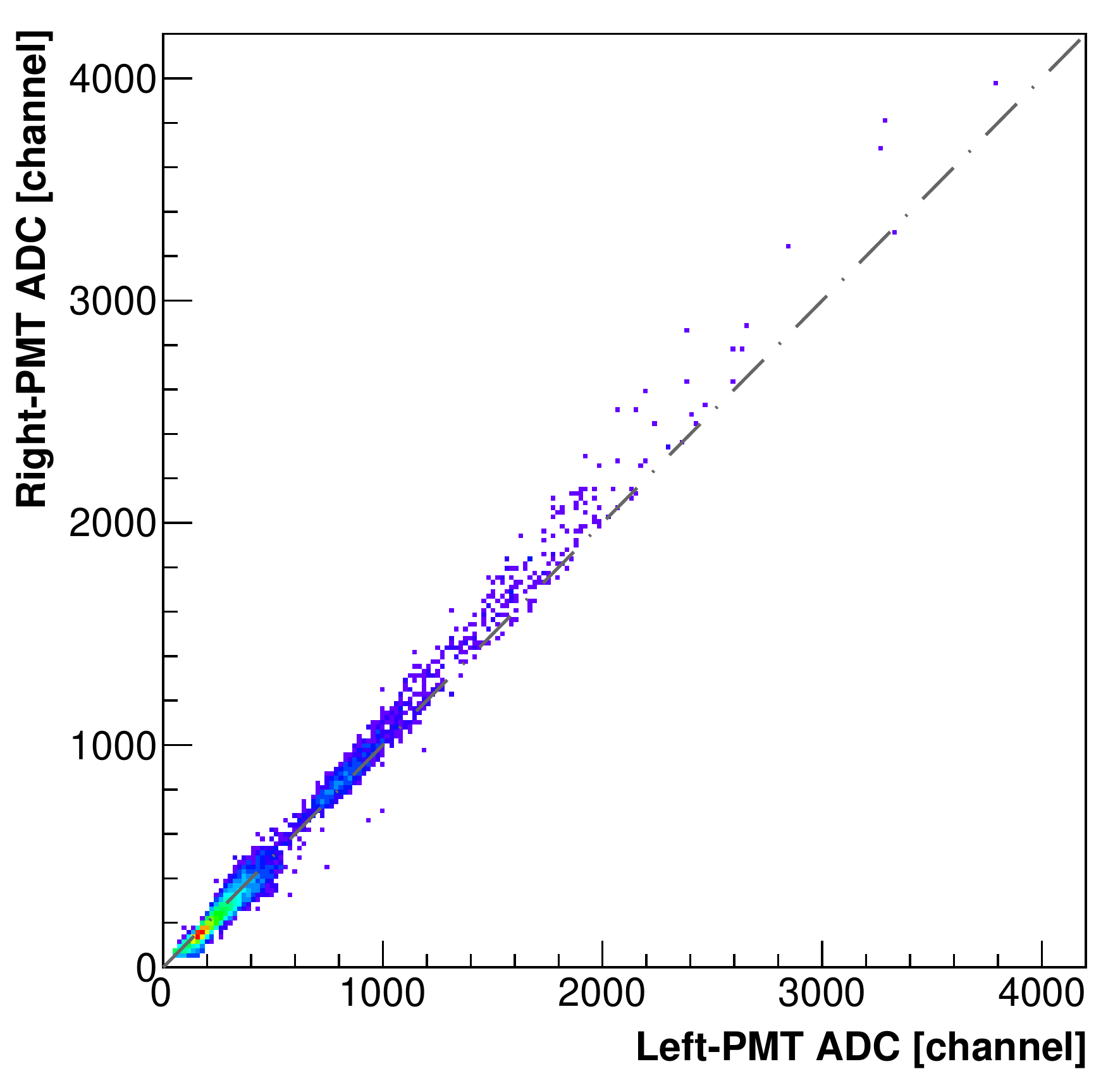}
\caption{The raw ADC readings from two PMTs mounted on the left and right side
of the E-plane scintillation paddle No.~10 before (left) and after (right) first 
calibration of the ADCs. Plots show only events that hit the center of the 
scintillation bar. Before any adjustments were made, the HV for the right PMT was set too high,
causing the signals to reach the saturation limit of the ADC module at $4096$. On the 
other hand, the HV for the left PMT could be increased in order to extend the digital ADC readouts 
beyond channel $1200$. After the changes to the HVs were made, the yields from both PMTs 
become almost identical and free of overflows. The strong peak at 
channel $\approx 200$ corresponds to the pedestal values. 
\label{fig_BBADC_RawADCReadings}}
\end{center}
\end{figure}

When changing the HV on the PMTs one had to 
be careful to keep the voltage under the maximum allowed value of 
$|U_{\mathrm{max}}| = 1700\,\mathrm{V}$. Setting the HV to above this limit would cause
a PMT to draw to much current which could permanently damage the tube. Furthermore, 
although it was desirable to have large PMT gains for better resolution, they had to be kept  
below a saturation limit of the ADC modules. In the opposite case the ADC readout was pushed to the 
maximum value of $4096$ and the overflow bit was set. This is demonstrated 
in Fig.~\ref{fig_BBADC_RawADCReadings}.

\begin{figure}[!ht]
\begin{center}
\begin{minipage}[t]{0.6\textwidth}
\hrule height 0pt
\includegraphics[width=1\textwidth]{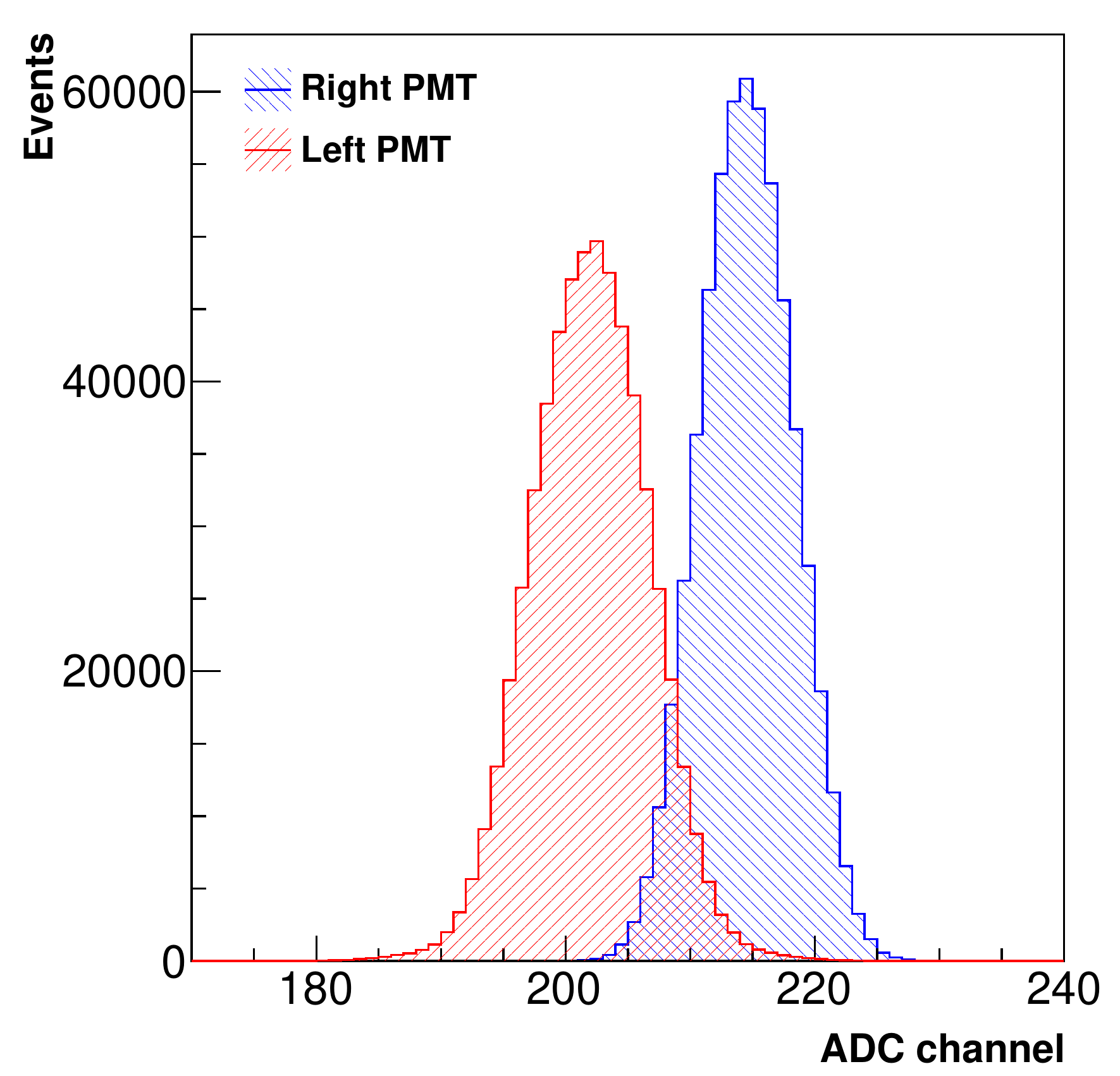}
\end{minipage}
\hfill
\begin{minipage}[t]{0.38\textwidth}
\hrule height 0pt
\caption{Pedestals of the two ADC channels corresponding to the 
PMT signals  for E-plane paddle No.~10.
\label{fig_BBADC_Pedestals}}
\end{minipage}
\end{center}
\end{figure}

A more detailed calibration needed for the final analysis was
performed after the experiment, by performing appropriate transformations
on the raw measured signals in the analysis programming scripts. 
The applicable ADC values $A_i^{\mathrm{Corr}}$ for each detected channel are 
calculated from the raw ADC readouts $A_i^{\mathrm{Raw}}$ by using
\begin{eqnarray}
 A_i^{\mathrm{Corr}} = G_i\left( A_i^{\mathrm{Raw}}  - A_i^{\mathrm{Ped}}\right)
    \,,\qquad i = 0\,,\dots\,,96\,, \label{eq_BBADC_Acorr}
\end{eqnarray}
where $G_i$ and $A_i^{Ped}$ are the relative gains and pedestal values
for each channel, respectively. The pedestals represent the values
returned by the ADC modules  when no signal is present at
their input. When valid signals are present at the input of the ADCs, the corresponding 
digitalized output values are positioned relative to the starting point, determined 
by the pedestals. Hence, pedestals need to be subtracted from the raw ADC value to 
obtain the clean ADC readout which corresponds to the size of the pulse at the input.
The pedestal values vary among different ADC channels and need to be determined 
for each channel separately. For that purpose, a dedicated set of data with 
scintillation detectors turned off was considered. Without high-voltage the PMTs 
produce no measurable signal and consequently the ADC modules  return only 
pedestal values. An example of such measurement is shown in Fig.~\ref{fig_BBADC_Pedestals}.
The pedestal peaks were typically located around ADC-bin 200 with a mean 
width of $\approx 3.2\,\mathrm{bins}$. 

\begin{figure}[!ht]
\begin{center}
\includegraphics[width=0.49\linewidth]{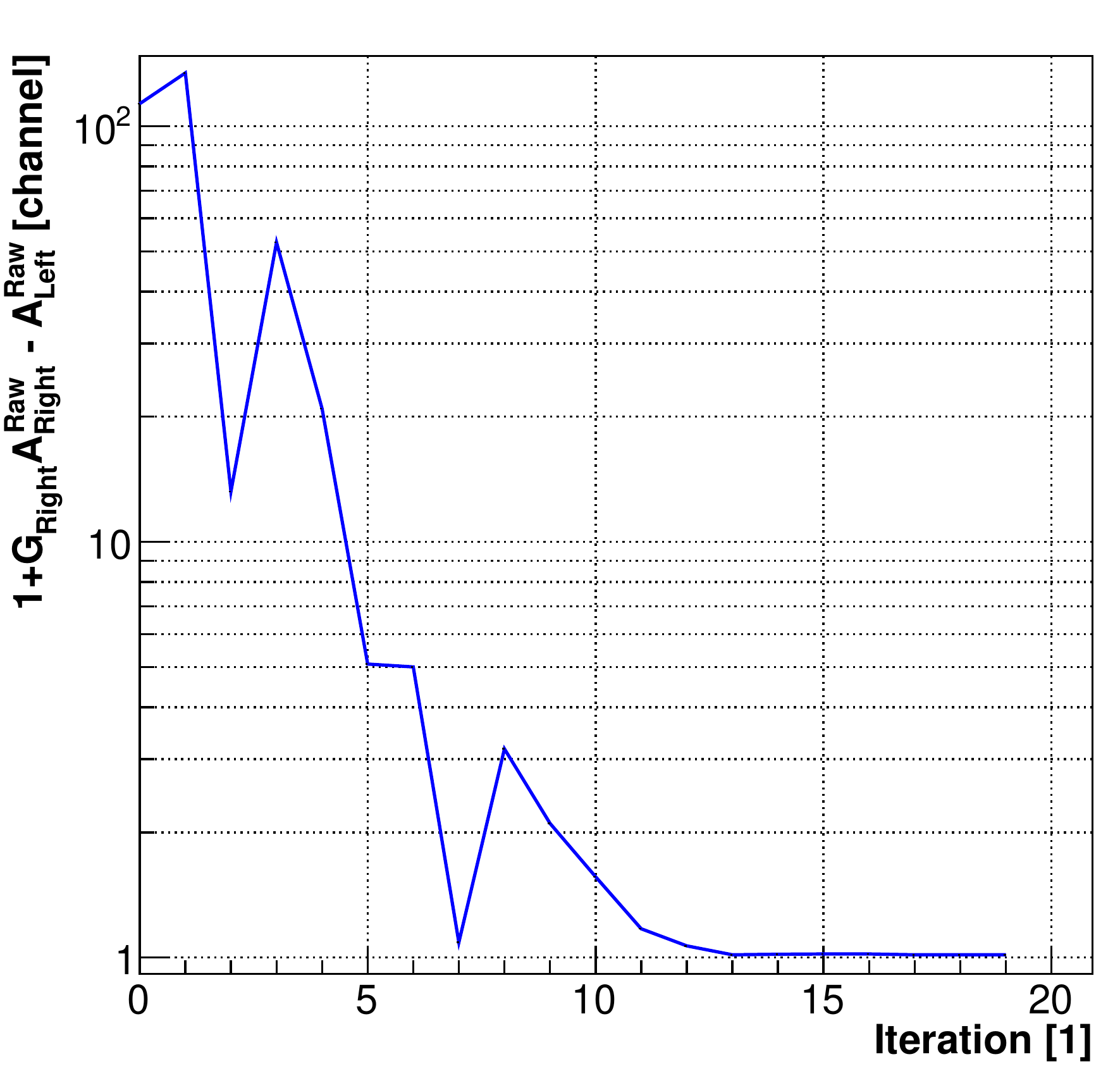}
\includegraphics[width=0.49\linewidth]{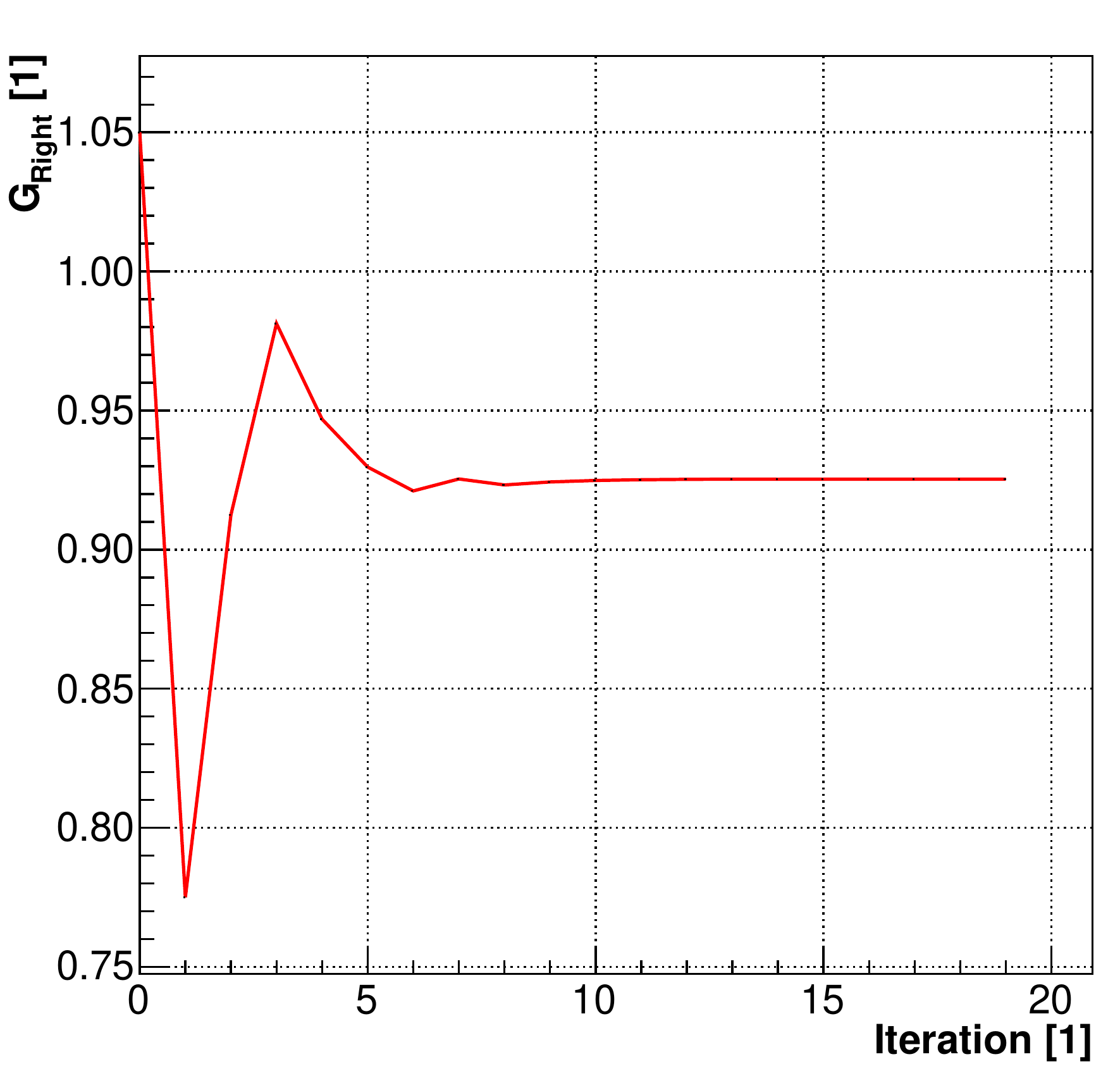}
\caption{The search for the relative gain factor for the right ADC signal for 
paddle No.~10. Bisection was used to minimize the difference
between signals detected in the left and right PMT for the light pulses generated 
at the center of the paddle. The plots show that values converge fast and that only 
few iterations of the algorithm are required to obtain the final result. 
\label{fig_BBADC_Bisection}}
\end{center}
\end{figure}

The relative gains $G_i$ were determined by a two-step procedure very similar
to that performed before the experiment, where the PMT HVs were set. If the first
step, a calibration of each scintillation paddle was performed independently to 
match the ADC yields from the left and right PMTs.  For that purpose, gains for 
the left PMTs were kept fixed at $G_{i} = 1$, while the the gains for the 
right PMTs were modified to provide effectively the same ADC readings from both
PMTs for events hitting the middle of the bar. The gain factors were determined
by an automated program, by minimizing the difference  
$(A_{\mathrm{Right}}^{\mathrm{Corr}} - A_{\mathrm{Left}}^{\mathrm{Corr}})$ using
bisection. An example of performance of this program is 
demonstrated in Fig.~\ref{fig_BBADC_Bisection}. This time, the MWDCs were utilized 
for the horizontal positioning  of the particle track at the scintillation bar. 

After the gain calibration in the individual paddles was finished, the mean ADC values 
from all bars were compared to each other. 
The gains for left and right ADC channels for each paddle were 
adjusted simultaneously in order to align the yields of neighboring paddles. Again, only 
events hitting the center of the scintillation bars were considered.  
For the visual monitoring of the changes, caused by the modifications to the gain 
factors, two dimensional plots of energy deposit in the overlapping dE- and E-paddles 
were employed. The final values of the gain factors for all
channels are summarized in Tables~\ref{table_BBADC_Eparameters} 
and~\ref{table_BBADC_dEparameters}.

\begin{figure}[!hb]
\begin{center}
\includegraphics[width=0.49\linewidth]{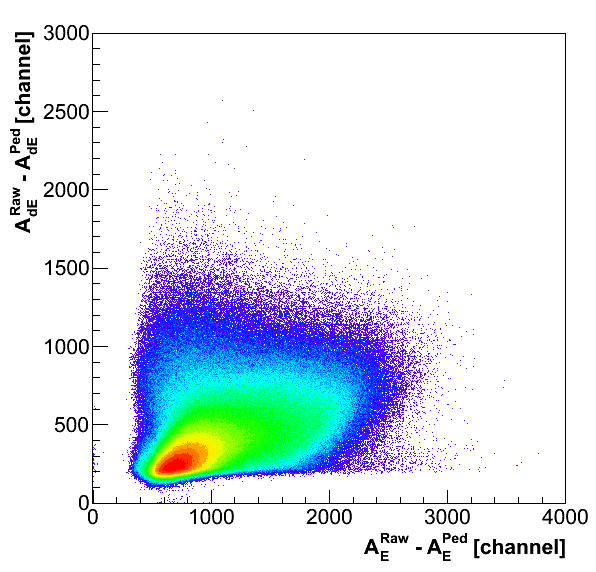}
\includegraphics[width=0.49\linewidth]{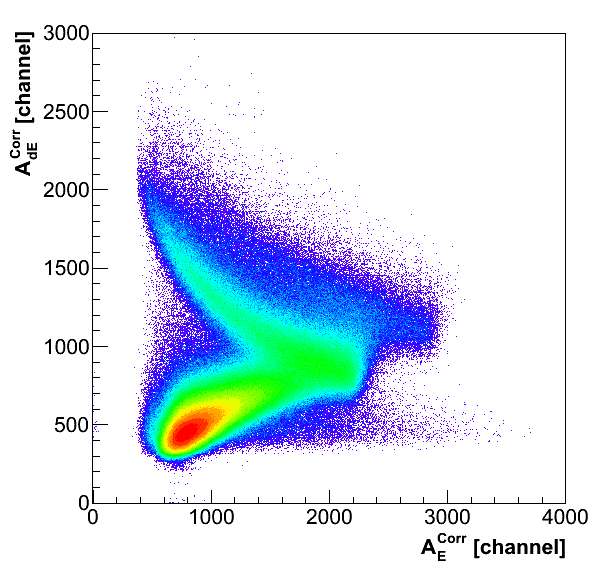}
\includegraphics[width=0.49\linewidth]{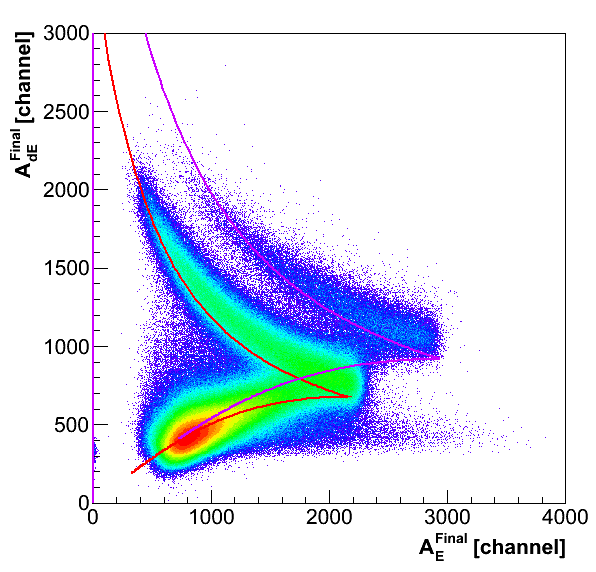}
\includegraphics[width=0.49\linewidth]{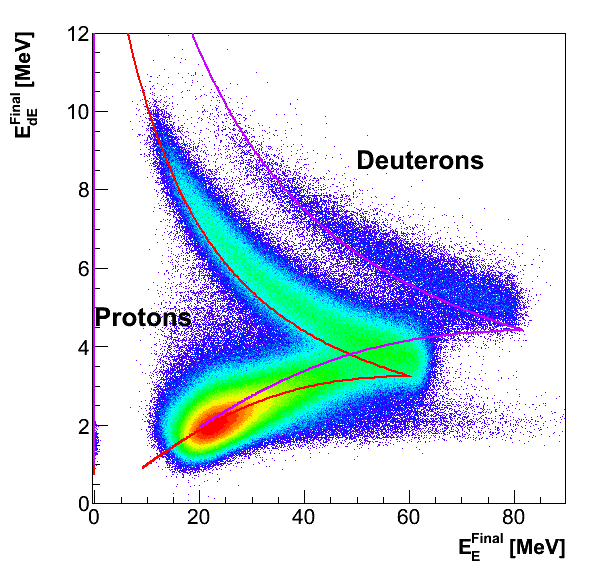}
\caption{Energy losses in the thin scintillator dE-plane versus
energy losses in the thicker E-plane at different stages of the BigBite ADC
calibration. [Top Left] The sum of the ADC signals for all 24 paddles before
gain matching. Here only pedestals were subtracted. [Top Right] The sum of 
the ADC signals in both planes after gains were matched according to 
Eq.~(\ref{eq_BBADC_Acorr}). [Bottom Left] The sum of the ADC signals after the
final step of the calibration, described by Eq.~(\ref{eq_BBADC_finalcorrection}),
was performed. The simulated energy losses for protons (red) and deuterons (violet)
inside the scintillators, calculated by using the Bethe-Bloch equation, are also shown. 
[Bottom Right] The sum of the ADC readings in both scintillation 
planes transformed into the physical energy scale, 
using Eqs.~(\ref{eq_BBADC_setscale}). Particles lose much more energy in the thicker 
scintillator (E) than in the thinner one (dE). The events shown at the top of the plot
correspond to low energy particles, which lose all their energy in the scintillators. 
High momentum particle accumulate at the bottom of the plot. The deuterons can be 
distinguished from the protons. The punch-through points, at which the protons 
and deuterons have just enough energy to penetrate both scintillation planes, are 
also clearly visible. 
\label{fig_BBADC_EdE}}
\end{center}
\end{figure}

The gain factors obtained with this calibration are stored in the \texttt{db\_BB.tp.dat}
database file  and are automatically considered in the analysis 
of all measured datasets. We are particularly 
interested in the two-dimensional ``dE/E`` plots where energies deposited by 
the incident particle inside the dE- and E-plane scintillators are shown. 
The structures observed in these plots depend on the kinetic energy and type 
of a particle, and can therefore be exploited for particle identification. 
A typical example is shown in 
Fig.~\ref{fig_BBADC_EdE}. The calibration significantly improved the
resolution of these plots and disentangled deuterons and protons.   

To further improve the PID, we decided not only to align the centers of the
paddles but also to correct for the $y$-dependence of the ADC reading of
each paddle. As described in the simple model, the mean ADC values 
depend on the position of the hit in the bar and consequently blur the empty
band between protons and deuterons. Therefore we decided to correct the mean
ADC values in each paddle for their positional dependence. 

\begin{figure}[!ht]
\begin{center}
\includegraphics[width=0.49\linewidth]{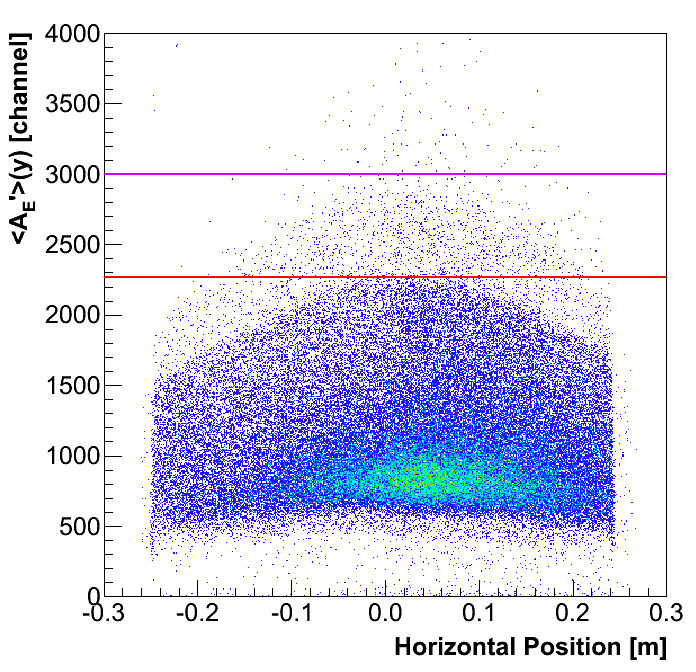}
\includegraphics[width=0.49\linewidth]{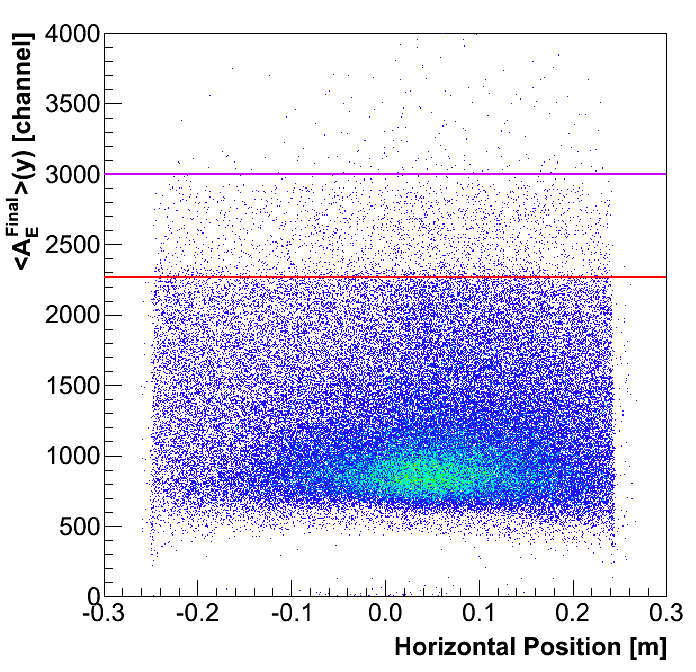}
\caption{ The mean values of the ADC readings in the E-plane paddle No.~10,
corrected for positional dependence using two considered approaches. 
[Left] The results when a correction derived from the 
simple attenuation model given by Eq.~(\ref{eq_BBADC_meanexp}) is
considered. The parameters $\lambda_{\mathrm{Left}}$ and 
$\lambda_{\mathrm{Right}}$ were obtained from 
 Fig.~\ref{fig_BBADC_MeanDifference}. This approach over-corrects 
the ADC spectrum. [Right] The results determined with the phenomenological 
approach given by Eq.~(\ref{eq_BBADC_finalcorrection}), where an exponential 
correction is subtracted from the mean ADC value in order to remove the 
positional dependence. The red and violet line 
denote the proton and deuteron punch-through points, respectively.
\label{fig_BBADC_MeanGoodBad}}
\end{center}
\end{figure}

In the first attempt we considered Eq.~(\ref{eq_BBADC_meanvalue}) and tried to 
extract the corrected mean ADC value $\left\langle A' \right\rangle$ by dividing 
$\left\langle A \right\rangle(y)$ with the attenuation factor:
\begin{eqnarray}
  \left\langle A' \right\rangle = \frac{2\left\langle A \right\rangle(y)}
     { e^{\lambda_{\mathrm{Left}}y} +  
      e^{-\lambda_{\mathrm{Right}}y}}. \label{eq_BBADC_meanexp}
\end{eqnarray}
The distribution shown in Fig.~\ref{fig_BBADC_RawADCReadings} (left)  was sliced 
in bins of $y$ (horizontal coordinate). The ADC distribution for each $y$ bin
was fitted by a Landau curve. From the positions of the maxima of these curves, 
the parameters $\lambda_{\mathrm{Left}}$ and $\lambda_{\mathrm{Right}}$ were 
determined, by fitting Eq.~(\ref{eq_BBADC_meanvalue}) to these maxima. 
This approach did not work well. In many cases the method failed to properly 
correct the $y$-dependence of the ADC signals. It turned out that the $y$-dependence 
of the positions of the maxima of the Landau curves, does not properly describe
the $y$-dependence of the whole ADC distribution. It happened that after the corrections
were applied, the ADC spectra (especially at the edges) got over-corrected, causing
the bending of the ADC spectra in the opposite directions. This is clearly
demonstrated in Fig.~\ref{fig_BBADC_MeanGoodBad}. Because of these problems, this 
type of correction was abandoned and another option was considered. 

In the second attempt of removing the $y$-dependence from the dE- and E-plane
ADC spectra we decided to subtract the $\cosh{\left(\lambda_\mathrm{Left}y\right)}$
instead of dividing with it. This phenomenological type of correction was easier
to control and turned out to be a good choice which produced much better results. 
The final value of the ADC signal for each channel $i$ in the dE- and E-plane was
calculated from 
\begin{eqnarray}
  \left\langle A^{\mathrm{Final}}_{i} \right\rangle = 
     \left\langle A_i \right\rangle(y) - {C'}_i \left[
      \frac{C_i}{2}\left( e^{\lambda_{\mathrm{Left}}y} +  e^{-\lambda_{\mathrm{Right}}y}
        -2\right)\right]\,. \label{eq_BBADC_finalcorrection}
\end{eqnarray}
Here $C_i$ represents the amplitude of the  correction for each paddle. Its values 
are typically of order $\approx 1500$ for E-plane and $\approx 300$ for dE-plane.
The corrections are introduced in a way that leave the ADC signals from 
the center of the paddle intact, while the readings away from the center 
are lowered according to Eq.~(\ref{eq_BBADC_finalcorrection}).
This results in a flat ADC distribution for each paddle, as desired. 

The paddles in the dE-plane are not aligned with the paddles in the E-plane, but 
are shifted by half the paddle width. Hence, the $i_{\mathrm{E}}$-th paddle in the 
E-plane overlaps with the $i_{\mathrm{dE}}$-th and $(i_\mathrm{dE}+1)$-th 
paddle in the dE-plane.  In the process of gain matching by looking at the dE/E-plots 
only paddles with the same indices $(i_\mathrm{E} = i_\mathrm{dE})$ were considered. 
Although that yields from all such pairs were matched, this does not mean that 
pairs $(i_\mathrm{E} = i_\mathrm{dE}-1)$ become matched as well. It turned out that
adjustments are required to align them with the rest. For that purpose an additional
multiplication factor ${C'}_i$ was introduced to  correct for that discrepancy. 
For the E-plane, the ${C'}_{i}$ is a priori set to one. For the dE-plane, the 
factor ${C'}_{i}$ was set to unity when $(i_\mathrm{E} = i_\mathrm{dE})$, while
it was set to value
different from unity when $(i_\mathrm{E} = i_\mathrm{dE} -1)$. The final values of 
the parameters appearing in Eq.~(\ref{eq_BBADC_finalcorrection}) for all paddles
are listed in Tables~\ref{table_BBADC_Eparameters} and~\ref{table_BBADC_dEparameters}. 

The corrections modeled by Eq.~(\ref{eq_BBADC_finalcorrection}) are not included in the
standard BigBite library (\texttt{bigbitelib}), but were coded in a separate library called
Hadron Detector Package (\texttt{HadrDetPack}) which inherits from the standard library. This way
we were able to keep the original version of the BigBite Library intact 
but still use all desirable corrections.

\begin{table}[!ht]
\begin{center}
\caption{Parameters for the ADC calibration of the E-plane scintillation 
detectors (Eqs.~(\ref{eq_BBADC_Acorr}) and~(\ref{eq_BBADC_finalcorrection})). 
\vspace*{1.5mm}
\label{table_BBADC_Eparameters}}
\begin{tabular}{llllllllll}
\toprule
i & $A_{\mathrm{Left}}^{\mathrm{Ped}}$ & $\sigma_{\mathrm{Left}}^{\mathrm{Ped}}$ 
  & $A_{\mathrm{Right}}^{\mathrm{Ped}}$ & $\sigma_{\mathrm{Right}}^{\mathrm{Ped}}$ 
   & $G_{\mathrm{Left}}$ & $G_{\mathrm{Right}}$ & $\lambda_{\mathrm{Left}}$ 
    & $\lambda_{\mathrm{Right}}$ & $C$\\
\midrule
0 & 182.677 & 2.36 & 225.599 & 3.33 & 0.922 & 0.930 & 2.124 & 2.516 & 980.607\\
1 & 152.78 & 1.49 & 206.255 & 2.08 & 1.077 & 0.950 & 2.531 & 3.01 & 1494.527\\
2 & 124.161 & 1.97 & 211.177 & 2.59 & 1.034 & 1.151 & 1.735 & 2.049 & 1978.991\\
3 & 137.427 & 2.14 & 237.898 & 2.11 & 1.140 & 1.164 & 2.08 & 1.84 & 2188.242\\
4 & 125.021 & 1.50 & 229.474 & 2.31 & 1.099 & 1.140 & 1.724 & 1.385 & 2185.218\\
5 & 146.594 & 2.30 & 219.869 & 3.73 & 1.180 & 1.164 & 1.982 & 1.476 & 2182.772\\
6 & 153.248 & 2.08 & 201.265 & 3.10 & 1.181 & 1.116 & 1.733 & 1.627 & 2183.715\\
7 & 126.654 & 2.09 & 131.56 & 1.65 & 1.131 & 1.053& 1.408 & 1.789 & 2142.507\\
8 & 152.689 & 2.87 & 128.451 & 2.58 & 1.0853 & 1.151 & 1.930 & 1.908 & 2236.151\\
9 & 124.788 & 2.59 & 222.03 & 3.58 & 1.146 & 1.161 & 1.754 & 1.904 & 2164.355\\
10 & 201.621 & 4.60 & 214.192 & 3.68 & 1.221 & 1.124 & 1.723 & 1.774 & 2240.818\\
11 & 200.381 & 2.91 & 166.908 & 3.86 & 1.133 & 1.149 & 1.852 & 1.666 & 2147.581\\
12 & 192.873 & 3.49 & 228.053 & 4.44 & 1.148 & 1.141 & 1.928 & 1.556 & 2190.922\\
13 & 141.47 & 3.045 & 213.533 & 4.45 & 1.172 & 1.088 & 1.841 & 1.457 & 2200.376\\
14 & 147.638 & 2.98 & 209.356 & 3.81 & 1.179 & 1.091 & 1.551 & 1.266 & 2174.675\\
15 & 196.307 & 3.08 & 174.49 & 3.38 & 1.210 & 1.102 & 2.394 & 2.495 & 893.722\\
16 & 286.174 & 2.33 & 258.997 & 2.39 & 1.208 & 1.068 & 2.095 & 2.58 & 860.803\\
17 & 238.374 & 2.88 & 256.831 & 3.55 & 1.156 & 1.101 & 1.225 & 1.385 & 2193.279\\
18 & 187.152 & 2.77 & 301.935 & 2.91 & 1.104 & 1.144 & 0.299 & 0.867 & 901.551\\
19 & 227.13 & 2.93 & 322.729 & 3.80 & 0.963 & 1.185 & 0.759 & 0.587 & 891.529 \\
20 & 163.785 & 3.05 & 342.693 & 2.96 & 1.009 & 1.276 & 1.881 & 3.014 & 840.171 \\
21 & 224.934 & 2.38 & 321.323 & 3.41 & 0.906 & 1.061 & 1.312 & 2.45 & 884.276\\
22 & 232.066 & 2.53 & 304.617 & 3.17 & 1.118 & 0.956 & 0.641 & 1.133 & 866.364\\
23 & 223.115 & 2.99 & 301.143 & 4.73 & 1 & 0.842 & 0 & 0 & 0\\
\bottomrule
\end{tabular}
\end{center}
\end{table}

At the end, after all the ADC signals were properly calibrated, the energy scale
was also set, in order to transform the measured ADC signals 
$\left\langle A^{\mathrm{Final}}_{i} \right\rangle$ to the physically 
relevant energy $\left\langle E^{\mathrm{Final}}_{i} \right\rangle$, 
deposited in the scintillation material by the particle flying through the detectors:
\begin{eqnarray}
      \left\langle E^{\mathrm{Final}}_{\mathrm{E}} \right\rangle = \frac{\left\langle 
        A^{\mathrm{Final}}_{\mathrm{E}} \right\rangle + O_{\mathrm{E}}}{k_{\mathrm{E}}}\,,\qquad 
      \left\langle E^{\mathrm{Final}}_{\mathrm{dE}} \right\rangle = \frac{\left\langle 
        A^{\mathrm{Final}}_{\mathrm{dE}} \right\rangle + O_{\mathrm{dE}} }{k_{\mathrm{dE}}}\,, \label{eq_BBADC_setscale}
\end{eqnarray}
where $O_{\mathrm{E}}$, $ O_{\mathrm{dE}}$, $k_{\mathrm{E}}$ and 
$ k_{\mathrm{dE}}$ are the calibration constants. 
A simulation using Bethe-Bloch formula was used to determine the energy 
deposit in both scintillation planes by particles with different kinetic energies. 
By comparing the results of the calculations to the measured ADC values, the 
calibration constants for both detector planes were obtained:
\begin{eqnarray}
         k_{\mathrm{E}} = 35.868\,\mathrm{\frac{channel}{MeV}}\,,\qquad
         O_{\mathrm{E}} = 32.0\,\mathrm{channel}\,, \nonumber\\
         k_{\mathrm{dE}} = 208.3125\,\mathrm{\frac{channel}{MeV}}\,,\qquad 
         O_{\mathrm{dE}} = 0.0\,\mathrm{channel}\,. \nonumber 
\end{eqnarray}
The reference points considered in this calibration were the two proton punch-through 
points. The first one corresponds to the protons with just
enough energy to penetrate through the first, thin layer of the scintillator. 
Similarly, the second one represents the point where the protons have enough 
energy also to exit the second layer (E-plane) of the scintillator. Hence, 
both points represent the maximum energy deposit that could be generated
by protons in the dE- and E-plane. They are clearly visible in 
Fig.~\ref{fig_BBADC_EdE}, as the sharp edges, where energy-losses curves change
their course.

\begin{table}[!ht]
\begin{center}
\caption{Parameters for the ADC calibration of the dE-plane scintillation 
detectors (Eqs.~(\ref{eq_BBADC_Acorr}) and~(\ref{eq_BBADC_finalcorrection})). 
\vspace*{1.5mm}
\label{table_BBADC_dEparameters}}
\begin{tabular}{lllllllllll}
\toprule
i & $A_{\mathrm{Left}}^{\mathrm{Ped}}$ & $\sigma_{\mathrm{Left}}^{\mathrm{Ped}}$ 
  & $A_{\mathrm{Right}}^{\mathrm{Ped}}$ & $\sigma_{\mathrm{Right}}^{\mathrm{Ped}}$ 
   & $G_{\mathrm{Left}}$ & $G_{\mathrm{Right}}$ & $\lambda_{\mathrm{Left}}$ 
    & $\lambda_{\mathrm{Right}}$ & $C$ & $C'$\\
\midrule
0 & 322.792 & 2.93 & 306.05 & 3.09 & 1.929 & 2.099 & 3.159 & 2.701 & 1469.103 & 1 \\
1 & 317.969 & 2.79 & 303.392 & 2.95 & 1.927 & 2.023 & 2.381 & 2.358 & 1310.953 & 1 \\
2 & 315.259 & 3.69 & 295.693 & 2.47 & 1.840 & 1.685 & 2.178 & 2.171 & 978.562 & 1.050 \\
3 & 308.33 & 3.39 & 305.935 & 3.81 & 1.618 & 1.733 & 4.540 & 3.586 & 738.350 & 0.915 \\
4 & 305.46 & 2.89 & 299.229 & 2.83 & 1.736 & 1.731 & 3.966 & 3.934 & 532.918 & 1 \\
5 & 323.829 & 3.03 & 314.015 & 2.48 & 1.864 & 1.902 & 4.911 & 4.963 & 485.122 & 1 \\
6 & 325.305 & 2.24 & 247.181 & 2.29 & 1.892 & 1.584 & 4.400 & 5.014 & 479.654 & 0.947 \\
7 & 298.384 & 3.79 & 305.914 & 3.29 & 1.672 & 1.542 & 4.473 & 5.292 & 469.829 & 1 \\
8 & 254.001 & 4.43 & 314.848 & 3.70 & 1.782 & 1.831 & 5.763 & 5.724 & 485.113 & 1 \\
9 & 327.316 & 4.23 & 336.343 & 3.50 & 2.018 & 1.931 & 5.208 & 5.734 & 496.651 & 1 \\
10 & 314.732 & 3.73 & 310.095 & 3.27 & 1.991 & 1.744 & 5.149 & 6.517 & 488.395 & 1 \\
11 & 321.1 & 4.02 & 324.128 & 3.27 & 1.930 & 1.644 & 4.367 & 4.609 & 464.356 & 1.011 \\
12 & 339.122 & 4.21 & 335.274 & 4.12 & 1.852 & 1.662 & 4.426 & 5.128 & 459.911 & 1 \\
13 & 345.066 & 4.21 & 322.456 & 3.76 & 1.681 & 1.723 & 5.434 & 5.268 & 460.990 & 1 \\
14 & 303.04 & 3.43 & 303.701 & 3.22 & 1.758 & 1.718 & 4.055 & 3.934 & 464.725 & 1 \\
15 & 317.361 & 4.12 & 307.136 & 4.73 & 1.545 & 1.695 & 4.442 & 3.983 & 455.136 & 1 \\
16 & 299.418 & 3.39 & 305.309 & 2.37 & 1.778 & 1.565 & 4.445 & 5.727 & 410.031 & 1.168 \\
17 & 313.009 & 3.02 & 332.018 & 3.68 & 1.473 & 1.932 & 3.974 & 4.381 & 436.072 & 0.975 \\
18 & 310.572 & 2.87 & 283.469 & 3.25 & 2.211 & 1.627 & 3.148 & 4.224 & 498.121 & 1.018 \\
19 & 310.759 & 3.45 & 296.797 & 2.37 & 1.396 & 1.933 & 4.043 & 3.413 & 452.242 & 1 \\
20 & 324.47 & 3.62 & 296.621 & 3.29 & 2.145 & 1.553 & 2.486 & 5.764 & 492.452 & 1 \\ 
21 & 306.902 & 4.47 & 319.811 & 2.37 & 1.825 & 1.479 & 4.028 & 5.47 & 468.191 & 0.949 \\
22 & 330.468 & 2.96 & 313.977 & 2.63 & 1.794 & 1.801 & 3.414 & 4.118 & 464.735 & 0.970 \\
23 & 309.46 & 4.43 & 328.881 & 3.79 & 1.092 & 1 & 0 & 0 & 0 & 1 \\
\bottomrule
\end{tabular}
\end{center}
\end{table}

\section{The Discriminator Threshold calibration}

The formation of BigBite triggers and readout of the TDCs is carried out
only when the amplitude of the analog signals from the BigBite scintillation 
detector PMTs exceed certain thresholds. They are set in the discriminator 
modules and can be controlled remotely from DAQ computers.

Three different threshold limits are considered in the BigBite electronics.
As described in Fig.~\ref{fig_BBT1trigger}, common discriminator modules, with 
the thresholds set to $U_{\mathrm{E}}$, are used to discriminate TDC pulses from 
the E-plane PMTs, as well as for setting the lower limit for the formation of the T1 trigger. 
On the other hand, two sets of discriminator modules are employed in the dE-plane 
circuit (see Fig.~\ref{fig_BBT2trigger}). The first set, with a threshold setting 
$U_{\mathrm{dE-TDC}}$, is used for discrimination of TDCs pulses, while the second set, 
with thresholds set to $U_{\mathrm{dE-Trigger}}$, is utilized for determining
the lower limit for formation of a T2 trigger. In principle, arbitrary values can be 
used for  $U_{\mathrm{dE-TDC}}$ and $U_{\mathrm{dE-Trigger}}$. Most often however, 
the thresholds were set to $U_{\mathrm{dE-Trigger}} \approx 2U_{\mathrm{dE-TDC}}$ in order 
to generate event detection conditions in the dE-plane that are similar to those 
used in the E-plane. 

\begin{figure}[!ht]
\begin{center}
\begin{minipage}[t]{0.6\textwidth}
\hrule height 0pt
    \includegraphics[width=\linewidth]{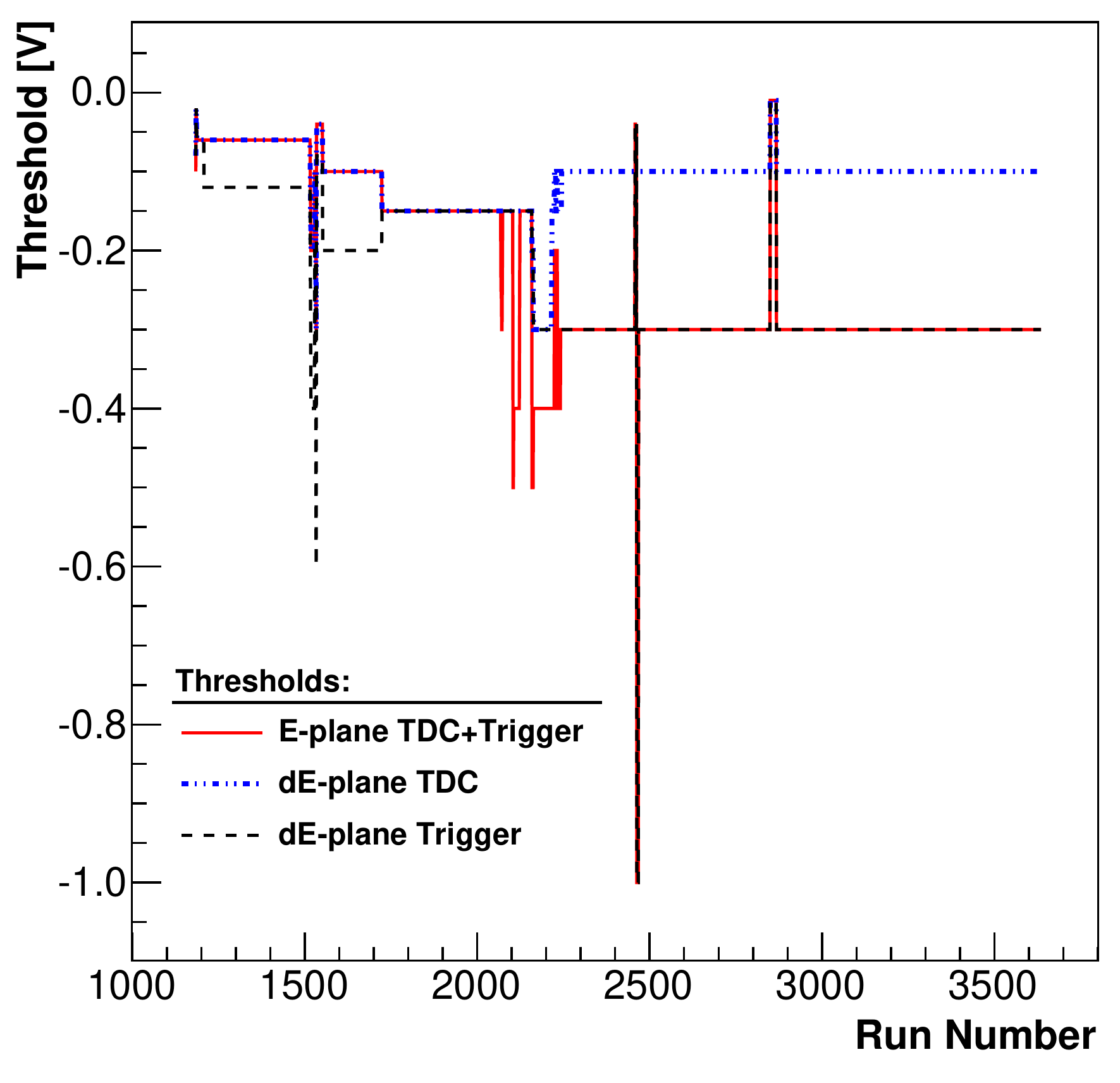}
\end{minipage}
\hfill
\begin{minipage}[t]{0.38\textwidth}
\hrule height 0pt
\caption{The discriminator threshold settings considered during the 
E05-102 experiment. In the commissioning phase (Runs 1200-2200) of the experiment, 
the thresholds were repeatedly modified for calibration purposes. During 
production running (Runs $\geq 2200$) the threshold settings were kept almost 
intact. Only minor changes were performed to improve detection efficiency.  
\label{fig_BBTDC_ThresholdSettings}}
\end{minipage}
\end{center}
\end{figure}

With the correct choice of the threshold settings we are able to select and trigger 
only on heavily ionizing protons and deuterons that leave strong signals in the BigBite 
scintillation detectors, while neglecting minimally  ionizing particles such as
electrons, muons and pions, which deposit only small amounts of energy.
The thresholds are set to the discriminator modules 
in units of $\mathrm{mV}$. The values considered in the experiment are listed 
in Fig~\ref{fig_BBTDC_ThresholdSettings}. Unfortunately these voltage settings can not 
be directly correlated to
the particle energy deposit in the scintillators, since the size of the detected analog 
pulses depends on many parameters, such as the type of the scintillator material and HVs 
applied to the PMTs. In order to transform threshold voltages to the physically meaningful
energy scale, a transformation constant is required. 

\begin{figure}[!ht]
\begin{center}
\includegraphics[width=0.9\linewidth]{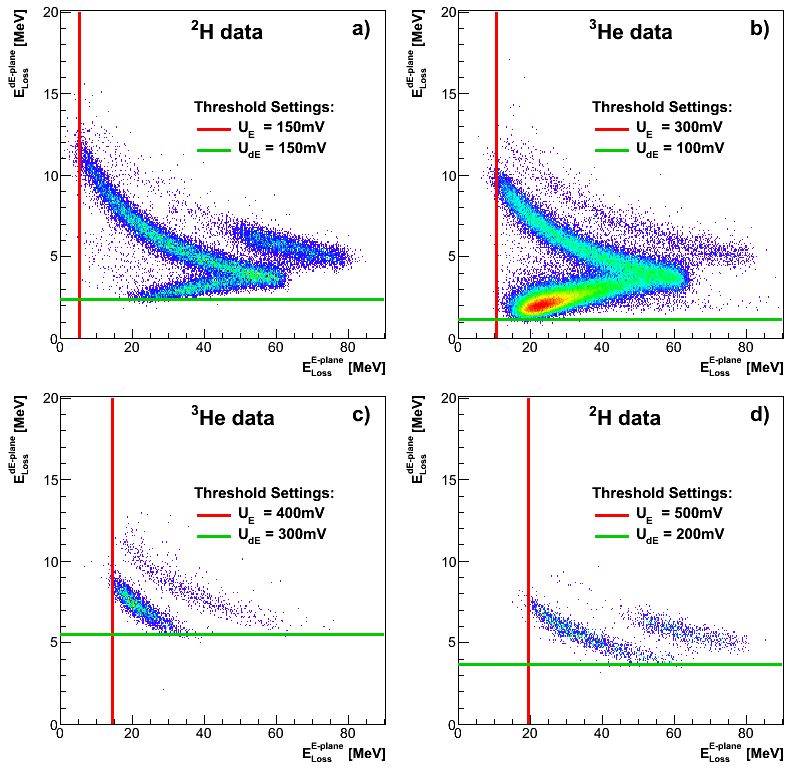}
\caption{Two-dimensional plots of the particle energy deposition inside dE- and
E-plane detectors for different TDC threshold settings. By increasing the threshold 
voltages $U_{\mathrm{E}}$ and $U_{\mathrm{dE-TDC}}$ in the  discriminator modules, 
the events that leave small amounts
of energy in the scintillation detectors are being gradually cut away.
These are either very low momentum particles, with barely enough energy to penetrate to
the E-plane and leave their whole remaining energy there, or very high momentum particles 
that deposit only a small portion of their 
energy. In this calibration procedure the production ${}^3\mathrm{He}$ 
and elastic ${}^2\mathrm{H}$ data were considered in order to study both proton and
deuteron energy losses. Since elastic deuterons all carry almost the same amount of 
energy, they generate only a small blob in the energy-deposition plots.
\label{fig_TDCThresholdEdE}}
\end{center}
\end{figure}

\begin{figure}[!ht]
\begin{center}
\begin{minipage}[t]{0.65\textwidth}
\hrule height 0pt
    \includegraphics[width=\linewidth]{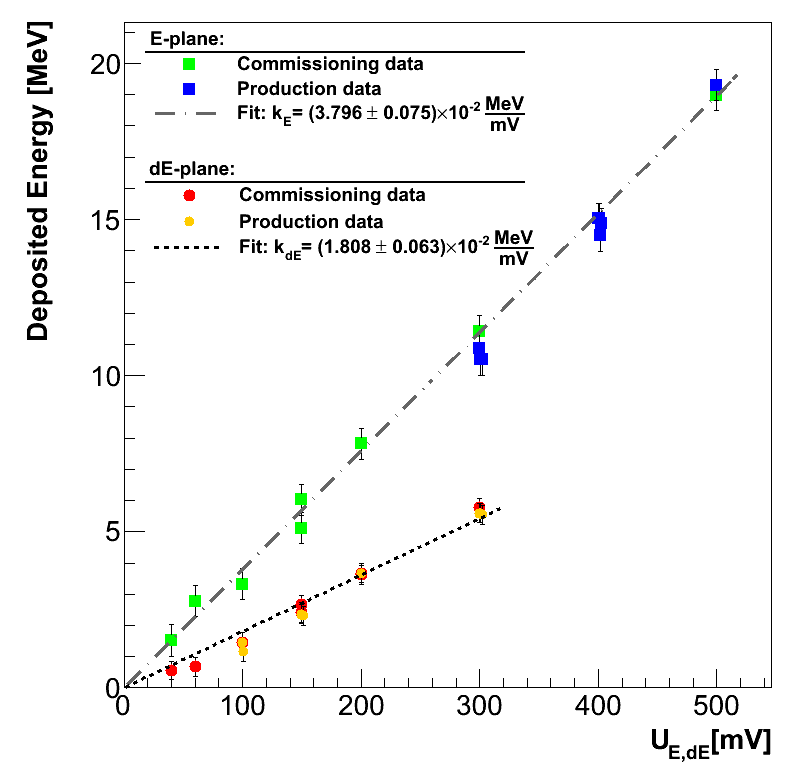}
\end{minipage}
\hfill
\begin{minipage}[t]{0.34\textwidth}
\hrule height 0pt
\caption{The minimal particle energy deposit that is detected in the 
scintillation detectors dE and E at various TDC threshold 
voltages $U_{\mathrm{E}}$ and $U_{\mathrm{dE_TDC}}$ set to
the discriminator modules. The results were obtained from the analysis 
of the two-dimensional E/dE plots shown in Fig.~\ref{fig_TDCThresholdEdE}.
The calibration measurements were performed during the 
commissioning phase of the experiment and also during the production 
running. The transformation constants $k_{\mathrm{E}}$ and $k_{\mathrm{dE}}$ 
can be used to directly transform threshold voltages to the physically 
meaningful energy deposit threshold by: 
$E_{\mathrm{E,dE}}^{_{\mathrm{Threshold}}} = k_{\mathrm{E,dE}}\cdot U_{\mathrm{E,dE}}$ 
\label{fig_TDCThresholdFit}}
\end{minipage}
\end{center}
\end{figure}

The transformation constants for both scintillation planes were obtained by inspecting 
the energy-deposit spectra (calibrated ADC spectra) at different threshold settings. 
When generating such spectra one had to be careful to read ADC information only from the 
paddles that also had a valid discriminated TDC hit. In the opposite case, changes in the 
ADC spectra related to the threshold settings would not be observed. 
Fig.~\ref{fig_TDCThresholdEdE} clearly 
shows how the lower limits of the observed energy-deposit spectra change with the threshold 
settings. Using this information,  transformation constants for both planes were obtained.
The results are shown in Fig.~\ref{fig_TDCThresholdFit}. Two sets of data were considered for this calibration. The 
first set was collected in the commissioning phase before the experiment, while the second one 
was obtained during production running. In the time between two tests, some modifications 
to the scintillation detectors were still performed, which may explain the observed small 
differences between the data sets. 

For production running, the threshold voltages were set to
$U_{\mathrm{E}} = -300\,\mathrm{mV}$, $U_{\mathrm{dE-TDC}} = -100\,\mathrm{mV}$
and $U_{\mathrm{dE-Threshold}} = -300\,\mathrm{mV}$. With these settings, the vast 
majority of the protons and deuterons were accepted and recorded, while the minimally 
ionizing particles were completely ignored. A typical E/dE spectrum obtained during 
the experiment is shown in Fig.~\ref{fig_TDCThresholdEdE} b). 
This was achieved by setting the E-plane 
threshold to a high value. On the other hand, the dE-plane TDC-threshold was set to 
a reasonably low value, for recording TDC information of high-momentum hadrons, located 
at the bottom of the E/dE spectrum. Furthermore, the dE-plane trigger threshold
(note that this is not the same as TDC threshold)
 was set to a very high value, $E_{\mathrm{Loss}}^{\mathrm{dE-plane}} = 5.55\,\mathrm{MeV}$.
Hence, when the T2 trigger is chosen, only low momentum protons and deuterons are 
accepted. However, we are mostly interested in events that deposit energy in both
scintillation planes and therefore use T1 as our primary trigger. Consequently, the high
threshold for trigger T2 does not represent a problem. The T2 trigger was
utilized mostly for detection of particles whose energy deposit is below the E-plane
threshold, and which can not form the T1 trigger. Fig.~\ref{fig_TDCThresholdEdE} shows 
that these events have energy deposits 
$E_{\mathrm{Loss}}^{\mathrm{dE-plane}} \geq 10\,\mathrm{MeV}$, which is well 
above the threshold for the T2 trigger. This way, by combining triggers T1 and T2,
no protons or deuterons are lost.

\section{Estimation of the particle energy losses}
\label{sec:energy_losses}
\begin{figure}[!ht]
\begin{center}
\includegraphics[width=1\linewidth]{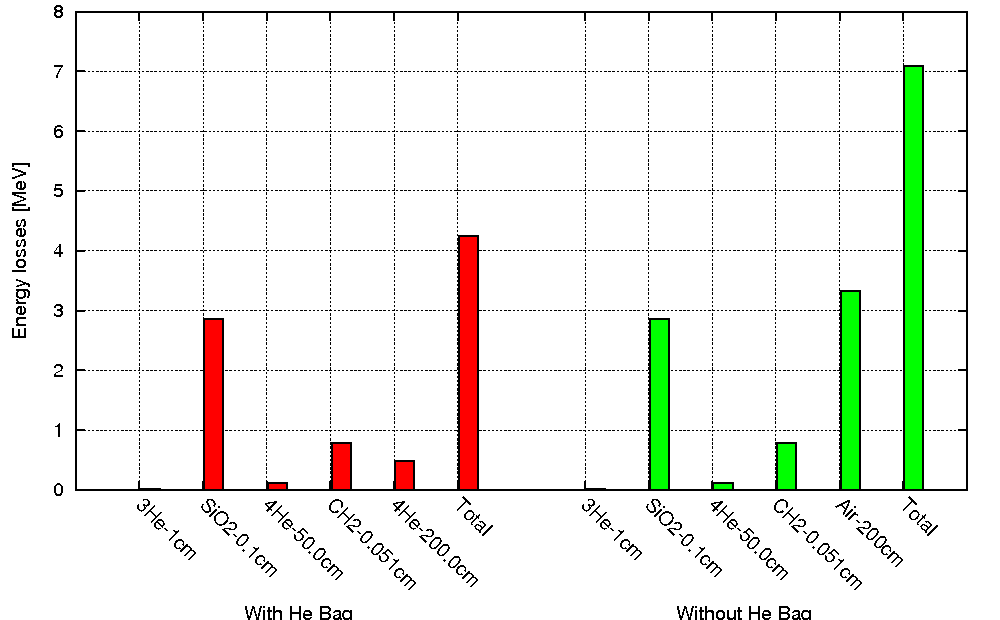}
\vspace*{-3mm}
\caption{ The predicted energy losses for deuterons with momentum $550\,\mathrm{MeV}/c$
(kinetic energy $\approx 80\,\mathrm{MeV}$) in  different materials from the target
to the detector package. The total energy losses are also shown. The total amount
of energy lost would be reduced for almost a factor of two if  helium bags were employed. 
\label{fig_EnLoss_HeBags}}
\end{center}
\end{figure}

Energy losses of particles in various materials were an important aspect of our 
experiment. First estimations of the losses needed to be performed already in the 
designing phase of the experiment to properly position the experimental equipment, 
so that the particles ejected from the target, could be detected. Very fast 
scattered electrons with energy $E_{\mathrm{e'}} \geq 1.2\,\mathrm{GeV}$ do not 
lose much energy on their path from the target to the detectors, especially
because there is vacuum inside the HRS-L spectrometer. The problematic ones are the
medium energy protons and deuterons, which may lose significant amounts of energy
on their way through the BigBite spectrometer. To reduce these losses, 
helium bags between the target and the detectors were fist foreseen. The estimated
corrections to the particle energy with and without helium bags are shown
in Fig.~\ref{fig_EnLoss_HeBags}. Later it was decided not to use them, because
their benefits were estimated to be smaller than the efforts required to install
and maintain the bags.

\begin{figure}[!hb]
\begin{center}
\includegraphics[width=0.49\textwidth]{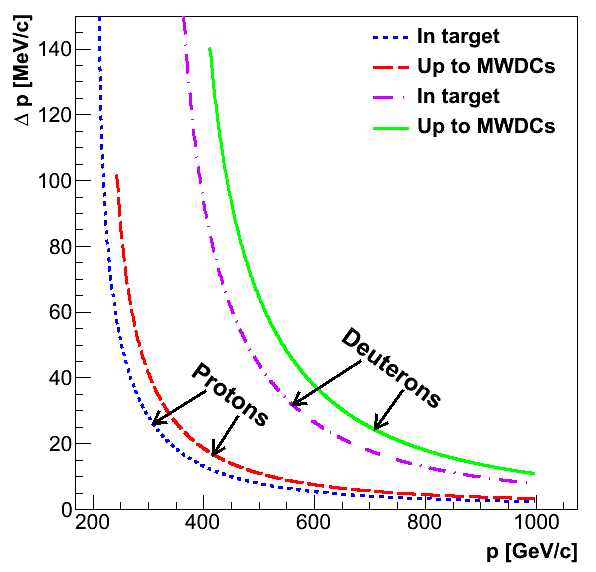}            
\includegraphics[width=0.49\textwidth]{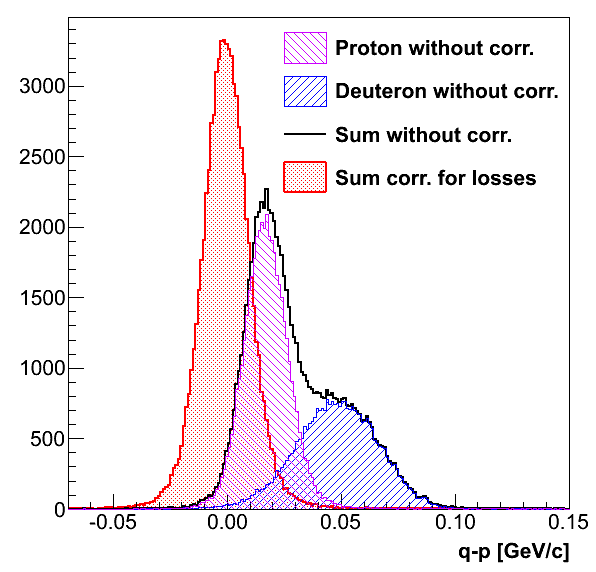}
\caption{[Left] Momentum losses of protons and deuterons
inside the target and the total momentum losses up to the MWDCs.
[Right] Quality of reconstructed momentum for elastic protons and deuterons. 
If energy losses are not taken into account, two peaks are visible (center
and right histograms summed to the full curve). With proper inclusion of 
energy losses both peaks merge into one (left histogram), resulting in
better momentum resolution.
\label{fig_EnLoss_Qvec}}
\end{center}
\end{figure}

\begin{figure}[!hb]
\begin{center}
\includegraphics[width=0.49\linewidth]{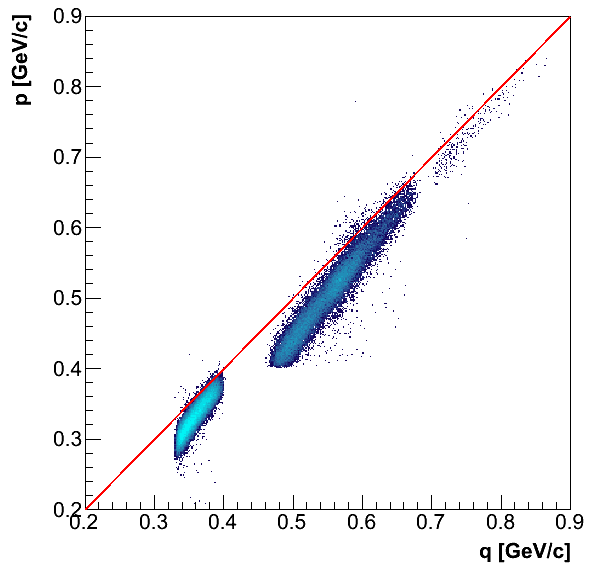}
\hfill
\includegraphics[width=0.49\linewidth]{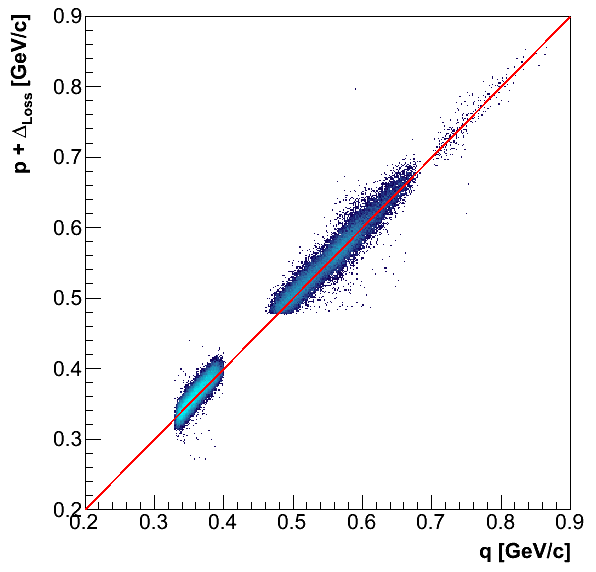}
\caption{ The comparison of the momentum transfer vector $|\vec q|$
to the reconstructed particle momentum $|\vec p|$ with (right) 
and without (left) consideration of energy-losses, for elastic scattering 
processes ${}^1\mathrm{H(e,e'p)}$ and ${}^2\mathrm{H(e,e'd)}$. In the 
analysis $\mathrm{1^\mathrm{st}}$-pass, $\mathrm{2^\mathrm{nd}}$-pass and
$\mathrm{3^\mathrm{rd}}$-pass data were utilized. In the absence of energy-losses,
the magnitudes of $\vec p$ and $\vec q$ should be identical. In the presence of 
energy-losses, particles traveling through various materials lose energy, causing
$|\vec p| \leq |\vec q|$.   
\label{fig_EnLoss_MomentumLosses}}
\end{center}
\end{figure}

A precise knowledge of the energy-losses has been very important also 
after the experiment for a proper analysis of the data, in particular
for the optical calibration of BigBite. 
The energy losses considered in the analysis were divided into 
two types. The first type are the transitional losses that occur 
during the transport of the particle from the target to the detector 
package and influence particle's momentum and direction. The particle travels 
through many layers of material. The amount of energy lost
in each section depends on the type of the material and its thickness.
Two largest contributors are the target glass and the air inside BigBite
(see Fig.~\ref{fig_EnLoss_Qvec}). The understanding of these losses is  
important for a correct reconstruction of the particle momentum 
and has a significant impact on the calculation of the derived kinematical
quantities, such as momentum, energy and mass of the undetected particle.
Figs.~\ref{fig_EnLoss_Qvec} and~\ref{fig_EnLoss_MomentumLosses} show 
the comparison of the magnitude of the electron momentum 
transfer vector $|\vec q|$ to the reconstructed momentum of 
the elastically scattered protons and deuterons. The agreement between
the two is significantly better when energy-losses are taken into account.

\begin{figure}[!ht]
\begin{center}
\begin{minipage}[t]{0.6\textwidth}
\hrule height 0pt
\includegraphics[width=1\textwidth]{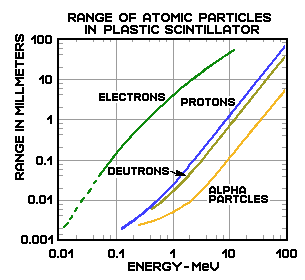}
\end{minipage}
\hfill
\begin{minipage}[t]{0.38\textwidth}
\hrule height 0pt
\caption{The intrinsic ranges of the particles in the plastic scintillator
EJ-204~\cite{gtech_EJ204, knoll}, that is considered for the dE- and E-plane detectors. 
In the scintillator deuterons loose their energy faster than protons. Consequently 
they require more energy to penetrate through a given thickness of material. 
In order to reach the other side of the dE/E scintillation detector 
with a designed thickness $3.3\,\mathrm{cm}$, protons and deuterons need momentum of 
$\approx 340\,\mathrm{MeV}/c$ and $\approx 576\,\mathrm{MeV}/c$, respectively.
\label{fig_EnLoss_EJ204}}
\end{minipage}
\end{center}
\end{figure}

\begin{figure}[!ht]
\begin{center}
\includegraphics[width=0.49\linewidth]{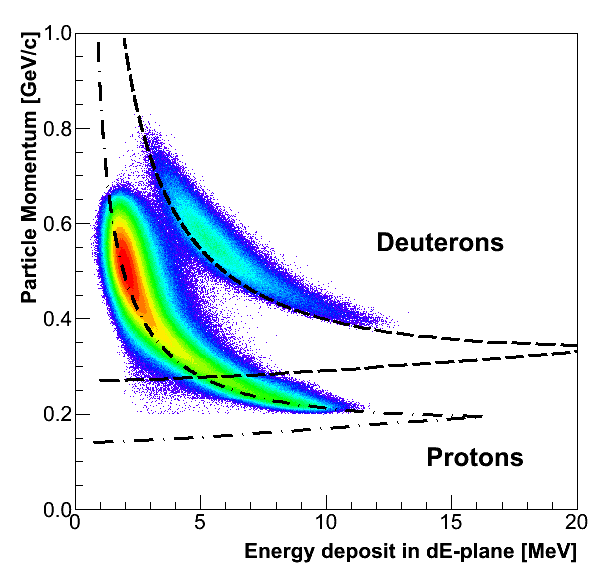}
\hfill
\includegraphics[width=0.49\linewidth]{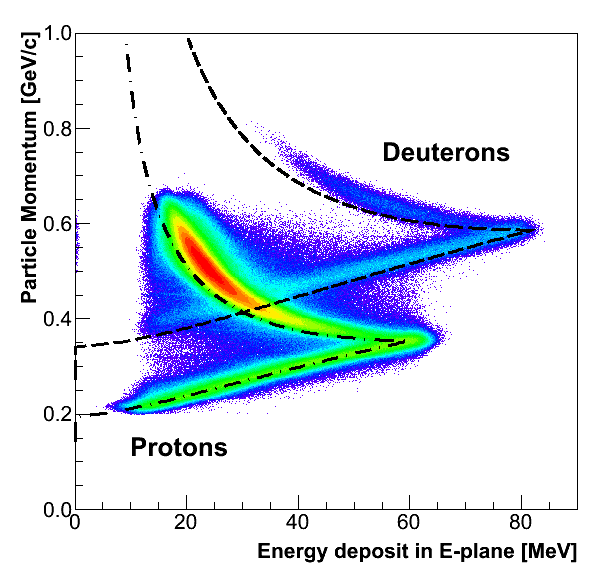}
\caption{The measured energy losses in the dE- and E- scintillation plane as 
a function of the particle momentum at the detector package. 
The measurements are compared to the
results of the Bethe-Bloch simulation. The dashed and dash-dotted lines 
represent simulated results for deuterons and protons, respectively. 
Initially, the amount of energy lost in 
each layer increases with the momentum, until particles manage to reach the other
end of the scintillation bar.  After that, the energy deposition inside the 
bars starts to decrease. For the E-plane, the proton and deuteron punch-through 
points are clearly visible. The punch-through points for the dE-plane are not observed,
since we are showing only particles that left measurable pulses in both scintillation
layers. 
\label{fig_EnLoss_EvsP}}
\end{center}
\end{figure}

The second type of losses are those that arise in the scintillation detectors,
and are utilized as part of the particle detection method. Hence, understanding 
of these losses is crucial for proper interpretation of the measured ADC 
spectra. Different types of particles lose different amounts of energy inside
the scintillators. The specific energy loss of deuterons is much larger than
the specific energy loss of protons.
Consequently, they stop much faster in the material and generate 
stronger ADC pulses (see Fig.~\ref{fig_EnLoss_EJ204}). The information 
about energy deposit inside these detectors can thus be considered for particle identification, 
as demonstrated in Figs.~\ref{fig_BBADC_EdE} and Fig.~\ref{fig_EnLoss_EvsP}.

The behavior of the observed spectra can be easily explained.  Slow hadrons are completely
stopped in the the first thin layer of the scintillation material and lose all their 
energy in it. By increasing the particle energy they lose more and more energy 
in the dE-plane, until they have enough energy to penetrate to the thicker E-layer
of the scintillator. This point is called the dE-plane punch-through point. To reach the
E-plane, the protons and deuterons require momenta of $\approx 185\,\mathrm{MeV}/c$ and 
$\approx 300\,\mathrm{MeV}/c$, respectively.  If particle 
momentum is further increased, the amount of energy deposited in the dE-plane starts 
to decrease, since the particle quickly passes through it, and then deposits all the 
remaining energy in the E-plane, where it stops. The amount of energy deposited in the E-plane 
again increases until the particle has enough energy to penetrate through both layers of scintillator 
material. This is known as the E-plane punch-through point. To reach the other
end of the scintillation detector, protons should have momentum $\approx 340\,\mathrm{MeV}/c$,
while deuterons need $\approx 576\,\mathrm{MeV}/c$. After that, the energy lost 
in both layers decreases quickly since the cross-section for the interaction 
of the particle with the material decreases rapidly with the particle momentum.

For the numerical description of the collisional energy losses of  heavy ions, 
the Bethe-Bloch formula was considered~\cite{leo}:
\begin{eqnarray}
  \frac{dE}{dx} = - R \rho \frac{Z}{A}\frac{z^2}{\beta^2}
    \left[ \ln\left( \frac{2m_e \gamma^2 v^2 W_{\mathrm{max}}}{I^2}\right) - 2\beta^2 - 
      \delta -2\frac{C}{Z} \right]\,, \label{eq_BetheBloch}
\end{eqnarray}
where $R = 0.1513\,\mathrm{\frac{MeV cm^2}{g}}$. The $\rho$, $Z$ and $A$ are 
the density, atomic number and atomic weight of the absorbing material, respectively,  
$m_e$ is the electron mass, $I$ is 
the mean excitation  potential of the material, $v$ represents the velocity of the incident particle, and 
$\beta = v/c$, $\gamma = 1/\sqrt{(1 - \beta^2)}$, while $W_{\mathrm{max}}$ represents 
the maximum energy transfer in a single collision 
of the incident particle with the electrons in the material. For our 
case  it can be approximated as $W_{\mathrm{max}} = 2m_e c^2 \beta^2 \gamma^2$. The quantities 
$C$ and $\delta$ represent the shell and density corrections to the Bethe-Bloch formula and are 
described in more detail in ref.~\cite{leo}. 

\begin{table}[!ht]
 \begin{center}
\caption{The relevant contributors to the energy losses of the hadrons
inside the target enclosure and BigBite in the sequence as experienced by the particle.
The materials gathered in this table 
were considered in the energy-loss simulation, which is utilizing Eq.~(\ref{eq_BetheBloch}). 
The parameters required in the Bethe-Bloch equation are taken from Ref.~\cite{sternheimer}.
The simulation considers the design thicknesses of all materials except of the air. The thickness
of the layer of the air was determined from the comparison of the simulation with the 
measured elastic proton and deuteron data. \vspace*{1.5mm}
\label{table_EnLoss_materials}}
\begin{tabular}{ccccc}
\toprule
Material & Thickness $\mathrm{[cm]}$ & $Z/A$  & $I~\mathrm{[eV]}$ & $\rho~\mathrm{[g/cm^3]}$ \\
\midrule
$\mathrm{SiO_2}$ & 0.17 & 0.499 & 139.2  &2.320 \\
Air & 83.0 & 0.499 & 85.7 & 1.205$\cdot 10^{-3}$ \\
$\mathrm{SiO_2}$ & 0.07812 & 0.499 & 139.2  &2.320 \\
Air & 200.0 & 0.499 & 85.7 & 1.205$\cdot 10^{-3}$ \\
Kapton & 0.003 & 0.513 & 79.6 & 1.42 \\
Mylar & 0.00712 & 0.521 & 78.7 & 1.4 \\
Kapton & 0.006 & 0.513 & 79.6 & 1.42 \\
Mylar & 0.00712 & 0.521 & 78.7 & 1.4 \\
Kapton & 0.003 & 0.513 & 79.6 & 1.42 \\
Vinyl Toluene & 0.3 & 0.542 & 64.7  &1.032 \\
Vinyl Toluene & 3.0 & 0.542 & 64.7  &1.032 \\
\bottomrule
 \end{tabular}
 \end{center}
\end{table}

\begin{figure}[!ht]
\begin{center}
\includegraphics[width=0.49\linewidth]{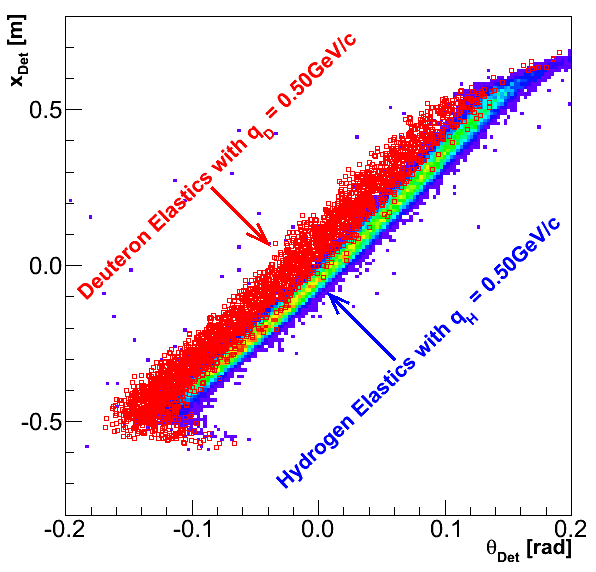}
\hfill
\includegraphics[width=0.49\linewidth]{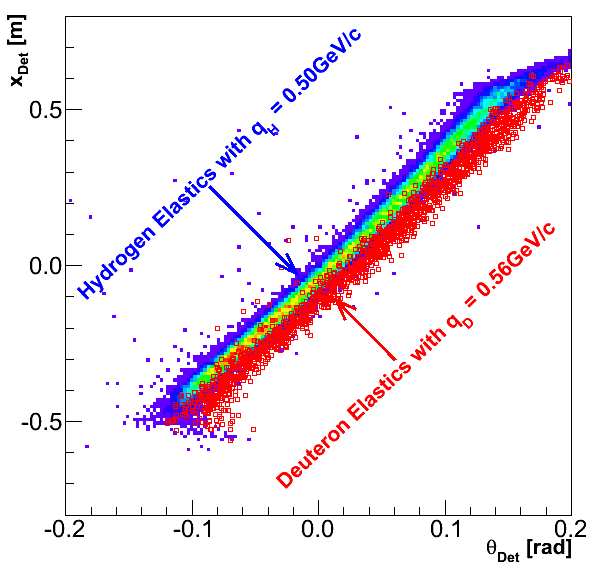}
\includegraphics[width=0.49\linewidth]{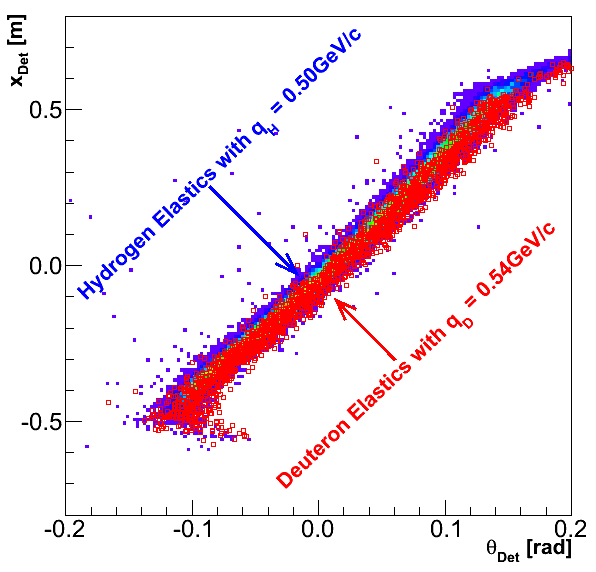}
\hfill
\includegraphics[width=0.49\linewidth]{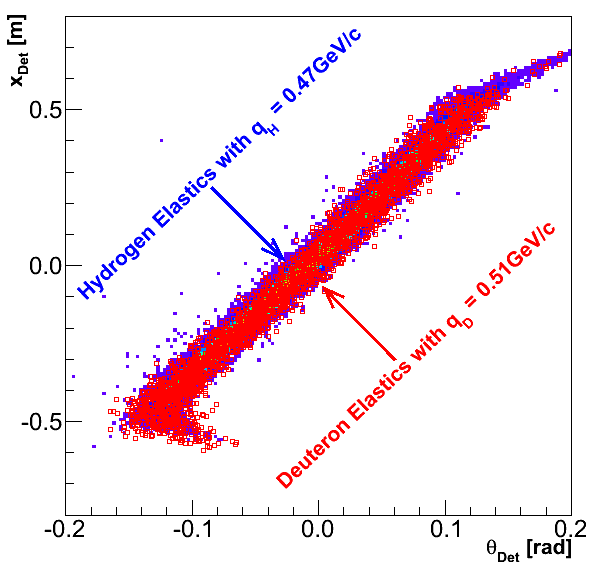}
\caption{The two-dimensional distributions ($x_{\mathrm{Det}}$ vs. $\theta_{\mathrm{Det}}$) of 
hits in the BigBite detector package, generated by elastic protons and deuterons
at different momentum transfers. Protons and deuterons with the
same $|\vec q|$ do not generate the same track patterns in the BigBite detector 
package due to the different energy-losses sustained inside the target and BigBite. 
Since deuterons lose their energy faster (due to their larger mass)
they need to have a larger momentum than protons in order to generate identical patterns
in the MWDCs. The required difference in the momenta is a measure of the difference
in their energy-losses.
\label{fig_EnLoss_DHComparison}}
\end{center}
\end{figure}

The materials considered in the simulation of the energy-losses are listed in 
Table~\ref{table_EnLoss_materials}, together with their fundamental properties and
design thicknesses. Incident particles may lose significant amount of their 
energies in these materials, especially inside the scintillators, where they can be stopped completely. 
Hence, the differential
approximation $\Delta E = - (dE/dx)\Delta x$ can no longer be utilized, hence the 
Bethe-Bloch formula needs to be integrated over the length of the material.
For the integration, simple trapezoidal
integration  was employed. Beside the the total amount
of energy lost by a hadron on its way from the target to the detectors, the simulation also
returns the losses in each layer separately. In particular, we are interested in the
description of the behavior of the energy-losses inside the two scintillation detectors. The 
comparison of the simulation to the measured energy deposits is shown in Figs.~\ref{fig_BBADC_EdE} 
and Fig.~\ref{fig_EnLoss_EvsP}. The simulation agrees well with the measurements.
In addition, the energy losses calculated for each material
were also found to agree well with the results from Ref.~\cite{ElossPloss}.

The losses obtained with the  simulation described above are not exact, mostly 
because some of the parameters are not well understood. The most uncertain  
was the thickness of the air, which was impossible 
to measure directly. It had to be obtained indirectly, by varying (within reasonable 
limits) its predicted thickness and comparing the simulation 
results to the true energy losses of protons 
and deuterons. For that purpose protons and deuterons 
from elastic processes ${}^1\mathrm{H(e,e'p)}$ and ${}^2\mathrm{H(e,e'd)}$ were utilized. 
By fixing the momentum transfer vector $\vec q$, the momentum of ejected hadrons in both
processes is also determined. In the absence of energy losses, protons and deuterons 
with the same momentum should create identical distributions of track-patterns in the MWDCs. 
However, in the presence of energy-losses, deuterons lose their energy faster 
than protons, resulting in different MWDC track-patterns, although both particles had identical
momenta at the target. 
Therefore, proton and deuteron hit-patterns for various momenta were compared, in order to determine 
those momenta for which MWDC track distributions for both particle types match. Examples of 
the conducted search are shown in Fig.~\ref{fig_EnLoss_DHComparison}. Once knowing the required 
momentum difference, the parameters of the Bethe-Bloch simulation (e.g. the thickness of the air layer) 
could be modified accordingly in order to reproduce the result.

\section{Insight into the Trigger Operation}
\label{sec:TriggerOperation}

The triggering system is the crucial part of the experiment. Its improper 
operation could have devastating consequences, since all detectors 
are read only when a valid triggering pulse arrives. Therefore, checks
need to be performed to test the trigger behavior. We need to make sure 
that coincidence events are really detected and that they can be separated 
from the random background. Understanding the trigger performance
is also important for particle identification. The relative time 
difference between electron (T3) and hadron triggers (T1, T2)  depends on
the particle type and its momentum. By inspecting the recorded trigger times
as a function of particle momenta, coincident protons can 
be well separated from the coincident deuterons.

\begin{figure}[!ht]
\begin{center}
\includegraphics[width=0.49\linewidth]{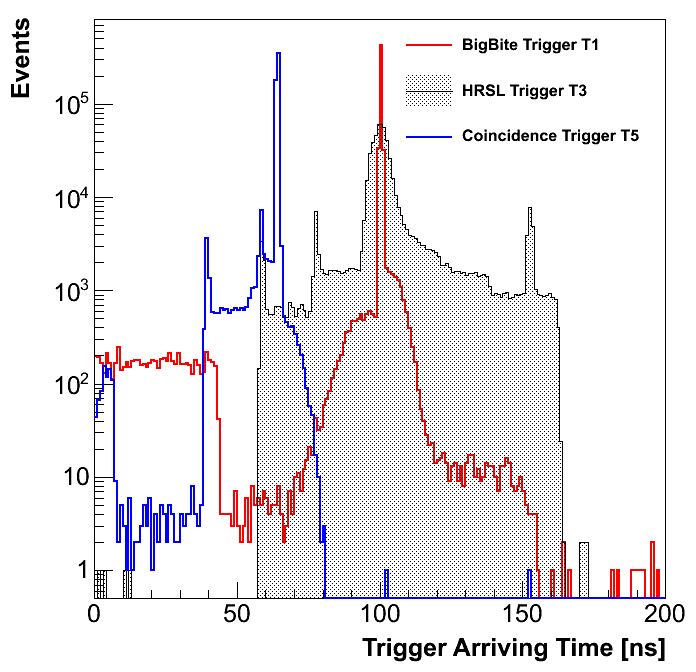}
\includegraphics[width=0.49\linewidth]{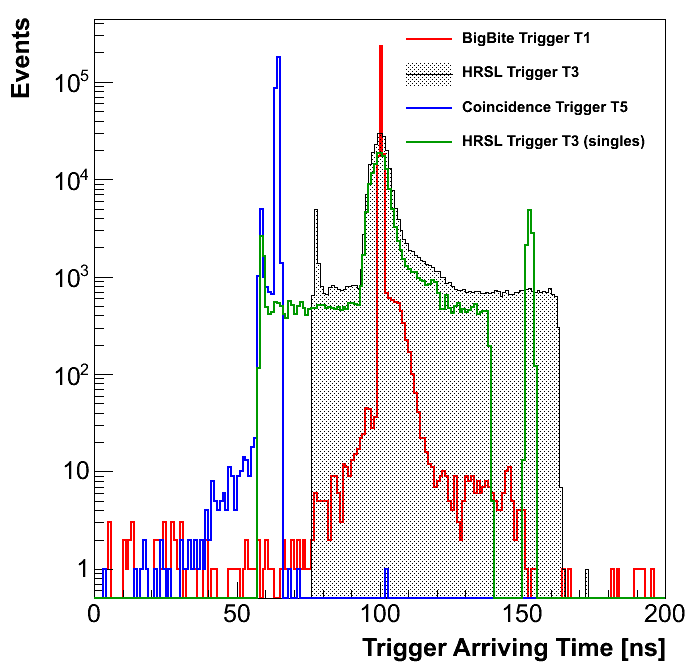}
\caption{ [Left] Recorded raw TDC information on the three most important
triggers in the experiment (T1, T3, T5). The arriving time is measured relative
to the BigBite re-timing pulse (see Fig.~\ref{fig_BBRetiming}) at 
time zero. Hence, greater arriving times mean earlier arrivals of the triggers.  
By design, BigBite T1 trigger should always come simultaneously with 
the BigBite re-timing pulse, while the T3 trigger should have the usual shape 
of a coincidence peak on a flat background. Due to the properties of the coincidence
trigger circuit, the T5 trigger should also be sharp. However, the structure of 
the obtained trigger information is far more complex due to 
the finite widths of the trigger pulses and 
un-compensated delays in the triggering circuit. 
[Right] The trigger TDC diagrams become much cleaner when only events that 
were accepted by the TS as coincidence events (T5s) are shown (gray histogram). 
With this limitation the second plateau in the T3 background disappears. 
The structure of triggers T1 and T5 also becomes clearer. 
The second step in the T3 background is caused
by the coincidence events that were accepted by the TS as T3 events (green
line). Since the T3 trigger comes to the TS earlier than T5, the background moves,
and becomes more varied when both types of events are joined.
\label{fig_Trigger_TDCs}}
\end{center}
\end{figure}

First tests of trigger operation were performed already during the construction of 
the triggering circuit, by using simulated EDTM pulses. A summary of that analysis 
is presented in Appendix~\ref{appendix:EDTM}. The analysis with production events 
is presented in this Section. During the experiment the trigger information was 
recorded via TDC modules and using this information, the performance of the triggering 
mechanism could be continuously monitored. By design, the TDC spectrum of the T1 trigger 
should have only one sharp peak, because it is self-timed, while the T3 trigger
should correspond to raw coincidence time, having a wide uniform background with
the coincidence peak in the middle.

\begin{figure}[!ht]
\begin{center}
\begin{minipage}[t]{0.49\textwidth}
\hrule height 0pt
\includegraphics[width=1\textwidth]{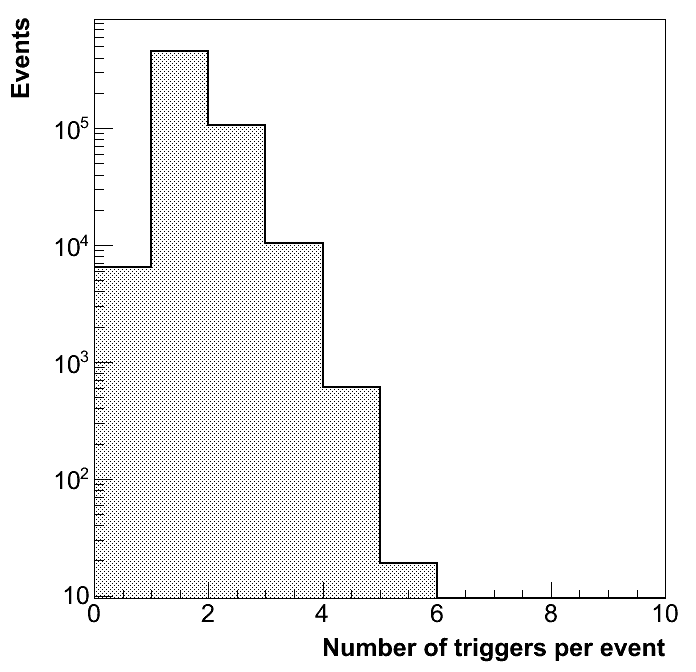}
\end{minipage}
\hfill
\begin{minipage}[t]{0.45\textwidth}
\hrule height 0pt
\caption{ The distribution of the number of T1 triggers per event, recorded by the 
TDC module. Only coincidence events are shown. For the  majority of the events 
trigger T1 is formed only once. Events without T1 trigger correspond to low momentum 
particles that are stopped in the dE-plane and are able to form only trigger T2. 
\label{fig_Trigger_T1number}}
\end{minipage}
\end{center}
\end{figure}

\begin{figure}[!ht]
\begin{center}
\includegraphics[width=0.49\linewidth]{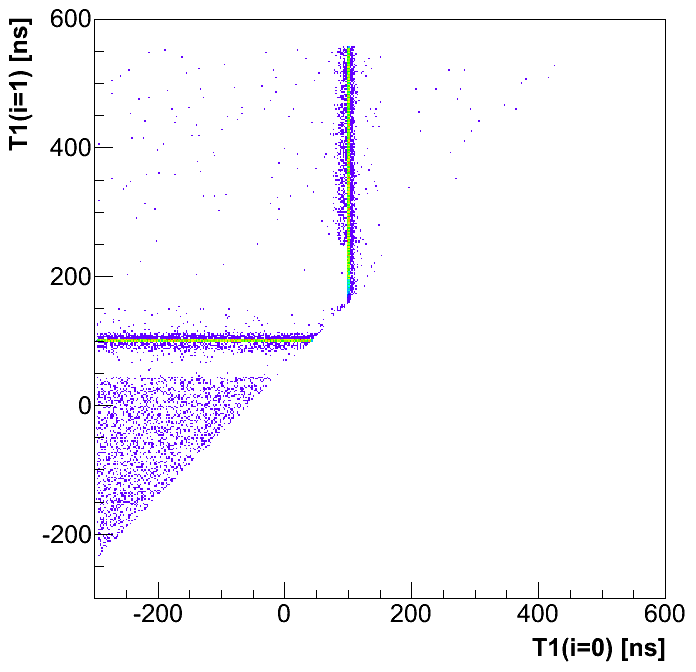}
\includegraphics[width=0.49\linewidth]{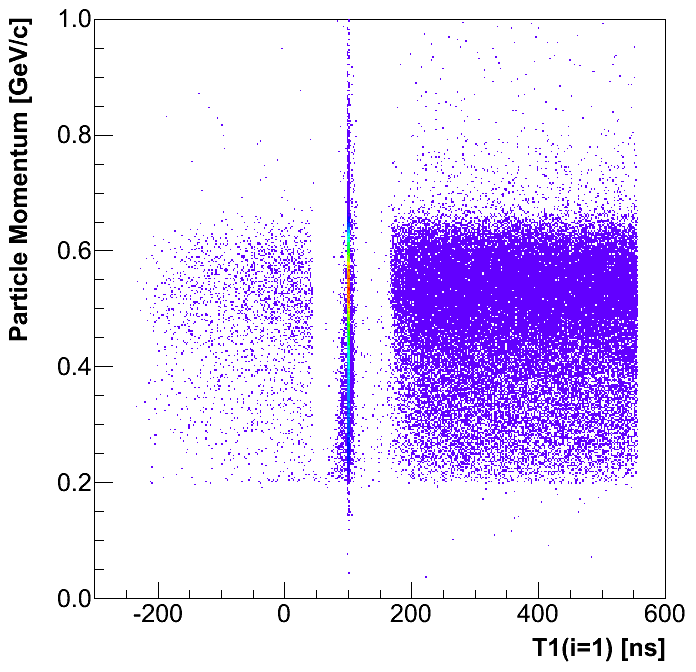}
\caption{[Left] The trigger timing information were recorded by the 
multi-hit TDC modules. Consequently more than one T1 trigger (or any other) 
can be recorded per event. However, only one of them is considered by the TS
to start the readout of the detectors. Which of the consecutive pulses is used depends
on each particular event. The chosen one can be clearly identified because it always 
comes at the same (relative) time. The rest of the triggers are smeared around it.  
The trigger arriving last has index $i=0$ while the first one has index $i=n-1$.  
Since $T1(i=1)$ always comes before $T1(i=0)$ all the points are gathered in the upper
corner of the plot. For instance, when $T1(i=1)$ is chosen to start the DAQ, $T1(i=0)$
is smeared, and vice-versa. The random points in the region $T1(i=1) < 50\,\mathrm{ns}$ 
correspond to the events where trigger $T1(i\geq2)$ was used to start the DAQ.
[Right] The recorded trigger $T1(i=1)$ as a function of the particle momentum. The sharp 
peak at the center corresponds to the events when  $T1(i=1)$ is selected to start the DAQ readout.
Due to self timing the T1 peak is independent the momentum. In the case that $T1(i=0)$ 
is selected to start the DAQ,  $T1(i=1)$ is spread and located at earlier times,
$T1(i=1) > 200\,\mathrm{ns}$. However, when the pulse $T1(i=2)$ is selected for triggering, 
$T1(i=1)$ comes later in time, $T1(i=1) < -50\,\mathrm{ns}$. The empty region around the
main peak is a consequence of the finite width of the T1 trigger pulses. Until one 
of the pulses is set, the next one can not be processed. 
\label{fig_Trigger_T1hits}}
\end{center}
\end{figure}

However, by looking at the measured TDC spectra shown in Fig.~\ref{fig_Trigger_TDCs}, 
one quickly realizes that the true anatomy of the trigger is far more complicated 
than we expected. 
The first complication arises from the fact that to record of the BigBite
T1 triggers, multi-hit TDC modules were considered that allow more than one trigger
hit per event to be recorded. This blurs our image and  
makes the interpretation of the TDC spectra difficult. According to 
Fig.~\ref{fig_Trigger_T1number} up to $6$ triggers were recorded. 
Since BigBite has a very large acceptance, 
high trigger rates have been anticipated. However, the trigger pulses in the 
electronics have a finite width and while one trigger is set  
to enabled, others are not accepted. This is clearly demonstrated in 
Fig.~\ref{fig_Trigger_T1hits}. Only after the previous pulse reverts  to 
zero, another trigger can be formed and recorded to TDCs. The number 
of recorded triggers per event therefore depends primarily on the  width of the TDC window
and the width of the trigger pulse.

In the TDC module, the recorded triggers are 
stored into an indexed stack memory. Although there is more than one trigger
recorded in the TDCs, only one of them is the right one, 
i.e. the one that  triggered the readout of
the electronics, and can be distinguished from the rest. It is always positioned 
at the same time $(t_1 \approx 100\,\mathrm{ns})$, regardless of its index in 
the stack. This is clearly visible in Fig.~\ref{fig_Trigger_T1hits}.
To get rid of the trigger ambiguity, we limited our trigger analysis only to the 
events with one recorded $T1$ trigger per event. With this limitation, the main 
T1 peak was left intact, while the uniform distribution in Fig.~\ref{fig_Trigger_TDCs} 
at $t_{\mathrm{T1}}\leq 45\,\mathrm{ns}$ disappeared.

\vspace*{-5mm}
A further complication which obscures the interpretation of the trigger TDC spectra is
related to the  operation of the trigger supervisor (TS). The TS prescale factors were 
set to accept only coincidence events (triggers T5 and T6) and HRS-L singles (trigger T3). 
All BigBite single events were rejected. We are mostly interested in the coincidence events, 
while the HRS-L single events were taken for  inclusive measurements. The true
left arm singles all accumulate in the most right peak (at $\approx 150\,\mathrm{ns}$) in the 
T3 TDC spectrum shown in Fig.~\ref{fig_Trigger_TDCs}. For these events there are no BigBite triggers,
and the trigger circuit needs to wait for the delayed L1A pulse (see Fig.~\ref{fig_BBRetiming})
before accepting them. Consequently, the events are self-timed, resulting in a sharp
peak, and due to the $90\,\mathrm{ns}$ delay, the peak is pushed to the far right side of the 
spectrum. 

Besides real HRS-L single events, coincidences may also be flagged and recorded as 
single events (T3 events). This happens, for instance, when the trigger supervisor,
according to the set prescale factors, decides to accept the coincidence event with a
single trigger (T3) instead of the T5 trigger. This is evidenced in Fig.~\ref{fig_Trigger_TDCs}. For these events, 
the structure of the T3 TDC spectrum (green line) is slightly different than the one for the primary 
coincidence events (gray histogram). 
This is a consequence of a trigger design flaw because of which the T5 trigger
arrives $\approx 30\,\mathrm{ns}$ after T1 and T3. This shortcoming was observed already during the EDTM test
(see Appendix~\ref{appendix:EDTM}, Fig.~\ref{fig_EDTM_TriggerT5}) performed before the experiment. 
Due to this delay, the events where the TS accepts the T3 trigger require less time to form the L1A pulse. HRS-L events 
that participate in the formation of the coincidence trigger can consequently come relatively late to
the trigger circuit and still be able to form coincidences with T1. However, the shorter T3 time 
delay affects only the acceptable range of the random background, which is shifted by
$\approx 30\,\mathrm{ns}$ to the left.

The position of the coincidence peak is predetermined by the underlying physics processes and 
is always relative to the T1. Therefore, coincidence peak stays at the same position, regardless 
of the delay issue.  By combining T3 TDC distributions for both  cases 
(green - T3 and gray - T5) shown in Fig.~\ref{fig_Trigger_TDCs} (right), 
the initial T3 TDC distribution shown in Fig.~\ref{fig_Trigger_TDCs} (left) is obtained.

By limiting our discussion only to the T5 coincidence events 
and neglecting (only for the purpose of this analysis) the T3 coincidence
events,  much cleaner TDC spectra are obtained (see Fig.~\ref{fig_Trigger_TDCs}). 
The only remaining issues
are the sharp peak near the edge of the T3 TDC spectrum at 
$t_{\mathrm{T3}} \approx 80\,\mathrm{ns}$, and the corresponding shoulder 
on the right side of the primary peak in the T1 TDC spectrum. These are 
random coincidence events caused by imperfections in the trigger electronics.
This happens when a very late T3 trigger manages to catch an 
enabled T1 pulse from a previous event, and forms a random coincidence.
Because the T3 trigger comes last, it fixes the time axis and is therefore
sharp, while the T1 trigger is smeared. 
The matching T1 triggers came earlier than usual (with respect to the 
BigBite re-timing pulse), hence all the events are gathered only on 
the right side of the main peak. 

\begin{figure}[!ht]
\begin{center}
\includegraphics[width=0.49\linewidth]{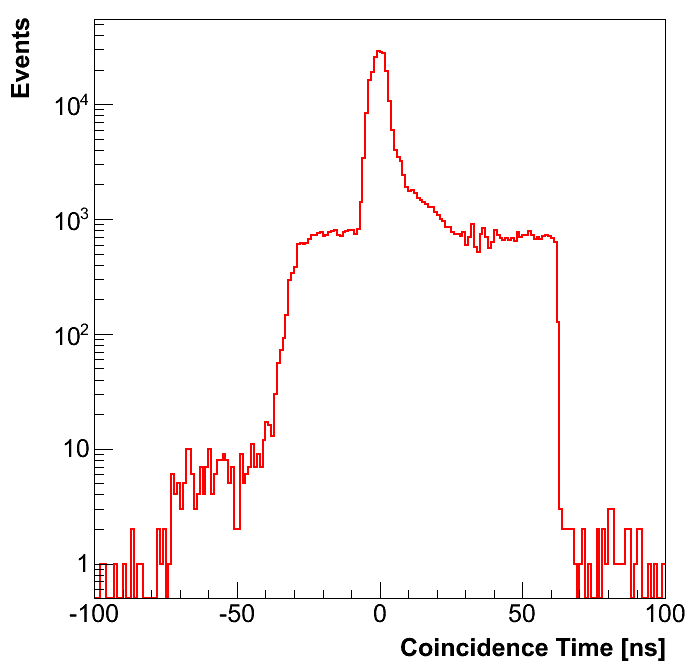}
\includegraphics[width=0.49\linewidth]{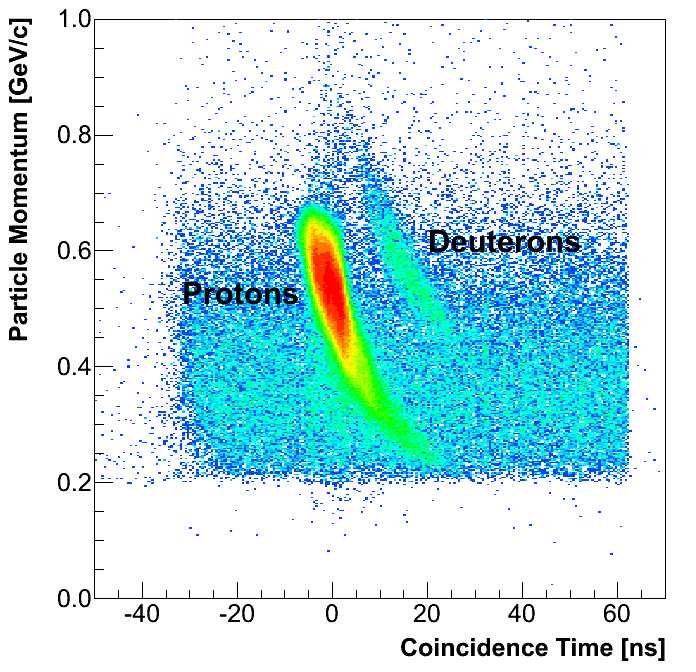}
\caption{ [Left] The coincidence time spectrum obtained as a difference between
the T3 and T1 triggers. In contrast to Fig.~\ref{fig_Trigger_TDCs}, 
a clean coincidence peak on top of a uniform background is obtained, without 
any undesirable sharp peaks near the edges. The main part of the peak corresponds
to the coincidence  protons while the shoulder on its right side contains 
coincidence deuterons. [Right] The coincidence deuterons can be
distinguished from the coincidence protons in the two-dimensional histograms 
of the particle momentum versus coincidence time. This clear separation 
can be exploited for the PID. Problems appear only in the low momentum region,
where the faint deuteron peak mixes with the random coincidence background. 
\label{fig_Trigger_coincidence}}
\end{center}
\end{figure}

In the electronics, this problem could be solved by introducing a veto signal to 
the module, forming the coincidence trigger (see Fig.~\ref{fig_BBCoincTrigger}),
in order to disable all input signals once the T1 trigger has arrived. This would
cut away all problematic events. However, the problem can also be resolved
in the off-line analysis. Instead of using the T3 trigger for measuring the raw coincidence,
a difference $t_{\mathrm{Coinc}} = t_{\mathrm{T3}} - t_{\mathrm{T1}}$ is formed.
The coincidence time $t_{\mathrm{Coinc}}$ properly considers the relative timing
issue between triggers T1 and T3, resulting in a perfect coincidence timing spectrum,
with uniform background and a nice coincidence peak. A typical coincidence timing 
spectrum is demonstrated in Fig.~\ref{fig_Trigger_coincidence}~(left). The majority 
of the events in the peak correspond to the coincidence protons from the 
$\mathrm{{}^3He(e,e' p)}$ reaction. The broadening of the peak to the right side
is explained by the momentum dependence of the coincidence time. While  
all electrons travel almost at the speed-of-light and require the same 
amount of time to come from the target to the HRS-L detector package, the protons of 
different momenta need different times to reach BigBite detectors. Low momentum 
particles travel longer than more energetic ones, resulting in the broadening 
of the coincidence peak. This phenomenon is clearly shown in 
Fig.~\ref{fig_Trigger_coincidence}~(right).

An additional small shoulder on the right
side of the peak (at $t_{\mathrm{Coinc}}\approx 20\,\mathrm{ns}$) corresponds
to the deuterons from the process $\mathrm{{}^3He(e,e' d)}$. The deuteron part of the peak
also has a very strong momentum dependence. Because of momentum broadening, the one-dimensional 
coincidence time histogram can not be used  to  distinguish deuterons from the 
dominant proton peak. The separation is much clearer when looking at the two-dimensional
histograms where the coincidence time is shown as a function of hadron momentum. 
This kind of histograms can be considered as an alternative to the 
PID technique based on the  ADC spectra. To get trustworthy results, 
the random background also needs to be properly subtracted.

\chapter{Magnetic Optics of Spectrometers}
\label{chapter:Optics}

This chapter is dedicated to the investigation of optical properties of the two
magnetic spectrometers, HRS-L and BigBite, that were employed in the  E05-102 experiment.
The content will be focused on the optical calibration of the BigBite, which was performed 
by the author. However, the results of the optical analysis for the 
HRS-L spectrometer that was contributed by Jin Ge~\cite{GeJinPhD}, will also be 
presented. 

\section{Overview}
The purpose of optical calibration is to establish the mapping
between the detector variables that are measured directly,
and the target variables corresponding to the actual physical 
quantities describing the particle at the reaction vertex.
In the detectors (VDCs or MWDCs), two position coordinates ($x_{\mathrm{Det}}$
and $y_{\mathrm{Det}}$) and two angles ($\theta_{\mathrm{Det}}$
and $\phi_{\mathrm{Det}}$) are measured.  From this information,
we wish to reconstruct the location of the interaction vertex 
($y_{\mathrm{Tg}}$), the in-plane and out-of-plane scattering
angles ($\phi_{\mathrm{Tg}}$ and $\theta_{\mathrm{Tg}}$),
and the particle momentum relative to the central momentum
($\delta_{\mathrm{Tg}}=(p_\mathrm{Tg}-p_\mathrm{c})/p_\mathrm{c}$).

\section{Optical Calibration of BigBite}

The transformation from the detector coordinates to the target
variables can be done in many ways. We have considered an analytical model
as well as a more sophisticated approach based on transport-matrix
formalism, with several means to estimate the reliability
of the results and the stability of the algorithms~\cite{miha_menu2010, miha_NIM}.

\begin{figure}[!ht]
\begin{center}
\includegraphics[height=0.62\linewidth]{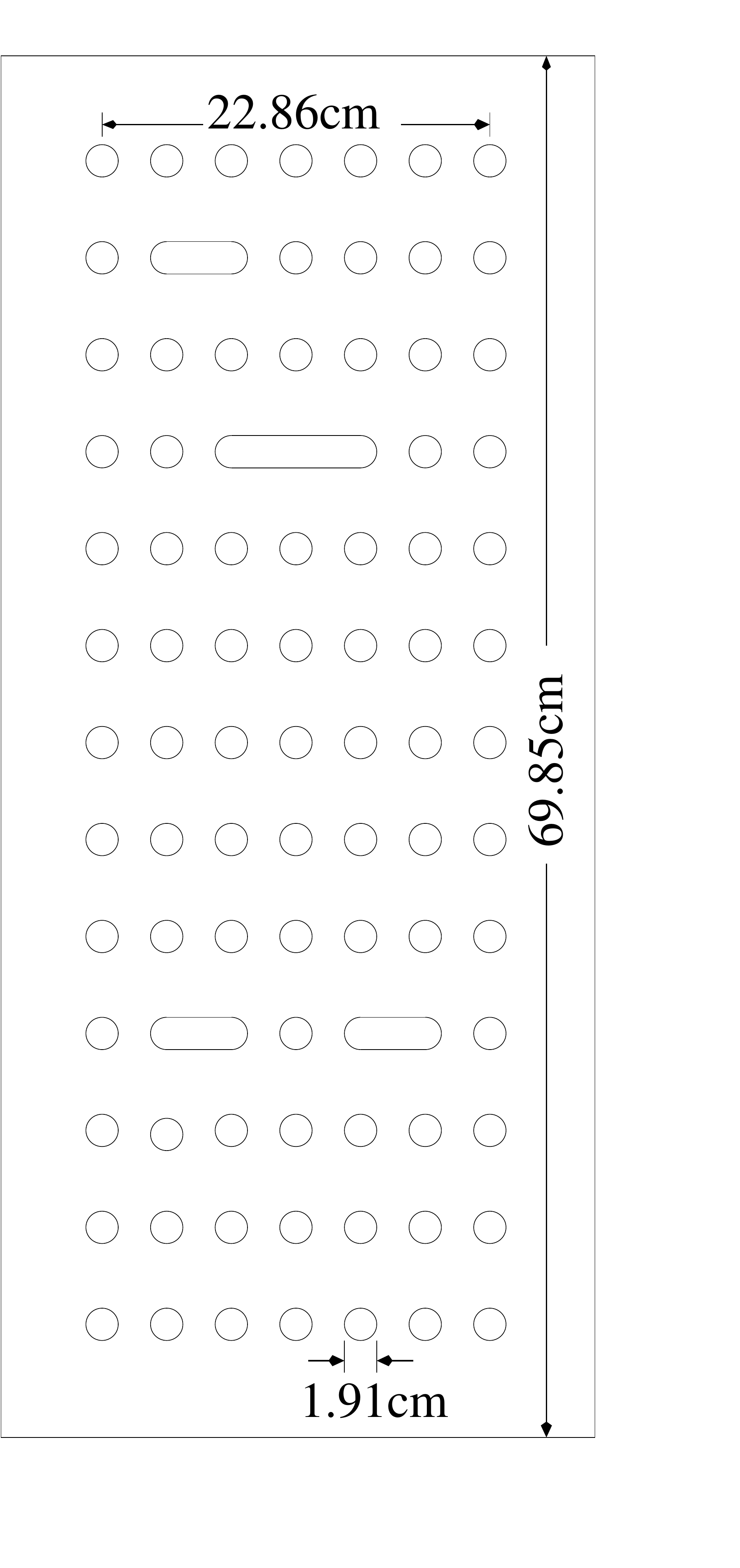}
\includegraphics[height=0.62\linewidth]{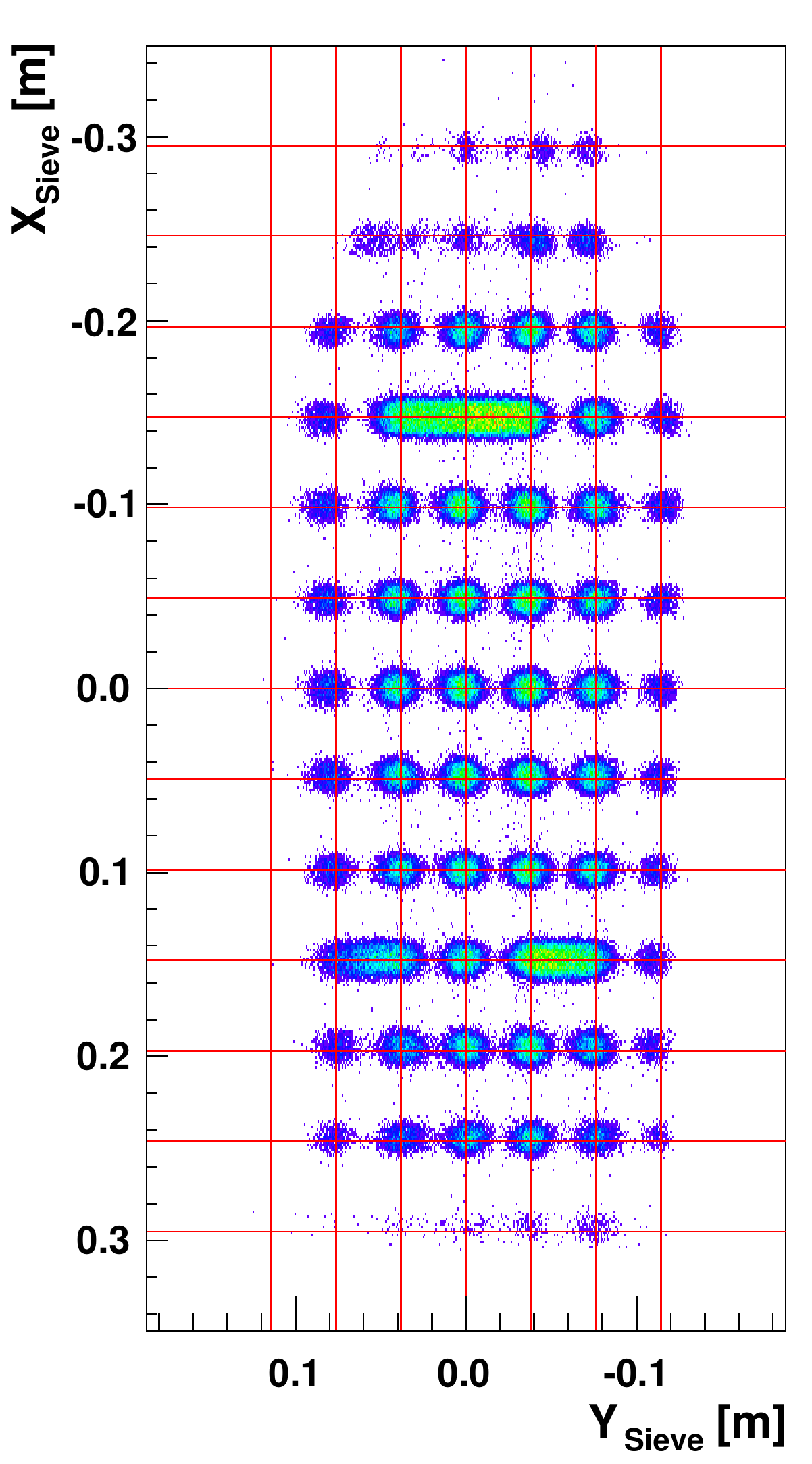}
\includegraphics[height=0.62\linewidth]{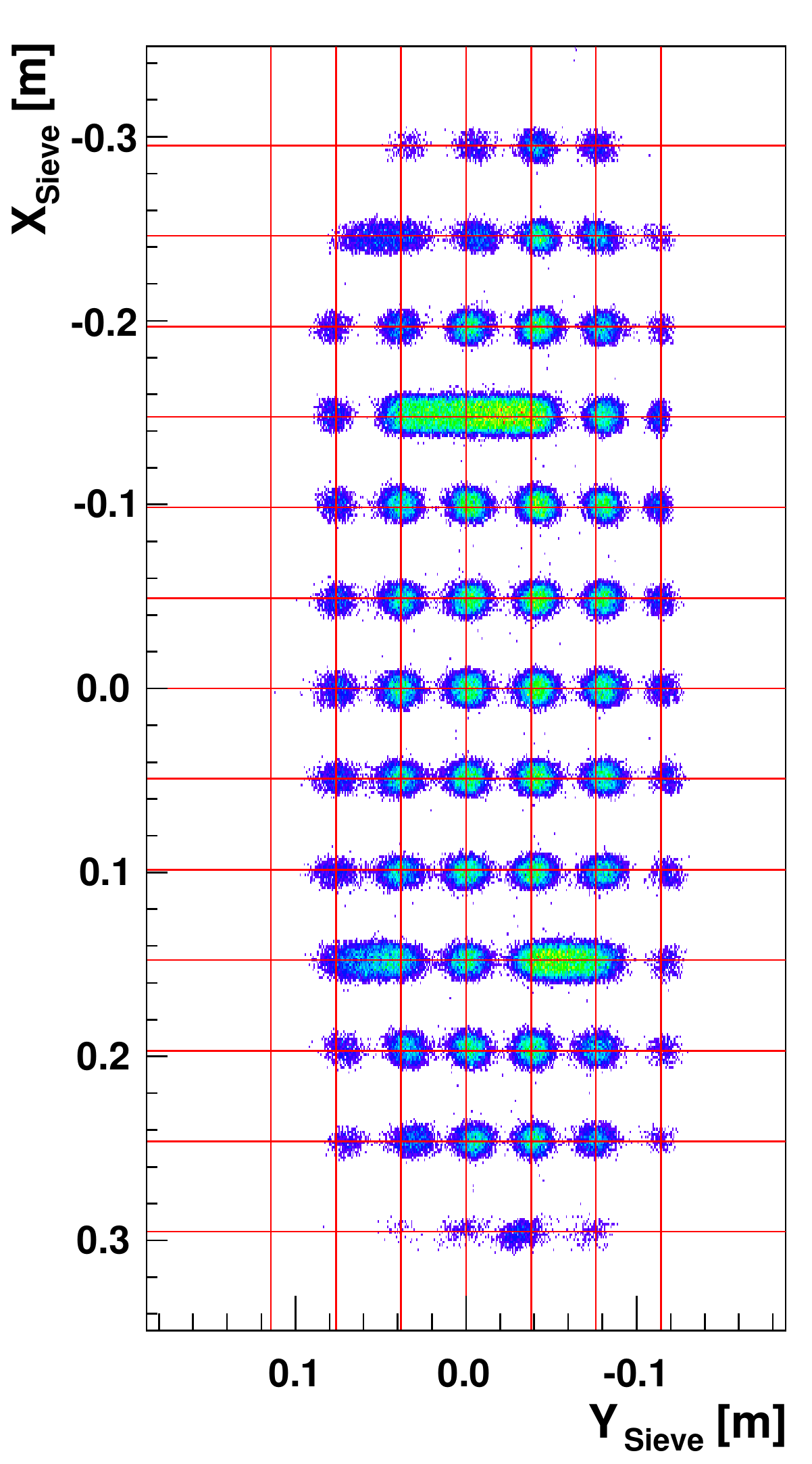}
\vspace*{-3mm}
\caption{[Left] Schematics of the BigBite sieve-slit collimator.
[Center, Right] Sieve pattern reconstruction by using the simplex method
and the SVD, respectively (see section~\ref{sub:nmsvd}).
The SVD technique resolves more holes and yields a much clearer pattern.
The holes at the left edge are missing due to geometrical obstacles
between the target and BigBite.\label{fig_optics_BBSieve}}
\vspace*{-1.8cm}
\end{center}
\end{figure}

For the purpose of optics calibration of BigBite a series of dedicated data 
sets were collected during the experiment using the seven-foil carbon target
and the reference target filled with various gases (see Sec.~\ref{sec:ref_targets}). 
In addition, a special set of measurements was performed with a $4\,\mathrm{cm}$-thick
lead sieve-slit collimator positioned at the entrance
to the BigBite magnet (see Fig.~\ref{BBSpectrometer}).
The sieve-slit collimator has $82$ circular holes that are
almost uniformly positioned over the whole acceptance
of the spectrometer, Fig.~\ref{fig_optics_BBSieve} (left).
The collimator also contains four elongated holes used
to remove ambiguities in horizontal and vertical orientations
and to allow for easier identification of the hole projections
at the detector package. 

\vspace*{5mm}
Prior to any optics analysis, a series of cuts were applied 
to the collected calibration data to eliminate the  background.
A HRS-BigBite coincidence trigger system was used to acquire
electron-proton and electron-deuteron coincidences,
at typical rates between $700\,\mathrm{Hz}$ and $1\,\mathrm{kHz}$.
True coincidences were selected by applying a cut on the raw
coincidence time.  The background was further reduced
by PID and HRS acceptance cuts.  Finally, only those events
that produce consistent hits in all BigBite detectors,
and could consequently be joined to form single particle tracks,
were selected.

Quasi-elastic protons from scattering on the seven-foil carbon 
target were used to calibrate $y_\mathrm{Tg}$; the same target
was also used to calibrate $\theta_\mathrm{Tg}$ and $\phi_\mathrm{Tg}$
when the sieve-slit collimator was in place.  In turn, elastic protons
and deuterons (from hydrogen and deuterium targets) were used
to calibrate $\theta_\mathrm{Tg}$, $\phi_\mathrm{Tg}$, and
$\delta_{\mathrm{Tg}}$. The $\delta_{\mathrm{Tg}}$ matrix elements 
could also be determined
by quasi-elastic events from $^3\mathrm{He}$ under the assumption
that the energy losses are well understood.

\subsection{The matrix formalism}
\label{sec:opticsmatrixformalism}
The study of BigBite began with the implementation of 
the analytical model described in 
Appendix~\ref{appendix:AnalyticalModel}.
In spite of its shortcomings, the analytical model 
was a good starting point.  Due to its simplicity,
it can be implemented and tested quickly, and lends itself well
to online estimation of the experimental data.
However, for the off-line analysis, a more sophisticated approach
based on transport matrix formalism is needed.  In this approach,
a prescription is obtained that transforms the detector variables
directly to the target variables.  Various parameterizations
of this transformation are possible.  We have adopted
a polynomial expansion of the form~\cite{bertozzi,nilangaTN}
\begin{eqnarray}
  \Omega_{\mathrm{Tg}} = \sum_{i,j,k,l} 
     a_{ijkl}^{\Omega_{\mathrm{Tg}}} \,\,
     x_\mathrm{Det}^i \,
     \theta_{\mathrm{Det}}^j \,
     y_{\mathrm{Det}}^k \,
     \phi_{\mathrm{Det}}^l \>, \qquad 
  \Omega_{\mathrm{Tg}} \in \left\{
     \delta_{\mathrm{Tg}}, \theta_{\mathrm{Tg}},
     \phi_{\mathrm{Tg}}, y_{\mathrm{Tg}} \right\} \>. \label{eq1}
\end{eqnarray}
Knowing the optics of a spectrometer is equivalent to
determining the expansion coefficients $a_{ijkl}^{\Omega_{\mathrm{Tg}}}$ 
(the so-called optical ``matrix'') and establishing
the limitations of such a parameterization. 

Ideally, one would like to obtain a single optical matrix
with full reconstruction functionality for all particle species
and momenta, with as few high-order terms as possible.
In a large-acceptance spectrometer like BigBite,
this represents a considerable challenge.  In particular,
one must clearly understand the contributions of the high-order elements.
Uncontrolled inclusion of these terms typically causes oscillations
of the reconstructed variables at the edges of the acceptance.
In the following we discuss the procedure of constructing
the optical matrix in which special attention is devoted
to checking the convergence of the method and estimating
the robustness of the matrix elements.

\subsubsection{Decoupled description}
\label{sub:direct}

The determination of the optical matrix starts with
a low-order analysis in order to estimate the dominant
matrix elements.  As in the analytical model, the BigBite magnet
is assumed to be an ideal dipole.  This assumption
decouples the in-plane and out-of-plane variables, resulting in
the simplification that $\delta_{\mathrm{Tg}}$ and $\theta_{\mathrm{Tg}}$
depend only on $x_{\mathrm{Det}}$ and $\theta_{\mathrm{Det}}$,
while $y_{\mathrm{Tg}}$ and $\phi_{\mathrm{Tg}}$ depend
only on $y_{\mathrm{Det}}$ and $\phi_{\mathrm{Det}}$.

Since each target coordinate depends only on two detector
coordinates, the matrix elements were estimated by
examining two-dimensional histograms of target coordinates
(as given by the HRS) versus BigBite detector variables,
using various detector-variable cuts.  Since BigBite
in this approximation does not bend horizontally,
only first-order polynomials were utilized to fit the data
for $y_{\mathrm{Tg}}$ and $\phi_{\mathrm{Tg}}$, while expansions
up to third-order were applied for $\delta_{\mathrm{Tg}}$
and $\theta_{\mathrm{Tg}}$:
\begin{eqnarray}
\delta_{\mathrm{Tg}}(x, \theta) &=& \left[a_{0000}^{\delta_{\mathrm{Tg}}} 
       +a_{1000}^{\delta_{\mathrm{Tg}}}x + 
        a_{2000}^{\delta_{\mathrm{Tg}}}x^2 \right]
       +\left[a_{0100}^{\delta_{\mathrm{Tg}}} +
        a_{1100}^{\delta_{\mathrm{Tg}}}x + 
        a_{2100}^{\delta_{\mathrm{Tg}}}x^2 \right]\theta \nonumber\\
      &+& \left[a_{0200}^{\delta_{\mathrm{Tg}}} 
       + a_{1200}^{\delta_{\mathrm{Tg}}}x \right] \theta^2
       + \left[a_{0300}^{\delta_{\mathrm{Tg}}} 
       + a_{1300}^{\delta_{\mathrm{Tg}}}x \right ]\theta^3\>, \nonumber  \\
\theta_{\mathrm{Tg}}(x, \theta) &=& \left[a_{0000}^{\theta_{\mathrm{Tg}}} 
       + a_{1000}^{\theta_{\mathrm{Tg}}}x + a_{2000}^{\theta_{\mathrm{Tg}}}x^2 \right]
       + \left[a_{0100}^{\theta_{\mathrm{Tg}}} + a_{1100}^{\theta_{\mathrm{Tg}}}x 
       + a_{2100}^{\theta_{\mathrm{Tg}}}x^2 \right] \theta\>, \nonumber  \\ 
\phi_{\mathrm{Tg}}(y,\phi) &=& a_{0000}^{\phi_{\mathrm{Tg}}} 
       + a_{0001}^{\phi_{\mathrm{Tg}}}\phi\>, \nonumber \\ 
y_{\mathrm{Tg}}(y, \phi) &=& \left[a_{0001}^{y_{\mathrm{Tg}}}
       + a_{0011}^{y_{\mathrm{Tg}}}y \right]\phi  + \left[ a_{0000}^{y_{\mathrm{Tg}}} 
       + a_{0010}^{y_{\mathrm{Tg}}}y\right]\,. \nonumber 
\end{eqnarray}
Some of the calculated matrix elements are shown in the second column
of Table~\ref{table_optics_table1}.  The $a_{0001}^{\phi_{\mathrm{Tg}}}$ matrix element 
was set to $1$ since there is no in-plane bending.
This approximation could not be used for further physics analysis
because higher-order corrections are needed.  However,
the low-order terms are very robust and do not change much when
more sophisticated models with higher-order terms are considered.
The results obtained by using this method serve
as a benchmark for more advanced methods, in particular
as a check whether the matrix elements computed by automated
numerical algorithms converge to reasonable values.

\begin{table}[!ht]
\caption{The dominant matrix elements of the BigBite
optics model (Eq.~(\ref{eq1})) determined by a decoupled
description (section~\ref{sub:direct}),
by simplex minimization (N\&M), and by singular
value decomposition (SVD, section~\ref{sub:nmsvd}).
\label{table_optics_table1}}
\begin{center}
\begin{tabular}{lrrr}
\toprule
 {\bf Matrix} & {\bf Decoupled} & {\bf N\&M} & {\bf SVD} \\[-2pt]
 {\bf element} & {\bf description} & & \\ 
\midrule
 & & & \\[-5mm]
$a_{0010}^{y_\mathrm{Tg}}\,[\mathrm{m/m}]$ 
  & $0.998$ & $1.024$ & $0.917$ \\ [1mm]
$a_{0001}^{y_\mathrm{Tg}}\,[\mathrm{m/rad}]$ 
  & $-2.801$ & $-2.839$ & $-2.766$ \\ [2mm]
\hline
 & & & \\[-4mm]
$a_{0001}^{\phi_\mathrm{Tg}}\,[\mathrm{rad/rad}]$ 
  & $1.000$ & $1.052$ &  $0.9517$ \\ [2mm]
\hline
 & & & \\[-4mm]
$a_{1000}^{\theta_\mathrm{Tg}}\,[\mathrm{rad/m}]$ 
  & $0.497$ & $0.549$ & $0.551$ \\[1mm]
$a_{0100}^{\theta_\mathrm{Tg}}\,[\mathrm{rad/rad}]$ 
  & $-0.491$ & $-0.490$ & $-0.484$ \\[2mm]
\hline
 & & & \\[-4mm]
$a_{1000}^{\delta_\mathrm{Tg}}\,[\mathrm{1/m}]$ 
  & $-0.754$ & $-0.716$ & $-0.676$ \\[1mm]
$a_{0100}^{\delta_\mathrm{Tg}}\,[\mathrm{1/rad}]$ 
  & $2.811$ & $2.881$ & $2.802$ \\
\bottomrule
\end{tabular}
\end{center}
\end{table}

\subsubsection{Higher order matrix formalism}
\label{sub:nmsvd}

For the determination of the optics matrix a numerical method
was developed in which matrix elements up to fourth order
were retained.  Their values were calculated by using
a $\chi^2$-minimization scheme, wherein the target
variables calculated by Eq.~(\ref{eq1}) were compared
to the directly measured values,
\begin{eqnarray}
  \chi^2\left(a_i^{\Omega_{\mathrm{Tg}}}\right) = 
     \sqrt{\left( \Omega_{\mathrm{Tg}}^{\mathrm{Measured}} -
     \Omega_{\mathrm{Tg}}^{\mathrm{Optics}}\left(x_{\mathrm{Det}},
      y_{\mathrm{Det}}, \theta_{\mathrm{Det}}, \phi_{\mathrm{Det}}; 
      a_i^{\Omega_{\mathrm{Tg}}}\right) \right)^2} \>,\>
      i = 1,2,\,\ldots\>,M\,. \label{eq2}
\end{eqnarray}
The use of $M$ matrix elements for each target variable means
that a global minimum in $M$-dimensional space must be found.
Numerically this is a very complex problem; two techniques
were considered for its solution.

Our first choice was the downhill simplex method developed
by Nelder and Mead \cite{nelder,nrc}.  The method tries
to minimize a scalar non-linear function of $M$ parameters
by using only function evaluations (no derivatives).
It is widely used for non-linear unconstrained optimization,
but it is inefficient and its convergence properties
are poorly understood, especially in multi-dimensional minimizations.
The method may stop in one of the local minima instead of the global
minimum \cite{lagarias,mckinnon}, so an additional examination
of the robustness of the method was required.

The set of functions $\Omega_{\mathrm{Tg}}$ is linear
in the parameters $a_i^{\Omega_{\mathrm{Tg}}}$.
Therefore, Eq.~(\ref{eq2}) can be written as
\begin{eqnarray}
  \chi^2 = \sqrt{ \, \left| A \, \vec{a} - \vec{b} \, \right |^2}\>,\label{eq3}
\end{eqnarray}
where the $M$-dimensional vector $\vec{a}$ contains
the matrix elements $a_i^{\Omega_{\mathrm{Tg}}}$,
and the $N$-dimensional vector $\vec{b}$ contains
the measured values of the target variable being considered.
The elements of the $N\times M$ matrix $A$ are various products
of detector variables ($x_\mathrm{Det}^i \theta_{\mathrm{Det}}^j
y_{\mathrm{Det}}^k \phi_{\mathrm{Det}}^l$) for each measured event.
The system $A\,\vec{a} = \vec{b}$ in Eq.~(\ref{eq3})
is overdetermined ($N>M$), thus the vector $\vec{a}$
that minimizes the $\chi^2$ can be computed by singular
value decomposition (SVD).  It is given by $A = UWV^\mathrm{T}$,
where $U$ is a $N\times M$ column-orthogonal matrix,
$W$ is a $M\times M$ diagonal matrix with non-negative
singular values $w_i$ on its diagonal, and $V$ is a $M\times M$
orthogonal matrix \cite{golub,nrc}.  The solution has the form
$$
\vec{a} = \sum_{i=1}^M\left(\frac{ \vec{U}_i\cdot\vec{b}}{w_i} \right)
    \vec{V}_i \>.
$$

The SVD was adopted as an alternative to simplex minimization
since it produces the best solution in the least-square sense,
obviating the need for robustness tests.  Another great advantage
of SVD is that it can not fail; the method always returns a solution,
but its meaningfulness depends on the quality of the input data.
The most important leading-order matrix elements computed
by using both techniques are compared in Table~\ref{table_optics_table1}.

\subsection{Calibration results for Vertex Position}

\begin{figure}[!ht]
\begin{center}
\includegraphics[width=0.49\textwidth]{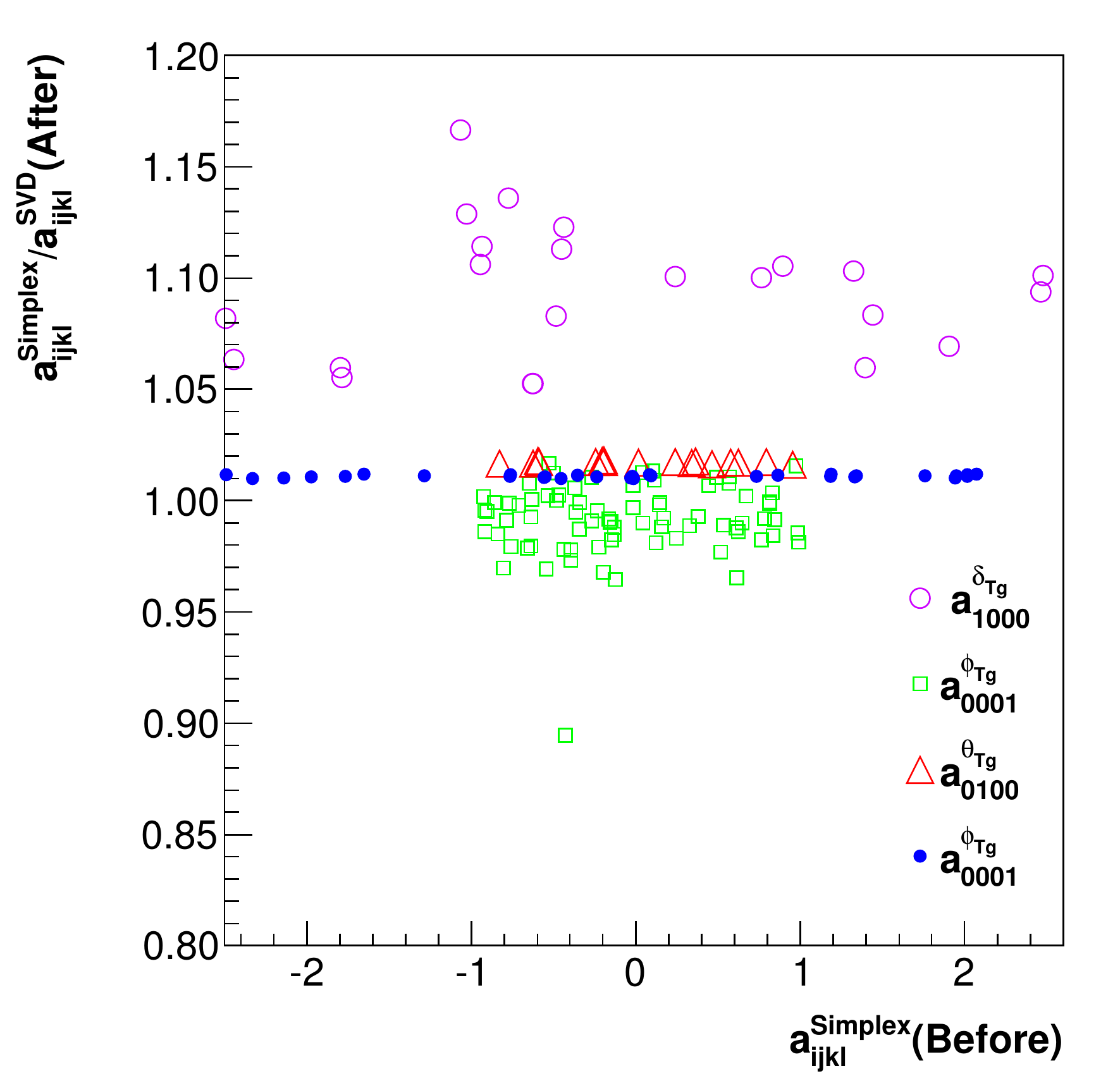}
\includegraphics[width=0.49\textwidth]{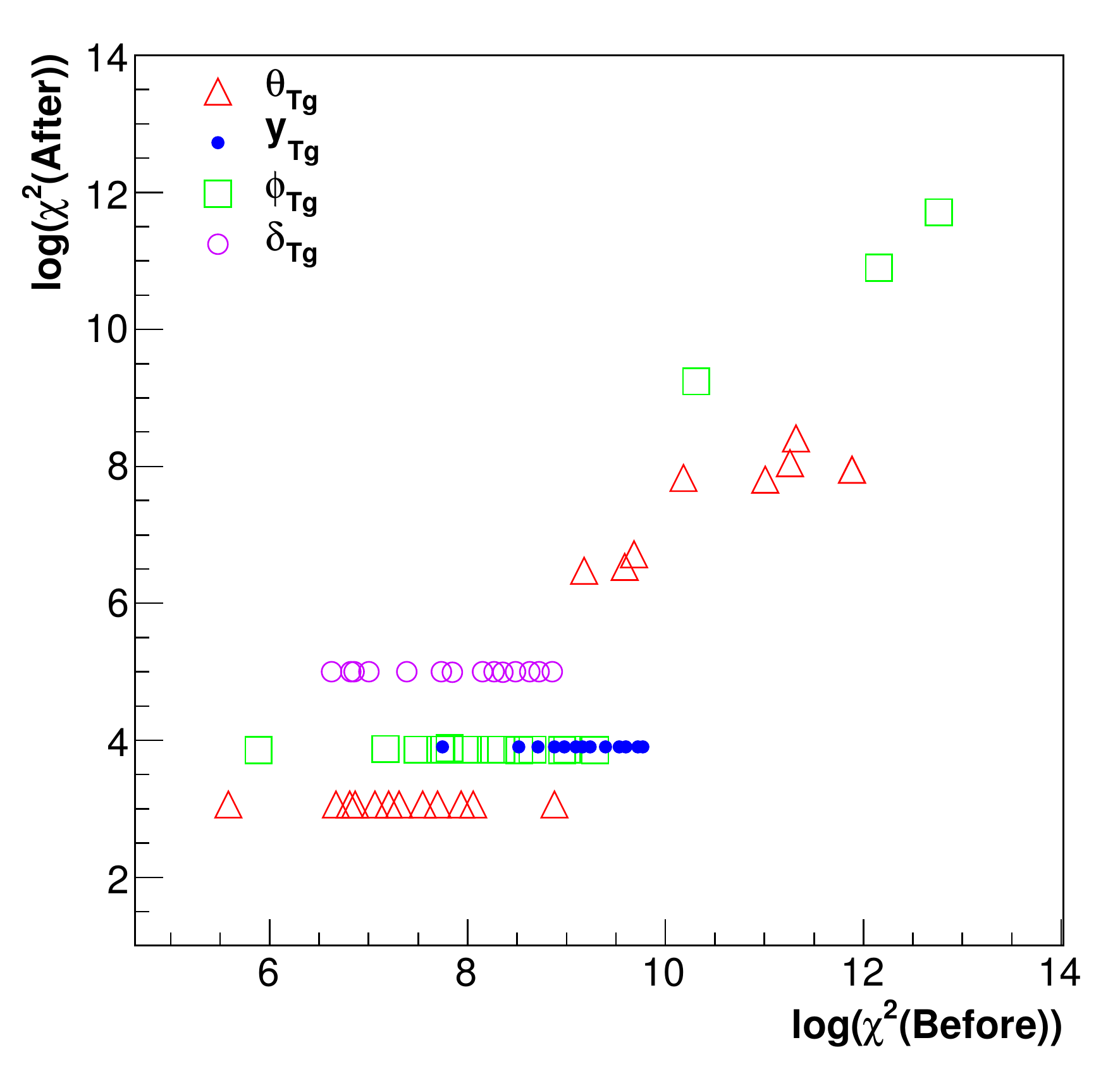}
\caption{[Left] Robustness checks of the simplex minimization method
for select matrix elements  $a_{ijkl}^{\Omega_{\mathrm{Tg}}}$.
The analysis was done for a large set of randomly chosen initial conditions
for each target coordinate.  The fact that the vast majority
of the initial conditions converge to a single value
is an indication of the robustness of the method.
[Right] The values of the $\chi^2$-function before and after
simplex minimization for all four target coordinates.
The method converges to a single $\chi^2$ value
for a wide range of initial conditions (note the log scales).
The solution with the smallest $\chi^2$ represents the result
used in the optical matrix. \label{fig_optics_TargetYConvergence} }
\end{center}
\end{figure}

The matrix for the vertex position variable $y_{\mathrm{Tg}}$
was obtained by analyzing the protons from quasi-elastic scattering
of electrons on the multi-foil carbon target.  The positions
of the foils were measured by a geodetic survey to sub-millimeter
accuracy, allowing for a very precise calibration of $y_{\mathrm{Tg}}$.
The vertex information from the HRS was used to locate
the foil in which the particle detected by BigBite originated.
This allowed us to directly correlate the detector variables
for each coincidence event to the interaction vertex.
When Eq.~(\ref{eq1}) is applied to $y_{\mathrm{Tg}}$,
a linear equation for each event can be written:
\begin{eqnarray}
 {y_\mathrm{Tg}}_{(n)}^{\mathrm{Measured}} = 
   {y_{\mathrm{Tg}}}_{(n)}^{\mathrm{Optics}} 
   &:=& a_{0000}^{y} + a_{0001}^{y}\phi_{(n)} + 
       a_{0002}^{y}\phi_{(n)}^2 + a_{0003}^{y}\phi_{(n)}^3 + 
       \cdots\ \nonumber\\
   &+& a_{0010}^{y}y_{(n)} + a_{0020}^{y}y_{(n)}^2 + 
       a_{0030}^{y}y_{(n)}^3 + a_{0040}^{y}y_{(n)}^4 + 
       \cdots\ \nonumber \\
   &+& a_{0100}^{y}\theta_{(n)} + a_{0200}^{y}\theta_{(n)}^2+ 
       a_{0300}^{y}\theta_{(n)}^3 + a_{0400}^{y}\theta_{(n)}^4 + 
       \cdots \nonumber \\
   &+& a_{1000}^{y}x_{(n)} + a_{2000}^{y}x_{(n)}^2 + 
       a_{3000}^{y}x_{(n)}^3 + a_{4000}^{y}x_{(n)}^4 + 
       \cdots\ \nonumber \\
   &+&   a_{1111}^{y}x_{(n)}\theta_{(n)} y_{(n)} \phi_{(n)}\>, 
        \label{TgYAnsatz}
\end{eqnarray}
where $n = 1,2,\ldots,N$, and $N$ is the number of coincidence events
used in the analysis. The overdetermined set of Eqs.~(\ref{TgYAnsatz})
represents a direct comparison of the reconstructed vertex position
$y_{\mathrm{Tg}}^{\mathrm{Optics}}$ to the measured value
$y_{\mathrm{Tg}}^{\mathrm{Measured}}$.  Initially a consistent
polynomial expansion to fourth degree ($i+j+k+l \leq 4$)
was considered, which depends on $70$ matrix elements $a_{ijkl}^{y}$.
Using this ansatz in Eq.~(\ref{eq2}) defines a $\chi^2$-minimization
function, which serves as an input to the simplex method.
To be certain that the minimization did not converge to one of
the local minima, the robustness of this method was examined
by checking the convergence of the minimization algorithm
for a large number of randomly chosen initial sets of parameters 
(see Fig.~\ref{fig_optics_TargetYConvergence}).

The results were considered to be stable if the $\chi^2$
defined by Eq.~(\ref{eq2}) converged to the same value
for the majority of initial conditions.
Small variations in $\chi^2$ were allowed: they are caused
by small matrix elements which are irrelevant for $y_{\mathrm{Tg}}$,
but have been set to non-zero values in order to additionally minimize $\chi^2$
in a particular minimization process.  These matrix elements
could be easily identified and excluded during the robustness checks
because they are unstable and converge to a different value
in each minimization. Ultimately only $25$ matrix elements that had the 
smallest fluctuations were kept for the $y_{\mathrm{Tg}}$ matrix.

The SVD method was used next.  To compute the matrix elements
for $y_{\mathrm{Tg}}$, the linear set of Eqs.~(\ref{TgYAnsatz})
first needs to be rewritten in the form $A\,\vec{a} = \vec{b}$
used in Eq.~(\ref{eq3}):
\begin{eqnarray}
 \left( 
  \begin{array}{cccc}
  1 & \phi_{(1)} & \cdots & x_{(1)}\theta_{(1)}y_{(1)}\phi_{(1)} \\ 
  1 & \phi_{(2)} & \cdots & x_{(2)}\theta_{(2)}y_{(2)}\phi_{(2)} \\ 
  1 & \phi_{(3)} & \cdots & x_{(3)}\theta_{(3)}y_{(3)}\phi_{(3)} \\ 
  \vdots & \vdots  & \ddots & \vdots \\ 
  1 & \phi_{(N-2)} & \cdots & x_{(N-2)}\theta_{(N-2)}y_{(N-2)}\phi_{(N-2)} \\ 
  1 & \phi_{(N-1)} & \cdots & x_{(N-1)}\theta_{(N-1)}y_{(N-1)}\phi_{(N-1)} \\ 
  1 & \phi_{(N)} & \cdots & x_{(N)}\theta_{(N)}y_{(N)}\phi_{(N)} \\ 
  \end{array}
  \right) 
 \left( 
  \begin{array}{c}
  a_{0000} \\ 
  a_{0001} \\ 
  \vdots \\
  a_{1111}
  \end{array}
 \right)  = 
 \left( 
  \begin{array}{c}
  {y_{\mathrm{Tg}}}_{(1)} \\ 
  {y_{\mathrm{Tg}}}_{(2)} \\ 
  {y_{\mathrm{Tg}}}_{(3)} \\ 
  \vdots \\
  {y_{\mathrm{Tg}}}_{(N-2)} \\
  {y_{\mathrm{Tg}}}_{(N-1)} \\
  {y_{\mathrm{Tg}}}_{(N)}
  \end{array}
 \right)\>, \nonumber
\end{eqnarray}
where $\vec{a}$ contains $M$ unknown matrix elements $a_{ijkl}^{y}$
to be determined by the SVD, $\vec{b}$ contains $N$ measured values
of $y_{\mathrm{Tg}}$, and $A$ is filled with the products
of detector variables accompanying the matrix elements
in the polynomial expansion of Eq.~(\ref{TgYAnsatz}) for each event.

\begin{figure}[!ht]
\begin{center}
\begin{minipage}[t]{0.6\textwidth}
\hrule height 0pt
\includegraphics[width=1\textwidth]{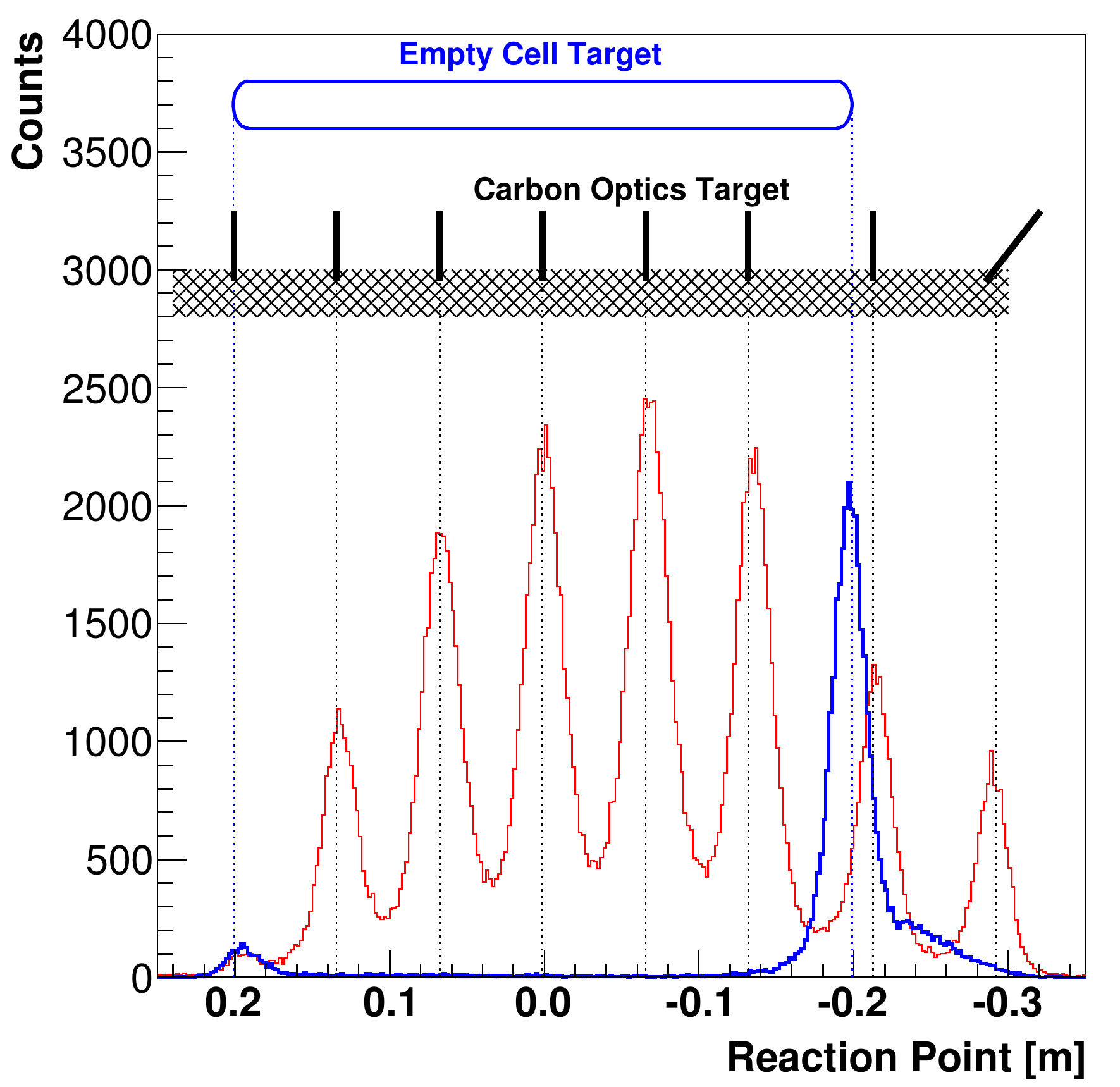}
\end{minipage}
\hfill
\begin{minipage}[t]{0.35\textwidth}
\hrule height 0pt
\caption{The reconstructed vertex position (reaction point)
for the multi-foil carbon target and the empty cell
of the production target, by using the SVD technique.
The vertical dashed lines indicate the actual positions
of the carbon foils and the empty-cell glass windows.
The small shoulder to the right of the reconstructed empty-cell
entry window is due to the jet of $^4\mathrm{He}$ gas used
to cool the window at the beam impact point. \label{fig_optics_TgYResult}} 
\end{minipage}
\end{center}
\end{figure}

The SVD analysis also began with $70$ matrix elements, 
but was not applied to one combined data set as in the 
simplex method in order to extract the most relevant ones.
Rather, it was used on each set of data separately. 
From the comparison of the matrix elements obtained
with different calibration data sets, only the elements 
fluctuating by less than $100\,\%$ were selected. Although
this choice appears to be arbitrary, the results do not change much by 
modifying this criterion, for example, by including elements
with as much as $\pm 1000\, \mathrm{\%}$ fluctuation.
The final set of matrix elements contained only $37$ of the best entries.
With these elements,  the entire analysis was repeated in order
to calculate their final values listed in 
Table~\ref{table_optics_matrixTgY}. The most relevant elements
are given in Table~\ref{table_optics_table1}.  The result of the calibration
of $y_\mathrm{Tg}$ is shown in Fig.~\ref{fig_optics_TgYResult}.

\begin{table}[!ht]
\begin{center}
\begin{minipage}[t]{0.6\textwidth}
\hrule height 0pt
\begin{tabular}{c|rrrr}
\toprule
j~k~l & \multicolumn{1}{c}{$i=0$} & \multicolumn{1}{c}{$i=1$} & \multicolumn{1}{c}{$i=2$} &
        \multicolumn{1}{c}{$i=3$}\\
\midrule
0~0~0 & $ 0.02318$ & $ 0.00000$ & $ 0.03641$ & $-0.13060$\\
0~0~1 & $-2.76580$ & $-0.24199$ & $-1.03855$ & $ 5.63012$\\
0~0~2 & $-2.33569$ & $ 4.94169$ & $ 0.00000$ & $ 0.00000$\\
0~0~3 & $ 19.9915$ & $-67.6181$ & $ 0.00000$ & $ 0.00000$\\
0~1~0 & $ 0.91701$ & $ 0.35879$ & $ 1.91131$ & $-8.24815$\\
0~1~1 & $ 1.98481$ & $-9.50365$ & $ 0.00000$ & $ 0.00000$\\
0~1~2 & $-30.8046$ & $ 177.272$ & $ 0.00000$ & $ 0.00000$\\
0~2~0 & $-0.63517$ & $ 4.41392$ & $-0.22726$ & $ 0.00000$\\
0~2~1 & $ 20.2199$ & $-168.771$ & $ 0.00000$ & $ 0.00000$\\
0~3~0 & $-6.60272$ & $ 49.9945$ & $ 0.00000$ & $ 0.00000$\\
1~0~0 & $ 0.03889$ & $ 0.00000$ & $ 0.00000$ & $ 0.00000$\\
1~0~1 & $-0.61325$ & $ 0.00000$ & $ 0.24360$ & $ 0.00000$\\
1~1~0 & $ 0.87685$ & $ 0.00000$ & $ 0.00000$ & $ 0.00000$\\
1~1~1 & $ 2.25219$ & $ 0.00000$ & $ 0.00000$ & $ 0.00000$\\
1~1~2 & $-28.7562$ & $ 0.00000$ & $ 0.00000$ & $ 0.00000$\\
1~2~0 & $-1.35994$ & $ 0.00000$ & $ 0.00000$ & $ 0.00000$\\
1~2~1 & $ 47.3592$ & $ 0.00000$ & $ 0.00000$ & $ 0.00000$\\
2~0~1 & $-0.87017$ & $ 0.00000$ & $ 0.00000$ & $ 0.00000$\\
2~1~0 & $ 1.61122$ & $ 0.00000$ & $ 0.00000$ & $ 0.00000$\\
2~2~0 & $ 4.62632$ & $ 0.00000$ & $ 0.00000$ & $ 0.00000$\\
\bottomrule
\end{tabular}
\end{minipage}
\hfill
\begin{minipage}[t]{0.35\textwidth}
\hrule height 0pt
\caption{The final list of matrix elements considered for the reconstruction of
$y_{\mathrm{Tg}}$. These parameters are introduced to Eq.~(\ref{eq1}) to calculate
the value of the variable from directly measured detector coordinates.
The units of the individual matrix elements are 
$[\mathrm{m}/\mathrm{m}^{i+k}\mathrm{rad}^{j+l}]$. 
\label{table_optics_matrixTgY}}
\end{minipage}
\end{center}
\end{table}

\subsection{Calibration results of Angular coordinates}

For the calibration of the angular variables $\theta_{\mathrm{Tg}}$
and $\phi_{\mathrm{Tg}}$, a set of quasi-elastic data on carbon
and deuterium targets taken with the sieve-slit collimator was analyzed.
The particles that pass through different holes can be well separated
and localized at the detector plane.

\begin{figure}[!ht]
\begin{center}
\begin{minipage}[t]{0.65\textwidth}
\hrule height 0pt
\includegraphics[width=1\textwidth]{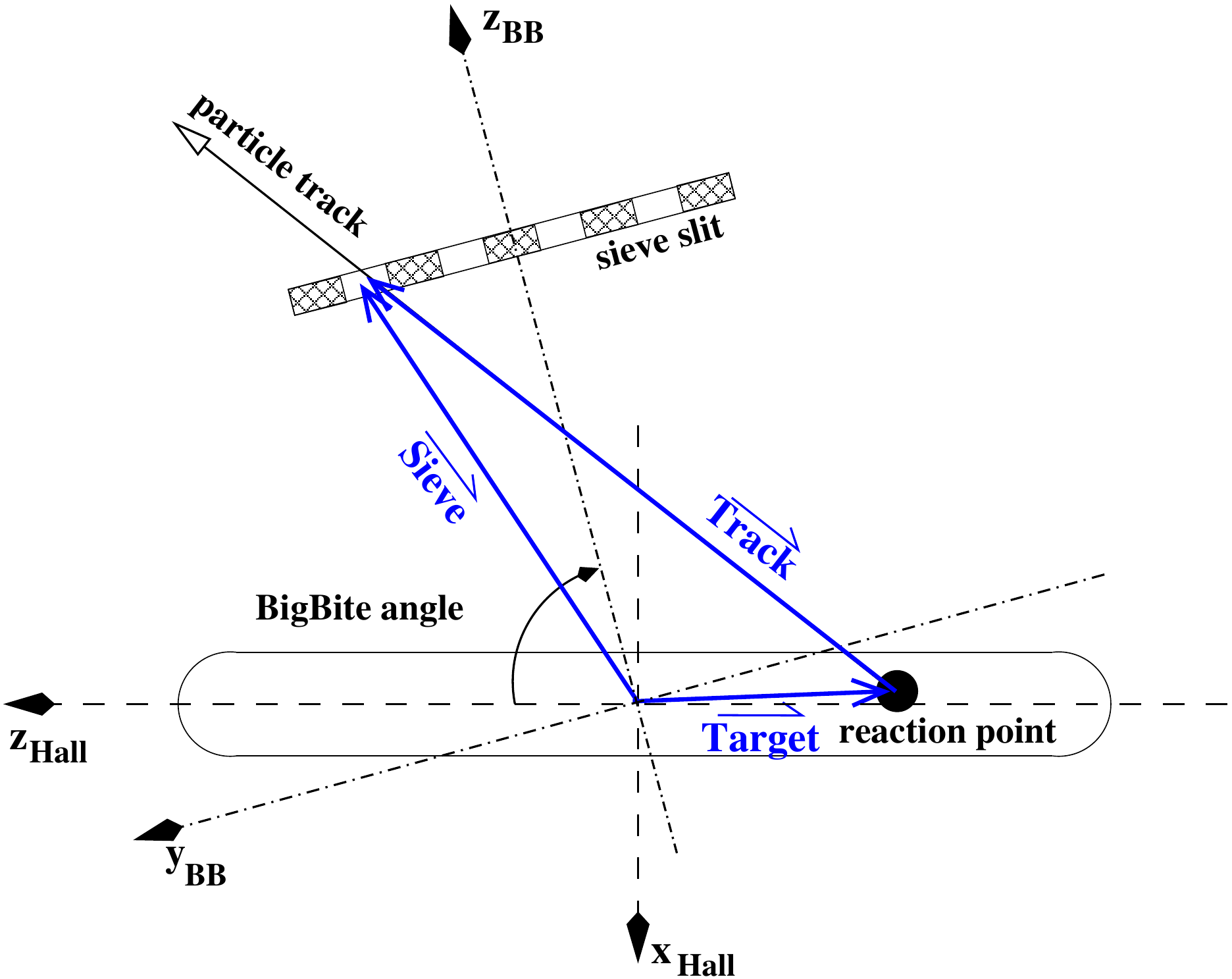}
\end{minipage}
\hfill
\begin{minipage}[t]{0.34\textwidth}
\hrule height 0pt
\caption{Position of the sieve-slit collimator relative to the target.
The vector of the particle track through a particular hole
in the sieve is the difference of the position vector at the hole
and the reaction-point vector.  BigBite is positioned
at $-75^\circ$ with respect to the beam direction.
\label{fig_optics_BBSieveSlitDiagram}}
\end{minipage}
\end{center}
\end{figure}

By knowing the detector coordinates and the accurate position
of the corresponding hole in the sieve, the target variables
can be calculated.  From the reaction point at the target
(see Fig.~\ref{fig_optics_BBSieveSlitDiagram}), $\theta_{\mathrm{Tg}}$
and $\phi_{\mathrm{Tg}}$ can be calculated:
\begin{eqnarray}
  \tan \phi_{\mathrm{Tg}} = \frac{y_{\mathrm{Sieve}} - 
    y_{\mathrm{Tg}}}{z_{\mathrm{Sieve}} - z_{\mathrm{Tg}}} \>,
    \qquad\tan \theta_{\mathrm{Tg}} = 
    \frac{x_{\mathrm{Sieve}} - 
    x_{\mathrm{Tg}}}{z_{\mathrm{Sieve}} - z_{\mathrm{Tg}}} \>. \nonumber
\end{eqnarray}

By using the values of the target variables, a set of linear
equations has been written for all measured events, and matrix
elements determined by using both numerical approaches described above.
In the simplex method, $30$ matrix elements for $\theta_{\mathrm{Tg}}$
and $68$ elements for $\phi_{\mathrm{Tg}}$ were retained.
Robustness checks for both angular variables were repeated
to ensure that the global minimum had been reached.

\begin{figure}[!ht]
\begin{center}
\begin{minipage}[t]{0.64\textwidth}
\hrule height 0pt
\includegraphics[width=1\textwidth]{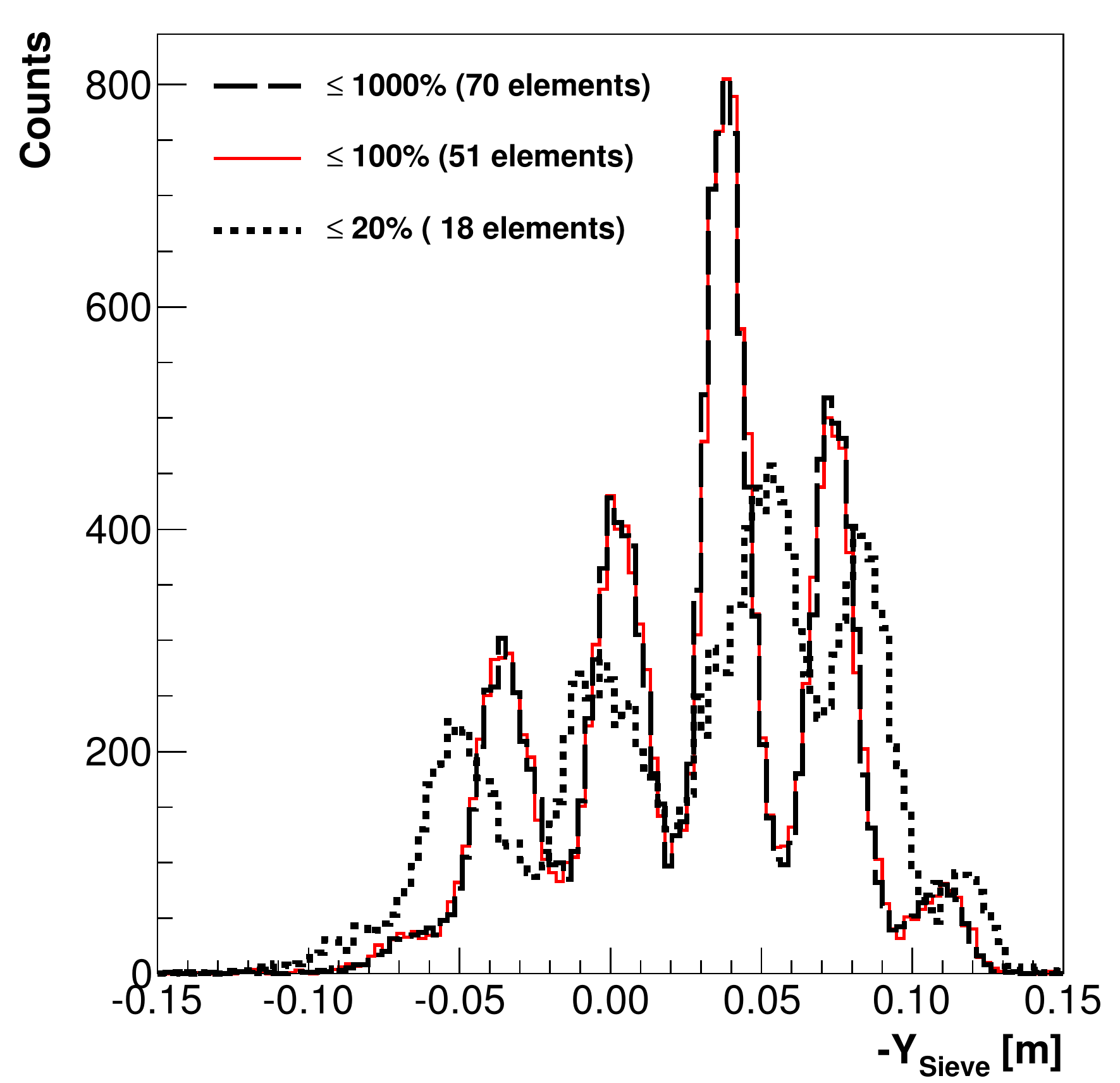}
\end{minipage}
\hfill
\begin{minipage}[t]{0.34\textwidth}
\hrule height 0pt
\caption{The reconstructed positions of the holes in the top row
of the sieve-slit collimator, computed from $\phi_\mathrm{Tg}$.
The quality of the reconstruction depends on the number of included
matrix elements.  There is almost no difference when the elements
fluctuating by up to $\pm 1000\,\%$ are retained ($70$ elements,
dashed lines) or only those that fluctuate by up to $\pm 100\,\%$
($51$ elements, full line).  The quality deteriorates if too many
elements are dropped (i.e.~keeping $18$ elements fluctuating
by less than $\pm 20\,\%$, dotted lines). \label{fig_optics_TgPhME}}
\end{minipage}
\end{center}
\end{figure}

The SVD analysis also started with $70$ matrix elements,
which were ultimately reduced to $37$ for $\theta_{\mathrm{Tg}}$
and $51$ for $\phi_{\mathrm{Tg}}$, again taking into account
only those elements that fluctuated by less than $100\,\mathrm{\%}$.
The resulting matrix elements for both angular variables
are gathered in Tables~\ref{table_optics_matrixTgPh} and~\ref{table_optics_matrixTgTh}.
Figure~\ref{fig_optics_BBSieve} (right) shows the reconstructed sieve pattern.
The majority of the holes are reconstructed, except those obscured
by parts of the experimental apparatus due to specific geometric
constraints during the experiment.  In order to demonstrate
the effect of gradually excluding redundant matrix elements,
Fig.~\ref{fig_optics_TgPhME} shows the reconstructed top row of the sieve-slit
collimator holes when the elements with up to $\pm 1000\,\%$,
$\pm 100\,\%$, and $\pm 20\,\%$ fluctuations are retained.
There is virtually no difference in the reconstructed pattern
when all elements exceeding the $\pm 100\,\%$ fluctuations are dropped,
while errors start to appear when those fluctuating by less than
$\pm 100\,\%$ are dropped.

\begin{table}[!ht]
\begin{center}
\begin{minipage}[t]{0.6\textwidth}
\hrule height 0pt
\begin{tabular}{c|rrrrr}
\toprule
j~k~l & \multicolumn{1}{c}{$i=0$} & \multicolumn{1}{c}{$i=1$} & \multicolumn{1}{c}{$i=2$} &
        \multicolumn{1}{c}{$i=3$} \\
\midrule
0~0~0 & $ 0.00731$ & $ 0.00904$ & $-0.03227$ & $-0.00085$\\ 
0~0~1 & $ 0.95166$ & $-0.12289$ & $ 1.09276$ & $-1.39597$\\
0~0~2 & $ 0.40387$ & $-1.40952$ & $-1.16763$ & $ 0.00000$\\
0~0~3 & $ 0.00000$ & $-6.86876$ & $ 0.00000$ & $ 0.00000$\\
0~1~0 & $ 0.10867$ & $-0.16805$ & $-1.53930$ & $ 2.06102$\\
0~1~1 & $-0.56894$ & $ 3.47503$ & $ 3.62306$ & $ 0.00000$\\
0~1~2 & $ 0.00000$ & $-15.6968$ & $ 0.00000$ & $ 0.00000$\\
0~2~0 & $ 0.27395$ & $-1.93917$ & $-1.50013$ & $ 0.00000$\\
0~2~1 & $ 0.00000$ & $ 53.7498$ & $ 0.00000$ & $ 0.00000$\\
0~3~0 & $ 1.13541$ & $-20.6053$ & $ 0.00000$ & $ 0.00000$\\
1~0~0 & $-0.04127$ & $-0.00523$ & $ 0.33872$ & $ 0.00000$\\
1~0~1 & $ 0.90995$ & $-0.41927$ & $-10.9548$ & $ 0.00000$\\
1~0~2 & $ 0.00000$ & $ 12.0085$ & $ 0.00000$ & $ 0.00000$\\
1~1~0 & $-0.89483$ & $ 0.00000$ & $ 13.8273$ & $ 0.00000$\\
1~1~1 & $-0.22486$ & $-16.9376$ & $ 0.00000$ & $ 0.00000$\\
1~2~0 & $ 0.43072$ & $ 4.53636$ & $ 0.00000$ & $ 0.00000$\\
1~2~1 & $-28.2799$ & $ 0.00000$ & $ 0.00000$ & $ 0.00000$\\
2~0~0 & $ 0.01202$ & $-0.25563$ & $ 0.00000$ & $ 0.00000$\\
2~0~1 & $ 0.17365$ & $ 13.0352$ & $ 0.00000$ & $ 0.00000$\\
2~0~2 & $-6.23897$ & $ 0.00000$ & $ 0.00000$ & $ 0.00000$\\
2~1~0 & $-0.54670$ & $-13.7950$ & $ 0.00000$ & $ 0.00000$\\
2~1~1 & $ 15.9660$ & $ 0.00000$ & $ 0.00000$ & $ 0.00000$\\
2~2~0 & $-8.13482$ & $ 0.00000$ & $ 0.00000$ & $ 0.00000$\\
3~0~1 & $-6.19519$ & $ 0.00000$ & $ 0.00000$ & $ 0.00000$\\
3~1~0 & $ 5.77147$ & $ 0.00000$ & $ 0.00000$ & $ 0.00000$\\
\bottomrule
\end{tabular}
\end{minipage}
\hfill
\begin{minipage}[t]{0.35\textwidth}
\hrule height 0pt
\caption{ The final list of matrix elements considered for the reconstruction of
$\phi_{\mathrm{Tg}}$. These parameters are introduced to Eq.~(\ref{eq1}) to calculate
the value of the variable from directly measured detector coordinates.
The units of the individual matrix elements are 
$[\mathrm{rad}/\mathrm{m}^{i+k}\mathrm{rad}^{j+l}]$.  
\label{table_optics_matrixTgPh}}
\end{minipage}
\end{center}
\end{table}

\begin{table}[!ht]
\begin{center}
\caption{ The final list of matrix elements considered for the reconstruction of 
$\theta_{\mathrm{Tg}}$. These parameters are introduced to Eq.~(\ref{eq1}) to calculate
the value of the variable from directly measured detector coordinates. 
The units of the individual matrix elements are 
$[\mathrm{rad}/\mathrm{m}^{i+k}\mathrm{rad}^{j+l}]$.  
\label{table_optics_matrixTgTh}}
\vspace*{2mm}
\begin{tabular}{c|rrrrr}
\toprule
j~k~l & \multicolumn{1}{c}{$i=0$} & \multicolumn{1}{c}{$i=1$} & \multicolumn{1}{c}{$i=2$} &
        \multicolumn{1}{c}{$i=3$} & \multicolumn{1}{c}{$i=4$}\\
\midrule
0~0~0 & $ 0.01714$ & $ 0.55084$ & $-0.06026$ & $-0.07013$ & $ 0.04245$\\
0~0~1 & $-0.00797$ & $-0.23834$ & $ 0.09164$ & $-0.14835$ & $ 0.00000$\\
0~0~2 & $-0.07563$ & $ 0.00000$ & $ 0.00000$ & $ 0.00000$ & $ 0.00000$\\
0~1~0 & $ 0.00721$ & $ 0.08022$ & $-0.13420$ & $ 0.20116$ & $ 0.00000$\\
0~1~1 & $ 0.06494$ & $ 0.00000$ & $ 0.27701$ & $ 0.00000$ & $ 0.00000$\\
0~2~0 & $ 0.05660$ & $ 0.00000$ & $-0.38301$ & $ 0.00000$ & $ 0.00000$\\
0~2~1 & $-0.35652$ & $ 0.00000$ & $ 0.00000$ & $ 0.00000$ & $ 0.00000$\\
0~3~0 & $ 0.21441$ & $ 0.00000$ & $ 0.00000$ & $ 0.00000$ & $ 0.00000$\\
1~0~0 & $-0.48367$ & $-0.04962$ & $ 0.17132$ & $ 0.15176$ & $ 0.00000$\\
1~0~1 & $ 0.19989$ & $ 0.00000$ & $ 0.70217$ & $ 0.00000$ & $ 0.00000$\\
1~1~0 & $-0.06061$ & $ 0.00000$ & $-0.38277$ & $ 0.00000$ & $ 0.00000$\\
1~3~0 & $ 0.28484$ & $ 0.00000$ & $ 0.00000$ & $ 0.00000$ & $ 0.00000$\\
2~0~0 & $ 0.04598$ & $-0.14871$ & $-0.41783$ & $ 0.00000$ & $ 0.00000$\\
2~0~1 & $ 0.00000$ & $-1.38744$ & $ 0.00000$ & $ 0.00000$ & $ 0.00000$\\
2~1~0 & $ 0.00000$ & $ 0.26537$ & $ 0.00000$ & $ 0.00000$ & $ 0.00000$\\
3~0~0 & $ 0.11351$ & $ 0.00000$ & $ 0.00000$ & $ 0.00000$ & $ 0.00000$\\
3~0~1 & $ 0.49991$ & $ 0.00000$ & $ 0.00000$ & $ 0.00000$ & $ 0.00000$\\
4~0~0 & $ 0.28425$ & $ 0.00000$ & $ 0.00000$ & $ 0.00000$ & $ 0.00000$\\
\bottomrule
\end{tabular}
\end{center}
\end{table}

\begin{figure}[!ht]
\begin{center}
\begin{minipage}[t]{0.6\textwidth}
\hrule height 0pt
\includegraphics[width=1\textwidth]{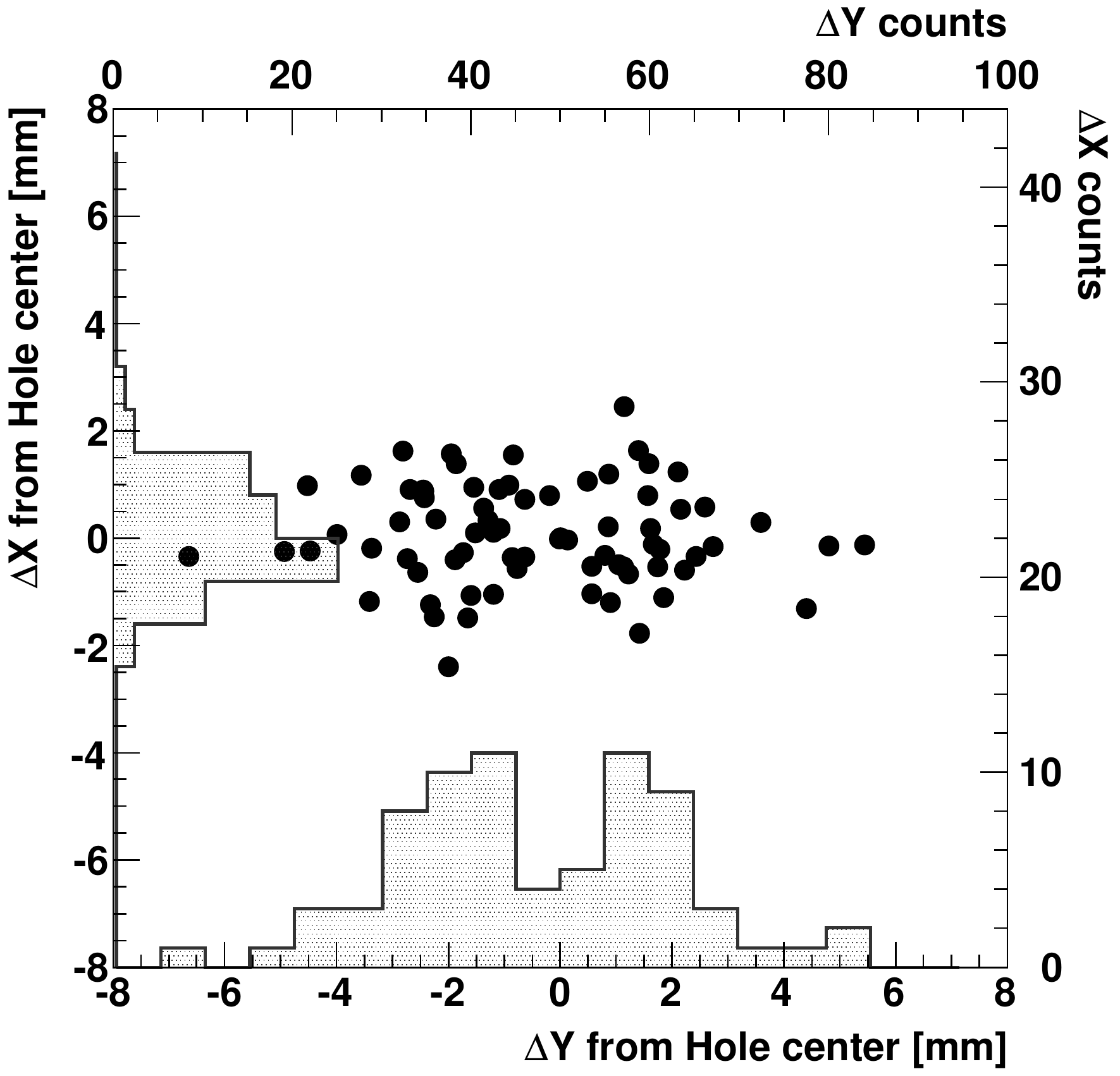}
\end{minipage}
\hfill
\begin{minipage}[t]{0.34\textwidth}
\hrule height 0pt
\caption{Distribution of vertical ($\Delta X$) and horizontal ($\Delta Y$)
deviation of the center of each reconstructed sieve-slit hole
from its true position. Observed 
deviations are much smaller than the diameter of a hole, which is  
$19.1\,\mathrm{mm}$.  The horizontal and vertical histograms (top and right
axis labels, respectively) represent the distributions
in the horizontal and vertical directions.
\label{fig_optics_ThPhCenterPositionError}}
\end{minipage}
\end{center}
\end{figure}

The quality of the sieve-pattern
reconstruction was examined by comparing the centers of the reconstructed
holes with their true positions.  Figure~\ref{fig_optics_ThPhCenterPositionError}
shows that, with the exception of a few holes near the acceptance edges,
these deviations are smaller than $2\,\mathrm{mm}$ in the vertical,
and smaller than $4\,\mathrm{mm}$ in the horizontal direction.
This is much less than the hole diameter, which is  $19.1\,\mathrm{mm}$. 
  
Once the sieve pattern was reconstructed, an absolute calibration
had to be performed to correct for any BigBite misalignment
and mispointing.  For that purpose hydrogen and deuterium elastic data 
were used. By comparing the direction of the momentum transfer vector
from the HRS to the calculated values of $\theta_{\mathrm{Tg}}$
and $\phi_{\mathrm{Tg}}$, the zero-order matrix elements could be
properly determined and the offsets corrected.  In addition,
the precise distance between the target and the sieve-slit collimator
was obtained, which we were not able to measure precisely
due to physical obstacles between the target and BigBite.
From this analysis, the sieve slit was determined to be 
positioned $1.13\,\mathrm{m}$ away from the target.

\subsection{Calibration results for Momentum}

The matrix elements for the $\delta_{\mathrm{Tg}}$ variable
were obtained by using data from elastic scattering of electrons
on hydrogen and deuterium for which the particle momentum
in BigBite should be exactly the same as the momentum 
transfer $\vec{q}$ given by the HRS-L.  We assumed that 
$\delta_{\mathrm{Tg}}$ depends only on $x_{\mathrm{Det}}$ and 
$\theta_{\mathrm{Det}}$, while the dependencies involving 
$y_{\mathrm{Det}}$ and $\phi_{\mathrm{Det}}$ were neglected.
Furthermore, the use of in-plane coordinates in the analysis
for $\delta_{\mathrm{Tg}}$ could result in an erroneous matrix due to 
the strong $\phi_{\mathrm{Tg}}$ dependence inherent
to elastic scattering (events strongly concentrated at one edge
of the acceptance).  Considering only $x_{\mathrm{Det}}$ and
$\theta_{\mathrm{Det}}$ matrix elements, $\delta_{\mathrm{Tg}}$ 
can be expressed as
\begin{eqnarray}
  \delta_{\mathrm{Tg}} = \frac{q_{\mathrm{HRS}} - \Delta_{\mathrm{Loss}}}
   {p_\mathrm{c}}-1 = 
   a_{0000}^\delta + a_{1000}^\delta x_{\mathrm{Det}} + 
   a_{0100}^\delta\theta_{\mathrm{Det}}+\cdots \,. 
   \label{momentumeq}
\end{eqnarray}
In order to obtain the optics matrix applicable to all types
of particles, energy losses $\Delta_{\mathrm{Loss}}$ for particle 
transport through the target enclosure and materials within the 
BigBite spectrometer had to be properly considered.
The energy losses were estimated with a procedure based on the 
Bethe-Bloch formula, as explained in Sec.~\ref{sec:energy_losses}.

The elastic data available for calibration (momentum range
approximately $0.45\,\mathrm{GeV}/c$ to $0.7\,\mathrm{GeV}/c$)
covered only about half of the BigBite momentum acceptance.
To calibrate the low-momentum region from $0.2\,\mathrm{GeV}/c$
to $0.45\,\mathrm{GeV}/c$, we used protons from quasi-elastic 
scattering on ${}^3\mathrm{He}$ by exploiting the information
from the scintillator dE- and E-planes; the deposited particle
energy in each plane was directly mapped to the particle momentum,
based on known properties of the scintillator material.
The punch-through point, corresponding to the particular
momentum at which the particle has just enough energy
to penetrate through the scintillators, served as a reference.

\begin{figure}[!ht]
\begin{center}
\begin{minipage}[t]{0.6\textwidth}
\hrule height 0pt
\includegraphics[width=1\textwidth]{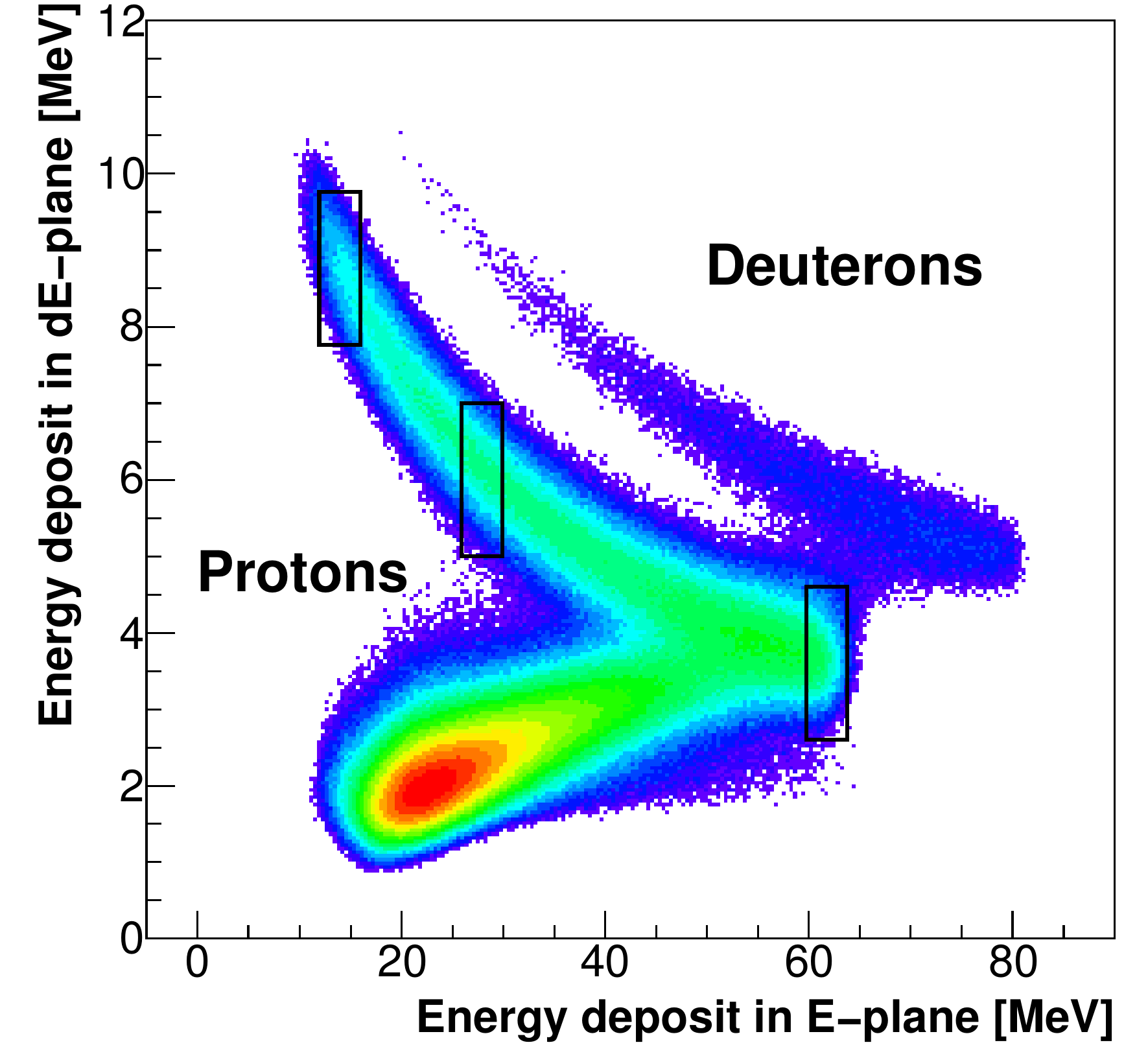}
\end{minipage}
\hfill
\begin{minipage}[t]{0.38\textwidth}
\hrule height 0pt
\caption{The energy losses in the thin ($3\,\mathrm{mm}$)
scintillator dE-plane versus the energy losses in the thicker
($3\,\mathrm{cm}$) E-plane.  The punch-through points, at which
the protons and deuterons have just enough energy to penetrate
both scintillation planes, are clearly visible.  The black boxes
show sections of events with precisely determined momenta
that were used in the $\delta_{\mathrm{Tg}}$ calibration.
\label{fig_optics_EdE}}
\end{minipage}
\end{center}
\end{figure}

Beside the proton punch-through point, two other points
with exactly known energy deposits in the dE- and E-planes
were identified, as illustrated in Fig.~\ref{fig_optics_EdE}.  With the 
additional information from these points, a complete momentum
calibration was possible.  To compute the $\delta_{\mathrm{Tg}}$
matrix elements, both numerical approaches described above
were used.  Since the available data were rather sparse,
the search for the most stable matrix elements was not performed
and a complete expansion to fifth order was considered
in both techniques.  Since only a two-variable dependency
was assumed, a complete description was achieved 
by using only $21$ matrix elements.

\begin{figure}[!ht]
\begin{center}
\begin{minipage}[t]{0.6\textwidth}
\hrule height 0pt
\includegraphics[width=1\textwidth]{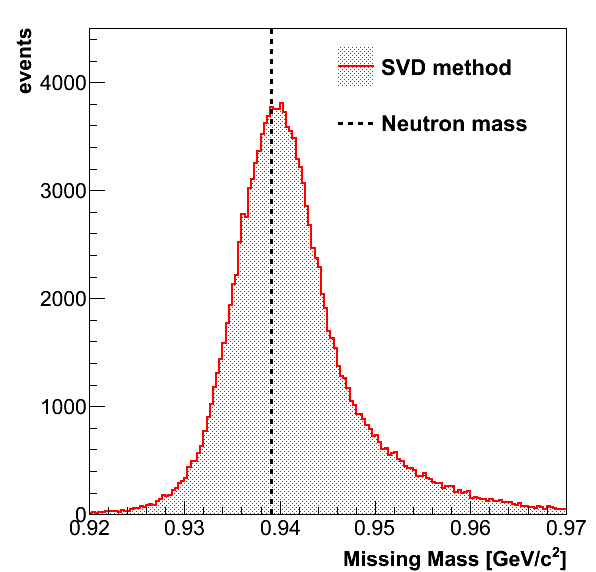}
\end{minipage}
\hfill
\begin{minipage}[t]{0.38\textwidth}
\hrule height 0pt
\caption{The reconstructed mass of the undetected neutron
(missing mass) from the process $\mathrm{{}^2H(e,e'p)n}$
by using matrix-formalism (SVD) approach.
The width of the peak (sigma) is $4\,\mathrm{MeV}/c^2$.
\label{fig_optics_MissingMassPlot}}
\end{minipage}
\end{center}
\end{figure}

The complete list of matrix elements considered for the reconstruction of 
 $\delta_{\mathrm{Tg}}$ is shown in Table~\ref{table_optics_matrixTgDelta}. 
The comparison of the most relevant matrix elements
obtained from both numerical approaches is shown
in Table~\ref{table_optics_table1}.  Figure~\ref{fig_EnLoss_EvsP} shows
that the $\delta_{\mathrm{Tg}}$ matrix is well under control.
The reconstructed momentum agrees with the simulation
of energy losses inside the scintillation planes
for the complete momentum acceptance of BigBite,
for both protons and deuterons.  Figure~\ref{fig_optics_MissingMassPlot}
shows the missing-mass peak for the $\mathrm{{}^2H( e, e'p)n}$ process.
The resolution of the reconstructed neutron mass is approximately
$4\,\mathrm{MeV}/c^2$. 

\begin{table}[!hb]
\begin{center}
\caption{ The final list of matrix elements considered for the reconstruction of 
$\delta_{\mathrm{Tg}}$. These parameters are introduced to Eq.~(\ref{eq1}) to calculate
the value of the variable from directly measured detector coordinates. 
The units of the individual matrix elements are 
$[1/\mathrm{m}^{i+k}\mathrm{rad}^{j+l}]$.  
\label{table_optics_matrixTgDelta}}
\vspace*{2mm}
\begin{tabular}{c|rrrrrr}
\toprule
j~k~l & \multicolumn{1}{c}{$i=0$} & \multicolumn{1}{c}{$i=1$} & \multicolumn{1}{c}{$i=2$} &
        \multicolumn{1}{c}{$i=3$} & \multicolumn{1}{c}{$i=4$} & \multicolumn{1}{c}{$i=5$} \\
\midrule
0~0~0 & $-0.08749$ & $-0.67627$ & $ 1.91361$ & $-3.11562$ & $-6.33754$ & $ 11.5908$ \\
1~0~0 & $ 2.80163$ & $-8.35861$ & $ 17.7113$ & $ 17.3145$ & $-56.5428$ & $ 0.00000$ \\
2~0~0 & $ 11.6524$ & $-46.0247$ & $ 7.91907$ & $ 119.282$ & $ 0.00000$ & $ 0.00000$ \\
3~0~0 & $ 44.3530$ & $-77.9449$ & $-99.1704$ & $ 0.00000$ & $ 0.00000$ & $ 0.00000$ \\
4~0~0 & $ 79.0134$ & $-4.96614$ & $ 0.00000$ & $ 0.00000$ & $ 0.00000$ & $ 0.00000$ \\
5~0~0 & $ 42.3238$ & $ 0.00000$ & $ 0.00000$ & $ 0.00000$ & $ 0.00000$ & $ 0.00000$ \\
\bottomrule
\end{tabular}
\end{center}
\end{table}

\subsection{Resolution}

\begin{figure}[!htp]
\begin{center}
\includegraphics[width=0.49\textwidth]{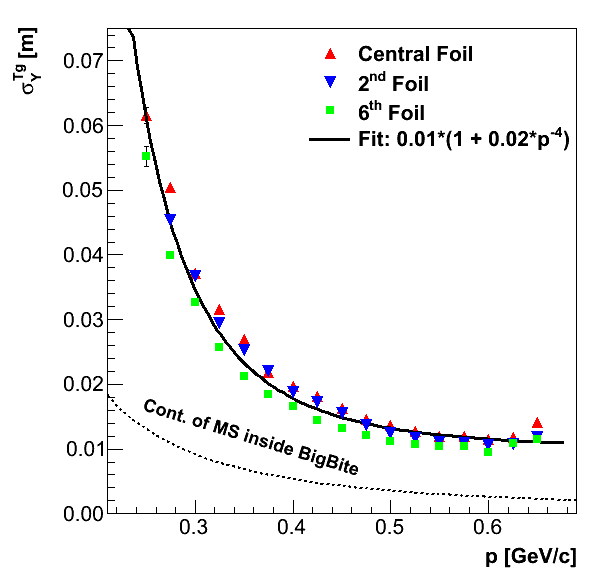}
\includegraphics[width=0.49\textwidth]{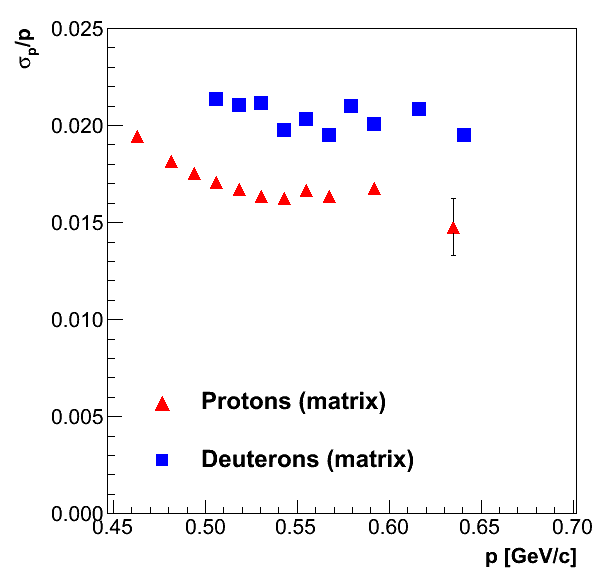}   
\vspace*{-3mm}
\caption{The absolute resolution of $y_{\mathrm{Tg}}$
and the relative momentum resolution as functions
of the momentum measured by BigBite, obtained by the SVD method. 
Irreducible multiple-scattering contributions, mostly
due to the air between the scattering chamber and MWDCs, 
are also shown. \label{fig_optics_ResolutionPlotsYDelta}}
\end{center}
\end{figure}

The quality of the BigBite optics was also studied.
The resolution of the vertex position was estimated
from the difference between the reconstructed $y_{\mathrm{Tg}}$
and the true position at the target by taking the width
(sigma) of the obtained distribution.  This part of the analysis 
was done by using $2$-pass ($2.425\,\mathrm{GeV}$ beam)
quasi-elastic carbon data. The results are shown in 
Fig.~\ref{fig_optics_ResolutionPlotsYDelta} (left). The extracted values
for the resolution of $y_\mathrm{Tg}$ in different momentum bins
can be parameterized as
$$
\sigma_{y_\mathrm{Tg}} \approx 0.01\left(1 + \frac{0.02}{p^{4}} \right)\>,
$$ 
where the particle momentum is in $\mathrm{GeV}/c$ and the result is in meters.
It is best at the upper limit of the accepted momentum range 
(about $p=0.7\,\mathrm{GeV}/c$) where it amounts to 
$\sigma_{y_{\mathrm{Tg}}}=1.1\,\mathrm{cm}$.
The deterioration of the resolution at lower momenta is due to
multiple scattering~\cite{leo} in the air between the scattering
chamber and the MWDCs.

\begin{figure}[!htp]
\begin{center}
\includegraphics[width=0.49\textwidth]{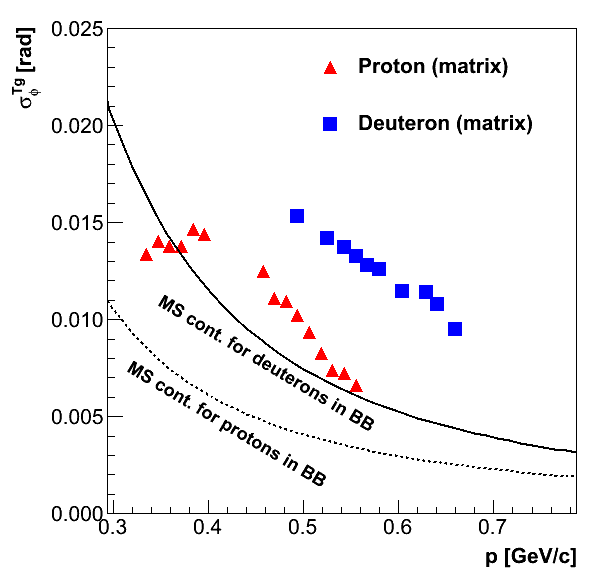}
\includegraphics[width=0.49\textwidth]{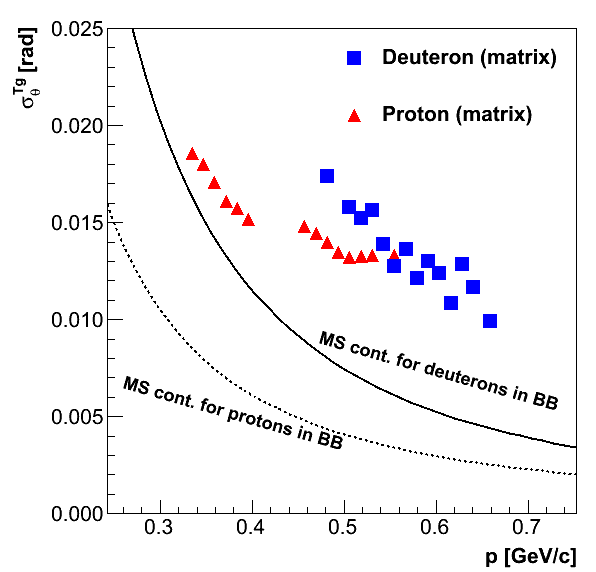}   
\vspace*{-3mm}
\caption{The absolute resolution of $\phi_{\mathrm{Tg}}$
and $\th_{\mathrm{Tg}}$ as functions
of the momentum measured by BigBite, obtained by the SVD method. 
Irreducible multiple-scattering contributions, mostly
due to the air between the scattering chamber and MWDCs, 
are shown by full and dashed lines
for deuterons and protons, respectively. 
\label{fig_optics_ResolutionPlotsPhTh}}
\end{center}
\end{figure}

The resolutions of $\theta_{\mathrm{Tg}}$ and $\phi_{\mathrm{Tg}}$ were
estimated by comparing them to the corresponding angles
as determined from the momentum transfer $\vec{q}$
in elastic scattering on hydrogen and deuterium.
The direction of $\vec{q}$ is given by the electron
kinematics and determined by the HRS-L spectrometer.
The corresponding HRS-L resolutions have been studied in \cite{GeJinPhD}.
Based on these values, the resolution of the reconstructed
$\vec{q}$ was estimated to be $6\,\mathrm{mrad}$
and $0.3\,\mathrm{mrad}$ for the vertical and horizontal angles,
respectively.  These contributions were subtracted in quadrature
from the calculated peak widths, yielding the final resolutions
attributable to BigBite.  The results for $\phi_{\mathrm{Tg}}$ 
and $\th_{\mathrm{Tg}}$ are shown in Fig.~\ref{fig_optics_ResolutionPlotsPhTh}.  
The strong momentum dependence of the resolution is again caused by multiple scattering
in the target and the spectrometer.  Different resolutions
for deuterons and protons occur because the peak broadening
in multiple scattering strongly depends on the particle mass
(at a given momentum).  As before, the biggest contributions come
from the air.  In a typical kinematics
of the E05-102 experiment, the resolutions of 
$\phi_{\mathrm{Tg}}$ and $\theta_{\mathrm{Tg}}$ are 
$\sigma_{\phi_\mathrm{Tg}} \approx 7\,\mathrm{mrad}$ and
$\sigma_{\theta_\mathrm{Tg}} \approx 13\,\mathrm{mrad}$
for $0.55\,\mathrm{GeV}/c$ protons, and approximately 
$\sigma_{\phi_\mathrm{Tg}} \approx 11\,\mathrm{mrad}$ and
$\sigma_{\theta_\mathrm{Tg}} \approx 13\,\mathrm{mrad}$
for $0.6\,\mathrm{GeV}/c$ deuterons.
(Due to multiple scattering, these resolutions are clearly
much larger than the intrinsic MWDC resolutions mentioned
in Sec.~\ref{sec:MWDCs}.)

The resolution of $\delta_{\mathrm{Tg}} = (p-p_\mathrm{c})/p_\mathrm{c}$
was also determined from elastic data by comparing the magnitude of $\vec{q}$
to the momentum reconstructed by BigBite.  The analysis
was done separately for the hydrogen and deuterium data sets.
Figure~\ref{fig_optics_ResolutionPlotsYDelta} (right) shows
the relative momentum resolution $\sigma_{p}/p$
as a function of momentum.  The  relative momentum 
resolution  is approximately 
$1.6\,\%$ for   $0.55\,\mathrm{GeV}/c$  protons, and 
$2\,\%$ for $0.6\,\mathrm{GeV}/c$   deuterons.

\begin{figure}[!hb]
\begin{center}
\begin{minipage}[t]{0.49\textwidth}
\hrule height 0pt
\includegraphics[width=1\textwidth]{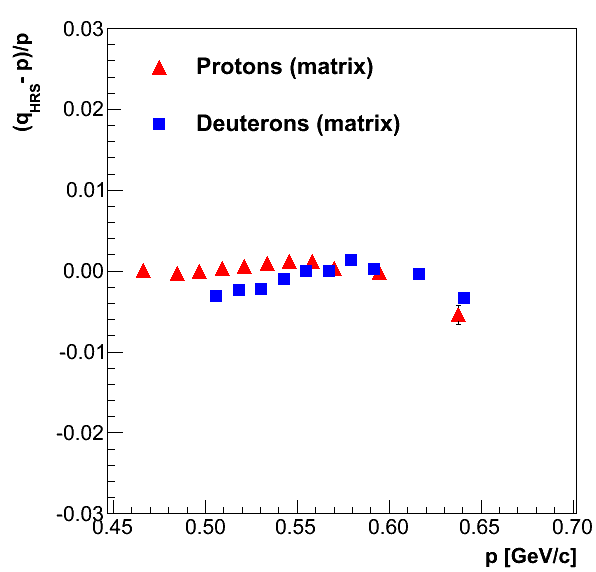}
\includegraphics[width=1\textwidth]{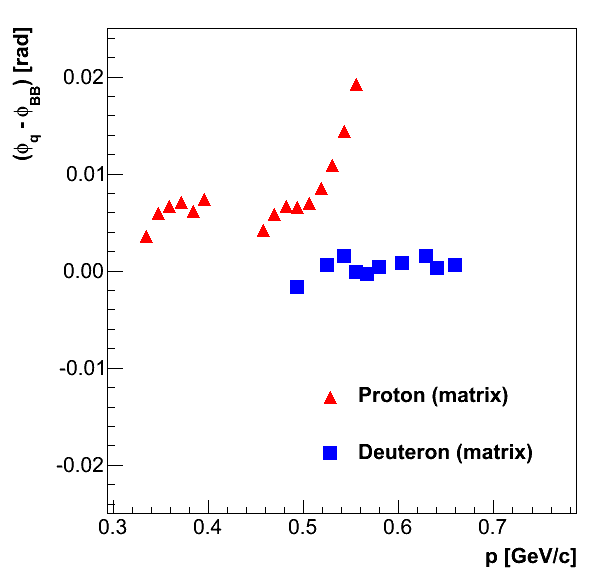}
\end{minipage}
\hfill
\begin{minipage}[t]{0.49\textwidth}
\hrule height 0pt
\includegraphics[width=1\textwidth]{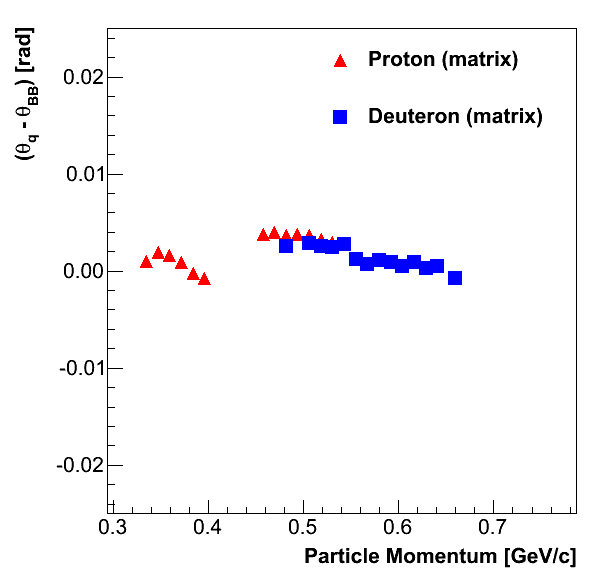}
\caption{The absolute calibrations of target variables $\delta_{\mathrm{Tg}}$, 
$\phi_{\mathrm{Tg}}$ and $\theta_{\mathrm{Tg}}$ as a function of the particle
momentum measured by BigBite. The analysis was done with both, elastic protons
and deuterons. Systematic offsets of the reconstructed variables are minimal 
and smaller than the corresponding resolutions (see 
Figs.~\ref{fig_optics_ResolutionPlotsYDelta} and \ref{fig_optics_ResolutionPlotsPhTh}).
Significant offset is observed only in the $\phi_{\mathrm{Tg}}$ for  elastic protons. 
It is believed, that this is a consequence of fact that analyzed elastic protons all 
come from the up-stream edge of the target, where the optics matrix starts to fail. 
On the other hand, elastic deuterons originate from the center of target and 
produce no systematic offsets. \label{fig_optics_Positions}}
\end{minipage}
\vspace*{-5mm}
\end{center}
\end{figure}

The absolute calibrations of the target variables $\phi_{\mathrm{Tg}}$, 
$\theta_{\mathrm{Tg}}$ and $\delta_{\mathrm{Tg}}$ are shown in 
Fig.~\ref{fig_optics_Positions}. The results do not show any significant systematic
offsets of the reconstructed variables, except for $\phi_{\mathrm{Tg}}$ in
the case of elastic protons. There a constant offset of $\approx 6\,\mathrm{mrad}$ 
is observed that increases rapidly when the momentum becomes greater than $0.53\,\mathrm{GeV}/c$.
This can be explained by the fact that all elastic protons, due to the 
Mott cross-section, come from the up-stream end of the target at maximum angles 
$\phi_{\mathrm{Tg}}$ (minimum scattering angles $\theta_{e}$).
The behavior of the optics matrix in that regime is expected to become imprecise.
On the other hand, elastic deuterons may also come from the center of the target, where
the optics works best, and consequently generate no systematic offsets. 

\vspace*{-2mm}
The quality of the absolute calibration for the $y_{\mathrm{Tg}}$ was not investigated, because
the absolute precision of the BigBite variable $y_{\mathrm{Tg}}$ is not crucial for the experiment 
where only coincidence events are considered, and the superior reconstruction of 
$y_{\mathrm{Tg}}$ by the HRS-L spectrometer can be used instead.

\section{Optical Calibration of HRS-L}

\begin{figure}[!ht]
\begin{center}
\begin{minipage}[t]{0.6\textwidth}
\hrule height 0pt
\includegraphics[width=1\textwidth]{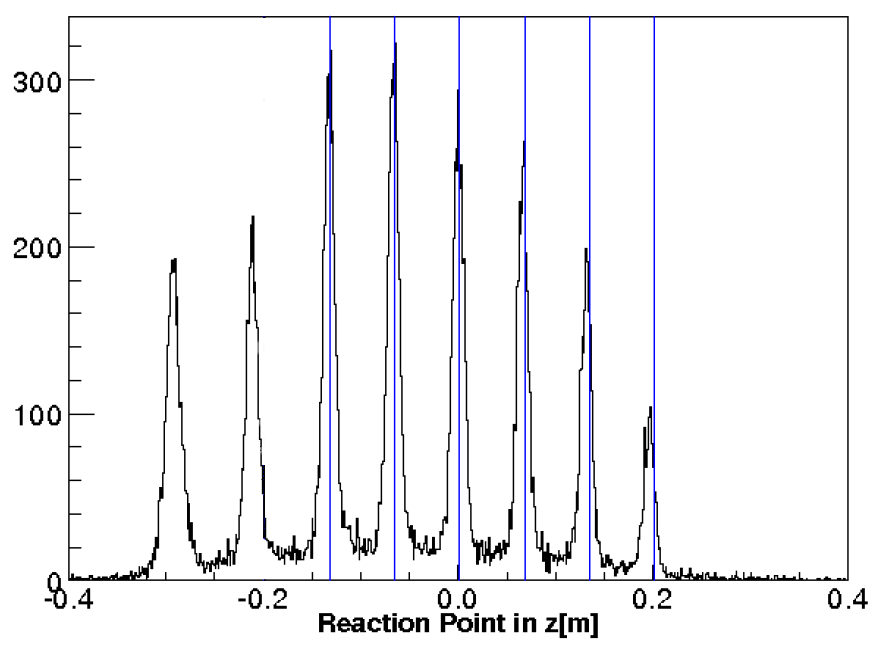}
\end{minipage}
\hfill
\begin{minipage}[t]{0.39\textwidth}
\hrule height 0pt
\caption{ The reconstruction of the seven-foil carbon target at $E_{\mathrm{Beam}} = 2.425\,\mathrm{GeV}$
and HRS-L angle of $14.5^\circ$. Note that the most upstream foil was mistakenly displaced 
in the experiment and its nominal position is not shown. The most left (eight) peak 
corresponds to the BeO window (see Sec.~\ref{sec:TargetSysyem}).
Figure is taken from Ref.~\cite{GeJinPhD}.
\label{fig_optics_HRSL_reactz}}
\end{minipage}
\end{center}
\end{figure}

For the description of the HRS-L magnetic optics, a polynomial parameterization identical to one
used for the BigBite spectrometer (see Sec.~\ref{sec:opticsmatrixformalism}), was considered. 
The optimization of spectrometer's optics matrix was performed by Ge Jin and is described 
in detail in Ref.~\cite{GeJinPhD}. The resolutions for all four target variables obtained 
with this process are given in Table~\ref{table_optics_HRSL_resolutions}.
\begin{table}[!ht]
\begin{center}
\begin{minipage}[t]{0.5\textwidth}
\hrule height 0pt
\begin{tabular}{lc}
\toprule
{\bf Reconstructed variable} & {\bf Resolution}\\
\midrule
Reaction Point $(z_{\mathrm{React}})$ & $6\,\mathrm{mm}$\\
Relative momentum $(\delta_{\mathrm{Tg}})$ & $2\times 10^{-4}$\\
Out-of-plane angle $(\theta_{\mathrm{Tg}})$ & $1.5\,\mathrm{mrad}$\\
In-plane angle $(\phi_{\mathrm{Tg}})$ & $0.5\,\mathrm{mrad}$\\
\bottomrule
\end{tabular}
\end{minipage}
\hspace*{0.5cm}
\begin{minipage}[t]{0.4\textwidth}
\hrule height 0pt
\caption{The resolutions of the reconstructed HRS-L target variables. 
Table is taken from Ref.~\cite{GeJinPhD}.
\label{table_optics_HRSL_resolutions}}
\end{minipage}
\end{center} 
\end{table}
The techniques considered in this calibration were very similar to those for the BigBite 
optics analysis. The calibration of the 
reaction point was performed by using the seven-foil carbon target. The final results
are shown in Fig.~\ref{fig_optics_HRSL_reactz}. 
The in-plane and out-of-plane target angles were calibrated using data sets, collected with the 
steel sheet with a pattern of 49 holes~\cite{alcorn} positioned in front of the HRS spectrometer. 
In the analysis of the sieve-slit data, those matrix elements for $\phi_{\mathrm{Tg}}^{\mathrm{HRS-L}}$ 
and $\theta_{\mathrm{Tg}}^{\mathrm{HRS-L}}$ were chosen that resulted in the best reconstructed sieve pattern. 

However, the matrix elements for the HRS-L momentum variable $\delta_{\mathrm{Tg}}^{\mathrm{HRS-L}}$
were determined differently than in the BigBite calibration. Here $\delta_{\mathrm{Tg}}$ could not be 
determined from the comparison of the particle momentum to the results from another spectrometer 
(i.e. comparison of the momentum to the momentum transfer vector as considered in the BigBite method).
Instead, a stand-alone calibration technique was employed  based on the formula for elastic 
electron-nucleus scattering, relating the momentum $p_{\mathrm{e'}}$ of the 
ejected electron to the scattering angle $\theta_{\mathrm{Tg}}^{\mathrm{HRS-L}}$ :
\begin{eqnarray}
  p_{\mathrm{e'}} = p_{\mathrm{c}}(1+\delta_{\mathrm{Tg}}) =  
        \frac{p_{\mathrm{e}}}{1+\frac{p_{\mathrm{e}}}{M}(1-\cos \theta_{\mathrm{Tg}}^{\mathrm{HRS-L}})}\,.\nonumber
\end{eqnarray}
Here, $p_{\mathrm{e}}$ is the momentum of the incident electron, $M$ is the mass of the target nucleus and
$p_{\mathrm{c}}$ is the central momentum of the spectrometer. Utilizing this formula, and knowing the 
scattering angle $\theta_{\mathrm{Tg}}$ and the target mass, the momentum of the scattered electron could be 
precisely determined. The comparison of calculated momenta for each event to the corresponding combination 
of directly measured detector coordinates then allows a precise momentum calibration of the spectrometer.

The optimization was performed by using hydrogen and deuterium elastic data for different momentum 
settings $p_{\mathrm{c}}$ of the spectrometer and for different beam energies. 
To improve the optics even further, the elastic scattering on carbon 
target was also utilized. With the carbon target, three additional peaks beside the elastic peak are visible 
in the momentum spectrum. They correspond to the first three excited states of the nuclei 
($\Delta E_1 = 4438.9\,\mathrm{keV}$,  $\Delta E_2 = 7654.2\,\mathrm{keV}$ and $\Delta E_3 = 9641.0\,\mathrm{keV}$) 
and represent a very rigid test of the quality of the magnetic transport optics.

\chapter{Data Analysis}
\label{chapter:DataAnalysis}

\section{Modus operandi}
\begin{figure}[!ht]
\begin{center}
\includegraphics[width=0.6\textwidth]{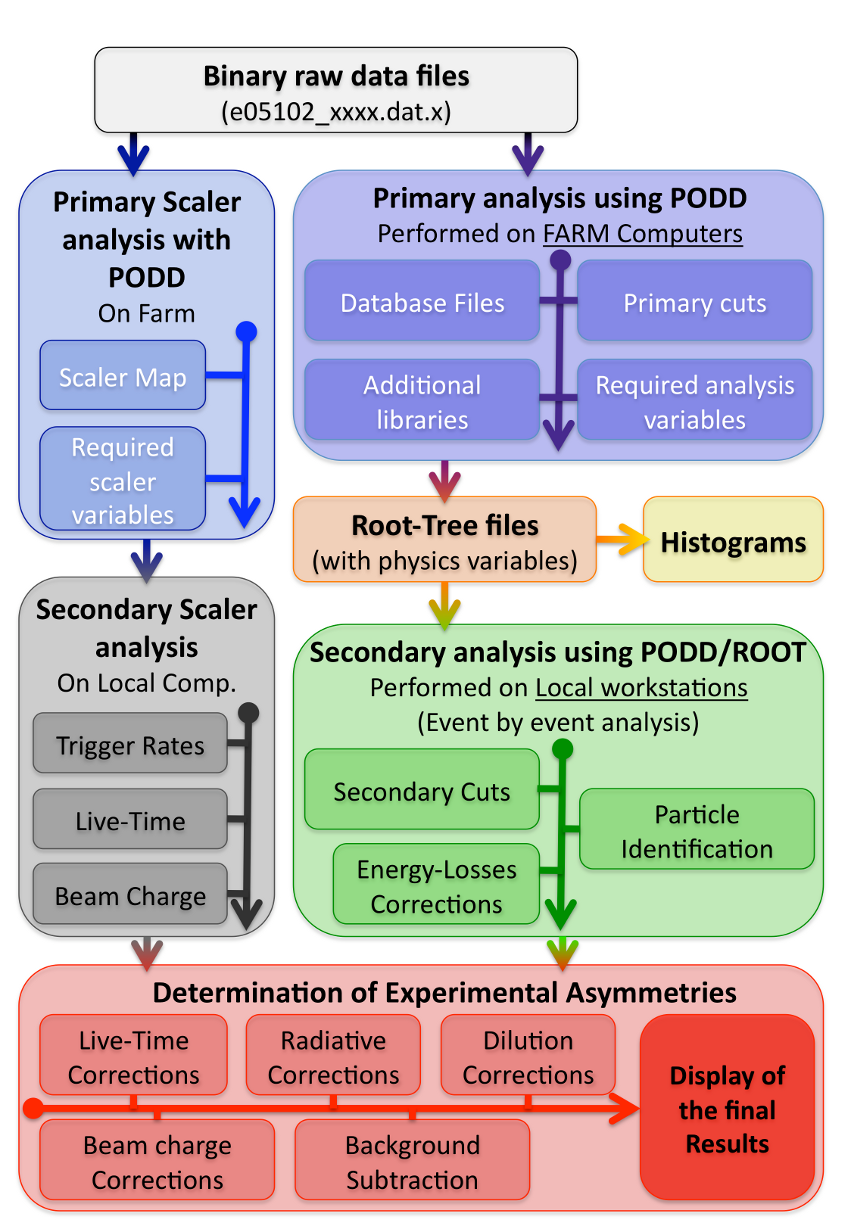}
\caption{ The data analysis flow chart.    
\label{fig_analysis_flowchart}}
\end{center}
\end{figure}

The analysis of the experimental data was accomplished in several steps, which are shown in 
Fig.~\ref{fig_analysis_flowchart}. The first part of the analysis was performed on the 
Jefferson Lab's FARM computers, where the raw experimental data-files were converted to the
root-tree files using the Hall A analysis framework Podd (see Sec.~\ref{sec:PODD}). 
For the successful analysis, Podd had to be adapted to the experimental configuration
of the E05-102 experiment by using correct analysis libraries and database (DB) files. The 
primary analysis was also furnished with a list of physics parameters required in succeeding steps
of the analysis. It was also given a list of primary cuts that were used for a first, rough filtering 
of events (see Sec.~\ref{sec:PrimaryCuts}).

The generated root-tree files contain physics variables such as 
particle momenta, spin orientations, beam helicity and momentum transfer vectors,
which are ready to be visualized with histograms. However, for extraction of experimental 
asymmetries, further, more detailed analysis is necessary. This secondary analysis requires
root-tree files as input and could be performed on a local workstation. Here, final 
cleaning of the data was performed, using secondary cuts (see Sec.~\ref{sec:SecondaryCuts}).
Particle identification (PID) was applied at this stage to isolate
deuteron channel (${}^3\mathrm{He(e, e'd)}$) from the proton channel (${}^3\mathrm{He(e, e'p)}$) events.
The details of PID are described in Sec.~\ref{sec:PID}. Once knowing the
identities of the particles detected by BigBite, their momenta could be 
appropriately corrected for energy losses. A the end of this analysis, two clean 
samples of events were obtained, one for each reaction channel. 

These samples  were then handled in the last part of the analysis, where
the experimental asymmetries were determined. The process of calculating the asymmetries
is described in Sec.~\ref{sec:Asymmetry}. The raw results had to be corrected for 
the dilutions caused by the finite target and beam polarizations, addition of 
nitrogen and presence of target cell-walls. Contributions to the asymmetry caused 
by the random coincidence background were also subtracted. 

Possible false asymmetries due to the beam charge asymmetry and dead-time asymmetry 
were investigated and considered in the results. These contributions were studied
through scaler analysis, which was performed independently of the main 
physics analysis. The first part of the scaler analysis was also performed on the 
FARM, while the second part, where the corrections to the asymmetry were determined, 
was done on the local computer. Findings of this analysis are presented in 
Sec.~\ref{sec:ScalerAnalysis}.

\section{Analysis software Podd}
\label{sec:PODD}
For the analysis of the experimental data, the Hall A physics analyzing tool Podd~\cite{Podd}
has been employed. It is considered for an on-line inspection of the data as well as for the 
final off-line analysis.  It was introduced in 2004 and replaced the old Fortran-based software 
ESPACE~\cite{alcorn}. It is an object-oriented code written in C++ and is built on top 
of the CERN Root analysis platform~\cite{CERNRoot}. Podd reads the information from 
raw data files and for each recorded event transforms detected electronic signals
into physically meaningful quantities. These are then filled into the Root 
trees for further analysis. The analyzer contains modules (classes) for the analysis of all 
standard Hall A instruments, such as beam quality monitors (BPMs and BCMs), HRS spectrometers, 
target and scaler counters. 

There have been continuous efforts among the developers to improve and update the Podd's code. 
A new version is released every year. For the analysis of the data collected in the experiment E05-102, 
version 1.5.12, released on March 12, 2010, has been utilized.

For the experiments utilizing BigBite and HAND in addition to HRSs,
the analyzer has been extended with additional libraries, to accommodate this additional
equipment. The complete interpretation of the signals detected with the BigBite and HAND 
is performed by the BigBite library (\texttt{libBigBite.so}), written by Ole Hansen and Jin 
Huang~\cite{OleThreeSearch,bigbitelib}. 
However, for the analysis of the E05-102 data, two more extra libraries were employed. 
The VertexTime library (\texttt{libVertexTime.so}) contains the implementation of the analytical model of 
the BigBite optics (see Appendix~\ref{appendix:AnalyticalModel}), which was exploited 
for the reconstruction of the BigBite vertex variables. On the other hand, the HadronDetectorPackage
library \\(\texttt{libBigBiteHadronDetectorPackageEdE.so}) was considered for more detailed calibration 
of the BigBite scintillation detectors, described in Sec.~\ref{sec:ADCCalibration}.

Podd can be installed and executed on any computer with a running ROOT. However, it turned
out that a single workstation computer is insufficient for the complete analysis of the
experimental data. In the experiment $\approx 1500$ data-sets were collected, each containing 
$\approx 10\,\mathrm{GB}$ of data, resulting in total amount of $\approx 15\,\mathrm{TB}$ of data,
which could not be stored on a single machine. Furthermore, additional space is required 
for the analyzed data files. Each analyzed ROOT file (containing only variables necessary for
the final analysis) requires $\approx5\,\mathrm{GB}$ of disk space. The reason for such large files
(in principle they could be smaller than $1\,\mathrm{GB}$) is related to the problem with the 
library (\texttt{gzip}) used to compress the ROOT files. Because the library was 
crashing uncontrollably, the option of compressing the ROOT files had to be abandoned.

\begin{figure}[hbtp]
\begin{center}
\includegraphics[width=\textwidth]{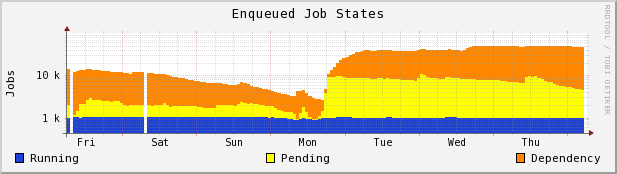} 
\vspace*{-3mm}
\caption{The batch farm load. The blue band shows the number of jobs
that are being simultaneously executed on the farm ($\approx 1000$).
The yellow band shows the number of jobs that are ready and waiting to be executed,
while the orange band shows the jobs  waiting in the queue, but were not
yet prepared for execution (i.e. the files are not yet loaded from the storage 
silo)~\cite{JLab_scicomp}.
\label{fig_FARMJobs}}
\end{center}
\end{figure}

The important limitation in the analysis was also the time required to analyze the raw data. 
Each data-set required $\approx 4\,\mathrm{hours}$ on a work-station computer to be 
analyzed. If all the data would be analyzed sequentially on a single computer, it would take 
$\approx 250\,\mathrm{days}$ to complete the analysis. 

Because of these constraints, the single workstation analysis was considered only 
during the development of the analysis scripts and for setting the proper values of cuts
used for event filtering. The complete analysis of the data was then 
performed on the Jefferson Lab's batch farm. The farm's cluster contains $\approx 100$ computing 
nodes, running CentOS 5.3. The cluster has the capacity of running $\approx 1000$ simultaneous 
batch jobs (see Fig.~\ref{fig_FARMJobs}), with the limitation of 256 
jobs per user at one time~\cite{JLab_scicomp}. The farm is also directly connected to 
the $7\,\mathrm{PB}$ storage silo where all raw data-files are stored.

With the use of farm computer the analysis of the E05-102 experimental data was 
accomplished in only a few weeks. The obtained root files containing the event trees
filled with only best coincidence events were then transfered from the farm to 
the workstation computer for further analysis. From that point on, the analysis 
scripts were not using raw signals any more (except if required), but dealt only with physical 
quantities, which made the search of the final experimental results much easier and
faster.

\section{Measurement of Double Polarized Asymmetries}
\label{sec:Asymmetry}
The experimental asymmetry $A_{\mathrm{exp}}$ is determined as the relative difference 
between the number of coincidence events, collected with positive and negative
beam helicity:
\begin{eqnarray}
A_\mathrm{exp} = { N^+ - N^- \over N^+ + N^- } \>. \label{eq_analysis_asym1}
\label{aexp}
\end{eqnarray}
Here $+/-$ represent the positive and negative beam helicity state, respectively. 
The total number of collected events is $N = N^+ + N^-$.  

Due to the experimental limitations, the number of electrons which hit the target 
with positive and negative spin direction is not the same. This can
introduce unwanted false asymmetries. To ensure that the 
measured asymmetry originates only in the given physical 
process, the amount of electric charge $Q^{\pm}$ that hit the target with helicity 
states $+$ and $-$, must be considered. In addition, the differences in dead-times 
for detecting events with each helicity state, $t_d^\pm$, also need to be accounted for.
Employing these corrections to Eq.~(\ref{eq_analysis_asym1}), the experimental 
asymmetry can be re-written as: 
\begin{eqnarray}
A_\mathrm{exp} = { N^+ / (Q^+ t_\mathrm{l}^+) - N^- / (Q^- t_\mathrm{l}^-) \over
N^+ / (Q^+ t_\mathrm{l}^+) + N^- / (Q^- t_\mathrm{l}^-) } = \frac{Y^+ - Y^-}{Y^+ + Y^-} \>,  \label{eq_analysis_asym2}
\end{eqnarray}
where $t_l^\pm = T^\pm - t_d^\pm$ is the live-time (the true data collecting
time for each helicity state).  $T^\pm$ is the total amount of time of beam having 
a positive (negative) beam helicity. The ratios $Y^\pm = \frac{N^\pm}{Q^\pm t_l^\pm}$ are defined as
yields for each helicity state. 

In the experiment, the asymmetries $A_{\mathrm{exp}}$ were measured with the target oriented
along the beam-line ($A_{\mathrm{exp}^\parallel}$) and with the target oriented perpendicularly
to the beam direction ($A_{\mathrm{exp}^\perp}$). With respect to the physics asymmetries
given with Eqs.~(\ref{eq_theory_finalasymmetry_pd}) and (\ref{eq_theory_finalasymmetry_ppn}), these 
experimental asymmetries are reduced due to dilutions present in the target (described by parameter $R$)
and because nor the target nor the beam are completely polarized $(P_\mathrm{t}, P_\mathrm{e} < 100\,\%)$. 
This way, the physics asymmetries can be expressed as:
\begin{eqnarray}
  A^\parallel = { A_\mathrm{exp}^\parallel \over P_\mathrm{e}^\parallel \cdot P_\mathrm{t}^\parallel \cdot R } \>,
\qquad
A^\perp = { A_\mathrm{exp}^\perp \over P_\mathrm{e}^\perp \cdot P_\mathrm{t}^\perp \cdot R } \>, \label{eq_analysis_phys_asym}
\end{eqnarray}

The  polarization of the target was measured every four hours (approximately after every four data-sets) 
using NMR polarimetry (see Sec.~\ref{sec:NMR}). For the data accumulated between two 
consecutive NMR measurements, the polarization was determined by using linear interpolation. This way, 
a target polarization was assigned precisely to each data-set considered in the analysis.

The polarization of the beam  was determined with the M\o{}ller polarimeter (see
Sec.~\ref{sec:Moller}). Since only four measurements were performed during the experiment, we decided to 
calculate the mean value of the beam polarization ($P_e = 84.3\,\mathrm{\%} \pm 2\,\mathrm{\%}$) and use it 
for the analysis of all production data.

The dilution factor $R$ describes how much the measured asymmetry $A_{\mathrm{exp}}$ is reduced due to the 
presence of unpolarized gases inside the polarized ${}^3\mathrm{He}$ target cell. The dominant contribution 
comes from the $\mathrm{N}_2$ gas, which is added ($\approx 1\,\mathrm{\%}$) 
to provide quenching of the de-excitation photons, in order to prevent the Rb atoms from 
depolarizing (see Sec.~\ref{sec:SEOP}). Hence, in the experiment, the incident electrons may 
scatter also off nitrogen nuclei. The unpolarized $N_2$ is expected to produce negligible false 
asymmetries. However, it will increase the total yield and dilute the asymmetry:
\begin{eqnarray}
  A_{\mathrm{exp}} = \frac{\left(Y^+ + \frac{Y_{\mathrm{N_2}}}{2}\right) - \left(Y^- + \frac{Y_{\mathrm{N_2}}}{2}\right)}
           {\left(Y^+ + \frac{Y_{\mathrm{N_2}}}{2}\right) + \left(Y^- + \frac{Y_{\mathrm{N_2}}}{2}\right)} = 
   \frac{Y^+ - Y^-}{\left(Y^+ + Y^-\right)R^{-1}} = R\,A\,, \nonumber
\end{eqnarray}
where the dilution factor $R \approx (1-Y_{\mathrm{N_2}}/Y_{\mathrm{Tot}})$, $Y_{\mathrm{N_2}}$ is the nitrogen yield
and $Y_{\mathrm{Tot}} = Y^+ + Y^-$ is the total ${}^3\mathrm{He}$ yield.

The analysis of the dilutions was performed by Ge Jin~\cite{GeJinPhD}. The dilution factor was determined 
through the analysis of the measurements utilizing a reference-cell filled with nitrogen gas. The obtained 
yield corresponds to the $N_2$ pressure inside the cell. Performing measurements for different
target pressures, the pressure curve was obtained, relating each target pressure to the yield.  Knowing 
the relative nitrogen pressure inside the production cell, this curve could be used to estimate 
the nitrogen yield. It was determined to be $R = 96\,\mathrm{\%} \pm 2\,\mathrm{\%}$.


Once knowing the asymmetries (Eqs.~(\ref{eq_analysis_phys_asym})), their statistical uncertainties can be expressed
as:
\begin{eqnarray}
	\Delta A = \frac{1}{P_e\,P_t\,R}\Delta A_{\mathrm{exp}}\,, \nonumber
\end{eqnarray}
where the uncertainty can be written in terms of $Y^\pm$ as:
\begin{eqnarray}
\Delta A_\mathrm{exp} 
  = \sqrt{ 
    \left( {\mathrm{d}A_\mathrm{exp}\over \mathrm{d}Y^+}\,\Delta Y^+\right)^2
  + \left( {\mathrm{d}A_\mathrm{exp}\over \mathrm{d}Y^-}\,\Delta Y^-\right)^2
    } \label{eq_analysis_eq1}
\end{eqnarray}
The derivatives are directly calculable:
\begin{eqnarray}
  \frac{\mathrm{d}A_\mathrm{exp}}{\mathrm{d}Y^\pm} = \frac{\pm 2Y^\mp}{(Y^+ + Y^-)^2} = \pm\frac{1\mp A_{\mathrm{exp}}}{Y^+ + Y^-}\,.\label{eq_analysis_eq2}
\end{eqnarray}
Considering that the $N^\pm$ obeys the Poissonian distribution, the error of the yield becomes
\begin{eqnarray}
	\Delta Y^\pm = \frac{1}{Q^\pm t_l^\pm}\Delta N^\pm = \frac{\sqrt{N^\pm}}{Q^\pm t_l^\pm} = \sqrt{\frac{Y^\pm}{Q^\pm t_l^\pm}}\,.\label{eq_analysis_eq3}
\end{eqnarray}
Using Eqs.~(\ref{eq_analysis_eq2}) and~(\ref{eq_analysis_eq3}) in Eq.~(\ref{eq_analysis_eq1}), the error of the experimental 
asymmetry assumes the form:
\begin{eqnarray}
  \Delta A_\mathrm{exp} = \sqrt{\frac{(1-A_{\mathrm{exp}})^2 \frac{Y^+}{Q^+ t_l^+} + (1+A_{\mathrm{exp}})^2 \frac{Y^-}{Q^- t_l^-}}{(Y^+ + Y^-)^2}}\,.
   \label{eq_analysis_eq4}
\end{eqnarray}
Let us assume that the dead-times do not depend on the helicity state $(t_l^+ = t_l^-)$ and that there is no beam charge 
asymmetry, $Q^+ = Q^-$. We also acknowledge that the expected asymmetries are small $(A_\mathrm{exp} \approx 0)$, which means that:
\begin{eqnarray}
	Y^+ \approx Y^- \approx \frac{Y_{Tot}}{2}\,. \nonumber
\end{eqnarray}
By applying all these assumptions to Eq.~(\ref{eq_analysis_eq4}), the statistical error for the experimental asymmetry 
can be estimated in a straightforward manner as:
\begin{eqnarray}
	\Delta A_\mathrm{exp} \approx \frac{1}{\sqrt{N}}\,.\nonumber
\end{eqnarray}

\section{Coordinate systems}
\begin{figure}[!ht]
\begin{center}
\begin{minipage}[t]{0.69\textwidth}
\hrule height 0pt
\includegraphics[width=1\linewidth]{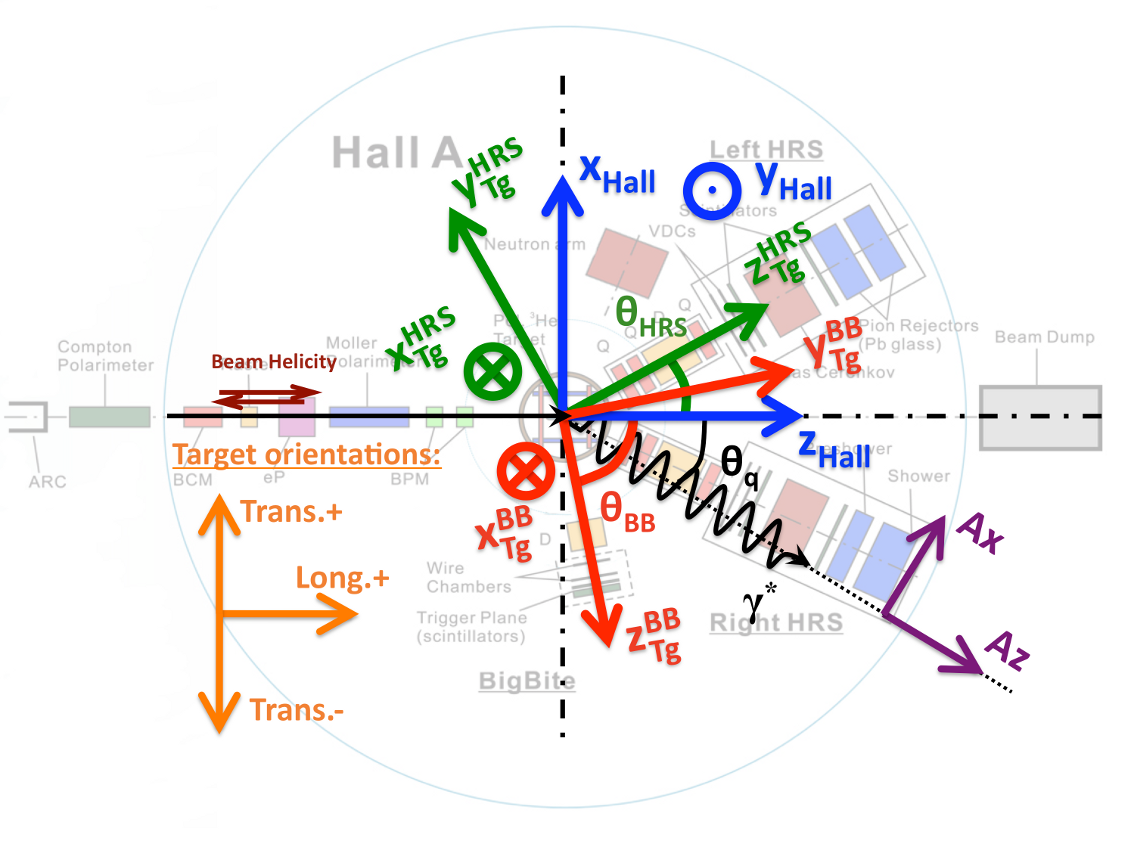} 
\end{minipage}
\begin{minipage}[t]{0.3\textwidth}
\hrule height 0pt
\caption{The coordinates systems considered in the analysis of the
experimental data. Hall coordinate system is oriented with respect to the 
direction of beam-line. The coordinate systems of two spectrometers (HRS and BB) are
aligned with respect to the spectrometer axes.\label{fig_Analysis_CS}}
\end{minipage}
\end{center}
\end{figure}

In the analysis of the experimental data, three coordinate systems were relevant. The first is the Hall A coordinate system (HCS).
The origin of this system is at the center of the experimental hall. As indicated in Fig.~\ref{fig_Analysis_CS}, its $\hat{z}_{Hall}$-axis 
is pointing along the direction of the beam, $\hat{x}_{Hall}$-axis is pointing to the left (in-plane) of the beam direction, and 
the $\hat{y}_{Hall}$-axis is pointing vertically upwards. This is the central coordinate system and is used in calculations of 
all direction-dependent physics quantities required by the analysis, such as the target-spin vector, momentum-transfer vector and 
the recoil momentum.

Beside the Hall coordinate system, each of the two spectrometers (HRS, BB) has its own target coordinate system (TCS). 
The $\hat{z}_{\mathrm{Tg}}$ axis is pointing along the spectrometer central trajectory. 
The $\hat{x}_{\mathrm{Tg}}$ is oriented in the spectrometer mid-plane, perpendicular to $\hat{z}_{\mathrm{Tg}}$ and pointing downwards. 
The $\hat{y}_{\mathrm{Tg}}$ is defined by $\hat{y}_{\mathrm{Tg}} = \hat{z}_{Tg} \times \hat{x}_{\mathrm{Tg}}$. In ideal circumstances
the origins of the TCSs coincide with the origin of the HCS. Any potential offsets from the center can be compensated through
the corrections to the spectrometer optics matrix. The primary purpose of the  spectrometer coordinate systems is to
present the particle's target coordinates $(y_{\mathrm{Tg}}, \theta_{\mathrm{Tg}}, \phi_{\mathrm{Tg}}, \delta_{\mathrm{Tg}})$, reconstructed
by the spectrometer optics (see Chapter~\ref{chapter:Optics}). 

The calculation of the physics variables, deducted from the particle coordinates and momenta, 
is performed in the HCS. The particle coordinates, measured by spectrometers, must therefore 
be transformed from the TCS to the HCS, by using rotations:
\begin{eqnarray}
\left(
\begin{array}{c}
x_{\mathrm{Hall}}\\
y_{\mathrm{Hall}}\\
z_{\mathrm{Hall}}
\end{array}
\right) = 
\left(
\begin{array}{rrr}
0 & \cos\Theta & \sin\Theta \\
-1 & 0 & 0 \\
0 & -\sin\Theta & \cos\Theta \\
\end{array}
\right)
\left(
\begin{array}{c}
x_{\mathrm{Tg}}\\
y_{\mathrm{Tg}}\\
z_{\mathrm{Tg}}
\end{array}
\right)\,. \nonumber
\end{eqnarray} 
When performing the rotation from the HRS-L TCS to the HCS,  $\Theta = \theta_{HRS} >0$.
However, when rotating from BigBite coordinate system,  $\Theta = \theta_{BB} <0$. 

Due to the limitations of the target laser system (see Sec.~\ref{sec:TargetSysyem}), the target could be 
polarized only along the beam line and perpendicularly to it (see Fig.~\ref{fig_Analysis_CS}). This way,  
only the asymmetries $A_{z_{\mathrm{Lab}}}$ and $A_{x_{\mathrm{Lab}}}$ corresponding to these target spin 
orientations could be measured. Physically more relevant are
the asymmetries $A_{z_{\vec{q}}}$  and $A_{x_{\vec{q}}}$, where the target spin is oriented parallel and 
perpendicular to $\vec{q}$. In the in-plane approximation ($\phi^* = 0^\circ$ or $180^\circ$), and assuming  
that the target spin is ideally aligned with the HCS axes, simple transformations  
can be deduced from Eq.~(\ref{eq_theory_finalasymmetry_pd}), to calculate 
$A_{z_{\vec{q}}}$  and $A_{x_{\vec{q}}}$ from $A_{z_{\mathrm{Lab}}}$ and $A_{x_{\mathrm{Lab}}}$:
\begin{eqnarray}
\left[
\begin{array}{c}
A_{z_{\vec{q}}}\\
A_{x_{\vec{q}}}
\end{array}
\right] = 
\left(
\begin{array}{rr}
\cos\left(|\theta_{\vec{q}}|\right) & -\sin\left(|\theta_{\vec{q}}|\right) \\
\sin\left(|\theta_{\vec{q}}|\right) & \cos\left(|\theta_{\vec{q}}|\right)
\end{array}
\right)
\left[
\begin{array}{c}
A_{z_{\mathrm{Lab}}}\\
A_{x_{\mathrm{Lab}}}
\end{array}
\right]\,. \label{eq_analysis_asymmetry_rotations}
\end{eqnarray}
Here the angle $\theta^*$ from Eq.~(\ref{eq_theory_finalasymmetry_pd}) was replaced by 
$\theta^* = |\theta_{\vec{q}}|$.

However, in the analysis of the experimental data, we prefer not to use 
rotations given by Eq.~(\ref{eq_analysis_asymmetry_rotations}) to interpret 
our results. We rather consider the theoretical predictions that were 
calculated for our experimental settings. This way we are able 
to examine also the out-of-plane configurations $\phi^* \neq 0^\circ, 180^\circ$
and consider small deviations of the target spin orientation from the 
selected direction (see Table~\ref{table_HoldingCoils}), in order to 
establish the most meaningful comparison between the measurements and the theory. 

\section{Identification of particles}
\label{sec:PID}
BigBite and HRS-L are able to detect various particles, also many unwanted
ones, which could pollute our data. However, the proper choice of kinematic 
settings and selection of coincidence events, reduces the contamination with 
other particles to a minimum. This way we are left only with electrons in 
HRS-L and protons and deuterons in BigBite. 

BigBite in principal detects also $\pi^+$, but they are suppressed by setting 
the triggering thresholds to high values, choosing only strongly-ionizing particles. 
The detection of heavier nuclei, like ${}^3\mathrm{He}$, is 
also highly unlikely with BigBite. Because of the enormous energy losses in the air, these particles 
stop inside the spectrometer before reaching the detector package.

In spite of that, proper particle identification (PID) is essential for the 
success of the E05-102 experiment. Since we are performing measurements of two 
reaction channels simultaneously, we need to be certain which particle was 
detected by the spectrometer. In the collected events there is approximately 
only $1-2\,\mathrm{\%}$ of deuterons. The rest are protons, which means that 
deuterons must be well distinguished from the protons, otherwise, the 
information about the deuteron channel will be lost. Two different approaches 
were utilized for the the BigBite PID: Coincidence time approach and 
the Energy-deposit approach.

\subsection{Coincidence time approach}
The first technique was based on the coincidence time information. After a nuclear
reaction, the ejected electron and hadron require times $t_e$ and $t_p$, respectively,
to reach the detector packages. The flight-paths of electrons inside the HRS-L can 
change maximally by $0.5\,\mathrm{m}$. Because electrons travel with almost the 
speed-of-light, the maximum spread in the electrons' time-of-flight 
is $\leq 2\,\mathrm{ns}$. Hence, the electron time-of-flight is almost constant. 
On the other hand, the hadron time-of-flight inside BigBite depends strongly on 
the particle momentum. The coincidence time 
can be expressed as:
\begin{eqnarray}
t_{\mathrm{Coinc}}(p_p, m_p) = t_p - t_e = \frac{s}{c_0}\sqrt{1 + \frac{m_p^2c^2}{p_p^2}} 
 - t_0\,, \label{eq_analysis_PID_TOF}
\end{eqnarray}
where $p_p$ and $m_p$ are the momentum and mass of a hadron detected by BigBite
and $c_0$ is the speed-of-light. The constants $s$ and $t_0$ are parameters 
of this equation. The $s$ represents the particle's effective flight-path inside
BigBite, while $t_0$ combines the electron's time-of-flight and some additional 
time offsets caused by the cabling. In the analysis the parameters were set
to $s = 3.15\,\mathrm{m}$ and $t_0 = 20.8\,\mathrm{ns}$. 

\begin{figure}[!ht]
\begin{center}
\includegraphics[width=0.49\linewidth]{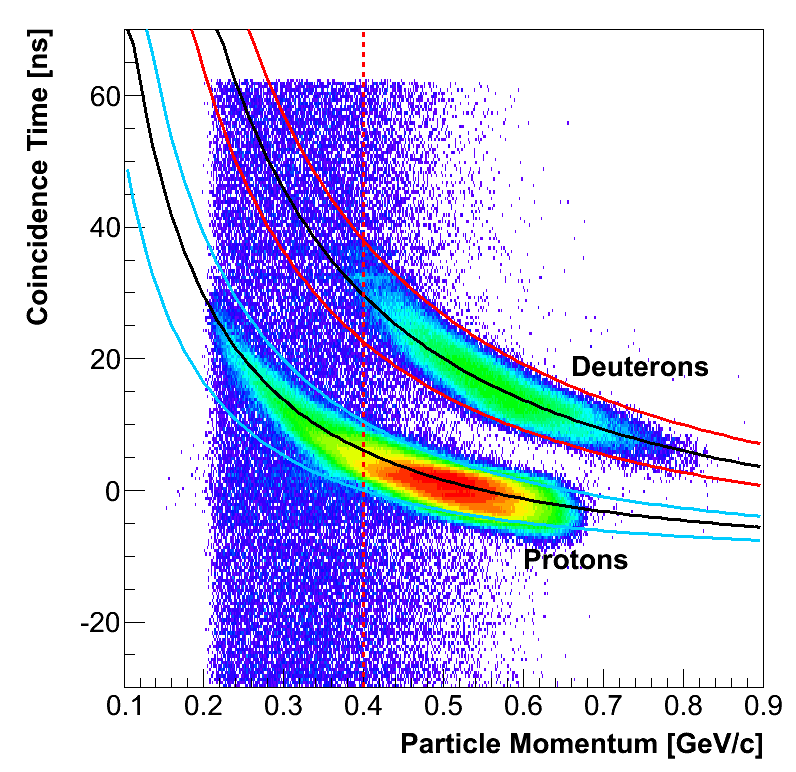} 
\includegraphics[width=0.49\linewidth]{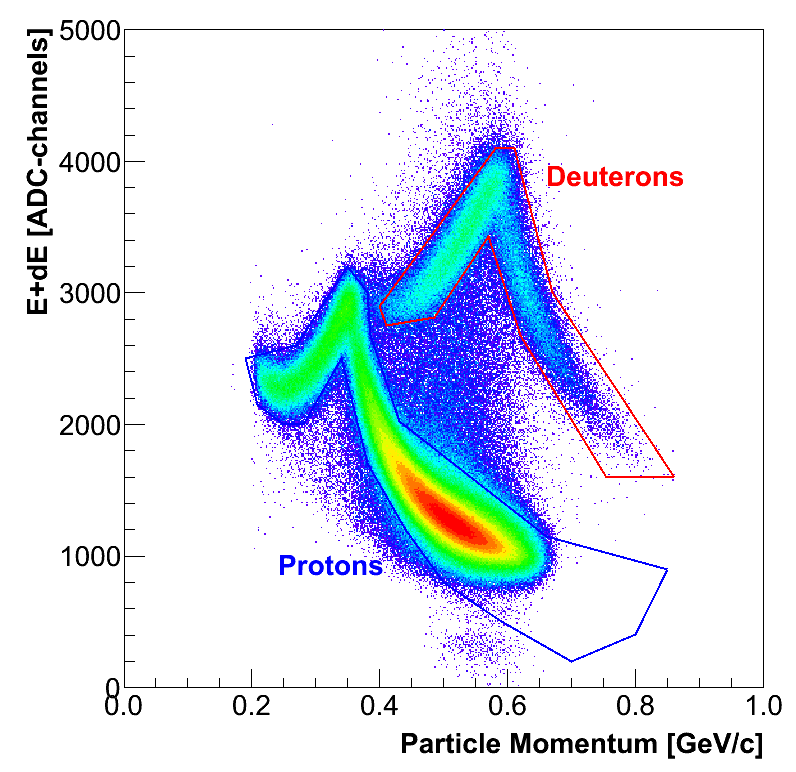} 
\caption{[Left] The coincidence time as a function of particle momentum. 
Deuterons have larger coincidence times than protons due to their larger mass. 
Consequently they can be clearly separated from the protons. 
Black lines represent calculated coincidence times using Eq.~(\ref{eq_analysis_PID_TOF}). 
Full red and blue lines show the limits of the PID. If the particle lies within the 
red lines it is recognized as a deuteron. If it is found within the blue lines, 
it is identified as a proton. The dashed red line represents an additional cut
for deuterons. 
[Right]  Energy deposit inside the scintillation
detector as a function of particle momentum. Because deuterons are slower, 
they lose more energy in the scintillators and generate stronger pulses.  
This can be used for PID. When a particle is found within the red area, it 
is recognized as deuteron. If it is found  inside the blue poly-line, it is 
identified as a proton.
\label{fig_analysis_PID1}}
\end{center}
\end{figure}

Equation~(\ref{eq_analysis_PID_TOF}) can be used to predict coincidence times 
for both protons and deute\-rons. The comparison with the measurements is 
presented in Fig.~\ref{fig_analysis_PID1}. The model curve (black) adequately 
describes both protons and deuterons. Due to the finite resolution 
the measured data are not gathered on a thin line, but generate 
a band with finite width. The limits of both bands were determined from the 
two-dimensional histogram shown in Fig.~\ref{fig_analysis_PID1}. To describe 
these limits, relation (\ref{eq_analysis_PID_TOF}) was considered,
but computed for slightly different particle masses. Four boundary curves
obtained by this procedure where employed for the PID. The following 
criteria were determined for protons and deuterons:
\begin{eqnarray}
\begin{array}{rc}
\mathrm{Protons:} & t_{\mathrm{Coinc}}(p_p, 0.68\,\mathrm{GeV}/c^2)\leq 
  t_{\mathrm{Coinc}}^{\mathrm{Protons}} \leq t_{\mathrm{Coinc}}(p_p, 1.12\,\mathrm{GeV}/c^2)\,, \\
 & \\
\mathrm{Deuterons:} & t_{\mathrm{Coinc}}(p_p, 1.6\,\mathrm{GeV}/c^2)\leq 
  t_{\mathrm{Coinc}}^{\mathrm{Deuterons}} \leq t_{\mathrm{Coinc}}(p_p, 2.2\,\mathrm{GeV}/c^2)\,.
\end{array}\label{eq_analysis_PID_TOFconditions}
\end{eqnarray}
To select deuterons, an additional cut was applied. From the calculations of energy-losses
it was determined (see Fig.~\ref{fig_EnLoss_EvsP}) that only deuterons with momenta 
$p_p \geq 0.34\,\mathrm{GeV}$ at the entrance to the detector package have enough 
strength to penetrate to the E-plane scintillators. In the analysis 
we select only events with valid hits in all detectors (full events). Hence, 
this momentum condition could be added to the PID, since below this limit no 
deuterons should exist. It also turned out that there are relatively 
few deuterons in that region with respect to the proton background. Therefore, 
it was decided to use an even more restrictive cut (shown with a
red dashed line in Fig.~\ref{fig_analysis_PID1}), and reject all particles 
with momenta $p_p \leq 0.4\,\mathrm{GeV}$.

The obtained method works well for  particles with momenta $p_p \geq 0.55\,\mathrm{GeV}/c$. 
Below this limit, the PID gets disturbed by the random coincidence background.
These events are mostly protons, and are not constrained by Eq.~(\ref{eq_analysis_PID_TOF}). 
Consequently they can appear inside the deuteron region limited by 
Eqs.~(\ref{eq_analysis_PID_TOFconditions}), and get mistakenly identified as deuterons.

\subsection{Energy-deposit approach}

The second PID method is based on the particle energy-losses inside the scintillation 
detectors. As demonstrated in Figs.~\ref{fig_BBADC_EdE} and~\ref{fig_EnLoss_EvsP}, 
deuterons lose more energy in the scintillation material than protons. This can be
exploited for PID. 

Several different combinations of detector information were considered in order to 
get a good separation of deuterons from protons.  In the end it turned out that the deuterons
and protons could be distinguished best when the energy lost in both detectors 
(E+dE) was compared to the particle momentum (see Fig.~\ref{fig_analysis_PID1}).
PID using only scintillation detector information (dE vs. E), 
fails for high momentum particles, where the deuteron band can no longer be separated 
from the proton band (Fig.~\ref{fig_BBADC_EdE}). The method, which
combined the energy deposit information from the E-plane with the particle momentum, 
performed better. However, this time the problems appeared in the low momentum 
region, where the proton and deuteron trails intersect (see Fig.~\ref{fig_EnLoss_EvsP}),  
which again makes PID impossible. As an alternative, 
the energy deposit inside the dE-plane in combination with particle momentum 
could be used for the PID. Here, the deuteron band always remains on the top of the 
proton band. However, it is always better to use the information from both detectors, 
rather than from a single one. Therefore, the first approach (dE+E vs. p) was chosen as
the primary PID, while the latter approach (dE vs. p) served as an additional 
constraint.  

The PID was managed by defining two regions: one for deuterons and one for protons.
As  shown in Fig.~\ref{fig_analysis_PID1}, the limits of both regions were 
determined by polygons whose edges were determined empirically.  To reduce
the chance of misidentification of low-momentum deuterons, restrictive cuts
were employed in the low-momentum region. However, to allow proper identification
of high momentum particles ($p_p \geq 0.7\,\mathrm{GeV}/c$), which could be obtained 
from the $3\,\mathrm{GeV}$ data, broader cuts were considered in the high-momentum 
region. The final points used to generate polygons for both types of particles 
are gathered in Table~\ref{table_analysis_PIDpoints}. 

\begin{table}[!ht]
\begin{flushleft}
\caption{Points defining two polygons used by the energy-loss PID method
for identifying protons and deuterons. The generated polygons are 
demonstrated in Fig.~\ref{fig_analysis_PID1}.
\label{table_analysis_PIDpoints}}
\vspace*{2mm}
\begin{tabular}{|c||c|c|c|c|c|c|c|c|c|c|}
\hline

\hline

\hline

\multicolumn{11}{|c|}{{\bf Deuterons}}\\
\hline
Point & 1 & 2 & 3 & 4 & 5 & 6 & 7 & 8 & 9 & 10 \\
\hline
$p_{\mathrm{Det}}$ & 0.4 & 0.41 & 0.485 & 0.57 & 0.62 & 0.755 & 0.86 & 0.67 & 0.61 & 0.58 \\
$[\mathrm{GeV}/c]$ & & & & & & & & & &\\
$ADC$ & 2900.0 & 2750 & 2810 & 3430 & 2680 & 1600 & 1600 & 3000 & 4100 & 4100 \\
$[\mathrm{channels}]$ & & & & & & & & & &\\
\hline

\hline

\hline
\end{tabular}
\end{flushleft}
\begin{flushleft}
\begin{tabular}{|c||c|c|c|c|c|c|c|c|c|c|}
\hline

\hline

\hline

\multicolumn{11}{|c|}{{\bf Protons}}\\
\hline
Point & 1 & 2 & 3 & 4 & 5 & 6 & 7 & 8 & 9 & 10 \\
\hline
$p_{\mathrm{Det}}$ & 0.19 & 0.208 & 0.250 & 0.280 & 0.34 & 0.380 & 0.44 & 0.5 & 0.59 & 0.7 \\
$[\mathrm{GeV}/c]$ & & & & & & & & & &\\
$ADC$ & 2500 & 2150 & 2000 & 2000 & 2500 & 1700 & 1200 & 800 & 500.0 & 200\\
$[\mathrm{channels}]$ & & & & & & & & & &\\
\hline

\hline

\hline
\end{tabular}
\end{flushleft}
\vspace*{-8mm}
\begin{flushleft}
\begin{tabular}{|c||c|c|c|c|c|c|c|c|}
\hline

\hline

\hline
Point & 11 & 12 & 13 & 14 & 15 & 16 & 17 & 18\\
\hline
$p_{\mathrm{Det}}$ & 0.8 & 0.85 & 0.66 & 0.430 & 0.385 & 0.38 & 0.35 & 0.270 \\
$[\mathrm{GeV}/c]$ & & & & & & & & \\
$ADC$ & 400 & 900 & 1150 & 2020 & 2680 & 3000 & 3200 & 2600 \\
$[\mathrm{channels}]$ & & & & & & & &\\
\hline

\hline

\hline
\end{tabular}

\end{flushleft}
\end{table}

By using this method, the identity of the detected particle
is determined by checking which polygon the particle is found in. If it is found 
inside the red polygon it is identified as a deuteron. If it is found inside the
blue polygon it is recognized as a proton. If the detected particle can not be found in 
any of the two regions, it is labeled as improper. With this constraint we mostly 
get rid of the protons that leave bogus signals in the scintillation detectors and
consequently form a proton tail in the region between two polygons. Unfortunately 
this tail reaches the deuteron polygon, where it causes false identifications. 
  
\subsection{Comparison of the PID methods}

\begin{figure}[!ht]
\begin{center}
\includegraphics[width=\linewidth]{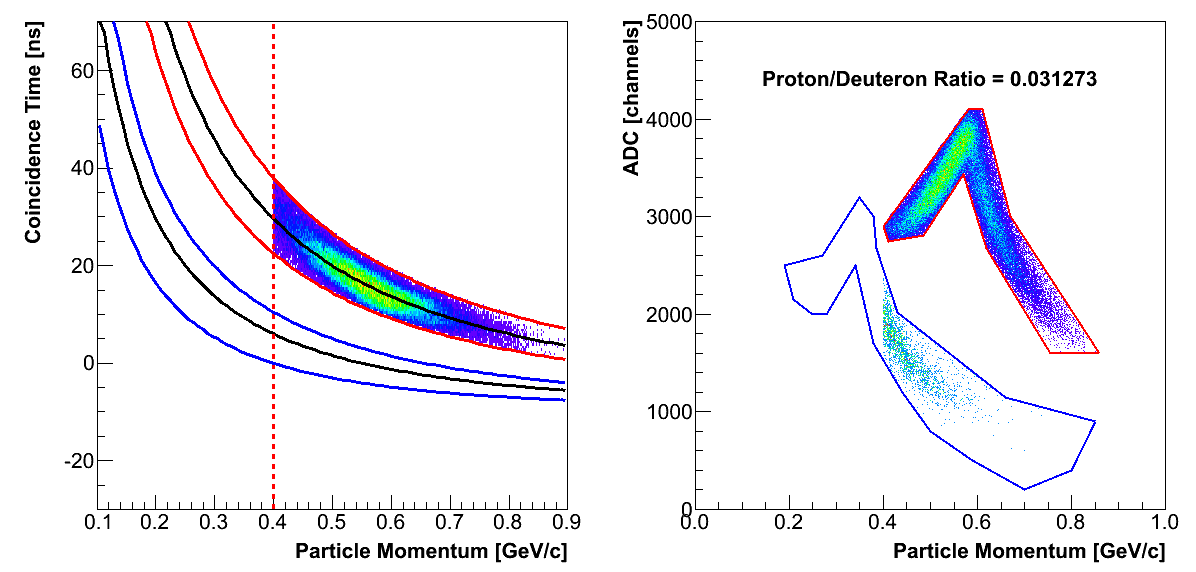} 
\vspace*{-7mm}
\caption{The efficiency of the Coincidence time PID for deuterons. The method is applied
to the production data to choose only deuterons (left figure). The selected events are then 
introduced to the Energy-loss PID (right figure) to check how many protons are mistakenly identified 
as deuterons. The obtained proton/deuteron ratio for the $Q^2 = 0.35\,(\mathrm{GeV}/c)^2$ data 
is estimated to $3.1\,\mathrm{\%}$.
\label{fig_analysis_PID_Comparison1}}
\end{center}
\end{figure}

\begin{figure}[!ht]
\begin{center}
\includegraphics[width=\linewidth]{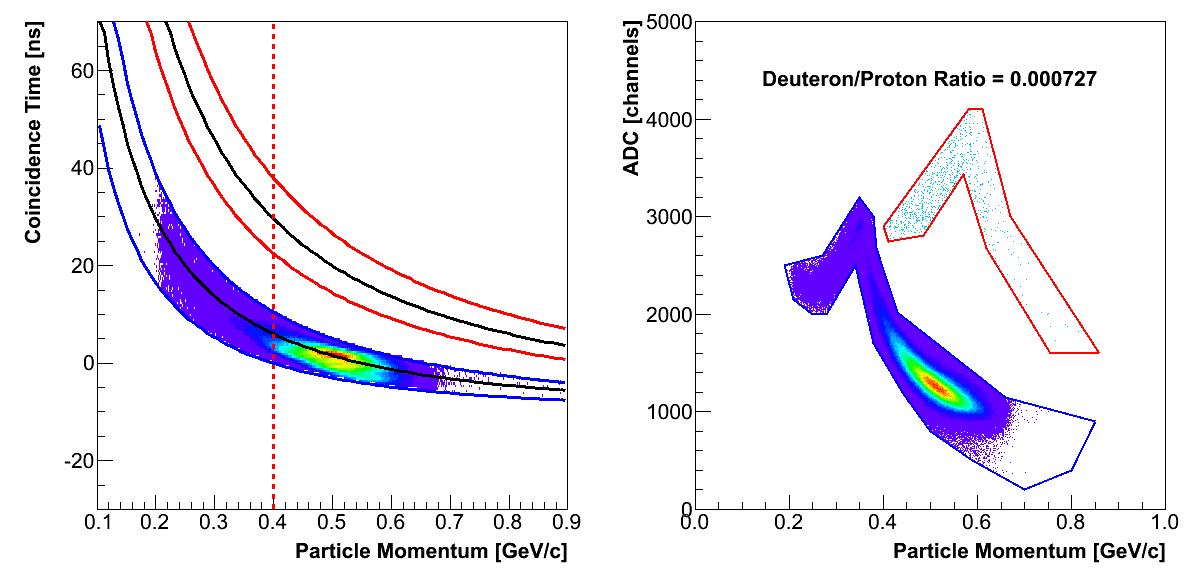} 
\vspace*{-7mm}
\caption{The efficiency of the Coincidence time PID for protons. The method is applied
to the production data to choose only protons (left figure). The selected events are then 
introduced to the Energy-loss PID (right figure) to check how many protons are mistakenly identified 
as deuterons. The obtained deuteron/proton ratio for the $Q^2 = 0.35\,(\mathrm{GeV}/c)^2$ data 
is estimated to $0.07\,\mathrm{\%}$.
\label{fig_analysis_PID_Comparison2}}
\end{center}
\end{figure}

\begin{figure}[!ht]
\begin{center}
\includegraphics[width=\linewidth]{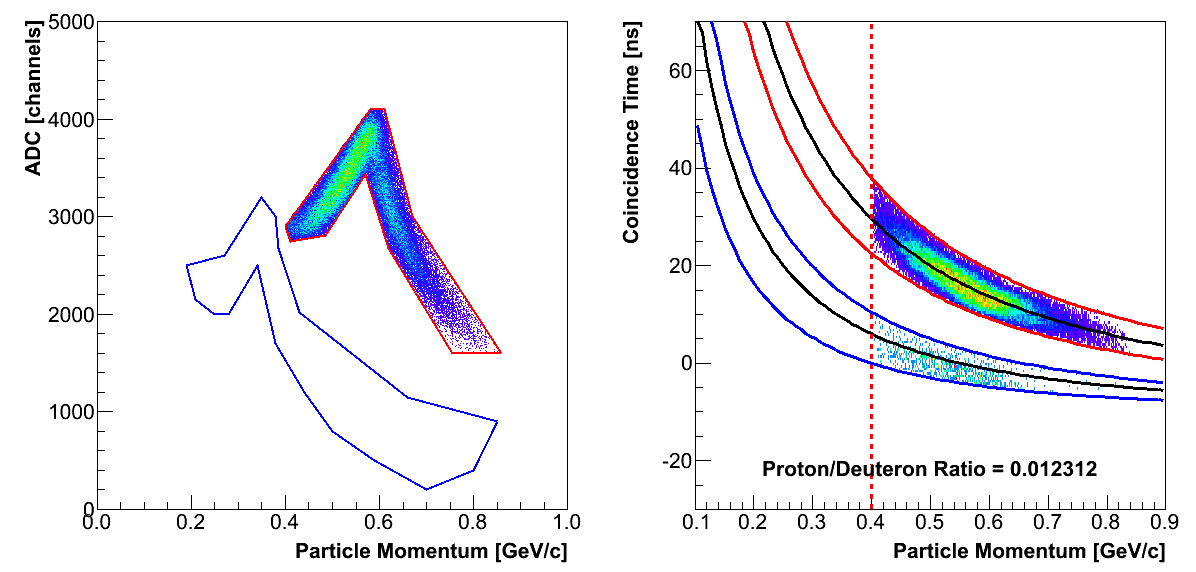} 
\vspace*{-7mm}
\caption{The efficiency of the Energy-Loss PID for deuterons. The method is applied
to the production data to choose only deuterons (left figure). The selected events are then 
introduced to the Coincidence time PID (right figure) to check how many deuterons are 
mistakenly identified as protons. The obtained proton/deuteron ratio for 
the $Q^2 = 0.35\,(\mathrm{GeV}/c)^2$ data is estimated to $1.2\,\mathrm{\%}$.
\label{fig_analysis_PID_Comparison3}}
\vspace*{-4mm}
\end{center}
\end{figure}

\begin{figure}[!ht]
\begin{center}
\includegraphics[width=\linewidth]{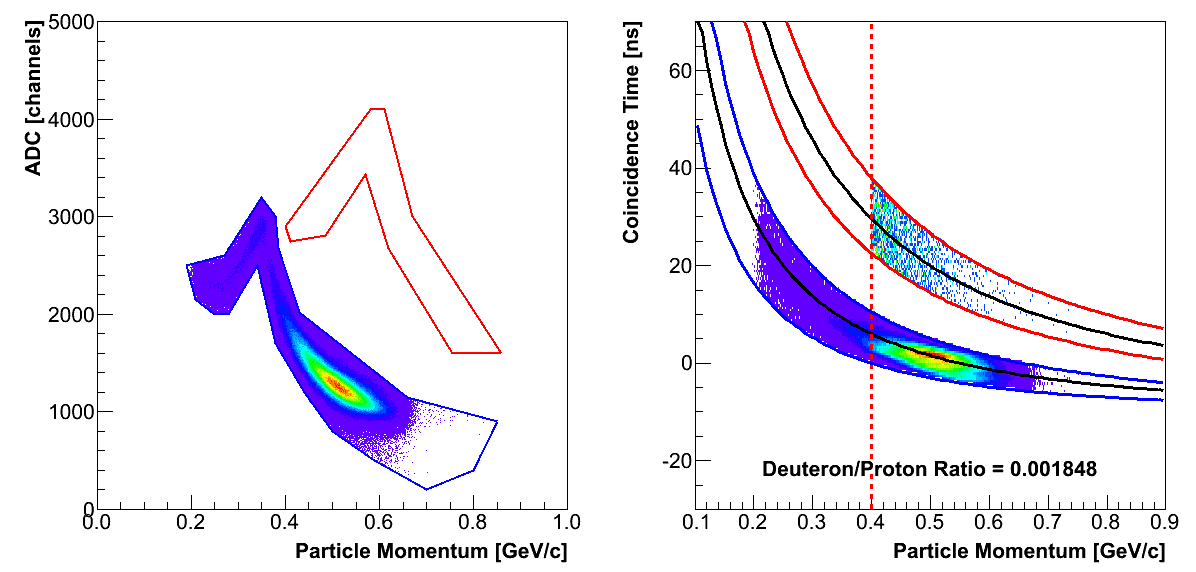} 
\vspace*{-7mm}
\caption{The efficiency of the Energy-Loss PID for protons. The method is applied
to the production data to choose only protons (left figure). The selected events are then 
introduced to the Coincidence time PID (right figure) to check how many deuterons are mistakenly 
identified as protons. The obtained deuteron/proton ratio for the $Q^2 = 0.35\,(\mathrm{GeV}/c)^2$ data 
is estimated to $0.18\,\mathrm{\%}$.
\label{fig_analysis_PID_Comparison4}}
\vspace*{-4mm}
\end{center}
\end{figure}

After the two PID approaches were developed, their efficiencies were investigated using a
two-step cross-examination test. In the first step, a chosen PID method was 
employed to identify protons and deuterons. In the second step, the other 
technique was used to check how many of the protons and deuterons, selected by the first 
method, were misidentified. 
The quality of the PID was then estimated from the ratio of misidentified with the properly
identified particles. Hence, a smaller ratio means a more reliable PID. The test were
performed for both PID techniques and for both experimental kinematic settings 
($Q^2 = 0.25\,(\mathrm{GeV}/c)^2$ and $Q^2 = 0.35\,(\mathrm{GeV}/c)^2$). The results are
gathered in Table~\ref{table_analysis_PID_efficancy}. For the $Q^2 = 0.35\,(\mathrm{GeV}/c)^2$ data, they
are also presented in Figs.~\ref{fig_analysis_PID_Comparison1} to~\ref{fig_analysis_PID_Comparison4}.

\begin{table}[!ht]
\begin{center}
\caption{Results of the two-step PID efficiency tests performed for both kinematical 
settings. The method declared in the first row was used to perform the PID. Then the other
method, stated in the brackets, was utilized to test how well the first method performed. 
The ratio between misidentified and properly recognized particles is a measure for the
quality of the PID. Test were made for both protons and deuterons. 
\label{table_analysis_PID_efficancy}}
\vspace*{2mm}
\begin{tabular}{|c||c|c|c|c|}
\hline

\hline

\hline
{\bf Kinematics} & \multicolumn{2}{|c|}{{\bf Energy-Losses}} & \multicolumn{2}{|c|}{{\bf Coincidence Time}} \\
                 & \multicolumn{2}{|c|}{(Coincidence Time)} & \multicolumn{2}{|c|}{(Energy-Losses)} \\[1mm]
\hline
 & & & &  \\[-4mm]
$Q^2$ & Proton PID & Deuteron PID & Proton PID & Deuteron PID \\ [-0.5mm]
$\left[(\mathrm{GeV}/c)^2\right]$ & $[\mathrm{\%}]$ & $[\mathrm{\%}]$ & $[\mathrm{\%}]$ & $[\mathrm{\%}]$\\ [1mm]
\hline

\hline
 & & & &  \\[-4mm]
$0.35$ & $\frac{\mathrm{Deuteron}}{\mathrm{Proton}} = 0.18$ &  
$\frac{\mathrm{Proton}}{\mathrm{Deuteron}} = 1.2$ &  $\frac{\mathrm{Deuteron}}{\mathrm{Proton}} = 0.07$ &
 $\frac{\mathrm{Proton}}{\mathrm{Deuteron}} = 3.1$ \\[2mm]
$0.25$ & $\frac{\mathrm{Deuteron}}{\mathrm{Proton}} = 0.10$ &  
$\frac{\mathrm{Proton}}{\mathrm{Deuteron}} = 1.6$ &  $\frac{\mathrm{Deuteron}}{\mathrm{Proton}} = 0.05$ &
 $\frac{\mathrm{Proton}}{\mathrm{Deuteron}} = 3.4$ \\ [1mm]

\hline

\hline

\hline
\end{tabular}
\end{center}
\end{table} 

For proton identification, a minimal contamination with deuterons is expected, since
deuterons represent only few percent of the total data. Even these small contamination ratios are 
probably overestimated.  The protons, that in the secondary step of  the test are
recognized  as deuterons could indeed be protons. However, because they come from the random coincidence 
background or they are part of the proton tail in the energy-loss plot, they are identified 
as deuterons.

A few percent contamination of deuterons with protons is contributed  mostly by the low energy 
particles. In the case of the coincidence time PID, the proton contamination originates in 
the random coincidences background, which are predominantly protons. In the energy-loss PID,
the protons which are recognized as deuterons, predominantly come from the proton tail, which reaches 
into the deuteron region and is strongest on the low-momentum side of the deuteron trail.
This also explains the higher proton/deuteron ratio for the $Q^2 = 0.25\,(\mathrm{GeV}/c)^2$
data. For this kinematic setting, the deuterons have smaller momenta and are gathered 
closer to the proton line, where they can collect more background protons. 

Since our main concern has been the misidentification of the protons as deuterons, we 
decided to check the behavior of the PID methods also with the data collected
with the hydrogen target. Ideally, no deuterons should be detected in this case. 
However, tests have revealed that $1-3\,\mathrm{\%}$ of protons were mistakenly
identified as deuterons, which is consistent with the results shown in 
Table~\ref{table_analysis_PID_efficancy}.

\begin{figure}[!ht]
\begin{center}
\begin{minipage}[t]{0.5\textwidth}
\hrule height 0pt
\includegraphics[width=1\textwidth]{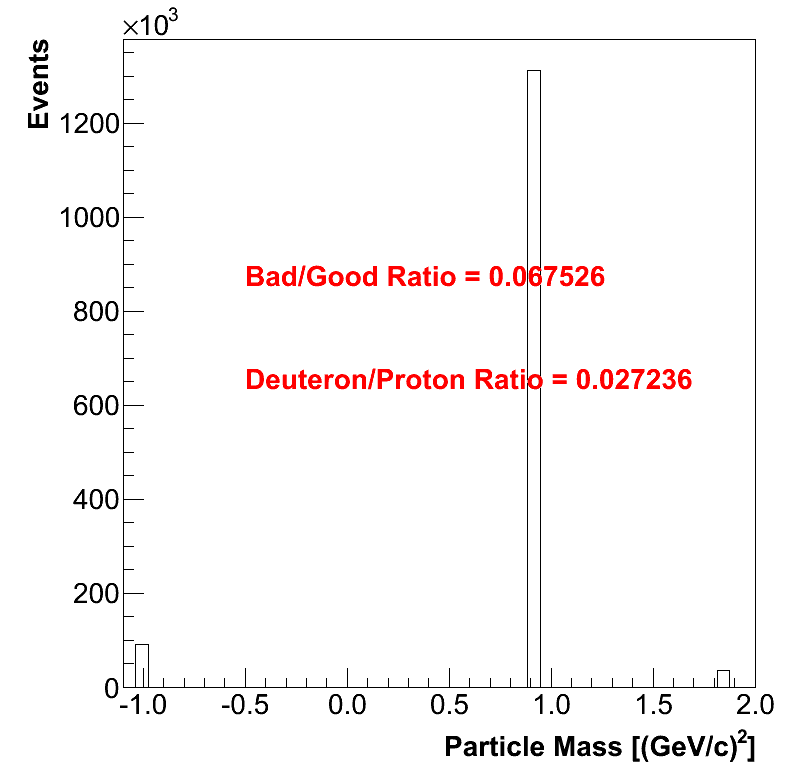}
\end{minipage}
\hspace*{1cm}
\begin{minipage}[t]{0.35\textwidth}
\hrule height 0pt
\caption{ Results of the energy-loss PID method for a typical production run.
Only about $3\,\mathrm{\%}$ of properly identified particles are deuterons. The rest 
are protons. Approximately $7\,\mathrm{\%}$ of events are left unidentified. These are 
the events that lie in between the proton and deuteron polygons and are rejected by 
the PID because of their bogus ADC readout. However,  the particles are most 
probably protons. 
\label{fig_analysis_PID_ratios}}
\end{minipage}
\end{center}
\end{figure}

Through the comparison of the two methods it was determined that the PID method based on the 
energy-losses distinguishes deuterons from protons better than the coincidence time 
technique. Therefore, it has been chosen as the primary PID method.
A result of the PID for a typical production run is shown in 
Fig.~\ref{fig_analysis_PID_ratios}. The coincidence time PID was used to monitor 
the PID efficiency or was in some cases considered as a supplementary PID method, 
to select the deuterons even more precisely.

\section{Selection of events}

Events accepted in the final analysis were filtered by a series of 
cuts that were applied to the data at two different stages of the analysis. 
Primary cuts were considered at the first step of the analysis, where raw data 
files get transformed into root-files, while the secondary cuts were put 
into use in a subsequent analysis of the created root files.

\subsection{Primary event cuts}
\label{sec:PrimaryCuts}
The primary cuts are listed in a \texttt{*.cdef} text file which is 
inserted as an input file to the Podd analysis software. Since the primary 
analysis is done massively on FARM computers, we choose these cuts to remove
all unphysical events in order to maximally reduce the size of the root-tree files
generated in this analysis. On the other hand, we do not want these cuts 
to be too restrictive in order to allow a flexible secondary analysis. For primary 
analysis, the following primary cuts were considered:
\begin{itemize}
\item A cut on the trigger variable event-type-bits 
(\texttt{DL.evtypebits}) was considered. For each detected event, this 
variables reports which triggers were present. Each bit of this variable 
corresponds to one trigger.  Since we are interested
in the coincidence events, we select only those events that have bits set
for coincidence triggers T5 and T6. Beside that, the HRS-L single events
also had to be selected, since these events can actually be coincidence events
due to the delay issues with the coincidence triggers (see Sec.~\ref{sec:TriggerOperation}).
In terms of programming code, this cut has the following form:
\begin{center}
\begin{verbatim}
(DL.evtypebits & 2^3)==2^3 || (DL.evtypebits & 2^5)==2^5 ||
(DL.evtypebits & 2^6)==2^6
\end{verbatim}
\end{center}

\item Only events with reconstructed tracks inside the BigBite MWDCs are selected. 
Without at least one valid track inside the BigBite wire-chambers, any further 
analysis would be impossible. The variable \texttt{BB.tr.n},
which reports the number of reconstructed tracks, must therefore be greater 
than zero,
\begin{center}
\begin{verbatim}
BB.tr.n > 0.
\end{verbatim}
\end{center}

\item 
Tracks reconstructed by the BigBite MWDCs must of course be consistent 
with hits in the adjacent scintillation detector. This is tested by 
the BigBite scintillation detector variable \texttt{BB.tp.trHitIndex}, which must
be greater than $-1$,  
\begin{center}
\begin{verbatim}
BB.tp.trHitIndex > -1.
\end{verbatim}
\end{center}
If its values are between $0$ and $99$, then 
a track is consistent with the hits in both layers of the scintillation detector.
If its values are between $100$ and $199$, the track agrees with the hits
in the E-plane. When values are $\geq 200$, the track agrees only with
the hits in the dE-plane.

\item 
Cuts on the nominal acceptance of the HRS-L were also employed.
The limiting values were taken from Table~\ref{table_HRSL}:
\begin{center}
\begin{verbatim}
L.gold.dp > -0.045 && L.gold.dp < 0.045
L.gold.th > -0.060 && L.gold.th < 0.060
L.gold.ph > -0.030 && L.gold.ph < 0.030.
\end{verbatim}
\end{center}

\item A loose constraint was applied also to the reaction point, which is 
described by the analysis variable \texttt{ReactPt\_L.z}. The absolute 
value of this variable had to be smaller than $0.3\,\mathrm{m}$,
\begin{center}
\begin{verbatim}
ReactPt_L.z > -0.30 && ReactPt_L.z < 0.30.
\end{verbatim}
\end{center}
Since  the targets are $0.4\,\mathrm{m}$ long ($\pm 0.2\,\mathrm{m}$ on 
each side of the center of the Hall), no real events can come from the 
region  beyond the set limit.

\end{itemize}

With the use of these cuts, the number of events recorded to the root-tree files
got reduced to $\approx 15\,\mathrm{\%}$ of the total number of events. This way
the created files became manageable in size and could be transfered from the FARM
computers to the local workstation and be successfully analyzed there. That 
would not be possible with the complete set of events.

\subsection{Secondary event cuts}
\label{sec:SecondaryCuts}
The secondary cuts represent a set of detailed constraints that are applied to 
the root-tree files, in order to perform a final filtering of events, which are 
then considered for the calculation of the experimental asymmetries. 
This part of the analysis has been performed 
on a local workstation, which has given us the possibility to modify these cuts 
repeatedly in order to obtain the best possible selection of events. In the end, 
the following set of cuts was implemented in the analysis scripts:

\begin{itemize}

\item Only real events are selected. Simulated EDTM events are excluded. This  is 
accomplished by rejecting events with a EDTM pulse present in the TDC modules.
The variable \texttt{Ndata.DL.edtmbb} must be zero:
\begin{center}
\begin{verbatim}
Ndata_DL.edtmbb == 0.
\end{verbatim}
\end{center}

\item The beam helicity state for each event is determined from the \texttt{g0hel.L.helicity}
variable. Only events with well defined helicity states $\pm1$
are accepted: 
\begin{center}
\begin{verbatim}
g0hel.L.helicity != 0. 
\end{verbatim}
\end{center}
Events with undefined helicity state (\texttt{g0hel.L.helicity=0}) are excluded.

\item We decided to consider only events with a single reconstructed track in 
each spectrometer. The number of reconstructed tracks in 
HRS-L and BigBite detector packages is monitored by the variables \texttt{L.tr.n} and 
\texttt{BB.tr.n}, respectively. This cut demands both variables to be $1$:
\begin{verbatim}
L.tr.n == 1 && BB.tr.n == 1. 
\end{verbatim}
This saves us the trouble of finding the best ("golden") track among all possible reconstructed 
tracks. This is not an issue for the HRS-L where the algorithm for finding the golden 
track works well, while the corresponding algorithm for the BigBite has not yet been written.
Fortunately the majority of events have only one track in each spectrometer.
Consequently, only a small portion of events is rejected by this cut.

\item Only real coincidence events are accepted. This cut is not performed by checking the 
event-type-bits variable (\texttt{DL.evtypebits}), since the trigger supervisor can miss
the coincidence trigger bits because of the $\approx30\,\mathrm{ns}$ delay present for the coincidence 
triggers T5 and T6. Instead, the trigger TDC information is exploited. For a valid coincidence event, 
the TDCs must record at least one T5 trigger. The number of recorded triggers per event is stored in the
variable \texttt{Ndata.DL.t5}. This gives us the ability to recognize coincidences even if they are not 
recognized in-time by the trigger supervisor. Furthermore, only events with a single 
trigger T1 in BigBite and a single trigger T3 inside the HRS-L are accepted. This 
additional constraint simplifies the calculation of the coincidence time. In order 
to consider events with multiple trigger hits, a correct pair (T1 and T3) of the TDC readouts 
would have to be chosen to correctly calculate the coincidence time. This is a non-trivial
problem which has not been addressed yet. 
By using these constrains $\approx 20\,\mathrm{\%}$ of events are lost in BigBite 
and $\approx 0.3\mathrm{\%}$ in the HRS-L. Note that Fig.~\ref{fig_Trigger_T1number} does not show the 
correct ratio between single and multiple T1-trigger events, because the use of the primary 
cuts significantly changes this ratio. In terms of programming code, these cuts are expressed as:
\begin{verbatim}
Ndata.DL.t5>0 && Ndata.DL.t3==1 && Ndata.DL.t1==1
\end{verbatim}

\item Two different cuts were applied to the coincidence time variable, which 
is defined as the difference between trigger-T3 and trigger-T1 TDC information:
\begin{eqnarray}
 t_{\mathrm{coinc}} = t_{\mathrm{T3-TDC}} - t_{\mathrm{T1-TDC}}\,. \label{eq_analysis_coinctime}
\end{eqnarray} 
The TDC timing information for the two triggers is accessible via 
variables \texttt{DL.t3} and \texttt{DL.t1}. When trying to extract good
coincidences accumulating in the  coincidence peak, the following 
cuts were considered:
\begin{eqnarray}
	-10\,\mathrm{ns} \leq t_\mathrm{coinc}\leq 35\,\mathrm{ns}\,. \nonumber
\end{eqnarray}
The chosen cut is visualized in Fig.~\ref{fig_analysis_coincidence_peak}. 
To estimate and subtract the contributions of the random coincidences, the events from 
the flat background of the coincidence time histogram were used. To get a 
reliable description of the background with as much statistics as possible, 
two identical sections of events were considered, 
one on each side of the coincidence peak:
\begin{eqnarray}
  -30\,\mathrm{ns} \leq t_\mathrm{coinc}\leq -18.75\,\mathrm{ns}\qquad \mathrm{and}\qquad
  45\,\mathrm{ns} \leq t_\mathrm{coinc}\leq 56.25\,\mathrm{ns}\,. \nonumber
\end{eqnarray}
The total width of the acknowledged background represents only one half of the width 
of the coincidence peak. Hence, the histograms corresponding to the random background 
must be multiplied by two before being subtracted from the primary histograms. 
An example of background subtraction in the particle-momentum histogram is demonstrated 
in Fig.~\ref{fig_analysis_coincidence_peak}.

\begin{figure}[!ht]
\begin{center}
\includegraphics[width=0.95\linewidth]{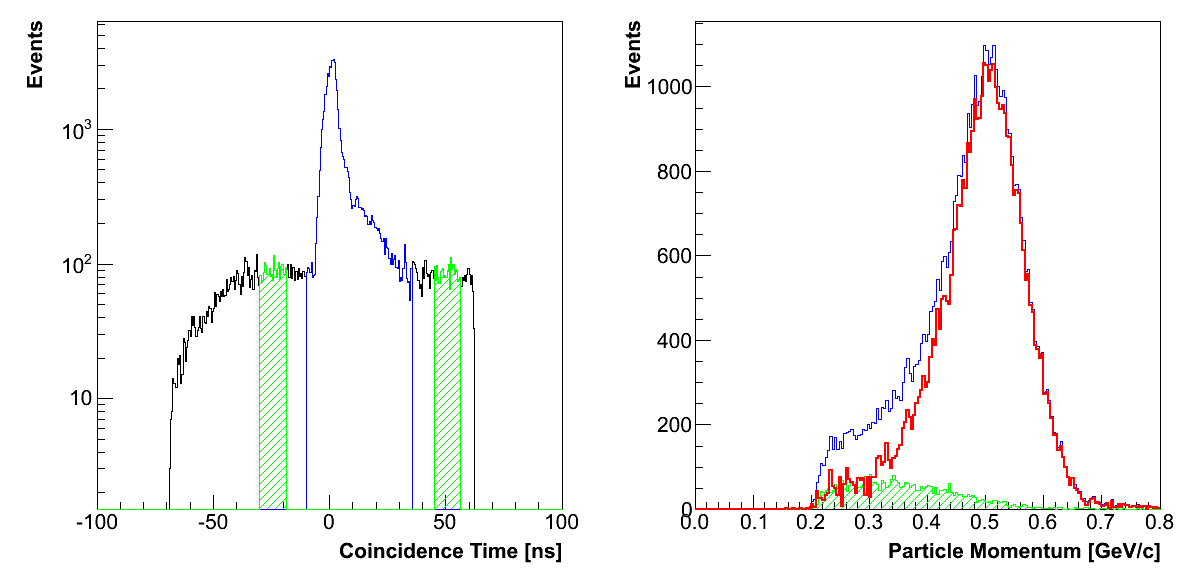} 
\caption{[Left] The coincidence time spectrum determined by using Eq.~(\ref{eq_analysis_coinctime}).
The blue line demonstrates the width of the cut used to select good events gathered in the
coincidence peak. The green bands represent the events used for the subtraction of 
background beneath the coincidence peak. [Right] Measured (by BigBite) distribution of 
particle momentum obtained with a cut on the coincidence peak. The blue and
red line show the momentum distribution before and after background subtraction. 
The momentum distribution of the background particles is shown by the green histogram. 
When subtracting, the background histogram must be considered with a factor two, since 
it contains only one half of the required statistics.
\label{fig_analysis_coincidence_peak}}
\end{center}
\end{figure}
\begin{figure}[!ht]
\begin{center}
\includegraphics[width=0.95\linewidth]{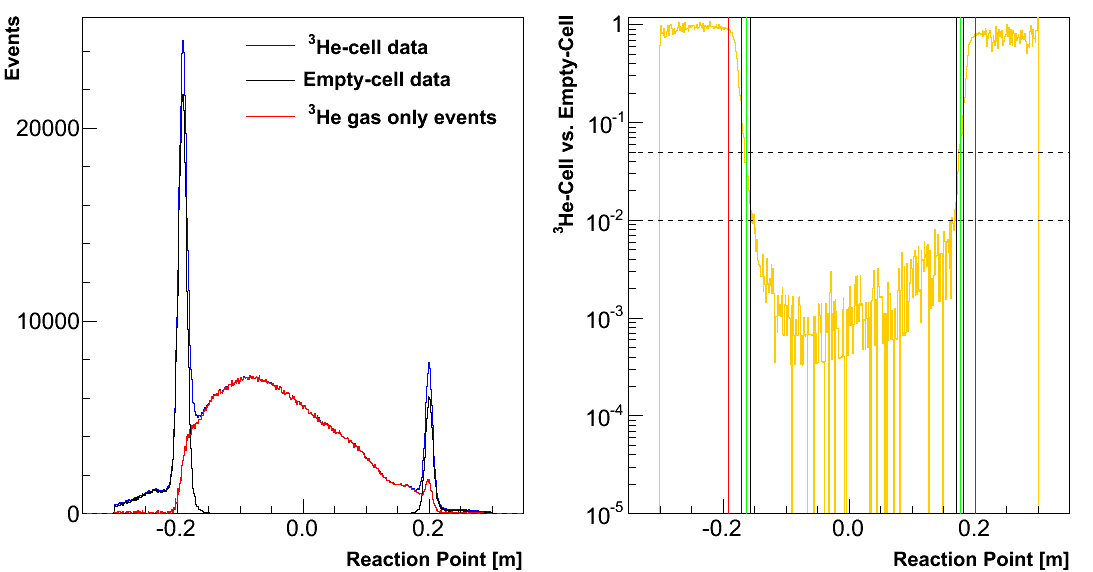} 
\caption{[Left] The reaction point variable for the ${}^3\mathrm{He}$-cell data (blue)
and empty-cell data (black).  The obtained results are normalized to the collected 
charge and corrected for pre-scale factors and dead-time. The red line shows the difference 
between the two data sets and represents the events arising  from the ${}^3\mathrm{He}$-gas.
[Right] The ratio between the empty-cell and ${}^3\mathrm{He}$-cell data. Red vertical lines
show locations of target windows. Blue, green and black lines show positions  of the 
$3\sigma$, $4\sigma$ and $5\sigma$ cuts, respectively.
\label{fig_analysis_targetcut1}}
\end{center}
\end{figure}

\begin{figure}[!ht]
\begin{center}
\begin{minipage}[t]{0.5\textwidth}
\hrule height 0pt
\includegraphics[width=1\textwidth]{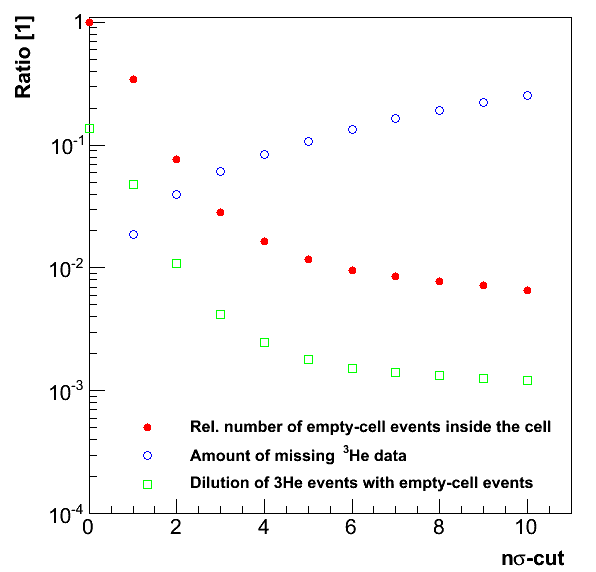}
\end{minipage}
\hfill
\begin{minipage}[t]{0.45\textwidth}
\hrule height 0pt
\caption{Analysis of the empty-cell and ${}^3\mathrm{He}$-cell
data. Red points show the number of collected events inside the empty-cell as
a function of the applied target cuts. In an ideal case, there should be no 
events coming from within the empty cell. Blue circles show the relative amount of 
${}^3\mathrm{He}$ events located in-between cell walls and set target cuts. 
The green circles show the relative contamination of 
pure ${}^3\mathrm{He}$-gas data with events from the glass-cell 
walls.
\label{fig_analysis_targetcut2}}
\end{minipage}
\end{center}
\end{figure}

\item The position of the reaction vertex along the beam direction is determined 
by the position variable $z_\mathrm{React}$ which is  measured
individually by both spectrometers. Cuts on this parameter
were applied in order to isolate events that arise from the ${}^3\mathrm{He}$-gas 
from those that happen inside the glass cell-walls. To determine the best possible 
position of the cuts, the ${}^3\mathrm{He}$-data were compared to the empty-cell data.
The results of the comparison are presented in Figs.~\ref{fig_analysis_targetcut1}
and~\ref{fig_analysis_targetcut2}. According to this analysis, the optimal position 
of the cuts is $\approx4\sigma$ away from the cell-walls, where $\sigma$ describes the 
widths of each cell-wall:
\begin{eqnarray}
 Z_{\mathrm{DS}}+4\sigma_{\mathrm{DS}} \leq z_\mathrm{React}^{\mathrm{BB,\>HRS-L}}\leq Z_{\mathrm{US}}-4\sigma_{\mathrm{US}}\,,
\label{eq_analysis_ReactPt_condition}
\end{eqnarray}
where $ Z_{\mathrm{US}} = 0.200\,\mathrm{m}$ and $ Z_{\mathrm{DS}} = -0.191\,\mathrm{m}$ 
represent the positions of the upstream and downstream cell windows, 
while $\sigma_{\mathrm{US}} = 0.006\,\mathrm{m}$ and $\sigma_{\mathrm{DS}} = 0.007\,\mathrm{m}$ 
are their reconstructed sigma-widths. With the $4\sigma$-cut more than $90\,\mathrm{\%}$ of the 
events on ${}^3\mathrm{He}$ are accepted, while keeping the cell-wall dilution below $0.3\,\mathrm{\%}$.

The condition (\ref{eq_analysis_ReactPt_condition}) of course needs to be satisfied 
for vertex positions $z_\mathrm{React}^{\mathrm{HRS-L}}$ and $z_\mathrm{React}^{\mathrm{BB}}$ 
measured by both HRS-L and BigBite. Furthermore, we demand that the vertex positions must agree
to within some specified precision:
\begin{eqnarray}
\left|z_\mathrm{React}^{\mathrm{HRS-L}} - z_\mathrm{React}^{\mathrm{BB}}\right|\leq 2.5\sigma_{y}^{\mathrm{BB}} = 3.2\,\mathrm{cm}\,. \nonumber
\end{eqnarray}
This constraint affects mostly BigBite events, since BigBite's resolution for 
vertex reconstruction ($\sigma_{y}^{\mathrm{BB}}$) is inferior to that of HRS-L.  
The $2.5\sigma_{y}^{\mathrm{BB}}$ limit corresponds to one half of the distance 
between two neighboring foils in the carbon optics target.

\item A cut was performed also on the quality of the tracks, reconstructed by the HRS-L. 
Best events hit between $3$ and $6$ wires in each wire plane. The number of hits in each plane 
are recorded by the  \texttt{L.vdc.*.nhit} variables. In the programming code, these cuts 
are expressed as:
\begin{verbatim}  
( L.vdc.u1.nhit > 2 && L.vdc.u1.nhit < 7 ) && 
( L.vdc.u2.nhit > 2 && L.vdc.u2.nhit < 7 ) &&
( L.vdc.v1.nhit > 2 && L.vdc.v1.nhit < 7 ) && 
( L.vdc.v2.nhit > 2 && L.vdc.v2.nhit < 7 ) &&    
( L.tr.chi2[0]/L.tr.ndof[0] < 100 )
\end{verbatim}
In addition, the $\chi^2/\mathrm{NDOF}$-value of the reconstructed track has also been tested. 
For this experiment it was decided that the $\chi^2/\mathrm{NDOF}$ must be smaller than 
$100$, in order for a track to be accepted.

\item All events with hits in the bottom-most scintillator bars of the dE and E-detectors are rejected. 
These paddles detect mostly very high momentum particles. We have 
not gathered enough statistics for their proper calibration, and it was decided to exclude them 
from the analysis:
\begin{verbatim}  
BB.HadrDetPack.E.hit_paddle[0] < 23 && 
BB.HadrDetPack.dE.hit_paddle[0] < 23. 
\end{verbatim}

\item Lastly, cuts on the BigBite event type (full/partial events) were performed. The type of each 
reconstructed event is saved in the Hadron-detector-package variable \texttt{GoodEvent}. In the 
programming code, the cut can be expressed as:
\begin{verbatim}  
BB.HadrDetPack.GoodEvent[0] == 1  && 
BB.HadrDetPack.GoodEvent[0] == 2
\end{verbatim}
In principle,
we are interested only in the full tracks (\texttt{GoodEvent==1}), which have MWDC tracks
consistent with TDC hits in both planes of the scintillation detector. However, we have also accepted  
partial events, which have valid TDC hits only in the E-plane(\texttt{GoodEvent==2}).  
It is possible that these partial events are actually full events reported as partials, 
because of the high TDC threshold settings for the dE-plane. In this case, they contain a valid ADC 
information and will be able to normally pass through the PID. If these events are really partial, 
they are missing also the ADC information in the dE-plane and will be eventually 
rejected by the PID, although they were accepted by this cut. Events with valid hits 
only in the dE-plane (\texttt{GoodEvent==3}) are always rejected.

\end{itemize}

\section{Scaler analysis}
\label{sec:ScalerAnalysis}

\begin{figure}[!ht]
\begin{center}
\includegraphics[width=\linewidth]{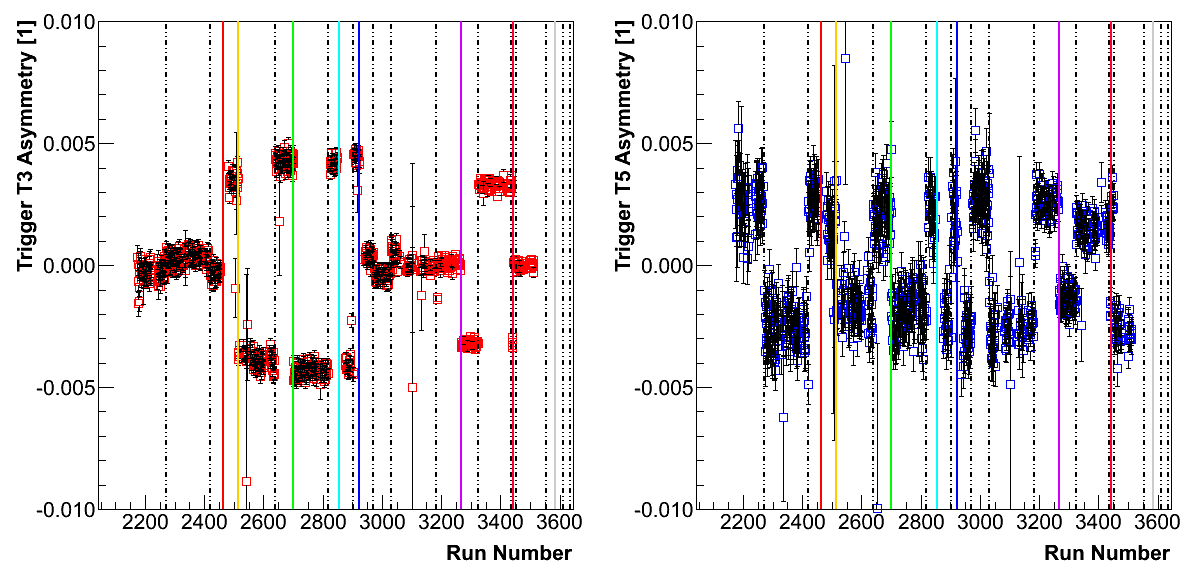} 
\caption{Raw scaler asymmetries for triggers T3 (left) and T5 (right), calculated 
for all collected data sets (runs). Dashed-dotted lines correspond to the 
insertion or removal of the beam HWP. The full vertical lines represent the reversals of 
the target orientations. The colors of the lines have the same meaning as those shown in 
Fig.~\ref{fig_TargetPolarizationPlot}. Both asymmetries must change sign after 
each vertical line. 
\label{fig_scalers_RawTriggerAsymmetries}}
\end{center}
\end{figure}

Scaler information was used to get a first glance at the measured asymmetries. By comparing the trigger
rates for both beam helicity states, mean trigger asymmetries for each collected data set 
could be determined:
\begin{eqnarray}
\overline{A}_{\mathrm{Ti}} = \frac{N_{\mathrm{h+}}^{\mathrm{Ti}}-N_{\mathrm{h-}}^{\mathrm{Ti}}}
{N_{\mathrm{h+}}^{\mathrm{Ti}}-N_{\mathrm{h-}}^{\mathrm{Ti}}}\,, \qquad i=1,2,\dots,7\,,\nonumber
\end{eqnarray} 
where $N_{\mathrm{h\pm}}^{\mathrm{Ti}}$ represents the number of scaler counts with helicity state $h\pm$ 
for each of the seven considered triggers $\mathrm{Ti}$. During the experiment, these asymmetries were 
calculated for data with half-wave plate (HWP) IN and HWP OUT and for data with opposite target orientations. 
This served for testing experimental data, since the asymmetry should flip sign when each of these 
parameters is reversed. The results of such analysis, performed for the relevant triggers T3 and T5, 
are shown in Fig.~\ref{fig_scalers_RawTriggerAsymmetries}. The behavior of the two asymmetries is 
very different since they are related to different physical processes. However, they 
both correctly change their sign after the reversal of the target orientation, or after HWP insertion/removal.

\begin{figure}[!ht]
\begin{center}
\includegraphics[width=0.49\linewidth]{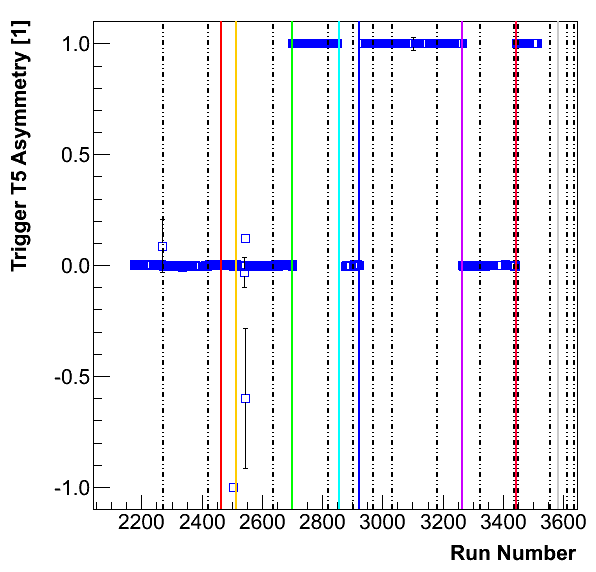} 
\includegraphics[width=0.49\linewidth]{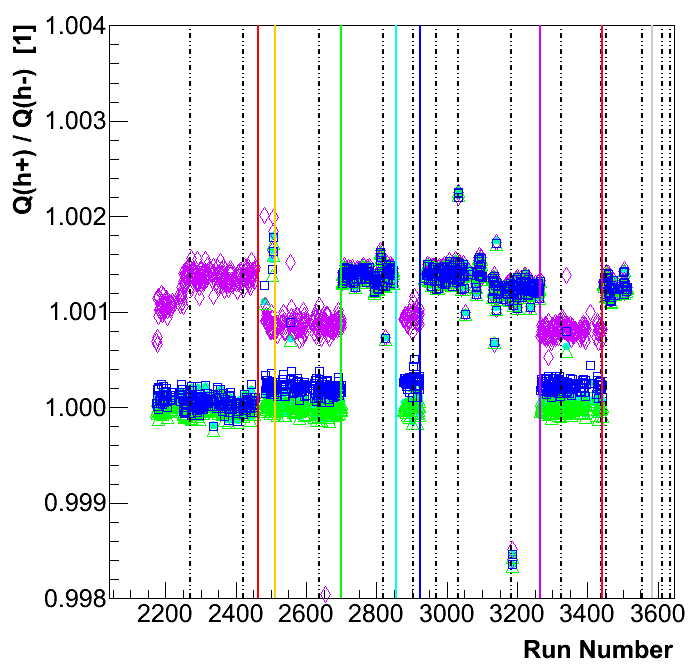} 
\caption{[Left] The raw T5 asymmetry determined with the \texttt{evbbite}
scalers. Since one of the modules that are gated with the positive target spin, 
is missing in the data-stream, the asymmetry equals one for regions
with positive target spin orientation. [Right] The ratio of the 
collected charges for positive and negative beam helicity. The cyan, violet, green 
and blue points represent the results obtained for U3, D3, U10 and D10 BCM monitors
using HRS-L scalers. Due to problems with the modules, bogus charge ratios
have been obtained. The presented results are corrected for the HWP position and 
target orientation. Dashed-dotted lines correspond to the 
insertion or removal of the HWP. The full vertical lines represent the reversals of 
the target orientation. The colors of the lines have the same meaning as those shown in 
Fig.~\ref{fig_TargetPolarizationPlot}.
\label{fig_scalers_ScalerIssues}}
\end{center}
\end{figure}

The main reason for performing the scaler analysis is to extract the information on the 
accumulated charge and dead time, which are important for the extraction of the
experimental asymmetries, as indicated in Eq.~(\ref{eq_analysis_asym2}). In the analysis 
\texttt{bbite} scalers were considered. The \texttt{evbbite} scaler could not be used
due to the absence of one of the gated scaler modules in the data stream (see 
Sec.~\ref{sec:ScalerMeasurements}). With one working pair of gated scalers, only data 
for negative target spin orientation could be properly analyzed (see 
Fig.~\ref{fig_scalers_ScalerIssues}).  Unfortunately, also neither of the HRS-L scalers
could be used. The analysis of both \texttt{left} and  \texttt{evleft} scalers has shown
that all modules had some kind of problems, resulting in bogus raw scaler asymmetries. 
Figure~\ref{fig_scalers_ScalerIssues} shows that the gated modules, corresponding to the 
negative target orientation, were returning unphysical results for one of the 
BCM current monitors. On the other hand, the modules gated with the 
positive target spin were returning false asymmetries in all recorded channels. This
suggests that one of the modules was counting too fast. This could be due to 
bad cabling or a faulty module. The observed deviations are small ($0.1\,\mathrm{\%}$),
which is the reason why they were overlooked during the experiment. 

\begin{figure}[!ht]
\begin{center}
\includegraphics[width=0.49\linewidth]{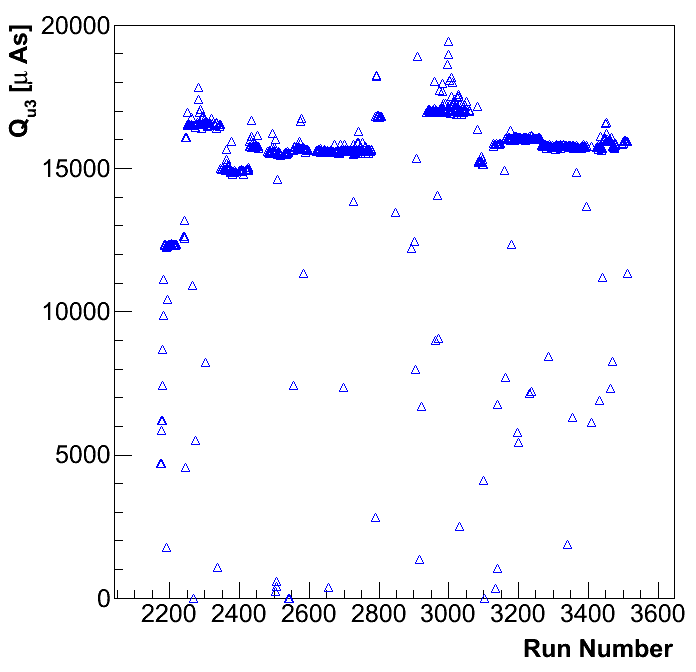} 
\includegraphics[width=0.49\linewidth]{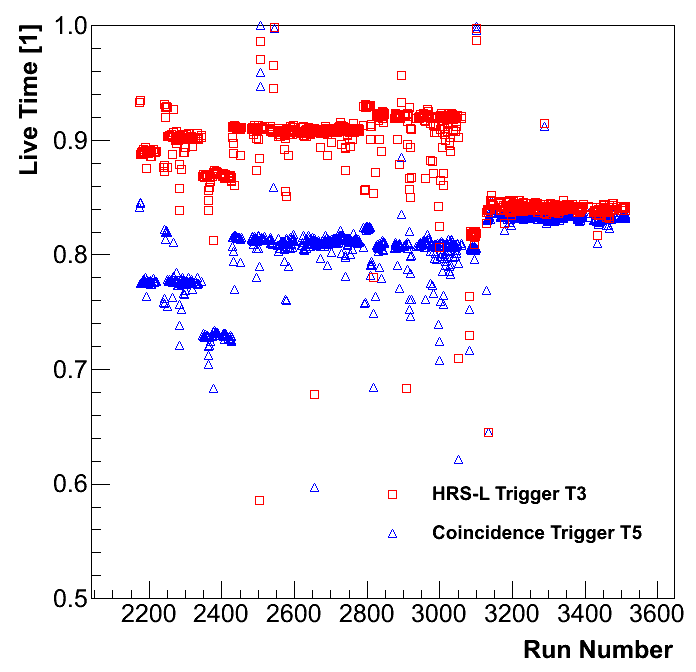} 
\caption{[Left] The accumulated charge for all datasets collected during the E05-102 experiment.
[Right] Live-times for triggers T3 (red) and T5 (blue). 
\label{fig_scalers_charges}}
\end{center}
\end{figure}

The accumulated charge for each dataset was determined from the scaler measurements 
of the BCM monitors U1, D1, U3, D3, U10 and D10. The raw scaler readings were then
transformed to the charge, by using Eq.(\ref{BCMCharge}).
The results for the most reliable current monitor U3 are shown in Fig.~\ref{fig_scalers_charges}.
The amount of charge collected during each run depends on the duration of each data set,
the intensity of the beam current, the number of beam trips and target type. 

The live time $t_l$ is determined as the ratio between the number of CODA accepted triggers
$N^{\mathrm{Ti}}(\mathrm{CODA})$ and the number of triggers recorded by scalers
$N^{\mathrm{Ti}}(\mathrm{Scalers})$:
\begin{eqnarray}
t_{l}^{\mathrm{Ti}} = \frac{ N^{\mathrm{Ti}}(\mathrm{CODA})\cdot PS^{\mathrm{Ti}}}
{N^{\mathrm{Ti}}(\mathrm{Scalers})}\frac{1}{f_{\mathrm{corr}}}\,.\label{eq_scalers_livetime}
\end{eqnarray}
The live time is calculated for each trigger separately, where the number of CODA events 
must be multiplied by the prescale factor ($PS^{\mathrm{Ti}}$), which determines the 
fraction of events accepted by the trigger supervisor. The factor $f_{\mathrm{corr}}$ corrects for
the non-synchronicity  of the scalers and CODA events, since \texttt{bbite} scalers are recorded 
to the data stream less frequently then CODA events. This is especially important 
for the analysis of small sections of runs.  With the analysis of full runs $f_{\mathrm{corr}}$
equals one. The factor $f_{\mathrm{corr}}$ is obtained from the ratio of the number of BigBite re-timing pulses
recorded by CODA and by the scalers:
\begin{eqnarray}
f_{\mathrm{corr}} = \frac{N^{\mathrm{BB\>re-time}}(\mathrm{CODA})}{N^{\mathrm{BB\>re-time}}(\mathrm{Scalers})} \approx 1\,.
\end{eqnarray}
The estimated live times for triggers T3 and T5 are shown in Fig.~\ref{fig_scalers_charges}.
The live time of trigger T3 is larger, as anticipated. The T3 events consist mostly of  HRS-L single
events and a small portion of coincidences. Hence, data for only one spectrometer need to be downloaded
to the data-stream. Consequently, CODA is not as occupied as during T5 events, where data from 
both spectrometers  need to be recorded.

The results for live-time and accumulated charge are presented in Fig.~\ref{fig_scalers_charges}.
They were obtained with the un-gated scalers, which recorded rates regardless of the target spin and 
beam helicity. However, to obtain the corrections required by Eq.~(\ref{eq_analysis_asym2}), both
live-time $t_{l}^{\pm}$ and charge $Q^{\pm}$ need to be determined for each helicity state separately, 
via the gated scalers. By defining the ratios:
\begin{eqnarray}
\frac{t_l^+}{t_l^-} = 1+\tau\,,\qquad \frac{Q^+}{Q^-} = 1+\delta\,, \nonumber
\end{eqnarray}
the asymmetry (\ref{eq_analysis_asym2}) can be expressed as:
\begin{eqnarray}
A_{\mathrm{exp}} = \frac{\frac{N^+}{t_l^-(1+\tau) Q^-(1+\delta)}  - \frac{N^-}{t_l^- Q^-}}
{\frac{N^+}{t_l^-(1+\tau) Q^-(1+\delta)}  + \frac{N^-}{t_l^- Q^-}} \approx \frac{N^+(1-\tau -\delta) - N^-}{N^+(1-\tau -\delta) + N^-}\,,\nonumber
\end{eqnarray}
where it is assumed that charge and live-time corrections, $\delta$ and $\tau$, are small. Because the
experimental asymmetries are also small, $\delta$ and $\tau$ have much stronger effects on
the numerator, where the difference is formed, than on the denominator:
\begin{eqnarray}
  A_\mathrm{exp} = \frac{N^+ - N^-}{N^+ + N^-} - (\tau + \delta)\frac{N^+}{N^+ + N^-}\,. \nonumber
\end{eqnarray} 
Considering also that $N^+/(N^+ + N^-)\approx 1/2$, the experimental asymmetry (\ref{eq_analysis_eq1}),
 corrected for the differences in live-times and charges for two helicity states, can finally be expressed as:
\begin{eqnarray}
  A_\mathrm{exp}^0 =   A_\mathrm{exp} +\frac{\tau}{2} + \frac{\delta}{2}\,. \label{eq_scalers_asymmetry}
\end{eqnarray}

\begin{figure}[!ht]
\begin{center}
\begin{minipage}[t]{0.5\textwidth}
\hrule height 0pt
\includegraphics[width=1\textwidth]{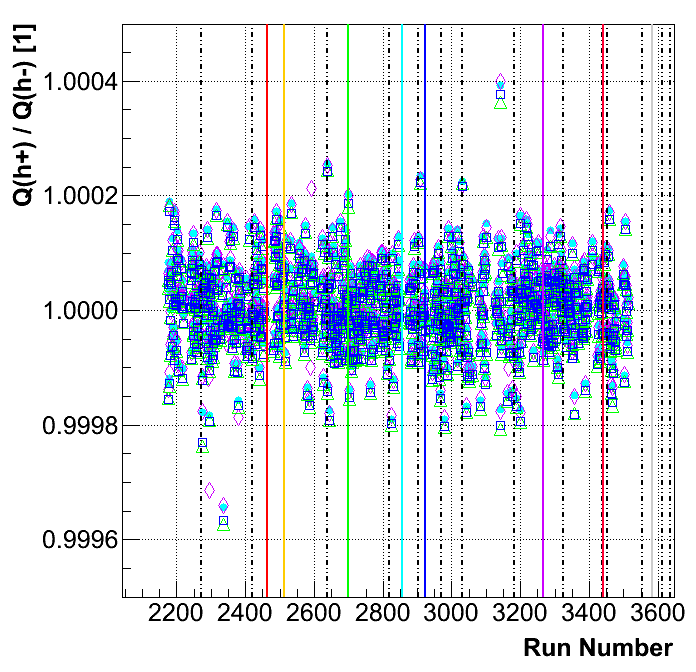}
\end{minipage}
\hfill
\begin{minipage}[t]{0.45\textwidth}
\hrule height 0pt
\caption{The ratio between accumulated charges with a positive and negative beam 
helicity. The cyan, violet, green and blue points represent the results obtained 
with U3, D3, U10 and D10 BCM monitors, by using \texttt{bbite} scalers.  All six
monitors return consistent results. Dashed-dotted lines correspond to the 
insertion or removal of the beam HWP. The full vertical lines represent the reversals of 
the target orientation. The colors of the lines have the same meaning as those shown in 
Fig.~\ref{fig_TargetPolarizationPlot}.
\label{fig_scalers_BCM_ratio}}
\end{minipage}
\end{center}
\end{figure}

\begin{figure}[!ht]
\begin{center}
\includegraphics[width=\linewidth]{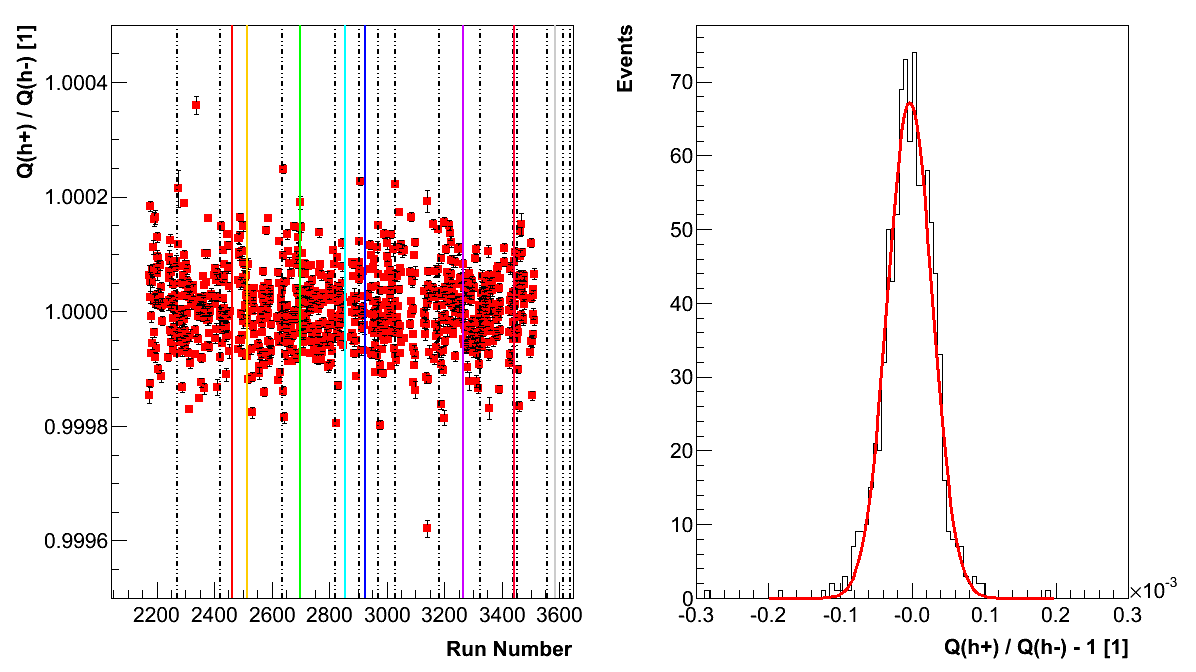} 
\caption{[Left] The ratio between accumulated charges $(1+\delta)$ with positive 
and negative beam helicity. Each point represents the mean value of the ratios, obtained 
with the most reliable monitors U3, U10 and D10. The values shown are properly corrected 
for the HWP position and target orientation. The description of the vertical lines is the same 
as in Fig.~\ref{fig_scalers_BCM_ratio}.
[Right] Spread of the charge correction factor $\delta$. The distribution was fitted
by a Gaussian with a mean value $\delta = -2.863\times 10^{-6}$ and width 
$\sigma_\delta = 3.1\times 10^{-5}$.
\label{fig_scalers_delta}}
\end{center}
\end{figure}

The charge correction factor $\delta$ was determined from the gated \texttt{bbite} scalers. 
Figure~\ref{fig_scalers_BCM_ratio} shows the results acquired from all six BCM monitors. They all 
give consistent results. For the final calculation of $\delta$, the mean value of the three 
most stable monitors U3, U10 and D10 was considered. The results are shown in Fig.~\ref{fig_scalers_delta}.
The values calculated for each run were corrected for the position of the beam HWP and the 
target orientation. This way all data could be analyzed together as a single data set. 
The mean value and the width (sigma) of the obtained distribution are:
\begin{eqnarray}
  \overline{\delta} = -2.863\times 10^{-6}\,,\qquad \sigma_\delta = 3.1\times 10^{-5}\,.\nonumber
\end{eqnarray}
One sees, that the systematic correction $\delta$ to the asymmetry (\ref{eq_scalers_asymmetry}), due to the
differences in the charges accumulated for the two helicity states, is minimal.

\begin{figure}[!hb]
\begin{center}
\includegraphics[width=\linewidth]{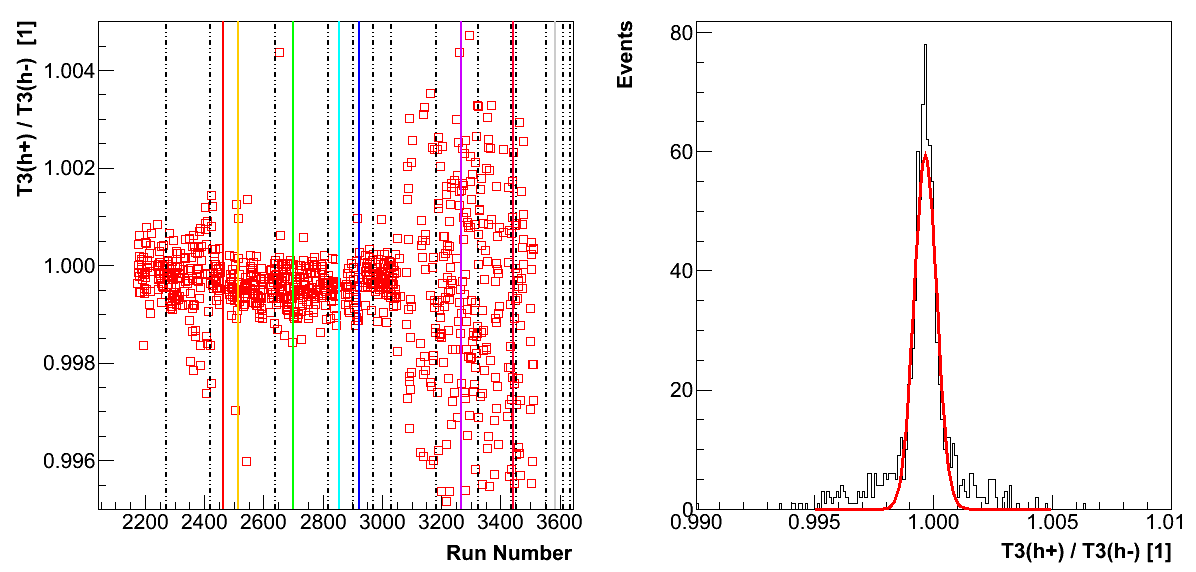} 
\caption{[Left] Trigger-T3 live-time ratio $(1+\tau)$ as a function of run number. Increased spread
of points for runs $\geq 3050$ is a consequence of lower CODA statistics due to the large prescale 
factor ($\mathrm{PS3}=200$) used for these runs. The shown results are corrected 
for the HWP position and target orientation. The description of the vertical lines is the same 
as in Fig.~\ref{fig_scalers_BCM_ratio}.
[Right] Spread of the live time ratio. The obtained distribution was fitted
by a Gaussian with a mean value $1+\tau = 0.99966$ and width $\sigma_\tau = 4.74\times 10^{-4}$.
\label{fig_scalers_T3livetime_ratio}}
\end{center}
\end{figure}

\begin{figure}[!hb]
\begin{center}
\includegraphics[width=\linewidth]{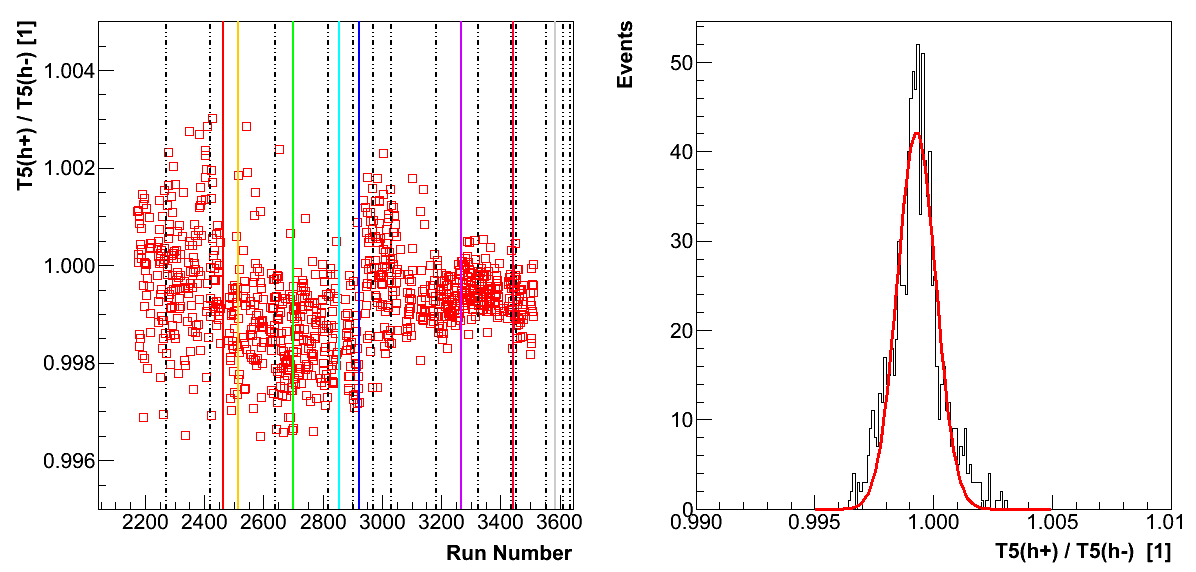} 
\caption{[Left] Trigger T5 live time ratio $(1+\tau)$ as a function of run number.  
The shown results have been corrected for the HWP position and target orientation. 
The description of the vertical lines is the same as in Fig.~\ref{fig_scalers_BCM_ratio}.
[Right] Spread of the live time ratio. The obtained distribution was fitted
by a Gaussian with a mean value $1+\tau = 0.99928$ and a width 
$\sigma_\tau = 8.12\times 10^{-4}$.
\label{fig_scalers_T5livetime_ratio}}
\end{center}
\end{figure}

The live-time correction factor $\tau$ was estimated separately for triggers T3 and T5. 
The analysis results are gathered in Figs.~\ref{fig_scalers_T3livetime_ratio} 
and~\ref{fig_scalers_T5livetime_ratio}. To determine the mean value and the width of the 
distributions of $\tau$, the same approach was considered as for the charge correction
factor $\delta$. The final results for $\tau$ are: 
\begin{eqnarray}
  \overline{\tau}_{\mathrm{T3}} = -3.39\times 10^{-4}\,,\qquad \sigma_\tau^{\mathrm{T3}} = 4.74\times 10^{-4}\,,\nonumber \\
  \overline{\tau}_{\mathrm{T5}} = -7.11\times 10^{-4}\,,\qquad \sigma_\tau^{\mathrm{T3}} = 8.12\times 10^{-4}\,.\nonumber
\end{eqnarray}
The corrections related to the different live-times for different helicity states are
also minimal and also (as $\delta$) do not have enough strength to significantly 
influence the experimental results for $A_\mathrm{exp}^0$.

\begin{figure}[!ht]
\begin{center}
\begin{minipage}[t]{0.5\textwidth}
\hrule height 0pt
\includegraphics[width=1\textwidth]{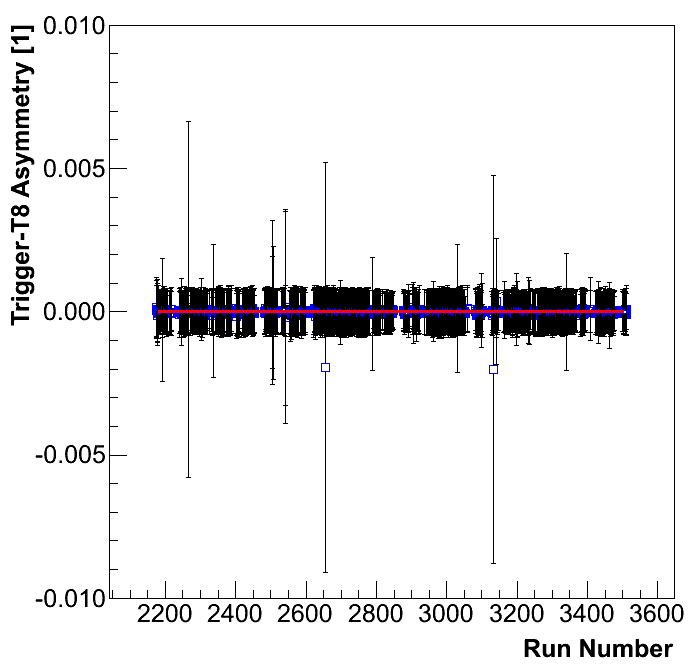}
\end{minipage}
\hfill
\begin{minipage}[t]{0.45\textwidth}
\hrule height 0pt
\caption{ Trigger T8 (pulser) asymmetry. The mean value was determined to be
$\overline{A}_{\mathrm{helicity}} = -2.1\times 10^{-6} \pm 2.5\times 10^{-5}$. 
Zero asymmetry demonstrates that the electron source at injector does not prefer 
any of the helicity states.
\label{fig_scalers_T8Asymmetry}}
\end{minipage}
\end{center}
\end{figure}

A false asymmetry could arise also from a different number of incident 
electrons with positive and negative helicity. Such occurrence would suggest 
that the electron source at the injector prefers electrons with 
one helicity state, while suppressing 
the other.  Equivalence of the positive/negative helicity states was also 
inspected by scaler analysis. For this purpose the T8 trigger was used, which 
corresponds to the $1024\,\mathrm{kHz}$ pulser. By comparing the pulser counts 
$N^{\pm}_{\mathrm{T8}}$ from the beam-helicity gated scalers, the helicity 
asymmetry could be calculated:
\begin{eqnarray}
 A_{\mathrm{helicity}} = \frac{N^{+}_{\mathrm{T8}} - N^{-}_{\mathrm{T8}}}{N^{+}_{\mathrm{T8}} + N^{-}_{\mathrm{T8}}}\,,\nonumber
\end{eqnarray}
which is a direct measure for such effects. Figure~\ref{fig_scalers_T8Asymmetry} shows 
that, to achievable accuracy, this asymmetry is consistent with zero.

\section{Extraction of asymmetries}
\label{sec:ExtractionOfAsymmetries}

The experimental measurements were performed for three different target
orientations (see Sec.~\ref{sec:TargetSysyem}), two positions of 
HRS-L $(\theta_{\mathrm{HRS-L}} = 12.5^\circ, 14.5^\circ)$, and two 
positions of BigBite $(\theta_{\mathrm{BB}} = -75.0^\circ, -82.0^\circ)$. The total number of 
runs collected for each kinematical setting is shown in Table~\ref{table_analysis_ListOfData}.
Cumulatively almost $1000$ production data-sets were created. Approximately half of them 
were measured with inserted beam HWP. 
\begin{table}[!ht]
\begin{center}
\caption{Assortment of runs considered in the data analysis. Each raw 
dataset contains between four and five million events. The spectrometer angles are 
stated in the Hall Coordinate System. The exact direction of the target spin for each 
target orientation is given in Table~\ref{table_CompassResults}.
\label{table_analysis_ListOfData}}
\vspace*{2mm}
\begin{tabular}{ccccccc}
\multicolumn{7}{c}{{\bf Production data on polarized ${}^3\mathrm{He}$ target}}\\
\hline

\hline

\hline
Data& { Target}  & $\theta_{\mathrm{HRS-L}}$ & $\theta_{\mathrm{BB}}$ & $|Q^2|$ & {Beam HWP } & { Number of } \\ 
Group& { Orientation} & $[{}^\circ]$ &  $[{}^\circ]$ & $[(\mathrm{GeV}/c)^2]$ &  { Position} & { datasets} \\
\hline

\hline

\hline
 & & & & & & \\[-4mm]
1 & Longitudinal $+$  & $14.5$ & $-75.0$ & $0.35$ & Out & $100$ \\
2 & Longitudinal $+$  & $14.5$ & $-75.0$ & $0.35$& In & $105$ \\
\hline
3 & Transverse $-$ &  $14.5$ & $-75.0$ & $0.35$ & Out & $78$ \\
4 & Transverse $-$ &  $14.5$ & $-75.0$ & $0.35$ & In & $59$ \\
5 & Transverse $+$ &  $14.5$ & $-75.0$ & $0.35$ & Out & $33$ \\
6 & Transverse $+$ &  $14.5$ & $-75.0$ & $0.35$ & In & $72$ \\
\hline
7 & Longitudinal $+$ & $12.5$ & $-75.0$ & $0.25$ & Out & $74$ \\
8 & Longitudinal $+$ & $12.5$ & $-75.0$ & $0.25$ & In & $93$ \\
\hline
9 & Transverse $-$ &  $12.5$ & $-75.0$ & $0.25$ & Out & $59$ \\
10& Transverse $-$ & $12.5$ & $-75.0$ & $0.25$ & In & $71$ \\
\hline
11& Longitudinal $+$ & $14.5$ & $-82.0$ & $0.35$ & Out & $55$ \\
12& Longitudinal $+$ & $14.5$ & $-82.0$ & $0.35$ & In & $51$ \\
\hline
13 & Transverse $-$ &  $14.5$ & $-82.0$ & $0.35$ & Out & $24$ \\
14 & Transverse $-$ &  $14.5$ & $-82.0$ & $0.35$ & In & $15$ \\
15 & Transverse $+$ &  $14.5$ & $-82.0$ & $0.35$ & Out & $24$ \\
16 & Transverse $+$ &  $14.5$ & $-82.0$ & $0.35$ & In & $27$ \\
\hline

\hline

\hline
\end{tabular}
\end{center}
\end{table}

\begin{figure}[!ht]
\begin{center}
\includegraphics[width=0.49\textwidth]{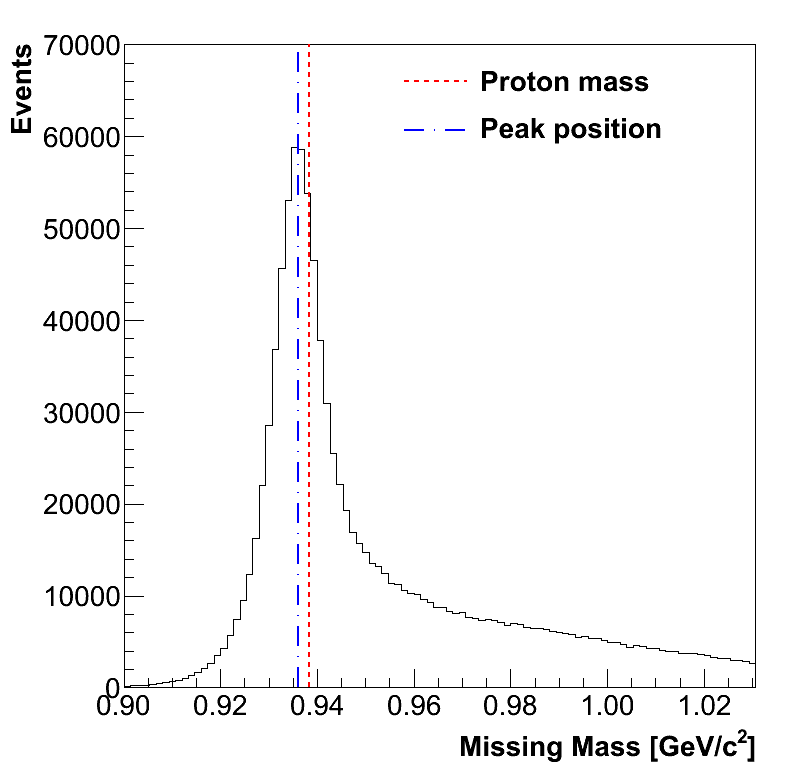}
\includegraphics[width=0.49\textwidth]{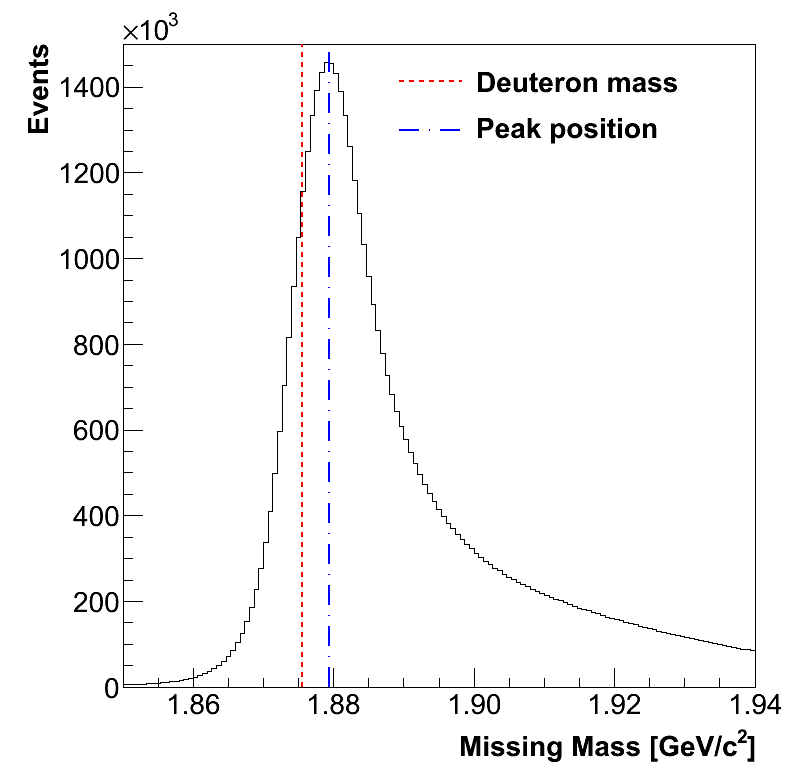}
\caption{The left and the right plot show the reconstructed missing-mass spectra for 
$(\mathrm{e,e'd})$  and $(\mathrm{e,e'p})$ reactions, respectively. 
In the proton channel, the mass of the undetected deuteron can not 
be distinguished from the mass of the proton-neutron pair because of the limited 
resolution of the spectrometers. Dashed blue lines show the positions of the reconstructed 
peaks. Dotted red lines indicate the  proton and deuteron masses.  A tail on the right side 
of the missing mass spectra is a consequence  of radiative losses.
HRS-L  was positioned at $\theta_{\mathrm{HRS-L}} = 12.5^\circ$.
\label{fig_analysis_MissingMass}}
\end{center}
\end{figure}

The extraction of the experimental asymmetries began with the individual analysis
of each collected data-set. The main benefit of such analysis is to precisely account
for the beam and target polarizations which were fluctuating 
during the experiment. After the completion of the secondary analysis 
(see Fig.~\ref{fig_analysis_flowchart}), two sets 
of events were generated from each run. The first set contained $\mathrm{(e,e'p)}$ events,
while the second one contained data from the $\mathrm{(e,e'd)}$ process. 
Figure~\ref{fig_analysis_MissingMass} shows the reconstructed
missing mass spectra for both reaction channels. The positions of the peaks 
correspond to the masses of undetected particles in each reaction channel. 
For the $\mathrm{(e,e'd)}$ channel, 
it should correspond to the mass of the undetected proton. For the 
$\mathrm{(e,e'p)}$ channel, the position of the missing-mass peak should agree 
with the mass of the undetected deuteron (in the case of two-body breakup) or 
combined mass of the undetected $pn$ pair. Unfortunately these two peaks, 
which are separated by the deuteron binding energy ($2.2\,\mathrm{MeV}/c^2$),
can not be distinguished because of the limited resolution of the apparatus. The bulk of 
the long tail on the right side of the observed missing mass spectra is caused by the 
radiative losses of the incident and scattered electron. In the deuteron channel, 
the tail may also include misidentified protons which appear as deuterons. 
However, it was estimated that this contribution is much smaller than 
the contribution of the radiative tail. 

\begin{figure}[!ht]
\begin{center}
\includegraphics[width=1\linewidth]{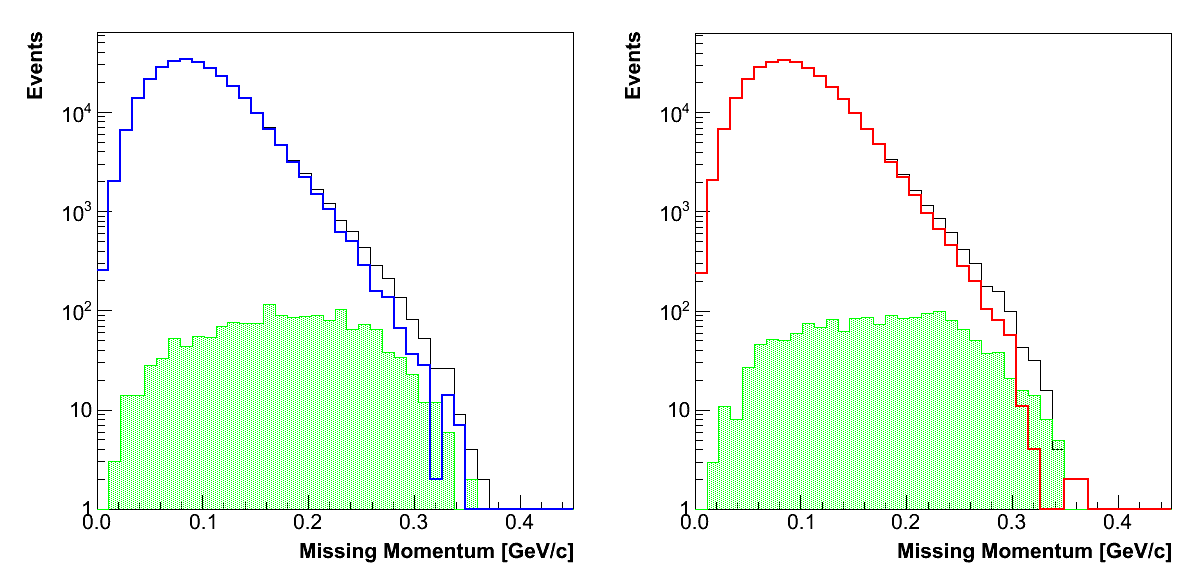}
\vspace*{-8mm}
\caption{Missing momentum distributions for the $(\mathrm{e,e'p})$ reaction, 
obtained for data from groups $7$ and $8$.
Distribution for events with positive and negative beam helicity are shown on the 
left and right plot, respectively. Black lines show uncorrected distributions. Green 
areas represent random coincidence events, obtained by utilizing cuts 
shown in Fig.~\ref{fig_analysis_coincidence_peak}. They need to be subtracted 
(with factor 2) from the main peak in order to determine the final missing-momentum 
distributions, which are shown with blue and red lines.
\label{fig_analysis_MissmMomBackSub}}
\vspace*{-7mm}
\end{center}
\end{figure}

\begin{figure}[!ht]
\begin{center}
\includegraphics[width=1\linewidth]{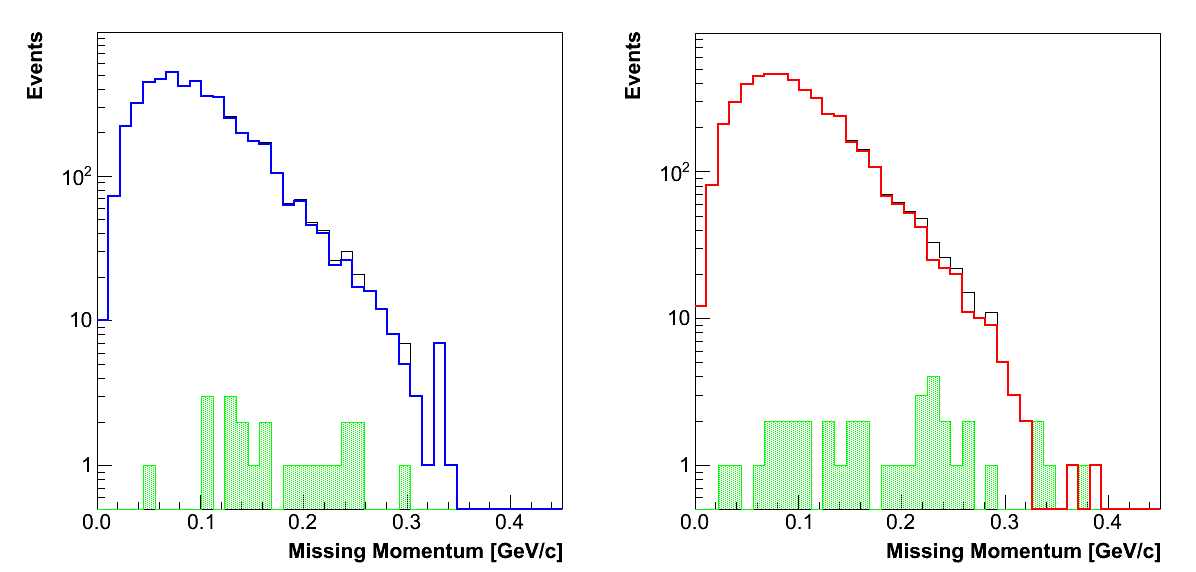}
\vspace*{-8mm}
\caption{Missing momentum distributions for the $(\mathrm{e,e'd})$ reaction, 
obtained for data from groups $7$ and $8$. 
Distribution for events with positive and negative beam helicity are shown on the 
left and right plot, respectively. Black lines show uncorrected distributions. Green 
areas represent random coincidence events, obtained by utilizing cuts 
shown in Fig.~\ref{fig_analysis_coincidence_peak}. They need to be subtracted 
(with factor 2) from the main peak in order to determine the final missing-momentum 
distributions, which are shown with blue and red lines.
\label{fig_analysis_DeuteronMissmMomBackSub}}
\vspace*{-7mm}
\end{center}
\end{figure}

Once the events for each reaction channel were isolated, they were binned 
in missing momentum. For this analysis, $40$ bins were chosen in the missing-momentum range between
$0\,\mathrm{MeV}/c$ and $0.45\,\mathrm{GeV}/c$. Separate histograms were created for 
events with positive and negative helicity. Examples for both reaction channels 
are demonstrated in Figs.~\ref{fig_analysis_MissmMomBackSub} and
~\ref{fig_analysis_DeuteronMissmMomBackSub}.
To subtract the backgrounds from the missing momentum histograms and preserve only 
proper coincidence events, cuts on coincidence-time spectrum were employed 
(see Sec.~\ref{sec:SecondaryCuts}).

\begin{figure}[!htbp]
\begin{center}
\includegraphics[angle=90,width=0.85\linewidth]{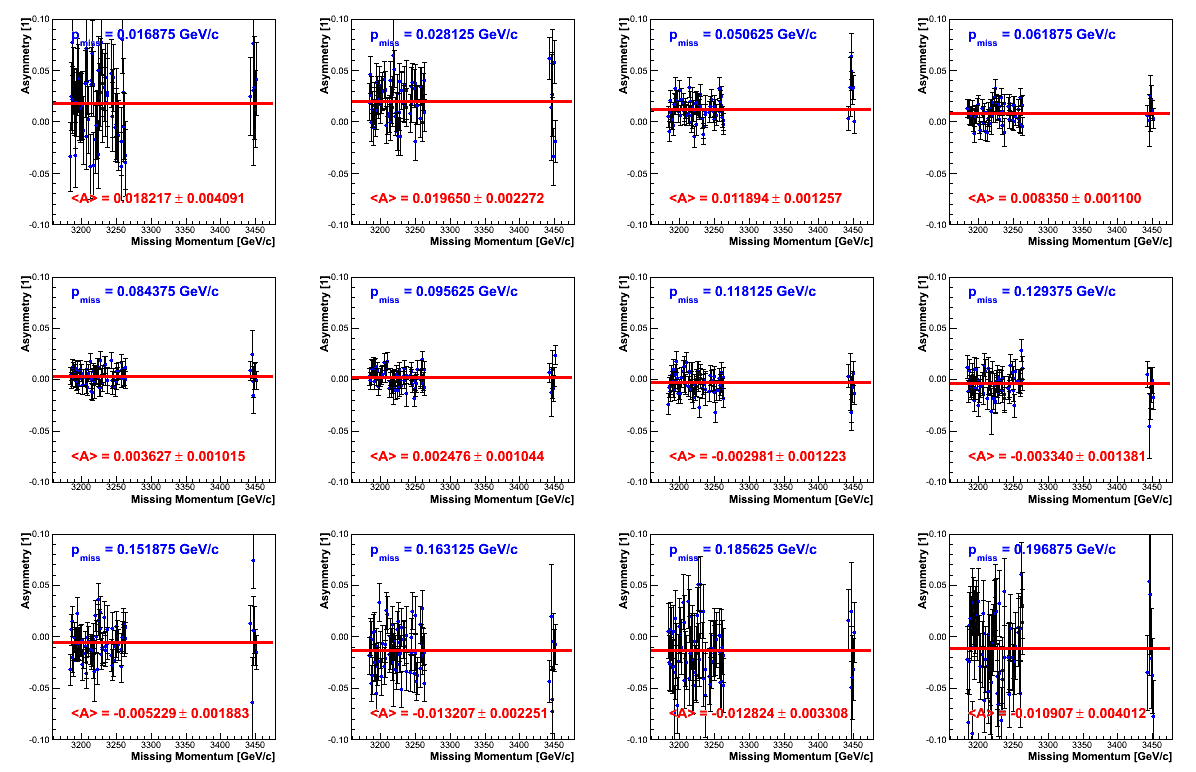} 
\caption{ Asymmetries in the $(\mathrm{e,e'p})$ reaction as function of run number for 
12 different missing momentum  ($p_{\mathrm{miss}}$) bins. 
The obtained results correspond to data from group $7$, where the target was 
polarized longitudinally, HRS-L was positioned at $\theta_{\mathrm{HRS-L}} = 12.5^\circ$ and 
beam HWP was removed. Data with this configuration were collected at two different time periods, 
which explains the empty gap between the two sections of data. Red lines show average 
asymmetries for each $p_{\mathrm{miss}}$ bin.
\label{fig_analysis_averaging1}}
\end{center}
\end{figure}

\begin{figure}[!htbp]
\begin{center}
\includegraphics[angle=90,width=0.85\linewidth]{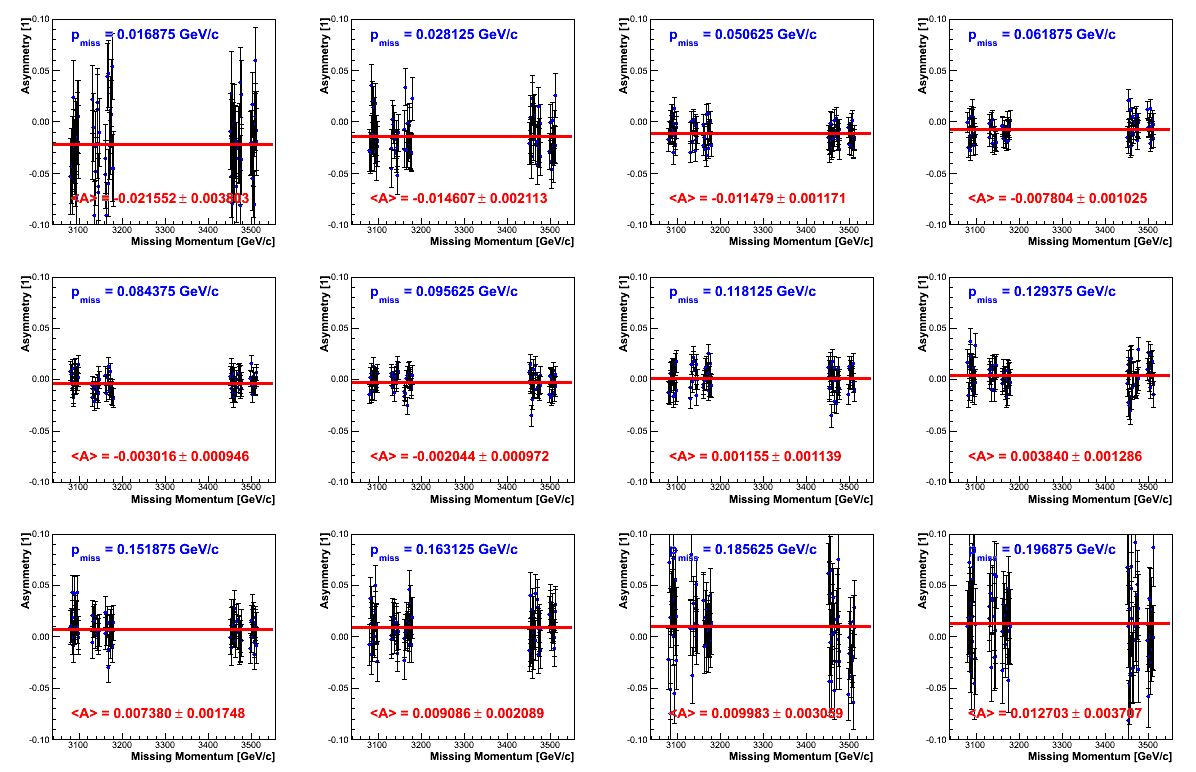} 
\caption{ Asymmetries in the $(\mathrm{e,e'p})$ reaction as function of run number for 
12 different missing momentum ($p_{\mathrm{miss}}$) bins. 
The obtained results correspond to data from group $8$, where the target was 
polarized longitudinally, HRS-L was positioned at $\theta_{\mathrm{HRS-L}} = 12.5^\circ$ and 
beam HWP was inserted. Data with this configuration were collected at two different time periods, 
which explains the empty gap between the two sections of data. Red lines show average 
asymmetries for each $p_{\mathrm{miss}}$ bin.
\label{fig_analysis_averaging2}}
\end{center}
\end{figure}

\begin{figure}[!htbp]
\begin{center}
\includegraphics[angle=90,width=0.85\linewidth]{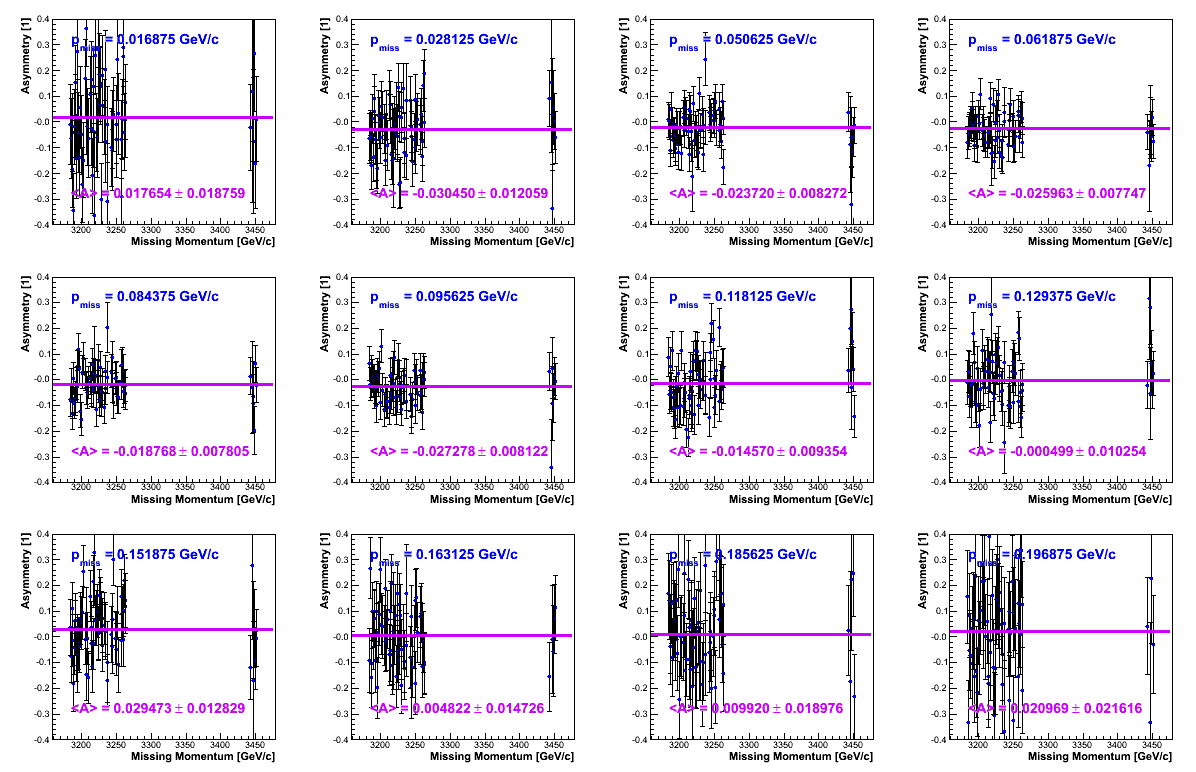} 
\caption{ Asymmetries in the $(\mathrm{e,e'd})$ reaction as function of run number for 
12 different missing-momentum  ($p_{\mathrm{miss}}$) bins. 
The obtained results correspond to data from group $7$, where the target was 
polarized longitudinally, HRS-L was positioned at $\theta_{\mathrm{HRS-L}} = 12.5^\circ$ and 
beam HWP was removed. Data with this configuration were collected at two different time periods, 
which explains the empty gap between the two sections of data. Red lines show average 
asymmetries for each $p_{\mathrm{miss}}$ bin.
\label{fig_analysis_averaging3}}
\end{center}
\end{figure}

\begin{figure}[!htbp]
\begin{center}
\includegraphics[angle=90,width=0.85\linewidth]{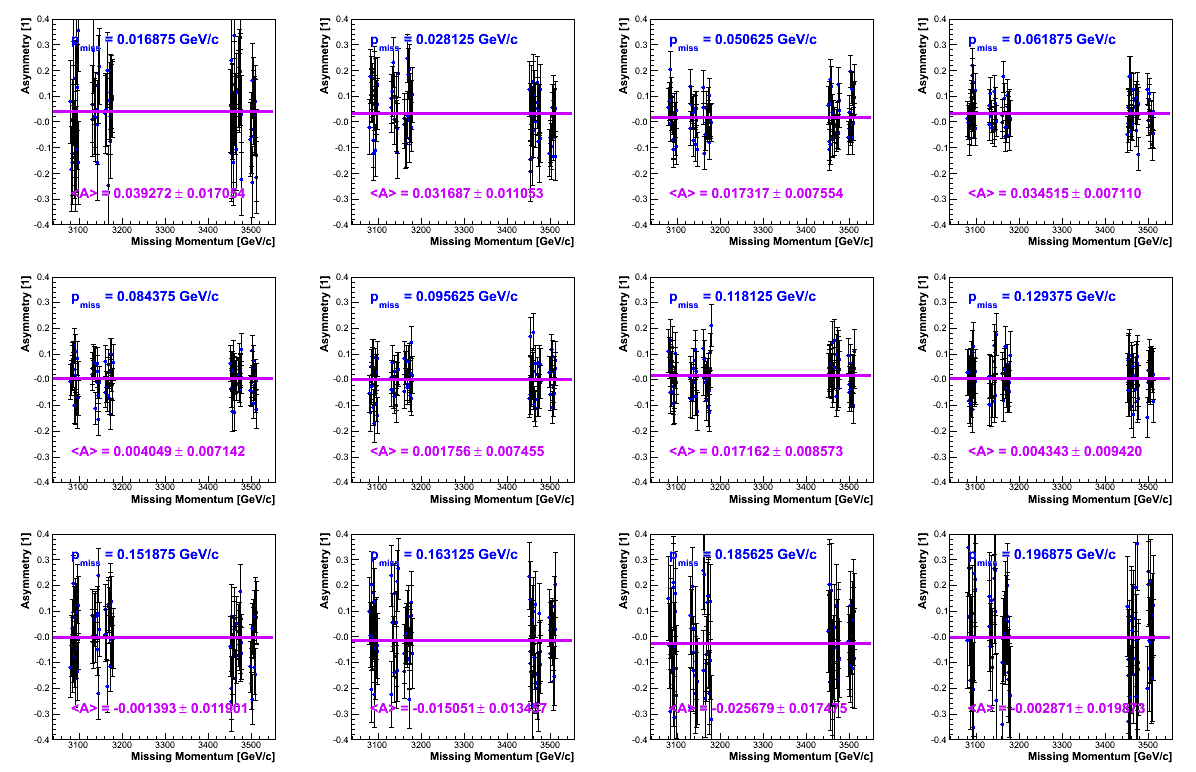} 
\caption{ Asymmetries in the $(\mathrm{e,e'd})$ reaction as function of run number for 
12 different missing momentum ($p_{\mathrm{miss}}$) bins. 
The obtained results correspond to data from group $8$, where the target was 
polarized longitudinally, HRS-L was positioned at $\theta_{\mathrm{HRS-L}} = 12.5^\circ$ and 
beam HWP was inserted. Data with this configuration were collected at two different time periods, 
which explains the empty gap between the two sections of data. Red lines show average 
asymmetries for each $p_{\mathrm{miss}}$ bin.
\label{fig_analysis_averaging4}}
\end{center}
\end{figure}

Next, the resulting histograms for two helicity states were 
joined. The counts in the matching missing-momentum bins were introduced to 
Eq.~(\ref{eq_analysis_asym1}) to calculate the raw asymmetry as a function of 
missing momentum. To retrieve the real physics asymmetry, the obtained raw results
then had to be corrected for the dilutions described in Eq.~(\ref{eq_analysis_asym2}).
This procedure was performed for all runs in  each data group. The results of 
the analysis performed for the data groups $7$ and $8$, where the target was polarized 
longitudinally and HRS-L was positioned at $\theta_{\mathrm{HRS-L}} = 12.5^\circ$,
are shown in Figs.~\ref{fig_analysis_averaging1} to~\ref{fig_analysis_averaging4}.

\begin{figure}[!ht]
\begin{center}
\includegraphics[width=0.49\textwidth]{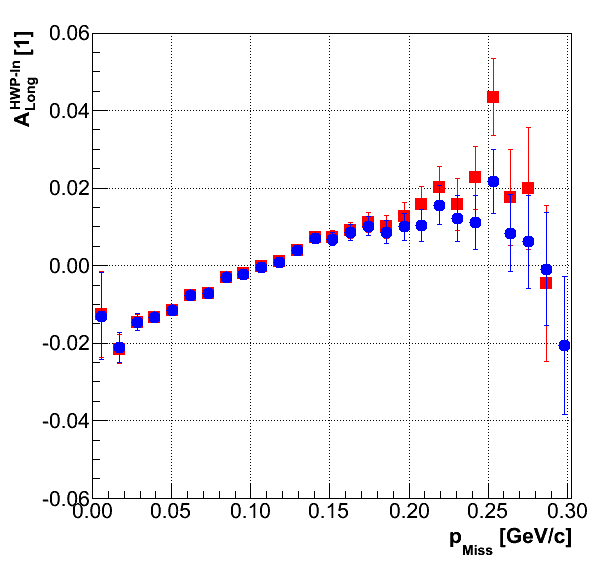}
\includegraphics[width=0.49\textwidth]{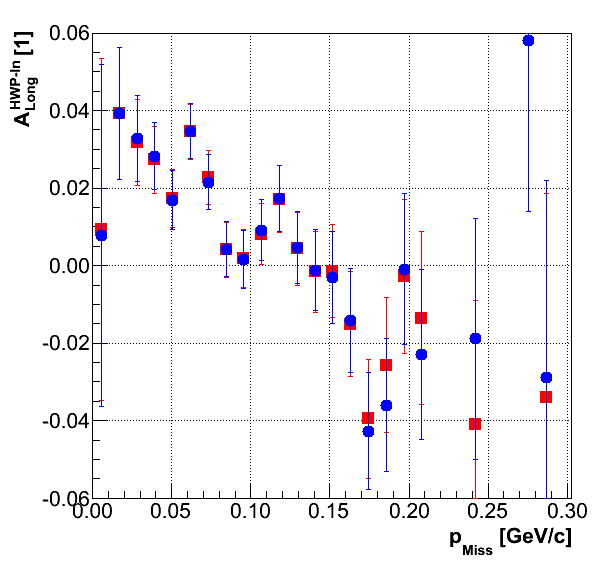}
\vspace*{-7mm}
\caption{ Asymmetry as a function of missing momentum ($p_{\mathrm{miss}}$), 
for $(\mathrm{e,e'p})$ (left) and $(\mathrm{e,e'd})$ (right) reactions.
The mean values and the uncertainties of the asymmetry were determined by calculating
the weighted average of all data in group $7$. Red squares 
and blue circles represent the results with and without random-coincidence background 
corrections. Significant differences appear only at high missing momenta.
\label{fig_analysis_AsymmetryBackSub}}
\vspace*{-7mm}
\end{center}
\end{figure}

By averaging the asymmetries over all runs within the same experimental setting 
(same group), the mean values for each missing momentum bin were determined. 
Averaged asymmetries for data-group $7$ are presented in 
Fig.~\ref{fig_analysis_AsymmetryBackSub}. The figure also shows the difference to the 
mean asymmetry after the subtraction of the random-coincidence background. 
Changes are visible only at high missing momenta, where the statistics of the 
peak and background distributions become comparable.

Once the experimental asymmetries for each group of data 
(see Table~\ref{table_analysis_ListOfData}) were extracted, comparable asymmetries 
could be shown to be consistent. For each kinematical setting, approximately 
half of the statistics were collected with the beam HWP removed and half with the HWP inserted.
The beam HWP (see Sec:~\ref{sec:HWP}) flips the orientation of the beam helicity. Hence, the
asymmetries obtained for the same kinematics setting, but with different HWP position, should 
differ only in sign.
Any inconsistencies in the observed results would be a direct indication of presence 
of false asymmetries.

\begin{figure}[!ht]
\begin{center}
\includegraphics[width=1\textwidth]{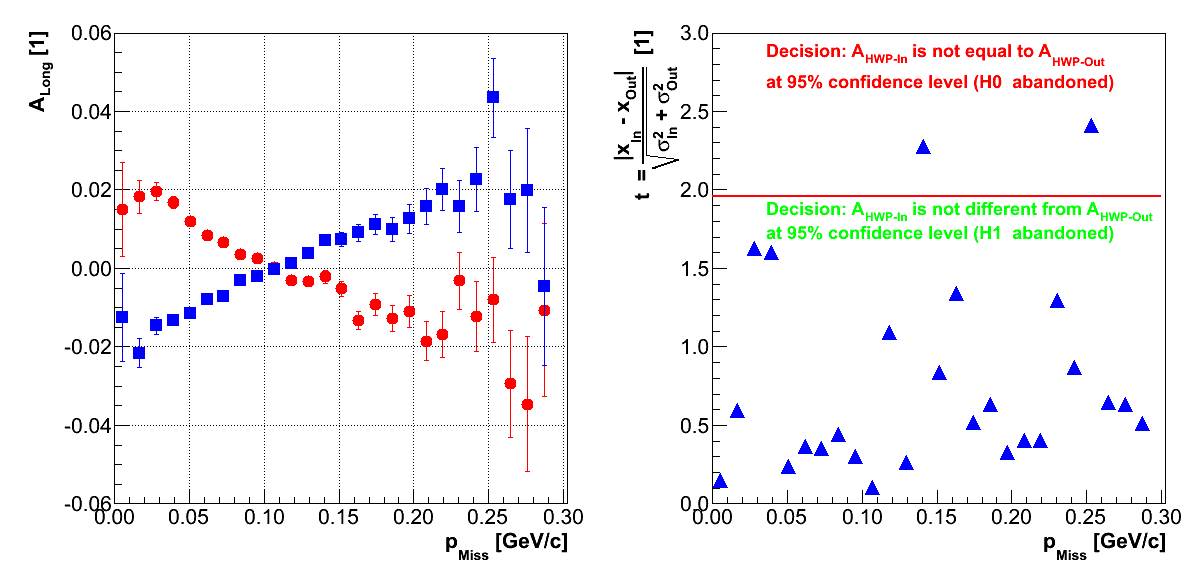}
\caption{[Left] Experimental asymmetries obtained with the data from  groups $7$ and $8$. Blue squares 
show asymmetries with HWP removed, while the red circles correspond to results with HWP inserted. 
[Right] Results of the Student's hypothesis test performed for data on the left. Red line corresponds 
to $t=1.96$. For values above this threshold, the hypothesis $H_0$ is rejected at
 $95\,\mathrm{\%}$ confidence level. 
\label{fig_analysis_StudentTest1}}
\end{center}
\end{figure}

Potential discrepancies between the data with HWP inserted and removed were pursued  via the Student's 
hypothesis test. In this test the null hypothesis $H_0$ claimed that the 
asymmetry with the HWP inserted, $A_{\mathrm{HWP-In}}$,
agrees with the negative value of asymmetry with HWP removed, $A_{\mathrm{HWP-Out}}$. 
The alternative hypothesis $H_1$ states that these two asymmetries do not match:
\begin{eqnarray}
  H_0:\> A_{\mathrm{HWP-In}} = -A_{\mathrm{HWP-Out}}\,, \qquad   H_1:\> A_{\mathrm{HWP-In}} \neq -A_{\mathrm{HWP-Out}}\,. \nonumber
\end{eqnarray}
The hypotheses were tested in terms of parameter $t$, which is defined as:
\begin{eqnarray}
 t = \frac{|A_{\mathrm{HWP-In}} + A_{\mathrm{HWP-Out}}|}{\sqrt{\sigma_{A_{\mathrm{HWP-In}}}^2 + \sigma_{A_{\mathrm{HWP-Out}}}^2}}\,,\nonumber
\end{eqnarray}
where $\sigma_{A_{\mathrm{HWP-Out}}}$ and $\sigma_{A_{\mathrm{HWP-In}}}$ represent statistical errors
of the determined asymmetries. Depending on the value of the parameter, the following decisions can be made:
\begin{itemize}
\item If $t \geq 1.960$, the hypothesis $H_0$ can be rejected at $95\,\mathrm{\%}$ confidence level. This 
means that there is a $95\,\mathrm{\%}$ probability that two asymmetries are inconsistent.

\item If $t < 1.960$, the hypothesis $H_1$ can be rejected at $95\,\mathrm{\%}$ confidence level, meaning
that there is a $95\,\mathrm{\%}$ probability that asymmetries are consistent.  
\end{itemize}
Student's test was performed for all kinematic settings considered in the analysis. The comparison was
done separately for each bin in missing momentum. The findings of the comparison done for the data from 
groups $7$ and $8$ are gathered in Fig.~\ref{fig_analysis_StudentTest1}.  The results show 
that, with the exception of two points, the asymmetries  $A_{\mathrm{HWP-Out}}$ and  $A_{\mathrm{HWP-In}}$ are not 
different. Similar results were obtained also for the rest of the experimental settings, all reporting
a good agreement between asymmetries with HWP inserted and removed. 

\begin{figure}[!htb]
\begin{center}
\includegraphics[width=1\textwidth]{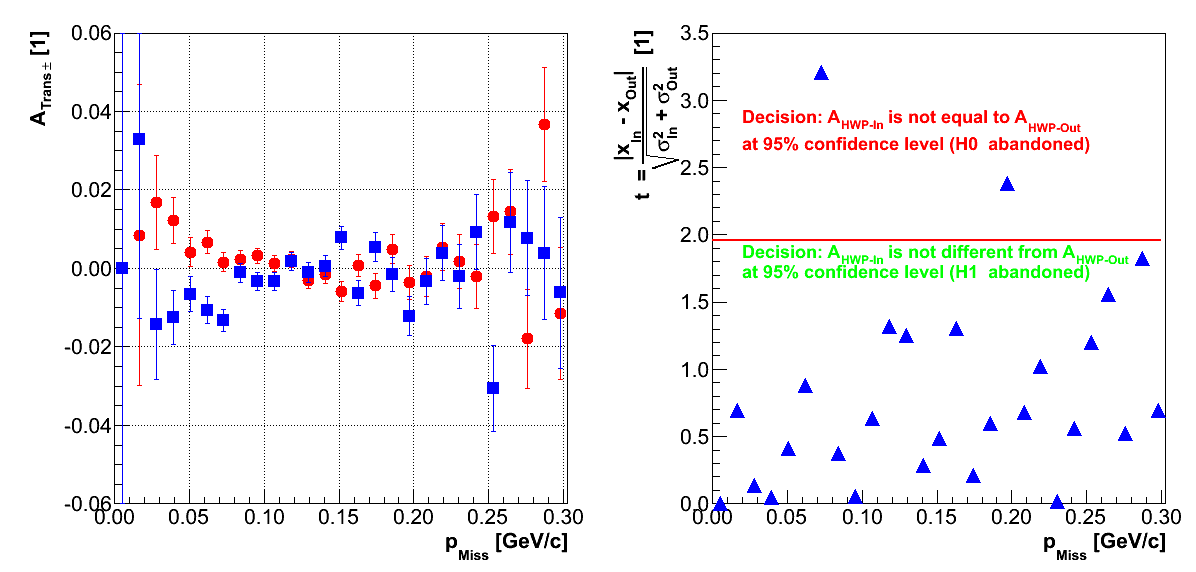}
\caption{[Left] Experimental asymmetries obtained with the data from groups $4$ and $6$. Blue squares 
show asymmetries with HWP removed, while the red circles correspond to results with HWP inserted. 
[Right] Results of the Student's hypothesis test performed for data on the left. Red line corresponds 
to $t=1.96$. For values above this threshold, the hypothesis $H_0$ is rejected at
 $95\,\mathrm{\%}$ confidence level.  
\label{fig_analysis_StudentTestTarget}}
\end{center}
\end{figure}

\begin{figure}[!htb]
\begin{center}
\begin{minipage}[t]{0.63\textwidth}
\hrule height 0pt
\includegraphics[width=1\textwidth]{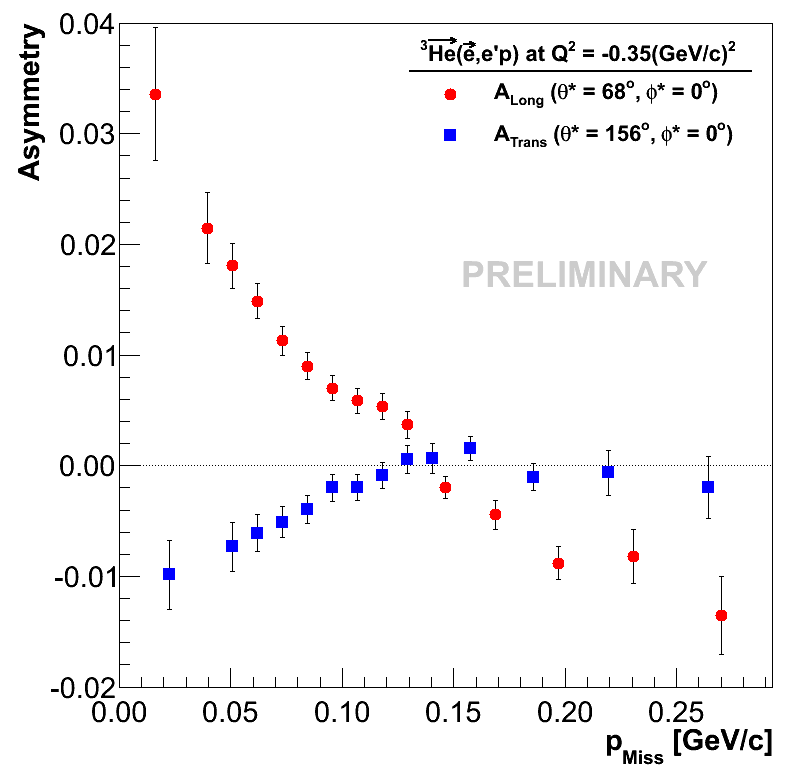}
\end{minipage}
\hfill
\begin{minipage}[t]{0.35\textwidth}
\hrule height 0pt
\caption{ Measured longitudinal (red circles) and transverse (blue squares) 
asymmetries for the ${}^3\vec{\mathrm{He}}(\vec{e},e'p)$  process 
as a function of $p_{\mathrm{Miss}}$, 
for  the setting, in which the target was polarized along  and perpendicularly 
to the beam direction, HRS-L was 
positioned at $\theta_{\mathrm{HRS-L}} = 14.5^\circ$ and BigBite 
at $\theta_{\mathrm{BB}} = -75^\circ$.  Angles $\theta^*$ and $\phi^*$ correspond to 
the mean direction of the momentum-transfer vector with respect to the target spin orientation. 
Error-bars represent statistical uncertainties only. Corresponding systematical 
uncertainties are  discussed in Sec.~\ref{Sec:Systematics}.
\label{fig_analysis_ProtonAsymmetry14}}
\end{minipage}
\end{center}
\end{figure}

Considering relation (\ref{eq_theory_finalasymmetry_pd}), equivalent tests could be performed
also by flipping the target orientation while keeping the beam-HWP in the same position. 
This offers an additional check to see if the asymmetry changes sign when the target polarization is
oriented in the opposite direction. These cross-checks could unfortunately be realized only for 
the data with transverse  target polarization and HRS-L at $\theta_{\mathrm{HRS-L}} = 14.5^\circ$. 
For the rest of the data (see Table~\ref{table_analysis_ListOfData}) only one target orientation was 
considered. The results of the test are shown Fig.~\ref{fig_analysis_StudentTestTarget}. With exception of 
a few points, the  data with opposite target orientations are found to be consistent.

\begin{figure}[!ht]
\begin{center}
\begin{minipage}[t]{0.63\textwidth}
\hrule height 0pt
\includegraphics[width=1\textwidth]{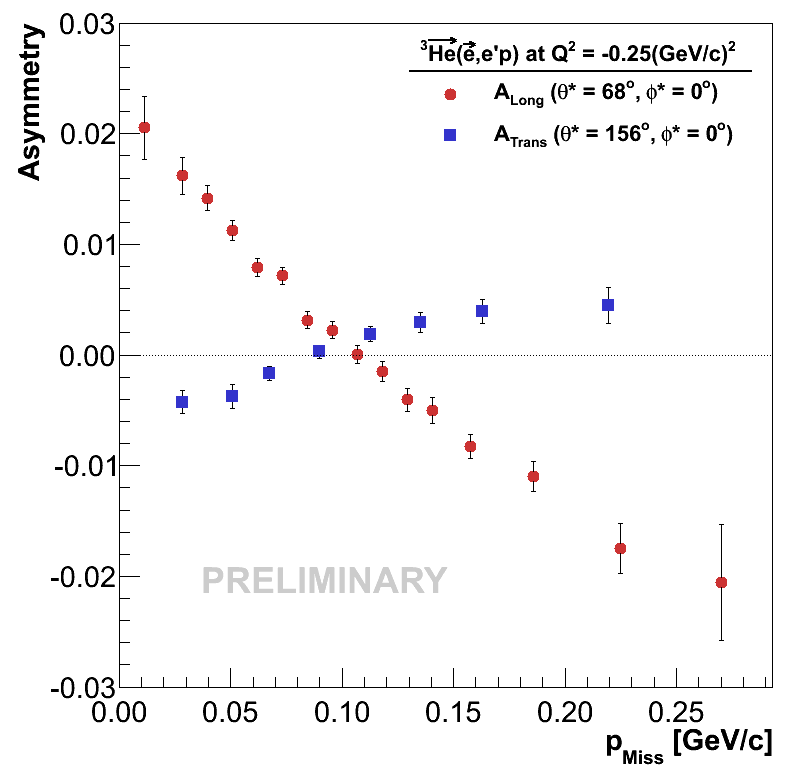}
\end{minipage}
\hfill
\begin{minipage}[t]{0.35\textwidth}
\hrule height 0pt
\caption{ Measured longitudinal (red circles) and transverse (blue squares) 
asymmetries for the ${}^3\vec{\mathrm{He}}(\vec{e},e'p)$  process
as a function of $p_{\mathrm{Miss}}$, 
for  the setting, in which the target was polarized along and perpendicularly 
to the beam direction, HRS-L was 
positioned at $\theta_{\mathrm{HRS-L}} = 12.5^\circ$ and BigBite was 
at $\theta_{\mathrm{BB}} = -75^\circ$. 
Notation as in Fig.~\ref{fig_analysis_ProtonAsymmetry14}.
\label{fig_analysis_ProtonAsymmetry12}}
\end{minipage}
\end{center}
\end{figure}

\begin{figure}[!ht]
\begin{center}
\begin{minipage}[t]{0.63\textwidth}
\hrule height 0pt
\includegraphics[width=1\textwidth]{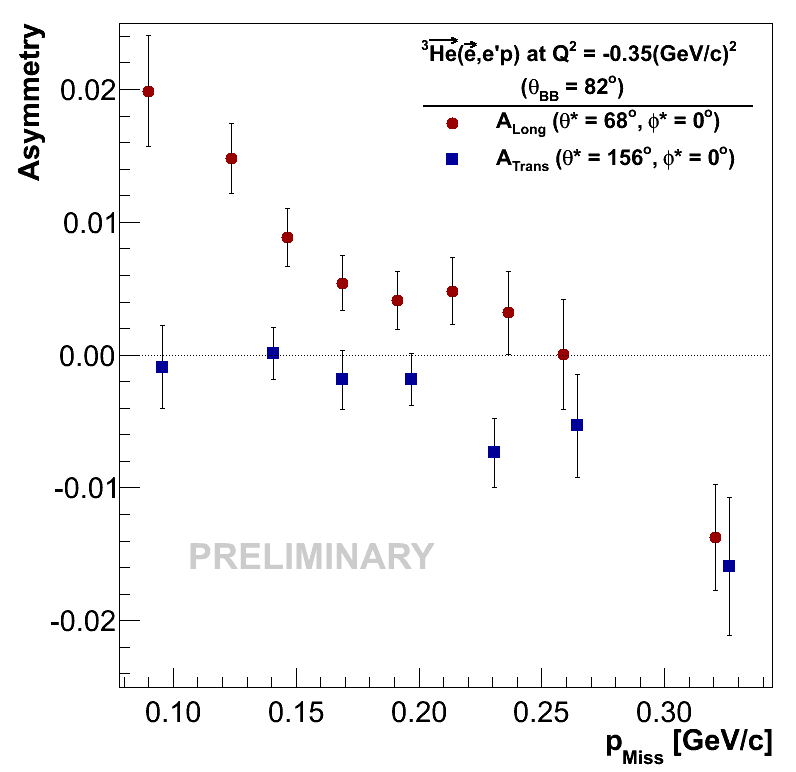}
\end{minipage}
\hfill
\begin{minipage}[t]{0.35\textwidth}
\hrule height 0pt
\caption{ Measured longitudinal (red circles) and transverse (blue squares) 
asymmetries for the ${}^3\vec{\mathrm{He}}(\vec{e},e'p)$  process
as a function of $p_{\mathrm{Miss}}$, 
for  the setting, in which the target was polarized along and perpendicularly 
to the beam direction, HRS-L was 
positioned at $\theta_{\mathrm{HRS-L}} = 14.5^\circ$ and BigBite was 
at $\theta_{\mathrm{BB}} = -82^\circ$. 
Notation as in Fig.~\ref{fig_analysis_ProtonAsymmetry14}.
\label{fig_analysis_ProtonAsymmetry82}}
\end{minipage}
\end{center}
\end{figure}

\begin{figure}[!ht]
\begin{center}
\begin{minipage}[t]{0.63\textwidth}
\hrule height 0pt
\includegraphics[width=1\textwidth]{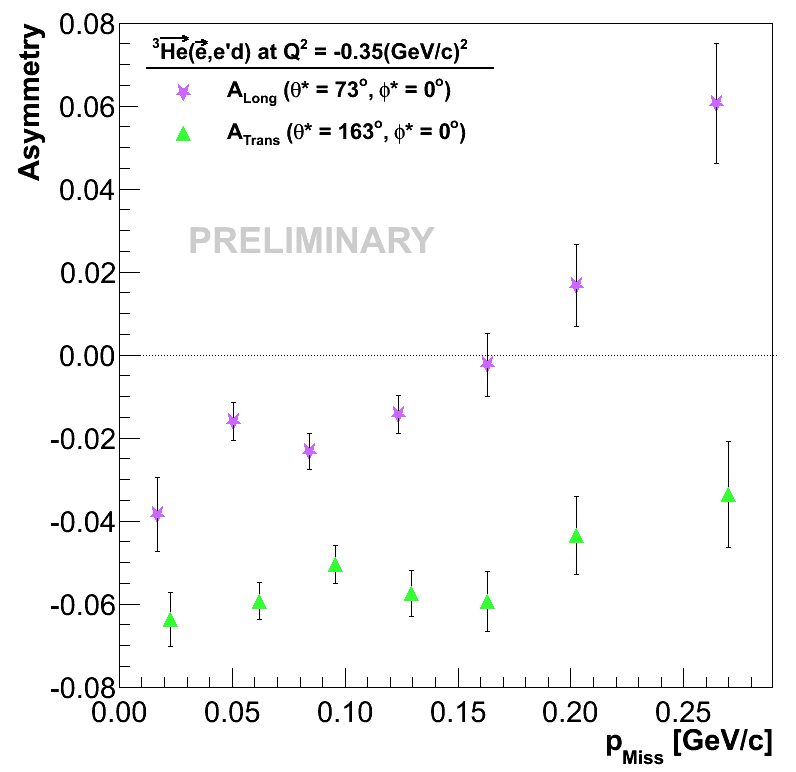}
\end{minipage}
\hfill
\begin{minipage}[t]{0.35\textwidth}
\hrule height 0pt
\caption{ Measured longitudinal (violet stars) and transverse (green triangles) 
asymmetries for the ${}^3\vec{\mathrm{He}}(\vec{e},e'd)p$  process
as a function of $p_{\mathrm{Miss}}$, 
for  the setting, in which the target was polarized along and perpendicularly 
to the beam direction, HRS-L was 
positioned at $\theta_{\mathrm{HRS-L}} = 14.5^\circ$ and BigBite was 
at $\theta_{\mathrm{BB}} = -75^\circ$. 
Notation as in Fig.~\ref{fig_analysis_ProtonAsymmetry14}.
\label{fig_analysis_DeuteronAsymmetry14}}
\end{minipage}
\end{center}
\end{figure}

\begin{figure}[!ht]
\begin{center}
\begin{minipage}[t]{0.63\textwidth}
\hrule height 0pt
\includegraphics[width=1\textwidth]{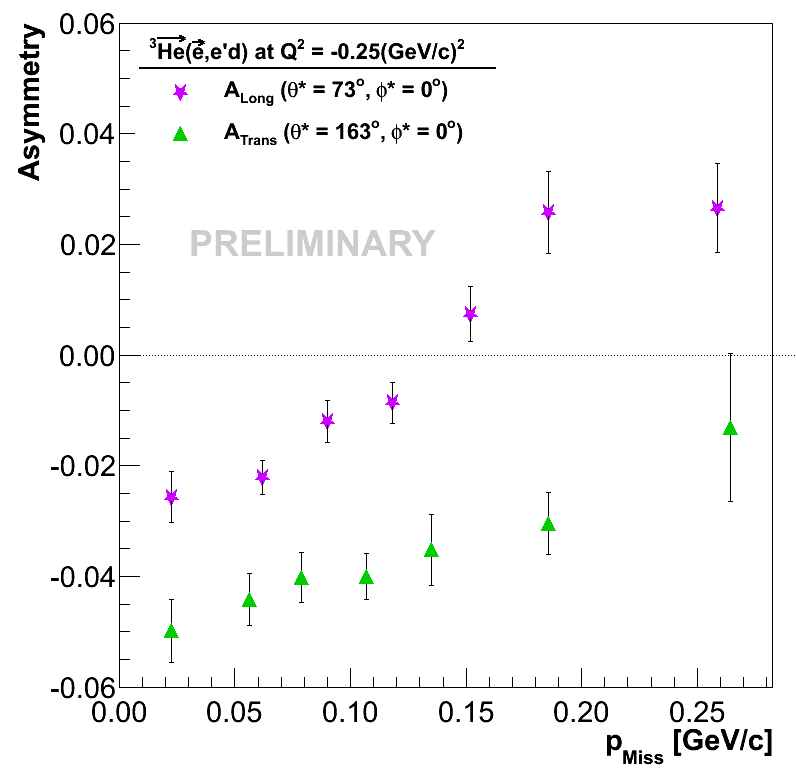}
\end{minipage}
\hfill
\begin{minipage}[t]{0.35\textwidth}
\hrule height 0pt
\caption{ Measured longitudinal (violet stars) and transverse (green triangles) 
asymmetries for the ${}^3\vec{\mathrm{He}}(\vec{e},e'd)p$  process
as a function of $p_{\mathrm{Miss}}$, 
for  the setting, in which the target was polarized along and perpendicularly 
to the beam direction, HRS-L was 
positioned at $\theta_{\mathrm{HRS-L}} = 12.5^\circ$ and BigBite was 
at $\theta_{\mathrm{BB}} = -75^\circ$. 
Notation as in Fig.~\ref{fig_analysis_ProtonAsymmetry14}.
\label{fig_analysis_DeuteronAsymmetry12}}
\end{minipage}
\end{center}
\end{figure}

\begin{figure}[!ht]
\begin{center}
\begin{minipage}[t]{0.63\textwidth}
\hrule height 0pt
\includegraphics[width=1\textwidth]{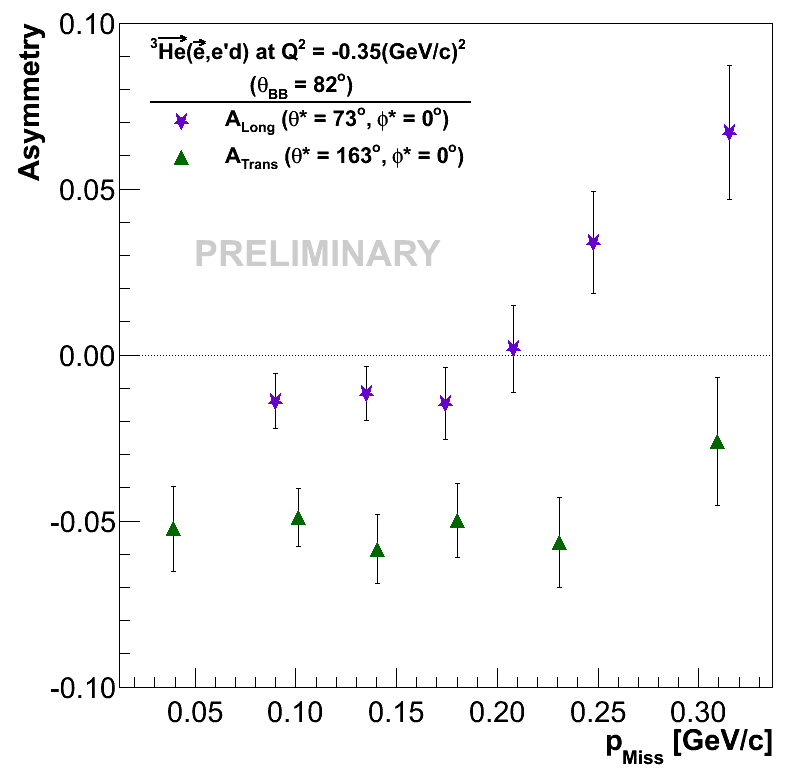}
\end{minipage}
\hfill
\begin{minipage}[t]{0.35\textwidth}
\hrule height 0pt
\caption{Measured longitudinal (violet stars) and transverse (green triangles) 
asymmetries for the ${}^3\vec{\mathrm{He}}(\vec{e},e'd)p$  process
as a function of $p_{\mathrm{Miss}}$, 
for  the setting, in which the target was polarized along and perpendicularly 
to the beam direction, HRS-L was 
positioned at $\theta_{\mathrm{HRS-L}} = 14.5^\circ$ and BigBite was 
at $\theta_{\mathrm{BB}} = -82^\circ$. 
Notation as in Fig.~\ref{fig_analysis_ProtonAsymmetry14}.
\label{fig_analysis_DeuteronAsymmetry82}}
\end{minipage}
\end{center}
\end{figure}

Once the consistency tests were successfully completed, comparable sets of data were merged
to increase the statistical accuracy. With the longitudinally
polarized target, the data with the HWP inserted and removed could be combined for each kinematical 
setting. Hence, group  $1$ was coupled with group $2$, $7$ with $8$, and $11$ with $12$. In the 
case of the transversely polarized target, the data with opposite spin directions could also been joined.
This way groups $3-6$ were fused, $9,\,10$, and $13-16$. In all these couplings, an appropriate
sign correction had to be applied for the data with the inserted HWP or with a negative orientation 
of the target spin. Using these combined sets of data, final asymmetries were calculated
for both target orientations and all three kinematical settings. Unfortunately, asymmetries could 
not be determined adequately for all 40 missing-momentum bins because of the limited statistics, 
especially at very low and very high missing momenta. Therefore, some 
of the bins were joined, resulting in an average asymmetry for selected range of momenta, with better 
statistical uncertainty. The final experimental results for the $(\mathrm{e,e'p})$ channel 
are shown in Figs.~\ref{fig_analysis_ProtonAsymmetry14},~\ref{fig_analysis_ProtonAsymmetry12} and
~\ref{fig_analysis_ProtonAsymmetry82}. The results for the $(\mathrm{e,e'd})$ reaction are
shown in Figs.~\ref{fig_analysis_DeuteronAsymmetry14},~\ref{fig_analysis_DeuteronAsymmetry12} and
~\ref{fig_analysis_DeuteronAsymmetry82}. 
Due to much smaller statistics, fever missing-momentum bins could be afforded for the 
$(\mathrm{e,e'd})$ reaction than for the $(\mathrm{e,e'p})$ reaction. Statistics was 
especially poor for configurations $11-16$ where BigBite was positioned at $\theta_{\mathrm{BB}} = -82^\circ$. 
In this kinematical setting, asymmetries at large missing momenta were measured. Regrettably,
only a small portion of experimental data was collected in this setting, resulting in larger uncertainties 
of the calculated asymmetries.

\section{Systematic uncertainties}
\label{Sec:Systematics}

A list of significant systematic uncertainties of the measured double-polarized asymmetries 
$A_{\mathrm{Long}}$ and $A_{\mathrm{Trans}}$ in the processes ${}^3\vec{\mathrm{He}}(\vec{e},e'd)$
and ${}^3\vec{\mathrm{He}}(\vec{e},e'p)$ is given in Table~\ref{table_analysis_SystematicErrors}.
The dominant part of the errors is contributed by the uncertainties in the target and beam polarization. 
Ambiguities in the nitrogen dilution factor also contribute significantly. 
The $(\vec{e},e'd)$ asymmetries are also affected by the misidentification 
of the protons as deuterons, which brings a few percent relative correction to the final results. 
For the $(\vec{e},e'p)$ channel, this contribution is negligible.

Uncertainties due to the  fluctuations of target density are also imperceptible. 
Chan\-ges in density can affect the particle detection rates, but it is believed
that these changes are much slower than the frequency of beam helicity flips, and  
therefore do not affect the asymmetry.

\begin{table}[!ht]
\begin{center}
\caption{Summary of the systematic uncertainties of the double-polarized asymmetries 
$A_{\mathrm{Long}}$ and $A_{\mathrm{Trans}}$ in processes ${}^3\vec{\mathrm{He}}(\vec{e},e'd)$
and ${}^3\vec{\mathrm{He}}(\vec{e},e'p)$.
\label{table_analysis_SystematicErrors}}
\vspace*{2mm}
\begin{tabular}{lcr}
\hline

\hline

\hline
{\bf Source} & {\bf Format} & {\bf Uncertainty} \\
\hline

\hline

\hline
 & &  \\[-4mm]
Beam polarization & $\left| \delta A_{\mathrm{Exp}}/A_{\mathrm{Exp}} \right|$ & $\sim 2\,\mathrm{\%}$\\[4pt]
Target polarization & $\left| \delta A_{\mathrm{Exp}}/A_{\mathrm{Exp}} \right|$ & $\sim 5\,\mathrm{\%}$\\[4pt]
Nitrogen dilution & $\left|\delta A_{\mathrm{Exp}}/A_{\mathrm{Exp}} \right|$ & $\sim 2.1\,\mathrm{\%}$ \\[4pt]
Target Wall dilution & $\left|\delta A_{\mathrm{Exp}}/A_{\mathrm{Exp}} \right|$ & $\lesssim 0.1\,\mathrm{\%}$ \\[4pt]
Live-time asymmetry & $\delta A_{\mathrm{Raw}}$ & $\sim 4\times 10^{-4}$ \\[3pt]
Beam-Charge asymmetry & $\delta A_{\mathrm{Raw}}$ & $\sim 1.5\times 10^{-5}$ \\[3pt]
Misidentification: & &  \\
\multicolumn{1}{l}{$\qquad(\mathrm{e,e'd})$} & $\left|\delta A_{\mathrm{Exp}}/A_{\mathrm{Exp}} \right|$ & $\lesssim 3\,\mathrm{\%}$ \\[4pt]
\multicolumn{1}{l}{$\qquad(\mathrm{e,e'p})$} & $\left|\delta A_{\mathrm{Exp}}/A_{\mathrm{Exp}} \right|$ & $\lesssim 0.1\,\mathrm{\%}$ \\[2mm]
Target orientation & $\left|\delta A_{\mathrm{Exp}}/A_{\mathrm{Exp}} \right|$ & $\sim 0.6\,\mathrm{\%}$ \\[4pt]
Target orientation & $\left|\delta A_{\mathrm{Exp}}/A_{\mathrm{Exp}} \right|$ & $\sim 0.6\,\mathrm{\%}$ \\[4pt]
BigBite Resolution: & &  \\
\multicolumn{1}{l}{$\qquad$Out-of-plane angle $\theta_{\mathrm{Tg}}$} & $\left|\delta A_{\mathrm{Exp}}\right|$ & $< 2\times10^{-5}$ \\[4pt]
\multicolumn{1}{l}{$\qquad$In-plane angle $\phi_{\mathrm{Tg}}$} & $\left|\delta A_{\mathrm{Exp}}\right|$ & $< 2\times10^{-4}$ \\[4pt]
\multicolumn{1}{l}{$\qquad$Position $y_{\mathrm{Tg}}$} & $\left|\delta A_{\mathrm{Exp}}\right|$ & $< 5\times10^{-4}$ \\[4pt]
\multicolumn{1}{l}{$\qquad$Momentum $\delta_{\mathrm{Tg}}$} & $\left|\delta A_{\mathrm{Exp}}\right|$ & $< 4\times10^{-4}$ \\[2mm]

\hline

\hline

\hline
 & &  \\[-4mm]
{\bf TOTAL $\quad(\mathrm{e,e'd})$} & $\left|\delta A_{\mathrm{Exp}}/A_{\mathrm{Exp}} \right|$ & $\lesssim 7\,\mathrm{\%}$\\[3pt]
{\bf TOTAL $\quad(\mathrm{e,e'p})$} & $\left|\delta A_{\mathrm{Exp}}/A_{\mathrm{Exp}} \right|$ & $\lesssim 6\,\mathrm{\%}$\\[3pt]
\end{tabular}
\end{center}
\end{table}

The uncertainties in the asymmetry due to differences in accumulated charges for both helicity states 
are minimal. Such are also the uncertainties caused by the asymmetric DAQ live-times. The detector
inefficiencies are also believed to be independent of the helicity state and can not modify the asymmetry.
Ambiguities caused by the elastic ${}^3\vec{\mathrm{He}}(\vec{e},e'{}^3\mathrm{He})$ events 
were also neglected. By choosing only coincidence events all elastic events got automatically
rejected, since  BigBite was unable to detect ${}^3\mathrm{He}$ nuclei.

The uncertainties in the asymmetries caused by the limitations of spectrometer optics 
have also been studied.  The dominant part is contributed by the BigBite spectrometer, whose optics  
resolution is inferior to the resolution of the HRS-L. The size of the correction was
estimated by individually shifting the values of the reconstructed target variables 
within the resolutions given in Table~\ref{table_BBChar}. Then, the asymmetries obtained with this
modification were compared to the asymmetries without it. The mean difference between both 
asymmetries was considered as an estimate for the systematic uncertainties. The obtained 
results are gathered in Table~\ref{table_analysis_SystematicErrors}. The systematic 
errors actually do not originate from the widths of the peaks, but from the systematic shift 
of the reconstructed coordinates. However, these shifts are always smaller than the typical 
widths of the reconstructed target coordinates. Therefore, our estimates represent
the upper limits of the systematic uncertainties caused by the spectrometer optics.

\section{Radiative corrections}

\begin{figure}[!ht]
\begin{center}
\includegraphics[width=0.49\textwidth]{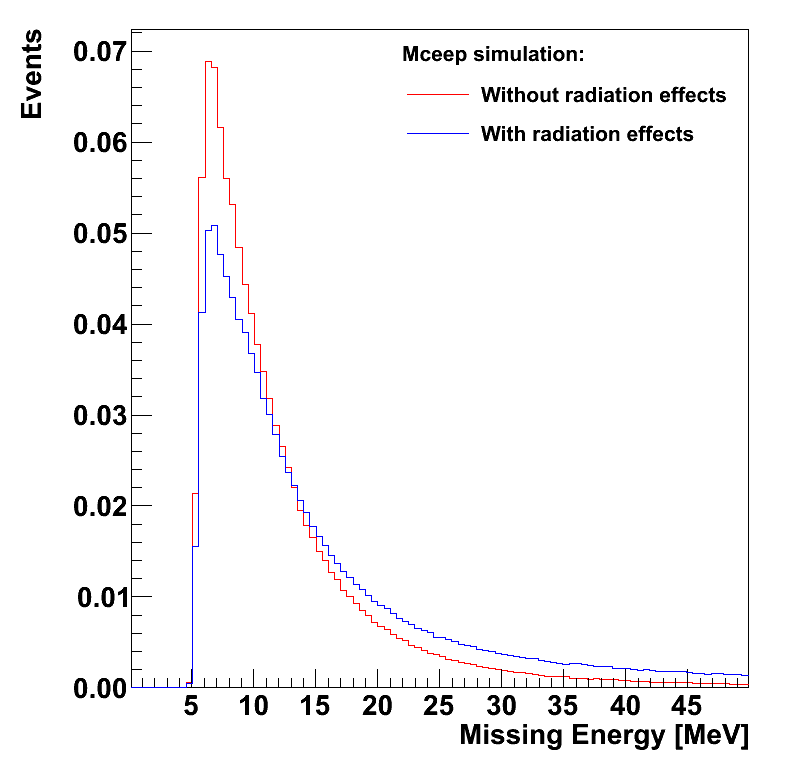}
\includegraphics[width=0.49\textwidth]{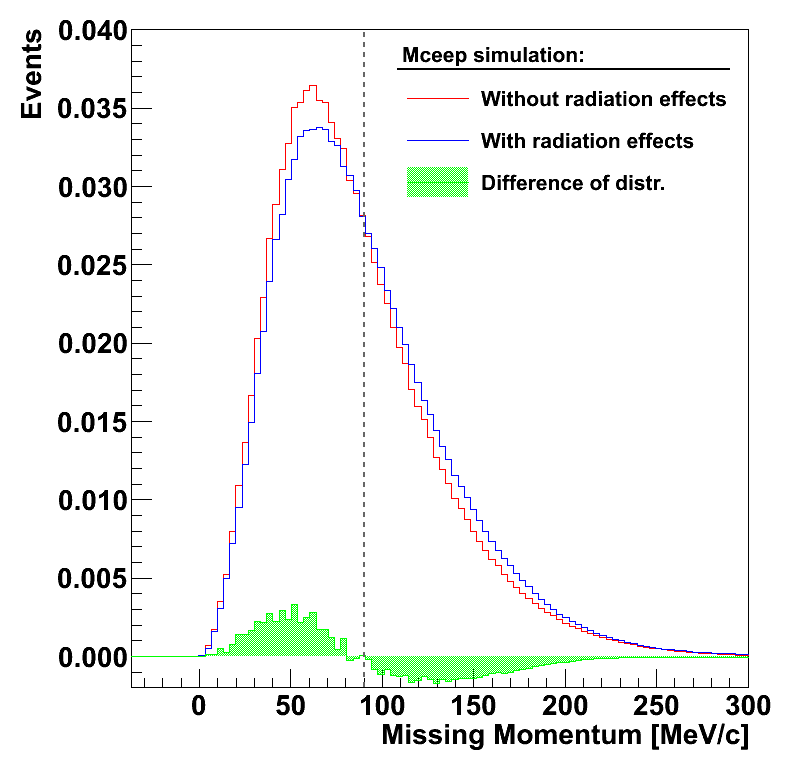}
\caption{[Left] The red and blue line show the simulated distributions of 
missing energy with and without radiative effects, respectively, for the
${}^3\mathrm{He(e,e'p)}$ process, by using MCEEP simulation.
[Right] Same as left, but for distributions of missing momentum. The green histogram
represents the difference between the two distributions. In the low momentum
region, the distribution without radiative smearing prevails, while the high 
momentum region is dominated by the distribution with the radiative effects,
causing the difference to be negative. The dashed
line represent the point, where the difference distribution changes sign.    
\label{fig_analysis_RadiativeCorr}}
\end{center}
\end{figure}

In the theory, the interaction of the electron with the ${}^3\mathrm{He}$ nucleus is
described by the exchange of a single virtual photon (see Fig.~\ref{fig_theory_Kinematics}).
In reality, the incoming and outgoing charged particles emit additional real and 
virtual photons~\cite{fatihaPhD}. The presence of these radiations changes the cross-section 
for the investigated reaction, as well as the distributions of energy-transfer, $\omega$, 
and momentum-transfer vector, $|\vec{q}|$. These changes usually result in 
long tails in the measured distributions (see Fig.~\ref{fig_analysis_RadiativeCorr}). 
Since the theoretical calculations do not include these effects,  they need to be 
considered before comparing the measurements to the 
theory~\cite{mo_tsai}. 

Effects of the radiative losses on the asymmetries, measured in the E05-102 experiment, 
have not been studied in detail yet. In order to obtain proper corrections to the 
asymmetries, one needs to understand the spin dependence of the radiation effects, 
beside the usual corrections to the  unpolarized cross-section, which significantly 
complicates the analysis. However, the majority of the radiation effects cancel 
with the calculation of the asymmetry.  Therefore we expect that the correction 
to the asymmetry will be substantially smaller than the effects on the  
measured cross-sections. 

In the first evaluation of the radiative corrections, we employed MCEEP~\cite{mceep} 
to simulate the unpolarized distributions of missing momentum with and without
use of radiative effects. The results are shown in Fig.~\ref{fig_analysis_RadiativeCorr} (right).
One sees that the distributions are not very different. 
To obtain a very conservative estimate for the correction we can assume 
that a portion of events ($\delta\lesssim 6\,\mathrm{\%}$) from the region 
below $p_{\mathrm{Miss}}\leq 80\,\mathrm{MeV}/c$ has moved to higher missing 
momenta ($p_{\mathrm{Miss}} > 80\,\mathrm{MeV}/c$). Since the asymmetry changes 
significantly with the missing momentum (see Fig.~\ref{fig_analysis_ProtonAsymmetry14}) 
the presence of the migrated events could alter the asymmetry in the high missing momentum 
region. Using the values of the asymmetries for the centers of both regions 
($A(p_{\mathrm{Miss}}=50\,\mathrm{MeV}/c) \approx 1.8\,\mathrm{\%}$ and
$A(p_{\mathrm{Miss}}=125\,\mathrm{MeV}/c) \approx 0.5\,\mathrm{\%}$), the following
correction to the asymmetry at high missing momenta was obtained:
\begin{eqnarray}
\Delta A(125\,\mathrm{MeV}/c) \approx -\delta A(125\,\mathrm{MeV}/c)
+ \delta A(50\,\mathrm{MeV}/c) = 8\times 10^{-4}\,,\nonumber
\end{eqnarray}
which effectively represents a $16\%$ relative correction to the asymmetry.
This estimation considers only  unpolarized radiative corrections, 
and  most probably significantly overestimates the size of the
necessary radiative corrections. Although we haven't  performed
the precise analysis of these effects yet, the results of this quick check
indicates that such effects do not have enough strength to
significantly change the experimental results shown in 
Figs.~\ref{fig_analysis_ProtonAsymmetry14} to~\ref{fig_analysis_DeuteronAsymmetry82}.

\chapter{Interpretation of Results}

This chapter presents the interpretation of the measured data. 
The obtained asymmetries, presented in section~\ref{sec:ExtractionOfAsymmetries}, will 
be confronted by theoretical predictions of the Bochum/Krakow group. The comparison 
with the calculations will be done separately for all three reaction channels 
${}^3\vec{\mathrm{He}}(\vec{e},e'd)p$, ${}^3\vec{\mathrm{He}}(\vec{e},e'p)d$ and 
${}^3\vec{\mathrm{He}}(\vec{e},e'p)pn$. Special attention will be devoted 
to the first two channels, which are currently better under control. The extraction and 
interpretation of the asymmetries for the latter channel is presently confined by 
an inaccurate separation of the three-body breakup events from the two-body breakup events. 
The results will also be put in the context of previous double-polarization asymmetry 
measurement from Mainz. In the end, the conclusions will be drawn, together with a summary 
of open problems and challenges for future work.

\section{The two-body breakup channel ${}^3\vec{\mathrm{He}}(\vec{e},e'p)d$}
\label{sec:2BBU}

The experimental asymmetries shown in Figs.~\ref{fig_analysis_ProtonAsymmetry14} to~
\ref{fig_analysis_ProtonAsymmetry82}, where the proton is detected by BigBite, 
are mixtures of the ${}^3\vec{\mathrm{He}}(\vec{e},e'p)d$  and 
${}^3\vec{\mathrm{He}}(\vec{e},e'p)pn$ asymmetries. The relative contribution of each reaction 
channel is governed by the ratio of cross-sections for the two processes.  To isolate the asymmetry 
corresponding to the reaction ${}^3\vec{\mathrm{He}}(\vec{e},e'p)d$, the two-body 
breakup events (2BBU) must be separated from the three-body breakup events (3BBU). 

\begin{figure}[!ht]
\begin{center}
\includegraphics[width=0.49\textwidth]{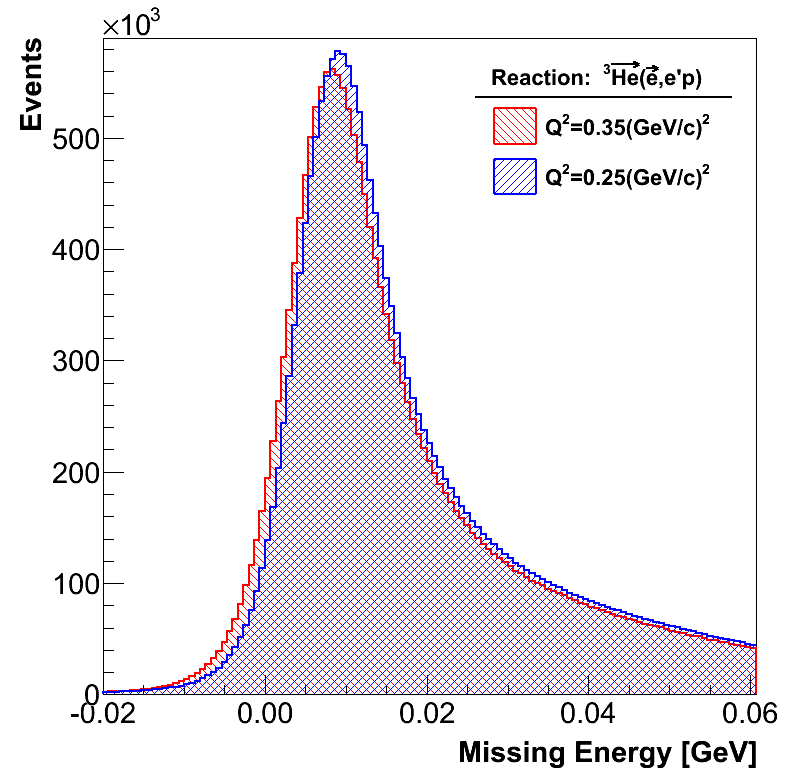}
\includegraphics[width=0.49\textwidth]{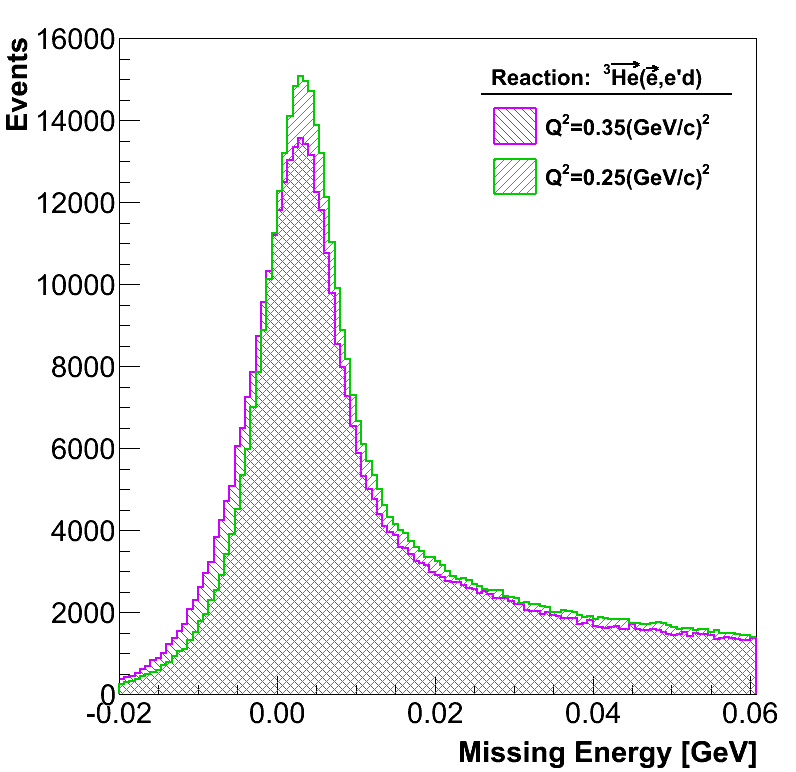}
\caption{The reconstructed missing energy ($E_{\mathrm{Miss}}$) distributions
for reactions ${}^3\vec{\mathrm{He}}(\vec{e},e'p)$ (left) and 
and ${}^3\vec{\mathrm{He}}(\vec{e},e'd)p$ (right). Histograms show results for 
both kinematical settings: $Q^2 \approx 0.35,\> 0.25\,(\mathrm{GeV/c})^2$. 
Due to radiative effects and spectrometer resolutions, the 2BBU and 
3BBU peaks in the proton channel can not be distinguished. 
\label{fig_analysis_MissingEnergy}}
\end{center}
\end{figure}

This is accomplished by inspecting the missing energy histogram, 
where 2BBU events generate a peak around $E_{\mathrm{Miss}} = 5.5\,\mathrm{MeV}$,
while 3BBU events gather around $E_{\mathrm{Miss}} = 7.7\,\mathrm{MeV}$. The 
obtained peaks are smeared by radiative processes and finite resolutions of the
spectrometers. Present analysis has shown (see Fig.~\ref{fig_analysis_MissingEnergy}) 
that for the E05-102 data, these effects are so large, that the 
two-body breakup peak can no longer be clearly distinguished from the three-body 
peak. This represents an important obstacle in the interpretation of our results
and requires the use of a Monte-Carlo simulation for a proper comparison of the theory to
the measured data.

\begin{figure}[!ht]
\begin{center}
\includegraphics[width=0.49\textwidth]{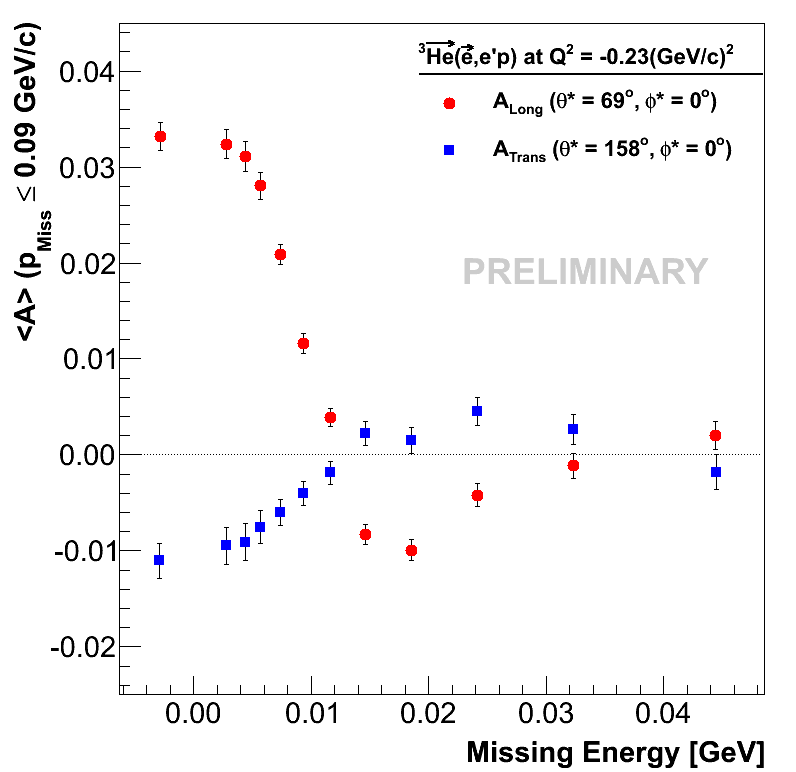}
\includegraphics[width=0.49\textwidth]{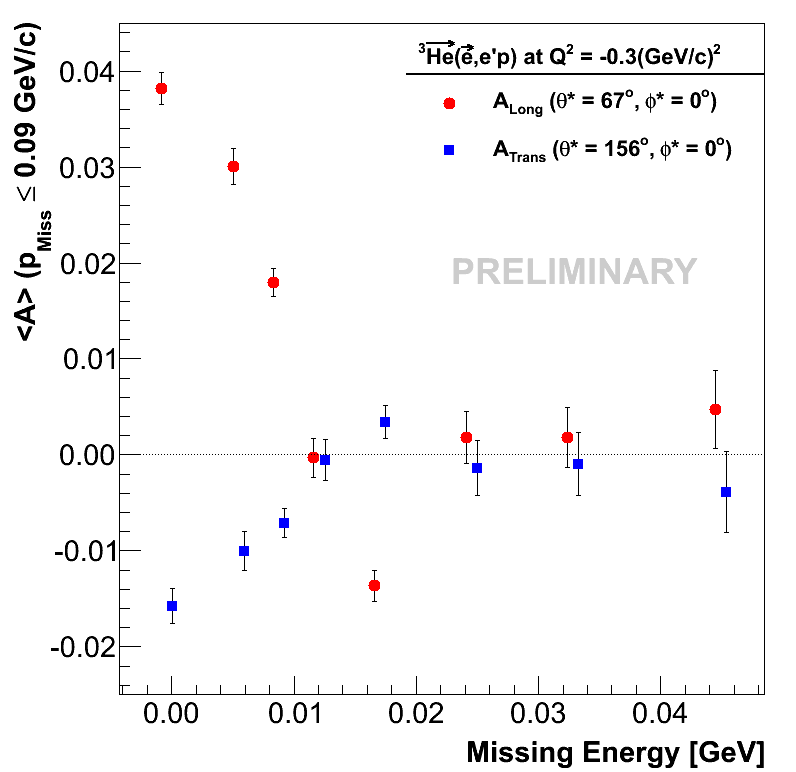}
\caption{Asymmetries as functions of missing energy ($E_{\mathrm{Miss}}$). 
Left and right plot show longitudinal and transverse asymmetries for kinematical 
settings $Q^2 = -0.23\,(\mathrm{GeV}/c)^2$ and $Q^2 = -0.3\,(\mathrm{GeV}/c)^2$ respectively. 
Each data point represents an average asymmetry for missing momenta in range of  
$p_{\mathrm{Miss}} = 0-90\,\mathrm{MeV}/c$. 
\label{fig_analysis_MissingEnergyAsymmetry}}
\end{center}
\end{figure}

Unfortunately a detailed simulation for the E05-102 experiment is not yet available. Instead, 
an approximate empirical approach was considered for this first extraction of  
the 2BBU asymmetries. In this procedure the measured $(e,e'p)$ asymmetries were plotted 
as a function of missing energy, shown in 
Fig.~\ref{fig_analysis_MissingEnergyAsymmetry}. Here only events with low missing momentum 
$p_{\mathrm{Miss}}\leq 90\,\mathrm{MeV}/c$ were retained. In this limit 
the S-state dominates the ${}^3\mathrm{He}$-wave function, and implies
a large asymmetry in the case of the 2BBU and almost a zero asymmetry for the 
3BBU (see Sec.~\ref{sec:nagorny} for more detail). The measured asymmetry agrees well
with this hypothesis. A large positive asymmetry has been observed in the region of small
$E_\mathrm{Miss}$, where 2BBU dominates. When moving to higher $E_\mathrm{Miss}$,
where the 3BBU is expected to prevail,
the asymmetry eventually decreases towards zero. 
A negative asymmetry in the middle region ($E_\mathrm{Miss}\approx 15\,\mathrm{MeV}$) 
is the result of interactions between nucleons. The Mainz experiment~\cite{achenbach2008} 
has described this in terms of the FSI that are expected to generate a 
strong effect in the 3BBU channel at low $E_{\mathrm{Miss}}$. This also explains the 
rapid drop of the asymmetry at $E_\mathrm{Miss}\approx 6\mathrm{MeV}$, where the 
3BBU process starts to contribute. At higher $E_{\mathrm{Miss}}$
the strength of the FSI weakens and the asymmetry approaches zero.

\begin{figure}[!ht]
\begin{center}
\includegraphics[width=0.49\textwidth]{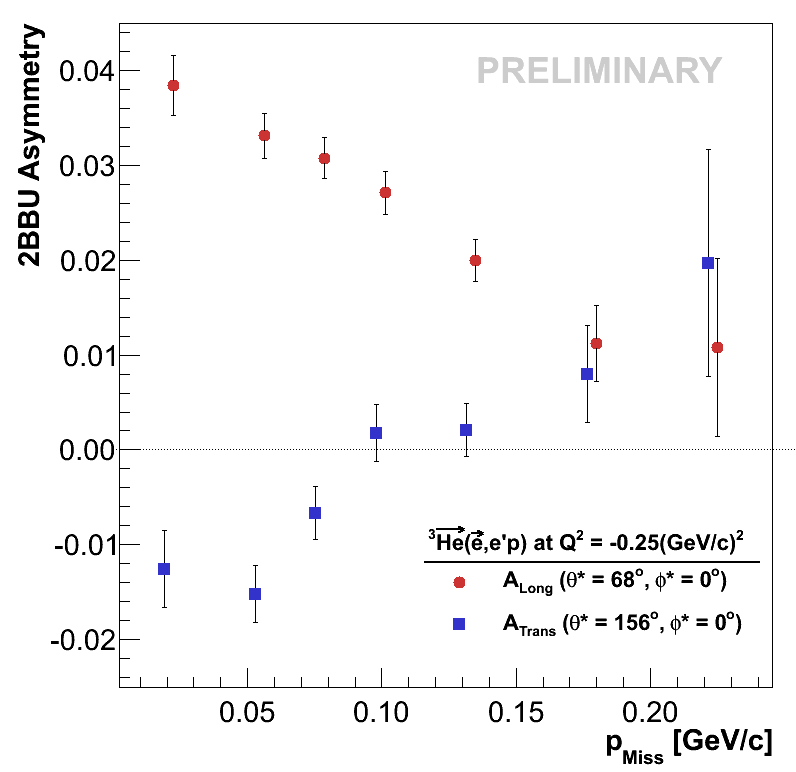}
\includegraphics[width=0.49\textwidth]{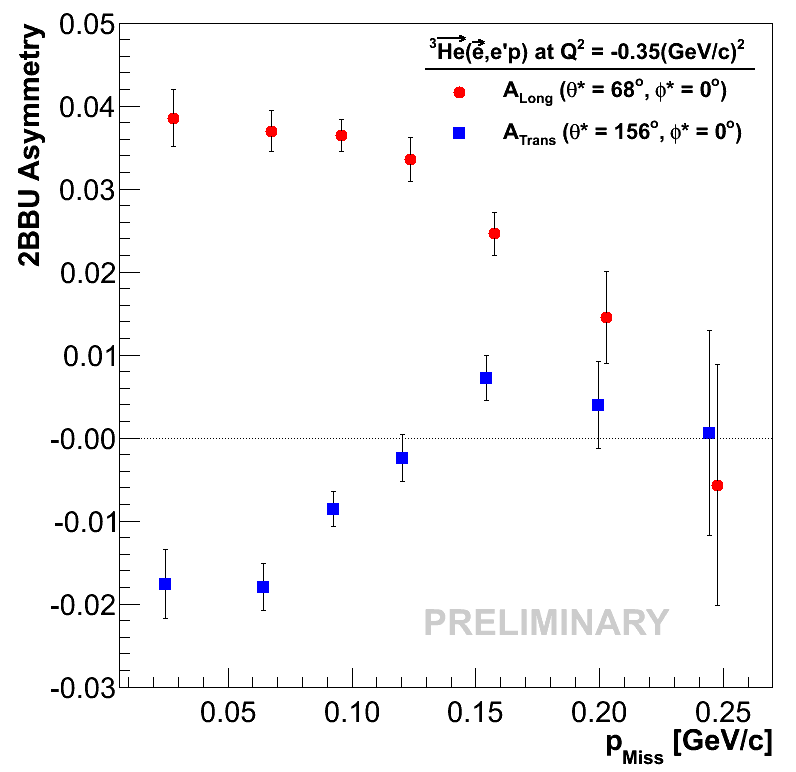}
\caption{The longitudinal and transverse  ${}^3\vec{\mathrm{He}}(\vec{e},e'p)$ 
asymmetries at $Q^2 = -0.23\,(\mathrm{GeV}/c)^2$ (left) and 
$Q^2 = -0.3\,(\mathrm{GeV}/c)^2$ (right). The asymmetries were determined
by selecting $E_{\mathrm{Miss}}\leq 1.6\,\mathrm{MeV}$; with this assumption,
the shown asymmetries pertain to the 2BBU channel only. 
\label{fig_analysis_2BBUAsymmetry}}
\end{center}
\end{figure}

An almost flat asymmetry at very low missing energies indicates a dominance of the 2BBU
reaction in that region. Relying on this assumption, the 2BBU asymmetries
were extracted from the measurements by selecting only events with 
$E_\mathrm{Miss}\leq1.6\,\mathrm{MeV}$. Both longitudinal and transverse asymmetries 
were obtained. The results for both kinematical settings are gathered 
in Fig.~\ref{fig_analysis_2BBUAsymmetry}. 

\begin{table}[!hb]
\begin{center}
\caption{The list of eleven kinematic points considered in the theoretical 
calculations. The points are selected to cover the kinematical acceptance 
for the setting when HRS-L is positioned at $\theta_{\mathrm{HRS-L}} = 12.5^\circ$.
\label{table_golak_kinpoints}}
\vspace*{2mm}
\begin{tabular}{rrrrrr}
\multicolumn{6}{c}{{\bf Kinematic Points for Theory}} \\
\hline

\hline

\hline

$i$ &  $E\>[\mathrm{MeV}]$  &    $E'\>[\mathrm{MeV}]$  &   
$\theta_e\>[\mathrm{deg}]$ & $q\>[\mathrm{MeV}/c]$ & $\omega\>[\mathrm{MeV}]$ \\[0.5mm]
\hline

\hline
1 & 2425.5 &  2235.0 & 11.35 & 498.2 & 190.5 \\
2 & 2425.5 &  2268.0 & 11.35 & 488.0 & 157.5\\
3 & 2425.5 &  2285.0 & 11.35 & 485.0 & 140.5\\
4 & 2425.5 &  2302.0 & 11.35 & 485.0 & 123.5\\
5 & 2425.5 &  2335.0 & 11.35 & 480.0 & 90.5\\
6 & 2425.5 &  2235.0 & 12.45 & 538.7 & 190.5\\
7 & 2425.5 &  2285.0 & 12.45 & 526.9 & 140.5\\
8 & 2425.5 &  2335.0 & 12.45 & 519.8 & 90.5\\
9 & 2425.5 &  2235.0 & 13.55 & 579.8 & 190.5\\
10 & 2425.5 &  2285.0 & 13.55 & 570.7 & 140.5\\
11 & 2425.5 &  2335.0 & 13.55 & 567.7 & 90.5\\
\hline

\hline

\hline
\end{tabular}
\end{center}
\end{table}

\begin{figure}[!hb]
\begin{center}
\includegraphics[width=0.49\textwidth]{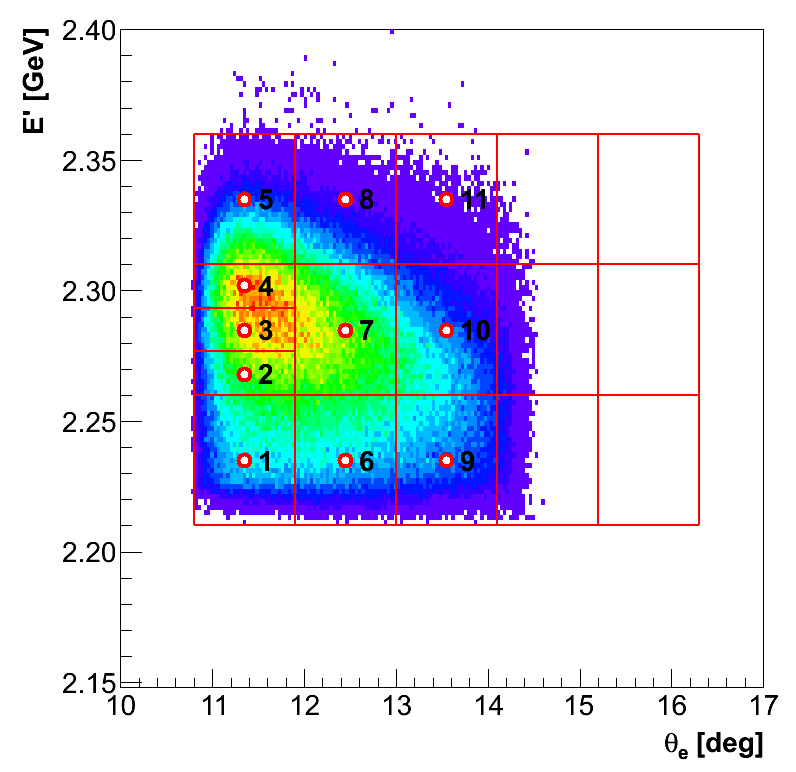}
\includegraphics[width=0.49\textwidth]{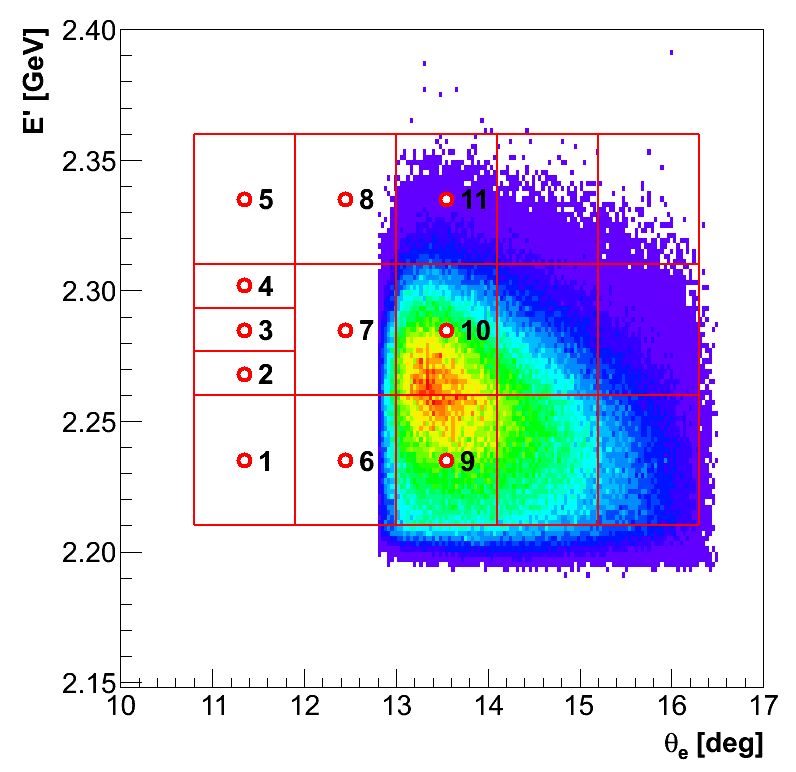}
\caption{ The electron kinematics accessible when  HRS-L was positioned at 
$\theta_{\mathrm{HRS-L}} = 12.5^\circ$ (left) and at $14.5^\circ$ (right).
The whole kinematical coverage was divided into 17 bins denoted by red squares. 
Theoretical calculations were presently performed for the centers of the leftmost
11 bins (denoted by circles). 
\label{fig_analysis_kinpoints}}
\end{center}
\end{figure}

The determined approximate 2BBU asymmetries can now be compared to the theoretical 
predictions. The calculations were performed by the Bochum/Krakow group~\cite{golak_private_2012}.
Due to extreme computational demands, 
they were able to calculate the asymmetries for only eleven different kinematics points
(Table~\ref{table_golak_kinpoints} and Fig.~\ref{fig_analysis_kinpoints}). 
Hence, a set of points was selected to cover the most important portions of the
kinematical acceptance for the setting with HRS-L
positioned at $\theta_{\mathrm{HRS-L}} = 12.5^\circ$. The bin with the highest 
statistics was divided even further into three smaller bins. 
The kinematical points at $Q^2 > 0.3\,(\mathrm{GeV}/c)^2$ that were accessible 
when HRS-L was positioned at
$\theta_{\mathrm{HRS-L}} = 14.5^\circ$ were not considered. The theory will therefore be tested 
mostly with the $Q^2 = -0.25\,(\mathrm{GeV}/c)^2$ data. However, since the
kinematical acceptances of the two experimental setups overlap in the region around 
$Q^2 = -0.3\,(\mathrm{GeV}/c)^2$, some checks could also be performed with the data 
taken with $\theta_{\mathrm{HRS-L}} = 14.5^\circ$.

\begin{sidewaysfigure}[!htbp]
\begin{center}
\includegraphics[angle=0,width=0.85\linewidth]{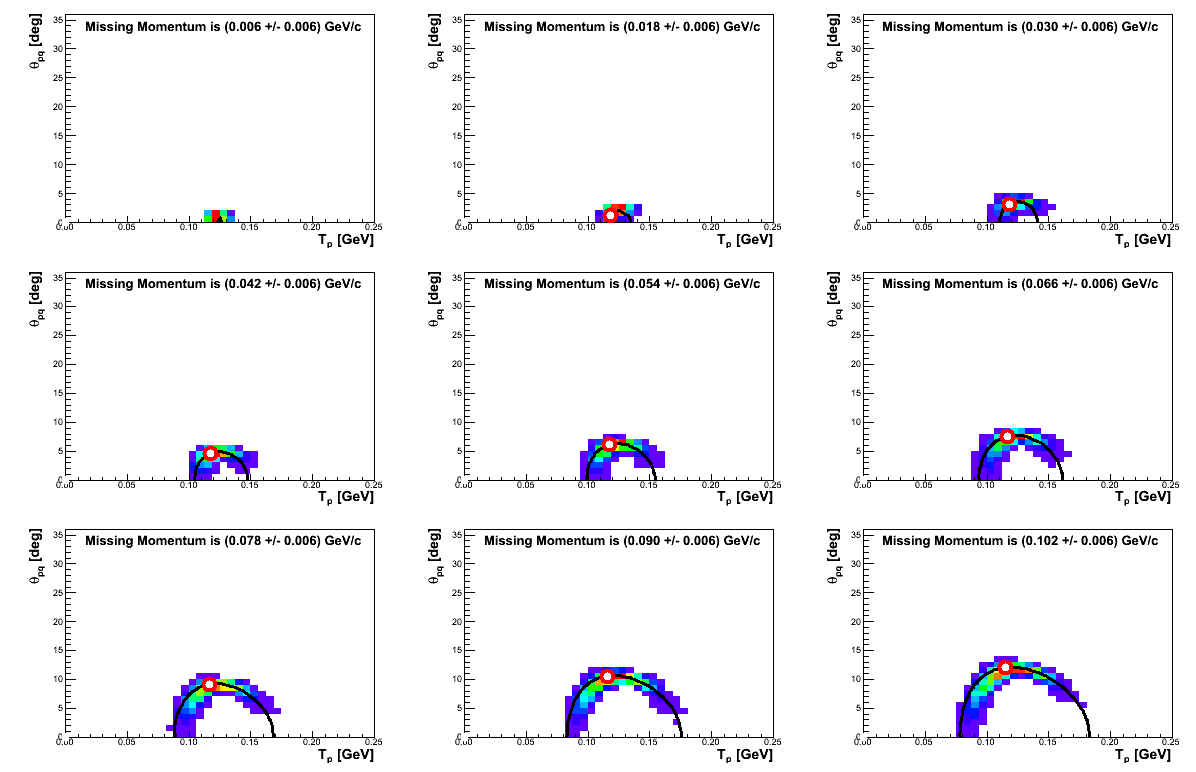}  
\vspace*{-3mm}
\caption{ Two-dimensional histograms showing the relation between the proton angle 
$\theta_{p}$ and its kinetic energy $T_p\approx p^2/2M_{p}$, for selected bins 
in $p_{\mathrm{Miss}}$. In these plots only the data from the $4^{\mathrm{th}}$ 
kinematics bin were considered (see Fig.~\ref{fig_analysis_kinpoints}).
The lengths of the obtained bands are controlled by the remaining spread in 
$\omega$ and $\vec{q}$. Black lines show the solutions of Eq.~(\ref{eq_momentum_conservation}) 
for a given $p_{\mathrm{Miss}}$ and $|\vec{q}| = 485.0\,\mathrm{MeV}/c$. Circles 
show $(\theta_{p}, \tilde{p})$ pairs considered in the theoretical calculations. 
For the $4^{\mathrm{th}}$ kinematic point, events with very low missing momenta are not accessible.
Hence, the theoretical point is missing in the histogram for $p_{\mathrm{Miss}} = 6\,\mathrm{MeV}$.
\label{fig_analysis_KinBananas1}}
\end{center}
\end{sidewaysfigure}

\begin{sidewaysfigure}[!htbp]
\begin{center}
\includegraphics[angle=0,width=0.85\linewidth]{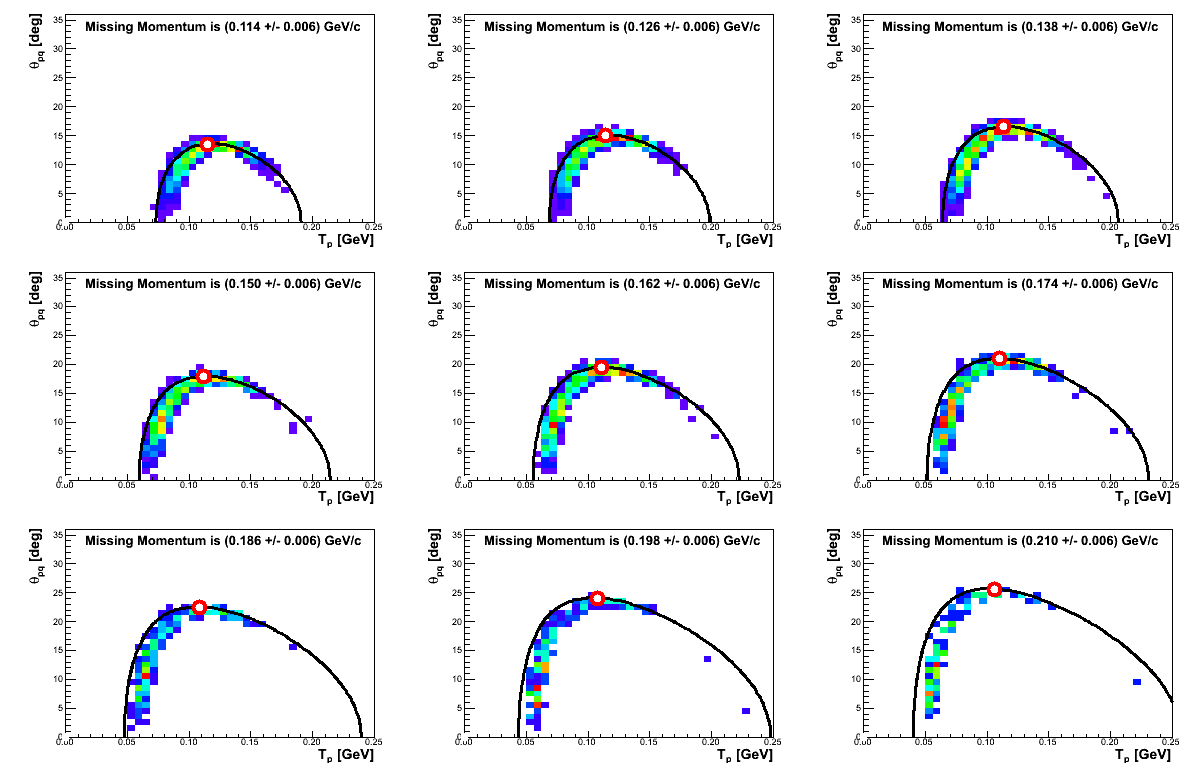}  
\caption{(continued): The remaining range in $p_{\mathrm{Miss}}$.
\label{fig_analysis_KinBananas2}}
\end{center}
\end{sidewaysfigure}

Beside the information on the electron kinematics and target spin orientation
$(\theta^*, \phi^*)$, the theoretical calculations require
also the momentum of the detected proton, $\tilde{p}$, and the polar angle $\theta_{p}$ for each 
selected bin in missing momentum $p_{\mathrm{Miss}}$  as an input. Here, $\theta_p$ represents the angle 
between the momentum transfer vector $\vec{q}$ and proton momentum $\vec p$ 
(see Fig.~\ref{fig_theory_planes}). 
Considering the conservation of energy and momentum in the non-relativistic limit, theoreticians 
use  $\theta_{p}$ to calculate the momentum of the detected proton $p$, independently 
of the input parameter $\tilde{p}$:
\begin{eqnarray}
&&\mathrm{Conservation\>of\>Energy }:\quad \omega + M_{{}^3\mathrm{He}} = M_p + M_{\mathrm{Miss}} + \frac{\vec{p}^2}{2M_p} + \frac{\vec{p}_{\mathrm{Miss}^2}}{2M_{\mathrm{Miss}}}\,,
 \label{eq_energy_conservation} \\
&&\mathrm{Conservation\>of\>Momentum }:\quad \vec{p}_{\mathrm{Miss}}^2 = \vec{q}^2 + \vec{p}^2 - 2|\vec{p}|\,|\vec{q}|\cos{\theta_{p}}\,. 
\label{eq_momentum_conservation}
\end{eqnarray}
Here, $p_{\mathrm{Miss}}$ and $M_{\mathrm{Miss}}$ are the momentum and the mass of the undetected 
deuteron. Inserting Eq.~(\ref{eq_momentum_conservation}) into Eq.~(\ref{eq_energy_conservation}), 
a quadratic equation for the proton momentum $p$ is obtained:
\begin{eqnarray}
  p = \frac{(2M_{p}q\cos{\theta_{pq}})\pm \sqrt{(2M_{p}q\cos{\theta_{p}})^2-4(M_{p}+M_{\mathrm{Miss}})(M_{p}q^2-2M_p M_{\mathrm{Miss}}H)} }
{2(M_{p}+M_{\mathrm{Miss}})}\,,\nonumber
\end{eqnarray}
where $H=\omega+M_{{}^3\mathrm{He}}-M_p-M_{\mathrm{Miss}}$. The equation has two solutions, and 
the algorithm chooses the one closest to the input momentum $\tilde{p}$, whose only role is to select
the physical solution. This procedure  is utilized to prevent non-physical combinations of $\tilde{p}$  
and $\theta_{p}$. In spite of this safety precaution, only proper combinations
of the two parameters should be introduced to the code. Non-matching combinations of $\tilde{p}$ and 
$\theta_{p}$ would result in asymmetry calculations for different $p_{\mathrm{Miss}}$ than desired.

The correct $(\theta_{p}, \tilde{p})$ pairs for each $p_{\mathrm{Miss}}$ bin were
obtained from the corresponding two-dimensional histograms. The analysis was done 
separately for each kinematic bin. The obtained distributions for the $4^{\mathrm{th}}$ bin are 
demonstrated in Figs.~\ref{fig_analysis_KinBananas1} and~\ref{fig_analysis_KinBananas2}. 
In spite of the tight kinematical cuts, the accepted events still have some freedom in 
$\omega$ and $\vec{q}$. Consequently, the data for each $p_\mathrm{Miss}$ bin are not gathered in 
a single point, but form a band.  The shape of the band is dictated by  
Eq.~(\ref{eq_momentum_conservation}), while its length is governed by the spread in $\omega$ 
and $\vec{q}$. The $(\theta_{p}, \tilde{p})$ pairs considered in the asymmetry calculations are 
labeled by circles. They also represent points where all the data would 
accumulate if the chosen kinematical bin were reduced to an infinitesimally small 
section around the chosen point. 

\begin{figure}[!htpb]
\begin{center}
\includegraphics[width=0.9\textwidth]{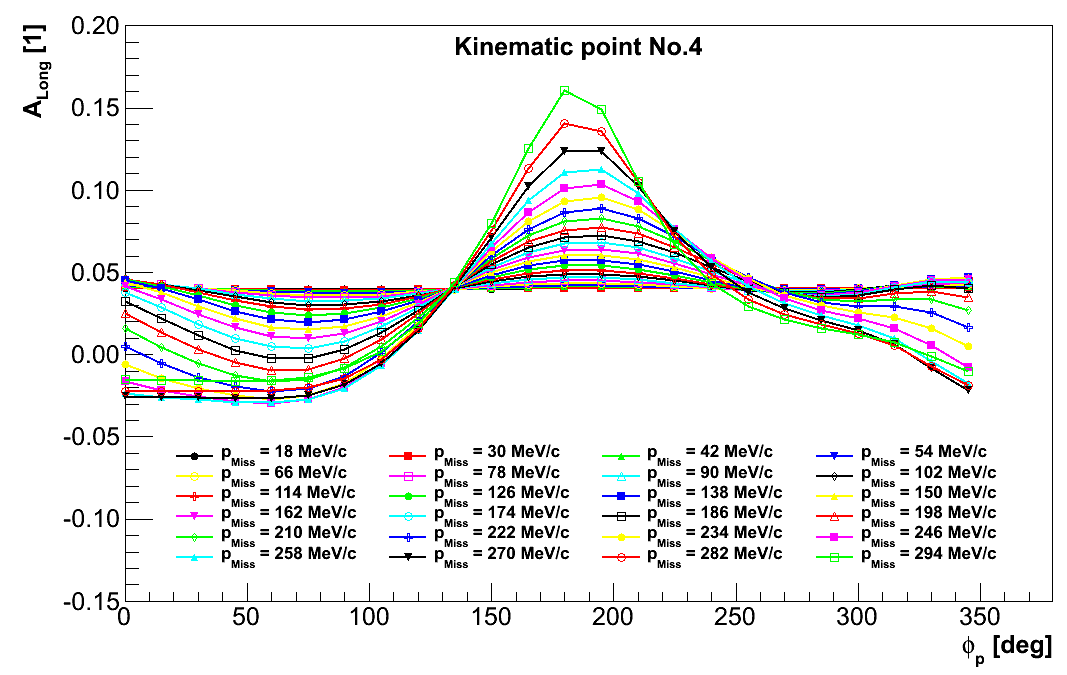}
\caption{The theoretical predictions for the longitudinal ${}^3\vec{\mathrm{He}}(\vec{e},e'p)d$ asymmetry 
$A_{\mathrm{Long}} = A(\theta^*=68^\circ, \phi^* = 0^\circ)$ as a function of angle $\phi_p$, 
for various missing momenta up to $p_{\mathrm{Miss}}\leq 300\,\mathrm{MeV}$.
Presented asymmetries were determined for the $4^{\mathrm{th}}$ kinematic point. Calculations were 
provided by the Bochum/Krakow group~\cite{golak_private_2012}.  
\label{fig_analysis_TheoryLongKin4}}
\end{center}
\end{figure}

\begin{figure}[!htpb]
\begin{center}
\includegraphics[width=0.9\textwidth]{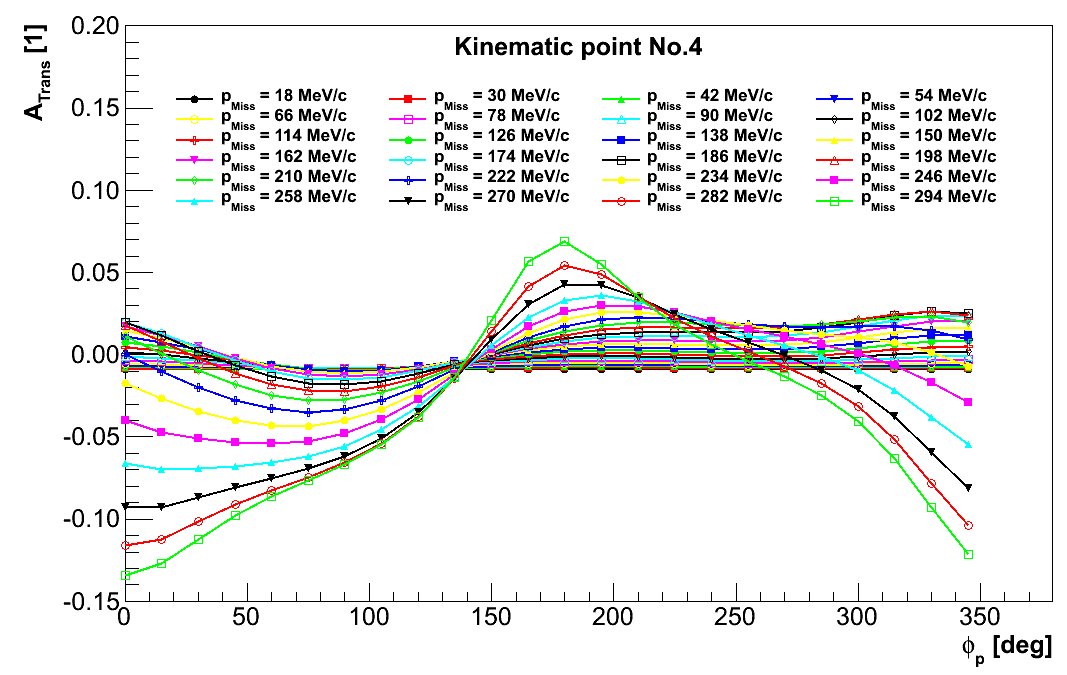}
\caption{ The theoretical predictions for the transverse ${}^3\vec{\mathrm{He}}(\vec{e},e'p)d$ asymmetry 
$A_{\mathrm{Trans}} = A(\theta^*=156^\circ, \phi^* = 0^\circ)$ as a function of angle $\phi_p$, 
for various missing momenta up to $p_{\mathrm{Miss}}\leq 300\,\mathrm{MeV}$.
Presented asymmetries were determined for the $4^{\mathrm{th}}$ kinematic point. Calculations were 
provided by the Bochum/Krakow group~\cite{golak_private_2012}. 
\label{fig_analysis_TheoryTransKin4}}
\end{center}
\end{figure}

For each target orientation and each bin in missing momentum, asymmetries for all eleven kinematic 
settings were generated as functions of the angle $\phi_{p}$. The calculated longitudinal and transverse 
asymmetries for the $4^{\mathrm{th}}$ kinematic point are shown in Figs.~\ref{fig_analysis_TheoryLongKin4} 
and~\ref{fig_analysis_TheoryTransKin4}, respectively.

\begin{figure}[!htb]
\begin{center}
\begin{minipage}[t]{0.5\textwidth}
\hrule height 0pt
    \includegraphics[width=\linewidth]{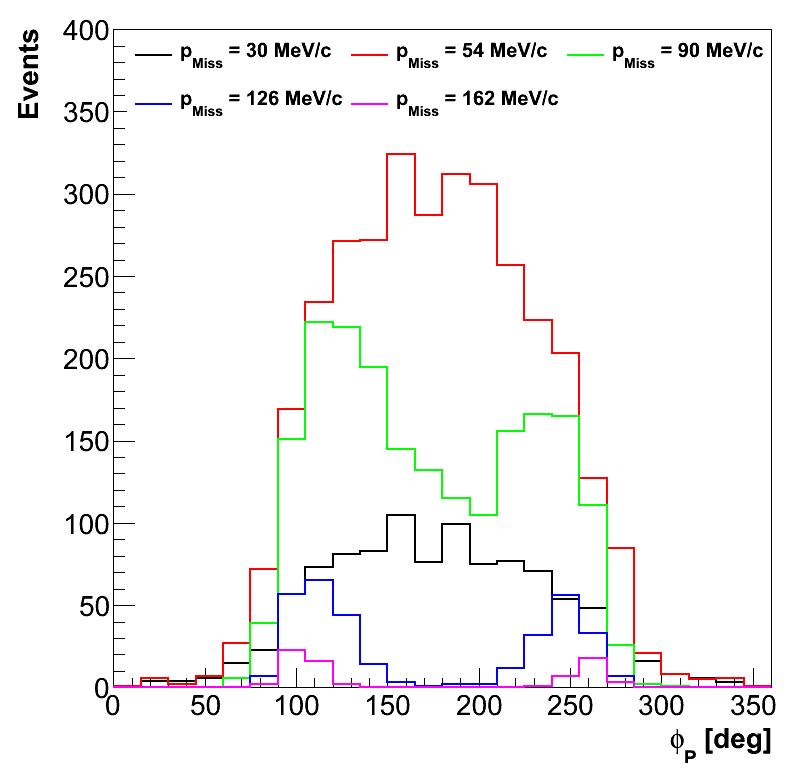}
\end{minipage}
\hspace*{1cm}
\begin{minipage}[t]{0.32\textwidth}
\hrule height 0pt
\caption{The distributions of $\phi_{p}$ at different $p_{\mathrm{Miss}}$, determined
for the events gathered around the $4^{\mathrm{th}}$ kinematic point. At low missing momenta, 
the angles near $\phi_{p} = 180^\circ$ dominate. At higher missing momenta, 
events with $\phi_p \approx 90^\circ, 270^\circ$ prevail. 
\label{fig_analysis_PhiPQAveraging}}
\end{minipage}
\end{center}
\end{figure}

The experimental results have not been binned in $\phi_{p}$. The theoretical calculations
must therefore be averaged over $\phi_{p}$ in order to compare them to the measured
asymmetries. A proper averaging over the $\phi_{p}$ is crucial for correct interpretation of the calculations,
since the theoretical asymmetries at $p_\mathrm{Miss}\gtrsim 100\,\mathrm{MeV}/c$ have a strong angular dependence.
This procedure is not trivial, since the $\phi_{p}$ distribution depends strongly on both 
the selected kinematical point and  $p_{\mathrm{Miss}}$. Figure~\ref{fig_analysis_PhiPQAveraging} shows
the $\phi_{p}$ distributions for various $p_{\mathrm{Miss}}$,  obtained for the $4^{\mathrm{th}}$ kinematic point.
In the region of low missing momenta, angles around $\phi_{p}\approx 180^\circ$ dominate. When moving
towards the higher missing momenta, the events with $\phi_{p} \approx 90^\circ, 270^\circ$ predominate. 

An appropriate averaging of the calculated asymmetries was achieved by generating the $\phi_p$ histograms
for each $p_{\mathrm{Miss}}$ in all eleven kinematic bins. The obtained distributions were
then considered as weights in the weighted average formula that was used to average the
asymmetries:
\begin{eqnarray}
\overline{A}(p_{\mathrm{Miss}}) = \frac{\sum_{\phi_{p}^{i}}A(p_{\mathrm{Miss}}, \phi_{p}^i) N_{\phi_{p}^i}}
{\sum_{\phi_{p}^i}N_{\phi_{p}^i}}\,, \nonumber
\end{eqnarray}
where $\phi_{p}^{i}$ goes over all bins in the $\phi_{p}(p_{\mathrm{Miss}})$ distribution and $N_{\phi_{p}^{i}}$
represents the number of events in $i$-th bin. $A(p_{\mathrm{Miss}}, \phi_{p}^i)$ represent the calculated asymmetries 
shown in Figs.~\ref{fig_analysis_TheoryLongKin4} and~\ref{fig_analysis_TheoryTransKin4}, while 
$\overline{A}(p_{\mathrm{Miss}})$ is the resulting average asymmetry for a chosen $p_{\mathrm{Miss}}$.

After the average asymmetries were calculated for all $p_{\mathrm{Miss}}$ available for a selected kinematic point,
they could be compared to the measured asymmetries. Separate comparisons were done for each kinematic point. 
Such comparisons are only approximate since each of the eleven calculated asymmetries describes only one 
section of data, while the experimental asymmetries represent an average over the whole acceptance. Hence, for a 
rigorous comparison, averaging over the whole acceptance has to be performed, combining the 
theoretical asymmetries of all eleven kinematic points. This requires an understanding of the asymmetry 
behavior in regions between two calculated points.  The interpolation of the calculated asymmetries to the 
whole kinematic acceptance has not been addressed yet and represents one of the challenges for future work. 

\begin{figure}[!htb]
\begin{center}
\includegraphics[width=0.49\linewidth]{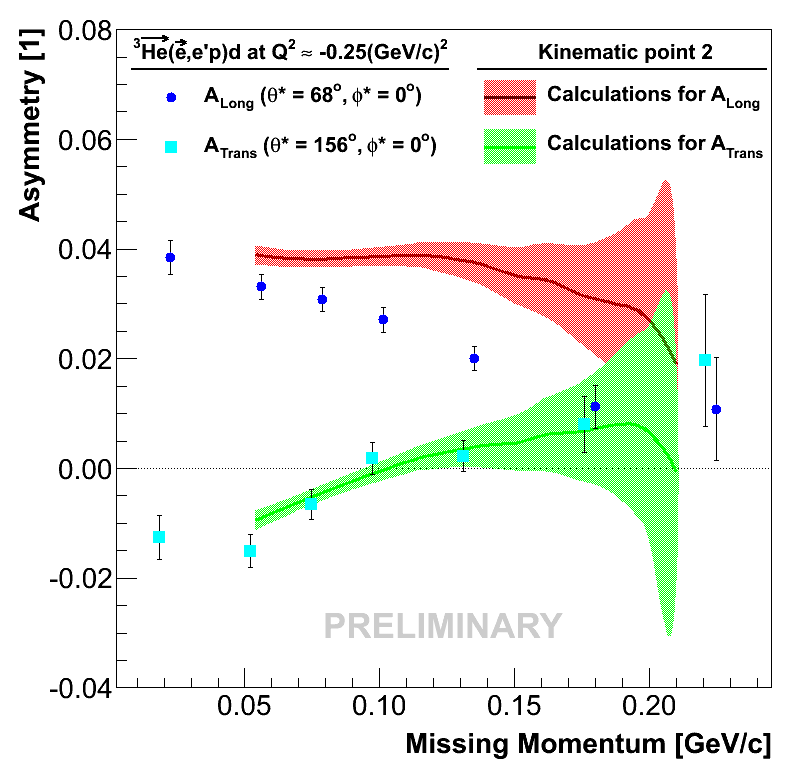}
\includegraphics[width=0.49\linewidth]{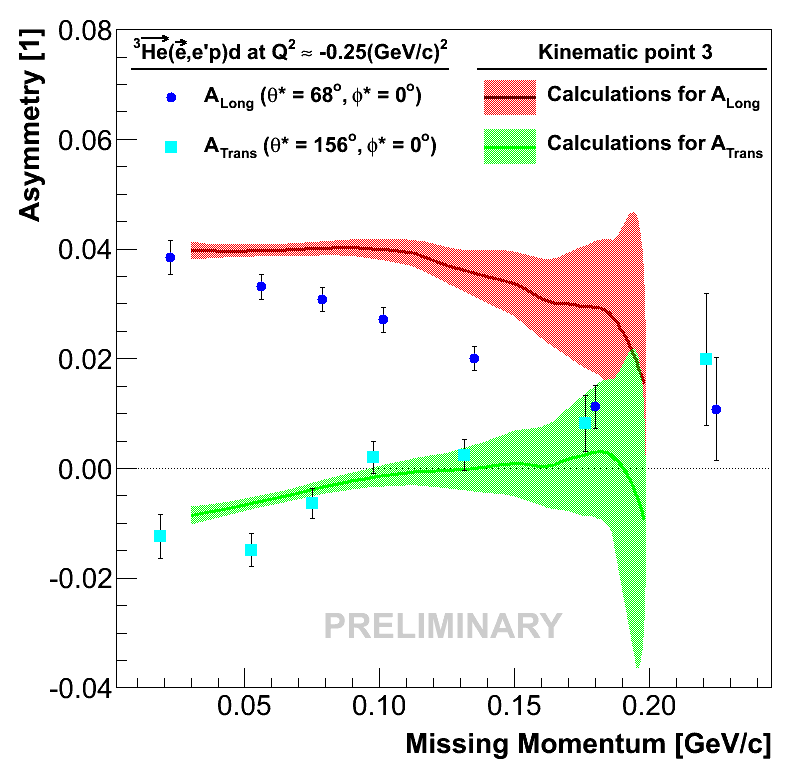}
\includegraphics[width=0.49\linewidth]{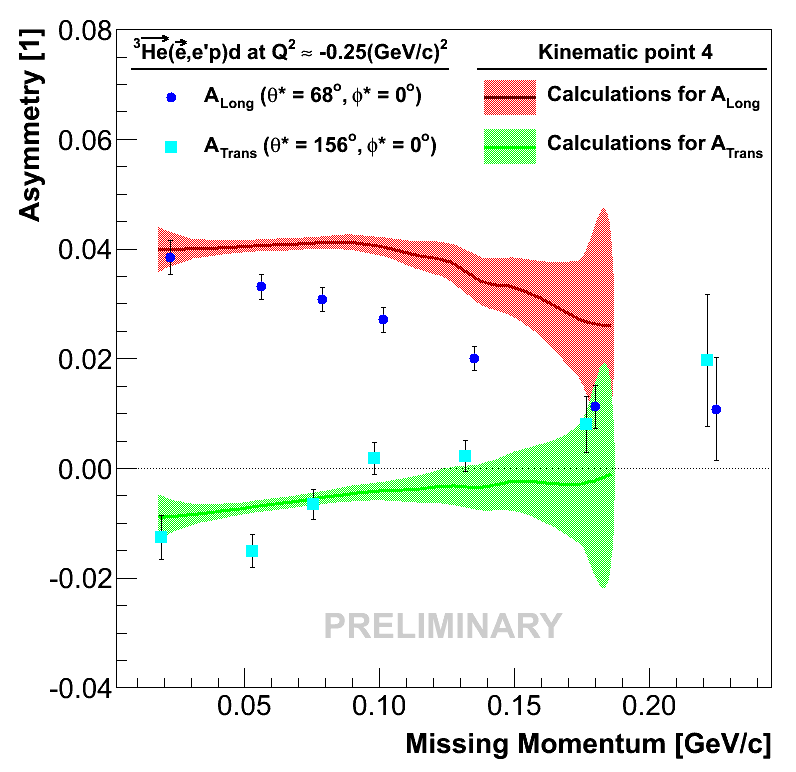}
\includegraphics[width=0.49\linewidth]{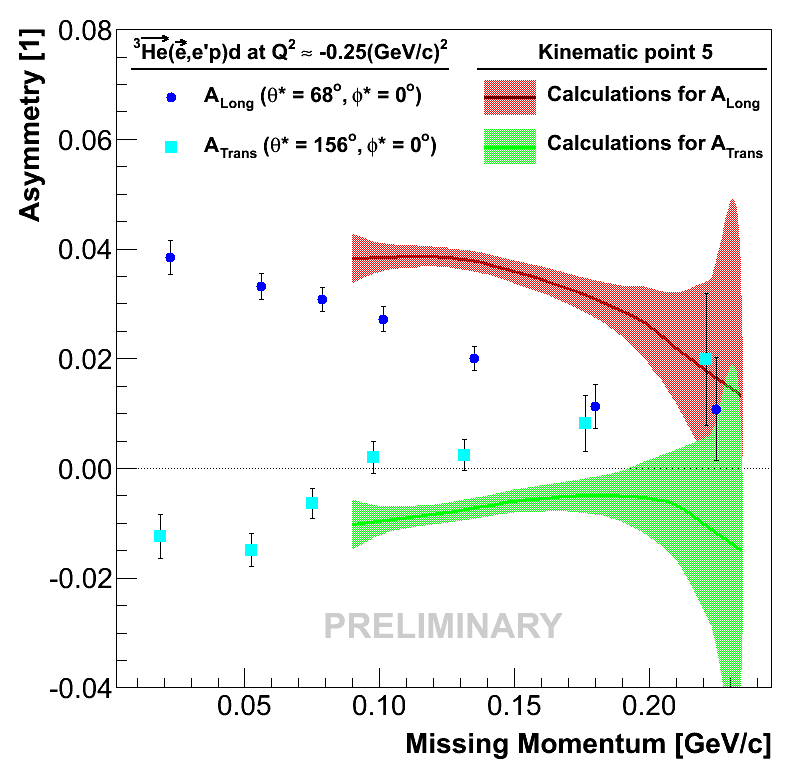}
\caption{ Comparison of the calculated ${}^3\vec{\mathrm{He}}(\vec{e},e'p)d$  
asymmetries for the $2^{\mathrm{nd}}$ $3^{\mathrm{rd}}$, $4^{\mathrm{th}}$ and 
$5^{\mathrm{th}}$ kinematic bin with the extracted experimental asymmetries at 
$Q^2 = -0.25\,(\mathrm{GeV}/c)^2$. The theoretical asymmetries are shown with 
full lines. The error bands demonstrate the uncertainties of the procedure used to average
the theoretical asymmetries, and are governed by the statistics  
in the $\phi_{p}$-histograms (see Fig.~\ref{fig_analysis_PhiPQAveraging}).
\label{fig_analysis_2BBUResult1}}
\end{center}
\end{figure}

In spite of these open problems, a comparison of the calculations, corresponding to individual kinematic 
points, to the data, already provide some important findings. Figure~\ref{fig_analysis_2BBUResult1} shows 
the results for the most populated $4^{\mathrm{th}}$ kinematical point. This is also 
the only point which brings information on the asymmetries at lowest missing momenta. All other
points provide data only at higher missing momenta. With the exception of the points at lowest missing momenta,
the calculated asymmetries do not agree with the data, with the important caveat that the present
level of disagreement could be reduced or might even vanish by applying a more 
refined averaging procedure. At the moment, our kinematics grid is too coarse to 
allow for such refinement.

\begin{figure}[!ht]
\begin{center}
\includegraphics[width=0.49\linewidth]{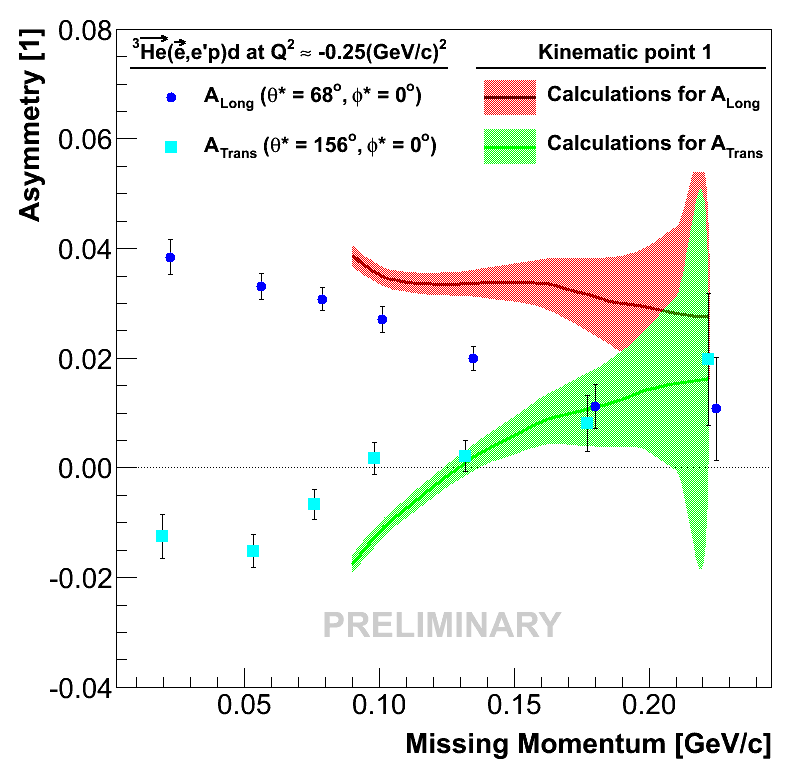}
\includegraphics[width=0.49\linewidth]{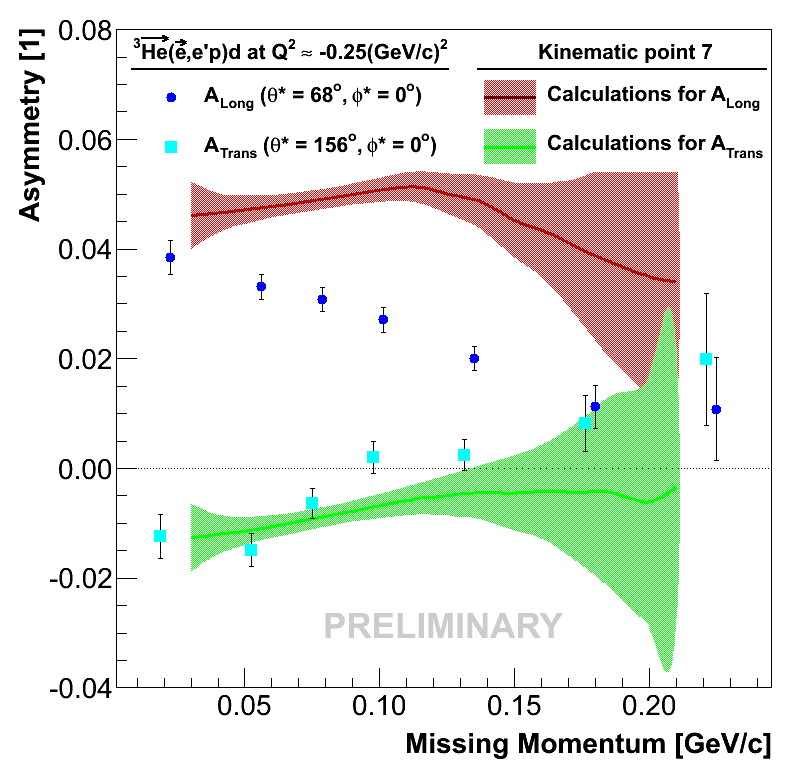}
\vspace*{-3mm}
\caption{ Comparison of the calculated asymmetries for the $1^{\mathrm{st}}$ 
and $7^{\mathrm{th}}$ kinematic bin with the 
extracted experimental ${}^3\vec{\mathrm{He}}(\vec{e},e'p)d$ asymmetries at 
$Q^2 = -0.25\,(\mathrm{GeV}/c)^2$. The theoretical asymmetries are shown with 
full lines. The error bands demonstrate the uncertainties of the procedure used to average
the theoretical asymmetries, and are governed by the statistics  
in the $\phi_{p}$-histograms (see Fig.~\ref{fig_analysis_PhiPQAveraging}). 
\label{fig_analysis_2BBUResult2}}
\vspace*{-5mm}
\end{center}
\end{figure}

The measured and acceptance-averaged theoretical asymmetries have consistent signs, and also have 
similar trends, but the absolute values are very different. For example, the experimental asymmetry 
$A_{\mathrm{Long}}$ seems to be decreasing much faster towards zero than the calculated one, which  
remains at values $\approx 4\,\mathrm{\%}$. The inspection of the rest of the calculations 
has shown similar behavior of the predicted asymmetries also in all other kinematical bins. Some examples are 
shown in Figs.~\ref{fig_analysis_2BBUResult1} and~\ref{fig_analysis_2BBUResult2}.

\begin{figure}[!ht]
\begin{center}
\includegraphics[width=0.49\linewidth]{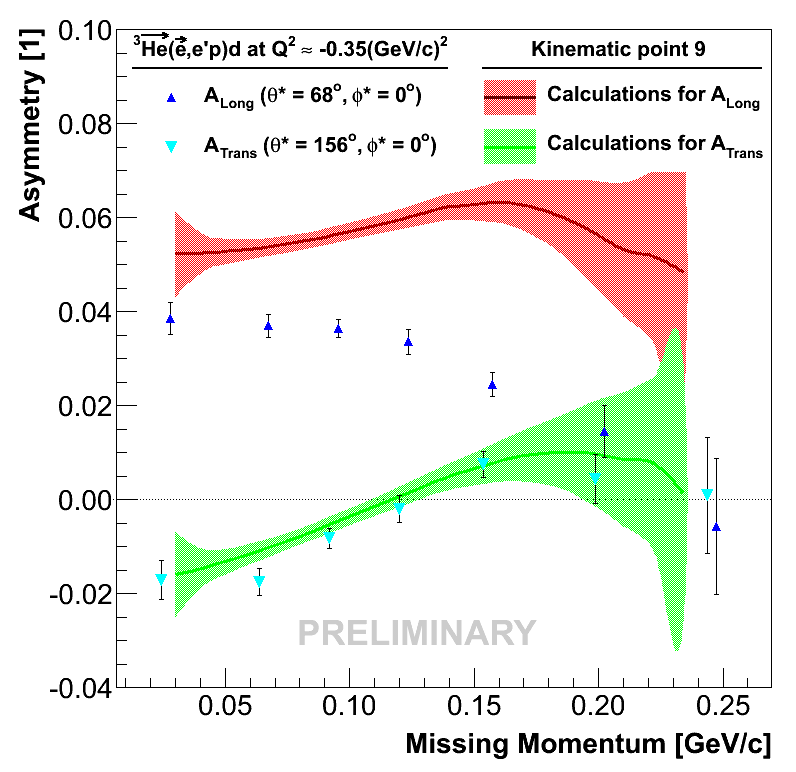}
\includegraphics[width=0.49\linewidth]{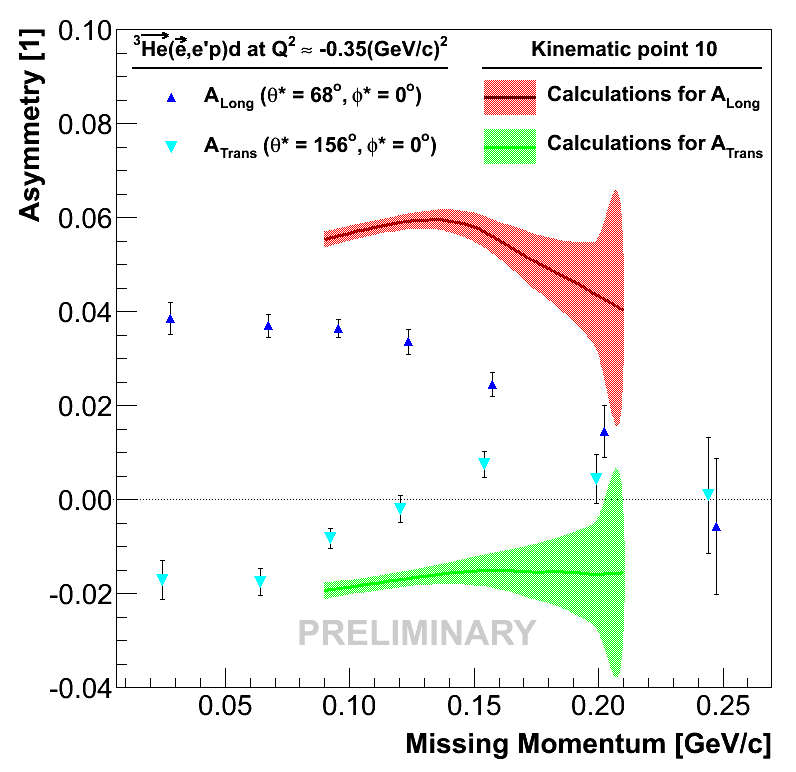}
\vspace*{-3mm}
\caption{ Comparison of the calculated asymmetries for the $9^{\mathrm{th}}$ 
and $10^{\mathrm{th}}$ kinematic bin with the 
extracted experimental ${}^3\vec{\mathrm{He}}(\vec{e},e'p)d$ asymmetries at 
$Q^2 = -0.35\,(\mathrm{GeV}/c)^2$. The theoretical asymmetries are shown with 
full lines. The error bands demonstrate the uncertainties of the procedure used to average
the theoretical asymmetries, and are governed by the statistics  
in the $\phi_{p}$-histograms (see Fig.~\ref{fig_analysis_PhiPQAveraging}).
\label{fig_analysis_2BBUResult3}}
\vspace*{-5mm}
\end{center}
\end{figure}

Identical problems appear also with the comparison of the calculations 
for the $9^{\mathrm{th}}$, $10^{\mathrm{th}}$ and $11^{\mathrm{th}}$ 
kinematic point to the $Q^2 = -0.35\,(\mathrm{GeV}/c)^2$ measurements. These results are shown 
in Fig.~\ref{fig_analysis_2BBUResult3}. This persisting discrepancy needs to be resolved in the future. 

\section{Relation to elastic scattering on $\vec{p}$}
\label{sec:ProtonElastic}

In a very simple picture the ${}^3\mathrm{He}$ ground-state can be imagined as a bound 
state of a deuteron and a proton. In this case the spin-part of the ${}^3\mathrm{He}$
wave-function can be expressed in terms of Clebsh-Gordan coeficients as:
\begin{eqnarray}
 \left|J=1/2, m_{J}=1/2\right\rangle_{{}^3\mathrm{He}} &=& 
\sqrt{\frac{2}{3}}\left|J=1, m_{J}=1\right\rangle_d \left|J=1/2, m_{J}=-1/2\right\rangle_{p}\nonumber\\
&-&\sqrt{\frac{1}{3}}\left|J=1, m_{J}=0\right\rangle_d \left|J=1/2, m_{J}=1/2\right\rangle_p\,,
\label{eq_analysis_HeCG}
\end{eqnarray}
where $J$ and $m_{J}$ represent the spin of a particle and its third component, respectively. 
The expression (\ref{eq_analysis_HeCG}) can be used to estimate the polarization of the
proton inside the nucleus. When the  ${}^3\mathrm{He}$ nucleus is polarized along the $z$-axis, the proton 
polarization $P_{p}$ can be written as:
\begin{eqnarray}
P_{p} &=&  {}_{{}^3\mathrm{He}}\left\langle 1/2, 1/2\right|\> 2\,\hat{\sigma}_{z}^{(p)}\, 
          P_{{}^3\mathrm{He}}\>\left|1/2, 1/2\right\rangle_{{}^3\mathrm{He}} 
      = P_{{}^3\mathrm{He}}\left[ \frac{2}{3}\left(-\frac{2}{2}\right) +  \frac{1}{3}\left(\frac{2}{2}\right) \right] = 
         -\frac{1}{3}P_{{}^3\mathrm{He}}\,,\nonumber
\end{eqnarray}
where $P_{{}^3\mathrm{He}}$ is the effective polarization of the ${}^3\mathrm{He}$ nucleus, and $\hat{\sigma}_{z}^{(p)}$ 
is the Pauli matrix operating on proton part of the wave-function.
When ${}^3\mathrm{He}$ is $100\,\mathrm{\%}$ polarized,
the proton polarization is $P_{p} \approx -33.3\,\mathrm{\%}$. The negative sign of the polarization 
means that the average proton spin is  opposite to the nuclear spin. 

\begin{figure}[!ht]
\begin{center}
\includegraphics[width=0.49\textwidth]{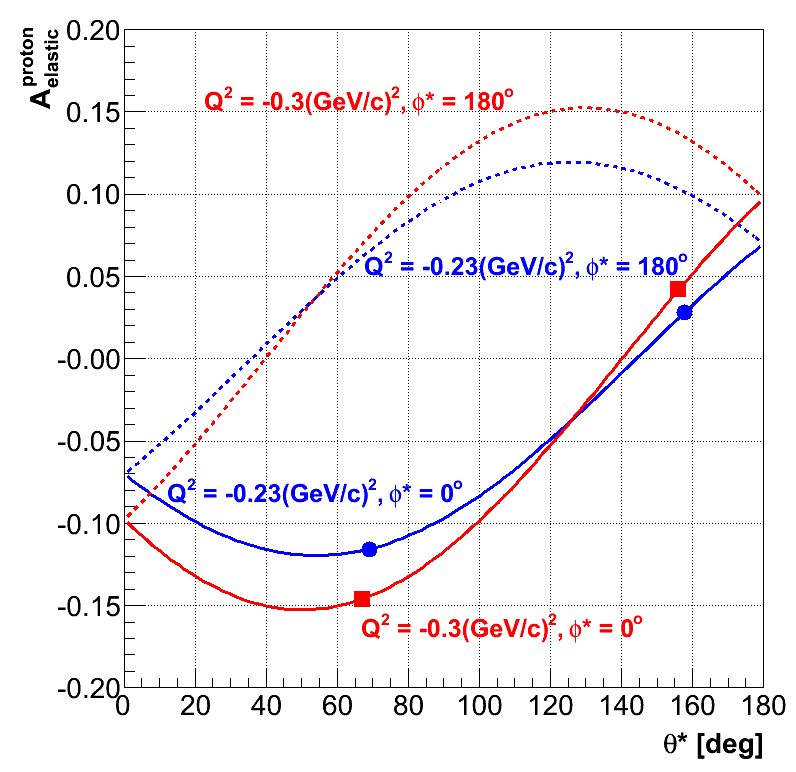}
\includegraphics[width=0.49\textwidth]{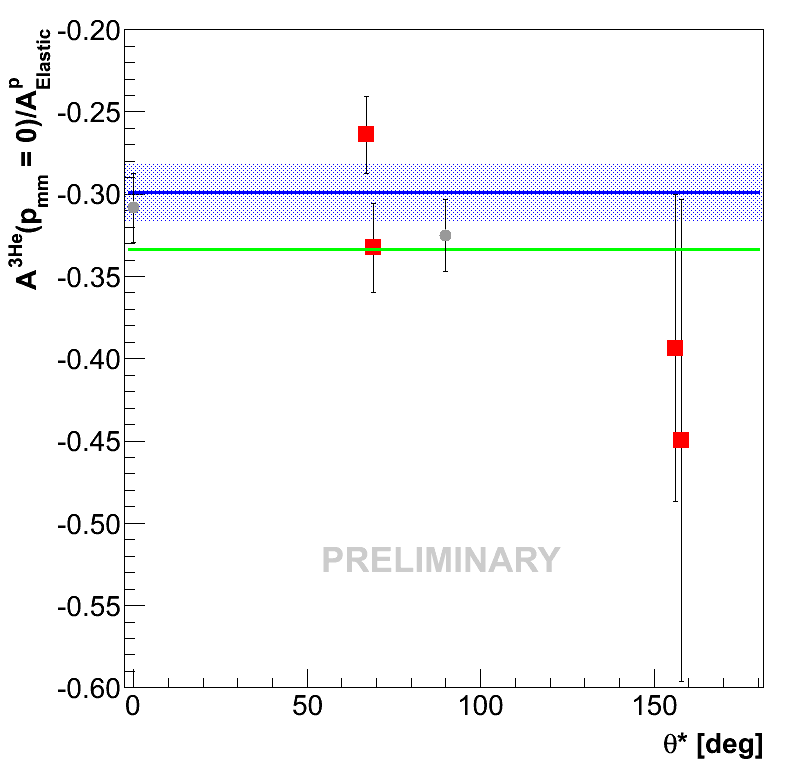}
\caption{[Left] The asymmetries $A^{\vec{e}\vec{p}}$ for elastic scattering of polarized electrons
on polarized protons as functions of $\theta^*$, obtained from Eq.~(\ref{eq_theory_ElasticProtonAsymmetry}).
Blue and red lines represent results for $Q^2 = -0.25\,(\mathrm{GeV}/c)^2$  and  
$Q^2 = -0.35\,(\mathrm{GeV}/c)^2$, respectively. Full and dotted lines distinguish between 
calculations performed for $\phi^* = 0^\circ$ and $\phi^* = 180^\circ$. Squares and circles
show four points considered for the comparison with the measured helium asymmetries. 
[Right] Ratios (red squares) between the extracted 2BBU asymmetries at $p_{\mathrm{Miss}}\approx 0$
and the corresponding elastic proton asymmetries. The green line represents the polarization of a proton
in the ${}^3\mathrm{He}$ nucleus,  $P_{p} = -33.3\,\mathrm{\%}$, predicted by the naive model. 
The blue line and the error band show the average of the four data points 
and its uncertainty. Gray circles represent the data-points
obtained by the Mainz experiment~\cite{achenbach2008}.
\label{fig_analysis_ProtonElasticAsymmetry}}
\end{center}
\end{figure}

This naive model of ${}^3\mathrm{He}$ can be further used to approximately describe the two-body 
electrodisintegration process ${}^3\vec{\mathrm{He}}(\vec{e},e'p)d$ at low missing momenta. In this limit, 
the virtual photon interacts only with a proton, while leaving the deuteron as a spectator at rest 
(see Sec.~\ref{sec:nagorny}). 
By neglecting any interaction between the proton and the deuteron, this process can be approximated by elastic
scattering of electrons on a polarized proton, $\vec{p}(\vec{e},e'p)$. This means that the 
extracted ${}^3\vec{\mathrm{He}}(\vec{e},e'p)d$  asymmetries at $p_{\mathrm{Miss}}\approx 0$ should 
agree with the  elastic proton asymmetry $A_{\vec{e}\vec{p}}$, corrected for the effective proton 
polarization inside ${}^3\mathrm{He}$:
\begin{eqnarray}
A_{\mathrm{2BBU}}(p_{\mathrm{Miss}} = 0, \theta^*, \phi^*)  \approx -\frac{1}{3}A^{\vec{e}\vec{p}}(\theta^*, \phi^*)\,.
  \label{eq_analysis_EffProtonPolarization}
\end{eqnarray}

To test this hypothesis, the asymmetry ratios $A_{\mathrm{2BBU}}/A^{\vec{e}\vec{p}}$  were calculated for 
four data points closest to $p_{\mathrm{Miss}} = 0$ (see Fig.~\ref{fig_analysis_MissingEnergyAsymmetry}).
The elastic asymmetries corresponding to the selected data-points were calculated using 
Eq.~(\ref{eq_theory_ElasticProtonAsymmetry}), and are presented in 
Fig.~\ref{fig_analysis_ProtonElasticAsymmetry} (left). The determined ratios are shown in 
Fig.~\ref{fig_analysis_ProtonElasticAsymmetry} (right). The results are nicely gathered around the predicted
value (green line). They also agree with the data-points determined by the Mainz 
experiment~\cite{achenbach2008}. By calculating the average value of the four data points (blue line), 
the effective proton polarization was estimated to be:
\begin{eqnarray} 
\langle P_p\rangle = -0.299 \pm  0.017\,,\nonumber
\end{eqnarray}
which agrees well with the value predicted in Eq.~(\ref{eq_analysis_EffProtonPolarization}).

\section{The three-body breakup channel ${}^3\vec{\mathrm{He}}(\vec{e},e'p)pn$ }

\begin{figure}[!htpb]
\begin{center}
\includegraphics[width=0.9\textwidth]{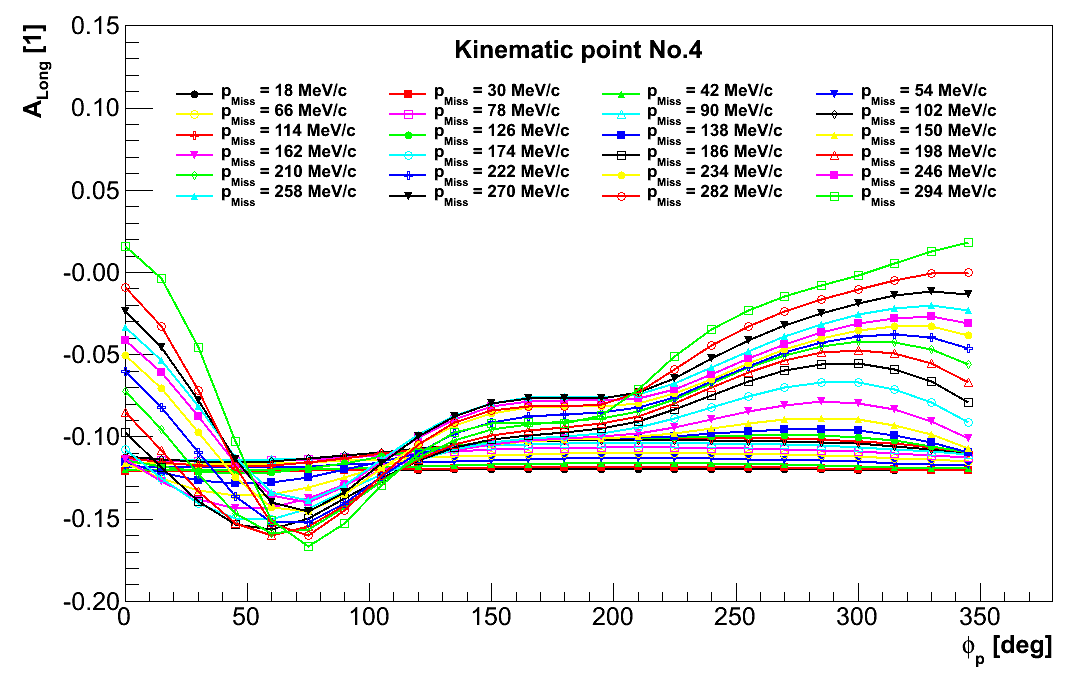}
\vspace*{-3mm}
\caption{The theoretical predictions for the longitudinal ${}^3\vec{\mathrm{He}}(\vec{e},e'p)pn$ asymmetry 
$A_{\mathrm{Long}} = A(\theta^*=68^\circ, \phi^* = 0^\circ)$ as a function of an angle $\phi_p$, 
for missing momenta up to $p_{\mathrm{Miss}}\leq 300\,\mathrm{MeV}$.
Presented are the asymmetries for the $4^\mathrm{th}$  kinematic bin.
Calculations were provided by the Bochum/Krakow group~\cite{golak_private_2012}.  
\label{fig_analysis_TheoryLongKin4ppn}}
\vspace*{-3mm}
\end{center}
\end{figure}

\begin{figure}[!htpb]
\begin{center}
\includegraphics[width=0.9\textwidth]{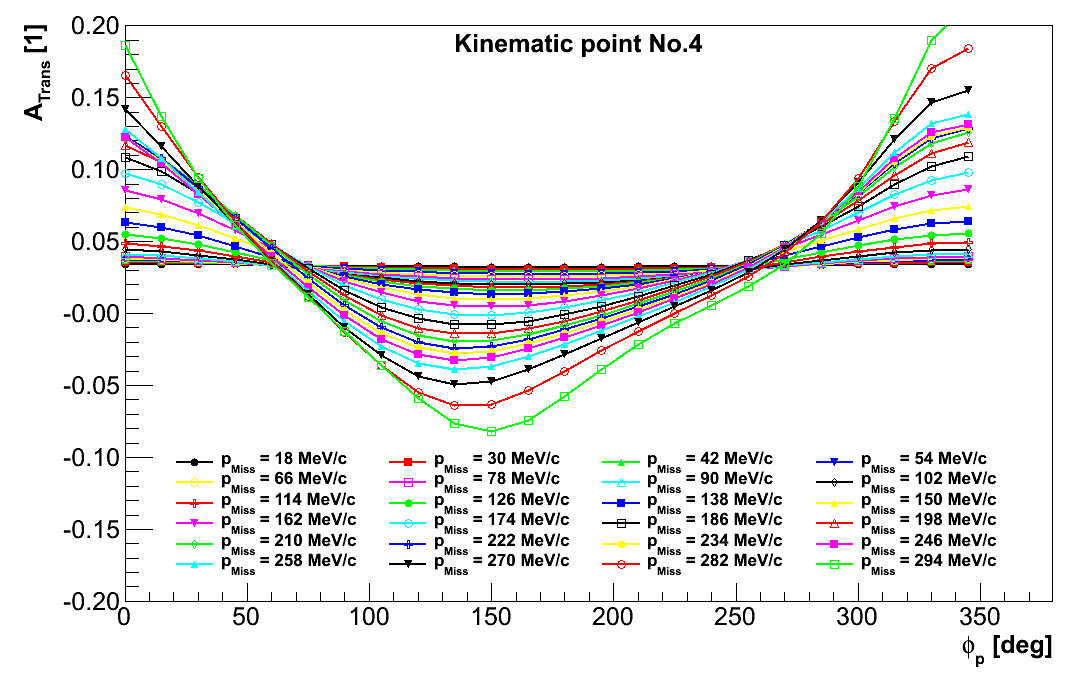}
\vspace*{-3mm}
\caption{ The theoretical predictions for the transverse ${}^3\vec{\mathrm{He}}(\vec{e},e'p)pn$ asymmetry 
$A_{\mathrm{Trans}} = A(\theta^*=156^\circ, \phi^* = 0^\circ)$ as a function of an angle $\phi_p$, 
for missing momenta up to $p_{\mathrm{Miss}}\leq 300\,\mathrm{MeV}$.
Presented are the asymmetries for the $4^\mathrm{th}$  kinematic bin.
Calculations were provided by the Bochum/Krakow group~\cite{golak_private_2012}.
\label{fig_analysis_TheoryTransKin4ppn}}
\vspace*{-3mm}
\end{center}
\end{figure}

The Bochum/Krakow group has provided calculations also for the three-body breakup channel 
${}^3\vec{\mathrm{He}}(\vec{e},e'p)pn$. The predicted longitudinal and transverse asymmetries, 
which were obtained with an identical procedure as described in Sec.~\ref{sec:2BBU}, are presented 
in Figs.~\ref{fig_analysis_TheoryLongKin4ppn} and~\ref{fig_analysis_TheoryTransKin4ppn}.
To be able to compare these calculations to the measured asymmetries, a separation of the 
3BBU data from the 2BBU data is essential. As indicated already in Sec.~\ref{sec:2BBU}, this 
can not be done without an additional input from a Monte-Carlo simulation. 

First such attempt was performed by using MCEEP (Monte Carlo for (e,e'p)). It was designed by 
P.~E.~Ulmer~\cite{mceep} to simulate coincidence (e,e'X) experiments by averaging theoretical models
over the experimental acceptance. It offers several different cross-section parameterizations for
${}^3\vec{\mathrm{He}}(\vec{e},e'p)$ reactions.  Unfortunately the standard version contains only
implementations for the HRS spectrometers. To use it for the E05-102 experiment, the
acceptance of the HRS-R spectrometer was broadened to emulate BigBite, but in such a simplified
view, only physical quantities at the target were accessible.

\begin{figure}[!ht]
\begin{center}
\includegraphics[width=0.49\textwidth]{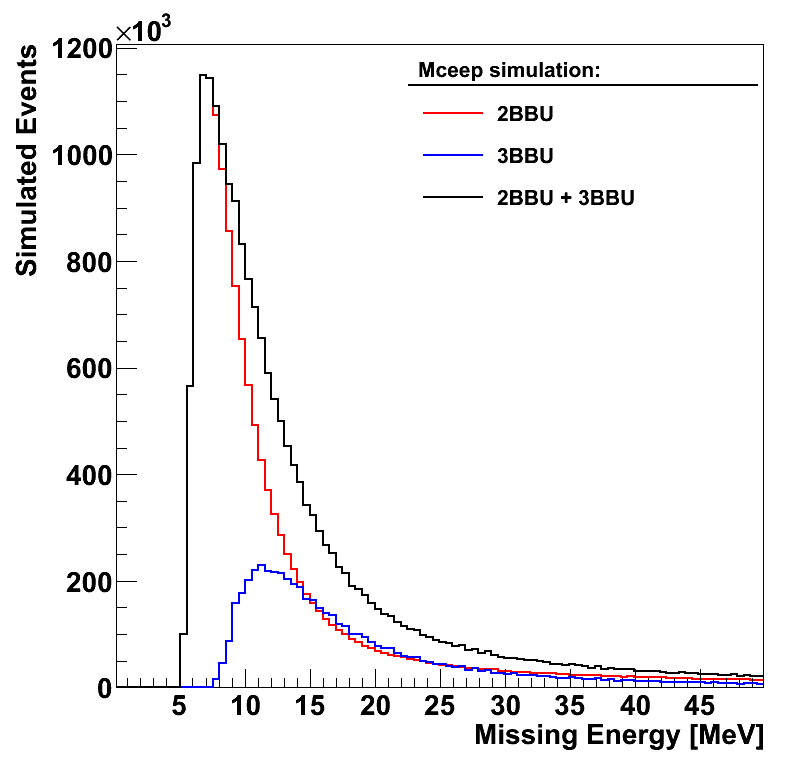}
\includegraphics[width=0.49\textwidth]{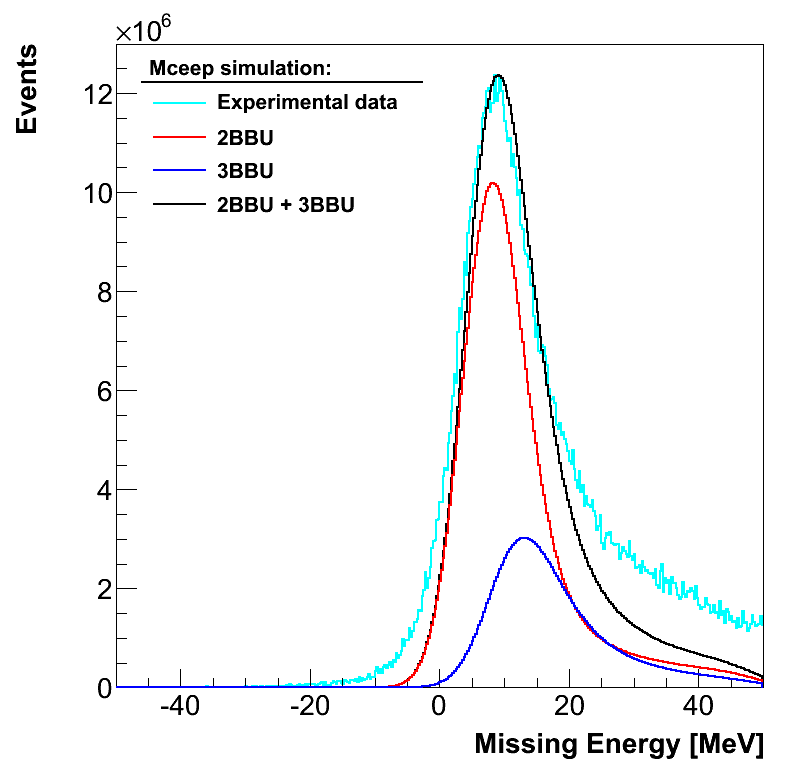}
\caption{[Left] The results of the MCEEP simulation for the ${}^3\vec{\mathrm{He}}(\vec{e},e'p)d$
and ${}^3\vec{\mathrm{He}}(\vec{e},e'p)pn$ reactions. The dominant part of the missing-energy 
peaks is contributed by the two-body reaction process. Even at very high missing energies, where 
the dominance of the 3BBU is expected, the 2BBU still contributes significantly.
The simulation also reveals that, even without any additional 
worsening of the resolution, the 3BBU contribution can not be distinguished from the 
main 2BBU peak. [Right] The results of the MCEEP simulation after 
the convolution with a Gaussian function ($\sigma_{\mathrm{Gauss}} = 4.3\,\mathrm{MeV}$).
For comparison, the measured data are shown (cyan line).
\label{fig_analysis_Mceep}}
\end{center}
\end{figure}

Having these limitations in mind, MCEEP was run for our experimental conditions. 
Simulations for the two- and three-body breakup were run separately for the same amount of 
accumulated charge. The events from both simulations were then joined in  
a combined missing energy spectrum that can be directly compared to the data.
The results of the simulation are shown in Fig.~\ref{fig_analysis_Mceep} (left). By comparing these
to the experimental results shown in Fig.~\ref{fig_analysis_MissingEnergy} 
one can see that MCEEP significantly 
underestimates the width of the missing energy spectrum, even with the consideration 
of the radiative losses.

The observed disagreement  was attributed to the incomplete treatment
of the spectrometer resolutions. To compensate for the difference in 
the widths, we decided to artificially broaden the generated peaks. This was accomplished by 
convoluting both missing energy peaks (2BBU and 3BBU) with the same Gaussian function. 
The width of the Gaussian distribution ($\sigma_{\mathrm{Gauss}} = 4.3\,\mathrm{MeV}$) was chosen for 
the combined missing energy spectrum to agree best with the measured data. 
See Fig.~\ref{fig_analysis_Mceep} (right). 
However, even with this correction, MCEEP is still unable to properly describe the strong, 
long tail present in the data. This can be contributed to a known, but unsolved 
issue~\cite{doug_private2012}, that MCEEP underestimates the cross-section for 
the ${}^3\vec{\mathrm{He}}(\vec{e},e'p)pn$ reaction. A correction to the cross-section for 
this process would raise the 3BBU peak in the simulated missing energy spectrum, but would still 
require additional broadening of the peak.

\begin{figure}[!htb]
\begin{center}
\begin{minipage}[t]{0.5\textwidth}
\hrule height 0pt
    \includegraphics[width=\linewidth]{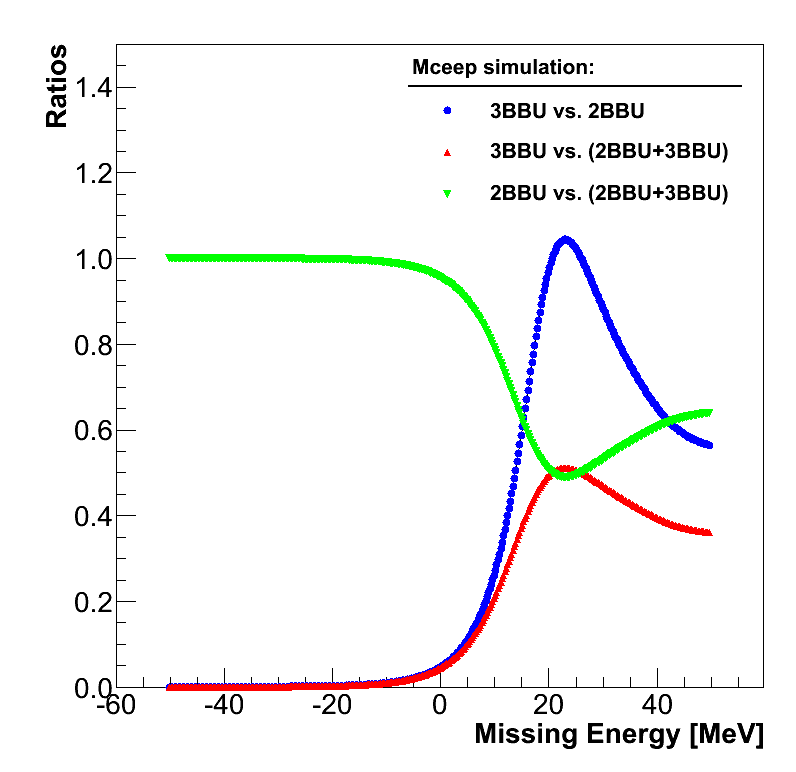}
\end{minipage}
\hfill
\begin{minipage}[t]{0.45\textwidth}
\hrule height 0pt
\caption{The relative contributions of the 2BBU and 3BBU channel as a function 
of missing energy. The ratio between the two-body and three-body 
breakup is also shown. The results were obtained from the comparisons of the
convoluted MCEEP results.  In the region of negative missing energies, 
two-body breakup dominates. Three-body breakup starts contributing at 
positive missing energies, prevailing in the region of 
$20\leq E_{\mathrm{Miss}}\leq 40$. At very high $E_{\mathrm{Miss}}$, where 
3BBU loses its strength, the contribution of the 2BBU tail again becomes
comparable to the 3BBU part.  
\label{fig_analysis_Mceep_Ratios}}
\end{minipage}
\end{center}
\end{figure}

The comparison of the broadened 2BBU and 3BBU missing energy peaks reveals that 
the coarse resolution causes 3BBU events to appear also at missing energies below the 
theoretical threshold $E_{\mathrm{Miss}} = 7.7\,\mathrm{MeV}$. Figure~\ref{fig_analysis_Mceep_Ratios}
shows the simulated ratio between the two-body and three-body breakup strength as
a function of the missing energy. Although the contamination of 2BBU events 
with 3BBU events at $E_{\mathrm{Miss}}\leq 2\,\mathrm{MeV}$ seems to be small,
their contribution to the 2BBU asymmetry can not be neglected, because 
the  ${}^3\vec{\mathrm{He}}(\vec{e},e'p)pn$ asymmetries are predicted to be large.
Since the sign of the 3BBU asymmetries is opposite to the sign of the 2BBU asymmetries, 
such corrections could explain the unresolved discrepancy between the theory and measurements 
for the ${}^3\vec{\mathrm{He}}(\vec{e},e'p)d$ reaction. 

\begin{figure}[!htb]
\begin{center}
\includegraphics[width=0.49\linewidth]{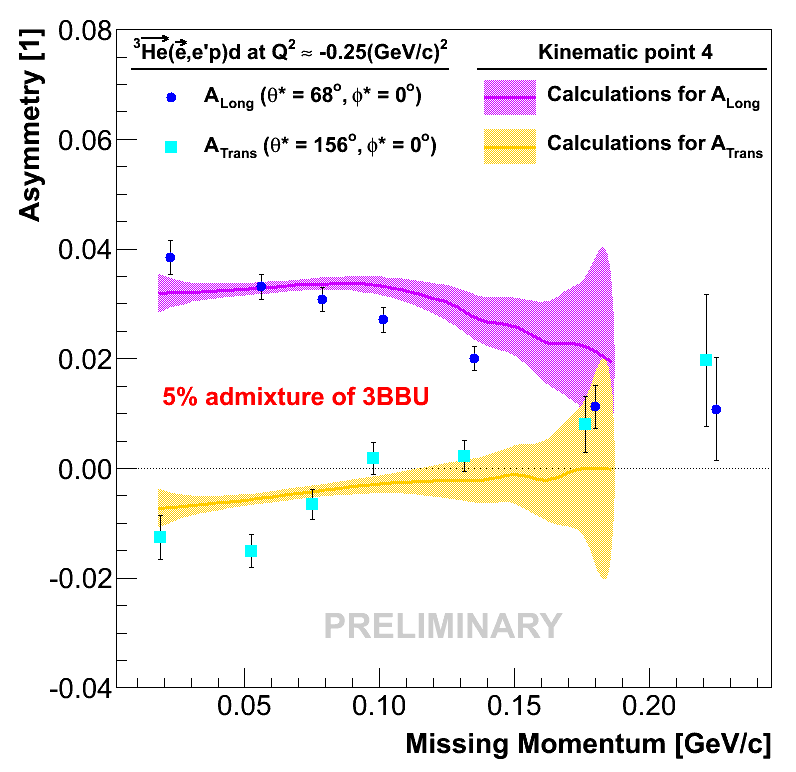}
\includegraphics[width=0.49\linewidth]{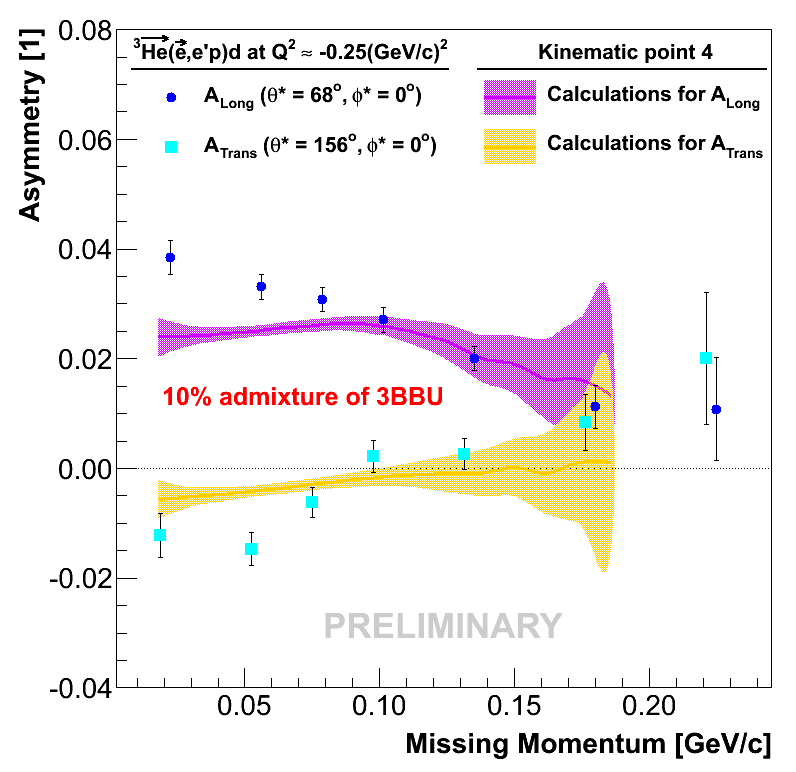}
\caption{ Comparison of the theoretical 2BBU asymmetries, calculated  for the $4^{\mathrm{th}}$ 
kinematic bin, to the extracted experimental asymmetries at $Q^2 = -0.25\,(\mathrm{GeV}/c)^2$, 
when  $5\,\mathrm{\%}$ (left) and $10\,\mathrm{\%}$ admixture (right) of the 
${}^3\vec{\mathrm{He}}(\vec{e},e'p)pn$ asymmetry is  added to the 
calculations for the ${}^3\vec{\mathrm{He}}(\vec{e},e'p)d$ asymmetry. 
Labels as in Fig.~\ref{fig_analysis_2BBUResult1}
\label{fig_analysis_3BBUResult1}}
\end{center}
\end{figure}

\begin{figure}[!htb]
\begin{center}
\begin{minipage}[t]{0.5\textwidth}
\hrule height 0pt
\includegraphics[width=\linewidth]{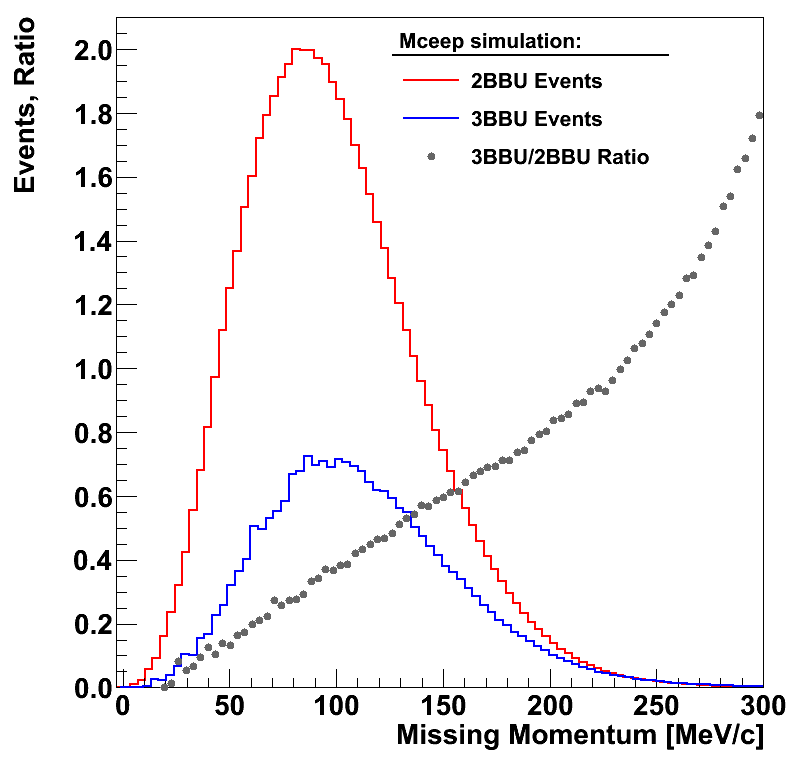}
\end{minipage}
\hfill
\begin{minipage}[t]{0.45\textwidth}
\hrule height 0pt
\caption{Missing-momentum distributions for the 2BBU and 3BBU channels, obtained with
the MCEEP simulation. Both distributions were scaled to $1/2$ of the two-body breakup 
peak. No cuts on $E_{\mathrm{Miss}}$ were applied and no corrections 
to the resolution were considered.  The ratio 3BBU/2BBU shows that the contribution 
of the 3BBU grows with increasing missing momentum. Hence, the 3BBU correction to the 2BBU 
asymmetries  is largest at high
missing momenta. 
\label{fig_analysis_MCEEP_MissingMomentum}}
\end{minipage}
\end{center}
\end{figure}

To test this assumption, a $5\,\mathrm{\%}$ and $10\,\mathrm{\%}$ admixture of 
the ${}^3\vec{\mathrm{He}}(\vec{e},e'p)pn$ asymmetries was added to
the theoretical asymmetries for 
the ${}^3\vec{\mathrm{He}}(\vec{e},e'p)d$ reaction. 
The modified asymmetries are presented in Fig.~\ref{fig_analysis_3BBUResult1}.
As anticipated, a small contamination with the 3BBU asymmetries has caused a sizable 
change to the 2BBU asymmetries. Note that such a correction with a fixed 3BBU/2BBU ratio
can be used only for demonstrative purposes. 
For a detailed analysis, an individual corrections to each bin in missing momentum is 
required, since the ratio has a strong missing-momentum dependence
(see Fig.~\ref{fig_analysis_MCEEP_MissingMomentum}).
This again emphasizes an urgent need for a better and more trustworthy Monte-Carlo 
simulation, which could be used to adequately estimate the 3BBU/2BBU ratios.

\begin{figure}[!htb]
\begin{center}
\begin{minipage}[t]{0.5\textwidth}
\hrule height 0pt
    \includegraphics[width=\linewidth]{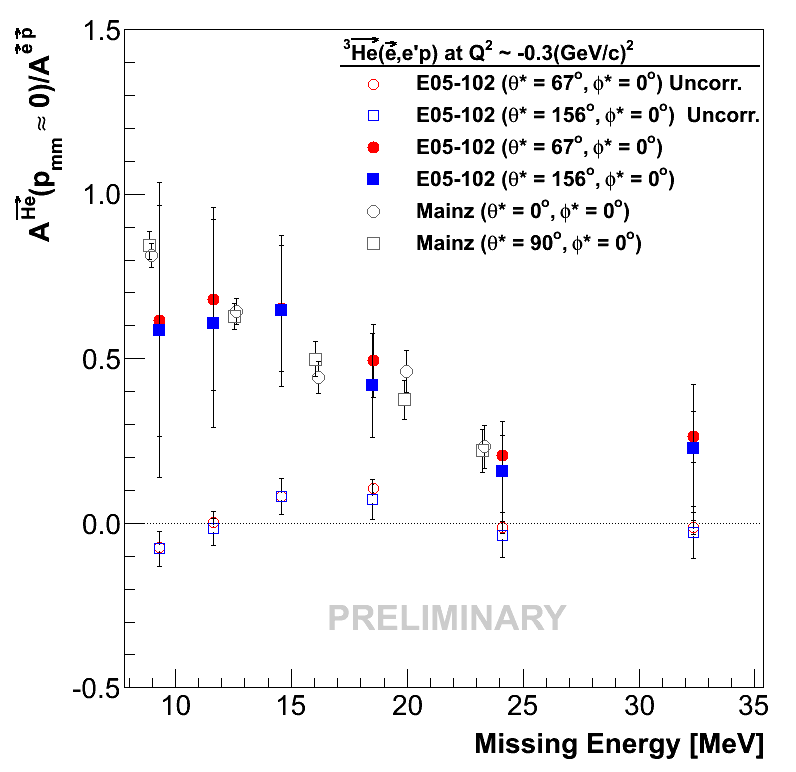}
\end{minipage}
\hfill
\begin{minipage}[t]{0.48\textwidth}
\hrule height 0pt
\caption{Ratios of the  ${}^3\vec{\mathrm{He}}(\vec{e},e'p)pn$ 
asymmetries with the elastic proton asymmetries for the $Q^2\approx -0.35\,(\mathrm{GeV}/c)^2$
data. Hollow red and blue data-points show the 3BBU results before corrections for the 2BBU
contamination. Full red and blue points show the corrected ratios. The results of 
the Mainz experiment~\cite{achenbach2008} are demonstrated with gray hollow points.
The significant worsening of the uncertainty for points at 
$E_{\mathrm{Miss}}\approx 10\,\mathrm{MeV}$ is predominantly caused by the 
small values of the denominator in Eq.~(\ref{eq_analysis_2BBU_3BBU}) 
near the threshold for the 3BBU (see Fig.~\ref{fig_analysis_Mceep_Ratios}).
The second most important contribution to the error is the estimated $20\,\mathrm{\%}$
uncertainty of the two-body contamination factor $p_{\mathrm{2BBU}}$, which
is determined from MCEEP simulation.
\label{fig_analysis_3BBU_Extraction}}
\end{minipage}
\end{center}
\end{figure}

In spite  of the imperfections of MCEEP, we decided to use it in our pursuit of extracting 
information on the three-body breakup asymmetries. For this trial we have 
selected longitudinal and transverse data at $Q^2 = -0.35\,(\mathrm{GeV}/c)^2$.
We have concentrated only on the events at low missing momenta
$(p_{\mathrm{Miss}}\leq 90\,\mathrm{MeV}/c)$ shown in 
Fig.~\ref{fig_analysis_MissingEnergyAsymmetry}. High missing momentum data were not yet 
analyzed. We selected only points with $E_{\mathrm{Miss}}\geq 10\,\mathrm{MeV}$ 
because one expects that 3BBU will be most clearly accessible in that region. 
However, due to a large contamination with the 2BBU asymmetry, established in 
Fig.~\ref{fig_analysis_Mceep_Ratios}, the measured asymmetries $A_{\mathrm{Exp}}$ must be 
properly corrected for the admixture of 2BBUs.  Assuming that the two-body breakup 
asymmetry $A_{\mathrm{2BBU}}$ is under control, the three-body asymmetry 
$A_{\mathrm{3BBU}}$ can be determined  via~\cite{achenbach2008}:
\begin{eqnarray}
A_{\mathrm{3BBU}}(E_{\mathrm{E_\mathrm{Miss}}}) = \frac{A_{\mathrm{Exp}}(E_{\mathrm{Miss}}) - A_{\mathrm{2BBU}}\>p_{\mathrm{2BBU}}(E_{\mathrm{Miss}})}
{1-p_{\mathrm{2BBU}}(E_{\mathrm{Miss}})}\,,\label{eq_analysis_2BBU_3BBU}
\end{eqnarray}
where $p_{\mathrm{2BBU}}$ represents the fraction of 2BBU events in a particular 
$E_\mathrm{Miss}$ bin. The obtained results are gathered in Fig.~\ref{fig_analysis_3BBU_Extraction},
which shows the ratios of extracted 3BBU asymmetries 
with the corresponding elastic proton asymmetries (see Fig.~\ref{fig_analysis_ProtonElasticAsymmetry}).
One can see that the applied corrections have a strong effect on the result and considerably 
increase the value of the asymmetry. This is consistent with the theoretical predictions 
of Bochum/Krakow group (see Figs.~\ref{fig_analysis_TheoryLongKin4ppn} 
and~\ref{fig_analysis_TheoryTransKin4ppn}), which predict 
large negative asymmetries $(\sim -10\,\mathrm{\%})$.  

\begin{figure}[!htb]
\begin{center}
\begin{minipage}[t]{0.5\textwidth}
\hrule height 0pt
    \includegraphics[width=\linewidth]{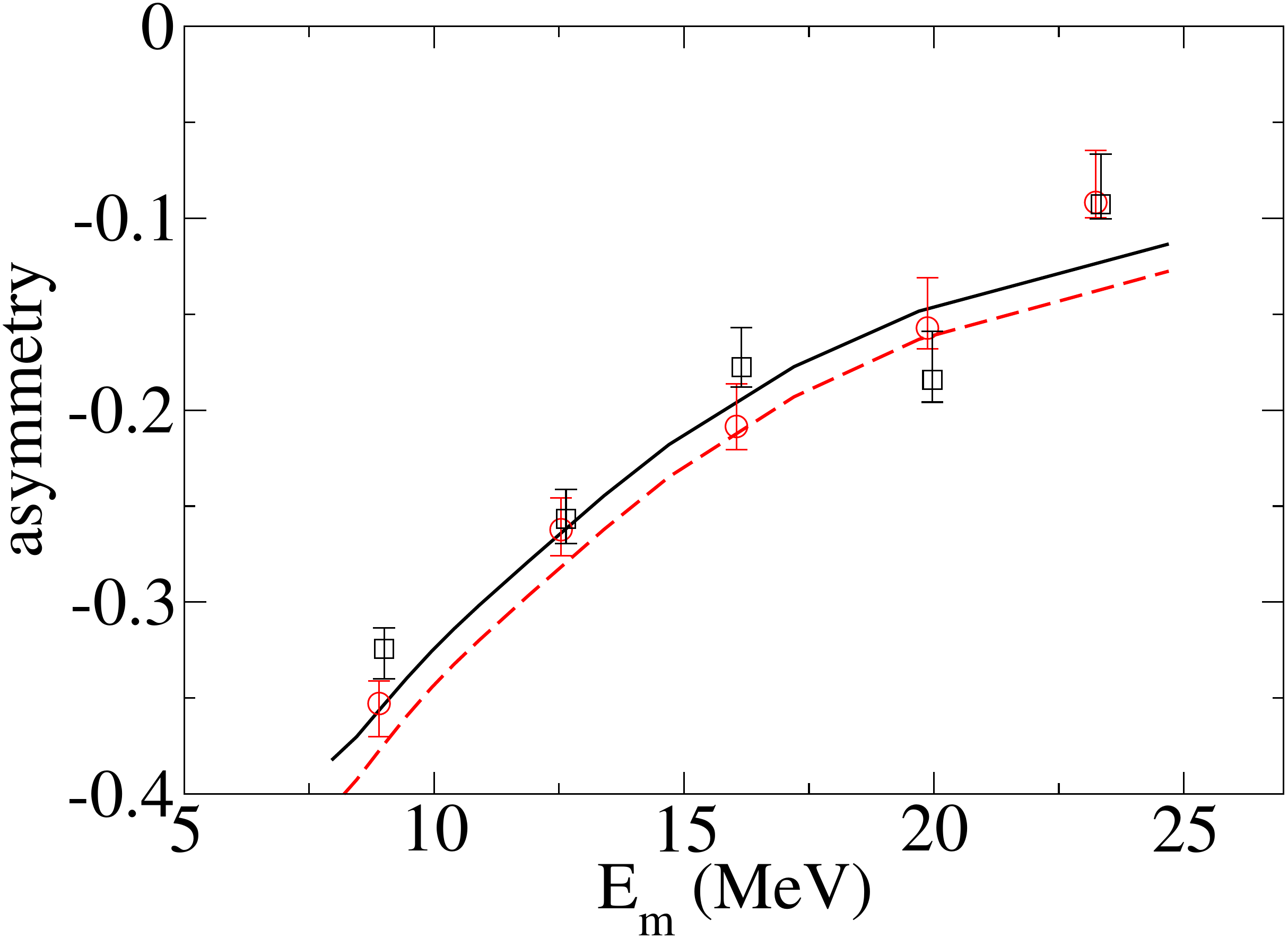}
\end{minipage}
\hfill
\begin{minipage}[t]{0.48\textwidth}
\hrule height 0pt
\caption{ The Mainz results for the parallel and perpendicular 
asymmetry as a function of missing energy $E_{m}$ in the 3BBU-channel. 
$A(\theta^* = 0^\circ,\phi^*= 0^\circ)$: data (squares), theory by 
Krakow/Bochum group (solid line). 
$A(\theta^* = 90^\circ,\phi^*= 0^\circ)$: data (circles), theory by 
Krakow/Bochum group (dashed line).
~\cite{achenbach2008}
\label{fig_analysis_Mainz_3BBU}}
\end{minipage}
\end{center}
\end{figure}

The interpretation can be compared to the results from the Mainz 
experiment~\cite{achenbach2008}. Their measurements were performed at very similar 
kinematic conditions ($\omega = 135\,\mathrm{MeV}$, $|\vec{q}| = 570\,\mathrm{MeV}/c$, 
$Q^2\approx -0.3\,(\mathrm{GeV}/c)^2$, 
$p_{\mathrm{Miss}}\approx 40\,\mathrm{MeV}/c$). However, their data were collected for different
target spin orientations ($\theta^* = 0^\circ\,,\>90^\circ$), resulting in much larger 
absolute values of the asymmetries (see Fig.~\ref{fig_analysis_Mainz_3BBU}). To make 
both results comparable, the normalization to the elastic proton asymmetry was chosen. 
The Mainz ratios show good agreement with our results. A more detailed comparison to the
data at $p_{\mathrm{Miss}}>0$ will become possible with a more sophisticated 
simulation, which remains one of our future tasks.  

\section{The deuteron channel ${}^3\vec{\mathrm{He}}(\vec{e},e'd)p$}
\begin{figure}[!ht]
\begin{center}
\includegraphics[width=0.49\textwidth]{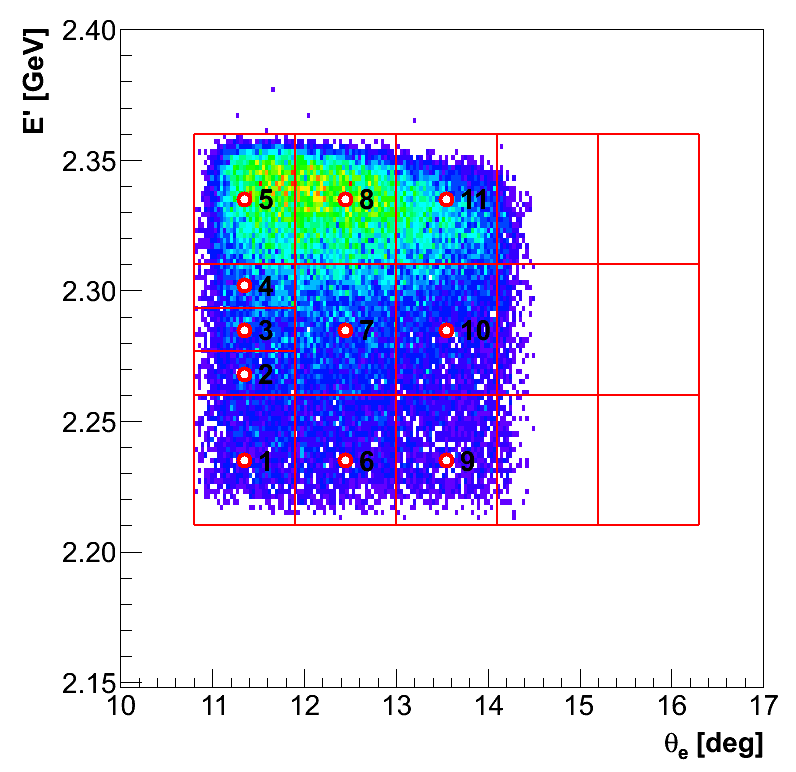}
\includegraphics[width=0.49\textwidth]{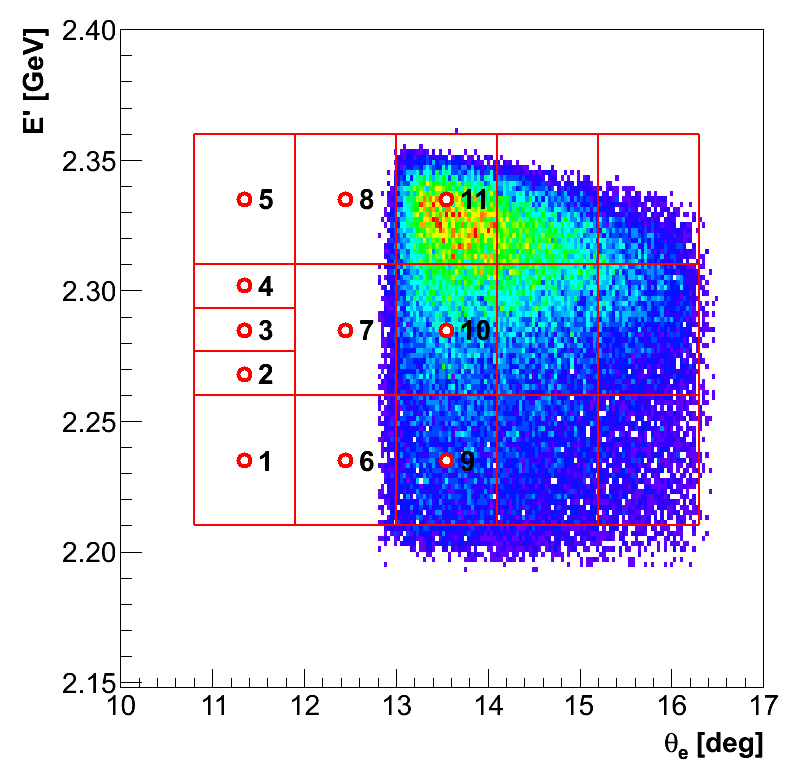}
\caption{ The available electron kinematics for the ${}^3\vec{\mathrm{He}}(\vec{e},e'd)p$ 
reaction. The left and right plots show the results when HRS-L was positioned at 
$\theta_{\mathrm{HRS-L}} = 12.5^\circ$ and $14.5^\circ$, respectively. 
The whole kinematical coverage was divided into 17 bins denoted by red squares. 
Theoretical calculations were performed for the centers of the leftmost 11 bins 
(circles). 
\label{fig_analysis_deuteron_kinpoints}}
\end{center}
\end{figure}

The comparison of the measured ${}^3\vec{\mathrm{He}}(\vec{e},e'd)p$ asymmetries with the 
theoretical calculations of Bochum/Krakow group was carried out with an 
approach identical to the one used for the interpretation of the proton channels. 
The asymmetries for each calculated kinematic point were again examined individually.
Since the majority of the events for this reaction channels are gathered inside the top
three kinematic bins (see Fig.~\ref{fig_analysis_deuteron_kinpoints}), we limited 
our present analysis to kinematic points $5$, $8$ and $11$.

\begin{sidewaysfigure}[!htp]
\begin{center}
\includegraphics[angle=0,width=0.85\linewidth]{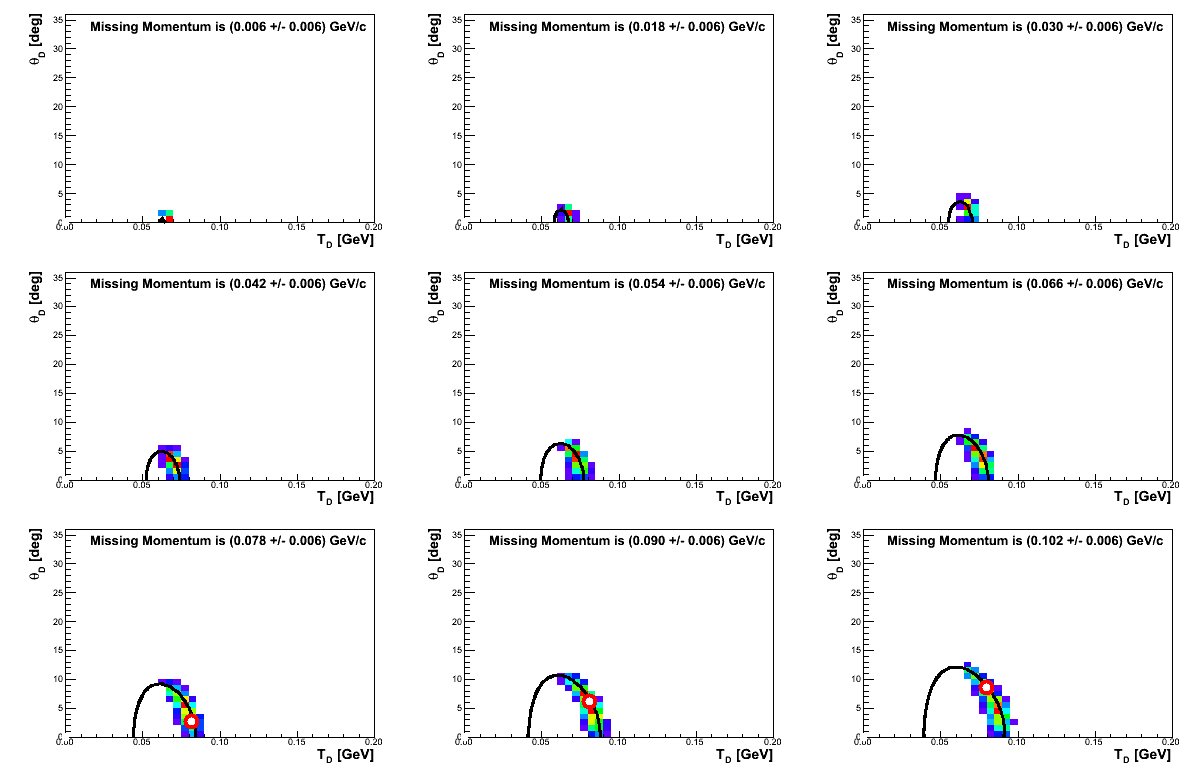} 
\vspace*{-3mm}
\caption{ Two-dimensional histograms showing the relation between the deuteron angle 
$\theta_{d}$ and its kinetic energy $T_d\approx p_d^2/2M_{d}$ for selected bins 
in $p_{\mathrm{Miss}}$. In these plots the  data surrounding the $5^{\mathrm{th}}$ 
central kinematics point were considered (see Fig.~\ref{fig_analysis_kinpoints}).
The lengths of the obtained bands are controlled by the remaining spread in 
$\omega$ and $\vec{q}$. Black lines show the solutions of Eq.(\ref{eq_momentum_conservation}) 
for a given $p_{\mathrm{Miss}}$ and $|\vec{q}| = 485.0\,\mathrm{MeV}/c$. Circles 
show $(\theta_{d}, \tilde{p}_d)$ pairs considered in the theoretical calculations. 
In a selected kinematic point, theoretical points at very low missing momenta are not 
permitted. Therefore, theoretical points are missing in first six histograms.
\label{fig_analysis_Deuteron_KinBananas1}}
\end{center}
\end{sidewaysfigure}

\begin{sidewaysfigure}[!htp]
\begin{center}
\includegraphics[angle=0,width=0.85\linewidth]{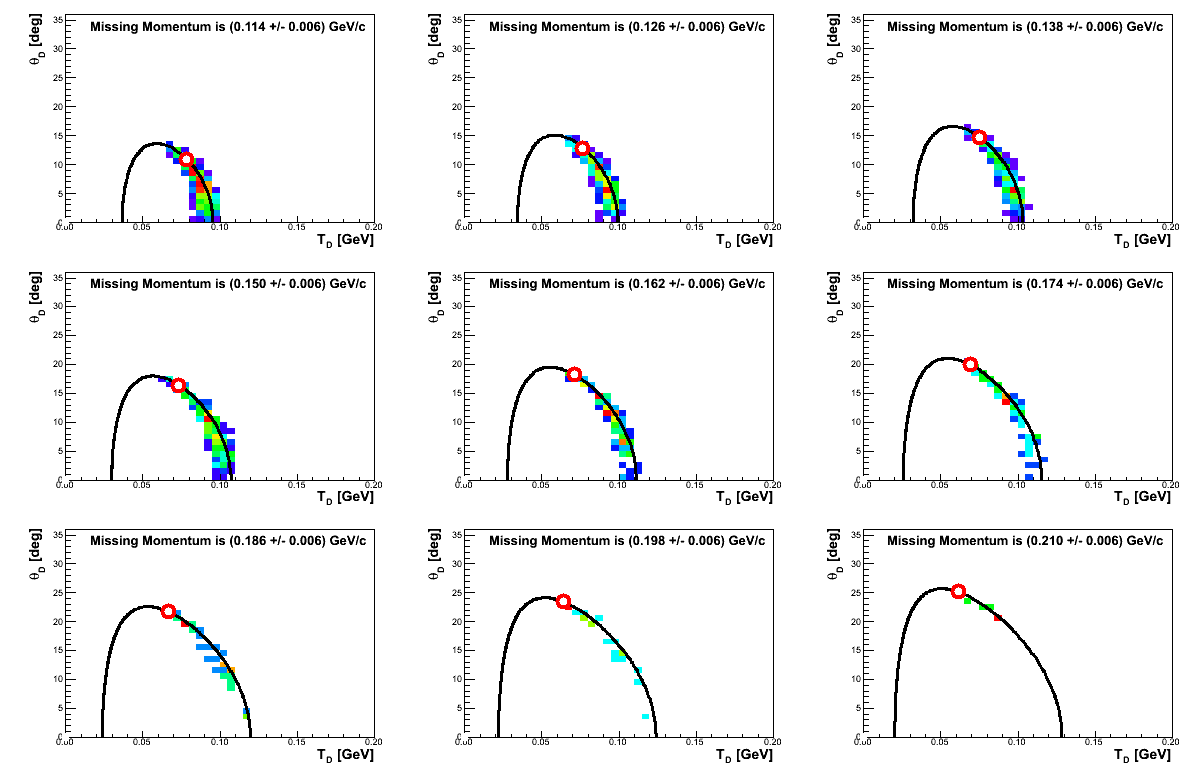} 
\caption{
(continued): The remaining range in $p_{\mathrm{Miss}}$.
\label{fig_analysis_Deuteron_KinBananas2}}
\end{center}
\end{sidewaysfigure}

Similarly as for the proton channel, correct pairs of deuteron kinetic energies $T_{d}$ and 
polar angle $\theta_{d}$ had to be input to the code, in order for the 
theoretical calculations to be executed for a desirable set of missing momenta. 
Selected points for the $5^{\mathrm{th}}$ kinematic bin are shown in 
Figs.~\ref{fig_analysis_Deuteron_KinBananas1} and~\ref{fig_analysis_Deuteron_KinBananas2}.
Again, not all missing momenta are accessible with each kinematic point. Some may be
prohibited by the equations (\ref{eq_energy_conservation}) and (\ref{eq_momentum_conservation}).
Histograms belonging to $p_{\mathrm{Miss}} < 78\,\mathrm{MeV}/c$ therefore do not 
contain theoretical points. 
On the other hand, the experimental data are not limited to a single kinematic point. They 
are smeared over the selected kinematic bin (red squares in Fig.~\ref{fig_analysis_deuteron_kinpoints}).
This gives them enough freedom to appear also in histograms with smaller missing momenta.

\begin{figure}[!hb]
\begin{center}
\includegraphics[width=0.9\textwidth]{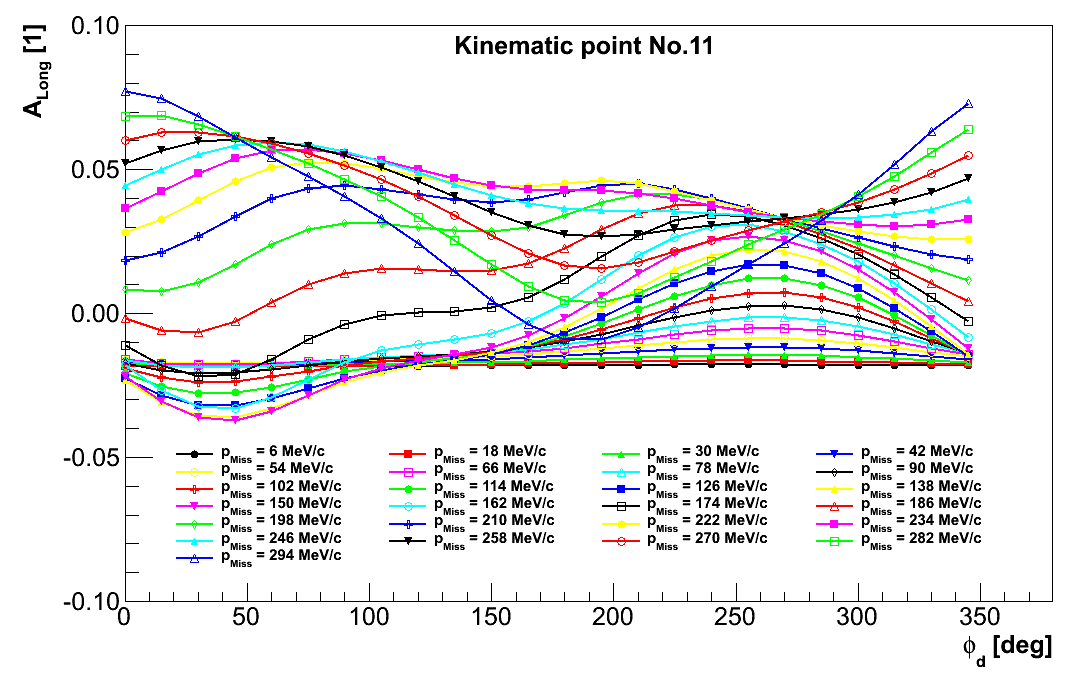}
\vspace*{-3mm}
\caption{The theoretical predictions for the longitudinal ${}^3\vec{\mathrm{He}}(\vec{e},e'd)p$ asymmetry 
$A_{\mathrm{Long}} = A(\theta^*=73^\circ, \phi^* = 0^\circ)$ as a function of the angle $\phi_d$,
for missing momenta up to $p_{\mathrm{Miss}}\leq 300\,\mathrm{MeV}$.
Presented asymmetries were obtained for the $11^\mathrm{th}$ kinematic bin.
Calculations were  provided by the Bochum/Krakow group~\cite{golak_private_2012}.  
\label{fig_analysis_Deuteron_TheoryLongKin4ppn}}
\vspace*{-5mm}
\end{center}
\end{figure}

\begin{figure}[!ht]
\begin{center}
\includegraphics[width=0.9\textwidth]{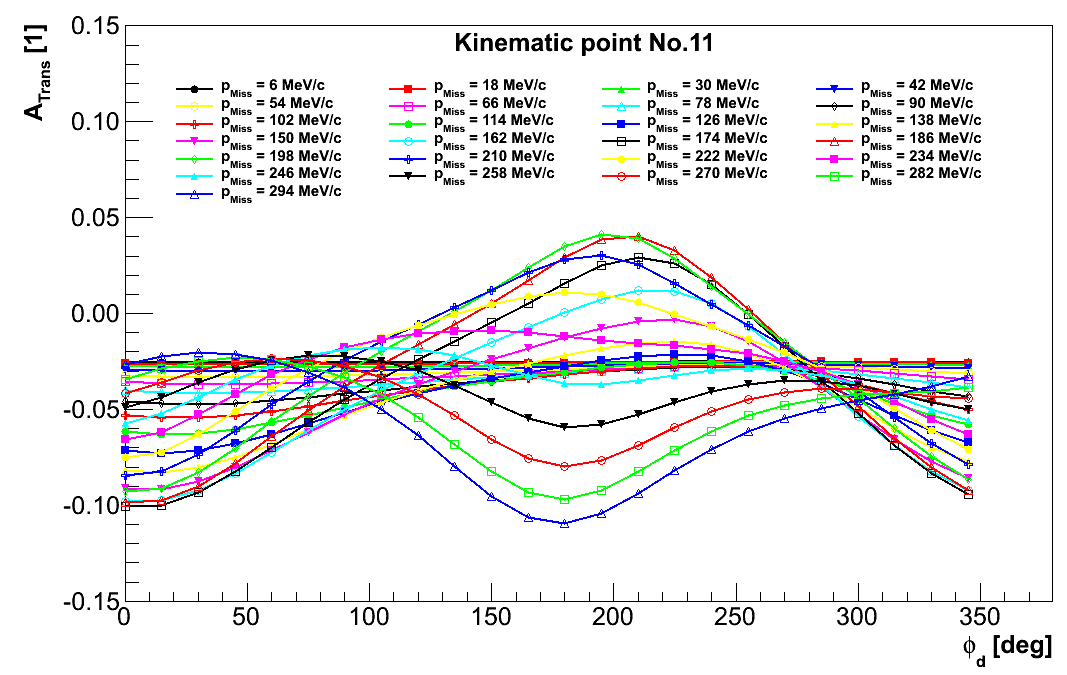}
\vspace*{-3mm}
\caption{ The theoretical predictions for the transverse ${}^3\vec{\mathrm{He}}(\vec{e},e'd)p$ asymmetry 
$A_{\mathrm{Trans}} = A(\theta^*=163^\circ, \phi^* = 0^\circ)$ as a function of the angle $\phi_d$,
for missing momenta up to $p_{\mathrm{Miss}}\leq 300\,\mathrm{MeV}$.
Presented asymmetries were obtained for the $11^\mathrm{th}$ kinematic bin.
Calculations were  provided by the Bochum/Krakow group~\cite{golak_private_2012}. 
\label{fig_analysis_Deuteron_TheoryTransKin4ppn}}
\vspace*{-5mm}
\end{center}
\end{figure}

\begin{figure}[!htb]
\begin{center}
\begin{minipage}[t]{0.5\textwidth}
\hrule height 0pt
    \includegraphics[width=\linewidth]{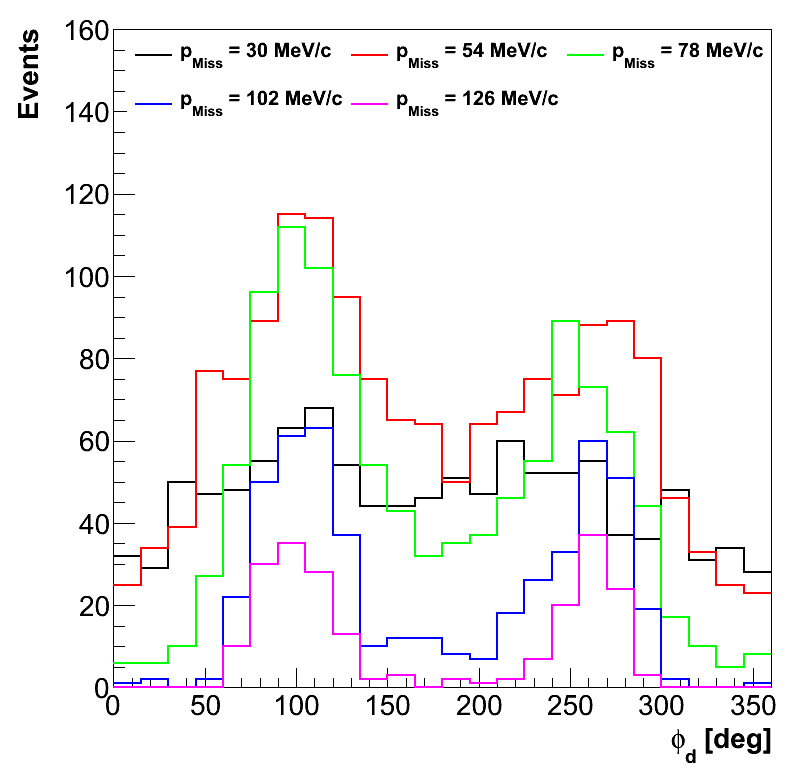}
\end{minipage}
\hspace*{1cm}
\begin{minipage}[t]{0.32\textwidth}
\hrule height 0pt
\caption{The distributions of $\phi_{d}$ at different $p_{\mathrm{Miss}}$, determined
for the events gathered around the $11^{\mathrm{th}}$ kinematic point. 
At low missing momenta the events are uniformly distributed over the whole angular range.
At high missing momenta, events with $\phi_d \approx 90^\circ, 270^\circ$ dominate. 
\label{fig_analysis_Deuteron_PhiDQAveraging}}
\end{minipage}
\end{center}
\end{figure}

In the sense of missing momenta the $11^{\mathrm{th}}$ kinematic 
point is the most interesting one, 
because it is the only theoretical point where the asymmetries at very low missing momenta are 
accessible. The calculated longitudinal and transverse asymmetries for this kinematic
point are gathered in Figs.~\ref{fig_analysis_Deuteron_TheoryLongKin4ppn} 
and~\ref{fig_analysis_Deuteron_TheoryTransKin4ppn}. In order to compare these computations
to the experimental asymmetries, an averaging over the $\phi_d$ angle needs to 
be performed for each $p_{\mathrm{Miss}}$, analogously as it was done for the 
proton channel. Here $\phi_d$ represents the angle between the scattering
plane and the reaction plane, which is  determined by the deuteron momentum 
and  $\vec{q}$ (see Fig.~\ref{fig_theory_planes}). Examples of the $\phi_d$ distributions
for various missing momenta are presented in Fig.~\ref{fig_analysis_Deuteron_PhiDQAveraging}.

\begin{figure}[!ht]
\begin{center}
\includegraphics[width=0.49\linewidth]{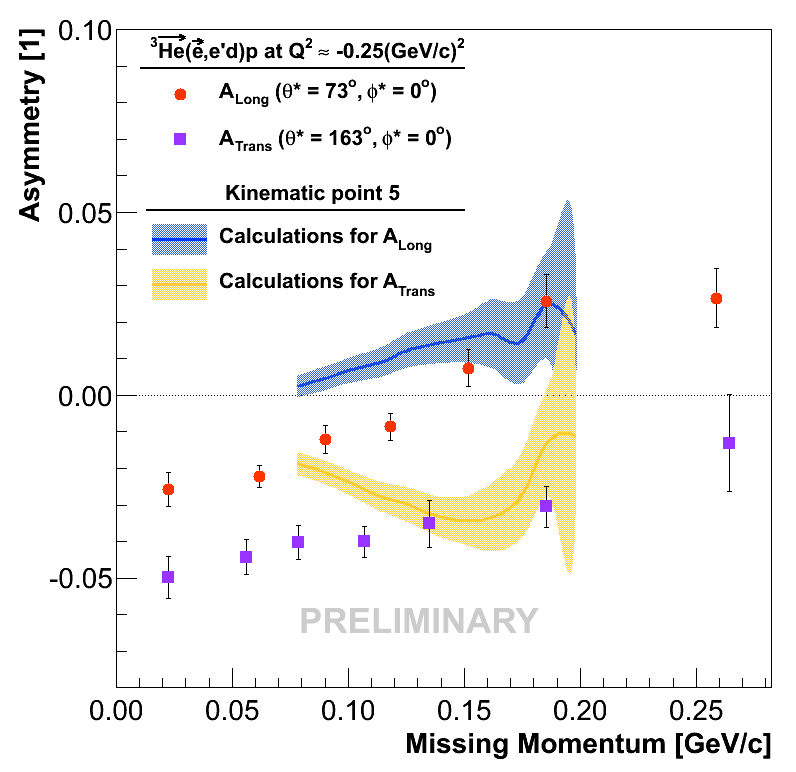}
\includegraphics[width=0.49\linewidth]{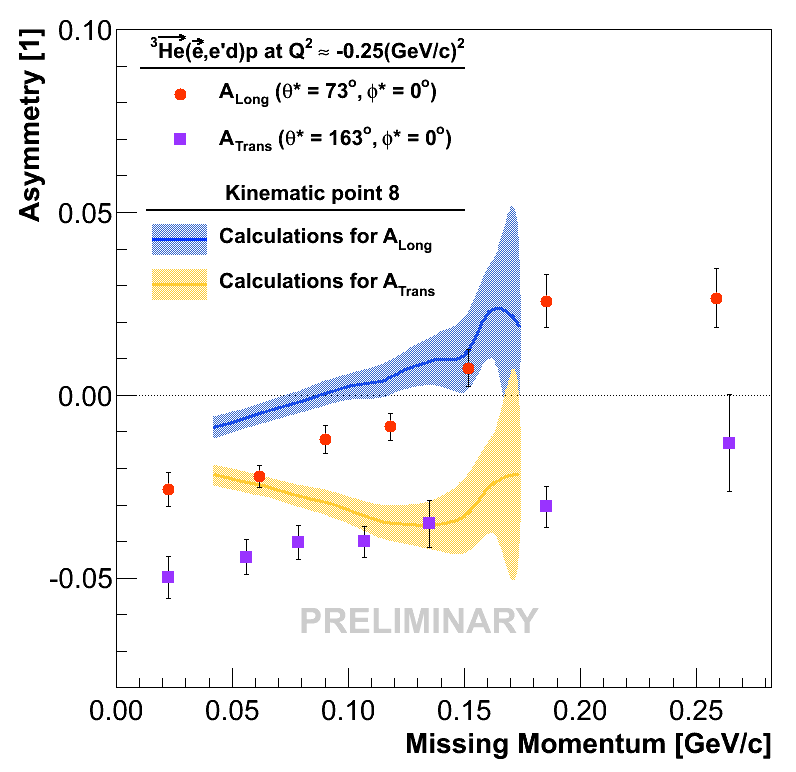}
\vspace*{-3mm}
\caption{ Comparison of the theoretical ${}^3\vec{\mathrm{He}}(\vec{e},e'd)p$ asymmetries, 
calculated for the $5^{\mathrm{th}}$ (left) and the $8^{\mathrm{th}}$ (right) kinematic bin, with the 
experimental  asymmetries at $Q^2 = -0.25\,(\mathrm{GeV}/c)^2$. The theoretical asymmetries are 
shown with full lines. The error bands demonstrate the uncertainties of the procedure used 
to average the theoretical asymmetries. Errors are governed by the statistics  of the $\phi_{d}$-histograms 
(see Fig.~\ref{fig_analysis_Deuteron_PhiDQAveraging}).
\label{fig_analysis_Deuteron_Comparison1}}
\vspace*{-5mm}
\end{center}
\end{figure}

\begin{figure}[!ht]
\begin{center}
\includegraphics[width=0.49\linewidth]{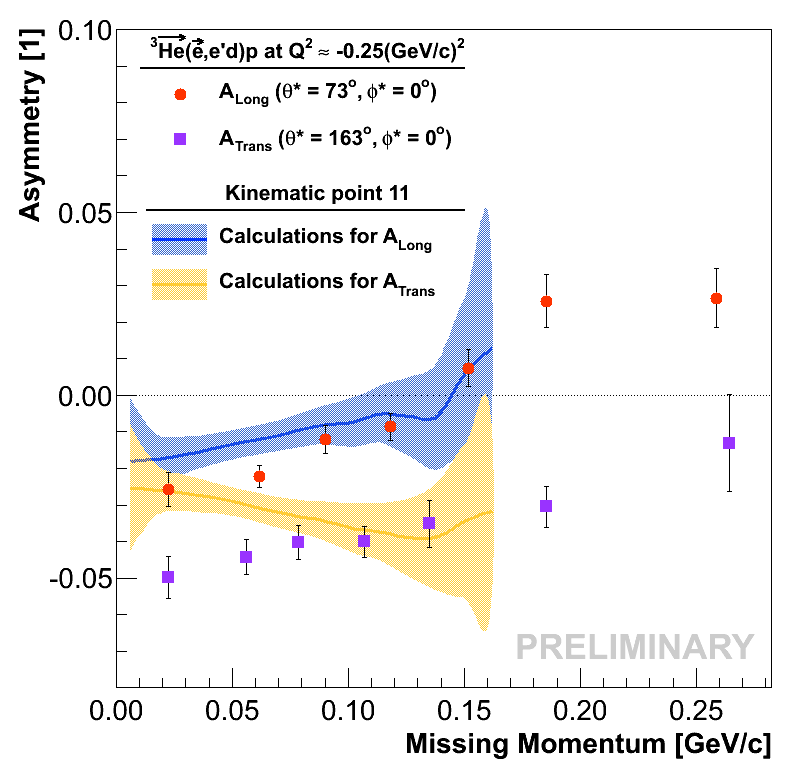}
\includegraphics[width=0.49\linewidth]{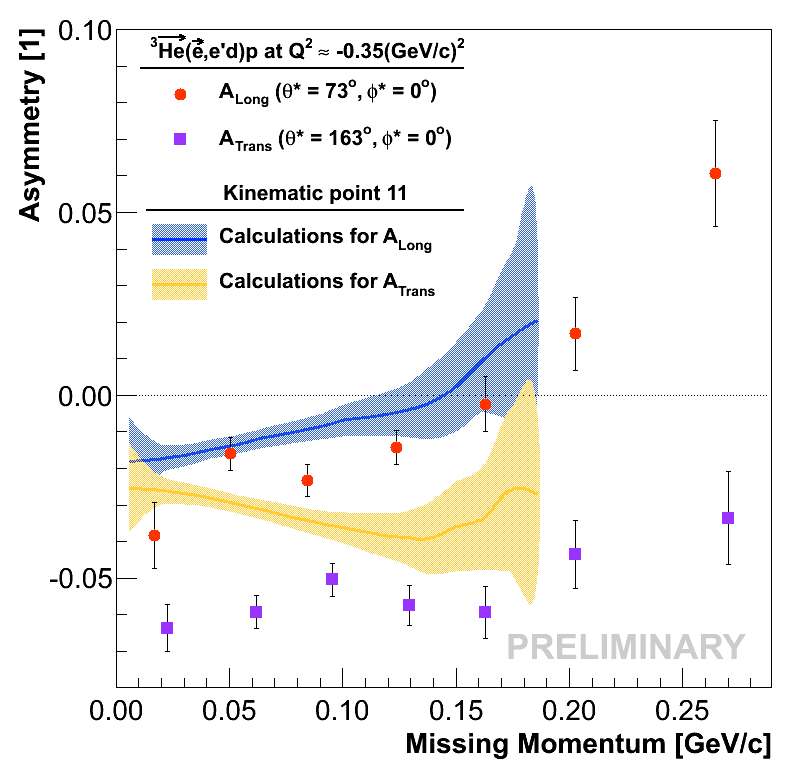}
\vspace*{-3mm}
\caption{ Comparison of the theoretical ${}^3\vec{\mathrm{He}}(\vec{e},e'd)p$ asymmetries, 
calculated for the $11^{\mathrm{th}}$  kinematic bin, with the experimental 
asymmetries at $Q^2 = -0.25\,(\mathrm{GeV}/c)^2$ (left) and 
$Q^2 = -0.35\,(\mathrm{GeV}/c)^2$ (right). The theoretical asymmetries are 
shown with full lines. The error bands demonstrate the uncertainties of the procedure used 
to average the theoretical asymmetries. Errors are governed by the statistics  
of the $\phi_{d}$-histograms 
(see Fig.~\ref{fig_analysis_Deuteron_PhiDQAveraging}).
\label{fig_analysis_Deuteron_Comparison2}}
\vspace*{-5mm}
\end{center}
\end{figure}

Once the theoretical calculations were properly averaged over $\phi_d$, 
they could be compared to the measured data. The results for the three 
considered theoretical bins are shown in Figs.~\ref{fig_analysis_Deuteron_Comparison1}
and~\ref{fig_analysis_Deuteron_Comparison2}. Both $Q^2 = -0.25\,(\mathrm{GeV}/c)^2$ and 
$Q^2 = -0.35\,(\mathrm{GeV}/c)^2$ data were put to the test. 
The measured and predicted asymmetries have
consistent signs. They also agree in the position of the zero-crossing point. 

In other regions the agreement is worse, with apparently opposing slopes for 
the transverse asymmetries. Even at very low missing momenta, where the 
best consistency was expected, the acceptance-averaged theory predicts a much smaller 
transverse asymmetry $(\approx 2.5\,\mathrm{\%})$ than it was measured 
$(\approx 5\,\mathrm{\%})$. The inconsistencies are observed at both $Q^2$.
Similar discrepancies are observed for the longitudinal asymmetry. 
One should again note that the primary reason for these mismatches could 
be the incomplete kinematic averaging procedure (see Sec.~\ref{sec:2BBU}).

\vspace*{-5mm}
For a correct interpretation of the deuteron channel one needs to resort to  
full theoretical calculations. Simple models,
analogous to the one considered for the proton channel (see Sec.~\ref{sec:ProtonElastic}),
can not be applied. For example, it was shown by previous 
experiments (see Fig.~\ref{fig_Tripp}) as well as the  
theory (see Sec.~\ref{sec:nagorny}), that the deuteron pole diagram alone  does not provide 
a satisfactory description of the ${}^3\vec{\mathrm{He}}(\vec{e},e'd)p$ process at 
low missing momenta. A complete calculation in all sophistication is needed for a 
meaningful comparison. 

\section{Conclusions}

Inconsistencies between the theory and the measurements in this first 
iteration of the analysis are not unexpected. The measurements presented 
in this thesis are the first of its kind. For the first time, we have measured 
double-polarization asymmetries in all exclusive nucleon knockout channels on
${}^3\mathrm{He}$, at approximately the same value of $Q^2$, and as a function of 
$p_{\mathrm{Miss}}$. The theoreticians so far were never faced with a data set so rich
and comprehensive, and thus had no proper reference point for a precise calibration of 
their theories. These state-of-the-art theories have been able to describe a large
body of un-polarized data astonishingly well, but significant discrepancies 
remain, but could be even more pronounced in the case of double-polarized 
observables. 

For example, the Krakow/Bochum calculations exhibit remarkable differences
to certain observables (see e.g.~Fig.~\ref{fig_spaltro}) and it is unclear which 
ingredient may be responsible. The disagreement seems to originate in the 
longitudinal part of the cross-section. The use of the Sachs form-factors 
$G_\mathrm{E}^\mathrm{p}$ and $G_\mathrm{E}^\mathrm{n}$ instead of the Pauli 
form-factors  $F_\mathrm{1}^\mathrm{p}$ and $F_\mathrm{1}^\mathrm{n}$, the 
three-nucleon force, as well as the inclusion of MEC in the charge-density 
operator are some candidates  to resolve the issue~\cite{e05102}. 

Recently we have also acquired the Faddeev calculations of the two-body 
breakup processes on ${}^3\mathrm{He}$ by the Hannover/Lisbon Group,
which significantly depart from both our data and the Krakow/Bochum results
(see Fig.~\ref{fig_HannoverKrakowComparison}). The Hannover calculations also include FSI and 
MEC, and the theoretical apparatus  is presumed to be comparable to the one 
used in Krakow~\cite{yuan02a, yuan02b, deltuva03, deltuva04a, deltuva04b, BKHprivate}. 
However, they  add the $\Delta$-isobar as an active 
degree of freedom providing a mechanism  for an effective three-nucleon 
force and for exchange currents.

One of our remaining tasks is to devise a better acceptance averaging procedure
which may have incorrectly modified the theoretical asymmetries. When this is 
accomplished, our data will provide any theory with a handle for careful 
fine-tunning. This will open the way to resolve the possible discrepancies
discussed above.

\appendix

\chapter{Analytical Optics Model for BigBite}
\label{appendix:AnalyticalModel}

The analytical model of the BigBite optics was utilized for the 
reconstruction of the BigBite target variables during the data-taking
phase of the experiment, when the final method for BigBite optics was
still not available. It is based on the assumption that the BigBite
magnet can be approximated by an ideal dipole and requires only few 
geometrical parameters for a successful implementation. 
In spite of its simplicity the model can reconstruct the target variables well 
enough to be adequately used in the on-line analysis of the data. 

\begin{figure}[!ht]
\begin{center}
\includegraphics[width=\linewidth]{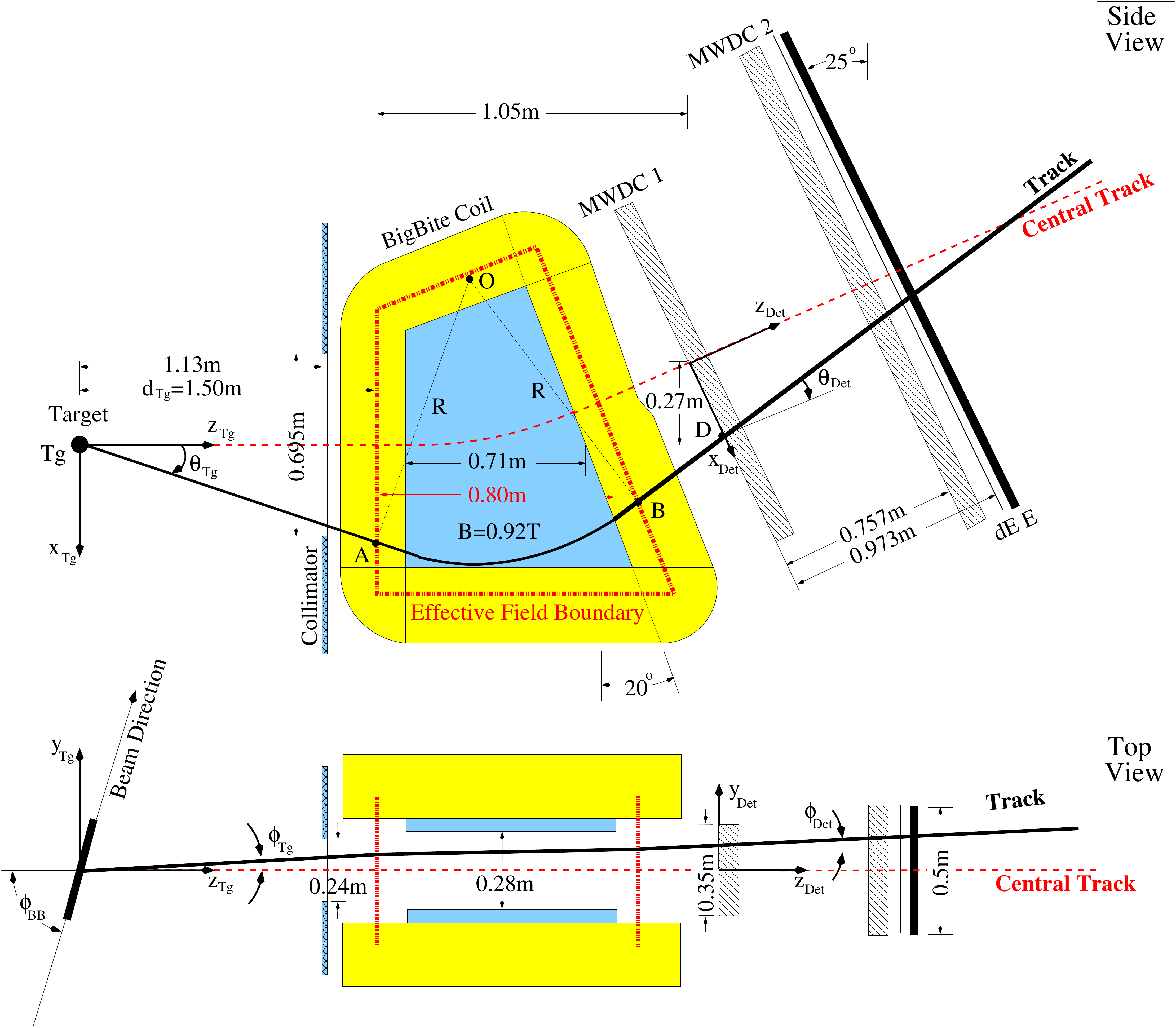} 
\caption{The schematic of the dispersive (top) and non-dispersive
(bottom) planes of the BigBite spectrometer.  Small angular
deflections in the non-dispersive plane occur if the particle
trajectory is not perpendicular to the effective field boundary
\cite{lange-optics,penner,brown}.  At the entrance to the magnet,
they are at most $18\,\mathrm{mrad}$ (close to the acceptance
boundaries in the dispersive direction).  At the exit field
boundary, the effect acts in the opposite sense and partially
cancels the deflection at the entrance.\label{fig_BBAnalyticalModel_BBScheme}}
\end{center}
\end{figure}

The magnetic field of the BigBite magnet is oriented in the
$y_{\mathrm{Tg}}$ direction (see Fig.~\ref{fig_BBField}).
Field mapping has shown \cite{lange-general} that the field density
is almost constant inside the magnet, with fringe fields
that decrease exponentially outside of the magnet.
In the analytical model, the true field was approximated
by a constant field within the effective field boundaries,
while edge effects were neglected.  Under these assumptions
all target coordinates were calculated by applying
a circular-arc approximation~\cite{ShneorPhD} of the track
inside the field. The particle transport was divided into
free motion (drift) in the $(y,z)$ plane and circular motion
in the $(x,z)$ plane (see Fig.~\ref{fig_BBAnalyticalModel_BBScheme}), 
described by the Lorentz equation
\begin{eqnarray}
p_{y} = \mathrm{const} \>, \qquad p_{xz}= e R B_y \>. \label{eq_BBAnal0}
\end{eqnarray}
To determine the momentum, the radius $R$ of the trajectory needs
to be calculated first.  This can be done by using the track
information obtained from the detector package, combined
with the geometrical properties of BigBite. 
A few reference points are needed, as shown in Fig.~\ref{fig_BBAnalyticalModel_BBScheme}.
The point $\mathrm{Tg}$ represents the position of the particle
at the target, and $D$ corresponds to the point where the particle
hits the detector package.  The point $B$ at which the particle
exits the magnet is the intersection between the extrapolated
particle track through the detector package and the effective 
exit face of the magnet.  Similarly, the point $A$ lies
at the intersection of the effective entrance face
of the magnet and the particle track from the target.
The point $O$ is the center of the circular trajectory. 
In order for all these points to correspond to a single
particle  track through the spectrometer, the following conditions
must be satisfied:
\begin{eqnarray}
   \overline{A\mathrm{{Tg}}} \perp \overline{AO} \>, \qquad 
   \overline{OB} \perp \overline{BD} \>, \qquad
   |\overline{AO}| = |\overline{BO}| = R \>, \nonumber
\end{eqnarray}
In the target coordinate system, these conditions can be expressed as:
\begin{eqnarray}
  R^2 = (x_B - x_O)^2 + (z_B - z_O)^2 &=& (x_O - X_A)^2 + (z_O - z_A)^2 \,,\label{eq_BBAnal1}
\end{eqnarray}
\begin{eqnarray}
  \frac{x_O - x_A}{z_O - z_A} &=& -\frac{z_{\mathrm{Tg}} - z_A}{x_{\mathrm{Tg}} - x_A} 
                              = \frac{d_{\mathrm{Tg}}}{x_{\mathrm{Tg}} - x_A}\,,\label{eq_BBAnal2}\\
  \frac{x_B - x_O}{z_B - z_O} &=& - \frac{z_{D} - z_{B}}{x_{D} - x_{B}}\,.\label{eq_BBAnal3}
\end{eqnarray}
The coordinates $x_B$ and $z_B$ of $B$, and the coordinates $x_D$
and $z_D$ of $D$ can be directly calculated from the information 
obtained by the detector package.  The position of the target 
$(x_{\mathrm{Tg}},z_{\mathrm{Tg}})$ is known. Since we are using only
elongated targets (dimensions along beam line much longer than transverse dimensions),  
the $x_{\mathrm{Tg}}$ is approximated to be zero. 
It turns out that this constraint significantly simplifies the model.
The coordinate $z_A$ of $A$ corresponds to the known distance $d_{\mathrm{Tg}}$
between the target center and the effective field boundary at the 
entrance to the magnet. The remaining coordinates $x_O$, $z_O$ and
$x_A$ are unknown and will be obtained as results of the analytical 
model. Using Eqs.~(\ref{eq_BBAnal2}) and (\ref{eq_BBAnal3}),
Eq.~(\ref{eq_BBAnal1}) can be written as:

\begin{eqnarray}
   R^2 &=& (z_O - z_A)^2\left[1 + \left(\frac{d_{\mathrm{Tg}}}{x_A} \right)^2 \right]  
        = (z_B - z_O)^2\left[1 
        + \left(\frac{z_D - z_B}{x_D - x_B} \right)^2 \right]\,.\label{eq_BBAnal4}
\end{eqnarray}

An additional relation between the coordinates can also be obtained for the
intersection $O$ between the line segments $\overline{AO}$ and $\overline{OB}$:
\begin{eqnarray}
   x_O &=& -\frac{d_{\mathrm{Tg}}}{x_A}(z_O-z_A) + x_A  
        = -\frac{z_D - z_B}{x_D - x_B}(z_O - z_B) 
        + x_B\>, \label{eq_BBAnal5}
\end{eqnarray}
By expressing $z_O$ from Eq.~(\ref{eq_BBAnal5}) and inserting
it into Eq.~(\ref{eq_BBAnal4}), a cubic equation for $x_A$ is obtained:
\begin{eqnarray}
  x_{A}^3 &+& A_2 x_{A}^2 + A_1 x_{A} + A_0 = 0\,,\nonumber\\
  A_2 &=& \frac{x_D - x_B}{z_D - z_B}(2z_B+d_{\mathrm{Tg}}) - 2x_B\,,\nonumber \\
  A_1 &=& x_3^2 - 2\frac{x_D - x_B}{z_D - z_B}x_B(z_B + z_{\mathrm{Tg}}) -2z_Bz_{\mathrm{Tg}}-z_{B}^2\,,\label{eq_BBAnal6}\\
  A_0 &=& \frac{x_D - x_B}{z_D - z_B}z_{\mathrm{Tg}}(x_B^2 - z_B^2) + 2x_B z_B z_{\mathrm{Tg}}\,.\nonumber
\end{eqnarray}
Equation.~(\ref{eq_BBAnal6}) has three complex solutions in general.  
The physically meaningful result for $x_A$ should be real and lie within
the effective field boundaries.  Two additional physical
constraints are applied.  The particle track should always
represent the shortest possible arc of the circle
(the arc between $A$ and $B$ in Fig.~\ref{fig_BBAnalyticalModel_BBScheme}).
Moreover, the track should bend according to the polarity
of the particle and orientation of the magnetic field.
The procedure of finding the physically meaningful 
solution is described in Ref.~\cite{ShneorPhD}.
The determined solution for $x_A$ can then be used to calculate the 
position of the point $O$:
\begin{eqnarray}
 z_O = \frac{\frac{x_D - x_B}{z_D - z_B}x_A(x_B-x_A) + z_B x_A}{x_A - \frac{x_D - x_B}{z_D - z_B} d_{\mathrm{Tg}}}\,,\qquad
 x_O = x_B + \frac{z_D - z_B}{x_D - x_B}(z_B - z_0)\,. \nonumber
\end{eqnarray}
Introducing these results to Eqs.~(\ref{eq_BBAnal0}) and (\ref{eq_BBAnal4}),
the radius $R$ and the momentum $p_{xz}$ can be calculated.  The particle trajectory length $l_{xz}$
in the $(x,z)$ plane can also be calculated by using
the cosine formula for the angle $\beta = \measuredangle AOB$,
\begin{eqnarray}
  l_{xz} &=&  \sqrt{x_A^2 + d_{\mathrm{Tg}}^2} + R\beta + \sqrt{(x_D-x_B)^2 
          + (z_D - z_B)^2}\>, \nonumber \\
  \cos \beta &=& \frac{\vec{OA}\cdot \vec{OB}}{|\vec{OA}||\vec{OB}|} = \frac{(x_A-x_O)(x_B-x_O) 
          + (z_A-z_O)(z_B-z_O)}{R^2} \> . \nonumber
\end{eqnarray}
By using this information, all target coordinates can finally be expressed as:
\begin{eqnarray}
  \phi_{\mathrm{Tg}} &=& \phi_{\mathrm{Det}} \>,\nonumber \\
  \theta_{\mathrm{Tg}} &=& \arctan\left(\frac{x_A}{d_{\mathrm{Tg}}}\right)  \>,\nonumber \\  
  y_{\mathrm{Tg}} &=& y_{\mathrm{Det}} 
    - l_{xz}\tan\phi_{\mathrm{Det}}\>,  \nonumber \\
  \delta_{\mathrm{Tg}} &=& \frac{p_{xz}}{p_\mathrm{c}}\frac{\sqrt{1
    +\tan^2\phi_{\mathrm{Tg}} + \tan^2\theta_{\mathrm{Tg}}}}{\sqrt{1
    + \tan^2\theta_{\mathrm{Tg}}}} - 1\>,\nonumber \\	
  L &=&l_{xz}\sqrt{1 + \tan^2\phi_{\mathrm{Tg}}}\>, \nonumber	
\end{eqnarray}
where $p_\mathrm{c}$ is the central momentum and $L$ is the 
total flight-path of the particle.

The described analytical method requires just a few geometry parameters,
but these need to be known quite accurately. They could be obtained by the
geodetic survey of the spectrometer and the target. Unfortunately, no such
survey was performed for our experiment. Some geodetic measurements were 
conducted, but none could be exploited in this model. Instead,
the sizes of spectrometer components and the distances between them were 
measured by hand,  and were used to get first estimates of the necessary parameters. 
The final numbers were obtained by calibrating the model 
with elastic events.  However, the solution is not unique.
Different combinations of parameters have been shown to yield
almost identical results for the target variables, while only one 
combination is correct. The final values of the parameters, considered in 
my implementation of the model are gathered in Table~\ref{table_BBAnal}.
\begin{table}[!ht]
\begin{center}
\caption{Parameters required by the analytical model of the BigBite
magnetic optics. The final values of the parameters were obtained
from the calibration with the $2\,\mathrm{GeV}$ elastic hydrogen data.
See also Fig.~\ref{fig_BBAnalyticalModel_BBScheme}.
\label{table_BBAnal}}
\vspace*{2mm}
\begin{tabular}{lc}
\toprule
{\bf Parameter description} & {\bf Value} \\
\midrule
Magnetic field density & $0.92\,\mathrm{T}$ \\
Distance from the target to the collimator & $1130\,\mathrm{mm}$ \\
Distance from target to magnet entrance-face  & $1500\,\mathrm{mm}$ \\
Position of magnet exit-face with respect to the entrance-face & $(795.0, 0.0)\,\mathrm{mm}$\\
Length of BigBite magnet & $1000\,\mathrm{mm}$ \\
Height of BigBite magnet & $1200\,\mathrm{mm}$ \\
Inclination of the magnet exit-face & $20.0^o$ \\
Inclination of the MWDCs & $25.0^o$ \\
Position of the first MWDC with respect to 
the entrance-face  & $(1050, 271)\,\mathrm{mm}$ \\
Distance from the first MWDC to the second MWDC & $757.0\,\mathrm{mm}$ \\
Distance from the first MWDC to the Scintillation planes & $973.0\,\mathrm{mm}$ \\
\bottomrule
\end{tabular}
\end{center}
\end{table} 

\begin{figure}[!ht]
\begin{center}
\includegraphics[width=0.49\textwidth]{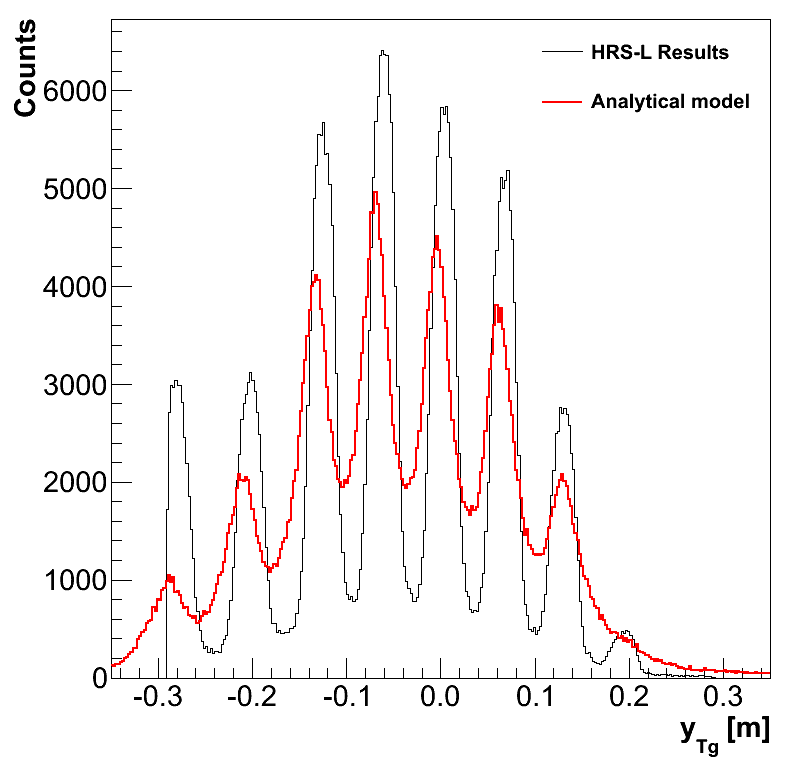}
\includegraphics[width=0.49\textwidth]{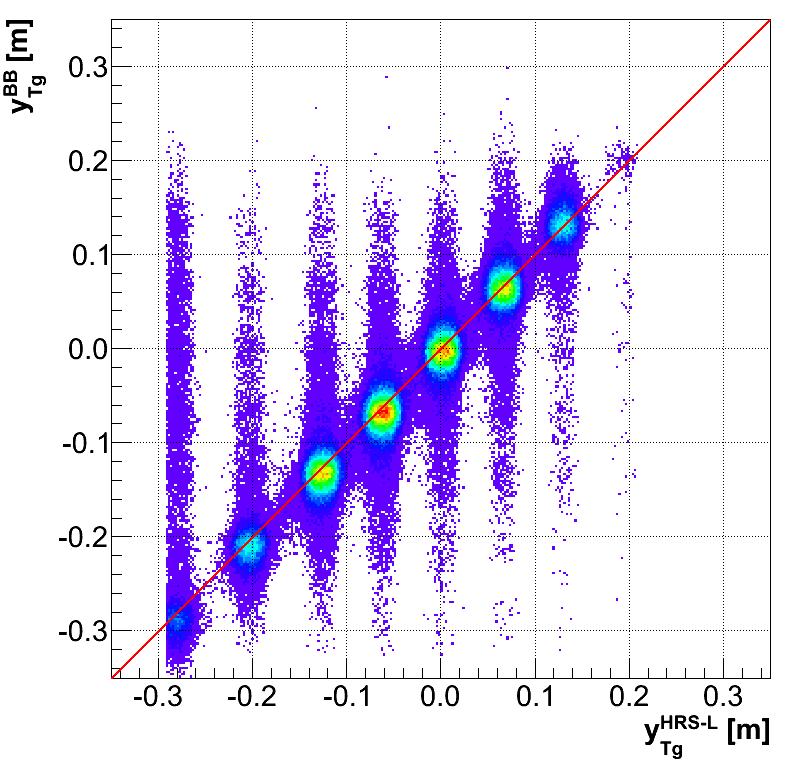}
\vspace*{-3mm}
\caption{[Left] The BigBite target variable $y_{\mathrm{Tg}}$ (the vertex position  
stated in the BigBite coordinate system) reconstructed with the  analytical model 
of BigBite. The peaks of the carbon-optics target can be well 
distinguished. The estimated resolution is $\sigma_{y}^{\mathrm{Tg}} \approx 1.7\mathrm{cm}$. 
For comparison, $y_{\mathrm{Tg}}$ reconstructed by the HRS-L spectrometer is shown. 
[Right] The comparison of the reconstructed $y_{\mathrm{Tg}}$ by using the BigBite 
analytical model, and $y_{\mathrm{Tg}}$ reconstructed by the HRS-L spectrometer 
(rotated from the HRS-L to the BigBite 
coordinate system).\label{fig_BBAnal_TgY}}
\vspace*{-3mm}
\end{center}
\end{figure}

\begin{figure}[!ht]
\begin{center}
\includegraphics[height=0.55\linewidth]{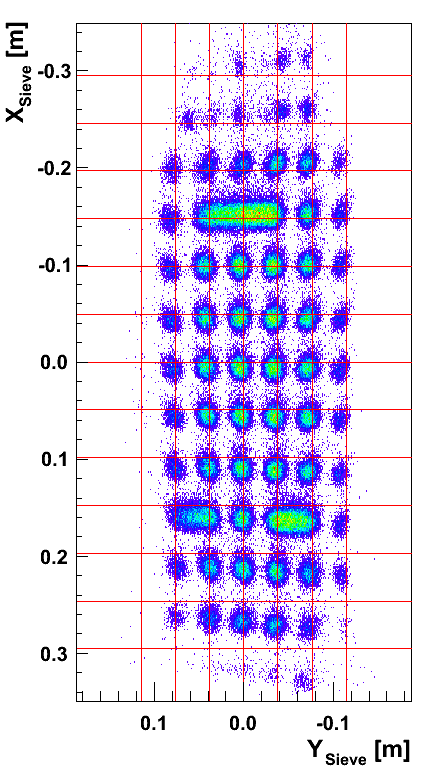}
\hspace*{1cm}
\includegraphics[height=0.55\linewidth]{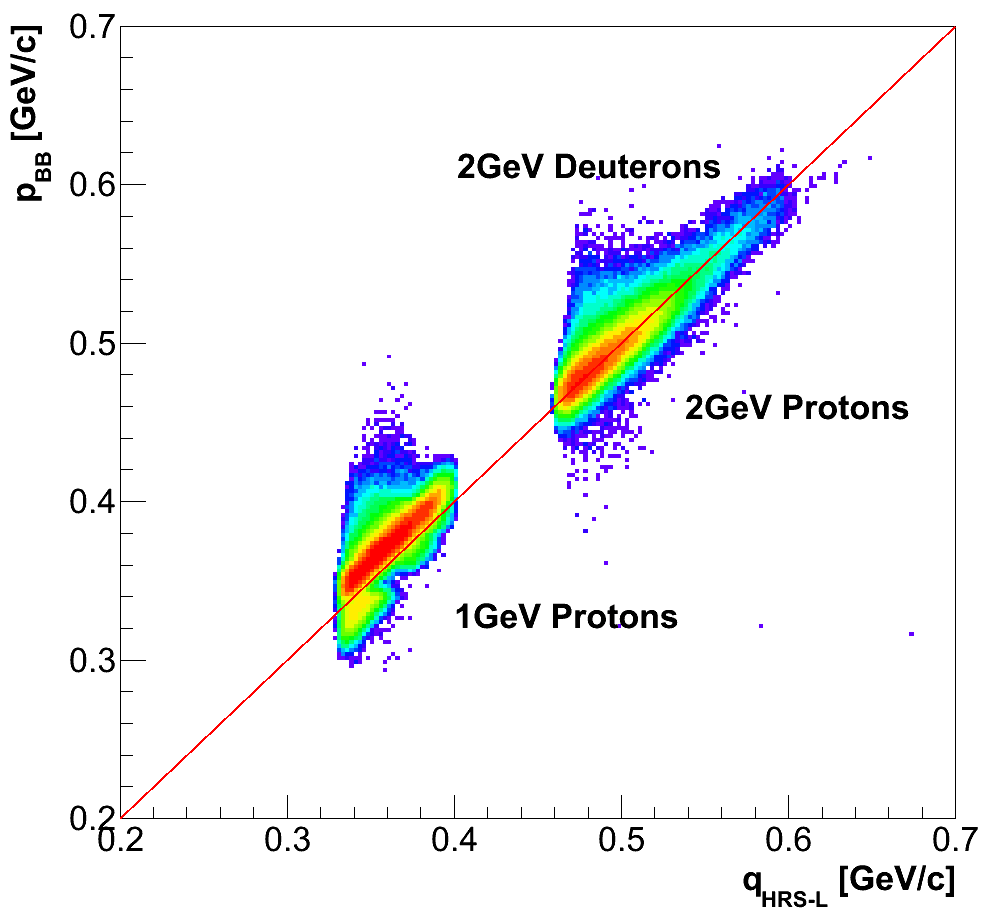}
\caption{[Left] The sieve pattern reconstruction by using the analytical model for
BigBite optics. [Right] The reconstructed BigBite momentum $p_{\mathrm{BB}}$ using the 
analytical model versus the size of the momentum transfer vector $q_{\mathrm{HRS-L}}$
obtained from the HRS-L spectrometer, for elastically scattered protons
and deuterons. The analytical model seems to work well for the $2\,\mathrm{GeV}$
data, while it generates a constant offset when applied to $1\,\mathrm{GeV}$
protons.
 \label{fig_BBAnal_BBSieve}}
\end{center}
\end{figure}

\begin{figure}[!hb]
\begin{center}
\includegraphics[width=0.49\textwidth]{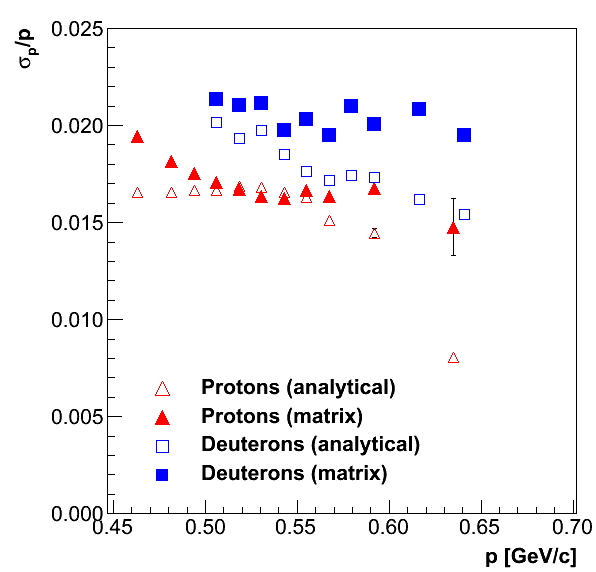}
\includegraphics[width=0.49\textwidth]{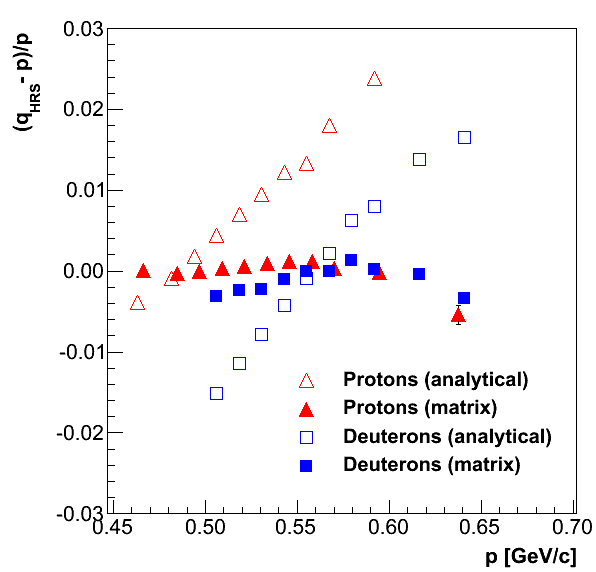}
\caption{
[Left] The relative momentum resolution as function of the momentum
measured by BigBite, obtained by the analytical model. The resolutions obtained
by the matrix approach are also shown. [Right] The absolute calibration of 
$\delta_{\mathrm{Tg}}$ as a function of the particle momentum determined by BigBite.
The relative resolution of $\delta_{\mathrm{Tg}}$
is a bit better in the analytical model than in the matrix method,
but the absolute momentum calibration is inferior to the
matrix approach, except in the narrow region around $p \approx 0.55\,\mathrm{GeV}/c$.
\label{fig_BBAnal_TgDelta}}
\end{center}
\end{figure}

Typical results for the target variable $y_{\mathrm{Tg}}$ obtained with this model 
are shown in Fig.~\ref{fig_BBAnal_TgY}. The resolution 
of $\sigma_y^{\mathrm{Tg}} \approx 1.7\,\mathrm{cm}$
was achieved. The analysis was done for data with multi-foil carbon target and 
considering only coincidence events. The calibration results for 
$\delta_{\mathrm{Tg}}$ are gathered in Figs.~\ref{fig_BBAnal_BBSieve} and 
\ref{fig_BBAnal_TgDelta}. The relative resolution obtained with this model is 
comparable or even better than the resolution obtained by the matrix 
approach (see Sec.~\ref{sec:opticsmatrixformalism}). 
The resolution for both protons and deuterons was estimated
to be better than $\sigma_p/p  = 2\,\mathrm{\%}$. However, the absolute 
momentum calibration determined with the analytical model is inferior to the 
matrix approach. Instead of being constant the difference $(q_{\mathrm{HRS-L} - p})/p$ 
drifts over the whole momentum acceptance. 

\begin{figure}[!ht]
\begin{center}
\includegraphics[width=0.49\textwidth]{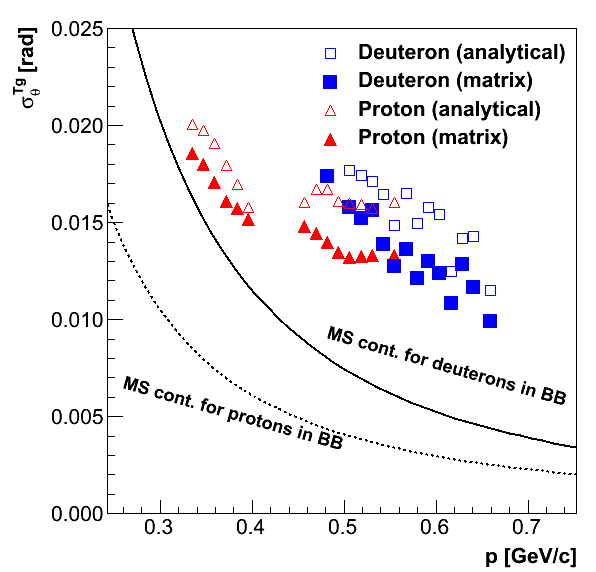}
\includegraphics[width=0.49\textwidth]{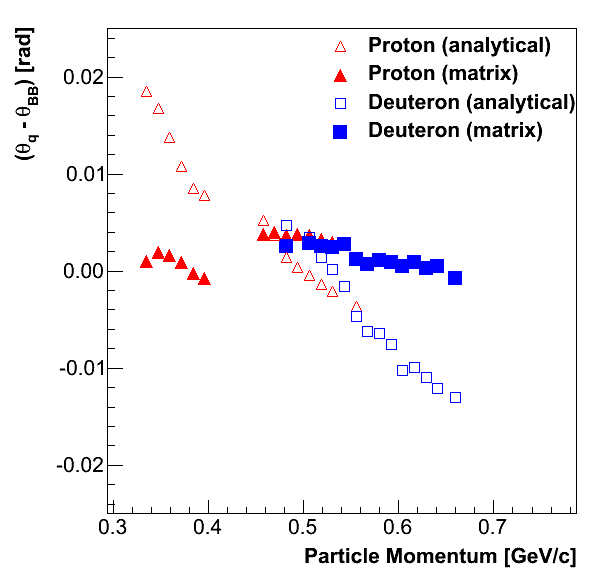}
\caption{[Left] The absolute resolution (sigma) of $\theta_{\mathrm{Tg}}$  as a function
of the momentum measured by BigBite, obtained by the analytical model. For
comparison, the resolutions obtained by the SVD method are also shown. Irreducible
multiple-scattering contributions, mostly due to the air within BigBite, are shown by 
full and dashed lines for deuterons and protons, respectively.
[Right] The absolute calibration of $\theta_{\mathrm{Tg}}$ as a function of the particle
momentum measured by BigBite. The figure shows that the absolute calibration 
obtained by the analytical model is inferior to the matrix approach.
\label{fig_BBAnal_TgTh}}
\end{center}
\end{figure}

\begin{figure}[!hb]
\begin{center}
\includegraphics[width=0.49\textwidth]{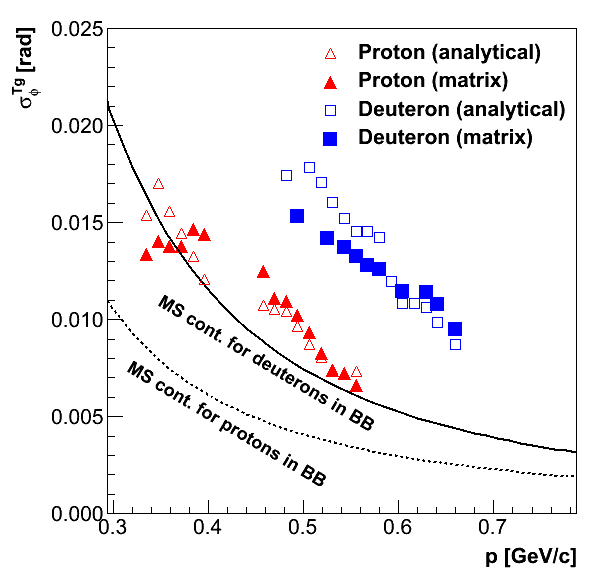}
\includegraphics[width=0.49\textwidth]{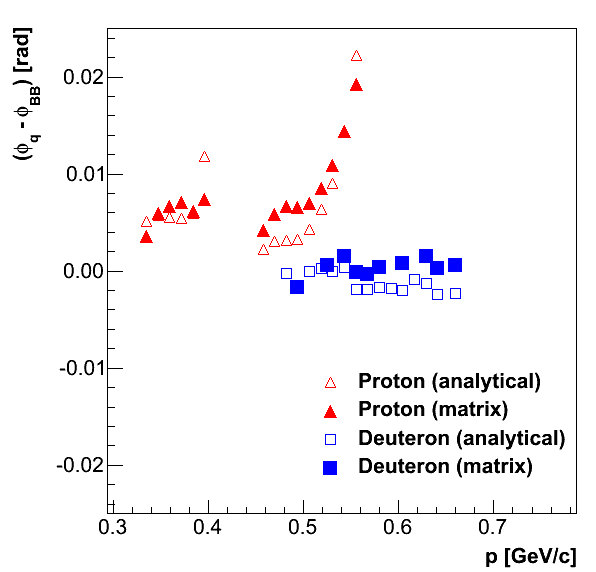}
\caption{[Left] The absolute resolution (sigma) of $\phi_{\mathrm{Tg}}$  as a function
of the momentum measured by BigBite, obtained by the analytical model. For
comparison, the resolutions obtained by the SVD method are also shown. Irreducible
multiple-scattering contributions, due to the air within BigBite, are also shown.
[Right] The absolute calibration of $\phi_{\mathrm{Tg}}$ as a function of the particle
momentum measured by BigBite. A rapid deterioration of the proton resolution in both
approaches at momenta $p \geq 0.5\,\mathrm{GeV}/c$ is due to the target edge effects:
the optics is expected to become faulty for events coming from the ends of the target. 
\label{fig_BBAnal_TgPh}}
\end{center}
\end{figure}

The calibration results for the in-plane angle $\phi_{\mathrm{Tg}}$ and the 
out-of-plane angle $\theta_{\mathrm{Tg}}$
are shown in Figs.~\ref{fig_BBAnal_BBSieve}, 
\ref{fig_BBAnal_TgTh} and \ref{fig_BBAnal_TgPh}. The absolute resolutions for
both angles increase with the increasing momentum, resulting in widths
$\sigma_{\theta}^{\mathrm{Tg}} \approx 17\,\mathrm{mrad}$ and
$\sigma_{\phi}^{\mathrm{Tg}} \approx 10\,\mathrm{mrad}$ for $0.5\,\mathrm{GeV}/c$
protons. The absolute calibration is also not stable. Similar 
behavior is observed as for the $\delta_{\mathrm{Tg}}$ variable.
Additionally, Fig.~\ref{fig_BBAnal_TgPh} shows a a rapid deterioration 
of the absolute calibration for protons with momenta 
$p \geq 0.5\,\mathrm{GeV}/c$. The same behavior is observed for analytical model
as well as for the matrix approach (see Sec.~\ref{sec:opticsmatrixformalism}). 
This happens because elastic protons 
mainly come from the up-stream end 
of the target, where the optical calibration is expected to start failing. 
This is mostly due to the effects of fringe-fields, which are strongest on the 
edges of the BigBite magnet. A great influence of the fringe-fields on the 
edges of the magnet is evident also 
in the reconstructed sieve-pattern, shown in Fig.~\ref{fig_BBAnal_BBSieve}.
The elastic deuterons of the same momenta, on the other hand, are limited by 
different kinematics conditions and are ejected from the center of the target. 
There the optics works best, which results in a stable absolute 
calibration for the deuterons.

\begin{figure}[!ht]
\begin{center}
\begin{minipage}[t]{0.6\textwidth}
\hrule height 0pt
\includegraphics[width=1\textwidth]{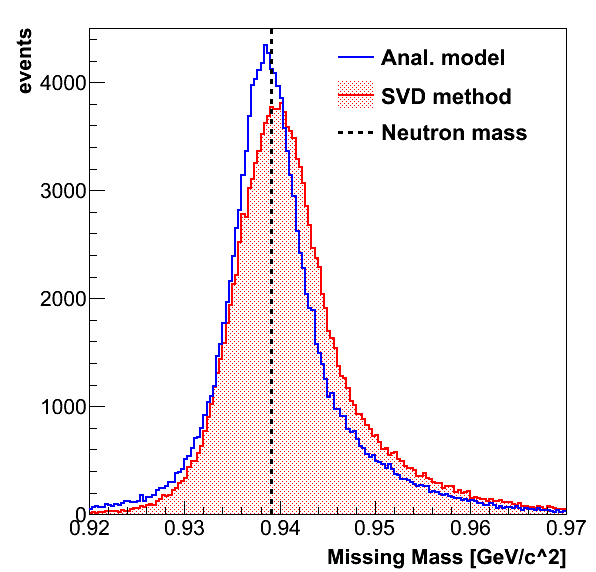}
\end{minipage}
\hfill
\begin{minipage}[t]{0.35\textwidth}
\hrule height 0pt
\caption{The reconstructed mass of the undetected neutron
(missing mass) from the process $\mathrm{{}^2H(e,e'p)n}$
by using the analytical model and the matrix-formalism (SVD) approach
described in Sec.~\ref{sec:opticsmatrixformalism} (see also Fig.~\ref{fig_optics_MissingMassPlot}). 
The width (sigma) of the peak determined with the analytical model
is $3.3\,\mathrm{MeV}/c^2$ (corresponding to $0.35\,\mathrm{\%}$
relative resolution).  The width of the peak reconstructed
by the SVD method is $4\,\mathrm{MeV}/c^2$.
 \label{AnalMissingMassPlot}}
\end{minipage}
\end{center}
\end{figure}

By the analytical approximation of  BigBite,  
resolutions of a few percent can be achieved, but they deteriorate
when moving towards the edges of the acceptance where the fringe fields begin 
to affect the optics. This is particularly true for $\phi_{\mathrm{Tg}}$.
Figure~\ref{AnalMissingMassPlot} (left) shows the reconstructed mass of 
the neutron from the process $\mathrm{{}^2H(e,e'p)n}$, obtained by using
the analytical model.  The relative resolution is $0.35\,\mathrm{\%}$.

\vspace*{5cm}
{}

\chapter{EDTM and Cosmics Checks}
\label{appendix:EDTM}

The simulated pulses generated by the Event Dead Time Monitors (see
Sec.~\ref{sec:EDTM}) were exploited to check the basic properties
and settings of the trigger system before the real experiment started. 
In particular, we were interested in time differences between triggers
entering the Trigger Supervisor (TS) and scaler modules, the widths and positions
of the coincidence windows, and the performance of the BigBite re-timing
circuit. To test all these properties, we utilized a two-channel Tektronix 
Oscilloscope TDS2012B. It supports an Ethernet connection and can be controlled
remotely via HTML interface. In this manner we were able to record the plots
of the pulses measured at different locations in the trigger circuit. From the
comparison of the results obtained from these plots with the original trigger 
schemes shown in Figs.~\ref{fig_BBT1trigger}-~\ref{fig_BBRetiming} we were 
then able to determine if the trigger system operates properly. 
\begin{figure}[!ht]
\begin{center}
\includegraphics[width=0.48\linewidth]{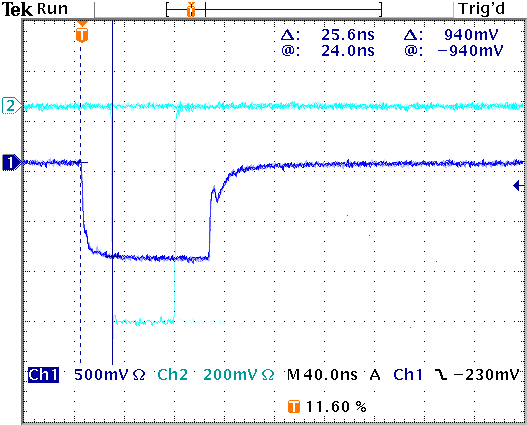}
\hfill
\includegraphics[width=0.48\linewidth]{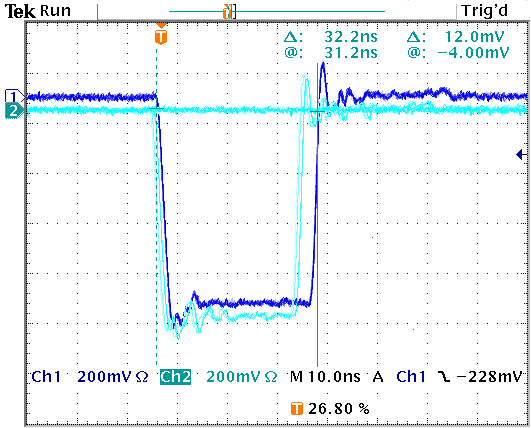}
\caption{[Left] The time difference between the T1 trigger (cyan) and the T3 trigger (blue) at the 
input to the coincidence circuit. The shape of the T3 pulse is deformed by the long cable used
to transport T3 from HRS-L to BigBite weldment. The T3 pulse is refreshed before
forming the coincidence window. [Right] The time difference between T1 (cyan) and T3 (blue) 
at the input to the TS and scaler modules. 
\label{fig_EDTM_TriggerT1}}
\end{center}
\end{figure}

Figure~\ref{fig_EDTM_TriggerT1} (left) shows the comparison of T1 and T3 triggers
at the input to the coincidence circuit 
(see Fig.~\ref{fig_BBCoincTrigger}). The T3 comes to the  circuit
$\approx 26\,\mathrm{ns}$ before T1. This gives it enough time to form a 
coincidence window, and wait for T1 to form the coincidence trigger T5.  
The T5 trigger will therefore be timed relative to T1.
Figure~\ref{fig_EDTM_TriggerT1} (right) shows the T1 and T3 triggers at the input to 
the TS and scalers. Both triggers come to the modules simultaneously 
(time difference is smaller than $1\,\mathrm{ns}$). This is required
for the proper operation of the trigger supervisor. On the contrary
the trigger coming first to the TS would  be in a privileged position 
to be accepted.
\begin{figure}[!ht]
\begin{center}
\includegraphics[width=0.48\linewidth]{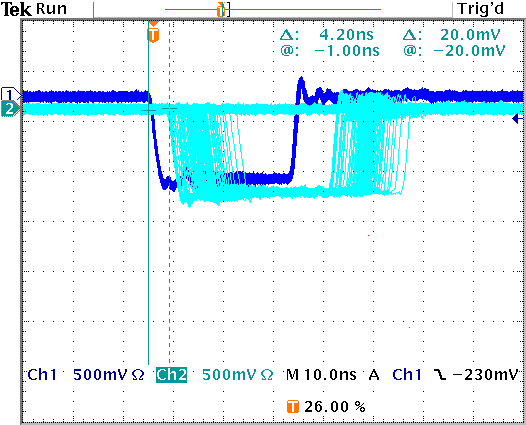}
\hfill
\includegraphics[width=0.48\linewidth]{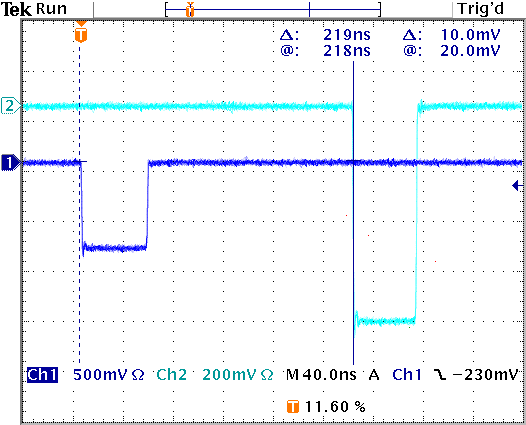}
\caption{[Left] The time difference between T1 (blue) and T2 (cyan) at the input
to the scaler modules. [Right] The time difference between two pulses
represents the time delay considered in the T1 circuit. The delay
was set to $219\,\mathrm{ns}$. 
\label{fig_EDTM_TriggerT2}}
\end{center}
\end{figure}

In addition to EDTM pulses, cosmic rays were considered for testing. 
Cosmic rays can be only used to test single triggers (T1, T2, T3), since
they can not produce systematic coincidence events. 
Figure~\ref{fig_EDTM_TriggerT2} (left) demonstrates the time difference between
the T1 and T2 triggers  at the input to the scaler modules, obtained 
with the use of cosmic rays passing the BigBite detector package. 
By design, T2 comes to the modules after the T1. This is 
accomplished by properly setting the delay and guarantees that 
the primary trigger (T1) will be taken for timing. The time difference between the 
triggers is not constant but ranges from $4.2\,\mathrm{ns}$ 
to $18\,\mathrm{ns}$. This spread is caused by different amounts of time
required by cosmic particles with different momenta to pass both scintillation
detectors.

After it has been formed, the T1 trigger is taken through some additional delay  
to wait for the T3 trigger to arrive from HRS-L. See Sec.~\ref{sec:TriggerSetup}
for detailed explanation.  The precise amount of delay needed is obtained by
 delay cables and two programmable delay modules. 
Figure~\ref{fig_EDTM_TriggerT2} (right) shows that the total amount 
of delay was  $219\,\mathrm{ns}$. 
\begin{figure}[!ht]
\begin{center}
\includegraphics[width=0.48\linewidth]{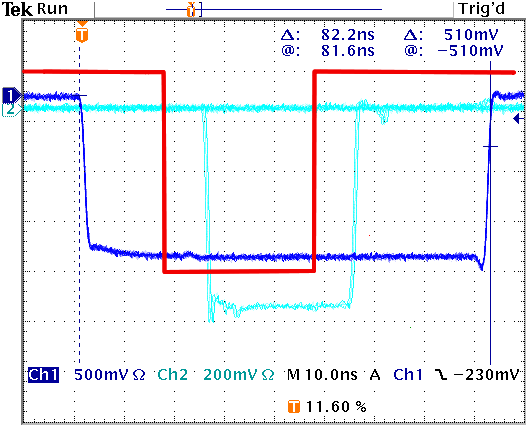}
\hfill
\includegraphics[width=0.48\linewidth]{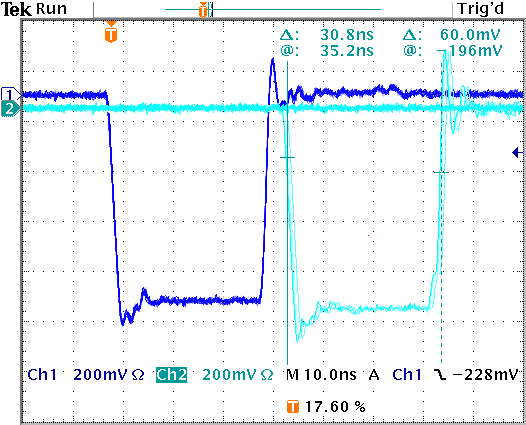}
\caption{[Left] The coincidence trigger T5 (cyan) relative to the coincidence window 
made of T3 (blue). The red line represents the T1 trigger. Its position was
obtained from Fig.~\ref{fig_EDTM_TriggerT1} and T1 circuit diagram presented 
in Fig.~\ref{fig_BBT1trigger}. [Right] The time difference between T3 (blue)
and T5 (cyan) at the input to the TS and scalers. T5 comes to the trigger supervisor
$\approx 34\,\mathrm{ns}$ after the T3. 
\label{fig_EDTM_TriggerT5}}
\end{center}
\end{figure}

The performance of the coincidence trigger T5 is shown in 
Fig.~\ref{fig_EDTM_TriggerT5} (left). The T3 trigger first opens a coincidence window
that is $\approx 82\,\mathrm{ns}$ wide. 
According to EDTM tests, T1 comes $18\,\mathrm{ns}$ later
and together they form the coincidence trigger T5 which appears 
$\approx 26\,\mathrm{ns}$ after the T3 window opens. This agrees well with
the coincidence trigger design shown in Fig.~\ref{fig_BBCoincTrigger}. 
T1 comes to the coincidence circuit $\approx 26\,\mathrm{ns}$ after T3 (see
Fig.~\ref{fig_EDTM_TriggerT1}). Considering also some additional electronics that is
necessary to form the coincidence window and logic AND between the T1 and T3 triggers, 
this results in $26-8-1+8 = 25\,\mathrm{ns}$ delay of T5 with respect to T3.
The results of the EDTM tests for the secondary coincidence trigger T6 are almost 
identical to those presented here.

\begin{figure}[!ht]
\begin{center}
\includegraphics[width=0.48\linewidth]{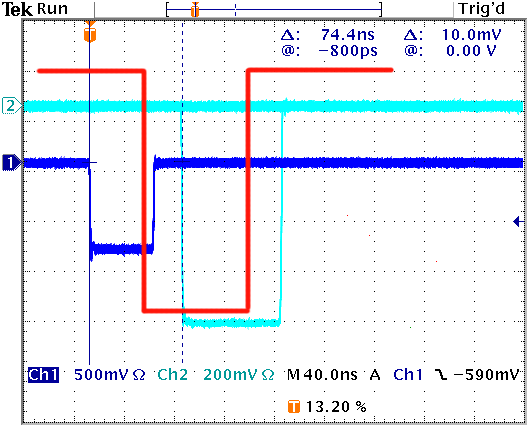}
\hfill
\includegraphics[width=0.48\linewidth]{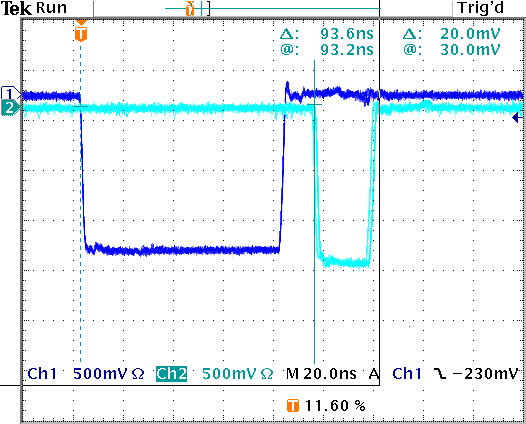}
\caption{[Left] Time difference between T1 (blue) and
L1A pulse (cyan) for the case when Trigger Supervisor (TS) accepts the coincidence
trigger T5. The position of the L1A pulse when TS accepts
T1 or T3 is shown in red. [Right] The time difference between the L1A pulse 
and the delayed L1A pulse which is used as a supplementary re-timing pulse in the case
when neither T1 nor T2 are present. The time delay for the secondary pulse is set 
to $94\,\mathrm{ns}$.
\label{fig_EDTM_TriggerL1A}}
\end{center}
\end{figure}

After the coincidence triggers are created, they are coupled to the trigger supervisor
and scaler modules, together with the rest of the triggers. However, due to the additional
electronics, these two triggers come to the TS $\approx 35\,\mathrm{ns}$ after the
single-arm triggers. This additional delay is presented in 
Fig.~\ref{fig_EDTM_TriggerT5} (right) and is a result of an additional electronics necessary
to form the coincidence triggers. The delayed triggers also lead to the delayed L1A pulse returned 
by the trigger supervisor. This is demonstrated in Fig~\ref{fig_EDTM_TriggerL1A} (left).

The triggers at the input to the TS could in principle be aligned by adding more
delay to the single-arm triggers. However, we decided not to do that, because too much 
delay could cause the ADC gate to open too late. Fortunately these time differences between 
the single-arm and coincidence triggers do not cause serious problems. 
The only consequence is that some of the coincidence events are recorded to the data
stream as single-arm events. This happens because T5 may come too late for the TS to 
recognize the coincidence event,  although both T1 and T3 are present. 
However, we can overcome this problem
by looking at the trigger TDC spectra, where all trigger information is stored. 
Using that information we can reconstruct such "hidden" coincidences.

For events (HRS-L single arm (T3) or pulser (T8) events), where BigBite triggers  
are not present, the L1A pulse can not be timed off T1 or T2. For these events the re-timing circuit 
(see Fig.~\ref{fig_BBRetiming}) uses the delayed L1A pulse to open the ADC gate and read the TDCs. 
According to the EDTM check presented in Fig.~\ref{fig_EDTM_TriggerL1A}, the primary L1A
pulse is $\approx 80\,\mathrm{ns}$ long. This means that if TS accepts T3, we are willing 
to wait for approximately $\approx 80\,\mathrm{ns}$ for BigBite triggers to come. 
If they are still missing after $94\,\mathrm{ns}$ (time difference between primary 
and delayed L1A pulse), the circuit stops waiting, accepts the delayed L1A pulse, 
and starts reading the ADCs and the TDCs.
\begin{figure}[!ht]
\begin{center}
\begin{minipage}[t]{0.6\textwidth}
\hrule height 0pt
    \includegraphics[width=\linewidth]{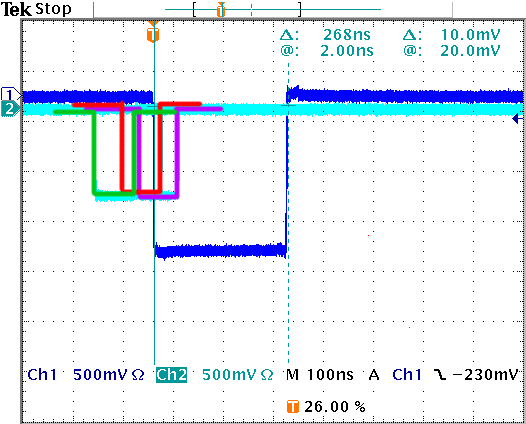}
\end{minipage}
\hfill
\begin{minipage}[t]{0.32\textwidth}
\hrule height 0pt
\caption{The time difference between BigBite re-timing pulse (blue) and L1A (cyan). Combination 
of different triggers and different decisions of the TS result in three L1A peaks displayed 
in violet, red and green. The width of the BigBite re-timing window, i.e. the width of 
the ADC gate, is $\approx 268\,\mathrm{ns}$. 
\label{fig_EDTM_BBRetime}}
\end{minipage}
\end{center}
\end{figure}

At the output from the re-timing circuit, which is explained in Fig.~\ref{fig_BBRetiming}, a
BigBite re-time pulse if formed. This pulse is then used directly as the gate for ADC and 
TDC modules. The width of the gate signal was set to be $\approx 268\,\mathrm{ns}$. 
As demonstrated in this chapter, different triggers come to the TS at different times, 
resulting in more than one position of the L1A with respect to the BigBite re-time pulse. 
This can also be clearly seen in Fig.~\ref{fig_EDTM_BBRetime}. The EDTM test results in 
three different L1A peaks. The red peak corresponds to the events where 
the trigger supervisor accepts the single-arm triggers (T1, T2, T3, T4). The position of the  
closest, violet peak agrees with the events where TS accepts coincidence triggers. Since
T5 and T6 come $\approx 35\,\mathrm{ns}$ after single-arm triggers, this results in a smaller 
time difference between the L1A and the BigBite re-time pulse (which is relative to T1). The 
leftmost, green peak corresponds to HRS-L single events, where the circuit needs to wait for 
the delayed L1A signal in order to create the BigBite re-time pulse.

\chapter{All about EVe}

\section{Introduction}

The {\bf E}vent {\bf V}i{\bf e}wer (or EVe) was written in 2008 to visualize events detected by the 
BigBite spectrometer. It is based on the CERN Root data analysis framework~\cite{CERNRoot},
but it is not a part of the standard Root's event viewer Eve. It is a separate code which
only uses Root's graphics packages (the Geometry package and GUI classes). 

\begin{figure}[!ht]
\begin{center}
\includegraphics[width=\linewidth]{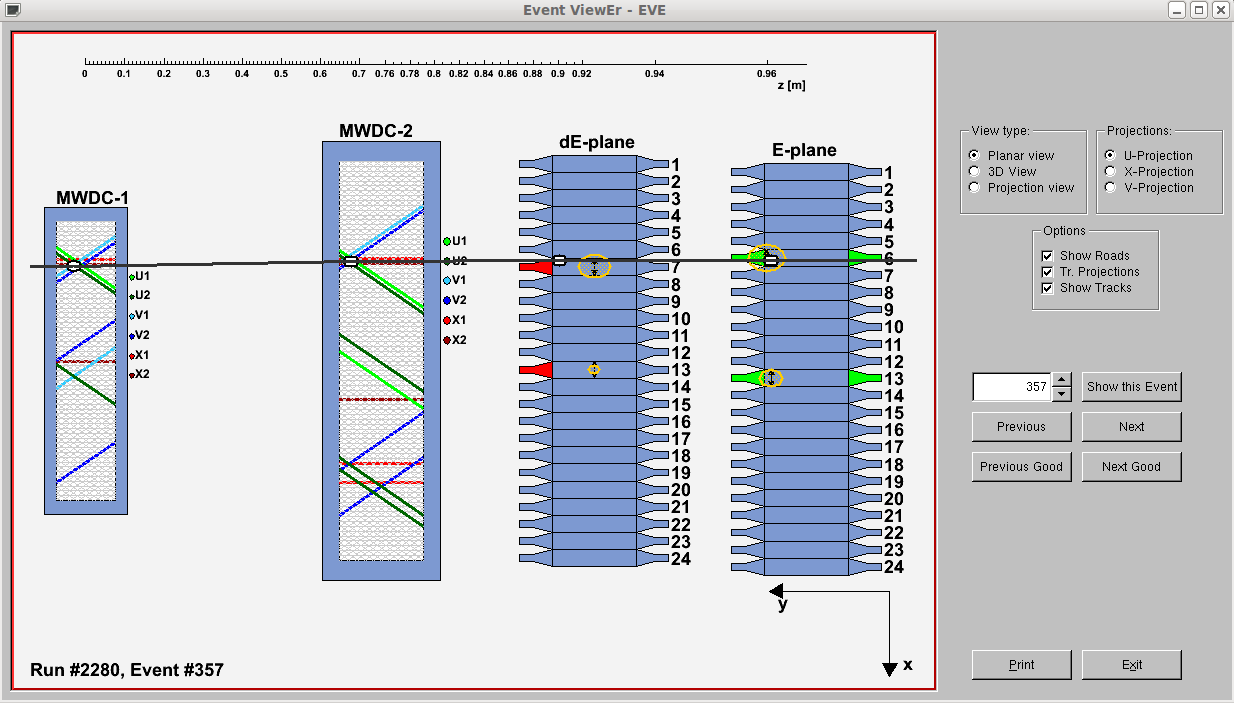}
\caption{The graphical user interface (GUI) of the BigBite event display EVe, showing 
the two-dimensional (planar) view of the Hadron detector package. The figure shows hits 
in the MWDCs and scintillation detectors E and dE for a selected event. The reconstructed 
track through the detectors is also shown. 
\label{fig_EVe_GUI}}
\end{center}
\end{figure}

EVe was developed to help debugging the BigBite tracking algorithm~\cite{OleThreeSearch} 
in its development phase. Bugs in the code could be discovered much easier if the hits 
in the MWDCs could be visualized together
with the corresponding reconstructed particle tracks. The event display also turned out to
be useful in the commissioning phase of the spectrometer for finding errors in the 
operation of the spectrometer and for adjusting the parameters of the reconstruction code. 

Presently EVe has the ability to show hit wires in the MWDCs and hits in the 
dE and E scintillation planes (see Fig.~\ref{fig_EVe_GUI}). Besides coloring the 
PMTs with non-zero signals, it also shows the position of the hit at the scintillation
detector, which is determined from the time difference between two hits. Additionally
it also displays the amount of accumulated charge in each plane, which is closely related 
to the particle momentum. For events possessing consistent hits in all detectors,
the reconstructed particle tracks through the detector package are shown.

\begin{figure}[!ht]
\begin{center}
\includegraphics[height=0.5\linewidth]{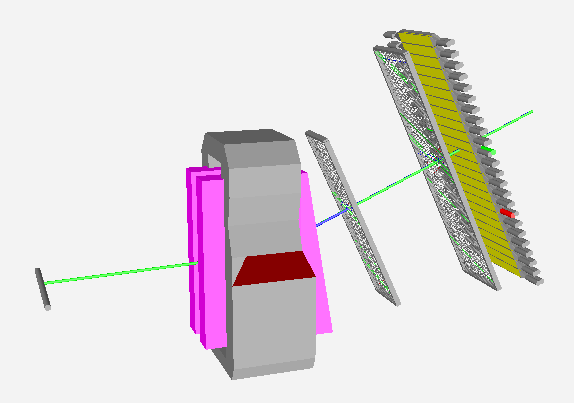}
\hfill
\includegraphics[height=0.5\linewidth]{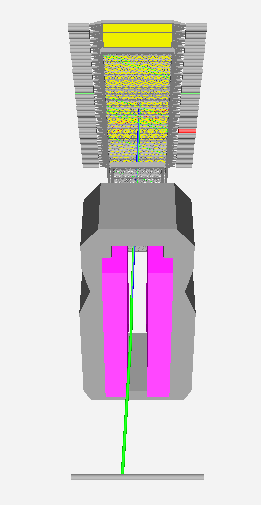}
\caption{The three-dimensional view of the BigBite event display
EVe, showing a fully reconstructed particle track from the target
to the detector package. Hit wires in the MWDCs and hit scintillation
paddles are illuminated distinctly according to the hit pattern.
\label{fig_EVe_3DView}}
\end{center}
\end{figure}

EVe supports three-dimensional and two-dimensional (planar) views of the detector package
(see Fig.~\ref{fig_EVe_3DView}), which can be selected and controlled  
through the graphical user interface (GUI). It also features a projection view, 
to display roads and track projections in the MWDCs in all three wire orientations ($U$, $X$, $V$).
This option was used for the debugging of the BigBite analysis library, and for each 
event provides a detailed information on the formation of the particle track 
(see Sec.~\ref{sec:MWDCs}).

\begin{figure}[!ht]
\begin{center}
\begin{minipage}[!ct]{0.6\textwidth}
    \includegraphics[width=\linewidth]{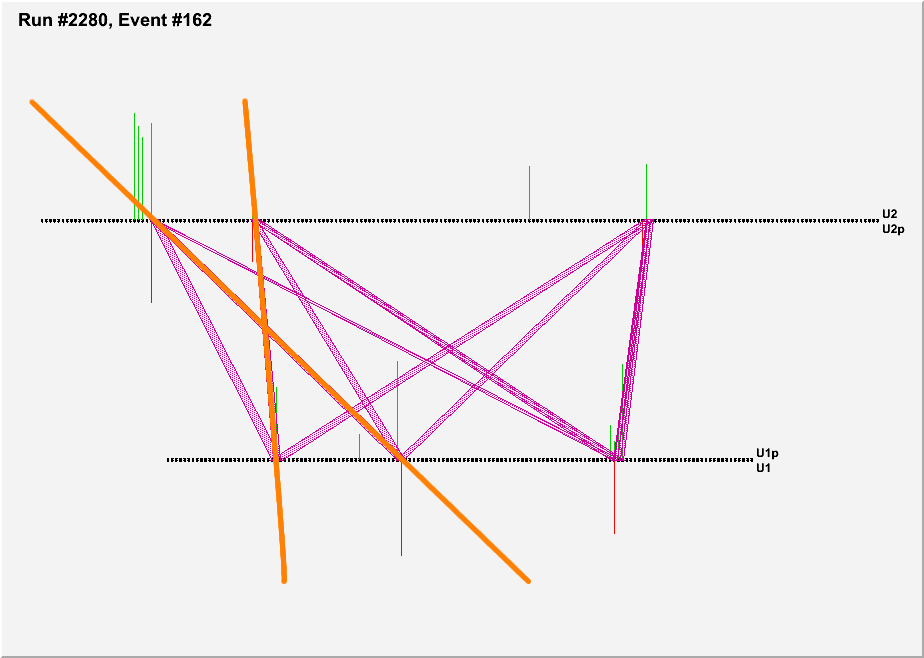}
\end{minipage}
\hfill
\begin{minipage}[!ct]{0.32\textwidth}
\caption{The planar view of the BigBite event display EVe, showing the 
u-projection of the MWDCs. Each projection combines four wire planes 
(e.g. u1, u1p, u2, u2p). The wire hits in each plane are shown with red 
and green vertical lines. The length of a line corresponds to the 
distance from the wire. Hits from all wires are introduced to the 
Pattern Match Tree Search algorithm (see Sec.~\ref{sec:MWDCs})
which finds possible one-dimensional tracks (orange), also 
known as roads. The roads from all three projections (x, u, v) 
are then combined into three-dimensional 
tracks.
\label{fig_EVe_Projection}}
\end{minipage}
\end{center}
\end{figure}

\section{Using EVe}
The source codes of the event display can be downloaded from the link in Ref.~\cite{EVe}.
The package \texttt{EVe.tgz} contains all necessary files. The program can be built by
using the command \texttt{make} inside the directory that contains the extracted source files. 
After the successful installation, a shared library \texttt{libEVe.so} is generated,  
which contains all objects. To run the event display, either 
JLab's analyzer or CERN Root can be used. EVe uses replayed root files obtained from 
the raw datafile analysis. Hence, it does not require any additional libraries 
(e.g. \texttt{libBigBite.so}) to operate. However, the CERN Root library 
\texttt{libGeom.so} must be loaded before using EVe. An analysis file \texttt{simEVe.C} is 
is added to the \texttt{EVe.tgz} package, which already contains the minimal required root script
and can be immediately used to display events. A typical analysis script should have the
following structure:
\begin{verbatim}
#include <TGClient.h>
gSystem->Load("libGeom");
gSystem->Load("libEVe.so");
EVe *sim = new Eve(gClient->GetRoot(),400,700);
sim->initRun(rootfile); 
\end{verbatim} 
In order to display hits in BigBite detectors, the \texttt{rootfile} introduced to the 
event display needs to contain all  necessary detector variables. 

\section{Modifying the code}

EVe is a collection of C++ classes employed to describe BigBite particle detectors (e.g.
Scintillation detector ) or their parts (e.g. Scintillator paddle). They are then merged
in the main class \texttt{EVe.cxx}, which reads all necessary data from the 
root files and sets the variables of detector objects accordingly so that all hits in a 
given event get shown on the screen. This class is not universal, but depends of the 
detector configuration. The current version of the event display is dedicated for the 
experiments that utilize the hadron detector package (e.g. E05-102 and E04-007).
To use EVe with the electron detector package, the file \texttt{EVe.cxx} must be modified 
to define the corresponding detectors. 
The modular design of the code allows EVe to be easily accommodated also for any other 
BigBite detector configuration or the HRS's detector packages, since these spectrometers 
possess very similar detectors. At the moment only scintillation and wire-chamber detectors 
are implemented in the code. 
The geometrical properties of the detectors together with their 
positions and orientations are stored in the database file \texttt{EVe\_DB.dat}. When changing 
the spectrometer configuration this file needs to be modified as well.


\chapter{Povzetek v slovenskem jeziku}

V doktorskem delu sem raziskoval spinsko-izospinsko zgradbo polariziranih 
jeder ${}^3\mathrm{He}$ z meritvijo dvojnopolarizacijskih asimetrij v kvazielasti\v{c}nih 
procesih 
${}^3\vec{\mathrm{He}}\left(\vec{\mathrm{e}},\mathrm{e'} \mathrm{d }\right)$ in 
${}^3\vec{\mathrm{He}}\left(\vec{\mathrm{e}},\mathrm{e'} \mathrm{p}\right)$, ki smo 
jo izvedli leta 2009 v okviru eksperimenta E05-102~\cite{e05102} v Thomas Jefferson 
National Accelerator Facility (TJNAF) v Zdru\v{z}enih dr\v{z}avah Amerike.  



Osnovne lastnosti protona so danes dobro znane in natan\v{c}no izmerjene. Strukturo nevtrona 
poznamo bistveno slab\v{s}e. Porazdelitev naboja, magnetnega momenta in spina nevtrona so 
znani le pribli\v{z}no, saj neposredna meritev ni mogo\v{c}a, ker nevtronske tar\v{c}e ni. 
Lastnosti nevtrona zato dolo\v{c}amo posredno. Najve\v{c}krat za raziskovanje njegove strukture 
uporabimo sipanje elektronov na devteronu, kjer privzamemo, da je zaradi majhne vezavne 
energije devterona nevtron skoraj prost. Kot efektivno pola\-rizirano nevtronsko tar\v{c}o pa 
lahko uporabimo tudi polarizirana jedra ${}^3\mathrm{He}$. Do kolik\v{s}ne mere in s kak\v{s}nimi popravki 
tak pribli\v{z}ek velja, je odvisno od razumevanja zgradbe ${}^3\mathrm{He}$~\cite{e05102}. 

Natan\v{c}no poznavanje strukture ${}^3\mathrm{He}$ je odlo\v{c}ilnega pomena 
predvsem za prihajajo\v{c}e eksperimente 
posve\v{c}ene raziskovanju nevtrona. Slika~\ref{fig_A1n} prikazuje 
predvidene napake eksperimenta E12-06-122~\cite{e12-06-122}, kjer bodo vnovi\v{c} merili spinsko asimetrijo 
nev\-trona, z \v{z}eljo po bolj\v{s}em razumevanju njegove kvarkovske strukture. 
Ocenjena statisti\v{c}na negotovost meritve je primerljiva ali celo manj\v{s}a od negotovosti
polarizacije protona in nevtrona znotraj ${}^3\mathrm{He}$. Uspe\v{s}nost
eksperimenta je tako neposredno odvisna od tega, kako dobro razumemo ${}^3\mathrm{He}$.

Jedro ${}^3\mathrm{He}$ pa je zelo zanimivo tudi samo po sebi. \v{C}eprav sestoji le iz treh nukleonov, 
je njegova struktura dovolj bogata, da je z njim mo\v{c} raziskovati pojave, ki so navadno 
prisotni pri jedrskih reakcijah s te\v{z}jimi jedri. Po drugi strani pa je \v{s}e vedno dovolj 
preprost, da ga je mogo\v{c}e ra\v{c}unsko obvladati, kar omogo\v{c}a, da teoreti\v{c}ne 
napovedi o strukturi jedra primerjamo z izmerjenimi podatki. Tako lahko natan\v{c}no 
preverimo na\v{s}e razumevanje jedrskih sil med nukleoni ter zgradbe ${}^3\mathrm{He}$
~\cite{sircaWL2009, golak2005}.

Teoreti\v{c}ni izra\v{c}uni vezanega stanja dveh protonov in nevtrona ka\v{z}ejo, da v 
valovni funkciji osnovnega stanja ${}^3\mathrm{He}$ prevladujejo tri 
komponente (glej sliko~\ref{fig_He3States})~\cite{blankleider, afnan}. 
Prva je prostorsko simetri\v{c}no stanje $S$, kjer sta spina protonov antiparalelna. 
V tem stanju je spin ${}^3\mathrm{He}$ dolo\v{c}en s spinom nevtrona. To je prevladujo\v{c}a 
komponenta osno\-vnega stanja, z verjetnostjo okrog $90\,\%$. Dodatnih $8\,\%$ k valovni 
funkciji prispeva stanje D, ki je posledica tenzorske sile med nukleoni. 
V tem primeru so spini vseh treh nukleonov orientirani v nasprotno smer kot spin 
jedra ${}^3\mathrm{He}$. Preostala $2\,\%$ v valovni funkciji predstavlja stanje z me\v{s}ano 
simetrijo $S'$, ki se pojavi, ker so dvodel\v{c}ne jedrske sile med nukleoni za razli\v{c}ne 
spine in izospinski stanji $T=1$ in $T=0$ razli\v{c}ne. Posledica tega je me\v{s}ana simetrija 
stanja tako v prostorskem  kot tudi v spinsko-izospinskem delu valovne funkcije~\cite{blatt}. 
Razumevanje vpliva stanj $D$ in $S'$ v ${}^3\mathrm{He}$ je pomembno pri razumevanju ve\v{c}del\v{c}nih 
sistemov.

Razpolo\v{z}ljivi teoreti\v{c}ni izra\v{c}uni razli\v{c}nih teoretskih skupin slonijo na 
re\v{s}evanju zahtevnih nerelativisti\v{c}nih integralskih ena\v{c}b Faddeeva za vezano in 
kontinuumsko stanje~\cite{glockle2004, golak2005} ter upo\v{s}tevajo \v{s}irok spekter procesov, ki so 
prisotni in spremljajo interakcijo fotona in ${}^3\mathrm{He}$ ter klju\v{c}no vplivajo na rezultate 
pri majhnih prenosih gibalne koli\v{c}ine. V modelih so uporabljeni sodobni dvonukleonski 
potenciali (npr. AV18~\cite{wiringa}) v kombinaciji s trinukleonsko silo (npr. UrbanaIX).

Precej\v{s}nje zanimanje za raziskovanje ${}^3\mathrm{He}$ se 
odra\v{z}a v vrsti preteklih eksperimentov posve\v{c}enih
\v{s}tudiju tega jedra. Te je mo\v{c} v grobem razdeliti v dve skupini. Prvo tvorijo precizijske
meritve pri nizkih $Q^2$, izvedene v Mainzu~\cite{achenbach2008}, ki omogo\v{c}ajo 
jasno lo\v{c}itev dvodel\v{c}nih od tridel\v{c}nih 
razpadov (glej sliko~\ref{fig_MAINZeep}). Druga skupina pa zdru\v{z}uje meritve sipalnih presekov na 
\v{c}im \v{s}ir\v{s}em obmo\v{c}ju manjkajo\v{c}ih  gibalnih koli\v{c}in pri visokih $Q^2$ 
(glej sliko~\ref{fig_FatihaLaget}), narejene v TJNAF~\cite{marat05, fatiha05}.

Ob tem ne smemo pozabiti omeniti prelomnih 
poskusov na tem prodro\v{c}ju.  Prvi je eksperiment MIT-Bates~\cite{tripp96}, kjer so 
izmerili razmerje longitudinalnih in transverzalnih odzivnih funkcij 
za proces ${}^3\mathrm{He}\left(\mathrm{e},\mathrm{e'} \mathrm{d }\right)$ 
(glej sliko~\ref{fig_Tripp}) in prvi\v{c} pokazali, da je za pravilen opis 
reakcije potrebno upo\v{s}tevati tako izoskalarne kot tudi izovektorske tokove. Pomembno je k razumevanju 
${}^3\mathrm{He}$ prispeval tudi eksperiment iz IUCF~\cite{milner1996}. 
Tu so prvikrat opravili meritev dvojnopolarizacijskih asimetrij v procesih 
${}^3\vec{\mathrm{He}}\left(\vec{\mathrm{p}},\mathrm{2p}\right)$  in \\
${}^3\vec{\mathrm{He}}\left(\vec{\mathrm{p}},\mathrm{pn}\right)$ ter pokazali, da je
stanje $S$ prevladujo\v{c}e stanje valovne funkcije ${}^3\mathrm{He}$. S tem so potrdili, da 
lahko ${}^3\mathrm{He}$  uporabimo kot efektivno nevtronsko tar\v{c}o. Poleg tega je eksperiment
pokazal tudi na slabosti uporabe hadronskih \v{z}arkov za raziskovanje lastnosti jeder. S tem
je dal zagon pristopom, kjer namesto protonov uporabljajo elektrone, ki jih znamo 
eksaktno teoreti\v{c}no opisati in lahko zato z njimi  bolj natan\v{c}no raziskujemo 
jedra. Prvo meritev dvojnopolarizacijskih asimetrij z elektroni na ${}^3\mathrm{He}$ so izvedli v 
NIKHEF~\cite{poolman}. Z meritvijo asimetrije $A_z$ v kvazielasti\v{c}nih reakcijah 
${}^3\vec{\mathrm{He}}\left(\vec{\mathrm{e}},\mathrm{e'} \mathrm{p}\right)$ in 
${}^3\vec{\mathrm{He}}\left(\vec{\mathrm{e}},\mathrm{e'} \mathrm{n}\right)$ so ponovno potrdili
prevlado stanja S v osnovnem stanju valovne funkcije ${}^3\mathrm{He}$. Kljub slabi statistki
pa so uspeli tudi pokazati, da za veren teoreti\v{c}ni opis procesov ne zado\v{s}\v{c}a pribli\v{z}ek PWIA, 
pa\v{c} pa je potrebno uporabiti zahtevnej\v{s}o teorijo, ki sloni na re\v{s}evanju ena\v{c}b Faddeeva.


\v{S}tudije s polariziranimi tar\v{c}ami in polariziranimi \v{z}arki elektronov so pomemben 
korak naprej v raziskovanju zgradbe jeder in nukleonov, saj nam preko meritev asimetrij 
omogo\v{c}ajo vpogled v lastnosti jeder (na primer stanji $D$ in $S'$), 
ki jih prej v nepola\-riziranih pristopih ni bilo mo\v{c} izmeriti. 
Na \v{z}alost so vsi dosedanji dvojnopolarizacijski poskusi imeli
preslabo statistiko, da bi lahko z njimi natan\v{c}no preverili teoreti\v{c}ne napovedi 
o zgradbi ${}^3\mathrm{He}$. 
Tako je eksperiment E05-102 prvi, pri katerem bomo lahko preverili prisotnost stanj 
$S'$ in $D$ v valovni funckiji osnovnega stanja ${}^3\mathrm{He}$. S tem bomo bodisi 
potrdili bodisi ovrgli teoreti\v{c}ne napovedi o spinsko-izospinski zgradbi jedra, 
lastnostih mezonskih izmenjalnih tokov ter interakcij v kon\v{c}nem stanju in s 
svojimi ugotovitvami postavili pomemben mejnik v raziskovanju strukture jedra.

Kak\v{s}na so pri\v{c}akovanja? Vpliv stanja $S'$ je po teoreti\v{c}nih izra\v{c}unih 
najve\v{c}ji pri majhnih manjkajo\v{c}ih gibalnih koli\v{c}inah  delcev
(glej sliko~\ref{fig_eed_eep}). Tam prevladuje 
asime\-trija $A_x$, ki po napovedih zna\v{s}a $\approx -4\,\%$. Ta se nato z ve\v{c}anjem 
gibalne koli\v{c}ine po\v{c}asi ve\v{c}a proti $6\,\%$ pri $200\,\mathrm{MeV/c}$. 
Napovedana funkcijska odvisnost asimetrije $A_z$ je bolj divja. Pri majhnih manjkajo\v{c}ih 
gibalnih koli\v{c}inah naj bi bila majhna, z njenim ve\v{c}anjem pa naj bi nara\v{s}\v{c}ala 
do maksimalne vrednosti $\approx 2\,\%$ pri $100\,\mathrm{MeV}$. Nato naj bi za\v{c}ela 
hitro padati in pri nekaj manj kot $\approx 200~\mathrm{MeV}$ spremenila svoj predznak. 
Na tak\v{s}no obna\v{s}anje asimetrije $A_z$ vpliva predvsem komponenta $D$ valovne 
funkcije, ki z ve\v{c}anjem manjkajo\v{c}e gibalne koli\v{c}ine postaja vse pomembnej\v{s}a. 
Pre\v{c}kanje ni\v{c}le asimetrije je namre\v{c} znak, da v jedru postane pomembna 
komponenta $D$ valovne funkcije (relativna vrtilna koli\v{c}ina med nukleonoma $l=2$). 
Podobno je bilo \v{z}e opa\v{z}eno v raziskavah s polarizirano ${}^2\mathrm{H}$ 
tar\v{c}o (glej sliko~\ref{fig_Passchier}) v NIKHEF~\cite{passcier2002}, na\v{s} 
poskus pa je  prvi, ki to posku\v{s}a videti pri ${}^3\vec{\mathrm{He}}$.

V eksperimentu E05-102 smo uporabili polarizirano 
${}^3\mathrm{He}$ tar\v{c}o v kombinaciji s pola\-riziranim 
zveznim \v{z}arkom elektronov z energijo $2.4\,\mathrm{GeV}$. Sipane elektrone smo detektirali 
z visokolo\v{c}ljivim spektrometrom HRS-L v koincidenci s protoni in devteroni, ki smo jih 
zaznali s spektrometrom BigBite (glej sliko~\ref{fig_HALLA}).

Uporabljena visokotla\v{c}na polarizirana ${}^3\mathrm{He}$ tar\v{c}a je kompleksen sistem, 
ki jo tvorijo: steklena tar\v{c}na celica, napolnjena z me\v{s}anico ${}^3\mathrm{He}$, 
du\v{s}ika in par alkalnih kovin, tri Helmholtzove tuljave ter opti\v{c}ni laserski sistem
za polarizacijo. Tar\v{c}o polariziramo preko postopka SEOP (ang. Spin Exchange Optical 
Pumping), kjer z lasersko svetlobo najprej polariziramo atome Rb, ki nato preko hiperfine 
sklopitve svojo polarizacijo prenesejo na jedra ${}^3\mathrm{He}$.
Sestav treh Helmholtzovih tuljav na\v{c}eloma omogo\v{c}a vrtenje tar\v{c}ne polarizacije v katero koli smer,
vendar smo med eksperimentom zaradi omejitev opti\v{c}nega sistema lahko tar\v{c}o polarizirali le vzdol\v{z} \v{z}arka 
elektronov in v smereh pravo\-kotno na smer \v{z}arka (v vodoravni ravnini).
Stopnjo polarizacije  tar\v{c}e smo spremljali periodi\v{c}no vsake \v{s}tiri ure z meritvijo jedrske magnetne 
resonance. Ob\v{c}a\-sno smo izvedli tudi meritev elektronske spinske resonance. 
Velikosti tar\v{c}ne polarizacije, izmerjene med poskusom, so prikazane na sliki~\ref{fig_TargetPolarizationPlot}. 
Dosegli smo $\approx 60\,\mathrm{\%}$ stopnjo polarizacije, kar je izjemen dose\v{z}ek.

Spektrometer HRS-L je eden izmed dveh skoraj identi\v{c}nih spektrometrov, ki tvorita osnovno esperimentalno 
opremo kolaboracije Hall A. Sestoji iz treh kvadrupolnih magnetov in dipola v konfiguraciji QQDQ. Ima zelo veliko 
kotno in pozicijsko lo\v{c}ljivost, vendar za ceno majhne kotne in momentne sprejemljivosti. To\v{c}ne 
karakteristike spektrometra so zbrane v tabeli~\ref{table_HRSL}. Detektorski paket je name\v{s}\v{c}en na vrhu 
spektrometra (glej sliko~\ref{fig_HRS_layout}) in je sestavljen iz dveh ve\v{c}\v{z}i\v{c}nih komor (VDC), 
dveh scintilacijskih ravnin, detektorja sevanja \v{C}erenkova ter elektromagnetnega kalorimetra 
(glej sliko~\ref{fig_HRS_DetectorPackage}). 

BigBite~\cite{lange-general} je nefokusirajo\v{c} spektrometer z veliko kotno ($96\,\mathrm{msr}$) in 
momentno ($200-900\,\mathrm{MeV/c}$) sprejemljivostjo. Sestoji iz klasi\v{c}nega dipolnega magneta, 
ki mu sledi detektorski paket (glej sliko~\ref{BBSpectrometer}). 
V eksperimentu E05-102 smo uporabili hadronski 
detektorski sestav, ki ga tvorita dve ve\v{c}\v{z}i\v{c}ni komori (ang. Multi Wire Drift Chamber), 
ki smo ju uporabili za dolo\v{c}anje trajektorije in gibalne koli\v{c}ine delca, ter dve 
scintilacijski ravnini za identifikacijo delcev. Scintilatorji so hkrati slu\v{z}ili tudi kot 
pro\v{z}ilni mehanizem za zajemanje podatkov.

Poleg detektiranih delcev moramo za uspe\v{s}no izvedbo meritve natan\v{c}no
poznati tudi lastnosti vpadnih elektronov. Te smo dolo\v{c}ili z in\v{s}trumenti, ki so 
name\v{s}\v{c}eni na \v{z}arkovno 
linijo (glej sliko~\ref{fig_HALLA}). Lego \v{z}arka na tar\v{c}i in njegovo intenziteto 
smo dolo\v{c}ili z detektorji lege (ang. Beam Position Monitors) in detektorji toka 
(ang. Beam Current Monitors).  Energijo vpadnih elektronov smo ugotavljali z meritvijo odklona 
\v{z}arka v magnetnem polju (ang. Arc Method). Stopnjo polarizacije vpadnih elekronov pa smo 
merili s Comptonovim in M\o{}llerjevim polarimetrom.

Eden od klju\v{c}nih delov eksperimenta je pro\v{z}ilni mehanizem, katerega naloga je prepoznavanje 
veljavnih zadetkov v spektrometrih in spro\v{z}enje procesa zajemanja podatkov iz detektorjev.
Med eksperimentom E05-102 je bila moja osrednja naloga izdelava pro\v{z}ilnega mehanizma za 
spektrometer BigBite in koinciden\v{c}nega pro\v{z}ilega mehanizma s spektrometrom HRS-L. 
Lastnostim pro\v{z}ilnega mehanizma zato v svojem doktorskem delu namenim precej
pozornosti. Najprej opi\v{s}em elektronsko zgradbo mehanizma in na\v{c}in njegovega delovanja. 
Nato predstavim rezultate testov, s katerimi sem preveril delovanje sistema, na koncu pa 
omenim \v{s}e pomanjkljivosti sistema, ki smo jih odkrili po koncu 
meritve. Opa\v{z}ene napake razumemo, in jih je s pravimi postopki analize mo\v{c} zaobiti, 
tako da ne vplivajo na izid meritev.

Po kon\v{c}anem poskusu je sledila analiza zajetih podatkov. Najprej 
je bilo potrebno dokon\v{c}ati umeritev eksperimentalne opreme, saj tega zaradi \v{c}asovnih 
omejitev ni bilo mo\v{c} izvesti pred in med raziskavo. Potencialne najdene napake in
neto\v{c}ne kalibracijske konstante sem popravil s sekundarnimi korekcijskimi faktorji 
v programju za ana\-lizo podatkov. 
Najprej sem se posvetil umeritvi sintilacijskih detektrojev spektrometra BigBite, 
za katera sem skrbel \v{z}e med meritvijo. Oba 
scintilacijska detektorja (dE in E) sta zgrajena iz 24 plo\v{s}\v{c}ic plasti\v{c}nega 
scintilatorja, ki so zlo\v{z}ene druga ob drugi, svetlobo v njih pa detektiramo z 
dvema fotopomno\v{z}evalkama. Pravilna uglasitev teh detektorjev je klju\v{c}nega pomena 
za identifikacijo delcev, ki smo jo dolo\v{c}ali z meritvijo energijskih izgub v 
scintilatorjih.

Temu je sledila opti\v{c}na kalibracija spektrometra BigBite.
E05-102 je drugi poskus, kjer smo BigBite uporabili z danim hadronskim detektorskim 
paketom, vendar opti\v{c}na kalibracija zanj \v{s}e ni bila narejena. Zato sem 
po zgledu postopka, ki ga uporabljamo za HRS-L, zasnoval in implementiral postopek za 
dolo\v{c}itev transformacijske matrike. Z njo lahko koordinate delca v 
detektorjih, ki jih izmerimo,  neposredno transformiramo v fizikalno zanimive tar\v{c}ne 
koordinate. Z uporabo singularnega razcepa 
(ang. Singular Value Decomposition) sem uspe\v{s}no generiral enotne tranformacijske 
matrike za vse tar\v{c}ne koordinate in prvikrat pokazal, da je tak\v{s}en na\v{c}in
opti\v{c}ne kalibracije mogo\v{c} tudi za nefokusirajo\v{c}e spektrometre~\cite{miha_NIM}.

Po opravljeni kalibraciji sem za\v{c}el analizirati produkcijske podatke.
Pri analizi sem uporabil orodje Podd~\cite{Podd}, ki temelji na 
programskem paketu Root. Celotno analizo sem izvedel v dveh korakih. V prvem koraku
sem surove elektronske signale prevedel v fizikalne koli\v{c}ine. Ob tem sem izvedel tudi  prvo 
filtriranje podatkov in izmed vseh dogodkov izbral le fizikalno zanimive.
Zaradi obse\v{z}nosti podatkov in precej\v{s}nje ra\v{c}unske zahtevnosti njihove obdelave
sem ta del analize izvedel na FARM ra\v{c}unalnikih v TJNAF. 

Rezultate primarne analize sem nato prenesel na doma\v{c}i ra\v{c}unalnik, kjer sem 
izve\-del drugi del analize. Tu sem podatke \v{s}e dodatno filtriral ter jih lo\v{c}il na
tiste, ki pripadajo reakciji ${}^3\vec{\mathrm{He}}\left(\vec{\mathrm{e}},\mathrm{e'} \mathrm{d }\right)$, in 
tiste, ki pripadajo razpadnemu kanalu ${}^3\vec{\mathrm{He}}\left(\vec{\mathrm{e}},\mathrm{e'} \mathrm{d }\right)$. 
Dogodke sem delil glede na to, katere delce (p ali d) detektiram v spektrometru BigBite.
Za identifikacijo protonov in devteronov sem izdelal 
algoritem, ki temelji na energijskih izgubah delcev v scintilacijskih detektorjih spektrometra BigBite in 
se opira na dejstvo, da pri izbrani gibalni koli\v{c}ini devteroni izgubijo  
ve\v{c} energije kot protoni. 

Preden sem dolo\v{c}il kon\v{c}ne  eksperimentalne asimetrije sem se moral tudi prepri\v{c}ati,
da meritve niso obremenjene z la\v{z}nimi asimetrijami, ki bi lahko spremenile fizikalne 
rezultate. V eksperimentu E05-102 so glavna nevarnost asimetrije v 
orientaciji spina vpadnega elektrona. Te nastanejo, ker zaradi 
tehni\v{c}nih omejitev med meritvijo nismo zbrali enake koli\v{c}ine naboja za obe 
su\v{c}nosti, ter zato, ker je od su\v{c}nosti vpadnih elektronov odvisen tudi mrtvi \v{c}as 
detektorjev. La\v{z}ne asimetrije zaznamo s \v{s}tevci, ki merijo
tok in \v{s}tejejo sunke v detektorjih brez mrtvega \v{c}asa in neodvisno od preostalega 
sistema za zajemanje podatkov (DAQ). Izvle\v{c}ki analize \v{s}tevcev so prikazani na 
slikah~\ref{fig_scalers_BCM_ratio} do~\ref{fig_scalers_T8Asymmetry}
in ka\v{z}ejo, da so izmerjene la\v{z}ne asimetrije vsaj red velikosti 
manj\v{s}e od pri\v{c}akovanih  fizikalnih asimetrij.

Fizikalne asimetrije za oba reakcijska kanala sem dolo\v{c}il v odvisnosti od gibalne 
koli\v{c}ine nedetektiranih delcev ($p_{\mathrm{Miss}}$) pri $Q^2=-0.25\,(\mathrm{GeV}/c)^2$ in 
$Q^2=-0.35\,(\mathrm{GeV}/c)^2$. Upo\v{s}teval sem podatke vseh treh eksperimentalnih 
postavitev (glej tabelo~\ref{table_analysis_ListOfData}). Kon\v{c}ni rezultati so 
prikazani na slikah~\ref{fig_analysis_ProtonAsymmetry14} do~\ref{fig_analysis_DeuteronAsymmetry82}. 
Ovrednotil sem tudi najpomembnej\v{s}e sistematske napake in jih zbral v 
tabeli~\ref{table_analysis_SystematicErrors}.
K napaki najve\v{c} prispevata nedolo\v{c}enosti polarizacije tar\v{c}e ($\approx 5\,\mathrm{\%}$)
in \v{z}arka elektronov ($\approx 2\,\mathrm{\%}$). Asimetrija v procesu
${}^3\vec{\mathrm{He}}\left(\vec{\mathrm{e}},\mathrm{e'} \mathrm{d }\right)$
pa je obremenjena \v{s}e z napako zaradi napa\v{c}ne 
idenifikacije delcev ($\approx 3\,\mathrm{\%}$).

Izmerjene asimetrije sem soo\v{c}il z napovedmi teoretske skupine iz 
Krakova~\cite{golak_private_2012}. Primerjavo sem najprej naredil za 
protonski kanal. Zaradi velike ra\v{c}unske zahtevnosti so nam teoretiki
zaenkrat napovedi pripravili le za enajst to\v{c}k, ki 
pokrivajo celotno kinematsko sprejemljivost pri $Q^2\approx-0.25\,(\mathrm{GeV}/c)^2$
(glej sliko~\ref{fig_analysis_kinpoints}). 
Izra\v{c}uni so napravljeni za vsak reakcijski kanal posebej.
To pomeni, da moramo v izmerjenih podatkih dvodel\v{c}ne razpade
${}^3\vec{\mathrm{He}}\left(\vec{\mathrm{e}},\mathrm{e'} \mathrm{p }\right)d$ pravilno lo\v{c}iti
od tridel\v{c}nih ${}^3\vec{\mathrm{He}}\left(\vec{\mathrm{e}},\mathrm{e'} \mathrm{p}\right)pn$,
\v{c}e \v{z}elimo narediti verodostojno primerjavo. Delitev obi\v{c}ajno naredimo z rezi 
na manjkajo\v{c}o energijo ($E_{\mathrm{Miss}}$), kjer dvodel\v{c}ni razpadi
tvorijo vrh pri $5.5\,\mathrm{MeV}$, medtem ko so tridel\v{c}ni razpadi zbrani 
pri $7.7\,\mathrm{MeV}$. Zaradi omejene lo\v{c}ljivosti spektrometrov je v na\v{s}em 
eksperimentu tak\v{s}na separacija precej ote\v{z}ena, saj opisanih vrhov ni 
mogo\v{c}e lo\v{c}iti (glej sliko~\ref{fig_analysis_MissingEnergy}). 
Za pravilno interpretacijo meritev je zato potrebno uporabiti Monte-Carlo simulacijo, 
ki nam pove, kolik\v{s}en dele\v{z} dogodkov v
porazdelitvi $E_{\mathrm{Miss}}$ pripada dvodel\v{c}nim in  kolik\v{s}en tridel\v{c}nim razpadom. 

Na \v{z}alost za na\v{s} eksperiment ustrezna simulacija \v{s}e ni na voljo, 
zato sem se moral zate\v{c}i k pribli\v{z}kom. Iz analize asimetrij v odvisnosti od 
$E_{\mathrm{Miss}}$ sem fenomenolo\v{s}ko dolo\v{c}il mejo 
(glej sliko~\ref{fig_analysis_MissingEnergyAsymmetry}), do 
koder prevladujejo dvodel\v{c}ni razpadi. Nato sem ana\-lizo podatkov 
omejil le na izbrano obmo\v{c}je $E_{\mathrm{Miss}}\leq2\,\mathrm{MeV}$. Dobljene 
asimetrije naj bi ustrezale tistim za dvodel\v{c}ni razpad.

Teoreti\v{c}ni izra\v{c}uni so podani v odvisnosti od polarnega kota 
$\theta_{p}$ med smerjo detektiranega protona $p_{p}$ in smerjo vektorja prenosa 
gibalne koli\v{c}ine $\vec{q}$ (glej sliko~\ref{fig_theory_Kinematics}). 
To je zahtevalo dodaten trud, saj sem moral poiskati pravo zvezo med 
$p_{\mathrm{Miss}}$ in kotom $\theta_{p}$, da sem  lahko teorijo  pravilno 
primerjal z meritvami. Teoreti\v{c}ne izra\v{c}une sem povpre\v{c}il
tudi po kotu $\phi_{p}$ (glej sliko~\ref{fig_analysis_PhiPQAveraging}),
kar je bilo potrebno izvesti pazljivo, saj se porazdelitev $\phi_{p}$ mo\v{c}no 
spreminja s $p_{\mathrm{Miss}}$.

Primerjava dobljenih dvodel\v{c}nih asimetrij s teorijo poka\v{z}e (glej 
slike~\ref{fig_analysis_2BBUResult1}, ~\ref{fig_analysis_2BBUResult2} 
in~\ref{fig_analysis_2BBUResult3}), da se izra\v{c}uni ujemajo z meritvami
v predznaku in v splo\v{s}nem poteku krivulj, vendar, z izjemo to\v{c}k pri 
najni\v{z}jih $p_{\mathrm{Miss}}$, ne opi\v{s}ejo pravilno velikosti izmerjenih 
asime\-trij. Nekonsistence opazimo v vseh obravnavanih kinematskih to\v{c}kah  in 
pri obeh $Q^2$.

Opa\v{z}eno neujemanje pri velikih $p_{\mathrm{Miss}}$ lahko pripi\v{s}emo
prisotnosti majhnega dele\v{z}a tridel\v{c}nih razpadov, ki ga z izbranimimi 
rezi v $E_{\mathrm{Miss}}$ nismo uspeli odstraniti. 
Kolik\v{s}na je ta kontaminacija, sem poskusil oceniti z uporabo 
simulacije MCEEP~\cite{mceep}, ki gene\-rira porazdelitve $E_{\mathrm{Miss}}$
za nepolarizirane dvodel\v{c}ne in tridel\v{c}ne razpade. Na \v{z}alost MCEEP 
ni povsem prilagojen  na\v{s}i eksperimentalni konfiguraciji, zato lahko z njim 
tvorimo le ocene. Rezultati simulacije napovedo nekaj odstotno
($5$-$10\,\mathrm{\%}$) kontaminacijo dvodel\v{c}nih razpadov s tridel\v{c}nimi na 
obmo\v{c}ju $E_{\mathrm{Miss}}$, kjer smo pri\v{c}akovali le dvodel\v{c}ne razpade.  
Ker so teoreti\v{c}ne asimetrije za tridel\v{c}ni 
razpad (glej sliki~\ref{fig_analysis_TheoryLongKin4ppn} 
in~\ref{fig_analysis_TheoryTransKin4ppn}) bistveno ve\v{c}je od dvodel\v{c}nih
($\approx 10\,\mathrm{\%}$) in imajo nasproten predznak, lahko \v{z}e tak\v{s}ne majhne 
kontaminacije prispevajo dovolj velike popravke in 
zmanj\v{s}ajo vrzel med meritvami in teorijo. Slika~\ref{fig_analysis_3BBUResult1} 
ka\v{z}e, kako se spremeni izra\v{c}unana asimetrija za dvodel\v{c}ni razpad, \v{c}e ji 
dodamo $5\,\mathrm{\%}$ in $10\,\mathrm{\%}$ primes tridel\v{c}nih razpadov. 
Ob tem se moramo seveda zavedati, da lahko tak\v{s}ne konstantne popravke uporabimo zgolj 
za demonstracijo velikosti spremembe. V pravi analizi je namre\v{c} potrebno 
narediti lo\v{c}ene popravke za vsako to\v{c}ko posebej, saj je velikost popravka
mo\v{c}no odvisna od $p_{\mathrm{Miss}}$. Tak\v{s}ne popravke pa bomo lahko naredili
\v{s}ele tedaj, ko bo na razpolago natan\v{c}nej\v{s}a simulacija Monte Carlo.

Dolo\v{c}itev asimetrij za tridel\v{c}ni razpad je bistveno bolj zapletena, 
saj tu ne morem uporabiti enostavnih rezov v spektrih $E_{\mathrm{Miss}}$,
kot sem jih pri analizi dvodel\v{c}nih razpadov. Prevlado tridel\v{c}nih razpadov
bi pri\v{c}akovali v obmo\v{c}ju $E_\mathrm{Miss} >10\,\mathrm{MeV}$, vendar se izka\v{z}e, 
da zaradi kon\v{c}ne lo\v{c}ljivosti spektrometrov in sevalnih popravkov
v tem obmo\v{c}ju k celotni asimetriji znatno prispevajo  tudi
dvodel\v{c}ni razpadi. Razmerje obeh procesov v odvisnosti od $E_{\mathrm{Miss}}$
je prikazano na sliki~\ref{fig_analysis_Mceep_Ratios}. 
Dolo\v{c}itev asimetrij za proces 
${}^3\vec{\mathrm{He}}\left(\vec{\mathrm{e}},\mathrm{e'} \mathrm{p}\right)pn$
je tako mo\v{z}na le z uporabo ustrezno Monte Carlo simulacije, ki oceni
popravke k tridel\v{c}ni asimetriji, zaradi prisotnosti 
dvodel\v{c}nih razpadov. 

Prvi poskus dolo\v{c}itve tridel\v{c}nih asimetrij je prikazan na 
sliki~\ref{fig_analysis_3BBU_Extraction}. Tu sem se osredoto\v{c}il 
le na asimetrije pri 
$p_{\mathrm{Miss}}\leq 90\,\mathrm{MeV}/c$ in jih predstavil kot funkcijo 
$E_{\mathrm{Miss}}$. Razmerje med \v{s}tevilom dvodel\v{c}nih in tridel\v{c}nih razpadov
pri vsaki  $E_{\mathrm{Miss}}$ sem ocenil s simulacijo MCEEP. Rezultati 
jasno ka\v{z}ejo, da je izlu\v{s}\v{c}ena asimetrija za tridel\v{c}ni razpad bistveno ve\v{c}ja 
od prvotne izmerjene asimetrije,
kar dokazuje nujnost uporabe simulacije. Dobljene asimetrije sem primerjal tudi 
z rezultati eksperimenta, ki so ga pri skoraj enakih kinematskih pogojih 
izvedli v Mainzu~\cite{achenbach2008}. V okviru danih negotovosti sem opazil
dobro ujemanje obeh meritev.

S teoreti\v{c}nimi izra\v{c}uni sem soo\v{c}il tudi izmerjene asimetrije za proces
${}^3\vec{\mathrm{He}}\left(\vec{\mathrm{e}},\mathrm{e'} \mathrm{d }\right)p$.
Ravnal sem enako kot pri protonskem kanalu. Za vse izbrane to\v{c}ke v $p_{\mathrm{Miss}}$
sem poiskal pripadajo\v{c}e polarne kote $\theta_{p}$, ki so jih potem teoretiki uporabili
v svojih programih in izra\v{c}unali asimetrije 
(glej sliki~\ref{fig_analysis_Deuteron_TheoryLongKin4ppn} 
in~\ref{fig_analysis_Deuteron_TheoryTransKin4ppn}). Njihove izra\v{c}une sem nato \v{s}e povpre\v{c}il 
po kotu $\phi_d$ in jih na koncu primerjal z izmerjenimi vrednostmi. Rezultati primerjave
so prikazani na slikah~\ref{fig_analysis_Deuteron_Comparison1} 
in~\ref{fig_analysis_Deuteron_Comparison2}.
Tu zopet opazim slabo ujemanje meritev s povpre\v{c}enimi teoreti\v{c}nimi vrednostmi.
Celo pri najni\v{z}jih $p_{\mathrm{Miss}}$, kjer bi pri\v{c}akoval najbolj\v{s}e ujemanje med 
teorijo in meritvami, opazim znatne razlike, saj je izra\v{c}unana transverzalna asimetrija tam 
skoraj dvakrat manj\v{s}a od izmerjene.  Neujemanje opazim tako pri $Q^2=-0.25\,(\mathrm{GeV}/c)^2$
kot tudi pri $Q^2=-0.35\,(\mathrm{GeV}/c)^2$.

Neujemanje izmerjenih asimetrij s teoreti\v{c}nimi v tem prvem poskusu primerjave 
ni tako nepri\v{c}akovano. V eksperimentu E05-102 smo prvikrat izmerili 
dvojnopolarizacijske asimetrije v kvazielasti\v{c}nih procesih 
${}^3\vec{\mathrm{He}}\left(\vec{\mathrm{e}},\mathrm{e'} \mathrm{d }\right)$ in 
${}^3\vec{\mathrm{He}}\left(\vec{\mathrm{e}},\mathrm{e'} \mathrm{p}\right)$ v odvisnosti od
$p_{\mathrm{Miss}}$ pri fiksnih vrednostih $Q^2$. Do sedaj teoretiki \v{s}e niso imeli 
na voljo tako ob\v{s}irnih in celovitih podatkov, s katerimi bi lahko natan\v{c}no umerili svoje
napovedi. Razpolo\v{z}ljive teorije, ki temeljijo na re\v{s}evanju ena\v{c}b Faddeeva, so sicer do sedaj 
uspe\v{s}no opisale ve\v{c}ino razpolo\v{z}ljivih nepolariziranih podatkov, vendar ostajajo meritve,
ki se ne skla\-dajo z njimi. Dvojnopolarizacijske opazljivke pa predstavljajo \v{s}e stro\v{z}ji test
teorij kot nepolarizirani sipalni preseki, zato tu utemeljeno pri\v{c}akujemo se ve\v{c}jo razhajanje
 med meritvami in izra\v{c}uni. 

Za nastale razlike so seveda lahko krive tudi napake pri interpretaciji izmerjenih 
asimetrij. Menim, da utegne biti ujemanje meritev s teorijo bolj\v{s}e potem, ko bomo
izvedli natan\v{c}nej\v{s}e povpre\v{c}enje teoreti\v{c}nih asimetrij po sprejemljivosti. 
Trenutna mre\v{z}a teoretskih to\v{c}k je namre\v{c} pregroba, 
da bi omogo\v{c}ala tak\v{s}no podrobno analizo. Zatadi te omejitve sem se tudi odlo\v{c}il,
da primerjavo teoretskih asimetrij z meritvami naredim za vsako kinematsko 
to\v{c}ko lo\v{c}eno, in ne ra\v{c}unam povpre\v{c}ij, dokler tak\v{s}na analiza ni na voljo.

Pri analizi asimetrij v reaciji 
${}^3\vec{\mathrm{He}}\left(\vec{\mathrm{e}},\mathrm{e'} \mathrm{p}\right)$
dodatno oviro predstavlja tudi nenatan\v{c}na lo\v{c}itev dvodel\v{c}nega razpadnega kanala od 
tridel\v{c}nega.  Natan\v{c}nejsa delitev, ki bo na voljo z implementacijo ustrezne
simulacije Monte Carlo, bo gotovo pripomogla k to\v{c}nej\v{s}i intepretaciji 
rezultatov. 



\bibliographystyle{unsrt}

\newpage
\thispagestyle{empty}
\vspace*{10cm}
\begin{flushleft}
Spodaj podpisani Miha Mihovilovi\v{c} izjavljam, da sem avtor te doktorske disertacije.
\end{flushleft}

\end{document}